\documentclass[twoside]{cupbook}

\usepackage{graphicx}
\usepackage{amsmath}
\usepackage{amsfonts}
\usepackage{amssymb}
\usepackage{needspace}
\usepackage[T1]{fontenc}
\usepackage[french,german]{babel}

\renewcommand{\theequation}{\arabic{equation}}

\makeatletter
\def\flushboth{%
  \let\\\@normalcr
  \@rightskip\z@skip \rightskip\@rightskip
  \leftskip\z@skip
  \parindent 1em\relax}
\makeatother

\def\Xint#1{\mathchoice
   {\XXint\displaystyle\textstyle{#1}}%
   {\XXint\textstyle\scriptstyle{#1}}%
   {\XXint\scriptstyle\scriptscriptstyle{#1}}%
   {\XXint\scriptscriptstyle\scriptscriptstyle{#1}}%
   \!\int}
\def\XXint#1#2#3{{\setbox0=\hbox{$#1{#2#3}{\int}$}
     \vcenter{\hbox{$#2#3$}}\kern-.5\wd0}}

\def\dashint{\Xint-}

\makeatletter
\def\endnote{%
 \@ifnextchar[{\@xendnote
  }{\stepcounter{endnote}%
    \xdef\@theenmark{\theendnote}%
   }%
 \@endnotemark
 \@endnotetext
}
\makeatother

\makeatletter
\def\@makefnmark{\hbox{$^{\scriptsize \rm \@thefnmark}$}}
\makeatother

\makeatletter    

\newcommand{\unnumberedcaption}%
	{\@dblarg{\@unnumberedcaption\@captype}}

\newcommand{\@unnumberedcaption}{}
\long\def\@unnumberedcaption#1[#2]#3{\par
  \addcontentsline{\csname ext@#1\endcsname}{#1}{%
    \protect\numberline{}{\ignorespaces #2}%
    }%
  \begingroup
    \@parboxrestore
    \normalsize
    \@makeunnumberedcaption{\ignorespaces #3}\par
  \endgroup}

\newcommand{\@makeunnumberedcaption}[1]{%
  \vskip\abovecaptionskip
  \sbox\@tempboxa{#1}%
  \ifdim \wd\@tempboxa >\hsize
    #1\par
  \else
    \global \@minipagefalse
    \hbox to\hsize{\hfil\box\@tempboxa\hfil}%
  \fi
  \vskip\belowcaptionskip}

\@ifundefined{abovecaptionskip}{%
  \newlength{\abovecaptionskip}%
  \setlength{\abovecaptionskip}{10pt}%
}{}
\@ifundefined{belowcaptionskip}{%
  \newlength{\belowcaptionskip}%
  \setlength{\belowcaptionskip}{0pt}%
}{}

\makeatother

\makeatletter
\def\@schapter#1{%
 \ifSFB@multisty
  \markboth{#1}{#1}
  \gdef\SFB@endnotehead{\relax}
 \else
  \chaptermark{#1}
  \gdef\SFB@endnotehead{#1}
 \fi
 \if@twocolumn
  \@topnewpage[\@makeschapterhead{#1}]
 \else
  \@makeschapterhead{#1}
  \@afterheading
 \fi
 \normalsize
}
\makeatother

\makeatletter
\newlength{\beforeepigraphskip}
\newlength{\afterepigraphskip}
\newlength{\epigraphwidth}
\newlength{\epigraphrule}
\newcommand{\epigraphsize}{\small}
\newcommand{\epigraphflush}{flushright}
\newcommand{\textflush}{flushleft}
\newcommand{\sourceflush}{flushright}

\AtBeginDocument{%
  \@ifundefined{flushleftright}{%
    }%
    {\PackageWarningNoLine{epigraph}%
     {flushleftright has been previously defined.\MessageBreak
      Use flushepinormal for epigraphs instead}}}

\newcommand{\@epirule}{\rule[.5ex]{\epigraphwidth}{\epigraphrule}}
\newcommand{\@epitext}[1]{%
  \begin{minipage}{\epigraphwidth}\begin{\textflush} #1\\
    \ifdim\epigraphrule>\z@ \@epirule \else \vspace*{1ex} \fi
  \end{\textflush}\end{minipage}}
\newcommand{\@episource}[1]{%
  \begin{minipage}{\epigraphwidth}\begin{\sourceflush} #1\end{\sourceflush}
  \end{minipage}}
\newcommand{\epigraph}[2]{\vspace{\beforeepigraphskip}
  {\epigraphsize\begin{\epigraphflush}\begin{minipage}{\epigraphwidth}
    \@epitext{#1}\\ \@episource{#2}
    \end{minipage}\end{\epigraphflush}
    \vspace{\afterepigraphskip}}}
\newcommand{\qitem}[2]{{\raggedright\item \begin{minipage}{\epigraphwidth}
  \@epitext{#1}\\ \@episource{#2}
  \end{minipage}}}
\newcommand{\qitemlabel}[1]{\hfill}
\newenvironment{epigraphs}{%
  \vspace{\beforeepigraphskip}\begin{\epigraphflush}
  \epigraphsize
  \begin{minipage}{\epigraphwidth}
   \list{}%
    {\itemindent\z@ \labelwidth\z@ \labelsep\z@
     \leftmargin\z@ \rightmargin\z@
     }}%
  {\endlist\end{minipage}\end{\epigraphflush}
   \vspace{\afterepigraphskip}}
\providecommand{\cleartoevenpage}[1][\@empty]{%
  \clearpage%
  \ifodd\c@page\hbox{}#1\clearpage\fi}
\newcommand{\dropchapter}[1]{%
  \let\@epichapapp\@chapapp
  \renewcommand{\@chapapp}{\vspace*{#1}\@epichapapp}}
\newcommand{\undodrop}{\let\@chapapp\@epichapapp}
\newif\if@epirhs     \@epirhstrue
\newif\if@epicenter  \@epicentertrue
\newcommand{\@epipos}{
  \long\def\@ept{flushleft}
  \ifx\epigraphflush\@ept
    \@epirhsfalse \@epicenterfalse
  \else
    \long\def\@ept{center}
    \ifx\epigraphflush\@ept
      \@epirhsfalse \@epicentertrue
    \else
      \@epirhstrue  \@epicenterfalse
    \fi
  \fi}
\newcommand{\epigraphhead}[2][95]{%
  \def\@epitemp{\begin{minipage}{\epigraphwidth}#2\end{minipage}}
  \def\ps@epigraph{\let\@mkboth\@gobbletwo
    \@epipos
    \if@epirhs
      \def\@oddhead{\hfil\begin{picture}(0,0)
                         \put(0,-#1){\makebox(0,0)[r]{\@epitemp}}
                         \end{picture}}
    \else
      \if@epicenter
        \def\@oddhead{\hfil\begin{picture}(0,0)
                           \put(0,-#1){\makebox(0,0)[b]{\@epitemp}}
                           \end{picture}\hfil}
      \else
        \def\@oddhead{\begin{picture}(0,0)
                           \put(0,-#1){\makebox(0,0)[l]{\@epitemp}}
                           \end{picture}\hfil}
      \fi
    \fi
    \let\@evenhead\@oddhead
    \def\@oddfoot{\reset@font\hfil\thepage\hfil}
    \let\@evenfoot\@oddfoot}
  \thispagestyle{epigraph}}

\newcommand*\MakePerPage[2][\@ne]{%
  \expandafter\def\csname c@pchk@#2\endcsname{\c@pchk@{#2}{#1}}%
  \newcounter{pcabs@#2}%
  \@addtoreset{pchk@#2}{#2}}

\def\new@pagectr#1{\@newl@bel{pchk@#1}}

\def\c@pchk@#1#2{\z@=\z@
  \begingroup
  \expandafter\let\expandafter\next\csname pchk@#1@\arabic{pcabs@#1}\endcsname
  \addtocounter{pcabs@#1}\@ne
  \expandafter\ifx\csname pchk@#1@\arabic{pcabs@#1}\endcsname\next
  \else \setcounter{#1}{#2}\fi
  \protected@edef\next{%
    \string\new@pagectr{#1}{\arabic{pcabs@#1}}{\noexpand\thepage}}%
  \protected@write\@auxout{}{\next}%
  \endgroup\global\z@}
 
\makeatother

\begin{document}
\bibliographystyle{unsrt}
\title[Reconsidering the 1927 Solvay Conference]{Quantum Theory at the Crossroads}
\author{Guido Bacciagaluppi \and Antony Valentini}
\date{2006}
\maketitle

\MakePerPage{footnote}

\frontmatter

\thispagestyle{empty}
\hspace*{\fill} \\
\hspace*{\fill} \\
\hspace*{\fill} \\
\hspace*{\fill} \\
\hspace*{\fill} \\
\hspace*{\fill} \\
\hspace*{\fill}{\em To the memory of James~T.~Cushing}
\newpage
\thispagestyle{empty}

\tableofcontents

\chapter*{Preface}\markboth{{\it Preface}}{{\it Preface}}
\addcontentsline{toc}{schapter}{Preface}
\epigraph{And they said one to another: Go to, let us build us a tower, whose
top may reach unto heaven; and let us make us a name. And the Lord said: Go
to, let us go down, and there confound their language, that they may not
understand one another's speech.}{Genesis 11: 3--7}

\noindent Anyone who has taken part in a debate on the interpretation of quantum theory
will recognise how fitting is the above quotation from the book of Genesis,
according to which the builders of the Tower of Babel found that they could no
longer understand one another's speech. For when it comes to the
interpretation of quantum theory, even the most clear-thinking and capable
physicists are often unable to understand each other.

This state of affairs dates back to the genesis of quantum theory itself. In
October 1927, during the `general discussion' that took place in Brussels at
the end of the fifth Solvay conference, Paul Ehrenfest wrote the above
lines on the blackboard. As Langevin later remarked, 
the Solvay meeting in 1927 was the conference where `the confusion of 
ideas reached its peak'. 

Ehrenfest's perceptive gesture captured the essence of a situation that has
persisted for three-quarters of a century. According to widespread historical
folklore, the deep differences of opinion among the leading physicists of the
day led to intense debates, which were satisfactorily resolved by Bohr and
Heisenberg around the time of the 1927 Solvay meeting. But in fact, at the end
of 1927, a significant number of the main participants (in particular de
Broglie, Einstein, and Schr\"{o}dinger) remained unconvinced, and the deep
differences of opinion were never really resolved. 

The interpretation of quantum theory seems as highly controversial today as it
was in 1927. There has also been criticism --- on the part
of historians as well as physicists --- of the tactics used by Bohr and others
to propagate their views in the late 1920s, and a realisation that alternative
ideas may have been dismissed or unfairly disparaged. 
For many physicists, a sense of unease lingers over the whole
subject. Might it be that things are not as clear-cut as Bohr and Heisenberg
would have us believe? Might it be that their opponents had something
important to say after all?
Because today there is no longer an established interpretation of quantum mechanics,
we feel it is important to go back to the sources and re-evaluate them.

In this spirit, we offer the reader a return to a time just before the
Copen\-hagen interpretation was widely accepted, when the best physicists of the
day gathered to discuss a range of views, concerning many topics of
interest today (measurement,
determinism, nonlocality, subject\-ivity, interference, and so on), and when three distinct theories ---
de Broglie's pilot-wave theory, Born and Heisenberg's quantum mechanics, and
Schr\"{o}dinger's wave mechanics --- were presented and discussed on an equal footing.

\begin{center}
*
\end{center}

Since the 1930s, and especially since the Second World War, it has been common to dismiss questions about the 
interpretation of quantum theory as `metaphysical' or `just philosophical'. It will be clear from the lively 
and wide-ranging discussions of 1927 that at that time, for the most distinguished physicists of the day, the 
issues were decidedly \textit{physical}: Is the electron a point particle with a
continuous trajectory (de Broglie), or a wave packet (Schr\"{o}dinger), or neither (Born and Heisenberg)? 
Do quantum outcomes occur when nature makes a choice (Dirac), or when an observer
decides to record them (Heisenberg)? Is the nonlocality of quantum theory compatible 
with relativity (Einstein)? Can a theory with trajectories account for the recoil of a 
single photon on a mirror (Kramers, de Broglie)? Is indeterminism a fundamental limitation, 
or merely the outcome of coarse-graining over something deeper and deterministic (Lorentz)?

After 1927, the Copenhagen interpretation became firmly established. Rival views were marginalised, 
in particular those represented by de Broglie, Schr\"{o}dinger and Einstein, even though these scientists 
were responsible for many of the major developments in quantum physics itself. (This
marginalisation is apparent in most historical accounts written throughout the twentieth century.) 
From the very beginning, however, there were some notes of caution: for example, when Bohr's landmark 
paper of 1928 (the English version of his famous Como lecture) was published in \textit{Nature}, 
an editorial preface expressed dissatisfaction with the `somewhat vague 
statistical description' and ended with the hope that this would not be the `last word on the subject'. 
And there were a few outstanding alarm bells, in particular the famous paper by Einstein, Podolsky and Rosen in
1935, and the important papers by Schr\"{o}dinger (in the same year) on the cat paradox and on entanglement. 
But on the whole, the questioning ceased in all but a few corners. A general opinion arose that the questions 
had been essentially settled, and that a satisfactory point of view had been arrived
at, principally through the work of Bohr and Heisenberg. For subsequent
generations of physicists, `shut up and calculate' emerged as the working rule among the vast majority.

Despite this atmosphere, the questioning never completely died out, and some
very significant work was published, for example by Bohm in 1952, Everett in
1957, and Bell in 1964 and 1966. But attitudes changed very slowly. Younger
physicists were strongly discouraged from pursuing such questions. Those who
persisted generally had difficult careers, and much of the careful thinking
about quantum foundations was relegated to departments of philosophy.

Nevertheless, the closing decade of the twentieth century saw a resurgence of
interest in the foundations of quantum theory. At the time of writing, a range
of alternatives (such as hidden variables, many worlds, collapse models, among
others) are being actively pursued, and the Copenhagen interpretation can no
longer claim to be the dominant or `orthodox' interpretation.

The modern reader familiar with current debates and positions in quantum
foundations will recognise many of the standard points of view in the
discussions reproduced here, though expressed with a remarkable concision and
clarity. This provides a welcome contrast with the generally poor level of
debate today: as the distinguished cosmologist Dennis Sciama was fond of
pointing out, when it comes to the interpretation of quantum theory `the
standard of argument suddenly drops to zero'. We hope that the publication of
this book will contribute to a revival of sharp and informed debate about the
meaning of quantum theory.

\begin{center}
*
\end{center}

Remarkably, the proceedings of the fifth Solvay conference have not received
the attention they deserve, neither from physicists nor from histor\-ians, and
the literature contains numerous major misunderstand\-ings about what took place there.

The fifth Solvay conference is usually remembered for the clash that took
place between Einstein and Bohr over the uncertainty relations. It is
remarkable, then, to find that not a word of these discussions appears in the
published proceedings. It is known that Einstein and Bohr engaged in vigorous
informal discussions, but in the formal debates recorded in the proceedings
they were relatively silent. Bohr did contribute to the general discussion,
but this material was not published. Instead, at Bohr's request, it was
replaced by a translation of the German version of his Como lecture,
which appeared in \textit{Naturwissenschaften} in 1928. (We do not reproduce
this well-known paper here.) The appending of this translation to the
published proceedings  may be the cause of the common misunderstanding that
Bohr gave a report at the conference: in fact, he did not.

Born and Heisenberg present a number of unfamiliar viewpoints concerning, among
other things, the nature of the wave function and the role of time and of probability in quantum theory.
Particularly surprising is the seeming absence of a collapse postulate in their formulation, and the
apparently phenomenological status of the time-dependent Schr\"{o}dinger equation. Born and Heisenberg's 
`quantum mechanics' seems remarkably different from quantum mechanics (in the Dirac-von~Neumann
formulation) as we know it today. 

De Broglie's pilot-wave theory was the subject of extensive and varied
discussions. This is rather startling in view of the claim --- in Max Jammer's
classic historical study \textit{The Philosophy of Quantum Mechanics} --- that
de Broglie's theory `was hardly discussed at all' and that `the only serious
reaction came from Pauli' (Jammer 1974, pp.~110--11). Jammer's view is
typical even today. But in the published proceedings, at the end of de
Broglie's report there are 9 pages of discussion devoted to de Broglie's
theory, and of the 42 pages of general discussion, 15 contain
discussion of de Broglie's theory, with serious reactions and comments coming
not only from Pauli but also from Born, Brillouin, Einstein, Kramers, Lorentz,
Schr\"{o}dinger and others. Even the well-known exchange between
Pauli and de Broglie has been widely misunderstood.

Finally, another surprise is that in his report de Broglie proposed the many-body
pilot-wave dynamics for a system of particles, with the total configuration
guided by a wave in configuration space, and not just (as is generally
believed) the one-body theory in 3-space. De Broglie's theory is essentially
the same as that developed by Bohm in 1952, the only difference being that de
Broglie's dynamics (like the form of pilot-wave theory popularised by Bell)
is formulated in terms of velocity rather than acceleration.

\begin{center}
*
\end{center}

This work is a translation of and commentary on the proceedings of the fifth
Solvay conference of 1927, which were published in French in 1928 under the
title \textit{\'{E}lectrons et Photons}.

We have not attempted to give an exhaustive historical analysis of the fifth
Solvay conference. Rather, our main aims have been to present the material in
a manner accessible to the general physicist, and to situate the proceedings
in the context of current research in quantum foundations. We hope that the
book will contribute to stimulating and reviving serious debate about quantum
foundations in the wider physics community, and that making the proceedings
available in English will encourage historians and philosophers to reconsider 
their significance.

Part I begins with a historical introduction and provides essays on
the three main theories presented at the conference (pilot-wave theory,
quantum mechanics, wave mechanics). The lectures and discussions that took place 
at the fifth Solvay
conference contain an extensive range of material that is relevant to current
research in the foundations of quantum theory. In Part II, after a brief
review of the status of quantum foundations today, we summarise what seem to
us to be the highlights of the conference, from the point of view of current
debates about the meaning of quantum theory.
Part III of the book consists of translations of the reports, of the
discussions following them, and of the general discussion. Wherever possible,
the original (in particular English or German) texts have been used. We have
tacitly corrected minor mistakes in punctuation and spelling, and we have
uniformised the style of equations, references and footnotes. (Unless otherwise
specified, all translations of quotations are ours.)

Part I (except for chapter \ref{deBroglieEss}) 
and the reports by Compton, by Born and Heisenberg and by Schr\"{o}dinger, 
are principally the work of Guido Bacciagaluppi. Chapter \ref{deBroglieEss}, 
Part II, 
and the reports by Bragg and by de Broglie and the general 
discussion in Part III, are principally the work of Antony Valentini. 

Chapters~\ref{deBroglieEss}, \ref{meas-in-pwt} and \ref{Pilot-wave-in-retrospect} are 
based on a seminar, `The early history of Louis de Broglie's pilot-wave dynamics', given by Antony 
Valentini at the University of Notre Dame in September 1997, at a conference in 
honour of the sixtieth birthday of the late James T. Cushing.

\begin{center}
* 
\end{center}

To James T. Cushing, physicist, philosopher, historian and gentleman, we both owe 
a special and heartfelt thanks. It was he who brought us together on this project, 
and to him we are indebted for his encouragement and, above all, his example. This 
book is dedicated to his memory.


Guido Bacciagaluppi wishes to express his thanks to the Humboldt Foundation, which 
supported the bulk of his work in the form of an Alexander von Humboldt Forschungsstipendium, 
and to his hosts in Germany, Carsten Held and the Philosophisches Seminar I, University of 
Freiburg, and Harald Atmanspacher and the Institut f\"{u}r Grenzgebiete der Psychologie und 
Psychohygiene, Freiburg, as well as to Jacques Dubucs and the Institut d'Histoire et de Philosophie 
des Sciences et des Techniques (CNRS, Paris 1, ENS) for support during the final phase. He also wishes to 
thank Didier Devriese of the Universit\'{e} Libre de Bruxelles, who is in charge of the archives of the 
Instituts Internationaux de Physique et de Chimie Solvay, Universit\'{e} Libre de Bruxelles, for his kindness 
and availability, and Brigitte Parakenings (formerly Uhlemann) and her staff at the Philosophisches
Archiv of the University of Konstanz, for the continuous assistance with the Archive for the History of Quantum 
Physics. Finally, he should wish to thank Jeff Barrett for suggesting this project to him in Utrecht one day back 
in 1996, as well as Mark van Atten, Jennifer Bailey, Olivier Darrigol, Felicity Pors, Gregor Schiemann 
and many others for discussions, suggestions, correspondence, references and other help.

Antony Valentini began studying these fascinating proceedings while holding a
postdoctoral position at the University of Rome `La Sapienza' (1994--96), and
is grateful to Marcello Cini, Bruno Bertotti and Dennis Sciama for their
support and encouragement during that period. For support in recent
years, he is grateful to Perimeter Institute, and wishes to express a special 
thanks to Howard Burton, Lucien Hardy and Lee Smolin.

We are both grateful to Tamsin van Essen at Cambridge University Press for her support and encouragement
during most of the gestation of this book, and to Augustus College for support during the final
stages of this work.

\

Guido Bacciagaluppi

Antony Valentini\hfill Lake Maggiore, August 2006


\section*{Abbreviations}\markboth{{\it Abbreviations}}{{\it Abbreviations}}
\addcontentsline{toc}{section}{Abbreviations}

\noindent AEA: Albert Einstein Archives, Jewish National and University Library, Hebrew University of Jerusalem.

\noindent AHQP: Archive for the History of Quantum Physics.

\noindent AHQP-BSC: Bohr Scientific Correspondence, microfilmed from the
Niels Bohr Arkiv, Copenhagen.

\noindent AHQP-BMSS: Bohr Scientific Manuscripts, microfilmed from the Niels
Bohr Arkiv, Copenhagen.

\noindent AHQP-EHR: Ehrenfest collection, microfilmed from the
Rijksmuseum voor de Geschiedenis van de Natuurwetenschappen en van de
Geneeskunde `Museum Boerhaave', Leiden.

\noindent AHQP-LTZ: Lorentz collection, microfilmed from the Algemeen
Rijksarchief, Den Haag.

\noindent AHQP-RDN: Richardson Collection, microfilmed from the Harry Ransom
Humanities Research Center, University of Texas at Austin.

\noindent AHQP-OHI: Oral history interview transcripts.

\noindent IIPCS: Archives of the Instituts Internationaux de Physique et de
Chimie Solvay, Universit\'{e} Libre de Bruxelles.

\noindent\textit{Ann.\ d.\ Phys.\ }or \textit{Ann.\ der Phys.}: Annalen der Physik.

\noindent\textit{Bayr.\ Akad.\ d.\ Wiss.\ Math.\ phys.\ Kl.}: Sitzungsberichte der 
Ma\-the\-ma\-tisch-Phy\-si\-ka\-li\-schen Klasse der K\"{o}niglich-Bayerischen Akademie der 
Wissenschaften (M\"{u}nchen).

\noindent\textit{Berl.\ Ber.}: Sitzungsberichte der Preussischen Akademie
der Wissenschaften (Berlin).

\noindent\textit{Acad.\ Roy.\ Belg.\ }or 
\textit{Bull.\ Ac.\ R.\ Belg.\ }or \textit{Bull.\ Ac.\ roy.\ de Belgique\/} or
\textit{Bull.\ Ac.\ roy.\ Belgique\/} or \textit{Bull.\ Ac.\ roy.\ Belg.\ }or
\textit{Bull.\ Ac.\ R.\ Belg., Cl.\ des Sciences\/}: Bulletin de
l'Acad\'{e}mie Royale des Sciences, des Lettres et des Beaux-arts de Belgique.
Classe des Sciences.

\noindent\textit{Bull.\ Natl.\ Res.\ Coun.}: Bulletin of the National
Research Council (U.S.).

\noindent\textit{Comm.\ Fenn.}: Commentationes Physico-mathematicae, Societas Scientiarum Fennica.

\noindent\textit{C.\ R.\ }or \textit{C.\ R.\ Acad.\ Sc.\ }or \textit{Comptes
Rendus Acad.\ Sci.\ Paris\/}: Comptes Rendus Hebdomadaires des S\'{e}ances de
l'Acad\'{e}mie des Sciences (Paris).

\noindent\textit{G\"{o}tt.\ Nachr.}: Nachrichten der Akademie der
Wissenschaften in G\"{o}ttingen. II, Mathematisch-Physikalische
Klasse.

\noindent\textit{J.\ de Phys.\ }or \textit{Jour.\ de Phys.\ }or
\textit{Journ.\ Physique\/} or \textit{Journ.\ d.\ Phys.}:  
Journal de Physique (until 1919), then Journal de Physique et le Radium. 

\noindent\textit{Jour.\ Frank.\ Inst.}: Journal of the Franklin Institute.

\noindent\textit{Lincei Rend.}: Rendiconti Lincei.

\noindent\textit{Manchester Memoirs\/}: Manchester Literary and Philosophical Society, Memoirs and Proceedings.

\noindent\textit{Math.\ Ann.\ }or \textit{Mathem.\ Ann.}: Mathematische Annalen.

%
%
\noindent\textit{Naturw.\ }or \textit{Naturwiss.\ }or
\textit{Naturwissensch.\ }or \textit{Naturwissenschaften\/}: Die Naturwissenschaften.

\noindent\textit{Nat.\ Acad.\ Sci.\ Proc.\ }or \textit{Proc.\ Nat.\ Acad.\ Sci.\ }or 
\textit{Proc.\ Nat.\ Acad.}: Proceedings of the National Academy of Sciences (U.S.).

\noindent\textit{Phil.\ Mag.}: Philosophical Magazine.

\noindent\textit{Phil.\ Trans.\ }or \textit{Phil.\ Trans.\ Roy.\ Soc.}:
Philosophical Transactions of the Royal Society of London.

\noindent\textit{Phys.\ Rev.}: Physical Review.

\noindent\textit{Phys.\ Zeits.\ }or \textit{Phys.\ Zeitsch.\ }or
\textit{Physik.\ Zts.}: Physikalische Zeitschrift.

\noindent\textit{Proc.\ Camb.\ Phil.\ Soc.\ }or \textit{Proc.\ Cambr.\ Phil.\
Soc.\ }or \textit{Proc.\ Cambridge Phil.\ Soc.}: Proceedings of the Cambridge
Philosophical Society.

\noindent\textit{Proc.\ Phys.\ Soc.}: Proceedings of the Physical Society of London.

\noindent\textit{Proc.\ Roy.\ Soc.\ }or \textit{Roy.\ Soc.\ Proc.}: Proceedings
of the Royal Society of London.

\noindent\textit{Upsala Univ.\ \AA rsskr.}: Uppsala Universitets \AA rsskrift.

%
%
\noindent\textit{Z.\ f.\ Phys.\ }or \textit{Zts.\ f.\ Phys.\ }or \textit{Zeit.\
f.\ Phys.\ }or \textit{Zeits.\ f.\ Phys.\ }or \textit{Zeitsch.\ f.\ Phys.\ }or
\textit{Zeitschr.\ f.\ Phys.}: Zeitschrift f\"{u}r Physik.

\section*{Typographic conventions}\markboth{{\it Typographic conventions}}{{\it Typographic conventions}}
\addcontentsline{toc}{section}{Typographic conventions}

\noindent The following conventions have been used.

\noindent Square brackets [ ] denote editorial amendments or (in the translations) original wordings.

\noindent Curly brackets \{ \} denote additions (in original typescripts or manuscripts). 

\noindent Angle brackets $<$ $>$ denote cancellations (in original typescripts or manuscripts).

\section*{Note on the bibliography}\markboth{{\it Note on the bibliography}}{{\it Note on the bibliography}}
\addcontentsline{toc}{section}{Note on the bibliography}

The references  cited in Parts I and II, and in the endnotes 
and editorial footnotes to Part III, are listed in our bibliography. 
The references cited in the original Solvay volume are found
in the translation of the proceedings in Part III.

\section*{Permissions and copyright notices}\markboth{{\it Permissions and copyright notices}}{{\it Permissions and copyright notices}}
\addcontentsline{toc}{section}{Permissions and copyright notices}

\mainmatter

\part{\\ Perspectives on the 1927 Solvay conference}

\setcounter{endnote}{0}
\setcounter{equation}{0}

\chapter{Historical introduction}\label{HistEss}\chaptermark{Historical introduction}
  \begin{epigraphs}
    \qitem{Quantum reconciliation very [added, deleted] unpleasant [deleted] tendency
           [deleted] retrograde [deleted] questionable [added, deleted] idea [deleted] 
           flippant [deleted] title leads to misunderstanding.}{Ehrenfest, on the 
           conference plans\endnotemark\\ \ }
    \qitem{The conference was surely the most interesting scientific conference I have 
           taken part in so far.}{Heisenberg, upon receipt of the conference photograph\endnotemark}
  \end{epigraphs}
\addtocounter{endnote}{-1}
\endnotetext{Handwritten remark (by Ehrenfest) in the margin of Lorentz to Ehrenfest, 29 March 1926, AHQP-EHR-23 (in Dutch).}
\addtocounter{endnote}{1}
\endnotetext{Heisenberg to Lef\'{e}bure, 19 December [1927], IIPCS 2685 (in German).}

\noindent The early Solvay conferences were remarkable occasions, made possible by the generosity of Belgian industrialist Ernest 
Solvay and, with the exception of the first conference in 1912, planned and organised by the indefatigable Hendrik Antoon 
Lorentz. In this chapter, we shall first sketch 
the beginnings of the Solvay conferences, Lorentz's involvement and the situation
in the years leading up to 1927 (sections~\ref{Ernest} and \ref{war}). Then we shall describe specifically the planning
of the fifth Solvay conference, both in its scientific aspects (section~\ref{scientific}) and in its more practical 
aspects (section~\ref{details}). Section~\ref{day-by-day} presents the day-by-day progress of the conference as far
as it can be reconstructed from the sources, while section~\ref{editing} follows the making of the volume of proceedings,
which is the main source of original material from the fifth Solvay conference and forms Part III of this book.

\section{Ernest Solvay and the Institute of Physics}\label{Ernest}
Ernest Solvay had an extensive record of supporting scientific, educational and social initiatives,
as Lorentz emphasises in a two-page document written in September 1914, during the first months of the first world 
war:\endnote{AHQP-LTZ-12, talk X 23, `Ernest Solvay', dated 28 September 1914 (in English). Cf.\ also the French 
version of the same, X 10. (The two pages of the latter are in separate places on the microfilm.)}
  \begin{quote}
    I feel bound to say some words in these days about one of Belgium's noblest citizens, one of the men whom I admire 
    and honour most highly.
    Mr Ernest Solvay .... is the founder of one of the most flourishing industries of the world, the soda manufacture
    based on the process invented by him and now spread over Belgium, France, England, Germany, Russia and the United
    States. ....
    The fortune won by an activity of half a century has been largely used by Mr Solvay for the public benefit. In the
    firm conviction that a better understanding of the laws of nature and of human society will prove one of the most
    powerful means for promoting the happiness of mankind, he has in many ways and on a large scale encouraged and
    supported scientific research and teaching.
  \end{quote}
Part of this activity was centred around the project of the Cit\'{e} 
Scientifique, a series of institutes in Brussels founded and endowed by Ernest Solvay and by his brother Alfred Solvay,
which culminated in the founding of the Institutes of Physics and of Chemistry in 1912 and 1913.\footnote{The following 
material on the Cit\'{e} Scientifique is drawn mainly from Despy-Meyer and Devriese (1997).}

This project had originally developed through the chance encounter between Ernest Solvay and 
Paul H\'{e}ger, physician and professor of physiology at the Universit\'{e} Libre de Bruxelles (ULB), and involved a 
collaboration between Solvay, the ULB and the city of Brussels.
In June 1892, it was agreed that Solvay would construct and equip two Institutes of Physiology on land owned by 
the city in the Parc L\'{e}opold in Brussels.\footnote{One was to become property of the city and given in use 
to the ULB, while the other was to be leased for thirty years to and run by Solvay himself.} 
There soon followed in 1893--94 an Institute for Hygiene, 
Bacteriology and Therapy, funded mainly by Alfred Solvay, and a School of Political and Social Sciences, founded by 
Ernest Solvay in 1894, which moved to the Cit\'{e} Scientifique in 1901, and to which a School of Commerce was added 
in 1904.

The idea for what became known as the first Solvay conference in physics goes back to Wilhelm Nernst and Max 
Planck,\footnote{In the rest of this and in part of the following sections, we draw  
on an unpublished compilation of the contents of the Solvay archives by J.~Pelseneer.\endnotemark}\endnotetext{Pelseneer, J., 
{\em Historique des Instituts Internationaux de Physique et de Chimie Solvay depuis leur fondation jusqu'\`{a} la 
deuxi\`{e}me guerre mondiale}, [1962], 103~pp., AHQP-58, section 1 (hereafter referred to simply as `Pelseneer').} who around 1910 considered that the
current problems in the theory of radiation and in the theory of specific heats had become so serious that an international
meeting (indeed a `council') should be convened in order to attempt to resolve the situation. The further encounter between
Nernst and Solvay provided the material opportunity for the meeting, and by July 1910, Nernst was sending Solvay the 
detailed proposals. He had also secured the collaboration of Lorentz (who was eventually asked to preside), of Knudsen 
and naturally of Planck, who wrote:
   \begin{quote}
     .... anything that may happen in this direction will excite my greatest interest and .... I promise already
     my participation in any such endeavour. For I can say without exaggeration that in fact for the past 10 years
     nothing in physics has so continuously stimulated, excited and irritated me as much as these quanta of 
     action.\footnote{Exquisite ending in the original: `.... dass mich seit 10 Jahren im Grunde nichts 
     in der Physik so ununterbrochen an-, er-, und aufregt wie diese Wirkungsquanten'.\endnotemark}\endnotetext{Planck 
     to Nernst, 11 June 1910, Pelseneer, p.~7 (in German).}       
   \end{quote}

Lorentz set up a committee to consider questions relating to the new experimental research that had been deemed necessary 
during the conference (which took place between 30 October and 3 November 1911). This committee
included Marie Curie, Brillouin, Warburg, Kamerlingh Onnes, Nernst, Rutherford and Knudsen. Lorentz in turn was asked
to be the president. Further, at the end of the conference, Solvay proposed to Lorentz the idea of a scientific foundation. 
Lorentz's reply to Solvay's 
proposals, of 4 January 1912, includes extremely detailed suggestions on the functions and structure of the
foundation, all of which were put into practice and which can be summarised as follows.\endnote{Lorentz to Solvay, 
4 January 1912, Pelseneer, pp.~20--26 (in French). See also the reply, Solvay to Lorentz, 10 January 1912, AHQP-LTZ-12 (in French).} 

The foundation would be devoted principally to physics 
and physical chemistry, as well as to questions relating to physics from other sciences. It
would provide international support to researchers (`a Rutherford, a Lenard, 
a Weiss') in the form of money or loan of scientific instruments, and it would provide scholarships for young Belgian
scientists (both men and women) to work in the best laboratories or universities, mostly abroad. The
question of a link between the foundation and the `Conseil de physique' was left open, but Lorentz suggested to
provide meeting facilities if Solvay wished to link the two. Lorentz
suggested instituting an administrative board (consisting of a Solvay family member or appointee, an appointee of the 
King, and a member of the Belgian scientific establishment) and a scientific committee (which could initially be the
one he had formed during the first Solvay conference). Finally, Lorentz suggested housing the foundation in an annex 
of one of the existing institutes in the Cit\'{e} Universitaire.

During January, Solvay sent Paul H\'{e}ger to Leiden to work with Lorentz on the statutes of the foundation, which
Lorentz sent to Solvay on 2 February. Solvay approved them with hardly any modifications (only such as were required by the 
Belgian legislation of the time). The foundation, or rather the `Solvay International Institute of Physics', was officially 
established on 1 May 1912, which predates by several years the establishment of the comparable Belgian state institutions  
(Fondation Universitaire: 1920; Fonds National de la Recherche Scientifique: 1928). In this connection, Lorentz hoped `that 
governments would understand more and more the importance of scientific research and that in the long run one will 
arrive at a satisfactory organisation, independent of the individual efforts of private persons',\endnote{Lorentz 
to Solvay, 6 March 1912, Pelseneer, p.~27 (in French).} a sentiment echoed by Solvay himself.\footnote{`Mr Solvay 
also thinks that it is the role of the state to subsidise and organise scientific institutions, and he hopes that 
in thirty years the state will fulfill this duty better than it does
today.'\endnotemark}\endnotetext{H\'{e}ger to Lorentz, 16 February 1912, AHQP-LTZ-11 (in French).} 

The institute, which Solvay had endowed for thirty years, could soon boast of
remarkable activity in supporting scientific research. The numerous recipients of subsidies 
granted during the first two years until the first world war included Lebedew's laboratory,  
von Laue, Sommerfeld, Franck and Hertz, W.~L.~Bragg (who was later to become president of the 
scientific committee), Stark, and Wien. In 1913, an Institute of Chemistry followed suit, organised 
along similar lines to the Institute of Physics.

\section{War and international relations}\label{war}

The first meeting of the scientific committee, for the planning of the second Solvay conference, took place on 
30 September and 1 October 1912. The conference was held the following year, but the activities of the institute 
were soon disrupted by the start of the first world war, in particular the German invasion of Belgium.

Immediate practical disruption included the fear of requisitions, the difficulty of communication between the
international membership of the scientific committee and, with regard to the publication of the proceedings
of the second Solvay conference, the impossibility of sending Lorentz the proofs 
for correction and the eventual prospect of German censorship.\footnote{The proceedings of the first Solvay 
conference had had both a French and a German edition. Those of the 
second Solvay conference were printed in three languages in 1915, but never published in this form and later mostly 
destroyed. Only under the changed conditions after the war, in 1921, were the proceedings published in a French 
translation (carried out, as on later occasions, by J.-\'{E}.~Verschaffelt).} 

The war, however, had longer-term negative implications for international intellectual cooperation. In October 1914,
a group of 93 repres\-entatives of German science and culture signed the manifesto `An die Kulturwelt!', denying German 
responsibilities in the war.\footnote{The main claims of the manifesto were:~`.... {\em It is not true} that Germany 
is the cause of this war.~.... {\em It is not true} that we have wantonly [freventlich] infringed the neutrality of 
Belgium.~.... {\em It is not true} that the life and property of a single Belgian citizen has been touched by our soldiers,
except when utter self-defence required it.~.... {\em It is not true} that our troops have raged brutally against 
Leuven.~.... {\em It is not true} that our conduct of war disregards the laws of international right.~.... 
{\em It is not true} that the struggle against our so-called militarism is not a struggle against our culture~....' 
(translated from B\"{o}hme 1975, pp.~47--9).} Among the signatories were both Nernst and Planck. This manifesto was  
partly responsible for the very strong hostility of French and Belgian scientists and institutions towards 
renewal of scientific relations with Germany after the war. 

No Germans or Austrians were invited to the third Solvay conference of 1921. The only exception (which remained 
problematic until the last minute) was Ehrenfest, who was Austrian, but who had remained in Leiden throughout the war 
as Lorentz's successor. Similarly, no Germans participated in the fourth Solvay conference of 1924. 
French and Bel\-gian armies had occupied the Ruhr in January 1923, and the international situation was particularly tense.
Einstein had (temporarily) resigned from the League of Nations' Committee on Intellectual Cooperation, and wrote to
Lorentz that he would not participate in the Solvay conference because of the exclusion of the German scientists, 
and that he should please make sure that no invitation was sent.\endnote{Einstein to Lorentz, 16 August 1923, AHQP-LTZ-7 (in German).} 
Bohr also declined to participate in the conference apparently because of the continued exclusion of German scientists (Moore 1989,
p.~157). Schr\"{o}dinger, however, who was Austrian and working in Switzerland, was invited.\footnote{Van Aubel (a member of the
scientific committee) objected strongly in 1923 to the possibility of Einstein being invited to the fourth Solvay 
conference, and resigned when it was decided to invite him. It appears he was convinced to remain on the 
committee.\endnotemark}\endnotetext{Van Aubel to Lorentz, 16 April,
16 May and 19 July 1923, AHQP-LTZ-11 (in French).}

Einstein had distinguished himself by assuming a pacifist position during the war.\footnote{For instance, Einstein 
was one of only four signatories of the counter-manifesto `Aufruf an die Europ\"{a}er' (Nicolai 1917). Note also that 
Einstein had renounced his German citizenship and had become a Swiss citizen in 1901, although there was some 
uncertainty about his citizenship when he was awarded the Nobel prize (Pais 1982, pp.~45 and 503--4).} Lorentz was pointing 
out Einstein's exceptional case to Solvay already in January 1919:
   \begin{quote}
     However, in talking about the Germans, we must not lose sight of the
     fact that they come in all kinds of nuances. A man like Einstein, the great and profound physicist, is not `German'
     in the sense one often attaches to the word today; his judgement on the events of the past years will not differ
     at all from yours or mine.\endnote{Lorentz to Solvay, 10 January 1919, Pelseneer, p.~37 (in French).}
   \end{quote}
In the meantime, after the treaty of Locarno of 1925, Germany was going to join the League of Nations, but the 
details of the negotiations were problematic.\footnote{Lorentz to Einstein on 14 March 1926: 
`Things are bad with the League of Nations; if only one could yet find a way out until the day 
after tomorrow'.\endnotemark Negotiations provisionally broke down on 17 March, but Germany 
eventually joined the League in September 1926.}\endnotetext{Lorentz to Einstein, 14 March 1926, AHQP-86 (in German).}
As early as February 1926, one finds mention of the prospect of renewed inclusion of German scientists at the Solvay 
conferences.\endnote{Lef\'{e}bure to Lorentz, 12 February 1926, AHQP-LTZ-12 (in French).} In the same month, Kamerlingh 
Onnes died, and at the next meeting of the scientific committee, in early April (at which the fifth Solvay conference 
was planned), it was decided to propose both to invite Einstein to replace Onnes and to include again the German scientists. 

On 1 April, Charles Lef\'{e}bure, then secretary of the administrative commission, wrote to commission members
Armand Solvay and Jules 
Bordet,\footnote{Lef\'{e}bure was the appointee of the Solvay family to the administrative commission, and 
as such succeeded Eug\`{e}ne Tassel, who had died in October 1922 and had been a long-standing collaborator of Ernest Solvay 
since 1886. Armand Solvay was the son of Ernest Solvay, who had died on 26 May 1922. Bordet was the royal 
appointee to the commission, and had just been appointed in February 1926, following the death of Paul 
H\'{e}ger.\endnotemark}\endnotetext{Cf.\ Lef\'{e}bure to Lorentz, 12 February 1926, AHQP-LTZ-12 (in French).}
enquiring about the admissibility of `moderate figures like Einstein, 
Planck\footnote{According to Lorentz, Planck had always been helpful to him when he had tried 
to intervene with the German authorities during the war. Further, Planck had somewhat qualified his position 
with regard to the Kulturwelt manifesto in an 
open letter, which he asked Lorentz to publish in the Dutch newspapers in 1916. On the other hand, he explicitly ruled 
out a public disavowal of the manifesto in December 1923.\endnotemark}\endnotetext{Letter by Lorentz, 7 January 1919, 
Pelseneer, pp.~35--6 (in French), two letters from Planck to Lorentz, 1915, AHQP-LTZ-12 (in German), Planck to Lorentz, 
March 1916, Pelseneer, pp.~34--5 (in German), and Planck to Lorentz, 5 December 1923, AHQP-LTZ-9 (in German).}  and 
others'\endnote{From Lef\'{e}bure 
[possibly a copy for Lorentz], 1 April 1926, AHQP-LTZ-12 (in French). Obituary of Tassel, {\em L'\'{E}ventail}, 
15 October 1922, AHQP-LTZ-13 (in French).} (Bordet telegraphed 
back: `Germany will soon be League of Nations therefore no objection'\endnote{Cf.\ Lef\'{e}bure to Lorentz, 6 April 
1926, AHQP-LTZ-12 (in French).}). On 2 April, Lorentz himself had a long interview with the King, who gave his approval.

Thus, finally, Lorentz wrote to Einstein on 6 April, informing him of the unanimous decision by the members of the 
committee present at the meeting,\footnote{Listed as Marie Curie, Langevin, Richardson, 
Guye and Knudsen (with two members absent, W.~H.~Bragg and Van Aubel).} as well as of the whole administrative commission, 
to invite him to succeed Kamerlingh Onnes. The Solvay conferences were to readmit Germans, and if Einstein were a member 
of the committee, Lorentz hoped this would encourage the German scientists to accept the invitation.\endnote{Lorentz to 
Einstein, 6 April 1926, AHQP-86 (in German).} Einstein was favourably impressed by the positive Belgian attitude and glad to 
accept under the altered conditions.\endnote{Einstein to Lorentz, 14 April 1926 and 1 May 1927, AHQP-LTZ-11 (in German).}
Lorentz proceeded to invite the German scientists, `not because there should be such a great 
haste in the thing, rather to show the Germans as soon as possible our good will',\endnote{Lorentz to Einstein,
28 April 1926, AHQP-86 (in German).} and sent the informal invitations to Born, Heisenberg and Planck (as well as
to Bohr) in or around June 1926.\endnote{Compare the invitation of Lorentz to Bohr, 7 June 1926, AHQP-BSC-13 (in English), 
and the replies of Bohr to Lorentz, 24 June 1926, AHQP-LTZ-11 (in English), Planck to Lorentz, 13 June 1926, AHQP-LTZ-8 
(in German), Born to Lorentz, 19 June 1926, AHQP-LTZ-11 (in German), and Heisenberg to Lorentz, 4 July 1926, AHQP-LTZ-12 
(in German).} 

As late as October 1927, however, the issue was still a sensitive one. Van Aubel 
(who had not been present at the April 1926 meeting of the scientific committee) replied in the negative to the official 
invitation to the conference.\footnote{Lef\'{e}bure's comment was: `because there are Germans! Then why does he stay 
in the Institute of Physics?'\endnotemark}\endnotetext{Van Aubel to Lef\'{e}bure, 6 October 1927, IIPCS 2545 (in French), with 
Lef\'{e}bure's handwritten comment.} Furthermore, 
it was proposed to release the list of participants to the press only after the conference to avoid public demonstrations.
Lorentz travelled in person to Brussels on 17 October to discuss the matter.\endnote{Lef\'{e}bure to Lorentz, 14 October
1927, IIPCS 2534, and 15 October 1927, IIPCS 2536, telegramme Lorentz to Lef\'{e}bure, 15 October 1927, IIPCS 2535, and
Lef\'{e}bure to the King, 19 October 1927, IIPCS 2622 (all in French).}

Lorentz's own position during and immediately after the war, as a physicist from one of the neutral countries, had possibly 
been rather delicate. In the text on Ernest Solvay from which we have quoted at the beginning of this chapter, for instance,
he appears to be defending the impartiality of the policies of the Institute of Physics in the years leading up to the war. 
Lorentz started working for some form of reconciliation as soon as the war was over, writing as follows to Solvay in
January 1919: 
  \begin{quote}
    All things considered, I think I must propose to you not to exclude formally the Germans, that is, not to close
    the door on them forever. I hope that it may be open for a new generation, and even that maybe, in the course
    of the years, one may admit those of today's scholars who one can believe regret sincerely and honestly the
    events that have taken place. Thus German science will be able to regain the place that, despite everything, it
    deserves for its past.\endnote{Lorentz to Solvay, 10 January 1919, Pelseneer, p.~37 (in French).} 
  \end{quote}
It should be noted that Lorentz was not only the scientific organiser of the Solvay institute and the 
Solvay conferences, but also a prime mover behind efforts towards international intellectual cooperation, through his 
heavy involvement with the Conseil International de Recherches, as well as with the League of Nations' Committee 
on Intellectual Cooperation, of which he was a member from 1923 and president from 1925.\footnote{The Conseil 
International de Recherches (founded in 1919) has today become the International Council for Science (ICSU). 
The Committee on Intellectual Cooperation (founded in 1922) and the related International Institute of Intellectual 
Cooperation (inaugurated in Paris in 1926) were the forerunners of UNESCO.\endnotemark}\endnotetext{There is a large 
amount of relevant correspondence in AHQP-LTZ.} 

Lorentz's figure and contributions to the Solvay conferences are movingly recalled by Marie Curie in her obituary of 
Lorentz in the proceedings of the fifth Solvay conference (which opens Part III of this volume).

\section{Scientific planning and background}\label{scientific}
What was at issue in the remark that heads this chapter,\footnote{In the 
original: `Quantenverzoening $<$\{zeer\} antipathik$<$e$>>$ $<$tendentie$>$ $<$retrograde$>$ $<$\{bedenkelijk$<$e$>$\}$>$ 
$<$idee$>$ $<$loszinnige$>$ [?] titel wekt misverstand'. Many thanks to Mark van Atten for help with this passage.} 
scribbled by Ehrenfest in the margin of a letter from Lorentz, 
was the proposed topic for the fifth Solvay conference, namely `the conflict and the possible reconciliation between the 
classical theories and the theory of quanta'.\endnote{Lorentz to Ehrenfest, 
29 March 1926, AHQP-EHR-23 (in Dutch), with Ehrenfest's handwritten comments.} 
Ehrenfest found the phrasing objectionable in that it encouraged one to `swindle away the fruitful 
and suggestive harshness of the conflict by most slimy unclear thinking, quite in analogy with what happened also  
even after 1900 with the mechanical ether theories of the Maxwell equations', pointing out that 
`Bohr feels even more strongly than me against this slogan [Schlagwort], precisely because he takes it 
so particularly to heart to find the foundations of the future theory'.\endnote{Ehrenfest to Lorentz, 30 March 1926, 
AHQP-LTZ-11 (in German).}

Lorentz took Ehrenfest's suggestion into account, and dropped the reference to reconciliation
both from the title and from later descriptions of the focus of the
meeting.\footnote{To Bohr in June 1926: `.... the conflict between the classical theories and
the quantum theory .... '; to Schr\"{o}dinger in January 1927: ` .... the contrast between the current and the 
earlier conceptions [Auffassungen] and the attempts at development of a new mechanics'.\endnotemark}\endnotetext{Lorentz 
to Bohr, 7 June 1926, AHQP-BSC-13, section 3 (in English), Lorentz to Schr\"{o}dinger, 21 January 1927, AHQP-41, 
section 9 (in German).} 

The meeting of the scientific committee for the planning of the fifth Solvay conference took place in Brussels on
1 and 2 April 1926. Lorentz reported a few days later to Einstein: 
  \begin{quote}
    As the topic for 1927 we have chosen `The quantum theory and the classical theories of radiation', and we hope 
    to have the following reports or lectures:

    \

    \noindent{\em 1\/} W.~L.~Bragg. New tests of the classical theory.\\
    {\em 2\/} A.~H.~Compton. Compton effect and its consequences.\\
    {\em 3\/} C.~T.~R.~Wilson. Observations on photoelectrons and collision electrons by the condensation method.\\
    {\em 4\/} L.~de~Broglie. Interference and light quanta.\\
    {\em 5\/} (short note): Kramers. Theory of Slater-Bohr-Kramers and analogous theories.\\
    {\em 6\/} Einstein. New derivations of Planck's law and applications of statistics to quanta.\\
    {\em 7\/} Heisenberg. Adaptation of the foundations of dynamics to the quantum theory.\endnote{Lorentz to Einstein, 
    6 April 1926, AHQP-86 (in German and French).}
  \end{quote}
Another report, by the committee's secretary Verschaffelt,\endnote{Verschaffelt to Lef\'{e}bure, 8 April 1926, IIPCS 2573 (in French).} 
adds, concerning point {\em 5\/}: `(at least, if Mr Kramers judges that it is still useful)'; it further lists a few 
alternative speakers: Compton or Debye for {\em 2\/}, Einstein or Ehrenfest for {\em 6\/}, and Heisenberg or 
Schr\"{o}dinger for {\em 7}.\footnote{For details of the other participants, see the next section.}

Thus, the fifth Solvay conference, as originally planned, was to focus mainly on the theory of radiation and on light
quanta, including only one report on the new quantum theory of matter. The shift in focus between 1926 and 1927
was clearly due to major theoretical advances (for example by Schr\"{o}dinger and Dirac) and new experimental
results (such as the Davisson-Germer experiments), and it can be partly followed as the planning of the conference
progressed. 

Schr\"{o}dinger's wave mechanics was one of the major theoretical developments of the year 1926.
Einstein, who had been alerted to Schr\"{o}dinger's first paper by Planck (cf.~Przibram 1967, p.~23), 
suggested to Lorentz that Schr\"{o}dinger should talk at the conference instead of himself, on the basis 
of his new `theory of quantum states', which he described as a development of genius of de Broglie's 
ideas.\endnote{Einstein to Lorentz, 12 April 1926, AHQP-LTZ-11 (in German).} While
it is unclear whether Lorentz knew of Schr\"{o}dinger's papers by the time of the April meeting,\footnote{Cf.\ 
section~\ref{Schr-invitation}.} 
Schr\"{o}dinger was listed a week later as a possible substitute for Heisenberg, and Lorentz himself
was assuring Einstein at the end of April that Schr\"{o}dinger was already being considered, specifially as a 
substitute for the report on the new foundations of dynamics rather than for the report on quantum 
statistics.\footnote{Note that Schr\"{o}dinger (1926a) had written on `Einstein's gas theory' in a paper 
that is an immediate precursor to his series of papers on quantisation.} 

Lorentz closely followed the development of
wave mechanics, indeed contributing some essential critique in his correspondence with Schr\"{o}dinger from this period,
for the most part translated in Przibram (1967) (see chapter~\ref{SchrEss}, especially sections~\ref{Schr-packets}
and \ref{Schr-radiation}, for some more details on this correspondence).
Lorentz also gave a number of colloquia and lectures on wave mechanics (and on matrix mechanics) in the period leading
up to the Solvay conference, in Leiden, Ithaca and Pasadena.\endnote{See for instance the catalogue of Lorentz's manuscripts in AHQP-LTZ-11.} 
In Pasadena he also had the opportunity of discussing
with Schr\"{o}dinger the possibility that Schr\"{o}dinger may also give a report at the conference, as in fact he 
did.\footnote{Lorentz was at Cornell from 
September to December 1926, then in Pasadena until March 1927.\endnotemark On Schr\"{o}dinger's American voyage, see Moore 
(1989, pp.~230--33).}\endnotetext{See also Lorentz to Schr\"{o}dinger, 21 January and 17 June 1927, 
AHQP-41, section 9 (in German).}
Schr\"{o}dinger's wave mechanics had also made a great 
impression on Einstein, although he repeatedly expressed his unease to Lorentz at the use of wave 
functions on configuration space\label{mysterium} (`obscure',\endnote{Einstein to Lorentz, 1 May 1926, AHQP-LTZ-11 (in German).} 
`harsh',\endnote{Einstein to Lorentz, 22 June 1926, AHQP-LTZ-8 (in German).} a `Mysterium'\endnote{Einstein 
to Lorentz, 16 February 1927, AHQP-LTZ-11 (in German).}), and again during the general discussion 
(p.~\pageref{Einstein-config}).

One sees Lorentz's involvement with the recent developments also in his correspondence with Ehrenfest. In particular,
Lorentz appears to have been struck by Dirac's contributions to quantum mechanics.\footnote{This correspondence 
includes for instance a 15-page commentary by Lorentz on Dirac (1927a).\endnotemark}\endnotetext{Enclosed
with Lorentz to Ehrenfest, 3 June [1927, erroneously amended to 1925], AHQP-EHR-23 (in Dutch).} 
In June 1927, Lorentz invited Dirac to spend the following academic year in Leiden
,\endnote{Lorentz to Dirac, 9 June 1927, AHQP-LTZ-8 (in English)
.} and asked Born and Heisenberg to include a discussion of Dirac's work
in their report.\endnote{Cf.\ Born to Lorentz, 23 June 1927, AHQP-LTZ-11 (in German).} Finally, in late August, Lorentz
decided that Dirac, and also Pauli, ought to be invited to the conference, for indeed:
  \begin{quotation}
    Since last year, quantum mechanics, which will be our topic, has developed with an unexpected rapidity, and some 
    physicists who were formerly in the second tier have made extremely notable contributions. For this reason I would 
    be very keen to invite also Mr Dirac of Cambridge and Mr Pauli of Copenhagen. .... Their collaboration would be very 
    useful to us .... I need not consult the scientific committee because Mr Dirac and Mr Pauli were both on a list that 
    we had drawn up last year .... .\endnote{Lorentz to Lef\'{e}bure, 27 August 1927, IIPCS 2532A/B (in French). [There 
    appears to be a further item also numbered 2532.]}
  \end{quotation}
Lorentz invited Pauli on 5 September 1927 (Pauli 1979, pp.~408--9) 
and Dirac sometime before 13 September 
1927.\endnote{Dirac to Lorentz, 13 September 1927, AHQP-LTZ-11 (in English).} 

On the experimental side, some of the main achievements of 1927 were the experiments on matter waves. While originally 
de Broglie was listed to give a report on light quanta, the work he presented was about both light quanta and material 
particles (indeed, electrons and photons!), and Lorentz asked him explicitly to include some discussion of the recent 
experiments speaking in favour of the notion of matter waves, specifically discussing Elsasser's (1925)  
proposals, and the experimental work of Dymond (1927) and of Davisson and Germer (1927).\endnote{Cf.\ de Broglie 
to Lorentz, 22 June 1927, AHQP-LTZ-11 (in French), and Ehrenfest to Lorentz, 14 June 1927, AHQP-EHR-23 (in Dutch and 
German).} Thus, in the final programme of the conference, we find three reports on the foundations of a new mechanics, by
de Broglie, Heisenberg (together with Born) and Schr\"{o}dinger.

The talks given by Bragg and Compton, instead, reflect at least in part the initial orientation of the conference.
Here is how Compton presents the division of labour (p.~\pageref{accuser}):
  \begin{quote}
     Professor W.\ L.\ Bragg has just discussed a whole series of radiation phenomena in which the electromagnetic 
     theory is confirmed. .... I have been left the task of pleading the opposing cause to that of the electromagnetic 
     theory of radiation, seen from the experimental viewpoint. 
  \end{quote}

Bragg focusses in particular on the technique of X-ray analysis, as the `most direct way of analysing atomic and 
molecular structure' (p.~\pageref{Bragg-direct}), the development of which, as he had mentioned to Lorentz, was the `line 
in which [he had] been especially interested'.\endnote{W.~L.~Bragg to Lorentz, 7 February 1927, AHQP-LTZ-11 (in English).}
This includes in particular the investigation of the electronic charge distribution.
At Lorentz's request, he had also included a discussion of the refraction of X-rays (section 8 of his report), which is 
directly relevant to the discussion after Compton's 
report.\endnote{Cf.\ also W.~L.~Bragg to Lorentz, 27 June 1927, AHQP-LTZ-11 (in English).} As described by Lorentz
in June 1927, Bragg was to report `on phenomena that still somehow allow a classical description'.\endnote{Lorentz 
to Schr\"{o}dinger, 17 June 1927, AHQP-41, section 9 (in German).} A few more aspects of Bragg's 
report are of immediate relevance for the rest of the conference, especially to the discussion of Schr\"{o}dinger's
interpretation of the wave function in terms of an electric charge density (pp.~\pageref{Bragg-Schr1}, 
\pageref{Bragg-Schr2}, section~\ref{Schr-radiation}),
and so are some of the issues taken up further in the discussion (Hartree approximation, problems with waves in three 
dimensions), but it is fair to say that the report provides a rather distant background for what followed it. 

Compton's talk covers the topics of points {\em 2\/} and {\em 3\/} listed above. The explicit
focus of his report is the three-way comparison between the photon hypothesis, the Bohr-Kramers-Slater (BKS) theory of
radiation, and the classical theory of radiation. Note, however, that Compton introduces many of the topics of later 
discussions. For instance, he discusses
the problem of how to explain atomic radiation (section on `The emission of radiation', p.~\pageref{emission}), 
which is inexplicable from the point of view of the classical theory, given that the `orbital frequencies' in the atom
do not correspond to the emission frequencies. This problem was one of Schr\"{o}dinger's main concerns 
and one of the main points of conflict between Schr\"{o}dinger and, for instance, Heisenberg (see in particular the
discussion after Schr\"{o}dinger's report and, below, sections~\ref{Schr-radiation} and \ref{Schr-conflict}). Compton's 
discussion of the photon
hypothesis relates to the question of `guiding fields' (pp.~\pageref{Comp46} and \pageref{Comp67}) and of the 
localisation of particles or energy quanta within a wave (pp.~\pageref{Comp53} and \pageref{Comp61}). These in turn
are closely connected with some of de~Broglie's and Einstein's ideas (see below chapter~\ref{locality-and-incompleteness},
especially section~\ref{Einstein-at-Salzburg}, and chapter~\ref{guiding-fields-in-3-space} ), and with de~Broglie's 
report on pilot-wave theory and Einstein's remark about locality in the general discussion (p.~\pageref{Einstein-loc}). 

Bohr had been a noted sceptic of the photon hypothesis, and in 1924 Bohr, Kramers and Slater\label{forpage55} had 
developed a theory that was able to maintain a wave picture of radiation, by introducing a description of the atom 
based on `virtual oscillators' with frequencies equal to the frequencies of emission (Bohr, Kramers and Slater 
1924a,b).\footnote{As Darrigol (1992, p.~257) emphasises, while the free virtual fields obey the Maxwell equations, 
i.e.\ can be considered to be classical, the virtual oscillators and the interaction between the fields and the 
oscillators are non-classical in several respects.} 
A stationary state 
of an atom, say the $n$th, is associated with a state of excitation of the oscillators with frequencies corresponding to 
transitions from the energy $E_n$. Such oscillators produce a classical radiation field (a `virtual' one), 
which in turn determines the {\em probabilities} for spontaneous emission in the atom, that is, for the emission of 
energy from the atom and the jump to a stationary state of lower energy. The virtual field of one atom also interacts
with the virtual oscillators in other atoms (which in turn produce secondary virtual radiation) and influences the probabilities 
for induced emission and absorption in the other atoms. 
While the theory provides a mechanism for radiation consistent with the picture of stationary states (cf.\ Compton's
remarks, p.~\pageref{Comp50}), it violates energy and momentum conservation for single events, in that an emission in 
one atom is not connected directly to an absorption in another atom, but only indirectly through the virtual 
radiation field. Energy and momentum conservation hold only at a statistical level. The BKS proposal was 
short-lived, because the Bothe-Geiger and Compton-Simon experiments established the conservation laws for individual 
processes (as explained in detail by Compton in his report, pp.~\pageref{forpage13}~ff.). Thus, at the time of 
the planning of the fifth 
Solvay conference, the experimental evidence had ruled out the BKS theory (hence the above remark: `if Mr Kramers 
judges that it is still useful').\footnote{Bothe and Geiger had been working on their experiments since June 1924 
(Bothe and Geiger 1924), and provisional results were being debated by the turn of the year.   
For two differing views on the significance of these results for instance see 
Einstein to Lorentz, 16 December 1924 (the same letter in which he wrote to Lorentz about de Broglie's 
results)\endnotemark and the exchange of letters between Born and Bohr in January 1925 (Bohr 1984, pp.~302--6). 
By April 1925, Bothe and Geiger had clear-cut results 
against the BKS theory (Bothe and Geiger 1925a,b; see also the letters between Geiger and Bohr in Bohr 1984, 
pp.~352--4).}\endnotetext{Einstein to Lorentz, 16 December 1924, AHQP-LTZ-7 (in 
German).} The short note {\em 5}, indeed, dropped out of the programme altogether.\footnote{On the BKS theory and related
matters, see also chapter~\ref{BornEss} (especially sections~\ref{beforematrix} and \ref{BornBohr}), 
chapter~\ref{guiding-fields-in-3-space}, Darrigol (1992, chapter~9), the excellent introduction by Stolzenburg to 
Part I of Bohr (1984) and Mehra and Rechenberg (1982, section~V.2).}

The description of the interaction between matter and radiation, in particular the Compton effect, continued to 
be a problem for Bohr, and contributed to the development of his views on wave-particle dualism and complementarity.
In his contribution to the discussion after 
Compton's report (p.~\pageref{forpage170}, the longest of his published contributions in the Solvay 
volume\footnote{See below for the fate of his contribution to the general discussion.}), Bohr sketches 
the motivations behind the BKS theory, the conclusions to be drawn from the Bothe-Geiger
and Compton-Simon experiments and the further development of his views. 

Lorentz, in his report of the meeting to Einstein had mentioned `Slater-Bohr-Kramers and analogous theories'. This 
may refer to the further developments (independent of the validity of the BKS theory) that led in 
particular to Kramer's (1924) dispersion theory (and from there towards matrix mechanics), 
or to Slater's original ideas, which were roughly along the lines of guiding fields 
for the photons (even though the photons were dropped from 
the final BKS proposal).\footnote{Cf.\ Slater (1924) and Mehra and Rechenberg (1982, pp.~543--6). See also Pauli's
remark during the discussion of de Broglie's report (p.~\pageref{PaulionSlater}).} 
Note that Einstein at this time was also thinking about guiding fields (in three dimensions). Pais (1982, pp.~440--41) 
writes that, according to Wigner, Einstein did not publish these ideas because they also led to problems with the 
conservation laws.\footnote{Cf.\ Einstein's contribution to the general discussion (p.~\pageref{Einstein-disc}) and the 
discussion below in chapter~\ref{guiding-fields-in-3-space}.}

Einstein was asked by Lorentz to contribute a report on `New derivations of Planck's law and applications 
of statistics to quanta' (point {\em 6\/}), clearly referring to the work by Bose (1924) on Planck's law, championed 
by Einstein and applied by him to the theory of the ideal gas (Einstein 1924, 1925a,b). The second of these papers is also
where Einstein famously endorses de~Broglie's idea of matter waves. Einstein thought that his work on the subject was 
already too well-known, but he accepted after Lorentz repeated his invitation.\endnote{Einstein to Lorentz, 12 April 
1926, AHQP-LTZ-11, Lorentz to Einstein, 28 April 1926, AHQP-86, and Einstein to Lorentz, 1 May 1926, AHQP-LTZ-11 
(all in German).} On 17 June 1927, however, at about the time when Lorentz was sending detailed requests to the 
speakers, Einstein informed him in the following terms that he would not, after all, present a report:
  \begin{quote}
    I recall having committed myself to you to give a report on quantum statistics at the Solvay conference.
    After much reflection back and forth, I come to the conviction that I am not competent [to give] such a
    report in a way that really corresponds to the state of things. The reason is that I have not been able
    to participate as intensively in the modern development of quantum theory as would be necessary for this 
    purpose. This is in part because I have on the whole too little receptive talent for fully following the stormy
    developments, in part also because I do not approve of the purely statistical way of thinking on which the new
    theories are founded .... Up until now, I kept hoping to be able to contribute something of value in Brussels;
    I have now given up that hope. I beg you not to be angry with me because of that; I did not take this lightly
    but tried with all my strength .... (Quoted in Pais 1982, pp.~431--2)
  \end{quote}
Einstein's withdrawal may be related to the following circumstances. On 5 May 1927, during a meeting of the Prussian 
Academy of Sciences in Berlin, Einstein had read a paper on the 
question: `Does Schr\"{o}dinger's wave mechanics determine the motion of a system completely or only in the sense 
of statistics?'\endnote{`Bestimmt Schr\"{o}dingers Wellenmechanik die Bewegung eines Systems vollst\"{a}ndig oder 
nur im Sinne der Statistik?', AEA 2-100.00 (in German), available on-line at 
http://www.alberteinstein.info/db/ViewDetails.do?DocumentID=34338 .} 
As discussed in detail by Belousek (1996), the paper attempts to define deterministic particle motions from 
Schr\"{o}dinger's wave functions, but was also suddenly withdrawn on 21 May.\footnote{The news of Einstein's 
communication prompted an exchange 
of letters between Heisenberg and Einstein, of which Heisenberg's letters, of 19 May and 10 June, 
survive.\endnotemark The second of these is particularly interesting, because Heisenberg presents in some detail
his view of theories that include particle trajectories. Both Einstein's hidden-variables proposal and 
Heisenberg's reaction will be described in section~\ref{EinsteinHV}.}\endnotetext{Heisenberg to Einstein, 
19 May 1927, AEA 12-173.00, and 10 June 1927, AEA 12-174.00 (both in German).}

The plans for the talks were finalised by Lorentz around June 1927. An extract from his letter to Schr\"{o}dinger
on the subject reads as follows:
  \begin{quotation}
    [W]e hope to have the following reports [Referate] (I give them in the order in which we might discuss them):\\
    1. From Mr W.~L.~Bragg on phenomena that still somehow allow a classical description (reflexion of X-rays by 
    crystals, diffraction and total reflection of X-rays).\\
    2. From Mr Compton on the effect discovered by him and what relates to it.\\
    3. From Mr de Broglie on his theory. I am asking him also to take into account the application of his ideas to 
    free electrons (Elsasser, quantum mechanics of free electrons; Dymond, Davisson and Germer, scattering of 
    electrons).\\
    4. From Dr Heisenberg {\em or} Prof.~Born (the choice is left to them) on matrix mechanics, including 
    Dirac's theory.\\
    5. Your report [on wave mechanics].

    Maybe another one or two short communications [Berichte] on special topics will be added.\endnote{Lorentz 
    to Schr\"{o}dinger, 17 June 1927, AHQP-41, section 9 (in German).}
  \end{quotation}

This was, indeed, the final programme of the conference, with Born and Heisenberg deciding to contribute a joint 
report.\label{Mehrafootnote}\footnote{See section~\ref{Born-writing}. Note that, as we shall see below, while Bohr 
contributed significantly to the general discussion and reported the views he had developed in Como (Bohr 1949, p.~216, 
1985, pp.~35--7), he was unable to prepare an edited version of his comments in time and therefore suggested that a 
translation of his Como lecture, in the version for {\em Naturwissenschaften} (Bohr 1928), be included in the volume 
instead. This has given rise to a common belief that Bohr gave a report on a par with the other reports, and that
the general discussion at the conference was the discussion following it. See for instance Mehra (1975, p.~152), and 
Mehra and Rechenberg (2000, pp.~246 and 249), who appear further to believe that Bohr did not participate in the official 
discussion.}

\section{Further details of planning}\label{details}

In 1926--27 the scientific committee and the administrative commission of the Solvay institute were composed as follows.

Scientific committee: Lorentz (Leiden) as president, Knudsen (Copenhagen) as secretary, W.~H.~Bragg (until May 1927, 
London),\footnote{W.~H.~Bragg resigned due to overcommitment and was later replaced by Cabrera 
(Madrid).\endnotemark}\endnotetext{W.~H.~Bragg to Lorentz, 11 May 1927, AHQP-LTZ-11 (in English).} Marie Curie (Paris), 
Einstein (since April 1926, Berlin), Charles-Eug\`{e}ne Guye (Geneva),\footnote{In 1909 the university of Geneva 
conferred on Einstein his first honorary degree. According to Pais (1982, p.~152), this was probably due to Guye. 
Coincidentally, Ernest Solvay was honoured at the same time.} Langevin (Paris), Richardson (London), Edm.\ van Aubel (Gent). 

Administrative commission: Armand Solvay, Jules Bordet (ULB), Maurice Bourquin (ULB), \'{E}mile Henriot 
(ULB); the administrative secretary since 1922, and thus main correspondent of Lorentz and others, was Charles 
Lef\'{e}bure.

The secretary of the meeting was Jules-\'{E}mile. Verschaffelt (Gent), who had acted as secretary since the third Solvay 
conference.\endnote{Tassel to Verschaffelt, 24 February 1921, AHQP-LTZ-13 (in French).}

The first provisional list of possible participants (in addition to Ehrenfest) appears in Lorentz's letter to Ehrenfest 
of 29 March 1926: 
  \begin{quotation}
    Einstein, Bohr, Kramers, Born, Heisenberg (Jordan surely more mathematician), Pauli, Ladenburg (?), Slater, the young 
    Bragg (because of the `correspondence' to the classical theory that his work has often resulted in), J.~J.~Thomson, 
    another one or two Englishmen (Darwin? Fowler?), L\'{e}on Brillouin (do not know whether he has worked on this, he has
    also already been there a number of times), {\em Louis} de~Broglie (light quanta), one or two who have concerned 
    themselves with diffraction of X-rays (Bergen Davis?, Compton, Debye, Dirac (?)).\endnotemark
  \end{quotation}\endnotetext{Lorentz to Ehrenfest, 29 March 1926, AHQP-EHR-23 (in Dutch).}
Lorentz asked for further suggestions and comments, which Ehrenfest sent in a letter dated `Leiden 30 March 1926. Late 
at night':\addtocounter{footnote}{1}\footnotetext{For more on the special interest of the Ramsauer effect, 
see section~\ref{BornBohr} below.}\addtocounter{footnote}{-1}
  \begin{quotation}
    Langevin, Fowler, Dirac, J.~Fran[c]k (already for the experiments he devised by Hanle on the destruction of
    resonance polarisation through Larmor rotation and for the work he proposed by Hund on the Ramsauer 
    effect\footnotemark ---
    undisturbed passage of slow electrons through atoms and so on), Fermi (for interesting continuation of the experiments
    by Hanle), Oseen (possibly a wrong attempt at explanation of needle radiation and as sharpwitted critic), 
    Schr\"{o}dinger (was perhaps the first to give quantum interpretation of the Doppler effect, thus close to Compton 
    effect), Bothe (for Bothe-Geiger experiment on correlation of Compton quantum and electron, which destroys Bohr-Slater 
    theory, altogether a fine brain!) (Bothe should be considered perhaps {\em before} Schr\"{o}dinger), Darwin, Smekal 
    (is indeed a very deserving connoisseur of quantum finesses, only he writes so {\em frightfully} much).

    L\'{e}on Brillouin has published something recently on matrix physics, but I have not read it yet.\endnote{Ehrenfest 
    to Lorentz, 30 March 1926, AHQP-LTZ-11 (in German).}
  \end{quotation}

At the April meeting (as listed in the report by Verschaffelt\endnote{Verschaffelt 
to Lef\'{e}bure, 8 April 1926, IIPCS 2573 (in French).}) it was then decided to invite: Bohr, Kramers, Ehrenfest, 
two among Born, Heisenberg and Pauli, Planck, Fowler, W.~L.~Bragg, C.~T.~R.~Wilson, L.~de~Broglie, L.~Brillouin, 
Deslandres, Compton, Schr\"{o}dinger and Debye. Possible substitutes were listed as: M.~de~Broglie or Thibaud for 
Bragg, Dirac for Brillouin, Fabry for Deslandres, Kapitza for Wilson, Darwin or Dirac for Fowler, Bergen Davis  
for Compton, and Thirring for Schr\"{o}dinger.\footnote{A few days later, Guye suggested also Auger as a possible
substitute for Wilson.\endnotemark}\endnotetext{Guye to Lorentz, 14 April 1926, AHQP-LTZ-8 (in French).} 
The members of the scientific commitee would all take part {\em ex officio}, and invitations would be sent to the
professors of physics at ULB, that is, to Piccard, Henriot and De Donder\endnote{Cf.\ 
for instance Lorentz to Schr\"{o}dinger, 21 January 1927, AHQP-41, section 9 (in German).} (the latter apparently somewhat 
to Lef\'{e}bure's chagrin, who, just before the conference started, felt obliged to remind Lorentz that De Donder was 
`a paradoxical mind, loud [encombrant] and always ready to seize the word, often with great maladroitness'\endnote{Lef\'{e}bure
to Lorentz, 22 October 1927, AHQP-LTZ-12 (in French).}).

Both the number of actual participants and of observers was to be kept limited,\endnote{Lorentz to Einstein, 28 April 1926, 
AHQP-86 (in German), 
Lorentz to Lef\'{e}bure, 9 October 1927, IIPCS, 2530A/B (in French).} partly explaining why it was thought that
one should invite only two among Born, Heisenberg and Pauli. The choice initially fell on Born and Heisenberg (although
Franck was also considered as an alternative).\endnote{Lorentz to Einstein, 28 April 1926, AHQP-86 (in German), Einstein 
to Lorentz, 1 May 1926, AHQP-LTZ-11 (in German).} Eventually, as noted above, Pauli was also included, as was Dirac.\endnote{For the 
latter, cf.\ also Brillouin to Lorentz, 20 August 1927, AHQP-LTZ-11 (in French).} Lorentz was also keen to invite 
Millikan --- and possibly Hall ---, when he heard that Millikan would be in Europe anyway for the Como meeting (Einstein 
and Richardson agreed).\endnote{Lorentz to Einstein, 30 January 1927, AHQP-86 (in German), Einstein to Lorentz, 16 February 
1927, AHQP-LTZ-11 (in German), Richardson to Lorentz, 19 February 1927, AHQP-LTZ-12 (in English).} However,
nothing came of this plan. 

When Einstein eventually withdrew as a speaker, he suggested Fermi or Langevin as possible substitutes  (Pais 1982, 
p.~432). For a while it was not clear whether Langevin (who was anyway a member of the scientific committee) would be able 
to come, since he was in Argentina over the summer and due
to go on to Pasadena from there. Ehrenfest suggested F.~Perrin instead, in rather admiring tones. Langevin was needed 
in Paris in October, however, and was able to come to the conference.\endnote{Brillouin to Lorentz, 20 August 1927,
AHQP-LTZ-11 (in French), Ehrenfest to Lorentz 18 August 1927, AHQP-EHR-23 (in German).}
Finally, the week before the conference started, Lorentz extended the invitation to Irving Langmuir,\endnote{Telegramme 
Lorentz to Lef\'{e}bure, 19 October 1927, IIPCS 2541 (in French), with Lef\'{e}bure's note: `Oui'.} who would happen to
be in Brussels at the time of the conference.\footnote{To Langmuir we owe a fascinating `home movie' of the
conference; see the report in the {\em AIP Bulletin of Physics News}, number 724 (2005).}

Lorentz sent most of the informal invitations around January 1927.\endnote{Lorentz to Brillouin, 15 December 1926,
AHQP-LTZ-12 (in French), Lorentz to Ehrenfest, 18 January 1927, AHQP-EHR-23 (in Dutch), Lorentz to Schr\"{o}dinger, 
21 January 1927, AHQP-41, section 9 (in German); compare various other replies to Lorentz's invitation, in AHQP-LTZ-11, 
AHQP-LTZ-12 and AHQP-LTZ-13: Brillouin, 8 January 1927 (in French), de Broglie, 8 January 1927 (in French), W.~L.~Bragg, 
7 February 1927 (in English), Wilson, 11 February 1927 (in 
English), Kramers, 14 February 1927 (in German), Debye, 24 February 1927 (in Dutch), Compton, 3 April 1927 (in English) 
[late because he had been away for two months `in the Orient'].} In May 1927, he sent to Lef\'{e}bure the list of all the people he had 
`provisionally invited',\endnote{Lorentz to Lef\'{e}bure, 21 May 1927, IIPCS 2521A (in French).} including all the members 
of the scientific committee and the prospective invitees as listed above by Verschaffelt (that is, as yet without Pauli 
and Dirac). All had already replied and accepted, except Deslandres (who eventually replied much later declining the 
invitation\endnote{Deslandres to Lorentz, 19 July 1927, AHQP-LTZ-11 (in French).}). Around early July, 
Lorentz invited the physicists from the university,\endnote{See the letters of acceptance to Lorentz: De Donder, 
8 July 1927, AHQP-LTZ-11, Henriot, 10 July 1927, AHQP-LTZ-12, Piccard, 2 October 1927, AHQP-LTZ-12 (all in French).} 
and presumably sent a new invitation to W.~H.~Bragg, who thanked him but  
declined.\endnote{W.~H.~Bragg to Lorentz, 12 and 17 July 1927, AHQP-LTZ-11 (in English).} Formal letters of 
confirmation were sent out by Lef\'{e}bure shortly before the conference.\endnote{Copies in AHQP-LTZ-12 and IIPCS 2543.
Various replies: IIPCS 2544--51, 2553--6, 2558, 2560--3.} 

Around June 1927, Lorentz wrote to the planned speakers inviting them in the name of the scientific committee to 
contribute written reports, to reach him preferably by 1 September. The general guidelines were: to focus
on one's own work, without mathematical details, but rather so that `the principles are highlighted as clearly 
as possible, and the open questions as well as the connections [Zusammenh\"{a}nge] and contrasts are 
clarified'. The material in the reports did not have to be unpublished, and a bibliography would be 
welcome.\endnote{Cf.\ Lorentz to Schr\"{o}dinger, 17 June 1927, AHQP-41, section 9 (in German).} Compton wrote that
he would aim to deliver his manuscipt by 20 August, de Broglie 
easily before the end of August, Bragg, as well as Born and Heisenberg, by 1 September, and Schr\"{o}dinger presumably 
only in the second half of September.\endnote{Compare the answers by Bragg, 27 June 1927, and by Compton, 7 July 1927, 
both AHQP-LTZ-11 (in English), and the detailed ones by de Broglie, 27 June 1927, AHQP-LTZ-11 (in French) and by Born, 
23 June 1927, AHQP-LTZ-11 (in German); also Schr\"{o}dinger to Lorentz, 23 June 1927, AHQP-LTZ-13 (original 
with Schr\"{o}dinger's corrections), and AHQP-41 section 9 (carbon copy) (in German).} (For further details of the 
correspondence between some of the authors and 
Lorentz, see the relevant chapters below.) 

The written reports were to be sent to all participants in advance of the conference.\footnote{Mimeographed 
copies of Bragg's, Born and Heisenberg's and Schr\"{o}dinger's reports are to be found in the Richardson Collection, 
Harry Ransom Humanities Research Center, University of Texas at Austin.\endnotemark}\endnotetext{Microfilmed in AHQP-RDN, 
documents M-0059 (Bragg, catalogued as `unidentified author'), M-0309 (Born and Heisenberg, with seven pages of notes by 
Richardson) and M-1354 (Schr\"{o}dinger).} De Broglie's, which had been written directly 
in French, was sent by Lorentz to the publishers, Gauthier-Villars in Paris, before he left for the Como meeting. They 
hoped to send 35 proofs to Lorentz by the end of September.\label{forpageDEB22} In the meantime, Verschaffelt and Lorentz's son had the 
remaining reports mimeographed by the `Holland Typing Office' in Amsterdam, and Verschaffelt with the 
help of a student added in the formulas by hand, managing to mail on time to the participants at least Compton's 
and Born and Heisenberg's reports, if not all of them.\endnote{See in particular the already quoted Lorentz to 
Lef\'{e}bure, 27 August 1927, IIPCS 2532A/B (in French), as well as Lorentz to Lef\'{e}bure, 23 September 1927, IIPCS 
2523A/B, and 4 October 1927, IIPCS 2528 (both in French), Gauthiers-Villars to Lef\'{e}bure, 16 September 1927, IIPCS 2755 
(in French), Verschaffelt to Lorentz, 6 October 1927, AHQP-LTZ-13 (in Dutch), de Broglie to Lorentz, 29 August 1927,
AHQP-LTZ-8, and 11 October 1927, AHQP-LTZ-11 (both in French), and Verschaffelt to Lef\'{e}bure, 15 October 1927, 
IIPCS 2756 (in French).} Lorentz had further written to all speakers (except Compton) to ask them to bring reprints
of their papers.\endnote{Lorentz to Ehrenfest, 13 October 1927, AHQP-EHR-23 (in Dutch). See also Born 
to Lorentz, 11 October 1927 AHQP-LTZ-11 (in German), de Broglie to Lorentz, 11 October 1927, AHQP-LTZ-11 (in French), 
Planck to Lef\'{e}bure, 17 October 1927, IIPCS 2558, and Richardson to Lef\'{e}bure, IIPCS 2561.}      

Late during planning, a slight problem emerged, namely an unfortunate overlap of the Brussels conference with the 
festivities for the centenary of Fresnel in Paris, to be officially opened Thursday 27 October.
Lorentz informed Lef\'{e}bure of the clash writing from Naples after the Como meeting: neither the date of the 
conference could be changed nor that of the Fresnel celebrations, which had been fixed by the French President. The 
problem was compounded by the fact that de Broglie had accepted to give a lecture 
to the Soci\'{e}t\'{e} de Physique on the occasion.\footnote{In Lorentz's letter, the date of de Broglie's lecture
is mentioned as 28 October, but the official invitations state that it was Zeeman who lectured then, and 
de~Broglie the next evening, after the end of the Solvay conference. A report on de~Broglie's lecture, which was entitled 
`Fresnel's \oe uvre and the current development of physics', was published by Guye in the {\em Journal de Gen\`{e}ve} of 
16 and 18 April 1928.\endnotemark}\endnotetext{`Une crise dans la physique moderne I \&\ II', IIPCS 2750--51 (in French).} 
Lorentz suggested the compromise solution of a general invitation to attend the celebrations.
Those who wished to participate could travel to Paris on 27 October, returning to Brussels the next day, when sessions 
would be resumed in the afternoon. This was the solution that was indeed adopted.\endnote{Lorentz to Lef\'{e}bure, 29 September 1927, IIPCS 2523A/B, Brillouin to Lorentz, 
11 October 1927, AHQP-LTZ-11, Fondation Solvay to Lef\'{e}bure, IIPCS 2582. Invitation: IIPCS 2615 and 2619, AHPQ-LTZ-8. 
Replies: IIPCS 2617 and 2618. (All documents in French.)}

\section{The Solvay meeting}\label{day-by-day}

%

The fifth Solvay conference took place from 24 to 29 October 1927 in Brussels. As on previous occasions, the 
participants stayed at the H\^{o}tel Britannique, where a dinner invitation from Armand Solvay awaited 
them.\endnote{IIPCS 2530A/B, 2537.} Other meals were going to be taken at the institute, which was
housed in the building of the Institute of Physiology in the Parc L\'{e}opold; catering for 
50--55 people had been arranged.\endnote{IIPCS 2533, 2586A/B/C/D/E (the proposed menus from the Taverne Royale), 
2587A/B.} The participants were guests of the administrative commission and all travel expenses within Europe
were met.\endnote{Lorentz to Schr\"{o}dinger, 21 January 1927, AHQP-41, section 9 (in German).} From the 
evening of 23 October onwards, three seats were reserved in a box at the Th\'{e}atre de la Monnaie.\endnote{IIPCS 2340.}

The first session of the conference started at 10:00 on Monday 24 October. A tentative 
reconstruction of the schedule of the conference is as follows.\endnote{For the 
time and place of the first session, see IIPCS 2523A/B. For the reception at ULB, see IIPCS 2540 and 2629. There is a 
seating plan for the lunch with the royal couple, IIPCS 2627. For the dinner with Armand Solvay, see IIPCS 2533, 2624 
and 2625. See also Pelseneer, pp.~49--50.} We assume that the talks were given in the order 
they were described in the plans and printed in the volume, and that the reception by the university on the Tuesday 
continued throughout the morning. It is clear that the general discussion extended over at least two days, from the fact
that Dirac in his main contribution (p.~\pageref{Dirac-contribution}) refers explicitly to Bohr's comments of the day before.\footnote{In a 
letter to Verschaffelt, Kramers refers to `the general discussion of Thursday', but that in fact was the day 
of the Fresnel celebrations in Paris. The photograph of Lorentz included in the volume, according to the caption, 
was also taken on that day. Since the celebrations opened only at 8:30pm, it is conceivable that there was a 
first discussion session on Thursday morning. Pelseneer states, however, that sessions were 
suspended for the whole day.\endnotemark}\endnotetext{See 
Kramers to Verschaffelt, 23 March 1928, AHQP-28 (in Dutch), and Pelseneer, p.~50.}
  \begin{itemize}
    \item Monday 24 October, morning: W.~L.~Bragg's report, followed by discussion.
    \item Monday 24 October, afternoon: A.~H.~Compton's report, followed by discussion.
    \item Tuesday 25 October, starting 9:00 a.m.: reception offered by the ULB.
    \item Tuesday 25 October, afternoon: L.~de~Broglie's report, followed by discussion.
    \item Wednesday 26 October, morning: M.~Born and W.~Heisenberg's report, followed by discussion.
    \item Wednesday 26 October, afternoon: E.~Schr\"{o}dinger's report, followed by discussion.
    \item Thursday 27 October, all day: travel to Paris and centenary of Fresnel.\footnote{Most of the participants at the 
          Solvay conference, with the exception of Knudsen, Dirac, Ehrenfest, Planck, Schr\"{o}dinger, Henriot, Piccard 
          and Herzen, travelled to Paris to attend the inauguration of the celebrations, in the {\em grand 
          amphith\'{e}atre} of the Sorbonne.\endnotemark}\endnotetext{IIPCS 2621.} 
    \item Friday 28 October, morning: return to Brussels.
    \item Friday 28 October, afternoon: general discussion.
    \item Saturday 29 October, morning: general discussion,\footnote{The final session of the conference also included 
          a homage to Ernest Solvay's widow.\endnotemark}\endnotetext{See AHQP-LTZ-12 (draft), and presumably IIPCS 2667.} 
          followed by lunch with the King and Queen of the Belgians.
    \item Saturday 29 October, evening: dinner offered by Armand Solvay.
  \end{itemize}

The languages used were presumably English, German and French. Schr\"{o}dinger had volunteered to give his talk in 
English,\footnote{Note that Schr\"{o}dinger was fluent in English from 
childhood, his mother and aunts being half-English (Moore 1989, chapter 1).} 
while Born had suggested that he and Heisenberg could provide additional explanations in English (while
he thought that neither of them knew French).\endnote{Schr\"{o}dinger to Lorentz, 23 June 1927, 
AHQP-LTZ-13 and AHQP-41, section 9 (in German), Lorentz to Schr\"{o}dinger,
8 July 1927, AHPQ-41, section 9 (in German), Born to Lorentz, 23 June 1927, AHQP-LTZ-11 (in German).}
The phrasing used by Born referred to who should `explain orally the contents of the report', suggesting that the 
speakers did not present the exact or full text of the reports as printed. 

Multiplicity of languages had long been a characteristic of the Solvay conferences. A well-known letter by 
Ehrenfest\endnote{Ehrenfest to Goudsmit, Uhlenbeck and Dieke, 3 November 1927, AHQP-61 (in German).} tells us
of `[p]oor Lorentz as interpreter between the British and the French who were 
absolutely unable to understand each other. Summarising Bohr. And Bohr responding with polite despair' (as quoted in 
Bohr 1985, p.~38).\footnote{Both W.~H.~Bragg and Planck deplored in letters to Lorentz that they 
were very poor linguists. Indeed,
in a letter explaining in more detail why he would not participate in the conference, W.~H.~Bragg wrote: `I find it 
impossible to follow the discussions even though you so often try to make it easy for us', and Planck was in doubt 
about coming, particularly because of the language difficulties.\endnotemark}\endnotetext{W.~H.~Bragg to Lorentz,
17 July [1927], AHQP-LTZ-11 (in English), Planck to Lorentz, 2 February 1927, AHQP-LTZ-8 (in German).} 

On the last day of the conference, Ehrenfest went to the blackboard and evoked the image of the tower of Babel (presumably 
in a more metaphorical sense than the mere multiplicity of spoken languages), writing:
  \begin{quotation}
    And they said one to another: .... Go to, let us build us .... a tower, whose
    top may reach unto heaven; and {\em let us make us a name} .... And the Lord said: .... Go
    to, let us go down, and there confound their language, that they may not
    understand one another's speech. (Genesis 11: 3--7, reported by Pelseneer, his emphasis\footnote{This may 
    have been Ehrenfest's own emphasis. Note that, if not necessarily present at the sessions, Pelseneer had some 
    connection with the conference, having taken Lorentz's photograph reproduced in the 
    proceedings.\endnotemark}\endnotetext{See Pelseneer, pp.~50--51.})
  \end{quotation}

Informal discussions at the conference must have been plentiful, but information about them has to be
gathered from other sources. Famously, Einstein and Bohr engaged in discussions that were described in 
detail in later recollections by Bohr (1949), and vividly
recalled by Ehrenfest within days of the conference in the well-known letter quoted above (see also 
chapter~\ref{Beyond-Bohr-Einstein}).

Little known, if at all, is another reference by Ehrenfest to the discussions between Bohr and Einstein,
which appears to relate more directly to the issues raised by Einstein in the general discussion:
  \begin{quote}
    Bohr had given a very pretty argument in a conversation with Einstein, that one could not hope
    ever to master many-particle problems with three-dimensional Schr\"{o}dinger machinery. He said
    something like the following (more or less!!!!!!): a wave packet can never simultaneously determine 
    EXACTLY the position and the velocity of a particle. Thus if one has for instance TWO particles, then
    they cannot possibly be represented in {\em three-dimensional} space such that one can simultaneously 
    represent exactly their kinetic energy and the potential energy OF THEIR INTERACTION. Therefore.........
    (What comes after this therefore I already cannot reproduce properly.) In the multidimensional representation
    instead the potential energy of the interaction appears totally sharp in the relevant coefficients of the
    wave equation and one does [?] not get to see the kinetic energy at 
    all.\endnote{Ehrenfest to Kramers, 6 November 1927, AHQP-9, section 10 (in German). The words `bekommt man' [?] are
    very faint.}   
  \end{quote}

\section{The editing of the proceedings}\label{editing}
The editing of the proceedings of the fifth Solvay conference was largely
carried out by Verschaffelt, who reported regularly to Lorentz and to Lef\'{e}bure. During the last months of 1927 
Lorentz was busy writing up the lecture he had given at the Como meeting in September.\endnote{Lorentz 
to Lef\'{e}bure, 30 December 1927, IIPCS 2670A/B (in French).} He then died suddenly on 4 February, 
before the editing work was complete. 


The translation of the reports into French was carried out after the conference, except for de~Broglie's report which, 
as mentioned, was written directly in French. 
From Verschaffelt's letters we gather that by 6 January 1928 all the reports had been translated, Bragg's and Compton's 
had been sent to the publishers, and Born and Heisenberg's and Schr\"{o}dinger's were to be sent on that day or 
the next. Several proofs were back by the beginning of March.\endnote{Verschaffelt to Lef\'{e}bure, 6 January 1928, 
IIPCS 2609, 2 March 1928, IIPCS 2610 (both in French).} 

Lorentz had envisaged preparing with Verschaffelt an edited version of the discussions from notes taken during the 
conference, and sending the edited version to the speakers at proof stage.\footnote{According to D.~Devriese, curator 
of the IIPCS archives, the original notes have not survived.}   
In fact, stenographed notes appear to have been taken, typed up and sent to the 
speakers, who for the most part used them to prepare drafts of their contributions. From these, Verschaffelt then edited 
the final version, with some help from Kramers (who specifically completed two of Lorentz's 
contributions).\endnote{Lef\'{e}bure to Lorentz, [after 27 August 1927], IIPCS 2524 (in French), Bohr to Kramers,
17 February 1928, AHQP-BSC-13 (in Danish), Brillouin to Lorentz, 31 December 1927, AHQP-LTZ-8 (in French), and Kramers 
to Verschaffelt, 28 March 1928, AHQP-28, section 4 (in Dutch).} A copy of the galley proofs of the general discussion, 
dated 1 June 1928, survives in the Bohr archives in Copenhagen,\endnote{Microfilmed in AHQP-BMSS-11, section 5.} and 
includes some contributions that appear to have been still largely unedited at that time.\footnote{We have reproduced 
some of this material in the endnotes to the general discussion.}

By January, the editing of the discussions was proceeding well, and at the beginning of March it was almost 
completed. Some contributions, however, were still missing, most notably Bohr's. The notes sent by Verschaffelt
had many gaps; Bohr wanted Kramers's advice and help with the discussion contributions, and travelled to Utrecht 
for this purpose at the beginning of March.\endnote{Bohr to Kramers, 17 and 27 February 1928, AHQP-BSC-13 
(both in Danish).} At the end of March, 
Kramers sent Verschaffelt the edited version of Bohr's contributions to the discussion after Compton's report
(pp.~\pageref{forpage170} and \pageref{Bohr-disc2}) and after Born and Heisenberg's report (p.~\pageref{Bohr-BH}),
remarking that these were all of Bohr's contributions to the discussions during the first three days of the 
conference.\footnote{Note that Bohr also asked some brief questions after Schr\"{o}dinger's report 
(p.~\pageref{Bohr-Schr}). Kramers further writes\label{forpage54} that Bohr suggested to `omit the whole final Born-Heisenberg discussion 
(Nr.~18--23) and equally Fowler's remark 9'. Again, thanks to Mark van Atten for help with this 
letter.\endnotemark}\endnotetext{Kramers to Verschaffelt, 28 March 1928, AHQP-28, section 4 (in Dutch).} 

In contrast, material on Bohr's contributions to the general discussion survives only in the form of
notes in the Bohr archives.\footnote{This material is not microfilmed in AHQP. See also Bohr 
(1985, pp.~35--7, 100 and 478--9).} (Some notes by Richardson also relate to Bohr's contributions.\endnote{Included 
with the copy of Born and Heisenberg's report in AHQP-RDN, document M-0309.}) 
A translation of version of the Como lecture for {\em Naturwissenschaften} (Bohr 1928) was included instead, 
reprinted on a par with the other reports, and accompanied by the following footnote (p.~215 of the
published proceedings):
  \begin{quote}
    This article, which is the translation of a note published very recently in {\em Naturwissenschaften}, vol.~16, 1928, 
    p.~245, has been added at the author's request to replace the exposition of his ideas that he gave in the course of 
    the following general discussion. It is essentially the reproduction of a talk on the current state of quantum 
    theory that was given in Como on 16 September 1927, on the occasion of the jubilee festivities in honour of Volta.  
  \end{quote}

The last remaining material was sent to Gauthier-Villars sometime in September 1928, and the volume 
was finally published in early December of that year.\endnote{Gauthier-Villars to Verschaffelt, 6 December 
1928, IIPCS 2762 (in French), and Verschaffelt to Lef\'{e}bure, 11 [December] 1928, IIPCS 2761 (in French).}

\section{Conclusion}
The fifth Solvay conference was by any standards an important and memorable event. On this point all participants 
presumably agreed, as shown by numerous letters, such as Ehrenfest's letter quoted above (reproduced in Bohr 1985), 
Heisenberg's letter to Lef\'{e}bure at the head of this chapter, or various other letters of thanks addressed to the 
organisers after the conference:\endnote{IIPCS 2671 (in English), IIPCS 2672 (in German).}
  \begin{quote}
    I would like to take this opportunity of thanking you for your kind hospitality, and telling you how much I enjoyed 
    this particular Conference. I think it has been the most memorable one which I have attended for the subject which was 
    discussed was of such vital interest and I learned so much. (W.~L.~Bragg to Armand Solvay, 
    3 November 1927)

    It was the most stimulating scientific meeting I have ever taken part in. (Max Born to Charles Lef\'{e}bure, 
    8 November 1927)
  \end{quote}
Perceptions of the significance of the conference differed from each other, however. In the official history, the fifth 
Solvay conference went down (perhaps together with the Como meeting) as the occasion on which the interpretational issues 
were finally clarified. This was presumably a genuine sentiment on the part of 
Bohr, Heisenberg and the other physicists of the Copenhagen-G\"{o}ttingen school. We find it explicitly as early as 1929:
  \begin{quote}
    In relating the development of the quantum theory, one must in particular not forget the discussions at the Solvay
    conference in Brussels in 1927, chaired by Lorentz. Through the possibility of exchange [Aussprache] between the
    representatives of different lines of research, this conference has contributed extraordinarily to the clarification
    of the physical foundations of the quantum theory; it forms so to speak the outward completion of the quantum 
    theory~....~. (Heisenberg 1929, p.~495)
  \end{quote}
On the other hand, the conference was also described (by Langevin) as the one where `the confusion of ideas reached 
its peak'.\endnote{\ Quoted in Pelseneer, p.~50.} From a distance of almost 80 years, the beginnings of a more dispassionate 
evaluation should be possible. In the following chapters, we shall revisit the fifth Solvay conference, focussing in 
particular on the background and contributions relating to the three main `lines of research' into quantum
theory represented there: de Broglie's pilot-wave theory, Born and Heisenberg's quantum mechanics 
and Schr\"{o}dinger's wave mechanics.


\newpage

\renewcommand{\enoteheading}{\section*{Archival notes}}
\addcontentsline{toc}{section}{\em Archival notes}
\theendnotes

\setcounter{endnote}{0}
\setcounter{equation}{0}

\chapter{De Broglie's pilot-wave theory}\label{deBroglieEss}
\chaptermark{De Broglie's pilot-wave theory}

\section{Background}\label{background}

\epigraph{At a time when no single known fact supported this theory, Louis de Broglie
asserted that a stream of electrons which passed through a very small hole in
an opaque screen must exhibit the same phenomena as a light ray under the same 
conditions.}{Prof.\ C. W. Oseen, Chairman of the Nobel Committee for Physics, 
presentation speech, 12 December 1929 (Oseen 1999)}

\noindent In September 1923, Prince Louis de Broglie\footnote{The de Broglies had come
to France from Italy in the seventeenth century, the original name `Broglia'
eventually being changed to de Broglie. On his father's side, de Broglie's
ancestors included dukes, princes, ambassadors, and marshals of France. Nye
(1997) considers the conflict between de Broglie's pursuit of science and the
expectations of his aristocratic family. For a biography of de Broglie, see
Lochak (1992).} made one of the most astonishing predictions in the history of
theoretical physics: that material bodies would exhibit the wave-like
phenomena of diffraction and interference upon passing through sufficiently
narrow slits. Like Einstein's prediction of the deflection of light by the
sun, which was based on a reinterpretation of gravitational force in terms of
geometry, de Broglie's prediction of the deflection of electron paths by
narrow slits was made on the basis of a fundamental reappraisal of the nature
of forces and of dynamics. De Broglie had proposed that Newton's first law of
motion be abandoned, and replaced by a new postulate, according to which a
freely moving body follows a trajectory that is orthogonal to the surfaces of
equal phase of an associated guiding wave. The resulting `de Broglian
dynamics' --- or pilot-wave theory as de Broglie later called it --- was a new
approach to the theory of motion, as radical as Einstein's interpretation of
the trajectories of falling bodies as geodesics of a curved spacetime, and as
far-reaching in its implications. In 1929 de Broglie received the Nobel Prize,
`for his discovery of the wave nature of electrons'.

Strangely enough, however, even though de Broglie's prediction was confirmed
experimentally a few years later, for most of the twentieth century
single-particle diffraction and interference were routinely cited as evidence
\textit{against} de Broglie's ideas: even today, some textbooks on quantum
mechanics assert that such interference demonstrates that particle
trajectories cannot exist in the quantum domain (see section~\ref{prob-intrfce}). 
It is as if the deflection of light by the sun had come to be
widely regarded as evidence against Einstein's general theory of relativity.
This remarkable misunderstanding illustrates the extent to which de Broglie's
work in the 1920s has been underestimated, misrepresented, and indeed largely
ignored, not only by physicists but also by historians.

De Broglie's PhD thesis of 1924 is of course recognised as a landmark in the
history of quantum theory. But what is usually remembered about that thesis is
the proposed extension of Einstein's wave-particle duality from light to
matter, with the formulas $E=h\nu$ and $p=h/\lambda$ (relating energy and
momentum to frequency and wavelength) being applied to electrons or `matter
waves'. Usually, little attention is paid to the fact that a central theme of
de Broglie's thesis was the construction of a new form of dynamics, in which
classical (Newtonian or Einsteinian) laws are abandoned, and replaced by new
laws according to which particle \textit{velocities} are determined by guiding
waves, in a specific manner that unifies the variational principles of
Maupertuis and Fermat. Nor, indeed, have historians paid much attention to de
Broglie's later and more complete form of pilot-wave dynamics, which he
arrived at in a paper published in May 1927 in \textit{Journal de Physique},
and which he then presented in October 1927 at the fifth Solvay conference.

Unlike the other main contributors to quantum theory, de Broglie worked in
relative isolation, having little contact with the principal research centres
in Berlin, Copenhagen, G\"{o}ttingen, Cambridge and Munich. While Bohr,
Heisenberg, Born, Schr\"{o}dinger, Pauli and others visited each other
frequently and corresponded regularly, de Broglie worked essentially alone in
Paris.\footnote{In his typed `Replies to Mr Kuhn's questions' in the Archive
for the History of Quantum Physics,\endnotemark de Broglie writes
(p.~7): `Between 1919 and 1928, I worked \textit{very much in isolation}'
(emphasis in the original). Regarding his Ph.D.\ thesis de Broglie recalled (p.~9): 
`I worked very much alone and almost without any exchange of 
ideas'.}\endnotetext{AHQP-OHI, Louis de Broglie, `Replies to Mr Kuhn's questions'.} 
In France at the time, while pure mathematics was well represented, there was
very little activity in theoretical physics. In addition, after the first
world war, scientific relations with Germany and Austria were
interrupted.\footnote{For more details on de Broglie's situation in France at
the time, see Mehra and Rechenberg (1982a, pp.~578--84).} All this seems to
have suited de Broglie's rather solitary temperament. De Broglie's isolation,
and the fact that France was outside the mainstream of theoretical physics,
may account in part for why so much of de Broglie's work went relatively
unnoticed at the time, and has remained largely ignored even to the present day.

For some seventy years, the physics community tended to believe either that
`hidden-variables' theories like de Broglie's were impossible, or that such
theories had been disproven experimentally. The situation changed considerably
in the 1990s, with the publication of textbooks presenting quantum mechanics
in the pilot-wave formulation (Bohm and Hiley 1993; Holland 1993). Pilot-wave
theory --- as originated by de Broglie in 1927, and
elaborated by Bohm 25 years later (Bohm 1952a,b) --- is now accepted as an
alternative (if little used) formulation of quantum theory.

Focussing for simplicity on the nonrelativistic quantum theory of a system of
$N$ (spinless) particles with 3-vector positions $\mathbf{x}_{i}$ ($i=1,2,...,N$), 
it is now generally agreed that, with appropriate initial
conditions, quantum physics may be accounted for by the deterministic dynamics
defined by two differential equations, the Schr\"{o}dinger equation%
\begin{equation}
i\hbar\frac{\partial\Psi}{\partial t}=\sum_{i=1}^{N}-\frac{\hbar^{2}}{2m_{i}%
}\nabla_{i}^{2}\Psi+V\Psi\label{Sch}%
\end{equation}
for a `pilot wave' $\Psi(\mathbf{x}_{1},\mathbf{x}_{2},...,\mathbf{x}_{N},t)$
in configuration space, and the de Broglie guidance equation%
\begin{equation}
m_{i}\frac{d\mathbf{x}_{i}}{dt}=\mathbf{\nabla}_{i}S \label{deB}%
\end{equation}
for particle trajectories $\mathbf{x}_{i}(t)$, where the phase $S(\mathbf{x}%
_{1},\mathbf{x}_{2},...,\mathbf{x}_{N},t)$ is locally defined by
$S=\hbar\operatorname{Im}\ln\Psi$ (so that $\Psi=\left\vert \Psi\right\vert
e^{(i/\hbar)S}$).

This, as we shall see, is how de Broglie presented his dynamics in 1927.
Bohm's presentation of 1952 was somewhat different. If one takes the time
derivative of (\ref{deB}), then using (\ref{Sch}) one obtains Newton's law of
motion for acceleration%
\begin{equation}
m_{i}\mathbf{\ddot{x}}_{i}=-\mathbf{\nabla}_{i}(V+Q)\ , \label{B52}%
\end{equation}
where%
\begin{equation}
Q\equiv-\sum_{i}\frac{\hslash^{2}}{2m_{i}}\frac{\nabla_{i}^{2}\left\vert
\Psi\right\vert }{\left\vert \Psi\right\vert } \label{QQ}%
\end{equation}
is the `quantum potential'. Bohm regarded (\ref{B52}) as the law of motion,
with (\ref{deB}) added as a constraint on the initial momenta, a constraint
that Bohm thought could be relaxed (see section~\ref{hist-mis}). For de Broglie, 
in contrast, the law of motion (\ref{deB})
for velocity had a fundamental status, and for him represented the unification
of the principles of Maupertuis and Fermat. One should then distinguish
between de Broglie's first-order (velocity-based) dynamics of 1927, and Bohm's
second-order (acceleration-based) dynamics of 1952.

In this chapter, we shall be concerned with the historical origins of de
Broglie's 1927 dynamics defined by (\ref{Sch}) and (\ref{deB}). Some authors
have referred to this dynamics as `Bohmian mechanics'. Such terminology is
misleading: it disregards de Broglie's priority, and misses de Broglie's
physical motivations for recasting dynamics in terms of velocity; it also
misrepresents Bohm's 1952 formulation, which was based on (\ref{Sch}) and
(\ref{B52}). These and other historical misconceptions concerning de
Broglie-Bohm theory will be addressed in section~\ref{hist-mis}.

The two equations (\ref{Sch}), (\ref{deB}) define a deterministic (de Broglian
or pilot-wave) dynamics for a single multiparticle system: given an initial
wave function $\Psi(\mathbf{x}_{1},\mathbf{x}_{2},...,\mathbf{x}_{N},0)$ at
$t=0$, (\ref{Sch}) determines $\Psi(\mathbf{x}_{1},\mathbf{x}_{2}%
,...,\mathbf{x}_{N},t)$ at all times $t$; and given an initial configuration
$(\mathbf{x}_{1}(0),\mathbf{x}_{2}(0),...,\mathbf{x}_{N}(0))$, (\ref{deB})
then determines the trajectory $(\mathbf{x}_{1}(t),\mathbf{x}_{2}%
(t),...,\mathbf{x}_{N}(t))$. For an ensemble of systems with the same 
initial wave function $\Psi(\mathbf{x}_{1},\mathbf{x}_{2},...,\mathbf{x}%
_{N},0)$, and with initial configurations $(\mathbf{x}_{1}(0),\mathbf{x}%
_{2}(0),...,\mathbf{x}_{N}(0))$ distributed according to the Born rule%
\begin{equation}
P(\mathbf{x}_{1},\mathbf{x}_{2},...,\mathbf{x}_{N},0)=\left\vert
\Psi(\mathbf{x}_{1},\mathbf{x}_{2},...,\mathbf{x}_{N},0)\right\vert ^{2}\ ,
\label{Born0}%
\end{equation}
the statistical distribution of outcomes of quantum measurements will agree
with the predictions of standard quantum theory. This is shown by treating the
measuring apparatus, together with the system being measured, as a single
multiparticle system obeying de Broglian dynamics, so that $(\mathbf{x}%
_{1},\mathbf{x}_{2},...,\mathbf{x}_{N})$ defines the `pointer position' of
the apparatus as well as the configuration of the measured system. Given the
initial condition (\ref{Born0}) for any multiparticle system, the statistical
distribution of particle positions at later times will also agree with the
Born rule $P=\left\vert \Psi\right\vert ^{2}$. Thus, the statistical
distribution of pointer positions in any experiment will agree with the
predictions of quantum theory, yielding the correct statistical distribution
of outcomes for standard quantum measurements.

In his 1927 Solvay report, de Broglie gave some simple applications of
pilot-wave theory, with the assumed initial condition (\ref{Born0}). He
applied the theory to single-photon interference, to atomic transitions, and
to the scattering (or diffraction) of electrons by a crystal lattice. But a
detailed demonstration of equivalence to quantum theory, and in particular a
pilot-wave account of the general quantum theory of measurement, was not
provided until the work of Bohm in 1952.

How did de Broglie come to propose this theory in 1927? In this chapter, we
trace de Broglie's work in this direction, from his early work leading to his
doctoral thesis of 1924 (de Broglie 1924e, 1925), to his crucial paper of 1927
published in \textit{Journal de Physique} (de Broglie 1927b), and culminating
in his presentation of pilot-wave theory at the fifth Solvay conference. We examine in detail how de Broglie arrived at this new form of
particle dynamics, and what his attitude towards it was. Later, in chapter~\ref{meas-in-pwt}, 
we shall consider some of the discussions of de
Broglie's theory that took place at the conference, in particular the famous
(and widely misunderstood) clash between de Broglie and Pauli.

De Broglie's dynamics has the striking feature that electrons and photons are
regarded as both particles and waves. Like many scientific ideas, this
mingling of particle-like and wave-like aspects had precursors. In Newton's
\textit{Opticks} (first published in 1704), both wave-like and particle-like
properties are attributed to light. Newton's so-called `corpuscular' theory
was formulated on the basis of extensive and detailed experiments (carried out
by Grimaldi, Hooke, and Newton himself) involving what we would now call
interference and diffraction. According to Newton, light corpuscles --- or
light `Rays' as he called them\footnote{The opening definition of the
\textit{Opticks} defines `Rays' of light as `its least Parts'.} --- generate
`Waves of Vibrations' in an `Aethereal Medium', much as a stone thrown into
water generates water waves (Newton 1730; reprint, pp.~347--9). In addition,
Newton supposed that the waves in turn affect the motion of the corpuscles,
which `may be alternately accelerated and retarded by the Vibrations' (p.~348). 
In particular, Newton thought that the effect of the medium on the
motion of the corpuscles was responsible for the phenomena of interference and
diffraction. He writes, for example (p.~350):
  \begin{quotation}
    And doth not the gradual condensation of this Medium extend to some distance
    from the Bodies, and thereby cause the Inflexions of the Rays of Light, which
    pass by the edges of dense Bodies, at some distance from the Bodies?
  \end{quotation}
Newton understood that, for diffraction to occur, the motion of the
light corpuscles would have to be affected at a distance by the diffracting
body --- `Do not Bodies act upon Light at a distance, and by their action bend
its Rays .... ?' (p.~339) --- and his proposed mechanism involved waves in an
inhomogeneous ether. Further, according to Newton, to account for the coloured
fringes that had been observed by Grimaldi in the diffraction of white light
by opaque bodies, the corpuscles would have to execute an oscillatory motion
`like that of an Eel' (p.~339):

\begin{quotation}
Are not the Rays of Light in passing by the edges and sides of Bodies, bent
several times backwards and forwards, with a motion like that of an Eel? And
do not the three Fringes of colour'd Light above-mention'd arise from three
such bendings?
\end{quotation}
For Newton, of course, such non-rectilinear motion could be caused only
by a force emanating from the diffracting body.

It is interesting to note that, in the general discussion at the fifth Solvay
conference (p.~\pageref{oldp193}), de Broglie commented on this very
point, with reference to the `emission' (or corpuscular) theory, and pointed
out that if pilot-wave dynamics were written in terms of acceleration (as done
later by Bohm) then just such forces appeared:

\begin{quotation}
In the corpuscular conception of light, the existence of diffraction phenomena
occuring at the edge of a screen requires us to assume that, in this case, the
trajectory of the photons is curved. The supporters of the emission theory
said that the edge of the screen exerts a force on the corpuscle. Now, if in
the new mechanics as I develop it, one writes the Lagrange equations for the
photon, one sees appear on the right-hand side of these equations a term ....
[that] .... represents a sort of force of a new kind, which exists only ....
where there is interference. It is this force that will curve the trajectory
of the photon when its wave $\psi$ is diffracted by the edge of a screen.
\end{quotation}
The striking similarity between Newton's qualitative ideas and
pilot-wave theory has also been noted by Berry, who remarks that during
interference or diffraction the de Broglie-Bohm trajectories indeed `wriggle
like an eel' (Berry 1997, p.~42), in some sense vindicating Newton.

A mathematical precursor to de Broglian dynamics is found in the early
nineteenth century, in Hamilton's formulation of geometrical optics and
particle mechanics. As de Broglie points out in his Solvay report (pp.~\pageref{oldp86} 
and \pageref{oldp90}), Hamilton's theory is in fact the
short-wavelength limit of pilot-wave dynamics: for in that limit, the phase of
the wave function obeys the Hamilton-Jacobi equation, and de Broglie's
trajectories reduce to those of classical mechanics.

A physical theory of light as both particles and waves --- in effect a revival
of Newton's views --- emerged again with Einstein in 1905. It is less well
known that, after 1905, Einstein tried to construct theories of localised
light quanta coupled to vector fields in 3-space. As we shall see in 
chapter~\ref{guiding-fields-in-3-space}, Einstein's ideas in this vein show some
resemblance to de Broglie's but also differ from them.

It should also be mentioned that, in the autumn of 1923 (the same year in
which de Broglie first elaborated his ideas), Slater tried to develop a theory
in which the motion of photons was guided by the electromagnetic field. It
appears that Slater first attempted to construct a deterministic theory, but
had trouble defining an appropriate velocity vector; he then came to the
conclusion that photons and the electromagnetic field were related only
statistically, with the photon probability density being given by the
intensity of the field. After discussing his ideas with Bohr and Kramers in
1924, the photons were removed from the theory, apparently against Slater's
wishes (Mehra and Rechenberg 1982a, pp.~542--6). Note that, while de Broglie
applied his theory to photons, he made it clear (for example in the general
discussion, p.~\pageref{p192DRAFT1}) that in his theory the guiding `$\psi
$-wave' was distinct from the electromagnetic field.

In the case of light, then, the idea of combining both particle-like and
wave-like aspects was an old one, going back indeed to Newton. In the case of
ordinary matter, however, de Broglie seems to have been the first to develop a
physical theory of this form.

It is sometimes claimed that, for the case of electrons, ideas similar to de
Broglie's were put forward by Madelung in 1926. What Madelung proposed,
however, was to regard an electron with mass $m$ and wave function $\psi$ not
as a pointlike particle within the wave, but as a continuous fluid spread over
space with mass density $m\left\vert \psi\right\vert ^{2}$ (Madelung 1926a,b).
In this `hydrodynamical' interpretation, mathematically the fluid velocity
coincides with de Broglie's velocity field; but physically, Madelung's theory
seems more akin to Schr\"{o}dinger's theory than to de Broglie's.

Finally, before we examine de Broglie's work, we note what appears to be a
recurring historical opposition to dualistic physical theories containing both
waves and particles. In 1801--03, Thomas Young, who by his own account
regarded his theory as a development of Newton's ideas,\footnote{See, for
example, Bernard Cohen's preface to Newton's {\em Opticks} (1730, reprint).} 
removed the corpuscles from Newton's theory
and produced a purely undulatory account of light. In 1905, Einstein's dualist
view of light was not taken seriously, and did not win widespread support
until the discovery of the Compton effect in 1923. In 1924, Bohr and Kramers,
who regarded the Bohr-Kramers-Slater theory as a development of Slater's
original idea, insisted on removing the photons from Slater's theory of
radiation.\footnote{In 1925, Born and Jordan attempted to restore the photons,
proposing a stochastic theory reminiscent of Slater's original ideas; it
appears that they were dissuaded from publication by Bohr. See Darrigol (1992, p.~253) and
section~\ref{BornBohr}.} And in 1926,
Schr\"{o}dinger, who regarded his work as a development of de Broglie's ideas,
removed the trajectories from de Broglie's theory and produced a purely
undulatory `wave mechanics'.\footnote{Cf. section~\ref{Schr-deB}.}

\section{A new approach to particle dynamics: 1923--24}

In this section we show how de Broglie took his first steps towards a new form
of dynamics.\footnote{An insightful and general account of de Broglie's early
work, up to 1924, has been given by Darrigol (1993).} His aim was to explain
the quantum phenomena known at the time --- in particular the Bohr-Sommerfeld
quantisation of atomic energy levels, and the apparently dual nature of
radiation --- by unifying the physics of particles with the physics of waves.
To accomplish this, de Broglie began by extending Einstein's wave-particle
duality for light to all material bodies, by introducing a `phase wave'
accompanying every material particle. Then, inspired by the optical-mechanical
analogy,\footnote{Possibly, de Broglie was also influenced by the philosopher
Henri Bergson's writings concerning time, continuity and motion (Feuer 1974,
pp.~206--14); though this is denied by Lochak (1992).} de Broglie proposed
that Newton's first law of motion should be abandoned, and replaced by a new
principle that unified Maupertuis' variational principle for mechanics with
Fermat's variational principle for optics. The result was a new form of
dynamics in which the velocity $\mathbf{v}$ of a particle is determined by the
gradient of the phase $\phi$ of an accompanying wave --- in contrast with
classical mechanics, where accelerations are determined by forces. (Note that
de Broglie's phase $\phi$ has a sign opposite to the phase $S$ as we would
normally define it now.)

This new approach to dynamics enabled de Broglie to obtain a wave-like
explanation for the quantisation of atomic energy levels, to explain the
observed interference of single photons, and to predict for the first time the
new and unexpected phenomenon of the diffraction and interference of electrons.

As we shall see, the theory proposed by de Broglie in 1923--24 was, in fact, a
simple form of pilot-wave dynamics, for the special case of independent
particles guided by waves in 3-space, and without a specific wave equation.

\subsection{First papers on pilot-wave theory (1923)}\label{first-papers}

De Broglie's earliest experience of physics was closely tied to experiment.
During the first world war he worked on wireless telegraphy, and after the war
his first papers concerned X-ray spectroscopy. In 1922 he published a paper
treating blackbody radiation as a gas of light quanta (de Broglie 1922). In
this paper, de Broglie made the unusual assumption that photons had a very
small but non-zero rest mass $m_{0}$. He was therefore now applying Einstein's
relations $E=h\nu$ and $p=h/\lambda$ (relating energy and momentum to
frequency and wavelength) to \textit{massive} particles, even if these were
still only photons. It seems that de Broglie made the assumption $m_{0}\neq0$
so that light quanta could be treated in the same way as ordinary material
particles. It appears that this paper was the seed from which de Broglie's
subsequent work grew.\footnote{In a collection of papers by de Broglie and
Brillouin, published in 1928, a footnote added to de Broglie's 1922 paper on
blackbody radiation and light quanta remarks: `This paper .... was the origin
of the ideas of the author on wave mechanics' (de Broglie and Brillouin 1928,
p.~1).}

According to de Broglie's later recollections (L. de Broglie, AHQP interview,
7 January 1963, p.~1),\endnote{AHQP-OHI, Louis de Broglie, session 1 (tape 43a), 
7 January 1963, Paris; 0.75 h, in French, 13~pp.; by T. S. Kuhn, T. Kahan and A. George.} 
his first ideas concerning a pilot-wave theory of massive particles arose as follows.
During conversations on the subject of X-rays with his older brother Maurice
de Broglie,\footnote{Maurice, the sixth duc de Broglie, was a distinguished
experimental physicist, having done important work on the photoelectric effect
with X-rays --- experiments that were carried out in his private laboratory in
Paris.} he became convinced that X-rays were both particles and waves. Then,
in the summer of 1923, de Broglie had the idea of extending this duality to
ordinary matter, in particular to electrons. He was drawn in this direction by
consideration of the optical-mechanical analogy; further, the presence of
whole numbers in quantisation conditions suggested to him that waves must be involved.

This last motivation was recalled by de Broglie (1999) in his Nobel lecture of 1929:

\begin{quotation}
.... the determination of the stable motions of the electrons in the atom
involves whole numbers, and so far the only phenomena in which whole numbers
were involved in physics were those of interference and of eigenvibrations.
That suggested the idea to me that electrons themselves could not be
represented as simple corpuscles either, but that a periodicity had also to be
assigned to them too.
\end{quotation}

De Broglie first presented his new ideas in three notes published (in French)
in the \textit{Comptes Rendus} of the Academy of Sciences in Paris (de Broglie
1923a,b,c), and also in two papers published in English --- one in
\textit{Nature} (de Broglie 1923d), the other in the \textit{Philosophical Magazine} 
(de Broglie 1924a).\footnote{It seems possible that
\textit{Comptes Rendus} was not widely read by physicists outside France, but
this certainly was not true of \textit{Nature} or the \textit{Philosophical Magazine}.}

The ideas in these papers formed the basis for de Broglie's doctoral thesis.
The paper in the \textit{Philosophical Magazine} reads, in fact, like
a summary of much of the material in the thesis. Since the thesis provides a
more systematic presentation, we shall give a detailed summary of it in the
next subsection; here, we give only a brief account of the earlier papers,
except for the crucial second paper, whose conceptual content warrants more
detailed commentary.\footnote{We do not always keep to de Broglie's original
notation.}

The first communication (de Broglie 1923a), entitled `Waves and quanta',
proposes that an `internal periodic phenomenon' should be associated with
\textit{any} massive particle (including light quanta). In the rest frame of a
particle with rest mass $m_{0}$, the periodic phenomenon is assumed to have a
frequency $\nu_{0}=m_{0}c^{2}/h$. In a frame where the particle has uniform
velocity $v$, de Broglie considers the two frequencies $\nu$ and $\nu_{1}$,
where $\nu=\nu_{0}/\sqrt{1-v^{2}/c^{2}}$ is the frequency $\nu=mc^{2}/h$
associated with the relativistic mass increase $m=m_{0}/\sqrt{1-v^{2}/c^{2}}$
and $\nu_{1}=\nu_{0}\sqrt{1-v^{2}/c^{2}}$ is the time-dilated frequency. De
Broglie shows that, because $\nu_{1}=\nu(1-v^{2}/c^{2})$, a `fictitious' wave
of frequency $\nu$ and phase velocity $v_{\mathrm{ph}}=c^{2}/v$ (propagating
in the same direction as the particle) will remain in phase with the internal
oscillation of frequency $\nu_{1}$. De Broglie then considers an atomic
electron moving uniformly on a circular orbit. He proposes that orbits are
stable only if the fictitious wave remains in phase with the internal
oscillation of the electron. From this condition, de Broglie derives the
Bohr-Sommerfeld quantisation condition.

The second communication (de Broglie 1923b), entitled `Light quanta,
diffraction and interference', has a more conceptual tone. De Broglie begins
by recalling his previous result, that a moving body must be associated with
`a \textit{non-material} sinusoidal wave'. He adds that the particle velocity
$v$ is equal to the group velocity of the wave, which de Broglie here calls
`the phase wave' because its phase at the location of the particle is equal to
the phase of the internal oscillation of the particle. De Broglie then goes on
to make some very significant observations about diffraction and the nature of
the new dynamics that he is proposing.

De Broglie asserts that diffraction phenomena prove that light quanta cannot
always propagate in a straight line, even in what would normally be called
empty space. He draws the bold conclusion that Newton's first law of motion
(the `principle of inertia') must be abandoned (p.~549):

\begin{quotation}
The light quanta [atomes de lumi\`{e}re] whose existence we assume do not
always propagate in a straight line, as proved by the phenomena of
diffraction. It then seems \textit{necessary} to modify the principle of inertia.
\end{quotation}
De Broglie then suggests replacing Newton's first law with a new
postulate (p.~549):

\begin{quotation}
We propose to adopt the following postulate as the basis of the dynamics of
the free material point: `At each point of its trajectory, a free moving body
follows in a uniform motion the \textit{ray} of its phase wave, that is (in an
isotropic medium), the normal to the surfaces of equal phase'.
\end{quotation}
The diffraction of light quanta is then explained since, as de Broglie
notes, `if the moving body must pass through an opening whose dimensions are
small compared to the wavelength of the phase wave, in general its trajectory
will curve like the ray of the diffracted wave'.

In retrospect, de Broglie's postulate for free particles may be seen as a
simplified form of the law of motion of what we now know as pilot-wave
dynamics --- except for the statement that the motion along a ray be `uniform'
(that is, have constant speed), which in pilot-wave theory is true only in
special cases.\footnote{From Bohm's second-order equation (\ref{B52}) applied
to a single particle, for time-independent $V$ it follows that $d(\frac{1}%
{2}mv^{2}+V+Q)/dt=\partial Q/\partial t$ (the usual energy conservation
formula with a time-dependent contribution $Q$ to the potential). In free
space ($V=0$), the speed $v$ is constant if and only if $dQ/dt=\partial
Q/\partial t$ or $\mathbf{v}\cdot\mathbf{\nabla}Q=0$ (so that the `quantum
force' does no work), which is true only in special cases.} De Broglie notes
that his postulate respects conservation of energy but not of momentum. And
indeed, in pilot-wave theory the momentum of a `free' particle is generally
not conserved: in effect (from the standpoint of Bohm's Newtonian
formulation), the pilot wave or quantum potential acts like an `external
source' of momentum (and in general of energy too).\footnote{Again from
(\ref{B52}), in free space the rate of change of momentum $\mathbf{p}%
=m\mathbf{v}$ (where $\mathbf{v}=\mathbf{\nabla}S/m$) is $d\mathbf{p}%
/dt=-\mathbf{\nabla}Q$, which is generally non-zero. Further, in general
(\ref{B52}) implies $d(\frac{1}{2}mv^{2}+V)/dt=-\mathbf{v}\cdot\mathbf{\nabla
}Q$, so that the standard (classical) expression for energy is conserved if
and only if $\mathbf{v}\cdot\mathbf{\nabla}Q=0$. If, on the other hand, one
defines $\frac{1}{2}mv^{2}+V+Q$ as the `energy', it will be conserved if and
only if $\partial Q/\partial t=0$, which is true only for special cases (in
particular for stationary states, since for these $\left\vert \Psi\right\vert
$ is time-independent).} The abandonment of something as elementary as
momentum conservation is certainly a radical step by any standards. On the
other hand, if one is willing --- as de Broglie was --- to propose a
fundamentally new approach to the theory of motion, then the loss of classical
conservation laws is not surprising, as these are really properties of
classical equations of motion.

De Broglie then makes a remarkable prediction, that \textit{any} moving body
(not just light quanta) can undergo diffraction:

\begin{quotation}
.... any moving body could in certain cases be diffracted. A stream of
electrons passing through a small enough opening will show diffraction
phenomena. It is in this direction that one should perhaps look for
experimental confirmation of our ideas.
\end{quotation}

Next, de Broglie puts his proposals in a general conceptual and historical
perspective. Concerning the role of the phase wave, he writes (p.~549):

\begin{quotation}
We therefore conceive of the phase wave as guiding the movements of energy,
and this is what can allow the synthesis of waves and quanta.
\end{quotation}
Here, for the first time, de Broglie characterises the phase wave as a
`guiding' wave. De Broglie then remarks that, historically speaking, the
theory of waves `went too far' by denying the discontinuous structure of
radiation and `not far enough' by not playing a role in dynamics. For de
Broglie, his proposal has a clear historical significance (p.~549, italics in
the original):

\begin{quotation}
\textit{The new dynamics of the free material point is to the old dynamics
(including that of Einstein) what wave optics is to geometrical optics}. Upon
reflection one will see that the proposed synthesis appears as the logical
culmination of the comparative development of dynamics and of optics since the
seventeenth century.
\end{quotation}

In the second part of this note, de Broglie considers the explanation of
optical interference fringes. He assumes that the probability for an atom to
absorb or emit a light quantum is determined by `the resultant of one of the
vectors of the phase waves crossing each other there [se croisant sur lui]'
(pp.~549--50). In Young's interference experiment, the light quanta passing
through the two holes are diffracted, and the probability of them being
detected behind the screen will vary from point to point, depending on the
`state of interference' of the phase waves. De Broglie concludes that there
will be bright and dark fringes as predicted by the wave theories, no matter
how feeble the incident light.

This approach to optical interference --- in which interfering phase waves
determine the probability for interaction between photons and the atoms in the
detection apparatus --- is elaborated in de Broglie's thesis (see below). Soon
after completing his thesis (apparently), de Broglie abandoned this idea in
favour of a simpler approach, in which the interfering phase waves determine
the number density of photon trajectories (see section~\ref{opt-interf}).

In de Broglie's third communication (de Broglie 1923c), entitled `Quanta, the
kinetic theory of gases and Fermat's principle', part 1 considers the
statistical treatment of a gas of particles accompanied by phase waves. De
Broglie makes the following assumption:

\begin{quotation}
The state of the gas will then be stable only if the waves corresponding to
all of the atoms form a system of stationary waves.
\end{quotation}
In other words, de Broglie considers the stationary modes, or standing
waves, associated with a given spatial volume. He assumes that each mode `can
transport zero, one, two or several atoms', with probabilities determined by
the Boltzmann factor.\footnote{As remarked by Pais (1982, pp.~435--6), in this
paper de Broglie `evaluated independently of Bose (and published before him)
the density of radiation states in terms of particle (photon) language'.}
According to de Broglie, for a gas of nonrelativistic atoms his method yields
the Maxwell distribution, while for a gas of photons it yields the Planck
distribution.\footnote{As shown by Darrigol (1993), de Broglie made some
errors in his application of the methods of statistical mechanics.}

In part 2 of the same note, de Broglie shows how his new dynamical postulate
amounts to a unification of Maupertuis' principle of least action with
Fermat's principle of least time in optics. Let us recall that, in the
mechanical principle of Maupertuis for particle trajectories,%
\begin{equation}
\delta\int_{a}^{b}m\mathbf{v}\cdot d\mathbf{x}=0\ , \label{Mau}%
\end{equation}
the condition of stationarity determines the particle paths. (In (\ref{Mau})
the energy is fixed on the varied paths; at the end points, $\Delta
\mathbf{x}=0$ but $\Delta t$ need not be zero.) While in the optical principle
of Fermat for light rays,%
\begin{equation}
\delta\int_{a}^{b}d\phi=0\ , \label{Fer}%
\end{equation}
the stationary line integral for the phase change --- the stationary `optical
path length' --- provides a condition that determines the path of a ray
connecting two points, in space (for the time-independent case) or in
spacetime. Now, according to de Broglie's basic postulate: `The rays of the
phase waves coincide with the dynamically possible trajectories' (p.~632). The
rays are described by Fermat's principle (for the case of a dispersive
medium), which de Broglie shows coincides with Maupertuis' principle, as
follows: writing the element of phase change as $d\phi=2\pi\nu
dl/v_{\mathrm{ph}}$, where $dl$ is an element of path and $v_{\mathrm{ph}%
}=c^{2}/v$ is the phase velocity, and using the relation $\nu=E/h=mc^{2}/h$,
the element of phase change may be rewritten as $(2\pi/h)mvdl$, so that
(\ref{Fer}) coincides with (\ref{Mau}). As de Broglie puts it (p.~632):

\begin{quotation}
In this way the fundamental link that unites the two great principles of
geometrical optics and of dynamics is brought fully to light.
\end{quotation}
De Broglie remarks that some of the dynamically possible trajectories
will be `in resonance with the phase wave', and that these correspond to
Bohr's stable orbits, for which $\int\nu dl/v_{\mathrm{ph}}$ is a whole number.

Soon afterwards, de Broglie introduces a covariant 4-vector formulation of his
basic dynamical postulate (de Broglie 1924a,b). He defines a 4-vector $w_{\mu
}=(\nu/c,-(\nu/v_{\mathrm{ph}})\mathbf{\hat{n}})$, where $\mathbf{\hat{n}}$ is
a unit vector in the direction of a ray of the phase wave, and assumes it to
be related to the energy-momentum 4-vector $p_{\mu}=(E/c,-\mathbf{p})$ by
$p_{\mu}=hw_{\mu}$. De Broglie notes that the identity of the principles of
Maupertuis and Fermat then follows immediately. We shall discuss this in more
detail in the next subsection.

\subsection{Thesis (1924)}\label{thesis}

\epigraph{He has lifted a corner of the great veil.}{Einstein, 
commenting on de Broglie's thesis\footnotemark}\footnotetext{Letter to Langevin, 16
December 1924 (quoted in Darrigol 1993, p.~355).}

\noindent De Broglie's doctoral thesis (de Broglie 1924e) was mostly based on the above
papers. It seems to have been completed in the summer of 1924, and was
defended at the Sorbonne in November. The thesis was published early in 1925
in the \textit{Annales de Physique} (de Broglie 1925).\footnote{An English
translation of extracts from de Broglie's thesis appears in Ludwig (1968). A
complete translation has been done by A. F. Kracklauer (currently online at 
http://www.ensmp.fr/aflb/LDB-oeuvres/De\_Broglie\_Kracklauer.htm~). 
All translations here are ours.}

When writing his thesis, de Broglie was well aware that his theory had gaps.
As he put it (p.~30):\footnote{Here and below, page references for de
Broglie's thesis correspond to the published version in \textit{Annales de
Physique} (de Broglie 1925).}

\begin{quotation}
.... the main aim of the present thesis is to present a more complete account
of the new ideas that we have proposed, of the successes to which they have
led, and also of the many gaps they contain.
\end{quotation}

De Broglie begins his thesis with a historical introduction. Newtonian
mechanics, he notes, was eventually formulated in terms of the principle of
least action, which was first given by Maupertuis and then later in another
form by Hamilton. As for the science of light and optics, the laws of
geometrical optics were eventually summarised by Fermat in terms of a
principle whose form is reminiscent of the principle of least action. Newton
tried to explain some of the phenomena of wave optics in terms of his
corpuscular theory, but the work of Young and Fresnel led to the rise of the
wave theory of light, in particular the successful wave explanation of the
rectilinear propagation of light (which had been so clear in the corpuscular
or `emission' theory). On this, de Broglie comments (p.~25):

\begin{quotation}
When two theories, based on ideas that seem entirely different, account for
the same experimental fact with equal elegance, one can always wonder if the
opposition between the two points of view is truly real and is not due solely
to an inadequacy of our efforts at synthesis.
\end{quotation}
This remark is, of course, a hint that the aim of the thesis is to
effect just such a synthesis. De Broglie then turns to the rise of
electrodynamics, relativity, and the theory of energy quanta. He notes that
Einstein's theory of the photoelectric effect amounts to a revival of Newton's
corpuscular theory. De Broglie then sketches Bohr's 1913 theory of the atom,
and goes on to point out that observations of the photoelectric effect for X-
and $\gamma$-rays seem to confirm the corpuscular character of radiation. At
the same time, the wave aspect continues to be confirmed by the observed
interference and diffraction of X-rays. Finally, de Broglie notes the very
recent corpuscular interpretation of Compton scattering. De Broglie concludes
his historical introduction with a mention of his own recent work (p.~30):

\begin{quotation}
.... the moment seemed to have arrived to make an effort towards unifying the
corpuscular and wave points of view and to go a bit more deeply into the true
meaning of the quanta. That is what we have done recently ....
\end{quotation}
De Broglie clearly regarded his own work as a synthesis of earlier
theories of dynamics and optics, a synthesis increasingly forced upon us by
accumulating experimental evidence.

Chapter 1 of the thesis is entitled `The phase wave'. De Broglie begins by
recalling the equivalence of mass and energy implied by the theory of
relativity. Turning to the problem of quanta, he remarks (pp.~32--3):

\begin{quotation}
It seems to us that the fundamental idea of the quantum theory is the
impossibility of considering an isolated quantity of energy without
associating a certain frequency with it. This connection is expressed by what
I shall call the quantum relation:%
\[
\mathrm{energy}=h\times\mathrm{frequency}%
\]
where $h$ is Planck's constant.
\end{quotation}
To make sense of the quantum relation, de Broglie proposes that (p.~33)

\begin{quotation}
.... to each energy fragment of proper mass $m_{0}$ there is attached a
periodic phenomenon of frequency $\nu_{0}$ such that one has:%
\[
h\nu_{0}=m_{0}c^{2}%
\]
$\nu_{0}$ being measured, of course, in the system tied to the energy fragment.
\end{quotation}
De Broglie asks if the periodic phenomenon must be assumed to be
localised inside the energy fragment. He asserts that this is not at all
necessary, and that it will be seen to be `without doubt spread over an
extensive region of space' (p.~34).

De Broglie goes on to consider the apparent contradiction between the
frequency $\nu=mc^{2}/h=\nu_{0}/\sqrt{1-v^{2}/c^{2}}$ and the time-dilated
frequency $\nu_{1}=\nu_{0}\sqrt{1-v^{2}/c^{2}}$. He proposes that the
contradiction is resolved by the following `theorem of phase harmony' (p.~35):
in a frame where the moving body has velocity $v$, the periodic phenomenon
tied to the moving body and with frequency $\nu_{1}$ is always in phase with a
wave of frequency $\nu$ propagating in the same direction as the moving body
with phase velocity $v_{\mathrm{ph}}=c^{2}/v$. This is shown by applying the
Lorentz transformation to a rest-frame wave $\sin\left(  \nu_{0}t_{0}\right)
$, yielding a wave%
\begin{equation}
\sin\left[  \nu_{0}\left(  t-vx/c^{2}\right)  /\sqrt{1-v^{2}/c^{2}}\right]
\end{equation}
of frequency $\nu=\nu_{0}/\sqrt{1-v^{2}/c^{2}}$ and phase velocity $c^{2}/v$.
Regarding the nature of this wave de Broglie says that, because its velocity
is greater than $c$, it cannot be a wave transporting energy: rather, `it
represents the spatial distribution of the \textit{phases} of a phenomenon; it
is a \textquotedblleft phase wave\textquotedblright' (p.~36). De Broglie shows
that the group velocity of the phase wave is equal to the velocity of the
particle. In the final section of chapter 1 (`The phase wave in spacetime'),
he discusses the appearance of surfaces of constant phase for differently
moving observers, from a spacetime perspective.

Chapter 2 is entitled `Maupertuis' principle and Fermat's principle'. The aim
is to generalise the results of the first chapter to non-uniform,
non-rectilinear motion. In the introduction to chapter 2 de Broglie writes (p.~45):

\begin{quotation}
Guided by the idea of a deep unity between the principle of least action and
that of Fermat, from the beginning of my investigations on this subject I was
led to \textit{assume} that, for a given value of the total energy of the
moving body and therefore of the frequency of its phase wave, the dynamically
possible trajectories of the one coincided with the possible rays of the other.
\end{quotation}
De Broglie discusses the principle of least action, in the different
forms given by Hamilton and by Maupertuis, and also for relativistic particles
in an external electromagnetic field. He writes Hamilton's principle as%
\begin{equation}
\delta\int_{P}^{Q}p_{\mu}dx^{\mu}=0
\end{equation}
($\mu=0,1,2,3$, with $dx^{0}=cdt$), where $P$, $Q$ are points in spacetime and
$p_{\mu}$ is the canonical energy-momentum 4-vector, and notes that if $p_{0}$
is constant the principle becomes%
\begin{equation}
\delta\int_{A}^{B}p_{i}dx^{i}=0
\end{equation}
($i=1,2,3$), where $A$, $B$ are the corresponding points in space --- that is,
Hamilton's principle reduces to Maupertuis' principle.

De Broglie then discusses wave propagation and Fermat's principle from a
spacetime perspective. He considers a sinusoidal function $\sin\phi$, where
the phase $\phi$ has a spacetime-dependent differential $d\phi$, and writes
the variational principle for the ray in spacetime in the Hamiltonian form%
\begin{equation}
\delta\int_{P}^{Q}d\phi=0\ .
\end{equation}
De Broglie then introduces a 4-vector field $w_{\mu}$ on spacetime, defined by%
\begin{equation}
d\phi=2\pi w_{\mu}dx^{\mu}\ , \label{Kdef}%
\end{equation}
where the $w_{\mu}$ are generally functions on spacetime. (Of course, this
implies that $2\pi w_{\mu}=\partial_{\mu}\phi$, though de Broglie does not
write this explicitly.) De Broglie also notes that $d\phi=2\pi(\nu
dt-(\nu/v_{\mathrm{ph}})dl)$ and $w_{\mu}=(\nu/c,-(\nu/v_{\mathrm{ph}%
})\mathbf{\hat{n}})$, where $\mathbf{\hat{n}}$ is a unit vector in the
direction of propagation; and that if $\nu$ is constant, the principle in the
Hamiltonian form%
\begin{equation}
\delta\int_{P}^{Q}w_{\mu}dx^{\mu}=0
\end{equation}
reduces to the principle in the Maupertuisian form%
\begin{equation}
\delta\int_{A}^{B}w_{i}dx^{i}=0\ ,
\end{equation}
or%
\begin{equation}
\delta\int_{A}^{B}\frac{\nu}{v_{\mathrm{ph}}}dl=0\ ,
\end{equation}
which is Fermat's principle.

De Broglie then discusses an `extension of the quantum relation' (that is, an
extension of $E=h\nu$). He states that the two 4-vectors $p_{\mu}$ and
$w_{\mu}$ play perfectly symmetrical roles in the motion of a particle and in
the propagation of a wave. Writing the `quantum relation' $E=h\nu$ as
$w_{0}=\frac{1}{h}p_{0}$, de Broglie proposes the generalisation%
\begin{equation}
w_{\mu}=\frac{1}{h}p_{\mu}\ , \label{relgeq}%
\end{equation}
so that%
\begin{equation}
d\phi=2\pi w_{\mu}dx^{\mu}=\frac{2\pi}{h}p_{\mu}dx^{\mu}\ .
\end{equation}
Fermat's principle then becomes%
\begin{equation}
\delta\int_{A}^{B}p_{i}dx^{i}=0\ ,
\end{equation}
which is the same as Maupertuis' principle. Thus, de Broglie arrives at the
following statement (p.~56):

\begin{quotation}
Fermat's principle applied to the phase wave is identical to Maupertuis'
principle applied to the moving body; the dynamically possible trajectories of
the moving body are identical to the possible rays of the wave.
\end{quotation}
He adds that (p.~56):

\begin{quotation}
We think that this idea of a deep relationship between the two great
principles of Geometrical Optics and Dynamics could be a valuable guide in
realising the synthesis of waves and quanta.
\end{quotation}

De Broglie then discusses some particular cases: the free particle, a particle
in an electrostatic field, and a particle in a general electromagnetic field.
He calculates the phase velocity, which depends on the electromagnetic
potentials. He notes that the propagation of a phase wave in an external field
depends on the charge and mass of the moving body. And he shows that the group
velocity along a ray is still equal to the velocity of the moving body.

For the case of an electron of charge $e$ and velocity $v$ in an electrostatic
potential $\varphi$, de Broglie writes down the following expressions for the
frequency $\nu$ and phase velocity $v_{\mathrm{ph}}$ of the phase wave (p.~57):%
\begin{equation}
\nu=(mc^{2}+e\varphi)/h\ ,\ \ \ \ v_{\mathrm{ph}}=(mc^{2}+e\varphi)/mv
\label{elec}%
\end{equation}
(where again $m=m_{0}/\sqrt{1-v^{2}/c^{2}}$). He shows that $v_{\mathrm{ph}}$
may be rewritten as the free value $c^{2}/v$ multiplied by a factor
$h\nu/(h\nu-e\varphi)$ that depends on the potential $\varphi$. The
expressions (\ref{elec}) formed the starting point for Schr\"{o}dinger's work
on the wave equation for de Broglie's phase waves (as reconstructed by Mehra
and Rechenberg (1987, pp.~423--5), see section~\ref{towards-a-complete}).

While de Broglie does not explicitly say so in his thesis, note that from the
definition (\ref{Kdef}) of $w_{\mu}$, the generalised quantum relation
(\ref{relgeq}) may be written in the form%
\begin{equation}
p_{\mu}=\hslash\partial_{\mu}\phi\ . \label{relgeq2}%
\end{equation}
This is what we would now call a relativistic guidance equation, giving the
velocity of a particle in terms of the gradient of the phase of a pilot wave
(where here de Broglie defines the phase $\phi$ to be dimensionless). In other
words, the extended quantum relation is a first-order equation of motion. In
the presence of an electromagnetic field, the canonical momentum $p_{\mu}$
contains the 4-vector potential. For a free particle, with $p_{\mu
}=(E/c,-\mathbf{p})$, the guidance equation has components%
\begin{equation}
E=\hslash\dot{\phi},\;\;\;\mathbf{p}=-\hslash\mathbf{\nabla}\phi\ ,
\label{relgeq3}%
\end{equation}
where the spatial components may also be written as%
\begin{equation}
m\mathbf{v}=\frac{m_{0}\mathbf{v}}{\sqrt{1-v^{2}/c^{2}}}=-\hslash
\mathbf{\nabla}\phi\ . \label{relgeq4}%
\end{equation}
For a plane wave of phase $\phi=\omega t-\mathbf{k}\cdot\mathbf{x}$, we have%
\begin{equation}
E=\hslash\omega,\;\;\;\mathbf{p}=\hslash\mathbf{k}\ .
\end{equation}

Thus, de Broglie's unification of the principles of Maupertuis and Fermat
amounts to a new dynamical law, (\ref{relgeq}) or (\ref{relgeq2}), in which
the phase of a guiding wave determines the particle velocity. This new law of
motion is the essence of de Broglie's new, first-order dynamics.

Chapter 3 of de Broglie's thesis is entitled `The quantum conditions for the
stability of orbits'. De Broglie reviews Bohr's condition for circular orbits,
according to which the angular momentum of the electron must be a multiple of
$\hslash$, or equivalently $\int_{0}^{2\pi}p_{\theta}d\theta=nh$ ($p_{\theta}$
conjugate to $\theta$). He also reviews Sommerfeld's generalisation, $\oint
p_{i}dq_{i}=n_{i}h$ (integral $n_{i}$) and Einstein's invariant formulation
$\oint\sum_{i=1}^{3}p_{i}dq_{i}=nh$ (integral $n$). De Broglie then provides
an explanation for Einstein's condition. The trajectory of the moving body
coincides with one of the rays of its phase wave, and the phase wave moves
along the trajectory with a constant frequency (because the total energy is
constant) and with a variable speed whose value has been calculated. To have a
stable orbit, claims de Broglie, the length $l$ of the orbit must be in
`resonance' with the wave: thus $l=n\lambda$ in the case of constant
wavelength, and $\oint(\nu/v_{\mathrm{ph}})dl=n$ ($n$ integral) generally. De
Broglie notes that this is precisely the integral appearing in Fermat's
principle, which has been shown to be equal to the integral giving the
Maupertuisian action divided by $h$. The resonance condition is then identical
to the required stability condition. For the simple case of circular orbits in
the Bohr atom de Broglie shows, using $v_{\mathrm{ph}}=\nu\lambda$ and
$h/\lambda=m_{0}v$, that the resonance condition becomes $\oint m_{0}vdl=nh$
or $m_{0}\omega R^{2}=n\hslash$ (with $v=\omega R$), as originally given by
Bohr.\footnote{De Broglie also claims to generalise his results from closed
orbits to quasi-periodic (or multi-periodic) motion: however, as shown by
Darrigol (1993), de Broglie's derivation is faulty.} (Note that the simple
argument commonly found in textbooks, about the fitting of whole numbers of
wavelengths along a Bohr orbit, originates in this work of de Broglie's.)

De Broglie thought that his explanation of the stability or quantisation
conditions constituted important evidence for his ideas. As he puts it (p.~65):

\begin{quotation}
This beautiful result, whose demonstration is so immediate when one has
accepted the ideas of the preceding chapter, is the best justification we can
give for our way of attacking the problem of quanta.
\end{quotation}
Certainly, de Broglie had achieved a concrete realisation of his
initial intuition that quantisation conditions for atomic energy levels could
arise from the properties of waves.

In his chapter 4, de Broglie considers the two-body problem, in particular the
hydrogen atom. He expresses concern over how to define the proper masses,
taking into account the interaction energy. He discusses the quantisation
conditions for hydrogen from a two-body point of view: he has two phase waves,
one for the electron and one for the nucleus.

The subject of chapter 5 is light quanta. De Broglie suggests that the
classical (electromagnetic) wave distribution in space is some sort of time
average over the true distribution of phase waves. His light quantum is
assigned a very small proper mass: the velocity $v$ of the quantum, and the
phase velocity $c^{2}/v$ of the accompanying phase wave, are then both very
close to $c$.

De Broglie points out that radiation is sometimes observed to violate
rectilinear propagation: a light wave striking the edge of a screen diffracts
into the geometrical shadow, and rays passing close to the screen deviate from
a straight line. De Broglie notes the two historical explanations for this
phenomenon --- on the one hand the explanation for diffraction given by the
wave theory, and on the other the explanation given by Newton in his emission
theory: `Newton assumed [the existence of] a force exerted by the edge of the
screen on the corpuscle' (p.~80). De Broglie asserts that he can now give a
unified explanation for diffraction, by abandoning Newton's first law of
motion (p.~80):

\begin{quotation}
.... the ray of the wave would curve as predicted by the theory of waves, and
the moving body, for which the principle of inertia would no longer be valid,
would suffer the same deviation as the ray with which its motion is bound up
[solidaire] ....
\end{quotation}
De Broglie's words here deserve emphasis. As is also very clear in his
second paper of the preceding year (see section~\ref{first-papers}),
de Broglie regards his explanation of particle diffraction as based on a new
form of dynamics in which Newton's first law --- the principle that a free
body will always move uniformly in a straight line --- is abandoned. At the
same time, de Broglie recognises that one can always adopt a
classical-mechanical viewpoint if one wishes (pp.~80--81):

\begin{quotation}
.... perhaps one could say that the wall exerts a force on it [the moving
body] if one takes the curvature of the trajectory as a criterion for the
existence of a force.
\end{quotation}
Here, de Broglie recognises that one may still think in Newtonian
terms, if one continues to identify acceleration as indicative of the presence
of a force. Similarly, as we shall see, in 1927 de Broglie notes that his
pilot-wave dynamics may if one wishes be written in Newtonian form with a
quantum potential. But de Broglie's preferred approach, throughout his work in
the period 1923--27, is to abandon Newton's first law and base his dynamics on
velocity rather than on acceleration.

After considering the Doppler effect, reflection by a moving mirror, and
radiation pressure, all from a photon viewpoint, de Broglie turns to the
phenomena of wave optics, noting that (p.~86):

\begin{quotation}
The stumbling block of the theory of light quanta is the explanation of the
phenomena that constitute wave optics.
\end{quotation}
Here it becomes apparent that, despite his understanding of how
non-rectilinear particle trajectories arise during diffraction and
interference, de Broglie is not sure of the details of how to explain the
observed bright and dark fringes in diffraction and interference experiments
with light. In particular, de Broglie did not have a precise theory of the
assumed statistical relationship between his phase waves and the
electromagnetic field. Even so, he went on to make what he called `vague
suggestions' (p.~87) towards a detailed theory of optical interference. De
Broglie's idea was that the phase waves would determine the probability for
the light quanta to interact with the atoms constituting the equipment used to
observe the radiation, in such a way as to account for the observed fringes
(p.~88):

\begin{quotation}
.... the probability of reactions between atoms of matter and atoms of light
is at each point tied to the resultant (or rather to the mean value of this)
of one of the vectors characterising the phase wave; where this resultant
vanishes the light is undetectable; there is interference. One then conceives
that an atom of light traversing a region where the phase waves interfere will
be able to be absorbed by matter at certain points and not at others. This is
the still very qualitative principle of an explanation of interference .... .
\end{quotation}
As we shall see in the next section, after completing his thesis de
Broglie arrived at a simpler explanation of optical interference fringes.

The final section of chapter 5 considers the explanation of Bohr's frequency
condition $h\nu=E_{1}-E_{2}$ for the light emitted by an atomic transition
from energy state $E_{1}$ to energy state $E_{2}$. De Broglie derives this
from the assumption that each transition involves the emission of a single
light quantum of energy $E=h\nu$ (together with the assumption of energy conservation).

De Broglie's chapter 6 discusses the scattering of X- and $\gamma$-rays.

In his chapter 7, de Broglie turns to statistical mechanics, and shows how the
concept of statistical equilibrium is to be modified in the presence of phase
waves. If each particle or atom in a gas is accompanied by a phase wave, then
a box of gas will be `criss-crossed in all directions' (p.~110) by the waves.
De Broglie finds it natural to assume that the only stable phase waves in the
box will be those that form stationary or standing waves, and that only these
will be relevant to thermodynamic equilibrium. He illustrates his idea with a
simple example of molecules moving in one dimension, confined to an interval
of length $l$. In the nonrelativistic limit, each phase wave has a wavelength
$\lambda=h/m_{0}v$ and the `resonance condition' is $l=n\lambda$ with $n$
integral. Writing $v_{0}=h/m_{0}l$, one then has $v=nv_{0}$. As de Broglie
notes (p.~112): `The speed will then be able to take only values equal to
integer multiples of $v_{0}$'. (This is, of course, the well-known
quantisation of momentum for particles confined to a box.) De Broglie then
argues that a velocity element $\delta v$ corresponds to a number $\delta
n=(m_{0}l/h)\delta v$ of states of a molecule (compatible with the existence
of stationary phase waves), so that an element $m_{0}\delta x\delta v$ of
phase space volume corresponds to a number $m_{0}\delta x\delta v/h$ of
possible states. Generalising to three dimensions, de Broglie is led to take
the element of phase space volume divided by $h^{3}$ as the measure of the
number of possible states of a molecule, as assumed by Planck.

De Broglie then turns to the photon gas, for which he obtains Wien's law. He
claims that, in order to get the Planck law, the following further hypothesis
is required (p.~116):

\begin{quotation}
If two or several atoms [of light] have phase waves that are exactly
superposed, of which one can therefore say that they are transported by the
same wave, their motions can no longer be considered as entirely independent
and these atoms can no longer be treated as separate units in calculating the probabilities.
\end{quotation}
In de Broglie's approach, the stationary phase waves play the role of
the elementary objects of statistical mechanics. De Broglie defines stationary
waves as a superposition of two waves of the form%
\begin{equation}%
\begin{array}
[c]{c}%
\sin\\
\cos
\end{array}
\left[  2\pi\left(  \nu t-\frac{x}{\lambda}+\phi_{0}\right)  \right]
\ \ \ \mathrm{and}\ \ \
\begin{array}
[c]{c}%
\sin\\
\cos
\end{array}
\left[  2\pi\left(  \nu t+\frac{x}{\lambda}+\phi_{0}\right)  \right]  \ ,
\end{equation}
where $\phi_{0}$ can take any value from $0$ to $1$ and $\nu$ takes one of the
allowed values. Each elementary wave can carry any number $0,1,2,...$ of
atoms, and the probability of carrying $n$ atoms is given by the Boltzmann
factor $e^{-nh\nu/kT}$. Applying this method to a gas of light quanta, de
Broglie claims to derive the Planck distribution.\footnote{Again, as shown by
Darrigol (1993), de Broglie's application of statistical mechanics contains
some errors.}

De Broglie's thesis ends with a summary and conclusions (pp.~125--8). The
seeds of the problem of quanta have been shown, he claims, to be contained in
the historical `parallelism of the corpuscular and wave-like conceptions of
radiation'. He has postulated a periodic phenomenon associated with each
energy fragment, and shown how relativity requires us to associate a phase
wave with every uniformly moving body. For the case of non-uniform motion,
Maupertuis' principle and Fermat's principle `could well be two aspects of a
single law', and this new approach to dynamics led to an extension of the
quantum relation, giving the speed of a phase wave in an electromagnetic
field. The most important consequence is the interpretation of the quantum
conditions for atomic orbits in terms of a resonance of the phase wave along
the trajectories: `this is the first physically plausible explanation proposed
for the Bohr-Sommerfeld stability conditions'. A `qualitative theory of
interference' has been suggested. The phase wave has been introduced into
statistical mechanics, yielding a derivation of Planck's phase volume element,
and of the blackbody spectrum. De Broglie has, he claims, perhaps contributed
to a unification of the opposing conceptions of waves and particles, in which
the dynamics of the material point is understood in terms of wave propagation.
He adds that the ideas need further development: first of all, a new
electromagnetic theory is required, that takes into account the discontinuous
structure of radiation and the physical nature of phase waves, with Maxwell's
theory emerging as a statistical approximation. The final paragraph of de
Broglie's thesis emphasises the incompleteness of his theory at the time:

\begin{quotation}
I have deliberately left rather vague the definition of the phase wave, and of
the periodic phenomenon of which it must in some sense be the translation, as
well as that of the light quantum. The present theory should therefore be
considered as one whose physical content is not entirely specified, rather
than as a consistent and definitively constituted doctrine.
\end{quotation}

As de Broglie's concluding paragraph makes clear, his theory of 1924 was
rather abstract. There was no specified basis for the phase waves (they were
certainly not regarded as `material' waves); nor was any particular wave
equation suggested. It should also be noted that at this time de Broglie's
waves were real-valued functions of space and time, of the form $\propto
\sin(\omega t-\mathbf{k}\cdot\mathbf{x})$, with a real oscillating amplitude.
They were \textit{not} complex waves $\propto e^{i(\mathbf{k}\cdot
\mathbf{x}-\omega t)}$ of uniform amplitude. Thus, de Broglie's `phase waves'
had an oscillating amplitude as well as a phase. (De Broglie seems to have
called them `phase waves' only because of his theorem of phase harmony.) Note
also that, in his treatment of particles in a box, de Broglie superposes waves
propagating in opposite directions, yielding stationary waves whose amplitudes
oscillate in time.

In his thesis de Broglie does not explicitly discuss diffraction or
interference experiments with electrons, even though in his second
communication of 1923 (de Broglie 1923b) he had suggested electron diffraction
as an experimental test. According to de Broglie's later recollections (L. de
Broglie, AHQP Interview, 7 January 1963, p.~6),\endnote{{\em Ibid.}} at his thesis defence on 25
November 1924:

\begin{quotation}
Mr Jean Perrin, who chaired the committee, asked me if my ideas could lead to
experimental confirmation. I replied that yes they could, and I mentioned the
diffraction of electrons by crystals. Soon afterwards, I advised Mr Dauvillier~.... 
to try the experiment, but, absorbed by other research, he did not do it.
I do not know if he believed, or if he said to himself that it was perhaps
very uncertain, that he was going to go to a lot of trouble for nothing ---
it's possible. .... But the following year it was discovered in America by
Davisson and Germer.
\end{quotation}

\subsection{Optical interference fringes: November 1924}\label{opt-interf}

On 17 November 1924, just a few days before de Broglie defended his thesis, a
further communication of de Broglie's was presented to the Academy of
Sciences: entitled `On the dynamics of the light quantum and interference',
and published in the \textit{Comptes Rendus}, this short note gave a new and
improved account of optical interference in terms of light quanta (de Broglie 1924d).

De Broglie began his note by recalling his unsatisfactory discussion of
optical interference in his recent work on the quantum theory (p.~1039):

\begin{quotation}
.... I had not reached a truly satisfying explanation for the phenomena of
wave optics which, in principle, all come down to interference. I limited
myself to putting forward a certain connection between the state of
interference of the waves and the probability for the absorption of light
quanta by matter. This viewpoint now seems to me a bit artificial and I tend
towards adopting another, more in harmony with the broad outlines of my theory itself.
\end{quotation}
As we have seen, in his thesis de Broglie was unsure about how to
account for the bright and dark fringes observed in optical interference
experiments. He did not have a theory of the electromagnetic field, which he
assumed emerged as some sort of average over his phase waves. To account for
optical fringes, he had suggested that the phase waves somehow determined the
probability for interactions between photons and the atoms in the apparatus.
Now, after completing his thesis, he felt he had a better explanation, that
was based purely on the spatial distribution of the photon trajectories.

De Broglie's note continues by outlining his `new dynamics', in which the
energy-momentum 4-vector of every material point is proportional to the
`characteristic' 4-vector of an associated wave, even when the wave undergoes
interference or diffraction. He then gives his new view of interference
fringes (p.~1040):

\begin{quotation}
The rays predicted by the wave theories would then be in every case the
possible trajectories of the quantum. In the phenomena of interference, the
rays become concentrated in those regions called `bright fringes' and become
diluted in those regions called `dark fringes'. In my first explanation of
interference, the dark fringes were dark because the action of fragments of
light on matter was zero there; in my current explanation, these fringes are
dark because the number of quanta passing through them is small or zero.
\end{quotation}
Here, then, de Broglie explains bright and dark fringes simply in terms
of a high or low density of photon trajectories in the corresponding regions.

When de Broglie speaks of the trajectories being concentrated and diluted in
regions of bright and dark fringes respectively, he presumably had in mind
that the number density of particles in an interference zone should be
proportional to the classical wave intensity, though he does not say this
explicitly. We can discern the essence of the more precise and complete
explanation of optical interference given by de Broglie three years later in
his Solvay report: there, de Broglie has the same photon trajectories, with a
number density specified as proportional to the amplitude-squared of the
guiding wave (see pp.~\pageref{alsooldp9192}~f., and our discussion in
section~\ref{interf-deB}).

In his note of November 1924, de Broglie goes on to illustrate his proposal
for the case of Young's interference experiment with two pinholes acting as
point sources. De Broglie cites the well-known facts that in this case the
surfaces of equal phase are ellipsoids with the pinholes as foci, and that the
rays (which are normal to the ellipsoidal surfaces) are concentrated on
hyperboloids of constructive interference where the classical intensity has
maxima. He then notes (p.~1040):

\begin{quotation}
Let $r_{1}$ and $r_{2}$ be the distances from a point in space to the two
holes and let $\psi$ be the function $\frac{1}{2}(r_{1}+r_{2})$, which is
constant on each surface of equal phase. One easily shows that the phase
velocity of the waves along a ray is equal to the value it would have in the
case of free propagation divided by the derivative of $\psi$ taken along the
ray; as for the speed of the quantum, it will be equal to the speed of free
motion multiplied by the same derivative.
\end{quotation}
De Broglie gives no further details, but these are easily
reconstructed. For an incident beam of wavelength $\lambda=2\pi/k$, each
pinhole acts as a source of a spherical wave of wavelength $\lambda$, yielding
a resultant wave proportional to the real part of%
\begin{equation}
\frac{e^{i(kr_{1}-\omega t)}}{r_{1}}+\frac{e^{i(kr_{2}-\omega t)}}{r_{2}}\ .
\end{equation}
If the pinholes have a separation $d$, then at large distances ($r_{1}$,
$r_{2}>>d$) from the screen the resultant wave may be approximated as%
\begin{equation}
2\frac{e^{i(kr-\omega t)}}{r}\cos\left(  \frac{kd}{2}\theta\right)
\end{equation}
where $\theta$ is the (small) angular deviation from the normal to the screen.
The amplitude shows the well-known interference pattern. As for the phase
$\phi=kr-\omega t$, with $r=\frac{1}{2}(r_{1}+r_{2})$, the surfaces of equal
phase are indeed the well-known ellipsoids. Further, the phase velocity is
given by%
\begin{equation}
v_{\mathrm{ph}}=\frac{\left\vert \partial\phi/\partial t\right\vert
}{\left\vert \mathbf{\nabla}\phi\right\vert }=\frac{\omega}{k}\frac
{1}{\left\vert \mathbf{\nabla}r\right\vert }=\frac{c}{\left\vert
\mathbf{\nabla}r\right\vert }\ ,
\end{equation}
while de Broglie's particle velocity --- given by (\ref{relgeq4}) --- has
magnitude%
\begin{equation}
v=\frac{\hslash\left\vert \mathbf{\nabla}\phi\right\vert }{m}=\frac
{\hslash\left\vert \mathbf{\nabla}\phi\right\vert }{(\hslash\omega/c^{2}%
)}=c^{2}\frac{k}{\omega}\left\vert \mathbf{\nabla}r\right\vert =c\left\vert
\mathbf{\nabla}r\right\vert
\end{equation}
(where $m$ is the relativistic photon mass), in agreement with de Broglie's assertions.

At the end of his note, de Broglie comments that this method may be applied to
the study of scattering.

In November 1924, then, de Broglie understood how interfering phase waves
would affect photon trajectories, causing them to bunch together in regions
coinciding with the observed bright fringes.

Note that in this paper de Broglie treats his phase waves as if they were a
direct representation of the electromagnetic field. In his discussion of
Young's interference experiment, he has phase waves emerging from the two
holes and interfering, and he identifies the interference fringes of his phase
waves with optical interference fringes. However, he seems quite aware that
this is a simplification,\footnote{De Broglie may have thought of this as
analogous to scalar wave optics, which predicts the correct optical
interference fringes by treating the electromagnetic field simply as a scalar
wave.} remarking that `the whole theory will become truly clear only if one
manages to define the structure of the light wave'. De Broglie is still not
sure about the precise relationship between his phase waves and light waves, a
situation that persists even until the fifth Solvay conference: there, while
he gives (pp.~\pageref{alsooldp9192}~f.) a precise account of optical
interference in his report, he points out (p.~\pageref{p192DRAFT1}) that
the connection between his guiding wave and the electromagnetic field is still unknown.

\section{Towards a complete pilot-wave dynamics: 1925--27}\label{towards-a-complete}

On 16 December 1924, Einstein wrote to Lorentz:\footnote{Quoted in Mehra and
Rechenberg (1982a, p.~604).}

\begin{quotation}
A younger brother of the de Broglie known to us [Maurice de Broglie] has made
a very interesting attempt to interpret the Bohr-Sommerfeld quantization rules
(Paris Dissertation, 1924). I believe that it is the first feeble ray of light
to illuminate this, the worst of our physical riddles. I have also discovered
something that supports his construction.
\end{quotation}
What Einstein had discovered, in support of de Broglie's ideas,
appeared in the second of his famous papers on the quantum theory of the ideal
gas (Einstein 1925a). Einstein showed that the fluctuations associated with
the new Bose-Einstein statistics contained two distinct terms that could be
interpreted as particle-like and wave-like contributions --- just as Einstein
had shown, many years earlier, for blackbody radiation. Einstein argued that
the wave-like contribution should be interpreted in terms of de Broglie's
matter waves, and he cited de Broglie's thesis. It was largely through this
paper by Einstein that de Broglie's work became known outside France.

In the same paper, Einstein suggested that a molecular beam would undergo
diffraction through a sufficiently small aperture. De Broglie had already made
a similar suggestion for electrons, in his second communication to the
\textit{Comptes Rendus} (de Broglie 1923b). Even so, in their report at the
fifth Solvay conference, Born and Heisenberg state that in his gas theory
paper Einstein `deduced from de Broglie's daring theory the possibility of 
``diffraction'' of material particles' (p.~\pageref{for-first-papers}), 
giving the incorrect impression that Einstein had
been the first to see this consequence of de Broglie's theory. It seems likely
that Born and Heisenberg did not notice de Broglie's early papers in the
\textit{Comptes Rendus}.

In 1925 Elsasser --- a student of Born's in G\"{o}ttingen --- read de
Broglie's thesis. Like most others outside France, Elsasser had heard about de
Broglie's thesis through Einstein's gas theory papers. Elsasser suspected that
two observed experimental anomalies could be explained by de Broglie's new
dynamics. First, the Ramsauer effect --- the surprisingly large mean free path
of low-velocity electrons in gases --- which Elsasser thought could be
explained by electron interference. Second, the intensity maxima observed by
Davisson and Kunsman at certain angles of reflection of electrons from metal
surfaces, which had been assumed to be caused by atomic shell structure, and
which Elsasser thought were caused by electron diffraction. Elsasser published
a short note sketching these ideas in \textit{Die Naturwissenschaften} (Elsasser
1925). Elsasser then tried to design an experiment to test the ideas further,
with low-velocity electrons, but never carried it out. According to
Heisenberg's later recollection, Elsasser's supervisor Born was sceptical
about the reality of matter waves, because they seemed in conflict with the
observed particle tracks in cloud chambers.\footnote{For this and further
details concerning Elsasser, see Mehra and Rechenberg (1982a, pp.~624--7).
See also the discussion in section~\ref{BornBohr}.}

On 3 November 1925, Schr\"{o}dinger wrote to Einstein: `A few days ago I read
with the greatest interest the ingenious thesis of Louis de Broglie ....
'.\footnote{Quoted in Mehra and Rechenberg (1987, p.~412).} Schr\"{o}dinger 
too had become interested in de Broglie's thesis by
reading Einstein's gas theory papers, and he set about trying to find the wave
equation for de Broglie's phase waves. As we have seen (section~\ref{thesis}), 
in his thesis de Broglie had shown that, in an
electrostatic potential $\varphi$, the phase wave of an electron of charge $e$
and velocity $v$ would have (see equation (\ref{elec})) frequency $\nu
=(mc^{2}+V)/h$ and phase velocity $v_{\mathrm{ph}}=(mc^{2}+V)/mv$, where
$m=m_{0}/\sqrt{1-v^{2}/c^{2}}$ and $V=e\varphi$ is the potential energy. These
expressions for $\nu$ and $v_{\mathrm{ph}}$, given by de Broglie, formed the
starting point for Schr\"{o}dinger's work on the wave equation.

Schr\"{o}dinger\label{forSchrEss} took de Broglie's formulas for $\nu$ and $v_{\mathrm{ph}}$ and
applied them to the hydrogen atom, with a Coulomb field $\varphi
=-e/r$.\footnote{Here we follow the analysis by Mehra and Rechenberg
(1987, pp.~423--5) of what they call Schr\"{o}dinger's
`earliest preserved [unpublished] manuscript on wave mechanics'. Similar
reasoning is found in a letter from Pauli to Jordan of 12 April 1926 (Pauli
1979, pp.~315--20).} Using the formula for $\nu$ to eliminate $v$,
Schr\"{o}dinger rewrote the expression for the phase velocity $v_{\mathrm{ph}%
}$ purely in terms of the frequency $\nu$ and the electron-proton distance
$r$:%
\begin{equation}
v_{\mathrm{ph}}=\frac{h\nu}{m_{0}c}\frac{1}{\sqrt{\left(  h\nu/m_{0}%
c^{2}-V/m_{0}c^{2}\right)  ^{2}-1}} \label{V}%
\end{equation}
(where $V=-e^{2}/r$). Then, writing de Broglie's phase wave as $\psi
=\psi(\mathbf{x},t)$, he took the equation for $\psi$ to be the usual wave
equation%
\begin{equation}
\nabla^{2}\psi=\frac{1}{v_{\mathrm{ph}}^{2}}\frac{\partial^{2}\psi}{\partial
t^{2}} \label{S0}%
\end{equation}
with phase velocity $v_{\mathrm{ph}}$. Assuming $\psi$ to have a time
dependence $\propto e^{-2\pi i\nu t}$, Schr\"{o}dinger then obtained the
time-independent equation%
\begin{equation}
\nabla^{2}\psi=-\frac{4\pi^{2}\nu^{2}}{v_{\mathrm{ph}}^{2}}\psi\label{S1}%
\end{equation}
with $v_{\mathrm{ph}}$ given by (\ref{V}). This was Schr\"{o}dinger's original
(relativistic) equation for the energy states of the hydrogen atom.

As is well known, Schr\"{o}dinger found that the energy levels predicted by
(\ref{S1}) --- that is, the eigenvalues $h\nu$ --- disagreed with experiment.
He then adopted a nonrelativistic approximation, and found that this yielded
the correct energy levels for the low-energy limit.

It is in fact straightforward to obtain the correct nonrelativistic limit.
Writing $E=h\nu-m_{0}c^{2}$, in the low-energy limit $\left\vert E\right\vert
/m_{0}c^{2}<<1$ and $\left\vert V\right\vert /m_{0}c^{2}<<1$, we have%
\begin{equation}
\frac{\nu^{2}}{v_{\mathrm{ph}}^{2}}=\frac{2m_{0}}{h^{2}}\left(  E-V\right)
\ ,
\end{equation}
so that (\ref{S1}) reduces to what we now know as the nonrelativistic
time-independent Schr\"{o}dinger equation for a single particle in a potential
$V$:%
\begin{equation}
-\frac{\hslash^{2}}{2m_{0}}\nabla^{2}\psi+V\psi=E\psi\ . \label{S2}%
\end{equation}

Historically speaking, then, Schr\"{o}dinger's original equation (\ref{S1})
for stationary states amounted to a mathematical transcription --- into the
language of the standard wave equation (\ref{S0}) --- of de Broglie's
expressions for the frequency and phase velocity of an electron wave. By
studying the eigenvalue problem of this equation in its nonrelativistic form
(\ref{S2}), Schr\"{o}dinger was able to show that the eigenvalues agreed
remarkably well with the observed features of the hydrogen spectrum. Thus, by
adopting the formalism of a wave equation, Schr\"{o}dinger transformed de
Broglie's elementary derivation of the quantisation of energy levels into a
rigorous and powerful technique.

The time-dependent Schr\"{o}dinger equation%
\begin{equation}
i\hslash\frac{\partial\psi}{\partial t}=-\frac{\hslash^{2}}{2m_{0}}\nabla
^{2}\psi+V\psi\ , \label{S3}%
\end{equation}
with a \textit{single} time derivative, was eventually obtained by
Schr\"{o}dinger in his fourth paper on wave mechanics, completed in June 1926
(Schr\"{o}dinger 1926g). The path taken by Schr\"{o}dinger in these four
published papers was rather tortuous. In the fourth paper, he actually began
by considering a wave equation that was of second order in time and of fourth
order in the spatial derivatives. However, he eventually settled on
(\ref{S3}), deriving it by the following argument: for a wave with time
dependence $\propto e^{-(i/\hslash)Et}$, one may write the term $E\psi$ in
(\ref{S2}) as $i\hslash\partial\psi/\partial t$ and thus obtain (\ref{S3}),
which must be valid for any $E$ and therefore for any $\psi$ that can be
expanded as a Fourier time series.\footnote{Actually, Schr\"{o}dinger
considered the time dependence $\propto e^{\pm(i/\hslash)Et}$, so that
$E\psi=\pm i\hslash\partial\psi/\partial t$, leading to \textit{two} possible
wave equations differing by the sign of $i$. He wrote: `\textit{We shall
require that the complex wave function }$\psi$\textit{ satisfy one of these
two equations}. Since at the same time the complex conjugate function
$\bar{\psi}$ satisfies the \textit{other} equation, one may consider the
real part of $\psi$ as a real wave function (if one needs it)'
(Schr\"{o}dinger 1926g, p.~112, original italics).} The price paid for having
an equation that was of only first order in time was that $\psi$ had to be complex.

In retrospect, the time-dependent Schr\"{o}dinger equation for a free particle
((\ref{S3}) with $V=0$) may be immediately derived as the simplest wave
equation obeyed by a complex plane wave $e^{i(\mathbf{k}\cdot\mathbf{x}-\omega
t)}$ with the nonrelativistic dispersion relation $\hslash\omega=(\hslash
k)^{2}/2m$ (that is, $E=p^{2}/2m$ combined with de Broglie's relations
$E=\hslash\omega$ and $p=\hslash k$). This derivation is, in fact, often found
in textbooks.

Not only did Schr\"{o}dinger adopt de Broglie's idea that quantised energy
levels could be explained in terms of waves, he also took up de Broglie's
conviction that classical mechanics was merely the short-wavelength limit of a
broader theory. Thus in his second paper on wave mechanics Schr\"{o}dinger
(1926c, p.~497) wrote:

\begin{quotation}
Maybe our classical mechanics is the \textit{complete} analogue of geometrical
optics and as such is wrong, not in agreement with reality; it fails as soon
as the radii of curvature and the dimensions of the trajectory are no longer
large compared to a certain wavelength, which has a real meaning in the
$q$-space. Then one has to look for an `undulatory mechanics' ---\footnote{Here
Schr\"{o}dinger adds a footnote: `Cf.\ also A.~Einstein, \textit{Berl.\ Ber.},  
pp.~9~ff. (1925)'.} and the most natural path towards this is surely the
wave-theoretical elaboration of the Hamiltonian picture.
\end{quotation}
Schr\"{o}dinger used the optical-mechanical analogy as a guide in the
construction of wave equations for atomic systems, particularly in his second
paper. Schr\"{o}dinger tended to think of the optical-mechanical analogy in
terms of the Hamilton-Jacobi equation and the equation of geometrical optics,
whereas de Broglie tended to think of it in terms of the principles of
Maupertuis and Fermat. These are of course two different ways of drawing the
same analogy.

While Schr\"{o}dinger took up and developed many of de Broglie's ideas, he did
\textit{not} accept de Broglie's view that the particle was localised within
an extended wave. In effect, Schr\"{o}dinger removed the particle trajectories
from de Broglie's theory, and worked only with the extended (non-singular)
waves. (For a detailed discussion of Schr\"{o}dinger's work, see chapter~\ref{SchrEss}.)

In the period 1925--26, then, many of the ideas in de Broglie's thesis were
taken up and developed by other workers, especially Schr\"{o}dinger. But what
of de Broglie himself? Unlike in his earlier work, during this period de
Broglie considered specific equations for his waves. Like Schr\"{o}dinger, he
took standard relativistic wave equations as his starting point. Unlike
Schr\"{o}dinger, however, de Broglie was guided by the following two ideas.
First, that particles are really small singular regions, of very large wave
amplitude, within an extended wave. Second, that the motion of the particles
--- or singularities --- must in some sense satisfy the condition%
\begin{equation}
\mathrm{Maupertuis\equiv Fermat\label{M=F}\ ,}%
\end{equation}
so as to bring about a synthesis of the dynamics of particles with the theory
of waves, along the lines already sketched in de Broglie's thesis. This work
came to a head in a remarkable paper published in May 1927, to which we now turn.

\subsection{`Structure': \textit{Journal de Physique}, May 1927}\label{Structure}

\epigraph{In the last number of the Journal de Physique, a paper by de Broglie has
appeared~.... de Broglie attempts here to reconcile the full determinism of
physical processes with the dualism between waves and corpuscles~.... even if
this paper by de Broglie is off the mark (and I hope that actually), still it
is very rich in ideas and very sharp, and on a much higher level than the
childish papers by Schr\"{o}dinger, who even today still thinks he may~....
abolish material points.}{Pauli, letter to Bohr, 6 August 1927 (Pauli 1979, pp.~404--5)}

\noindent What we now know as pilot-wave theory first appears in a paper by de Broglie
(1927b) entitled `Wave mechanics and the atomic structure of matter and of
radiation', which was published in \textit{Journal de Physique} in May 1927,
and which we shall discuss in detail in this section.\footnote{Again, we do
not always follow de Broglie's notation. Note also that de Broglie moves back
and forth between solutions of the form $f\cos\phi$ and solutions `written in
complex form' $fe^{i\phi}$.} We shall refer to this crucial paper as
`Structure' for short.

In `Structure', de Broglie presents a theory of particles as moving
singularities. It is argued, on the basis of certain assumptions, that the
equations of what we now call pilot-wave dynamics will emerge from this
theory. At the end of the paper de Broglie proposes, as a provisional theory,
simply taking the equations of pilot-wave dynamics as given, without trying to
derive them from something deeper. It is this last, provisional theory that de
Broglie presents a few months later at the fifth Solvay conference.

For historical completeness we should point out that, according to a footnote
to the introduction, `Structure' was `the development of two notes' published
earlier in \textit{Comptes Rendus} (de Broglie 1926, 1927a). The first, of 28
August 1926, considers a model of photons as moving singularities: arguments
are given (similar to those found in `Structure') leading to the usual
velocity formula, and to the conclusion that the probability density of
photons is proportional to the classical wave intensity. The second note, of
31 January 1927, sketches analogous ideas for material bodies in an external
potential, with the probability density now proportional to the
amplitude-squared of the wave function. Unlike `Structure', neither of these
notes contains the suggestion that pilot-wave theory may be adopted as a
provisional view.

We now turn to a detailed analysis of de Broglie's `Structure' paper. De
Broglie first considers a `material point of proper mass $m_{0}$' moving in
free space with a constant velocity $\mathbf{v}$ and represented by a wave
$u(\mathbf{x},t)$ satisfying what we would now call the Klein-Gordon equation%
\begin{equation}
\nabla^{2}u-\frac{1}{c^{2}}\frac{\partial^{2}u}{\partial t^{2}}=\frac{4\pi
^{2}\nu_{0}^{2}}{c^{2}}u\ , \label{KG}%
\end{equation}
with $\nu_{0}=m_{0}c^{2}/h$. De Broglie considers solutions of the form%
\begin{equation}
u(\mathbf{x},t)=f(\mathbf{x}-\mathbf{v}t)\cos\frac{2\pi}{h}\phi(\mathbf{x}%
,t)\ , \label{sing0}%
\end{equation}
where the amplitude $f$ is \textit{singular} at the location $\mathbf{x}%
=\mathbf{v}t$ of the moving body and the phase\footnote{For simplicity we
ignore an arbitrary constant in the phase, which was included by de Broglie.}%
\begin{equation}
\phi(\mathbf{x},t)=\frac{h\nu_{0}}{\sqrt{1-v^{2}/c^{2}}}\left(  t-\frac
{\mathbf{v}\cdot\mathbf{x}}{c^{2}}\right)  \label{phi0}%
\end{equation}
is equal to the classical Hamiltonian action of the particle. A single
particle is represented by a wave (\ref{sing0}), where the moving singularity
has velocity $\mathbf{v}$ and the phase $\phi(\mathbf{x},t)$ is extended over
all space.

De Broglie then considers an ensemble (or `cloud') of similar free particles
with no mutual interaction, all having the same velocity $\mathbf{v}$, and
with singular amplitudes $f$ centred at different points in space. According
to de Broglie, this ensemble of moving singularities may be represented by a
\textit{continuous} solution of the same wave equation (\ref{KG}), of the form%
\begin{equation}
\Psi(\mathbf{x},t)=a\cos\frac{2\pi}{h}\phi(\mathbf{x},t)\ , \label{cont0}%
\end{equation}
where $a$ is a constant and $\phi$ is given by (\ref{phi0}). This continuous
solution is said to `correspond to' the singular solution (\ref{sing0}). (De
Broglie points out that the continuous solutions $\Psi$ are the same as those
considered by Schr\"{o}dinger.)

The number density of particles in the ensemble is taken to have a constant
value $\rho$, which may be written as%
\begin{equation}
\rho=Ka^{2} \label{B0}%
\end{equation}
for some constant $K$. Thus, de Broglie notes, (\ref{cont0}) gives the
`distribution of phases in the cloud of points' as well as the `density of the
cloud'; and for a single particle with known velocity and unknown position,
$a^{2}d^{3}\mathbf{x}$ will measure the probability for the particle to be in
a volume element $d^{3}\mathbf{x}$.

Having discussed the free particle, de Broglie moves on to the case of a
single particle in a static external potential $V(\mathbf{x})$. The wave $u$
`written in complex form' now satisfies what we would call the Klein-Gordon
equation in an external potential\footnote{This follows from (\ref{KG}) by the
substitution $i\hslash\partial/\partial t\rightarrow i\hslash\partial/\partial
t+V$.}%
\begin{equation}
\nabla^{2}u-\frac{1}{c^{2}}\frac{\partial^{2}u}{\partial t^{2}}+\frac
{2i}{\hslash c^{2}}V\frac{\partial u}{\partial t}-\frac{1}{\hslash^{2}}\left(
m_{0}^{2}c^{2}-\frac{V^{2}}{c^{2}}\right)  u=0\ . \label{KG1}%
\end{equation}
De Broglie assumes that the particle begins in free space, represented by a
singular wave (\ref{sing0}), and then enters a region where $V\neq0$, into
which the free solution (\ref{sing0}) must be extended. De Broglie writes
(\ref{sing0}) `in complex form', substitutes into (\ref{KG1}), and takes the
real and imaginary parts, yielding two coupled partial differential equations
for $f$ and $\phi$. Writing%
\begin{equation}
\phi(\mathbf{x},t)=Et-\phi_{1}(\mathbf{x}) \label{Ph0}%
\end{equation}
(where $E=h\nu$ is constant), one of the said partial differential equations
becomes%
\begin{equation}
\hslash^{2}\frac{\square f}{f}=\left(  \mathbf{\nabla}\phi_{1}\right)
^{2}-\frac{1}{c^{2}}(E-V)^{2}+m_{0}^{2}c^{2} \label{pde1}%
\end{equation}
(where $\square\equiv\nabla^{2}-(1/c^{2})\partial^{2}/\partial t^{2}$). De
Broglie notes that if the left-hand side is negligible, (\ref{pde1}) reduces
to the `Jacobi equation' for relativistic dynamics in a static potential, so
that $\phi_{1}$ reduces to the `Jacobi function'. Deviations from classical
mechanics occur, de Broglie notes, when $\square f$ is non-zero.

De Broglie now remarks that, in the classical limit, the velocity of the
particle has the same direction as the vector $\mathbf{\nabla}\phi_{1}$; he
then explicitly \textit{assumes} that this is still true in the general case.
Here, the identity (\ref{M=F}) of the principles of Maupertuis and Fermat is
being invoked. With this assumption, de Broglie shows that the singularity
must have velocity%
\begin{equation}
\mathbf{v}=\frac{c^{2}\mathbf{\nabla}\phi_{1}}{E-V}\ , \label{G0}%
\end{equation}
and he adds that in the nonrelativistic approximation, $E-V\approx m_{0}c^{2}%
$, so that%
\begin{equation}
\mathbf{v}=\frac{1}{m_{0}}\mathbf{\nabla}\phi_{1}\ .
\end{equation}
De Broglie remarks that, in the classical approximation, $\phi_{1}$ becomes
the `Jacobi function' and that (\ref{G0}) agrees with the relativistic
relation between velocity and momentum. He adds (p.~230):

\begin{quotation}
The aim of the preceding arguments is to make it plausible that the relation
[(\ref{G0})] is strictly valid in the new mechanics.
\end{quotation}
The importance of the relation (\ref{G0}) for de Broglie is, of course,
that it embodies the identity of the principles of Maupertuis and Fermat.

As in the free case, de Broglie then goes on to consider an ensemble of such
particles, with each particle represented by a moving singularity. He assumes
that the phase function $\phi_{1}$ is the same for all the particles, whose
velocities are then given by (\ref{G0}). The (time-independent) particle
density $\rho(\mathbf{x})$ must then satisfy the continuity equation%
\begin{equation}
\mathbf{\nabla}\cdot(\rho\mathbf{v})=0\ .
\end{equation}
Once again, de Broglie introduces a representation of the ensemble by a
continuous solution of (\ref{KG1}). In the presence of the potential $V$, the
solution is taken to have the form%
\begin{equation}
\Psi(\mathbf{x},t)=a(\mathbf{x})\cos\frac{2\pi}{h}\phi^{\prime}(\mathbf{x}%
,t)=a(\mathbf{x})\cos2\pi\left(  \nu t-\frac{1}{h}\phi_{1}^{\prime}%
(\mathbf{x})\right)  \ .
\end{equation}
Again, writing $\Psi$ `in complex form', substituting into (\ref{KG1}) and
taking real and imaginary parts, de Broglie obtains two coupled partial
differential equations, now in $a$ and $\phi_{1}^{\prime}$. One of these reads%
\begin{equation}
\hslash^{2}\frac{\nabla^{2}a}{a}=\left(  \mathbf{\nabla}\phi_{1}^{\prime
}\right)  ^{2}-\frac{1}{c^{2}}(E-V)^{2}+m_{0}^{2}c^{2}\ . \label{pde2}%
\end{equation}
In this case de Broglie notes that if the left-hand side is negligible,
(\ref{pde2}) reduces to the equation of geometrical optics associated with the
wave equation (\ref{KG1}).

Comparing (\ref{pde1}) and (\ref{pde2}), de Broglie notes that when the
left-hand sides are negligible, one recovers both classical mechanics (for a
moving singularity) and geometrical optics (for a continuous wave). In this
limit, the functions $\phi_{1}$ and $\phi_{1}^{\prime}$ are identical, being
both equal to the `Jacobi function'.

At this point de Broglie makes a crucial assumption. He proposes to assume, as
a hypothesis, that $\phi_{1}$ and $\phi_{1}^{\prime}$ are \textit{always}
equal, regardless of any approximation (p.~231--2):

\begin{quotation}
We now make the essential hypothesis that $\phi_{1}$ and $\phi_{1}^{\prime}$
are still identical when [the left-hand sides of (\ref{pde1}) and
(\ref{pde2})] can no longer be neglected. Obviously, this requires that one
have:%
\[
\frac{\nabla^{2}a}{a}=\frac{\square f}{f}\ .
\]

We shall refer to this postulate by the name of `principle of the double
solution', because it implies the existence of two sinusoidal solutions of
equation [(\ref{KG1})] having the same phase factor, the one consisting of a
point-like singularity and the other having, on the contrary, a continuous amplitude.
\end{quotation}
This hypothesis expresses, in a more concrete form, de Broglie's
earlier idea of `phase harmony' between the internal oscillation of a particle
and the oscillation of an accompanying extended wave: the condition that
$\phi_{1}$ and $\phi_{1}^{\prime}$ should coincide amounts to a phase harmony
between the singular $u$-wave representing the point-like particle and the
continuous $\Psi$-wave.

Given this condition, de Broglie deduces that the ratio $\rho/a^{2}(E-V)$ is
constant along the particle trajectories. Since each particle is assumed to
begin in free space, where $V=0$ and (according to (\ref{B0})) $\rho=Ka^{2}$,
de Broglie deduces that in general%
\begin{equation}
\rho(\mathbf{x})=Ka^{2}(\mathbf{x})\left(  1-\frac{V(\mathbf{x})}{E}\right)
\ , \label{B1'}%
\end{equation}
where $K$ is a constant. The continuous wave then gives the ensemble density
at each point. In the nonrelativistic approximation, this becomes%
\begin{equation}
\rho(\mathbf{x})=Ka^{2}(\mathbf{x})\ . \label{B2}%
\end{equation}
De Broglie notes that each possible initial position for the particle gives
rise to a possible trajectory, and that an ensemble of initial positions gives
rise to an ensemble of motions. Again, de Broglie takes $\rho(\mathbf{x}%
)d^{3}\mathbf{x}$ as the probability for a single particle to be in a volume
element $d^{3}\mathbf{x}$. This probability is, as he notes, given in terms of
$\Psi$ by (\ref{B1'}) or (\ref{B2}). De Broglie adds that $\Psi$ also
determines the trajectories (p.~232):

\begin{quotation}
The form of the trajectories is moreover equally determined by knowledge of
the continuous wave, since these trajectories are orthogonal to the surfaces
of equal phase.
\end{quotation}
Here we see the first suggestion that $\Psi$ itself may be regarded as
determining the trajectories, an idea that de Broglie proposes more fully at
the end of the paper.

After sketching how the above could be used to calculate probabilities for
electron scattering off a fixed potential, de Broglie goes on to generalise
his results to the case of a particle of charge $e$ in a time-dependent
electromagnetic field $\left(  \mathcal{V}(\mathbf{x},t),\mathbf{A}%
(\mathbf{x},t)\right)  $. He writes down the corresponding Klein-Gordon
equation, and once again considers solutions of the form (\ref{sing0}) with a
moving singularity, following the same procedure as before. Because of the
time-dependence of the potentials, $\phi$ no longer takes the form
(\ref{Ph0}). Instead of (\ref{G0}), the velocity of the singularity is now
found to be%
\begin{equation}
\mathbf{v}=-c^{2}\frac{\mathbf{\nabla}\phi+\frac{e}{c}\mathbf{A}}{\dot{\phi
}-e\mathcal{V}}\ , \label{g1}%
\end{equation}
or, in the nonrelativistic approximation,%
\begin{equation}
\mathbf{v}=-\frac{1}{m_{0}}\left(  \mathbf{\nabla}\phi+\frac{e}{c}%
\mathbf{A}\right)  \ .
\end{equation}
Once again, de Broglie then considers an ensemble of such moving
singularities, with the same phase function $\phi(\mathbf{x},t)$. The density
$\rho(\mathbf{x},t)$ now obeys a continuity equation%
\begin{equation}
\frac{\partial\rho}{\partial t}+\mathbf{\nabla}\cdot(\rho\mathbf{v})=0\ ,
\end{equation}
with velocity field $\mathbf{v}$ given by (\ref{g1}). Again, de Broglie
proceeds to represent the motion of the ensemble by means of a continuous
solution $\Psi$ of the wave equation, this time of the form%
\begin{equation}
\Psi(\mathbf{x},t)=a(\mathbf{x},t)\cos\frac{2\pi}{h}\phi^{\prime}%
(\mathbf{x},t)\ ,
\end{equation}
with a time-dependent amplitude and a phase no longer linear in $t$. And
again, de Broglie assumes the principle of the double solution, that the phase
functions $\phi(\mathbf{x},t)$ and $\phi^{\prime}(\mathbf{x},t)$ are the same.
From this, de Broglie deduces that the ratio $\rho/a^{2}(\dot{\phi
}-e\mathcal{V})$ is constant along particle trajectories. Using once more the
expression $\rho=Ka^{2}$ for free space, where $\mathcal{V}=0$ and $\dot{\phi
}=E_{0}$, de Broglie argues that in general%
\begin{equation}
\rho(\mathbf{x},t)=\frac{K}{E_{0}}a^{2}(\mathbf{x},t)\left(  \dot{\phi
}-e\mathcal{V}\right)  =K^{\prime}a^{2}\left(  \dot{\phi}-e\mathcal{V}\right)
\ , \label{B3}%
\end{equation}
and that in the nonrelativistic limit $\rho$ is still proportional to $a^{2}$.
As before, the ensemble may be regarded as composed of all the possible
positions of a single particle, of which only the initial velocity is known,
and the probability that the particle is in the volume element $d^{3}%
\mathbf{x}$ at time $t$ is equal to $\rho(\mathbf{x},t)d^{3}\mathbf{x}$ and is
given in terms of $\Psi$ by (\ref{B3}).

De Broglie now shows how the above results for the motion of a particle in a
potential $V=e\mathcal{V}$ (ignoring for simplicity the vector potential
$\mathbf{A}$) may be obtained from \textit{classical} mechanics, with a
Lagrangian of the standard form%
\begin{equation}
L=-M_{0}c^{2}\sqrt{1-v^{2}/c^{2}}-V\ ,
\end{equation}
by assuming that the particle has a variable proper mass%
\begin{equation}
M_{0}(\mathbf{x},t)=\sqrt{m_{0}^{2}-\frac{\hslash^{2}}{c^{2}}\frac{\square
a}{a}}\ , \label{M}%
\end{equation}
where $a$ is the amplitude of the continuous $\Psi$ wave. Further, he
considers this point of view in the nonrelativistic limit. Writing
$M_{0}(\mathbf{x},t)=m_{0}+\varepsilon(\mathbf{x},t)$ with $\varepsilon$
small, the Lagrangian takes the approximate form%
\begin{equation}
L=-m_{0}c^{2}+\frac{1}{2}m_{0}v^{2}-\varepsilon c^{2}-V\ .
\end{equation}
As de Broglie remarks (p.~237):

\begin{quotation}
Everything then takes place as if there existed, in addition to $V$, a
potential energy term $\varepsilon c^{2}$.
\end{quotation}
This extra term coincides, of course, with Bohm's quantum potential $Q$
(equation (\ref{QQ})), since to lowest order%
\begin{equation}
\varepsilon c^{2}=-\frac{\hslash^{2}}{2m_{0}}\frac{\square a}{a}%
=-\frac{\hslash^{2}}{2m_{0}}\frac{\nabla^{2}a}{a}+\frac{\hslash^{2}}%
{2m_{0}c^{2}}\frac{\ddot{a}}{a}%
\end{equation}
and the second term is negligible in the nonrelativistic limit ($c\rightarrow
\infty$).

As we have already repeatedly remarked, for de Broglie the guidance equation
$\mathbf{v}\propto\mathbf{\nabla}\phi$ for \textit{velocity} is the
fundamental equation of motion, expressing as it does the identity of the
principles of Maupertuis and Fermat. Even so, de Broglie points out that one
\textit{can} write the dynamics of the particle in classical (Newtonian or
Einsteinian) terms, provided one includes a variable proper mass or, in the
nonrelativistic limit, an additional quantum potential. It is this latter,
Newtonian formulation that Bohm proposes in 1952.

De Broglie now turns to the case of a many-body system, consisting of $N$
particles with proper masses $m_{1},m_{2},...,m_{N}$. In the
nonrelativistic approximation, if the system has total (Newtonian) energy $E$
and potential energy $V(\mathbf{x}_{1},...,\mathbf{x}_{N})$, then following
Schr\"{o}dinger one may consider the propagation of a wave $u$ in the
$3N$-dimensional configuration space, satisfying the wave equation%
\begin{equation}
\sum_{i=1}^{N}-\frac{\hslash^{2}}{2m_{i}}\nabla_{i}^{2}u+Vu=Eu \label{SchOne}%
\end{equation}
(the time-independent Schr\"{o}dinger equation). De Broglie remarks that this
seems natural, because (\ref{SchOne}) is the obvious generalisation of the
one-body Schr\"{o}dinger equation, which follows from the nonrelativistic
limit of de Broglie's wave equation (\ref{KG1}). However, de Broglie
criticises Schr\"{o}dinger's interpretation, according to which a particle is
identified with an extended, non-singular wave packet, having no precise
position or trajectory. De Broglie objects that, without well-defined particle
positions, the coordinates $\mathbf{x}_{1},...,\mathbf{x}_{N}$ used to
construct the configuration space would have no meaning. Further, de Broglie
asserts that configuration space is `purely abstract', and that a wave
propagating in this space cannot be a physical wave: instead, the physical
picture of the system must involve $N$ waves propagating in 3-space.

`What then', asks de Broglie (p.~238), `is the true meaning of the
Schr\"{o}dinger equation?' To answer this question, de Broglie considers, for
simplicity, the case of two particles, which for him are singularities of a
wave-like phenomenon in 3-space. Neglecting the vector potential, the two
singular waves $u_{1}(\mathbf{x},t)$ and $u_{2}(\mathbf{x},t)$ satisfy the
coupled equations%
\begin{equation}
\begin{array}{lcl}
{\displaystyle \square u_{1}+\frac{2i}{\hslash c^{2}}V_{1}\frac{\partial u_{1}}{\partial
t}-\frac{1}{\hslash^{2}}\left(  m_{1}^{2}c^{2}-\frac{V_{1}^{2}}{c^{2}}\right)
u_{1}}  &  = & 0\ ,\\[1em]
{\displaystyle \square u_{2}+\frac{2i}{\hslash c^{2}}V_{2}\frac{\partial u_{2}}{\partial
t}-\frac{1}{\hslash^{2}}\left(  m_{2}^{2}c^{2}-\frac{V_{2}^{2}}{c^{2}}\right)
u_{2}} &  = & 0\ ,
\end{array}\label{2pdes}
\end{equation}
with potentials%
\begin{equation}
V_{1}=V(\left\vert \mathbf{x}-\mathbf{x}_{2}\right\vert )\ \ \ \ ,\ V_{2}%
=V(\left\vert \mathbf{x}-\mathbf{x}_{1}\right\vert )\ , \label{V12}%
\end{equation}
where $\mathbf{x}_{1}$, $\mathbf{x}_{2}$ are the positions of the two
particles (or singularities). The propagation of each wave then depends,
through $V$, on the position of the singularity in the other wave.

De Broglie is now faced with the formidable problem of solving the
simultaneous partial differential equations (\ref{2pdes}), in order to obtain
the motions of the two singularities. Not surprisingly, de Broglie does not
attempt to carry through such a solution. Instead he notes that, in classical
mechanics in the nonrelativistic limit, it is possible to find a `Jacobi
function' $\phi(\mathbf{x}_{1},\mathbf{x}_{2})$ for the two-particle system,
such that the particle momenta are given by%
\begin{equation}
m_{1}\mathbf{v}_{1}=-\mathbf{\nabla}_{1}\phi\ ,\ \ \ \ m_{2}\mathbf{v}%
_{2}=-\mathbf{\nabla}_{2}\phi\ . \label{G12}%
\end{equation}
De Broglie then asks (p.~238): `Can the new mechanics .... define such a
function $\phi$?'

De Broglie is asking whether, if one could solve the coupled equations
(\ref{2pdes}), the resulting motions of the singularities would (in the
nonrelativistic limit) satisfy the guidance equations (\ref{G12}) for some
function $\phi(\mathbf{x}_{1},\mathbf{x}_{2})$. He then asserts, on the basis
of an incorrect argument (to which we shall return in a moment), that this
will indeed be the case. Having concluded that a function $\phi(\mathbf{x}%
_{1},\mathbf{x}_{2})$ generating the motions of the singularities will in fact
exist, he goes on to identify $\phi(\mathbf{x}_{1},\mathbf{x}_{2})$ with the
phase of a continuous solution $\Psi(\mathbf{x}_{1},\mathbf{x}_{2},t)$ of the
Schr\"{o}dinger equation (\ref{SchOne}) for the two-particle system. While de
Broglie does not say so explicitly, in effect he \textit{assumes} that there
exists a continuous solution of (\ref{SchOne}) whose phase coincides with the
function $\phi(\mathbf{x}_{1},\mathbf{x}_{2})$ that generates the motions of
the singularities via (\ref{G12}). This seems to be de Broglie's answer to the
question he raises, as to the `true meaning' of the Schr\"{o}dinger equation:
solving the Schr\"{o}dinger equation (\ref{SchOne}) in configuration space
provides an effective means of obtaining the motions of the singularities
without having to solve the coupled partial differential equations
(\ref{2pdes}) in 3-space.

De Broglie's incorrect argument for the existence of an appropriate function
$\phi$, derived from the equations (\ref{2pdes}), proceeds as follows. He
first imagines that the motion of singularity 2 is already known, so that the
problem of motion for singularity 1 reduces to a case that has already been
discussed, of a single particle in a given time-dependent potential. For this
case, it has already been shown that the ensemble of possible trajectories may
be represented by a continuous wave%
\begin{equation}
a_{1}(\mathbf{x},t)e^{(i/\hslash)\phi_{1}(\mathbf{x},t)}\ . \label{Psi1}%
\end{equation}
Further, as was shown earlier, the equations of motion for the particle may be
written in Lagrangian form with an additional potential $\varepsilon
_{1}(\mathbf{x}_{1},t)c^{2}$. Similarly, considering the motion of singularity
1 as known, the motion of singularity 2 can be described by a Lagrangian with
an additional potential $\varepsilon_{2}(\mathbf{x}_{2},t)c^{2}$. The two sets
of (classical) equations of motion can be derived from a single Lagrangian, de
Broglie argues, only if $\varepsilon_{1}$ and $\varepsilon_{2}$ reduce to a
\textit{single} function $\varepsilon(\left\vert \mathbf{x}_{1}-\mathbf{x}%
_{2}\right\vert )$ of the interparticle separation. `If we assume this', says
de Broglie, the total Lagrangian for the system will be%
\begin{equation}
L=\frac{1}{2}m_{1}v_{1}^{2}+\frac{1}{2}m_{2}v_{2}^{2}-V(\left\vert
\mathbf{x}_{1}-\mathbf{x}_{2}\right\vert )-\varepsilon(\left\vert
\mathbf{x}_{1}-\mathbf{x}_{2}\right\vert )c^{2}\ ,
\end{equation}
and one will be able to deduce, in the usual fashion, the existence of a
function $\phi(\mathbf{x}_{1},\mathbf{x}_{2})$ satisfying the equations
(\ref{G12}). It is not entirely clear from the text if de Broglie is really
convinced that $\varepsilon_{1}$ and $\varepsilon_{2}$ will indeed reduce to a
single function $\varepsilon(\left\vert \mathbf{x}_{1}-\mathbf{x}%
_{2}\right\vert )$. With hindsight one sees, in fact, that this will
\textit{not} usually be the case. For it amounts to requiring that the quantum
potential $Q(\mathbf{x}_{1},\mathbf{x}_{2})$ for two particles should take the
form $Q(\mathbf{x}_{1},\mathbf{x}_{2})=Q(\left\vert \mathbf{x}_{1}%
-\mathbf{x}_{2}\right\vert )$, which is generally false.\footnote{One might
also question the meaning of de Broglie's wave function (\ref{Psi1}) for
particle 1, in the context of a system of two interacting particles, which
must have an entangled wave function $\Psi(\mathbf{x}_{1},\mathbf{x}_{2},t)$.}

After giving his incorrect argument for the existence of $\phi$, de Broglie
points out that, since the particles have definite trajectories, it is
meaningful to consider their six-dimensional configuration space, and to
represent the two particles by a single point in this space, with velocity
components given by (\ref{G12}). If the initial velocity components are given,
but the initial positions of the two particles are not, one may consider an
ensemble of representative points $(\mathbf{x}_{1},\mathbf{x}_{2})$, whose
density $\rho(\mathbf{x}_{1},\mathbf{x}_{2})$ in configuration space satisfies
the continuity equation,%
\begin{equation}
\mathbf{\nabla}\cdot(\rho\mathbf{v})=0\ , \label{Cont6}%
\end{equation}
where $\mathbf{v}$ is the (six-dimensional) velocity field given by (\ref{G12}).

De Broglie then considers, as we have said, a continuous solution of the
Schr\"{o}dinger equation (\ref{SchOne}), of the form%
\begin{equation}
\Psi(\mathbf{x}_{1},\mathbf{x}_{2},t)=A(\mathbf{x}_{1},\mathbf{x}%
_{2})e^{(i/\hslash)\phi}\ ,
\end{equation}
with the tacit assumption that the phase $\phi$ is the same function $\phi$
whose existence has been (apparently) established by the above (incorrect)
argument. He then shows, from (\ref{SchOne}), that $A^{2}$ satisfies the
configuration-space continuity equation (\ref{Cont6}). De Broglie concludes
that $A^{2}d^{3}\mathbf{x}_{1}d^{3}\mathbf{x}_{2}$ is the probability for the
representative point $(\mathbf{x}_{1},\mathbf{x}_{2})$ to be present in a
volume element $d^{3}\mathbf{x}_{1}d^{3}\mathbf{x}_{2}$ of configuration space.

Here, then, de Broglie has arrived at a higher-dimensional analogue of his
previous results: the wave function $\Psi$ in configuration space determines
the ensemble probability density through its amplitude, as well as the motion
of a single system through its phase.

De Broglie ends this section by remarking that it seems difficult to find an
equation playing a role similar to (\ref{SchOne}) outside of the nonrelativistic
approximation, and that Fermi's calculation concerning the scattering of
electrons by a rotator may be regarded as illustrating the above.

From the point of view of the history of pilot-wave theory, the most important
section of de Broglie's paper now follows. Entitled `The pilot wave' (`L'onde
pilote'), it begins by recalling the results obtained in the case of a single
particle in a time-dependent potential, which de Broglie states may be
summarised by `the two fundamental formulas' (\ref{g1}) and (\ref{B3}) for the
particle velocity and probability density respectively. De Broglie notes that
he arrived at the velocity formula (\ref{g1}) by invoking the principle of the
double solution --- a principle that is valid in free space but which `remains
a hypothesis in the general case' (p.~241). At this point, de Broglie makes a
remarkable suggestion: that instead of trying to derive the velocity field
(\ref{g1}) from an underlying theory, one could simply take it as a postulate,
and regard $\Psi$ as a physically-real `pilot wave' guiding the motion of the
particle. To quote de Broglie (p.~241):

\begin{quotation}
But if one does not wish to invoke the principle of the double solution, it is
acceptable to adopt the following point of view: one will assume the
existence, as distinct realities, of the material point and of the continuous
wave represented by the function $\Psi$, and one will take it as a postulate
that the motion of the point is determined as a function of the phase of the
wave by the equation [(\ref{g1})]. One then conceives the continuous wave as
guiding the motion of the particle. It is a pilot wave.
\end{quotation}

This is the first appearance in the literature of what we now know as
pilot-wave or de Broglian dynamics (albeit stated explicitly for only a single
particle). The pilot wave $\Psi$ satisfying the Schr\"{o}dinger equation, and
the material point, are regarded as `distinct realities', with the former
guiding the motion of the latter according to the velocity law (\ref{g1}).

De Broglie made it clear, however, that he thought such a dynamics could be
only provisional (p.~241):

\begin{quotation}
By thus taking [(\ref{g1})] as a postulate, one avoids having to justify it by
the principle of the double solution; but this can only be, I believe, a
provisional attitude. No doubt one will indeed have to \textit{reincorporate}
the particle into the wave-like phenomenon, and one will probably be led back
to ideas analogous to those that have been developed above.
\end{quotation}
As we shall discuss in the next section, de Broglie's proposal of
pilot-wave theory as a provisional measure has striking analogues in the early
history of Newtonian gravity and of Maxwellian electromagnetic theory.

De Broglie goes on to point out two important applications of the formulas
(\ref{g1}) and (\ref{B3}). First, applying the theory to light, according to
(\ref{B3}) `the density of the photons is proportional to the square of the
amplitude' of the guiding wave $\Psi$, yielding agreement with the predictions
of wave optics. Second, an ensemble of hydrogen atoms with a definite state
$\Psi$ will have a mean electronic charge density proportional to $\left\vert
\Psi\right\vert ^{2}$, as assumed by Schr\"{o}dinger.

De Broglie's `Structure' paper ends with some remarks on the pressure exerted
by particles on a wall (for example of a box of gas). He notes that according
to (\ref{g1}), because of the interference between the incident and reflected
waves, the particles will not actually strike the wall, raising the question
of how the pressure is produced. De Broglie claims that the pressure comes
from stresses in the interference zone, as appear in an expression obtained by
Schr\"{o}dinger (1927c) for the stress-energy tensor associated with the wave $\Psi$.

\subsection{Significance of de Broglie's `Structure' paper}\label{sigStruc}

The exceptional quality of de Broglie's `Structure' paper in \textit{Journal
de Physique} was noted by Pauli, in a letter to Bohr dated 6 August 1927,
already quoted as the epigraph to the last subsection: `.... it is very rich
in ideas and very sharp, and on a much higher level than the childish papers
by Schr\"{o}dinger ....' (Pauli 1979, pp.~404--5). In the same letter, Pauli
suggested that Bohr would have to refer to de Broglie's paper in his Como
lecture (in which Bohr developed the idea of complementarity between waves and
particles); and in fact, in what was to remain an unpublished version of the
Como lecture, Bohr did take an explicit stand against de Broglie's ideas (Bohr
1985, pp.~89--98). Bohr characterised de Broglie's work as attempting `to
reconcile the two apparently contradictory sides of the phenomena by regarding
the individual particles or light quanta as singularities in the wave field';
further, Bohr suggested that de Broglie's view rested upon `the concepts of
classical physics' and was therefore not `suited to help us over the
fundamental difficulties' (Bohr 1985, p.~92). On the whole, though, this paper
by de Broglie has been essentially ignored, by both physicists and historians.

De Broglie's `Structure' paper may be summed up as follows. De Broglie has a
model of particles as singularities of 3-space waves $u_{i}$, in which the
motion of the individual particles, as well as the ensemble distribution of
the particles, are determined by a continuous wave function $\Psi$. (We
emphasise that de Broglie has both the wave function $\Psi$ and the singular
$u$-waves.) The result is a first-order theory of motion, based on the
velocity law $m_{i}\mathbf{v}_{i}=\mathbf{\nabla}_{i}\phi$, in which the
principles of Maupertuis and Fermat are unified. Then, at the end of the
paper, de Broglie recognises that his singularity model of particles can be
dropped, and that the results he has obtained --- the formulas for velocity
and probability density in terms of $\Psi$ --- can be simply postulated,
yielding pilot-wave theory as a provisional measure.

A few months later, in October 1927, de Broglie presented this pilot-wave
theory at the fifth Solvay conference, for a nonrelativistic system of $N$
particles guided by a wave function $\Psi$ in configuration space. Before
discussing de Broglie's Solvay report, however, it is worth pausing to
consider the role played by the `Structure' paper in de Broglie's thinking.

From a historical point of view, the significance of de Broglie's `Structure'
paper inevitably depends on the significance one ascribes to the provisional
pilot-wave theory arrived at in that paper. Given what we know today --- that
pilot-wave theory provides a consistent account of quantum phenomena --- de
Broglie's `Structure' paper now seems considerably more significant than it
has seemed in the past.

From the point of view of pilot-wave theory as we know it today, the singular
$u$-waves played a similar role for de Broglie as the material ether did for
Maxwell in electromagnetic theory. In both cases, there was a conceptual
scaffolding that was used to build a new theory, and that could be dropped
once the results had been arrived at. De Broglie recognised at the end of his
paper that, if one took pilot-wave dynamics as a provisional theory, then the
scaffolding he had used to construct this theory could be dropped. At the same
time, de Broglie insisted that taking such a step was indeed only provisional,
and that an underlying theory was still needed, probably along the lines he
had been pursuing.

Similar situations have arisen before in the history of science. Abstracting
away the details of a model based on an older theory sometimes results in a
new theory in its own right, involving new concepts, where, however, the
author regards the new theory as only a provisional measure, and expects that
the model based on the older theory will eventually provide a proper basis for
the provisional theory. Thus, for example, Newton tried to explain gravitation
on the basis of action by contact, involving a material medium filling space
--- the same `Aethereal Medium', in fact, as he thought responsible for the
interference and diffraction of light (Newton 1730; reprint, pp.~350--53).
Newton regarded his theory of gravitation, with action at a distance, as
merely a provisional and phenomenological theory, that would later find an
explanation in terms of contiguous action. Eventually, however, the concept of
`gravitational action at a distance' became widely accepted in its own right
(though it was later to be overthrown, of course, by general relativity).
Similarly, Maxwell used mechanical models of an ether to develop his theory of
the electromagnetic field. Maxwell himself may or may not have recognised that
this scaffolding could be dropped (Hendry 1986). But certainly, many of his
immediate followers did not: they regarded working only with Maxwell's
equations as provisional and phenomenological, pending the development of an
underlying mechanical model of an ether. Again, the concept of
`electromagnetic field' eventually became widely accepted as a physical entity
in its own right.

In the case of de Broglie in 1927, in his `Structure' paper (and subsequently
at the fifth Solvay conference), he arrived at the new concept of `pilot wave
in configuration space', an entirely new kind of physical entity that,
according to de Broglie's (provisional) theory, guides the motion of material systems.

\section{1927 Solvay report: the new dynamics of quanta}\label{deB-1927-Solvay-report}

We now turn to a (brief) summary of and commentary on de Broglie's report at
the fifth Solvay conference.\footnote{Again, our notation sometimes departs
from de Broglie's.} As we have noted, the theory presented in this report is
pilot-wave theory as we know it today, for a nonrelativistic many-body system,
with a guiding wave in configuration space that determines the particle
velocities according to de Broglie's basic law of motion. De Broglie's ideas
about particles as singularities of 3-space waves are mentioned only briefly.
The theory de Broglie presents in Brussels in October 1927 is, indeed, just
the provisional theory he proposed a few months earlier at the end of his
`Structure' paper (though now explicitly applied to many-body systems as well).

De Broglie begins part I of his report by reviewing the results obtained in
his doctoral thesis. The energy $E$ and momentum $\mathbf{p}$ of a particle
are determined by the phase $\phi$ of an associated wave:\footnote{Again, de
Broglie's phase $\phi$ has a sign opposite to the phase $S$ as we would
normally define it now.}%
\begin{equation}
E=\partial\phi/\partial t\ ,\ \ \ \ \mathbf{p}=-\mathbf{\nabla}\phi\ .
\label{gu0}%
\end{equation}
(These are the relativistic guidance equations of de Broglie's early
pilot-wave theory of 1923--24.) It follows that, as de Broglie remarks, `the
principles of least action and of Fermat are identical' (p.~\pageref{identical-principles}). 
Quantisation conditions appear in a natural way, and there are far-reaching implications
for statistical mechanics. De Broglie ends the review of his early work with
some general remarks. He points out (p.~\pageref{always-trajectories}) that he `has
always assumed that the material point occupies a well-defined position in
space', and he asserts that as a result the wave amplitude must be singular,
or take very large values in a small region, somewhere within the extended
wave. Significantly, de Broglie adds: `But, in fact, the form of the amplitude
plays no role in the results reviewed above'. This seems to be a first hint
that the actual results, such as (\ref{gu0}), do not depend on the details of
any underlying model of the particles as moving singularities.

De Broglie then outlines the work of Schr\"{o}dinger. He notes that
Schr\"{o}dinger's wave equation is constructed in order that the phase $\phi$
of the wave function%
\begin{equation}
\Psi=a\cos\frac{2\pi}{h}\phi
\end{equation}
be a solution of the Hamilton-Jacobi equation in the geometrical-optics limit.
He points out that Schr\"{o}dinger identifies particles with localised wave
packets instead of with a small concentration within an extended wave, and
that for a many-body system Schr\"{o}dinger has a wave $\Psi$ propagating in
configuration space. For both the one-body and many-body cases, de Broglie
writes down the time-independent Schr\"{o}dinger equation only, with a static
potential energy function. As we shall see, de Broglie in fact considers
non-stationary wave functions as well. For the one-body case, de Broglie also
writes down a relativistic, and time-dependent, equation --- what we now know
as the Klein-Gordon equation in an external electromagnetic field.

De Broglie then raises two conceptual difficulties with Schr\"{o}dinger's work
(similar in spirit to the objections he raises in `Structure'): (1) He
questions how one can meaningfully construct a configuration space without a
real configuration, asserting that it `seems a little paradoxical to construct
a configuration space with the coordinates of points that do not exist' 
(p.~\pageref{paradoxical}). (2) He claims that the physical meaning of the wave
$\Psi$ cannot be compared with that of an ordinary wave in 3-space, because
the number of dimensions of the abstract space on which $\Psi$ is defined is
determined by the number of degrees of freedom of the system. (This second
point seems indeed a very effective way to make clear that $\Psi$ is quite
different from a conventional wave or field on 3-space.)

Part I of the report ends with some remarks on Born's statistical
interpretation, which de Broglie asserts is analogous to his own.

In part II of his report, de Broglie presents his own interpretation of the
wave function. Pilot-wave theory is clearly formulated, first for a single
particle, and then for a system of $N$ particles. Several applications are
outlined. It is interesting to see how de Broglie motivates his theory and
compares it with the contenders.

De Broglie begins by asking what the relationship is between particles and the
wave $\Psi$. He first considers a single relativistic particle in an external
electromagnetic field with potentials ($\mathcal{V}$, $\mathbf{A}$). De
Broglie notes that, in the classical limit, the phase $\phi$ of $\Psi$ obeys
the Hamilton-Jacobi equation, and the velocity of the particle is given by%
\begin{equation}
\mathbf{v}=-c^{2}\frac{\mathbf{\nabla}\phi+\frac{e}{c}\mathbf{A}}{\dot{\phi
}-e\mathcal{V}} \label{guid1}%
\end{equation}
(the same formula (\ref{g1}) discussed in `Structure'). De Broglie then
proposes that this velocity formula is valid even outside the classical limit
(p.~\pageref{oldp90}, italics in the original):

\begin{quotation}
We propose to assume by induction that this formula is still valid when the
old Mechanics is no longer sufficient, that is to say when [$\phi$] is no
longer a solution of the Jacobi equation. If one accepts this hypothesis,
which appears justified by its consequences, the formula [(\ref{guid1})]
completely determines the motion of the corpuscle \textit{as soon as one is
given its position at an initial instant}. In other words, the function
[$\phi$], just like the Jacobi function of which it is the generalisation,
determines a whole class of motions, and to know which of these motions is
actually described it suffices to know the initial position.
\end{quotation}

We emphasise that there is no appeal to singular $u$-waves anywhere in de
Broglie's report. No use is made of the principle of the double solution. As
he had suggested at the end of his `Structure' paper a few months earlier, de
Broglie simply postulates the basic equations of pilot-wave dynamics, without
trying to derive them from anything else. To motivate the guidance equation,
de Broglie simply generalises the classical Hamilton-Jacobi velocity formula
to the non-classical domain: he assumes `by induction' that the formula holds
even outside the classical limit. And de Broglie is quite explicit that he is
proposing a first-order theory of motion, based on velocities: given the wave
function, the initial position alone determines the trajectory.

So far, then, for a single particle de Broglie has a (relativistic) wave
function $\Psi=a\cos\frac{2\pi}{h}\phi$ whose phase $\phi$ determines the
particle velocity by the guidance equation (\ref{guid1}). De Broglie now goes
on to point out that, for an ensemble of particles guided by the velocity
field (\ref{guid1}), the distribution%
\begin{equation}
Ka^{2}\left(  \dot{\phi}-e\mathcal{V}\right)
\end{equation}
is preserved in time (that is, equivariant). He concludes that, if the initial
position of the particle is ignored, then the probability for the particle to
be present (at time $t$) in a spatial volume $d\tau$ is%
\begin{equation}
\pi d\tau=Ka^{2}\left(  \dot{\phi}-e\mathcal{V}\right)  d\tau\ . \label{prob1}%
\end{equation}
(This is the same probability formula (\ref{B3}) arrived at in `Structure'.)

De Broglie's expression `probabilit\'{e} de pr\'{e}sence' makes it clear that
we have to do with a probability for the electron \textit{being} somewhere,
and not merely with a probability for an experimenter \textit{finding} the
electron somewhere --- cf. Bell (1990, p.~29). De Broglie simply assumes that
the equivariant distribution is the correct probability measure for a particle
of unknown position. In fact, this distribution is only an equilibrium
distribution, analogous to thermal equilibrium in classical statistical
mechanics (see section~\ref{det-prob}).

De Broglie sums up his results so far (p.~\pageref{oldp9192}):

\begin{quotation}
In brief, in our hypotheses, each wave $\Psi$ determines a `class of
motions',\ and each one of these motions is governed by equation
[(\ref{guid1})] when one knows the initial position of the corpuscle. If one
ignores this initial position, the formula [(\ref{prob1})] gives the
probability for the presence of the corpuscle in the element of volume $d\tau$
at the instant $t$. The wave $\Psi$ then appears as both a \textit{pilot wave}
(F\"{u}hrungsfeld of Mr Born) and a \textit{probability wave}.
\end{quotation}
There are, de Broglie adds, no grounds for abandoning determinism, and
in this his theory differs from that of Born.

De Broglie adds that, in the nonrelativistic approximation (where $\dot{\phi
}-e\mathcal{V}\approx m_{0}c^{2}$), the guidance and probability formulas
(\ref{guid1}) and (\ref{prob1}) reduce to%
\begin{equation}
\mathbf{v}=-\frac{1}{m_{0}}\left(  \mathbf{\nabla}\phi+\frac{e}{c}%
\mathbf{A}\right)  \label{guid2}%
\end{equation}
and%
\begin{equation}
\pi=\mathrm{const}\cdot a^{2}\ . \label{prob2}%
\end{equation}
These are the standard pilot-wave equations for a single particle (in an
external electromagnetic field).

De Broglie then remarks on how the above formulas may be applied to the
scattering of a single particle by a fixed potential. The ensemble of incident
particles may be represented by a plane wave with a uniform probability
distribution. Upon entering a region of non-zero field, the behaviour of the
wave function may be calculated using Born's perturbation theory (for example,
for the Rutherford scattering of an electron by an atomic nucleus). De Broglie
draws an analogy between the scattering of the wave function $\Psi$ and the
classical scattering of light by a refracting medium.

As in his `Structure' paper, de Broglie remarks in passing that, for a
relativistic particle governed by the guidance equation (\ref{guid1}), one may
write down the equations of classical dynamics with a variable rest mass
$M_{0}$ given by equation (\ref{M}). As we have already noted, in the
nonrelativistic approximation this yields an additional potential energy term,
which is precisely Bohm's `quantum potential'. And, once again, we emphasise
that for de Broglie the equation $\mathbf{v}\propto\mathbf{\nabla}\phi$ for
\textit{velocity} is the fundamental equation of motion, expressing the
identity of the principles of Maupertuis and Fermat: de Broglie merely points
out that, if one wishes, the dynamics can be written in classical terms (as
Bohm did in 1952).

De Broglie then turns to the interpretation of interference and diffraction,
for the case of photons. Here, the guiding wave $\Psi$ is similar to but not
the same as a light wave.\footnote{De Broglie did not identify the photonic
pilot wave with the electromagnetic field. In the general discussion, on 
p.~\pageref{p192DRAFT1}, de Broglie explicitly states that in his theory the
wave $\Psi$ for the case of photons is distinct from the electromagnetic
field.} De Broglie considers scattering by fixed obstacles, in which case the
guiding wave may be taken to have a constant frequency $\nu$. The relativistic
equations (\ref{guid1}), (\ref{prob1}) then become%
\begin{equation}
\mathbf{v}=-\frac{c^{2}}{h\nu}\mathbf{\nabla}\phi,\;\;\;\;\pi=\mathrm{const}%
\cdot a^{2}\ .
\end{equation}
De Broglie points out that the second equation predicts the well-known
interference and diffraction patterns. He argues that the results will be the
same, whether the experiment is done with an intense beam over a short time or
with a feeble beam over a long time.

De Broglie remarks, as he did at the end of `Structure', on the stresses
appearing in Schr\"{o}dinger's expression for the stress-energy tensor of the
wave $\Psi$. According to de Broglie, these stresses provide an explanation
for the pressure exerted by light reflecting on a mirror --- despite the fact
that, according to (\ref{guid1}), the photons never actually strike the
surface of the mirror (as shown explicitly by Brillouin in the discussion
after de Broglie's report).

De Broglie then turns to the generalisation of the above dynamics to a
(nonrelativistic) \textit{many-body} system. Remarkably, many historians and
commentators have not noticed this proposal by de Broglie of a many-body
dynamics in configuration space, a proposal that is usually attributed to Bohm
(cf. section~\ref{hist-mis}.)

De Broglie begins by pointing out how the two difficulties he has raised
against Schr\"{o}dinger's wave mechanics might be solved. First, if a real
configuration exists at all times, one can meaningfully construct the
configuration space. As for the second difficulty, regarding the meaning of a
wave in the abstract configuration space, de Broglie makes the following
preliminary remark (p.~\pageref{otherwaves}):

\begin{quotation}
It appears to us certain that if one wants to \textit{physically} represent
the evolution of a system of $N$ corpuscles, one must consider the propagation
of $N$ waves in space, each of the $N$ propagations being determined by the
action of the $N-1$ corpuscles connected to the other waves.
\end{quotation}
This seems a clear reference to the theory of interacting singular
$u$-waves discussed a few months earlier in de Broglie's `Structure'
paper.\footnote{It might be thought that here de Broglie has in mind the fact
that $N$ moving particles may be associated with $N$ velocity fields in
3-space, which may be associated with $N$ (non-singular) guiding waves in
3-space. From a pilot-wave perspective, such $N$ guiding waves may be
identified with the $N$ `conditional' wave functions $\psi_{i}(\mathbf{x}%
_{i},t)\equiv\Psi(\mathbf{X}_{1},\mathbf{X}_{2},...,\mathbf{x}_{i}%
,...,\mathbf{X}_{N},t)$, where $\mathbf{x}_{i}$ ($i=1,2,...,N$) ranges over
all positions in 3-space and $\mathbf{X}_{j}$ is the \textit{actual} position
of the $j$th particle. Each $\psi_{i}(\mathbf{x}_{i},t)$ defines a wave in
3-space that determines the velocity of the $i$th particle (through the
gradient of its phase); and each $\psi_{i}(\mathbf{x}_{i},t)$ depends on the
positions of the other $N-1$ particles. However, given the context (in
particular the recent publication of `Structure'), it seems clear that here de
Broglie's $N$ waves are the singular $u$-waves of his double-solution theory.}
(The propagation of each wave must depend on the motion of the other $N-1$
particles, of course, in order to account for interactions between the
particles.) As we saw above (section~\ref{Structure}), in
`Structure' de Broglie explicitly considered the case of two particles, which
were represented by two singular waves $u_{1}(\mathbf{x},t)$ and
$u_{2}(\mathbf{x},t)$ satisfying a pair of coupled partial differential
equations (\ref{2pdes}), where the equation for each wave contained a
potential depending on the position of the singularity in the other wave.
Instead of trying to solve the equations, de Broglie assumed that the
resulting velocities of the moving singularities could be written as the
gradient of a function $\phi(\mathbf{x}_{1},\mathbf{x}_{2})$, which could be
identified with the phase of a solution $\Psi(\mathbf{x}_{1},\mathbf{x}%
_{2},t)$ of the (time-independent) Schr\"{o}dinger equation. In `Structure',
then, the wave function in configuration space appeared as an effective
description of the motions of the particles (or singularities). An echo of
this view is discernible in the Solvay report, which continues with the
following justification for introducing a guiding wave in configuration space
(p.~\pageref{andagain}):

\begin{quotation}
Nevertheless, if one focusses one's attention only on the corpuscles, one can
represent their states by a point in configuration space, and one can try to
relate the motion of this representative point to the propagation of a
fictitious wave $\Psi$ in configuration space.
\end{quotation}
De Broglie seems to be saying that, if one is concerned only with a
succinct mathematical account of particle motion (as opposed to a physical
representation), then this can be obtained by introducing a guiding wave in
configuration space. Certainly, the dynamics is much simpler with a single,
autonomous wave $\Psi$ in configuration space. As in his `Structure' paper,
this seems to be de Broglie's explanation for Schr\"{o}dinger's otherwise
mysterious configuration-space wave.

When de Broglie states that the wave $\Psi$ is `fictitious', he presumably
means that it has only mathematical, and not physical,
significance.\footnote{This is somewhat in contrast with de Broglie's
introduction of the pilot wave at the end of `Structure', where he refers to
the particle and the guiding wave as `distinct realities'. However, there de
Broglie explicitly proposed pilot-wave theory for a single particle only, and
it is likely that while he was comfortable with the idea of a physically real
pilot wave in 3-space (for the one-body case), he could not regard a pilot
wave in configuration space as having more than mathematical significance.} It
provides a convenient mathematical account of particle motion, pending a full
physical description by a more detailed theory. As we discussed in 
section~\ref{sigStruc}, the provisional introduction of a
`mathematical' description, pending the development of a proper `physical'
model, has distinguished precedents in the history of Newtonian gravity and
Maxwellian electrodynamics.

Having motivated the introduction of a guiding wave $\Psi$ in configuration
space, de Broglie suggests (p.~\pageref{singlematpoint}) that $\Psi$

\begin{quotation}
.... plays for the representative point of the system in configuration space
the same role of pilot wave and of probability wave that the wave $\Psi$ plays
in ordinary space in the case of a single material point.
\end{quotation}
Thus, in the nonrelativistic approximation, de Broglie considers $N$
particles with positions $\mathbf{x}_{1},\mathbf{x}_{2},...,
\mathbf{x}_{N}$. He states that the wave $\Psi$ determines the velocity of the
representative point in configuration space by the formula%
\begin{equation}
\mathbf{v}_{k}=-\frac{1}{m_{k}}\mathbf{\nabla}_{k}\phi\label{guid3}%
\end{equation}
(where $m_{k}$ is the mass of the $k$th particle). As in the case of a single
particle, notes de Broglie, the probability for the system to be present in a
volume element $d\tau$ of configuration space is%
\begin{equation}
\pi d\tau=\mathrm{const}\cdot a^{2}d\tau\ . \label{prob3}%
\end{equation}
These are the standard pilot-wave equations for a many-body system,
replacing the single-particle formulas (\ref{guid2}) and (\ref{prob2}). De
Broglie remarks that (\ref{prob3}) seems to agree with Born's results for
electron scattering by an atom and with Fermi's for scattering by a rotator.

Finally, at the end of his section on the many-body case, de Broglie notes
that it seems difficult to construct a wave $\Psi$ that can generate the
motion of a relativistic many-body system (in contrast with the relativistic
one-body case), a point that de Broglie had already noted in `Structure'. From
the very beginning, then, it was recognised that it would be difficult to
formulate a fundamentally Lorentz-invariant pilot-wave theory for a many-body
system, a situation that persists to this day.

Thus, in his report at the fifth Solvay conference, de Broglie arrived at
pilot-wave theory for a many-body system, with a guiding wave in configuration
space. Judging by de Broglie's comments about a physical representation
requiring $N$ waves in 3-space, and his characterisation of $\Psi$ as a
`fictitious' wave, it seems clear that he still regarded his new dynamics as
an effective, mathematical theory only (just as he had a few months earlier in
his `Structure' paper).

De Broglie's report then moves on to sketch some applications to atomic
theory. For stationary states of hydrogen, de Broglie notes that the electron
motion is circular, except in the case of magnetic quantum number $m=0$ for
which the electron is at rest. (Note that de Broglie expresses no concern that
the electron is predicted to be motionless in the ground state of hydrogen.
This result may seem puzzling classically, just as Bohr's quantised atomic
orbits seem puzzling classically: but both are natural consequences of de
Broglie's non-classical dynamics.)

De Broglie also points out that one can calculate the electron velocity during
an atomic transition $i\rightarrow j$, so that such transitions \textit{can}
be visualised. In this example, de Broglie's guiding wave is a solution of the
time-\textit{dependent} Schr\"{o}dinger equation, with a time-dependent
amplitude as well as phase. (The atomic wave function is taken to have the
form $\Psi=c_{i}\Psi_{i}+c_{j}\Psi_{j}$, where $\Psi_{i}$, $\Psi_{j}$ are
eigenfunctions corresponding to the atomic states $i$, $j$, and $c_{i}$,
$c_{j}$ are functions of time.) Clearly, then, de Broglie applied his
pilot-wave dynamics not just to stationary states, but to quite general wave
functions --- as he had also done (in the case of a single particle) in `Structure'.

De Broglie then outlines how, in his theory, one can obtain expressions for
the mean charge and current density for an ensemble of atoms. These
expressions are the same as those used by Schr\"{o}dinger and others. De
Broglie remarks that, `denoting by $\Psi$ the wave written in \textit{complex}
form, and by [$\Psi^{\ast}$] the conjugate function', in the nonrelativistic
limit the charge density is proportional to $\left\vert \Psi\right\vert ^{2}$:
the electric dipole moment then contains the correct transition frequencies.
By using these expressions as sources in Maxwell's equations, says de Broglie,
one can correctly predict the mean energy radiated by an atom. This is just
semiclassical radiation theory, which is still widely used today in quantum
optics.\footnote{Cf. section~\ref{Schr-radiation}.}

Part II of de Broglie's report ends with some general remarks. First and
foremost, de Broglie makes it clear that he regards pilot-wave theory as only
provisional (p.~\pageref{Draft1p.96}):

\begin{quotation}
So far we have considered the corpuscles as `exterior'\ to the wave $\Psi$,
their motion being only determined by the propagation of the wave. This is, no
doubt, only a provisional point of view: a true theory of the atomic structure
of matter and radiation should, it seems to us, \textit{incorporate} the
corpuscles in the wave phenomenon by considering singular solutions of the
wave equations.
\end{quotation}
De Broglie goes on to suggest that in a deeper theory one could `show
that there exists a correspondence between the singular waves and the waves
$\Psi$, such that the motion of the singularities is connected to the
propagation of the waves $\Psi$', just as he had suggested in `Structure'.
Part II ends by noting the incomplete state of the theory with respect to the
electromagnetic field and electron spin.

De Broglie's final part III contains a lengthy discussion and review of recent
experiments involving the diffraction, interference and scattering of
electrons (as had been requested by Lorentz\endnote{Cf.~de Broglie to Lorentz, 
27 June 1927, AHQP-LTZ-11 (in French).}). He regards the results as evidence for 
his `new Dynamics'. For the case of the diffraction of electrons by a crystal 
lattice, de Broglie points out that the scattered wave function $\Psi$ has maxima 
in certain directions, and notes that according to his theory the electrons should 
be preferentially scattered in these directions (p.~\pageref{selective}):

\begin{quotation}
Because of the role of pilot wave played by the wave $\Psi$, one must then
observe a selective scattering of the electrons in these directions.
\end{quotation}
What de Broglie is (briefly) describing here is the separation of the
incident wave function $\Psi$ into distinct (non-overlapping) emerging beams,
with each outgoing electron occupying one beam, and with an ensemble of
electrons being distributed among the emerging beams according to the Born
rule. This is relevant to a proper understanding of the de Broglie-Pauli
encounter, discussed in section~\ref{elastic-inelastic}.

Concerning the scattering maxima observed recently by Davisson and Germer, for
electrons incident on a crystal, de Broglie remarks (p.~\pageref{confirmation}):

\begin{quotation}
There is direct numerical confirmation of the formulas of the new Dynamics~....~.
\end{quotation}

De Broglie also discusses the inelastic scattering of electrons by atoms.
According to Born's calculations, one should observe maxima and minima in the
angular dependence of the differential scattering cross section. According to
de Broglie, it is premature to speak of an agreement with experiment. The
results of Dymond, for the inelastic scattering of electrons by helium atoms,
do however show maxima in the cross section, in qualitative agreement with the
predictions. De Broglie makes it quite clear that, at the time, theory was far
behind experiment.

The discussion following de Broglie's report was extensive, detailed, and
varied. A number of participants raised queries, and de Broglie replied to
most of them. Some of this discussion will be considered in detail in chapters~\ref{meas-in-pwt} 
and \ref{Pilot-wave-in-retrospect}. Here, we limit
ourselves to a brief summary of the questions raised.

Lorentz asked how, in the simple pilot-wave theory of 1924, de Broglie derived
quantisation conditions for the case of multiperiodic atomic orbits. Born
questioned the validity of de Broglie's guidance equation for an elastic
collision, while Pauli suggested that the key idea behind de Broglie's theory
was the association of particle trajectories with a locally conserved current.
Schr\"{o}dinger raised the question of an alternative velocity field different
from that assumed by de Broglie, while Kramers raised the question of how the
Maxwell energy-momentum tensor could arise from independently moving photons.
Lorentz, Ehrenfest and Schr\"{o}dinger asked about the properties of electron
orbits in hydrogen. Brillouin discussed at length the simple example of a
photon colliding with a mirror: in his Fig.~2, the incident and
emergent photon trajectories are located inside packets of limited extent (a
point relevant for the de Broglie-Pauli encounter in the general discussion).
Finally, Lorentz considered, in classical Maxwell theory, the near-field
attractive stress between two prisms, and claimed that this `negative
pressure' could not be produced by the motion of corpuscles (photons).

\section{Significance of de Broglie's work from 1923 to 1927}

In his papers and thesis of 1923--24, de Broglie proposed a simple form of a
new, non-classical dynamics of particles with velocities determined by the
phase gradient of abstract waves in 3-space. De Broglie constructed the
dynamics in such a way as to unify the mechanical principle of Maupertuis with
the optical principle of Fermat. He showed that it gave an account of some
simple quantum phenomena, including single-particle interference and quantised
atomic energy levels.

As we have seen, the scope and ambition of de Broglie's doctoral thesis went
far beyond a mere extension of the relations $E=h\nu$, $p=h/\lambda$ from
photons to other particles. Yet, the thesis is usually remembered solely for
this idea. The depth and inner logic of de Broglie's thinking is not usually
appreciated, neither by physicists nor by historians. An exception is Darrigol
(1993), who on this very point writes (pp.~303--4):

\begin{quotation}
For one who only knows of de Broglie's relation $\lambda=h/p$, two
explanations of his originality offer themselves. The first has him as a lucky
dreamer who hit upon a great idea amidst a foolish play with analogies and
formulas. The second has him a providential deep thinker, who deduced
unsuspected connections by rationally combining distant concepts. The first
explanation is the most popular, though rarely expressed in print. .... The
second explanation of Louis de Broglie's originality, though also extreme, is
certainly closer to the truth. Anyone who has read de Broglie's thesis cannot
help admiring the unity and inner consistency of his views, the inspired use
of general principles, and a necessary reserve.
\end{quotation}

In 1926, starting from de Broglie's expressions for the frequency and phase
velocity of an electron wave in an external potential, Schr\"{o}dinger found
the (nonrelativistic) wave equation for de Broglie's waves. It was de
Broglie's work, beginning in 1923, that initiated the notion of a wave
function for material particles. And it was de Broglie's view of the
significance of the optical-mechanical analogy, and of the role waves could
play in bringing about the existence of integer-valued quantum numbers, that
formed the basis for Schr\"{o}dinger's development of the wave equation and
the associated eigenvalue problem.

In 1927, de Broglie proposed what we now know as pilot-wave or de Broglian
dynamics for a many-body system, with a guiding wave in configuration space.
De Broglie regarded this theory as provisional: he thought it should emerge as
an effective theory, from a more fundamental theory in which particles are
represented by singularities of 3-space waves. Even so, he did propose the
theory, in the first-order form most commonly used today. Further, as we have
seen, de Broglie's view of the pilot wave as merely phenomenological is
strikingly reminiscent of (for example) late-nineteenth-century views of the
electromagnetic field. In retrospect, one may regard de Broglie as having
unwittingly arrived at a new and fundamental concept, that of a pilot wave in
configuration space.

De Broglie was unable to show that his new dynamics accounted for all quantum
phenomena. In particular, as we shall discuss in chapters~\ref{meas-in-pwt} 
and \ref{Pilot-wave-in-retrospect}, de Broglie did not understand how to
describe the process of measurement of arbitrary quantum observables in
pilot-wave theory: as shown in detail by Bohm in 1952, this requires an
application of de Broglie's dynamics to the measuring device itself. De
Broglie did, however, possess the fundamental dynamics in complete form.
Furthermore, de Broglie did understand how his theory accounted for
single-particle interference, for the directed scattering associated with
crystal diffraction, and for electron scattering by atoms (for the latter, see
section~\ref{elastic-inelastic}). Clearly, in 1927, many
applications of pilot-wave theory remained to be developed; just as, indeed,
many applications of quantum theory --- including the general quantum theory
of measurement --- remained to be developed and clarified.

In retrospect, de Broglian dynamics seems as radical as --- and indeed
somewhat reminiscent of --- Einstein's theory of gravity. According to
Einstein, there is no gravitational force, and a freely falling body follows
the straightest path in a curved spacetime. According to de Broglie, a massive
body undergoing diffraction and following a curvilinear path is not acted upon
by a Newtonian force: it is following the ray of a guiding wave. De Broglie's
abandonment of Newton's first law of motion in 1923, and the adoption of a
dynamics based on velocity rather than acceleration, amounts to a far-reaching
departure from classical mechanics and (arguably) from classical kinematics
too --- with implications for the structure of spacetime that have perhaps not
been understood (Valentini 1997). Certainly, the extent to which de Broglie's
dynamics departs from classical ideas was unfortunately obscured by Bohm's
presentation of it, in 1952, in terms of acceleration and a pseudo-Newtonian
quantum potential, a formulation that today seems artificial and inelegant
compared with de Broglie's (much as the rewriting of general relativity as a
field theory on flat spacetime seems unnatural and hardly illuminating). The
fundamentally second-order nature of classical physics is today embodied in
the formalism of Hamiltonian dynamics in phase space. In contrast, de
Broglie's first-order approach to the theory of motion seems more naturally
cast in terms of a dynamics in configuration space.

Regardless of how one may wish to interpret it, by any standards de Broglie's
work from 1923 to 1927 shows a remarkable progression of thought, beginning
from early intuitions and simple models of the relationship between particles
and waves, and ending (with Schr\"{o}dinger's help) with a complete and new
form of dynamics for nonrelativistic systems --- a deterministic dynamics that
was later shown by Bohm to be empirically equivalent to quantum theory (given
a Born-rule distribution of initial particle positions). The inner logic of de
Broglie's work in this period, his drive to unite the physics of particles
with the physics of waves by unifying the principles of Maupertuis and Fermat,
his wave-like explanation for quantised energy levels, his prediction of
electron diffraction, his explanation for single-particle interference, his
attempts to construct a field-theoretical picture of particles as moving
singularities, and his eventual proposal of pilot-wave dynamics as a
provisional theory --- all this compels admiration, all the more for being
largely unknown and unappreciated.

Today, pilot-wave theory is often characterised as simply adding particle
trajectories to the Schr\"{o}dinger equation. An understanding of de Broglie's
thought from 1923 to 1927, and of the role it played in Schr\"{o}dinger's
work, shows the gross inaccuracy of this characterisation: after all, it was
actually Schr\"{o}dinger who removed the trajectories from de Broglie's theory
(cf. section~\ref{hist-mis}). It is difficult to
avoid the conclusion that de Broglie's stature as a major contributor to
quantum theory has suffered unduly from the circumstance that, for most of the
twentieth century, the theory proposed by him was incorrectly regarded as
untenable or inconsistent with experiment. Regardless of whether or not de
Broglie's pilot-wave theory is closer to the truth than other interpretations,
the fact that it is a consistent and viable approach to quantum physics ---
which has no measurement problem, which shows that objectivity and determinism
are not incompatible with quantum physics, and which stimulated Bell to
develop his famous inequalities --- necessarily entails a reappraisal of de
Broglie's place in the history of twentieth-century physics.

\newpage
\renewcommand{\enoteheading}{\section*{Archival notes}}
\addcontentsline{toc}{section}{\em Archival notes}
\theendnotes

\setcounter{endnote}{0}
\setcounter{equation}{0}

\chapter{From matrix mechanics to quantum mechanics}\label{BornEss}
\chaptermark{From matrix to quantum mechanics}

The report by Born and Heisenberg on `quantum mechanics' may seem surprisingly difficult
to the modern reader. This is partly because Born and Heisenberg are describing various stages 
of development of the theory that are quite different from today's quantum mechanics. Among these, it should be noted 
in particular that the theory developed by Heisenberg, Born and Jordan in the years 1925--26 and known 
today as matrix mechanics (Heisenberg 1925b [1], Born and Jordan 1925 [2], Born, Heisenberg and Jordan 1926 
[4])\footnote{Throughout this chapter, numbers in square brackets refer to entries in the original 
bibliography at the end of Born and Heisenberg's report.}
differs from standard quantum mechanics in several important respects. At the same time, the interpretation of the
theory (the topic of section II of the report) also appears to have undergone important modifications, in particular regarding the notion 
of the state of a system. Initially, Born and Heisenberg insist on the notion that a system is always in a stationary state 
(performing quantum jumps between different stationary states). Then the notion of the wave function is introduced and 
related to probabilities for the stationary states. At a later stage probabilistic notions (in particular, what one now
calls transition probabilities) are extended to arbitrary observables, but it remains somewhat unclear whether the wave 
function itself should be regarded as a fundamental entity or merely as an effective one. This may reflect the different 
routes followed by Born and by Heisenberg in the development of their ideas. The common position presented by Born and 
Heisenberg emphasises the probabilistic aspect of the theory as fundamental, and the conclusion of the report expresses
strong confidence in the resulting picture.

The two main sections of this chapter, section~\ref{beyond} and section~\ref{stateprob}, will be devoted, 
respectively, to providing more details on the various stages of development of the theory, and to disentangling 
various threads of interpretation that appear to be present in Born and Heisenberg's report.  
Before that, we provide a summary (section~\ref{Born-summary}) and a few remarks on the authorship and writing 
of the various sections of the report (section~\ref{Born-writing}).

\section{Summary of Born and Heisenberg's report}\label{Born-summary}
Born and Heisenberg's report has four sections (together with an introduction and conclusion): I on formalism, II on 
interpretation, III on axiomatic formulations and on uncertainty, and IV on applications. The formalism that 
is described is initially that of matrix mechanics, which is then extended beyond the original framework (among other 
things, in order to make the connection with Schr\"{o}dinger's wave mechanics). Then, further developments of matrix 
mechanics are sketched. These allow one to incorporate a `statistical' interpretation. After a brief discussion of 
Jordan's (1927b,c [39]) axiomatic formulation, the uncertainty relations are used to justify the statistical element of 
the interpretation. A few applications of special interest and some brief final remarks conclude the report. 
As we discuss below, Born drafted sections I and II, and Heisenberg drafted the introduction, sections III and IV and 
the conclusion. 

%
%
Born and Heisenberg's introduction stresses the continuity of quantum mechanics with the old quantum theory of Planck, 
Einstein and Bohr, and touches briefly on such themes as discontinuity, observability in principle, `Anschaulichkeit'
(for which see section~\ref{Schr-conflict} below) and the statistical element in quantum mechanics.

%
%
Section I, `The mathematical methods of quantum mechanics', first sketches matrix mechanics roughly as 
developed in the `three-man paper' by Born, Heisenberg and Jordan (1926 [4]): the basic framework of position and 
momentum matrices and of the canonical equations, the perturbation theory, and the connection with the theory of 
quadratic forms (the latter leading to both discrete and continuous spectra).\footnote{Note that Born and Heisenberg 
use the term `quantum mechanics' to refer also to matrix mechanics, in keeping with the terminology of the original 
papers.  See also Mehra and Rechenberg (1982c, fn.~72 on pp.~61--2). Heisenberg expresses his dislike for the term 
`matrix physics' in Heisenberg to Pauli, 16 November 1925 (Pauli, 1979, pp.~255--6).} Next, 
Born and Heisenberg describe two generalisations of matrix mechanics: Dirac's (1926a [7]) q-number theory and what they 
refer to as Born and Wiener's (1926a [21], 1926b) operator calculus. As a matter of fact, they already sketch von Neumann's 
representation of physical quantities by operators in Hilbert space. This is identified as a mathematically rigorous 
version of the transformation theory of Dirac (1927a [38]) and Jordan (1927b,c [39]).\footnote{See below, 
sections~\ref{formex} and \ref{BW}, for some discussion of Born and Wiener's work, and section~\ref{transf-theory} for 
the transformation theory (mainly Dirac's). Note that von Neumann had published a series of papers on the Hilbert-space 
formulation of 
quantum mechanics before his well-known treatise of 1932. It is to two of these that Born and Heisenberg refer, Hilbert, 
von Neumann and Nordheim (1928 [41], submitted April 1927) and von Neumann (1927 [42]).} Born and Heisenberg note that, 
if one takes as the Hilbert space the appropriate function space, the problem of 
diagonalising the Hamiltonian operator leads to the time-independent Schr\"{o}dinger equation. Thus, in this (formal) 
sense, the Schr\"{o}dinger theory is a special case of the operator version of quantum mechanics. The eigenfunctions 
of the Hamiltonian can be associated with the unitary transformation that diagonalises it. Matrix mechanics, too, 
is a special case of this more general formalism, if one takes the (discrete) energy eigenstates as the basis for the 
matrix representation.

%
%
Section II on `Physical interpretation' is probably the most striking. It begins by stating that matrix mechanics describes 
neither the actual state 
of a system, nor when changes in the actual state take place, suggesting that this is related to the idea that matrix
mechanics describes only closed systems. In order to obtain some description of change, one must consider open systems. 
Two methods for doing this are introduced. First of all, following Heisenberg (1926b [35]) and Jordan (1927a 
[36]), one can consider the matrix mechanical description of two coupled systems that are in resonance. One can show 
that the resulting description can be interpreted in terms of quantum jumps between the energy levels of the two systems, 
with an explicit expression for the transition probabilities. The second method uses the generalised formalism of section I, 
in which time dependence can be introduced explicitly via the time-dependent Schr\"{o}dinger equation. Again, expressions 
can be found that can be interpreted as transition probabilities, and similarly the squared modulus of the wave function's 
coefficients in the energy basis is interpreted as the probability for the occurrence of the corresponding stationary 
state. Born and Heisenberg introduce the notion of interference of probabilities. Only at this stage are experiments
mentioned as playing a conceptually crucial role, namely in an argument why such interference `does not represent a 
contradiction with the rules of the probability calculus' (p.~\pageref{Born-contradiction}). Interference is then related to 
the wave theory of de Broglie and Schr\"{o}dinger, and the interpretation of the squared modulus of the wave function as 
a probability density for position is introduced. Born and Heisenberg also define the `relative' (i.e. conditional) 
position density given an energy value, as the squared modulus of the amplitude of a stationary state. Finally, in the 
context of Dirac's and Jordan's transformation theory, the notions of transition probability (for a single quantity) and 
of conditional probability (for a pair of quantities) are generalised to arbitrary physical quantities. 

%
%
Section III opens with a concise exposition of Jordan's (1927b,c [39]) axioms for quantum mechanics, which take the notion 
of probability amplitude as primary. Born and Heisenberg point out some formal drawbacks, such as the use of 
$\delta$-functions and the 
presence of unobservable phases in the probability amplitudes. They note that such drawbacks have been overcome by the 
formulation of von Neumann (which they do not go on to discuss further). Born and Heisenberg then proceed to discuss  
in particular whether the statistical element in the theory can be reconciled with macroscopic determinism. To this effect 
they first justify the necessity of using probabilistic notions by appeal to the notion of uncertainty (discussed using the 
example of diffraction of light by a single slit); then they point out that, while for instance in cases of diffraction the 
laws of propagation of the probabilities in quantum mechanics are very different from the classical evolution of a probability 
density, there are cases where the two coincide to a very good approximation; they write that this justifies the classical 
treatment of $\alpha$- and $\beta$-particles in a cloud chamber.

%
%
Section IV discusses briefly some applications, chosen for their `close relation to questions of principle' 
(p.~\pageref{closerelation}). Mostly, 
these applications highlight the importance of the generalised formalism (introduced in section I), by going beyond 
matrix mechanics proper or wave mechanics proper. The first example is that of spin, which, requiring finite 
matrices, is taken to be a problematic concept for wave mechanics (but not for matrix mechanics). The main example is 
given by the discussion of identical particles, the Pauli principle and quantum statistics. Born and Heisenberg note 
that the choice of whether the wave function should be symmetric or antisymmetric appears to be arbitrary. They note, 
moreover, that the appropriate choice arises naturally if one quantises the normal modes of a black body, leading 
to Bose-Einstein statistics, or, if one adopts Jordan's (1927d [54]) quantisation procedure, to Fermi-Dirac 
statistics.\footnote{The general discussion returns to these issues in more detail; see for instance Dirac's critical
remarks on Jordan's procedure, p.~\pageref{page187}.} Finally, Born and Heisenberg comment on Dirac's quantum electrodynamics (Dirac
1927b,c [51,52]), noting in particular that it yields the transition probabilities for spontaneous emission.

The conclusion discusses quantum mechanics as a `closed theory' (see section~\ref{foundmat} below) and addresses 
the question of whether indeterminism in quantum mechanics is fundamental. In particular, Born and Heisenberg state 
that the assumption of indeterminism agrees with experience and that the treatment of electrodynamics will not modify 
this state of affairs. They conclude by noting that the existing problems concern rather the development of a 
fully relativistic theory of quantum electrodynamics, but that there is progress also in this direction.

%
%
The discussion is comparatively brief. Dirac describes the analogy between the matrix method and the 
Hamilton-Jacobi theory. Then Lorentz makes a few remarks, in particular emphasising that the freedom to choose the 
phases in the matrices $q$ and $p$, (\ref{matrix1}) and (\ref{matrix2}) below, is not limited to a different choice 
of the time origin, a possibility already noted by Heisenberg (1925b [1]), but extends to arbitrary phase factors of 
the form $e^{i(\delta_m-\delta_n)}$, as noted by Born and Jordan (1925 [2]).\footnote{This corresponds of course to 
the choice of a phase factor for each energy state.} According to Lorentz, this fact suggests that the `true oscillators' 
are associated with the different stationary states rather than with the spectral frequencies.\endnote{Cf.\ also
Lorentz to Ehrenfest, 4 July 1927, AHQP-EHR-23 (in Dutch).} The last few and brief contributions address the question 
of the phases.\footnote{As mentioned in footnote on p.~\pageref{forpage54}, Bohr suggested to omit this entire discussion 
from the published proceedings.}

\section{Writing of the report}\label{Born-writing}
%
%
The respective contributions by Born and by Heisenberg to the report become clear from their correspondence with Lorentz. 
As mentioned in chapter~\ref{HistEss}, Lorentz had originally planned to have a report on quantum mechanics from either 
Born or Heisenberg, leaving to them the choice of who was to write it. In reply, Born and Heisenberg suggested that they 
would provide a joint report.\endnote{This and the following remarks are based on Born to Lorentz, 23 June 1927, 
AHQP-LTZ-11 (in German).} Heisenberg was going to visit Born in G\"{o}ttingen in early July 1927, on which occasion they 
would decide on the structure of the report and divide up the work, planning to meet again in August to finish it. The 
report (as requested by Lorentz) would include the results by Dirac, who was in G\"{o}ttingen at the time.\footnote{Dirac 
was in G\"{o}ttingen from February 1927 to the end of June, after having spent September to February in Copenhagen. 
Dirac's paper on transformation theory (1927a [38]) and his paper on emission and absorption (Dirac 1927b [51]) were 
written in Copenhagen, while the paper on dispersion (Dirac 1927c [52]) was written in G\"{o}ttingen. Compare Kragh (1990, 
ch.~2. pp.~37~ff.).} As Born explains, the report would `above all emphasise the viewpoints that are presumably taken less 
into account by Schr\"{o}dinger, namely the statistical conception of quantum mechanics'. Either author, or both, could 
orally present the report at the conference (Heisenberg, in a parallel letter, suggested that Born should be the appropriate 
choice, since he was the senior scientist\endnote{Heisenberg to Lorentz, 23 June 1927, AHQP-LTZ-12 (in German).}); Born and
Heisenberg could also give additional explanations in English. Born did not speak French, nor (Born thought) did Heisenberg.  

In a later letter,\endnote{Born to Lorentz, 29 August 1927, AHQP-LTZ-11 (in German).} Born reports that  
Heisenberg had sent him at the end of July the draft of introduction, conclusion and sections III and IV, upon which Born had written 
sections I and II, reworked Heisenberg's draft, and sent everything back to Heisenberg. Heisenberg in turn had made some 
further small changes. Due to illness and a small operation, Born had been unable to go to Munich in mid-August,  
so that he had not seen the changes. He trusted, however, that he would agree with the details of Heisenberg's final 
version.\footnote{Heisenberg had sent the final text to Lorentz on 24 August.\endnotemark}\endnotetext{Heisenberg 
to Lorentz, 24 August 1927, AHQP-LTZ-12 (in German).} Born also mentions that Heisenberg and himself would like 
to go through the text again, at least at proof stage.\footnote{The published version and the typescript show only minor 
discrepancies, and many of these are clearly mistakes in the published version (detailed in our endnotes to the report). 
This and the fact that Born and Heisenberg appear not to have spoken French suggests that there was in fact no 
proofreading on their part.} The presentation would be split between Born (introduction and sections I and II) and 
Heisenberg (sections III and IV and conclusion). Born concludes with the following words:\label{forpage255} `It would be 
particularly important for us to come to an agreement with Schr\"{o}dinger regarding the physical interpretation 
of the quantum formalisms'.

\section{Formalism}\label{beyond}
%
%
\subsection{Before matrix mechanics}\label{beforematrix}
In the old Bohr-Sommerfeld theory, electron orbits are described as classical Kepler orbits subject to additional 
constraints (the `quantum conditions'), yielding discrete stationary states. Such a procedure, in Born and Heisenberg's
introduction, is criticised as artificial.\footnote{The remark is rather brief, but note that Born and 
Heisenberg seem to consider the introduction of photons into the classical electromagnetic theory (a corpuscular 
discontinuity) to be as artificial as the introduction of discrete stationary states into classical mechanics.} An atom 
is assumed to `jump' between its various stationary states, and energy differences between stationary states are related 
to the spectral frequencies via Bohr's frequency condition. Spectral frequencies and orbital frequencies, however, appear 
to be quite unrelated. This is the crucial `radiation problem' of the old quantum theory (see also below, 
section~\ref{Schr-radiation}). 

The BKS theory of radiation\footnote{See also the description in section~\ref{scientific}.} 
includes a rather different picture of the atom, and arguably provides a solution to the radiation problem just 
mentioned. In the BKS theory, one associates to each atom a set of `virtual' oscillators, with frequencies equal to the 
spectral frequencies. Specifically, when the atom is in the stationary state $n$, the oscillators
that are excited are those with frequencies corresponding to transitions from the energy $E_n$ to both lower
and higher energy levels. These oscillators produce a virtual radiation field, which propagates (classically) 
according to Maxwell's equations and interacts (non-classically) with the virtual oscillators, in particular
influencing the probabilities for induced emission and absorption in other atoms and determining the probabilities 
for spontaneous emission in the atom itself. The actual emission or absorption of energy is associated with the 
corresponding transition between stationary states. Note that an emission in one atom is not connected directly 
to an absorption in another, so that energy and momentum are conserved only statistically. 

Several important results derived by developing correspondence arguments into translation rules from classical to 
quantum expressions were formulated within the context of the BKS theory, in particular Kramers' (1924) dispersion theory. 
Other examples were Born's (1924) perturbation formula, Kramers and Heisenberg's (1925) joint paper on dispersion and 
Heisenberg's (1925a) paper on polarisation of fluorescence radiation.\footnote{See Darrigol 
(1992, pp.~224--46) and Mehra and Rechenberg (1982b, sections III.4 and III.5).} The same arguments also led to a natural 
reformulation of the quantum conditions independent of the old `classical models' (Kuhn 1925, Thomas 1925). As 
first argued by Pauli (1925),\label{forpage14} such results were entirely independent of the presumed mechanism of 
radiation.\footnote{Cf.\ Darrigol (1992, pp.~244--5). Cf.\ also Jordan (1927e [63]), among others.} 

Indeed, after the demise of the BKS theory following the experimental verification of the conservation laws in the 
Bothe-Geiger experiments, these results and the corresponding techniques of symbolic translation formed the basis 
for Heisenberg's formulation of matrix mechanics in his famous paper of 1925 (Heisenberg 1925b [1]).
Heisenberg's stroke of genius was to give up altogether the 
kinematical description in terms of spatial coordinates, while retaining the classical form of the equations of motion, 
to be solved under suitably reformulated quantum conditions (those of Kuhn and Thomas). While Heisenberg had dropped both 
the mechanism of radiation and the older electronic orbits, there were  points of continuity with the BKS theory and with 
the old quantum theory; in particular, as we shall see, the new variables were at least formally related to the virtual 
oscillators, and the picture of quantum jumps was retained, at least for the time being.

\subsection{Matrix mechanics}\label{MM}
Heisenberg's original paper is rather different from the more
definitive presentations of matrix mechanics: the equations are given in Newtonian (not Hamiltonian) form, 
energy conservation is not established to all orders, and, as is well known, Heisenberg at first had not 
recognised that his theory used the mathematical machinery of matrices. Like Born and Heisenberg in their report, 
we shall accordingly follow in this section mainly the papers by Born and Jordan (1925 [2]) and Born, Heisenberg and 
Jordan (1926 [4]), as well as the book by Born (1926d,e [58]), supplementing the report with more details when useful.

Kinematically, the description of `motion' in matrix mechanics generalises that given by the set of the components of 
a classical Fourier series,
  \begin{equation}
    q_n e^{2\pi i n\nu t}
    \label{newyear1}
  \end{equation}
(and thus generalises the idea of a periodic motion). As frequencies, one chooses, instead of
$\nu (n)=n\nu$, the {\em transition frequencies} of the system under consideration, which are required to obey 
Ritz's combination principle, $\nu_{ij}=T_i-T_j$ (relating the frequencies to the spectroscopic terms $T_i=E_i/h$,
or, in Born and Heisenberg's notation, $T_i=W_i/h$). The components thus defined form a doubly infinite array:
  \begin{equation}
    q=\left(\begin{array}{ccc}
               q_{11}e^{2\pi i\nu_{11}t} &   q_{12}e^{2\pi i\nu_{12}t}   &  \ldots  \\
               q_{21}e^{2\pi i\nu_{21}t} &   q_{22}e^{2\pi i\nu_{22}t}   &  \ldots \\
               \vdots                             &   \vdots             &  \ddots
            \end{array}\right)\ .
     \label{matrix1}
  \end{equation}
From the combination principle for the $\nu_{ij}$, it follows that $\nu_{ij}=-\nu_{ji}$. Further, it is required of the 
(generally complex) amplitudes that
  \begin{equation}
    q_{ji}=q^*_{ij}\ ,
  \end{equation} 
by analogy with a classical Fourier series, so that the array is in fact a Hermitian matrix. 

In modern notation, the above matrix consists of the elements of the position observable (in Heisenberg picture) in the 
energy basis. For a closed system with Hamiltonian $H$, these indeed take the form
  \begin{equation}
    \begin{split}
    \left\langle E_{i}\right\vert Q(t)\left\vert E_{j}\right\rangle
    & =\left\langle E_{i}\right\vert e^{(i/\hbar)Ht}Q(0)e^{-(i/\hbar)Ht}\left\vert E_{j}\right\rangle  \\ 
    & =\left\langle E_{i}\right\vert Q(0)\left\vert E_{j}\right\rangle e^{(i/\hbar)(E_{i}-E_{j})t}\ .
    \end{split} 
  \label{newyear2}
  \end{equation}

Along with the position matrix $q$, one considers also a momentum matrix:
  \begin{equation}
    p=\left(\begin{array}{ccc}
               p_{11}e^{2\pi i\nu_{11}t} &   p_{12}e^{2\pi i\nu_{12}t}   &  \ldots  \\
               p_{21}e^{2\pi i\nu_{21}t} &   p_{22}e^{2\pi i\nu_{22}t}   &  \ldots \\
               \vdots                             &   \vdots             &  \ddots
            \end{array}\right)\ ,
     \label{matrix2}
  \end{equation}
as well as all other matrix quantities that can be obtained as functions of $q$ and $p$, defined as polynomials or 
as power series (leaving aside questions of convergence). Again from the combination principle, it follows that under 
multiplication of two such matrices, one obtains another matrix of the same form (i.e.\ with the same time-dependent 
phases). Thus one justifies taking these two-dimensional arrays as matrices. The most general physical quantity in matrix 
mechanics is a matrix whose elements (in modern terminology) are the elements of an arbitrary Heisenberg-picture observable 
in the energy basis. 

The values of these matrices (in particular the amplitudes $q_{ij}$ and the frequencies $\nu_{ij}$) will have to be 
determined by solving the equations of motion under some suitable quantum conditions. Specifically, the equations of 
motion are postulated to be the analogues of the classical Hamiltonian equations:
  \begin{equation}
    \frac{dq}{dt}=\frac{\partial H}{\partial p},\qquad\frac{dp}{dt}=-\frac{\partial H}{\partial q}\ ,
     \label{matrix3}
  \end{equation}
where $H$ is a suitable matrix function of $q$ and $p$. The quantum conditions are formulated as
  \begin{equation}
    pq-qp=\frac{h}{2\pi i}\cdot 1\ ,
      \label{matrix4}
  \end{equation}
where $1$ is the identity matrix. In the limit of large quantum numbers, (\ref{matrix4}) corresponds 
to the `old' quantum conditions.

Of course, differentiation with respect to $t$ and with respect to matrix arguments need to be suitably 
defined. The former is defined elementwise as
  \begin{equation}
    \dot{f}_{ij}=2\pi i\nu_{ij}f_{ij}\ ,
    \label{newyear3}
  \end{equation}
which can be written equivalently as
  \begin{equation}
    \dot{f}=\frac{2\pi i}{h}(Wf-fW)\ ,
    \label{newyear4}
  \end{equation}
where $W$ is the diagonal matrix with $W_{ij}=W_i\delta_{ij}$. 
Differentiation with respect to matrix arguments is defined (in slightly modernised form) as 
  \begin{equation}
    \frac{df}{dx}=\lim_{\alpha\rightarrow 0}\frac{f(x+\alpha 1)-f(x)}{\alpha}\ ,
    \label{newyear5}
  \end{equation}
where the limit is also understood elementwise.\footnote{This definition was agreed upon after some debate. See  
Mehra and Rechenberg (1982c, \mbox{pp.~69--71} and \mbox{97--100}).} 

Once the preceding scheme has been set up, the next question is how to go about solving the equations of motion. 
First, for the special case of differentiation with respect to $q$ or $p$,
one shows that
  \begin{equation}
    fq-qf=\frac{h}{2\pi i}\frac{\partial f}{\partial p}
    \label{newyear6a}
  \end{equation}
and
  \begin{equation}
    pf-fp=\frac{h}{2\pi i}\frac{\partial f}{\partial q}
    \label{newyear6b}
  \end{equation}
(by induction on the form of $f$ and using (\ref{matrix4})). 
From this and from (\ref{newyear4}), one can easily show that the equations of motion are equivalent to
  \begin{equation}
    Wq-qW=Hq-qH\  \label{newyear7a}
  \end{equation}
and
  \begin{equation} 
    Wp-pW=Hp-pH\ . \label{newyear7b}
  \end{equation}
Thus, $W-H$ commutes with both $q$ and $p$, and therefore also with $H$. It follows that
  \begin{equation}
    WH-HW=0\ ,
    \label{newyear8a}
  \end{equation}
which by (\ref{newyear4}) means that
  \begin{equation}
    \dot{H}=0\ ,
    \label{newyear8b}
  \end{equation}
that is, energy is conserved. In the non-degenerate case, it follows that $H$ is a diagonal matrix. 
Denoting its diagonal terms by $H_i$, (\ref{newyear7a}) yields further the Bohr frequency condition,
  \begin{equation}
    H_i-H_j=W_i-W_j=h\nu_{ij}\ ,
    \label{newyear9}
  \end{equation}
and fixing an arbitrary constant one can set $H=W$. The above proof shows also that conversely, 
(\ref{newyear8b}), (\ref{newyear9}) and the commutation relations imply the equations of motion. Therefore, 
the problem of solving the equations of motion essentially reduces to that of diagonalising the energy matrix. 

This discussion leads also to the notion of a `canonical transformation' (a transformation that leaves the equations 
of motion invariant) as a transformation that preserves the commutation relations. Obviously any
transformation of the form
  \begin{equation}
     Q=SqS^{-1},\qquad P=SpS^{-1}
    \label{matrix11}
  \end{equation}
(with $S$ invertible) will be such a transformation, and it was conjectured that these were the most general canonical 
transformations. At this stage, $S$ is required only to be invertible; indeed, the notion of unitarity\label{unitarity} 
has not been
introduced yet. If $H$ is a suitably symmetrised function of $q$ and $p$, however, it will also be a Hermitian matrix, 
and the problem of diagonalising it can be solved using the theory of quadratic forms of infinitely many variables (assuming 
that the theory as known at the time extends also to unbounded quadratic forms). In particular, it will be possible
to diagonalise $H$ using a unitary $S$, which also guarantees that the new coordinates $Q$ and $P$ will be Hermitian.

What does such a solution to the equations of motion yield? In the first place, one obtains the values 
of the frequencies $\nu_{ij}$ and of the energies $W_i$ (along with the values of other conserved quantities). 
In the second place, based on correspondence arguments, one can 
identify the diagonal elements of a matrix with `time average' in the respective stationary states,\footnote{The 
concept of `time averages' can also
be thought of as related to the BKS theory, as remarked by Heisenberg himself in a letter to Einstein of February 
1926.\endnotemark}\endnotetext{Heisenberg to Einstein, 18 February 1926, AEA 12-172.00 (in German).} and 
relate the squared amplitudes $|q_{ij}|^2$ to the intensities of spontaneous emission, or equivalently to the 
corresponding transition probabilities between stationary states.\footnote{The determination of intensities had
been a pressing problem since the work by Born (1924) and the `Utrecht observations' referred to in the report 
(p.~\pageref{utrecht}). See Darrigol (1992, pp.~234--5).} This, however, 
is not an argument based on first principles. Indeed, as Born
and Heisenberg repeatedly stress, matrix mechanics in the above form describes a closed, conservative system, in which
no change should take place. As Born and Heisenberg note in section II, actual transfer of energy is to be expected only 
when the atom is coupled to some other system, specifically the radiation field. Born and Jordan's paper (1925 [2]) and 
Born, Heisenberg and Jordan's paper (1926 [4]) include a first attempt at treating the radiation field quantum 
mechanically and at justifying the assumed interpretation of the squared amplitudes. A satisfactory treatment 
of spontaneous radiation was later given by Dirac (1927b [51]), as also mentioned in the report.\footnote{Note that by 
this time Dirac (1927a [38]) had already introduced a probabilistic interpretation based on the transformation theory.} In the third 
place, by setting up relations between the amplitudes 
$q_{ij}$ and the matrix elements of other conserved quantities such as angular momentum, one is able to identify which 
transitions are possible between states with the corresponding quantum numbers (derivation of `selection rules'). 
Expectation values, other than in the form of time averages for the stationary states, are lacking.

It should be clear that matrix mechanics in its historical formulation is not to be identified with today's 
quantum mechanics in the Heisenberg picture. The matrices allowed as solutions in matrix mechanics are 
basis-dependent. Moreover, quite apart from the fact that solutions to the equations of motion are hard to 
find,\footnote{In particular, one was unable to introduce action-angle variables to solve the Hamiltonian 
equations as in classical mechanics; cf.\ Born and Heisenberg's remarks on `aperiodic motions'.} the results 
obtained are relatively modest. In particular, there are no general expectation values for physical quantities. 
In discussing both the mathematical formalism and the physical interpretation, Born and Heisenberg 
suggest that the original matrix theory is inadequate and in need of extension.

\subsection{Formal extensions of matrix mechanics}\label{formex}
As presented in the report, the chief formal rather than interpretational difficulty for matrix mechanics is the 
failure to describe {\em aperiodic} quantities. As we have seen, the matrices were understood as a generalisation of 
classical Fourier series. Strictly speaking, if one follows this analogy, a periodic quantity is represented by a 
discrete (if doubly infinite) matrix, and one can envisage representing aperiodic quantities by continuous 
matrices --- analogously to the representation of classical quantities by Fourier integrals. Indeed, already Heisenberg 
(1925b [1]) points out that his quadratic arrays would have in general both a periodic part (discrete) and an 
aperiodic part (continuous). The connection between matrices and quadratic forms showed in fact that matrices in general
had also a continous spectrum (if the theory extended to the unbounded case). 
Still, continuous matrices are unwieldy, and in certain cases the matrix elements (which are always evaluated in the 
energy basis) will become singular, as when trying to describe a free particle. (This is not at all surprising, considering 
that also in the classical analogy the Fourier integral of the function $at+b$ fails to converge.) 

This problem was addressed by generalising the notion of a matrix to that of a q-number (Dirac 1926a [7]) and to that of
an operator (Born and Wiener 1926a [21], 1926b). Dirac's q-numbers are 
abstract objects that are assumed to form a noncommutative algebra, while Born and Wiener's operators are characterised 
not by their elements in some basis, but by their action on a space of functions. In both approaches, one can deal with
aperiodic quantities and quantities with a singular matrix representation. 

Born and Wiener's approach is the lesser-known of the two, and shall therefore be briefly 
sketched.\footnote{Born and Wiener published two very similar papers on their
operator theory, one in German, referred to in the report (1926a [21]), and one in English (1926b).} The operators 
in question are linear operators acting on a space of functions $x(t)$ (which are not given any specific physical 
interpretation). These functions are understood as generalising the functions having the form
  \begin{equation}
    x(t)=\sum_{n=1}^\infty x_n e^{\frac{2\pi i}{h}W_nt}\ .
    \label{newyear10}
  \end{equation}
That is, they generalise the functions that can be written in a Fourier-like series with frequencies equal to the spectral 
frequencies. Therefore, the operators generalise the matrices that act on in\-fi\-nite-di\-men\-sion\-al vectors indexed 
by the energy values, i.e.\ they generalise the matrices in the energy basis. By using these operators, Born and Wiener 
free themselves from 
the need to operate with the matrix elements, and they are able to solve the equations of motion explicitly for systems
more general than those treated in matrix mechanics until then. Their main example is the (aperiodic) one-dimensional 
free particle. 

Instead of presenting this formalism, however, in the report Born and Heisenberg present directly von Neumann's formalism 
of operators on Hilbert space (evidently but tacitly considering Born and Wiener's formalism as its natural precursor). 
They note that in this formalism it is possible to consider matrices in arbitrary, even 
continuous bases, and they use this fact to make the connection with Schr\"{o}dinger's theory. In particular, they 
point out that solving the (time-independent) Schr\"{o}dinger equation is equivalent to diagonalising the 
Hamiltonian quadratic form, in the sense that the set of Schr\"{o}dinger eigenfunctions $\varphi(q,W)$ (wave functions 
in $q$ indexed by $W$) yields the transformation matrix $S$ from the position basis to the energy basis. Thus one can 
use the familiar methods of partial differential equations to diagonalise $H$.\footnote{Cf.\ below, 
section~\ref{transf-theory}. The connection 
between matrix and wave mechanics was discovered independently by Schr\"{o}dinger (1926d), Eckart (1926 [22]) and Pauli 
(Pauli to Jordan, 12 April 1926, in Pauli 1979, pp.~315--20). For a description of the various contributions, including the 
work of Lanczos (1926 [23]), see Mehra and Rechenberg (1987, pp.~636--84). Muller (1997) argues that matrix mechanics and 
wave mechanics, as formulated and understood at the time, were nevertheless inequivalent theories. For concrete examples 
in which the two theories were indeed considered to yield different predictions, see below p.~\pageref{forpage62} (including the 
footnote) and the discussion on p.~\pageref{quadrupole}.}

Despite providing very useful formal extensions of matrix mechanics, the q-numbers and the operators (at least as 
presented in section I of the report) do not provide further insights into the physical interpretation of the theory. 
In the following section II, on `Physical interpretation', Born and Heisenberg discuss the problem of describing actual 
states and processes in matrix mechanics, suitably extended, and the surprising ramifications of this problem.

\section{Interpretation}\label{stateprob}
Born and Heisenberg's section II begins with the following statement (p.~\pageref{nothing}):
  \begin{quote}
    The most noticeable defect of the original matrix mechanics consists in 
    the fact that at first it appears to give information not about actual phenomena, 
    but rather only about possible states and processes. It allows one to calculate
    the possible stationary states of a system; further it makes a statement about the 
    nature of the harmonic oscillation that can manifest itself as a light wave in a quantum 
    jump. But it says nothing about when a given state is present, or when a change is to 
    be expected. The reason for this is clear: matrix mechanics 
    deals only with closed periodic systems, and in these there are indeed no changes. 
    In order to have true processes, as long 
    as one remains in the domain of matrix mechanics,  one must direct one's attention
    to a {\em part} of the system; this is no longer closed and enters into interaction
    with the rest of the system. The question is what matrix
    mechanics can tell us about this.
  \end{quote}
(This is again one of the sections originally drafted by Born.)
As raised here, the question to be addressed is how to incorporate into matrix mechanics the (actual) 
state of a system, and the time development of such a state. The discussion given in the report may give rise
to some confusion, because it arguably contains at least two, if not three, disparate approaches to what is a
state in quantum mechanics. The first, reflected in the above quotation, is the idea that a state of a 
system is always a stationary state, which stems from Bohr's quantum theory and which appears to have lived 
on through the BKS phase until well into the development of matrix mechanics, indeed at least as late as Heisenberg's
paper on resonance (Heisenberg 1926b [35]). The second is the idea that the state of a system is given by its wave 
function, but it is bound up with the question of whether the latter should be seen as a `spread-out' entity, a `guiding 
field', a `statistical state' or something else.\footnote{Cf.\ the next chapter.} Born's papers on collisions 
(Born 1926a,b [30]) can be said to contain elements of both these approaches. Yet a third 
approach may well be present in the report, an approach in which the notion of state would be purely an effective one. 
Some pronouncements by Heisenberg, in correspondence and in the uncertainty paper (Heisenberg 1927 [46]), 
may support this further (tentative) suggestion. 

It seems to us that Born and Heisenberg's statements become clearer if one is aware of 
the different backgrounds to their discussion. Accordingly, we shall discuss in turn, briefly,
various developments that appear to have fed into their conception of quantum mechanics,
in particular previous work by Born and Wiener on generalising matrix mechanics (section~\ref{BW}), by Born 
on guiding fields (section~\ref{BornBohr}), by Bohr as well as famously by Born on collision processes
(again section~\ref{BornBohr} and section~\ref{Bornsintroduction}), by Heisenberg on atoms in resonance 
(section~\ref{resonance}) and by Dirac on the transformation theory (section~\ref{transf-theory}). 
We shall then discuss the treatment of interpretational issues given in the report in sections~\ref{on-interference} 
and \ref{foundmat}. The latter includes some brief comments on the notion of a `closed theory', 
which is prominent in some of Heisenberg's later writings (e.g.\ Heisenberg 1948), and which appears to be used 
here for the first time. We do not claim to have settled the interpretational issues, and will return to several 
of them in Part II. Overall, the `physical interpretation' of the report requires careful reading and assessment.

\subsection{Matrix mechanics, Born and Wiener}\label{BW}
As we have seen, in the old quantum theory the only states allowed for atomic systems are stationary states,
understood in terms of classical orbits subject to the quantum conditions of Bohr and Sommerfeld. The same 
is true in the BKS theory, where stationary states become more abstract and are represented by the collection
of virtual oscillators corresponding to the transitions of the atom from a given energy level. In both theories,
discontinuous transitions between the stationary states, so-called quantum jumps, are assumed to occur. Although
in today's quantum theory one usually still talks about discontinuous transitions, these are associated with the
collapse postulate and with the concept of a measurement. The states performing these transitions are not necessarily
stationary states, i.e. eigenstates of energy (whether one describes them dynamically in the Schr\"{o}dinger picture 
or statically in the Heisenberg picture). 

This modern notion of state is strikingly absent also in matrix mechanics, 
as we have seen it formulated in the original papers. Indeed, the interpretation of the theory is still 
ostensibly in terms of stationary states and quantum jumps, but the formalism itself contains only matrices, 
which can at most be seen as a collective representation of all stationary states, in the following sense. 
The matrix (\ref{matrix1}) incorporates the oscillations corresponding to all 
possible transitions of the system. Each matrix element is formally analogous to a virtual 
oscillator with frequency $\nu_{ij}=(E_{i}-E_{j})/h$ (corresponding to an atomic transition 
$E_{i}\rightarrow E_{j}$). Therefore, each row (or column) contains all the frequencies corresponding 
to the transitions from (or to) a given energy level, much like a stationary state in the BKS theory. 
As in the above quotation, the matrix can thus be seen as the collective representation of all the stationary
states of a closed system.\footnote{A somewhat similar idea appears to be expressed by Dirac at the beginning of the discussion 
after the report (p.~\pageref{forpage65}), where he emphasises the parallel between matrix mechanics 
and classical Hamiton-Jacobi theory, with the latter also describing not single trajectories 
but whole families of trajectories.}

This analysis is further supported by examining Born and Wiener's work,
in which the picture of stationary states as the rows of the position matrix (now the position
operator) becomes even more explicit. While the functions $x(t)$ remain uninterpreted, 
Born and Wiener appear to associate the `rows' or, rather, the `columns' of their operators with (stationary) states 
of motion. This is clear from the following. Born and Wiener introduce a notion of `column sum', that is, 
a generalisation of the sum of the elements in the column of a matrix. In discussing their main example, the free
particle, they then show that, for the position operator, this generalised column sum $q(t,W)$ takes the form
  \begin{equation}
    q(t,W)=t\sqrt{\frac{2W}{\mu}} + F(W)\ , 
    \label{newyear11}
  \end{equation}
where $\mu$ is the mass of the particle and $F(W)$ is a complex-valued expression independent of $t$. 
They explicitly draw the conclusion that at least the real part of the generalised column sum represents a classical 
inertial motion with the energy $W$.

Note that this indicates not only that Born and Wiener associate the state of a system with a column of the position 
operator. It also suggests that, in their view, at least a limited spatio-temporal picture of 
particle trajectories in the absence of interactions is possible (analogously to the earlier limited use 
of spatio-temporal pictures in describing atomic states in the absence of transitions).

\subsection{Born and Jordan on guiding fields, Bohr on collisions}\label{BornBohr}
The early history of the guiding field idea, in connection with Einstein and with the BKS theory (in the case of photons), 
and in connection with de Broglie (in the case of both photons and material particles), is discussed mainly in 
chapters \ref{guiding-fields-in-3-space} and \ref{deBroglieEss}, respectively. Slater's original intention was in fact
to have the virtual fields to be guiding fields for the photons, which were to carry energy and momentum, 
but this aspect was not incorporated in the BKS theory. 
However, after this theory was rejected precisely because of the results on energy and momentum 
conservation in individual processes (detailed, for instance, in Compton's report), Slater's original idea was fleetingly 
revived. 

On 24 April 1925, after learning from Franck that Bohr had in fact given up the BKS theory, Born sent Bohr the description
of such a proposal (which, as he wrote, he had been working on for some weeks with Jordan). A manuscript
by Born and Jordan followed, entitled `Zur Strahlungstheorie', which is found today in the Bohr archives. \footnote{Cf.\ 
Darrigol (1992, p.~253). We are especially grateful to Olivier Darrigol for helpful correspondence on this matter.} The 
proposal combines the BKS idea of emission of waves while the atom is in a stationary state with the emission of a light 
quantum during an instantaneous quantum jump, in order to give a spacetime picture of radiation. The light quantum thus 
follows the rear end of the wave. It can be scattered or absorbed by other atoms (both processes depending on the dipole 
moment of the appropriate virtual oscillators in the atoms), in the latter case leaving a `dead' wave that has no further 
physical effects. Born and Jordan had applied this picture with some success to a few simple examples, and
were intending to publish the idea in {\em Naturwissenschaften}, provided Bohr or Kramers did not find fault with it 
(Bohr 1984, pp.~84--5 and 308--10). 

Bohr replied on 1 May, after receipt of both the letter and the manuscript, criticising the proposal, on the 
grounds, first, that the proposed mechanism did not guarantee that the trajectories of the light quanta would 
coincide with the propagation of the wave, and second, that the cross section for the absorption ought to be
a constant in order for the particle number density to be proportional to the intensity of the wave. Bohr reiterated
the beliefs he had expressed to Franck: that the coupling between state transitions in different atoms excluded a
description that used anschaulich pictures, and that he suspected the same conclusion to be likely in the case of 
collision phenomena (Bohr 1984, pp.~85 and 310--11).\footnote{See Bohr to Franck, 21 April
1925, in Bohr (1984, pp.~350--51).}

Bohr's work on collisions, to which he alluded here, was an extension of the BKS idea of merely statistical 
conservation laws (Bohr 1925).\footnote{Cf. also Darrigol 
(1992, pp.~249--51) and pp.~89--93 in Stolzenburg's introduction to Part I of Bohr (1984).} The idea, 
as paraphrased by Born, was `to regard the field of the particle passing by in the same way as the field of a 
light wave; thus, it only produces a probability for the absorption of energy, and this [absorption] only occurs when 
the ``collision'' lasts sufficiently long (the particle passes slowly)' (Born to Bohr, 15 January 1925, quoted in Bohr
1984, p.~73).\footnote{Another colourful paraphrase is in Bohr to Franck, 30 March 1925: `If two atoms have the 
possibility of settling their mutual account, it is, of course, simplest that they do so. However, when the invoices
cannot be submitted simultaneously, they must be satisfied with a running account' (quoted in Bohr 1984, p.~74).}
However, the Ramsauer effect --- the anomalously low cross section of atoms of certain gases  
for slow electrons --- could not be accommodated in Bohr's scheme. Bohr 
therefore was developing doubts about statistical conservation laws (and further, about the feasibility altogether 
of a spacetime picture of collisions\footnote{One of 
the most striking expressions of this is a passage in Bohr to Heisenberg, 18 April 1925: `Stimulated especially 
by talks with Pauli, I am forcing myself these days with all my strength to familiarise myself with the mysticism
of nature and am attempting to prepare myself for all eventualities, indeed even for the assumption of a coupling
of quantum processes in separated atoms. However, the costs of this assumption are so great that they cannot be 
estimated within the ordinary spacetime description' (Bohr 1984, pp.~360--61). For other qualms about such `quantum 
nonlocality', cf. also Jordan's habilitation lecture (1927f [62]), in which Jordan considers the idea of microscopic 
indeterminism to be comprehensible only if the elementary random events are independent. (This lecture 
was translated into English by Oppenheimer and published in {\em Nature} as Jordan (1927g).)}), even as he was
submitting his paper on collisions,\footnote{The paper was received by {\em Zeitschrift f\"{u}r Physik} on 30 March 1925;
on that very day Bohr was expressing his doubts in a letter to Franck (Bohr 1984, pp.~348--50).} that is, even before
the results of the Bothe-Geiger experiments were confirmed in April 1925. Bohr's paper was published only after Bohr 
included an addendum in July 1925, which draws the consequences from both the Bothe-Geiger experiments and the 
difficulties with the Ramsauer effect.  In the same month of July, an explanation for the Ramsauer effect was 
suggested by Elsasser (1925) in G\"{o}ttingen, on the basis of de Broglie's matter waves.

\subsection{Born's collision papers}\label{Bornsintroduction}
The above provides a useful backdrop for discussing Born's own work on collisions (Born 1926a,b [30]),\footnote{For 
a modern discussion of collisions, cf.\ section~\ref{scat-in-pwt}.}  which 
treats collision problems on the basis of Schr\"{o}dinger's wave mechanics.\footnote{The Ramsauer effect, however,
is excluded from Born's discussion (1926b [30], footnote on p.~824). A few passages might be seen to 
refer to Born's exchange with Bohr, in particular the remark that, at the price of 
dropping causality, the usual spacetime picture can be maintained (1926b [30], p.~826).} 

This work also reflects the idea that the states of a system are stationary states undergoing transitions. 
Indeed, Born presents the problem as that of including in matrix mechanics a description of the transitions between 
stationary states. The case of collisions, say between an atom and an electron, is chosen as the simplest for treating 
this problem (while still leading to interesting predictions), because it is natural to expect that in this case
the combined system is asymptotically in a stationary state for the atom and a state of uniform translational 
motion for the electron. (Note that this is connected to the treatment of the free particle by Born and Wiener.) If one 
can manage to describe the asymptotic behaviour of the combined system mathematically, this will give concrete 
indications as to the transitions between the initial and final asymptotic states. Born managed to find the solution 
specifically by wave mechanical methods.\footnote{Cf.\ Born to Schr\"{o}dinger, 16 May 1927: `the simple possibility 
of treating with it aperiodic processes (collisions) made me first believe that your conception was superior' (quoted 
in Mehra and Rechenberg 2000, p.~135).} 

Note that Born considers two conceptually distinct objects, the wave function on the one hand and the stationary 
states of the atom and the electron on the other, the connection between them being that the wave function 
defines a probability distribution over the stationary states. (Note also that he reserves the word `state' only for 
the stationary states.) 

Born solves by perturbation methods the time-independent Schr\"{o}dinger equation for the combined system
of atom and electron under the condition that asymptotically for $z\rightarrow\infty$, the solution has the form
$\psi_n(q)e^{\frac{2\pi}{\lambda}z}$ (a product of the $n$-th eigenstate of the atom with a plane wave coming from 
the $z$-direction), with energy $\tau$. Born's solution has the form: 
  \begin{equation}
    \sum_m\int_{\alpha x+\beta y+\gamma z >0} \Phi^\tau_{nm}(\alpha,\beta,\gamma)
    \psi_m(q)e^{k^\tau_{nm}(\alpha x+\beta y+\gamma z+\delta)}d\omega\ ,
    \label{newyear11b}
  \end{equation}
where the energy corresponding to the wave number $k^\tau_{nm}$ equals
  \begin{equation}
    E^\tau_{nm}=h\nu_{nm}+\tau\ ,
  \end{equation}
the $\nu_{nm}$ being the transition frequencies of the atom. The components of the superposition can thus be
associated with various, generally inelastic, collisions in which energy is conserved, and Born interprets 
$|\Phi^\lambda_{nm}(\alpha,\beta,\gamma)|^2$ as the probability for the atom to be in the stationary state 
$m$ and the electron to be scattered in the direction $(\alpha, \beta,\gamma)$.\footnote{A 
statistical interpretation for the modulus squared of the coefficients of
the wave function in the energy basis was introduced also by Dirac (1926c [37]), at the same time as and presumably 
independently of Born (cf.\ Darrigol 1992, p.~333). Note further that, even though it may be tempting to assume 
that each trajectory proceeds from the scattering centre, strictly speaking to each stationary state of the free 
particle corresponds a whole family of inertial trajectories. (Similarly, in the case treated by Born and Wiener,
each generalised column sum associated with a stationary state corresponds to two different inertial trajectories,
depending on the sign of the square root in (\ref{newyear11}).)} 

But now, crucially, since the initial wave function corresponds to a fully determined stationary state and inertial 
motion, this probability is also the probability for a quantum jump from the given initial state to the given final 
state, i.e. a transition probability.

This idea is linked to that of a guiding field. The link is made explicitly at the beginning of Born's second paper (which 
includes the details of the derivation and some quantitative predictions). While Born judges that in the context of optics 
one ought to wait until the development of a proper quantum electrodynamics, in the context of the quantum mechanics of 
material particles the guiding field idea can be applied already, using the de Broglie-Schr\"{o}dinger waves as guiding 
fields. The trajectories of material particles, however, are determined by the guiding field merely probabilistically 
(1926b [30], pp.~803--4). In his conclusion, Born regards the picture of the guiding field as fundamentally 
indeterministic. A deterministic completion, if possible, would not have any practical use. Born also expresses 
the hope that the `laws of motion for light quanta' will find a similar treatment to the one given for electrons, and 
refers to the difficulties `so far' of pursuing a guiding field approach in optics (pp.~826--7).\footnote{Born's 
views on quantum mechanics from this period are also presented in Born (1927), an expanded version 
(published March 1927) of a talk given by Born in August 1926.}

\subsection{Heisenberg on energy fluctuations}\label{resonance}
As discussed in the next chapter, Heisenberg in particular among matrix physicists was opposed to 
Schr\"{o}dinger's attempt to recast and reinterpret quantum theory on the basis of continuous wave functions.  
Schr\"{o}dinger's wave functions were meant 
from the start as descriptions of individual states of a physical system. Even though in general they are 
abstract functions (on configuration space), they can provide a picture of Bohr's stationary states. 
Furthermore, the solution of the time-dependent Schr\"{o}dinger equation appears to provide 
a generalisation of the state of a system as it evolves in time.

Heisenberg appears to have been disturbed
initially also by Born's use of Schr\"{o}dinger's theory in the treatment of collisions, an attitude reflected
in particular in his correspondence with Pauli.\footnote{`One sentence [of Born's paper] reminded me vividly of 
a chapter in the Christian creed: ``An electron {\em is} a plane wave...'' ' (Heisenberg to Pauli 28 July 1926, 
in Pauli 1979, p.~338, original emphasis).} In this connection, the fact that 
Pauli --- in his letter of 19 October (Pauli 1979, pp.~340--9)~--- was in effect able to sketch how one could reinterpret 
Born's results in terms of matrix elements, must have been of particular significance: `Your calculations have given me 
again great hope, because they show that Born's somewhat dogmatic viewpoint of the probability waves is only one of 
many possible schemes'.\footnote{Heisenberg to Pauli, 28 October 1926, in Pauli (1979, p.~350).} 

A few days later, Heisenberg sent Pauli the manuscript of his paper on fluctuation phenomena (Heisenberg 1926b [35]), 
in which he developed considerations similar to Pauli's in the context of a characteristic example, that of two atoms 
in resonance. By focussing on the subsystems of a closed system, Heisenberg was able to derive expressions for 
(transition) probabilities within matrix mechanics proper, without having to introduce the wave function as an 
external aid. A very similar result was derived at the same time by Jordan (1927 [36]), using two systems with 
a single energy difference in common.

We shall now sketch Heisenberg's reasoning. We adapt the presentation of the argument given by 
Heisenberg in {\em The Physical Principles of the Quantum Theory} (Heisenberg 1930b, pp.~142--7),\footnote{This 
very remarkable book is an expanded English edition of Heisenberg (1930a).} which is more general than the one 
given in the paper and clearer than the one given in Born and Heisenberg's report.

Take two systems, 1 and 2, that are
in resonance. Consider, to begin with, that the frequency of the transition $n_1\rightarrow m_1$ in system 1
corresponds to exactly one transition frequency in system 2, say 
  \begin{equation}
    \nu^{(1)}(n_1m_1)=\nu^{(2)}(n_2m_2)\ . 
  \end{equation}
If the systems are uncoupled, the combined system has the degenerate eigenvalue of energy 
  \begin{equation}
    W^{(1)}_{n_1}+W^{(2)}_{m_2}=W^{(1)}_{m_1}+W^{(2)}_{n_2}\ .
    \label{newyear12}
  \end{equation}
If we couple weakly the two systems, the degeneracy will be lifted. Let us label the new eigenstates of the combined
system by $a$ and $b$. We can now consider the
matrix $S$ that transforms the basis of eigenstates of energy of the coupled system to the (product) basis of eigenstates 
of energy of the uncoupled systems. In particular we can consider the submatrix
  \begin{equation}
    \left(\begin{array}{ll}
            S_{n_1m_2,a}   &    S_{n_1m_2,b}   \\
            S_{n_2m_1,a}   &    S_{n_2m_1,b}  
          \end{array}\right)\ .
    \label{newyear13}
  \end{equation}

Choose one of the stationary states of the combined system, say $a$. If the combined system
is in the state $a$, what can one say about the energy of the subsystems, for instance $H^{(1)}$? 
Heisenberg's answer, in the terminology and notation of his book (1930b), is that in the state $a$,
the time average of $H^{(1)}$ (which is no longer a diagonal matrix, thus no longer time-independent), 
or of any function $f(H^{(1)})$, is
  \begin{equation}
    \overline{f(H^{(1)})}=f(W^{(1)}_{n_1}) |S_{n_1m_2,a}|^2 + f(W^{(1)}_{m_1}) |S_{m_1n_2,a}|^2\ . 
    \label{newyear15}
  \end{equation}
Since $f$ is arbitrary, Heisenberg concludes that $|S_{n_1m_2,a}|^2$ is the probability that the state $n_1$ has 
remained the same (and the state $m_2$ has remained the same), and $|S_{m_1n_2,a}|^2$ is the probability that 
the state $n_1$ has jumped to the state $m_1$ (and $m_2$ has jumped to $n_2$). 

The associated transfer of energy between the systems appears to be instantaneous, in that a quantum jump 
in one system (from a higher to a lower energy level) is accompanied by a corresponding jump in the other 
system (from a lower to a higher energy level). The paper merely mentions, without 
elaborating further, that a light quantum (or better a `sound quantum') is exchanged over and over again 
between the two systems. The picture thus avoids the non-conservation of energy of the BKS theory, at 
the price of what appears to be an explicit correlation at a distance. 

In modern terms, Heisenberg has calculated the expectation value of the observable $f(H^{(1)})\otimes 1$ in 
the state $a$:
  \begin{equation}
    \begin{split}
      \langle a | f(H^{(1)})&\otimes 1 | a \rangle  =\\
      &
      \langle a \,| n_1m_2 \rangle    \langle n_1m_2 |\, f(H^{(1)})\otimes 1\, | n_1m_2 \rangle  \langle n_1m_2 |\, a \rangle + \\
      &
      \langle a \,| m_1n_2 \rangle    \langle m_1n_2 |\, f(H^{(1)})\otimes 1\, | m_1n_2 \rangle  \langle m_1n_2 |\, a \rangle = \\
      &
      \langle n_1 | f(H^{(1)}) | n_1 \rangle |\langle n_1m_2 | a \rangle |^2  + \\
      &
      \langle m_1 | f(H^{(1)}) | m_1 \rangle |\langle m_1n_2 | a \rangle |^2 \ .
    \end{split}
    \label{newyear14}
  \end{equation}
Note, however, that rather than focussing on the idea that 
  \begin{equation}
    |S_{n_1m_2,a}|^2=|\langle n_1m_2 | a \rangle |^2
    \label{traintoMilan}
  \end{equation}
is a conditional probability (in fact what we would today call a transition probability), Heisenberg is still
focussing on the transitions between the stationary states of the subsystems, as in Born's work on collisions.

\subsection{Transformation theory}\label{transf-theory}
Less than three weeks after completing his draft on fluctuation phenomena, we find Heisenberg reporting to Pauli
about Dirac's transformation theory, which generalises precisely the formal expression of a conditional probability
given by Heisenberg in terms of the transformation matrix: `Here [in Copenhagen] we have also been thinking 
more about the question of the meaning of the transformation function $S$ and {\em Dirac} has achieved an extraordinarily 
broad generalisation of this assumption from my note on fluctuations' (Pauli 1979, p.~357, original emphasis). 

Dirac indeed presents his results in his paper, significantly titled `The physical interpretation 
of quantum dynamics' (1927a [38]), as a generalisation of Heisenberg's approach. The main goal of the paper 
is the following.\footnote{In our presentation we shall partly follow the analysis by Darrigol (1992, pp.~337-45).}

Take any pair of conjugate matrix quantities $\xi$ and $\eta$, and any `constant of integration' 
$g(\xi,\eta)$.\footnote{This term is meant to include any value of a dynamical quantity at a specified time $t=t_0$ 
(Dirac 1927 [38], p.~623, footnote).} 
Given a value of $\xi$ as a c-number, find the fraction of $\eta$-space for which $g$ lies between any
two numerical values. If $\eta$ is assumed to be distributed uniformly,\footnote{This assumption may sound 
strange, especially since $\xi$ is assumed to have a definite value and $\xi$ and $\eta$ are canonically 
conjugate. Cf.\ however Dirac's remarks on interpretation below.} 
this result will yield the frequency of the given values of $g$ in an ensemble of systems. 
Equivalently, we can state Dirac's goal as that of finding the expectation value (or more precisely, the $\eta$-average) 
of any fixed-time observable $g$, given a certain value of $\xi$.  

The main part of the paper is devoted to developing a `transformation theory' that will allow Dirac to write the
quantity $g$ not in the usual energy representation but in an arbitrary $\xi$-representation. Dirac then suggests taking 
the $\eta$-averaged value of $g$, for $\xi$ taking the c-number value $\xi'$, as given by the diagonal element $g_{\xi'\xi'}$ of $g$ in this 
representation. This is in fact a natural if `extremely broad' generalisation, to an arbitrary pair of conjugate 
quantities $(\xi,\eta)$, of the assumption that the diagonal elements of a matrix in the energy representation 
(such as Heisenberg's $f(H^{(1)})$ from the previous section) are time averages, although the justification Dirac gives for 
this assumption is merely that `the diagonal elements .... certainly would [determine the average values] in the limiting 
case of large quantum numbers' (Dirac 1927 [38], p.~637). 

Dirac writes the elements of a general transformation matrix between two complete sets of variables $\xi$ and $\alpha$
(whether discrete or continuous) as $(\xi'/\alpha')$ (what we would now write $\langle\xi'|\alpha'\rangle$), so that 
the matrix elements transform as
  \begin{equation}
      g_{\xi'\xi''}=\int(\xi'/\alpha')g_{\alpha'\alpha''}(\alpha''/\xi'')d\alpha'd\alpha''
    \label{Dirac1}
  \end{equation}
(Dirac's notation is meant to include the possibility of discrete sums).

His main analytic tool is the manipulation of $\delta$-functions and their derivatives. Dirac shows in particular
that for the quantity $\xi$ itself,
  \begin{equation}
      \xi_{\xi'\xi''}=\xi'\delta(\xi'-\xi'')\ ,
    \label{Dirac2}
  \end{equation}
and that for the quantity $\eta$ canonically conjugate to $\xi$,    
  \begin{equation}
      \eta_{\xi'\xi''}=-i\hbar\delta'(\xi'-\xi'')\ .
    \label{Dirac3}
  \end{equation}
In a mixed representation, one has
  \begin{equation}
      \xi_{\xi'\alpha'}=\xi'(\xi'/\alpha')\ ,
    \label{Dirac4}
  \end{equation}
and
  \begin{equation}
      \eta_{\xi'\alpha'}=-i\hbar\frac{\partial}{\partial\xi'}(\xi'/\alpha')\ ,
    \label{Dirac5}
  \end{equation}
from which follows
  \begin{equation}
      g_{\xi'\alpha'}=g\Big(\xi,-i\hbar\frac{\partial}{\partial\xi'}\Big)(\xi'/\alpha')
    \label{Dirac6}
  \end{equation}
for arbitrary $g=g(\xi,\eta)$.

Choosing $\alpha$ such that $g$ is diagonal in the $\alpha$-representation, one has
  \begin{equation}
      g_{\xi'\alpha'}=g_{\alpha'}(\xi'/\alpha')\ ,
    \label{Dirac7}
  \end{equation}
where the $g_{\alpha'}$ are the eigenvalues of $g$. Therefore, (\ref{Dirac6}) becomes a differential
equation for the $(\xi'/\alpha')$ (seen as functions of $\xi'$), which generalises the time-independent
Schr\"{o}dinger equation.\footnote{Dirac also gives a generalisation of the time-dependent Schr\"{o}dinger equation.}

Once this equation is solved, one could obtain the desired $g_{\xi'\xi'}$  from the $g_{\xi'\alpha'}$ by the
appropriate transformation. The way Dirac states his final result, however, is by considering the matrix
$\delta(g-g')$. The numerical function $\delta(g-g')$, when integrated,
  \begin{equation}
      \int_a^b\delta(g-g')dg'\ ,
    \label{Dirac8}
  \end{equation}
yields the characteristic function of the set $a<g<b$. Therefore, in Dirac's proposed interpretation, the 
diagonal elements of the corresponding matrix in the $\xi$-representation yield, for each value $\xi=\xi'$,
the fraction of the $\eta$-space for which $a<g<b$. That is, if $\eta$ is assumed to be distributed uniformly, 
these diagonal elements yield the conditional probability for $a<g<b$ given $\xi=\xi'$. Thus, the diagonal 
elements of the matrix $\delta(g-g')$ yield the corresponding conditional probability density for $g$ given $\xi=\xi'$.
But now, e.g.\ since for any function $f(g)$ one has
  \begin{equation}
      f(g)_{\xi'\xi''}=\int(\xi'/g'')f(g'')(g''/\xi'')dg''\ ,
    \label{Dirac9}
  \end{equation}
one has in particular
  \begin{equation}
      \delta(g-g')_{\xi'\xi'}=\int(\xi'/g'')\delta(g''-g')(g''/\xi')dg''=(\xi'/g')(g'/\xi')\ .
    \label{Dirac10}
  \end{equation}
Therefore, the conditional probability density for $g$ given $\xi=\xi'$ is equal to $|(g'/\xi')|^2$, a
result that Dirac illustrates by discussing Heisenberg's example of transition probabilities in resonant atoms and 
Born's collision problem.

In parallel with Dirac's development of transformation theory, Jordan (1927b,c [39]) also arrived at a similar 
theory, following on directly from his paper on quantum jumps (1927a [36]) and from his earlier work on canonical 
transformations (1926a,b), to which Dirac also makes an explicit connection. Although Born and Heisenberg state in the 
report (p.~\pageref{completely-equivalent}) that the two methods are equivalent, Darrigol (1992, pp.~343--4)
points to some subtle differences, which are also related to Dirac's criticism in the general discussion of Jordan's 
introduction of anticommuting fields (p.~\pageref{page187}). Dirac also notes that his theory generalises the 
work by Lanczos (1926 [23]). The development of transformation theory from the idea of canonical transformations 
led further towards the realisation that quantum mechanical operators act on a Hilbert space and that the natural 
transformations are in fact unitary.\footnote{See Lacki (2004), who gives details also of London's (1926a,b) contributions to 
this development. Note also the connection between transformation theory and the work on the `equivalence' between 
matrix mechanics and wave mechanics.} 

Dirac concludes his paper with an intriguing suggestion of
  \begin{quote}
    .... a point of view for regarding quantum phenomena rather different from the usual ones.
    One can suppose that the initial state of a system determines definitely the state of the system at any subsequent 
    time. If, however, one describes the state of the system at an arbitrary time by giving numerical values to the
    co-ordinates and momenta, then one cannot actually set up a one-one correspondence between the values of these
    co-ordinates and momenta initially and their values at a subsequent time. All the same one can obtain a good deal
    of information (of the nature of averages) about the values at the subsequent time considered as functions of the
    initial values. The notion of probabilities does not enter into the ultimate description of mechanical processes:
    only when one is given some information that involves a probability ({\em e.g.}, that all points in $\eta$-space
    are equally probable for representing the system) can one deduce results that involve probabilities. (Dirac 1927a 
    [38], p.~641)
  \end{quote}
Here Dirac does not impute indeterminism to nature itself (the
matrix equations are after all deterministic), but instead apparently identifies the source of the statistical element
in the choice of probabilistic initial data.\footnote{This should be compared to the general discussion, 
in which Dirac (a) talks of `an irrevocable choice of nature' (p.~\pageref{Dirac-choice}) in relation to the outcomes 
of an experiment, (b) uses explicitly (perhaps for the first time) the notion of the state vector, when he affirms that 
`[a]ccording 
to quantum mechanics the state of the world at any time is describable by a wave function $\psi$, which normally varies 
according to a causal law, so that its initial value determines its value at any later time' (p.~\pageref{Dirac-state}), 
and in which (c) he describes
the initial data taken for quantum mechanical calculations as describing `acts of freewill', namely `the disturbances 
that an experimenter applies to a system to observe it' (p.~\pageref{Dirac-freewill}); see also section~\ref{det-prob}.}

According to Heisenberg, however, and despite the generality of the results and the `extraordinary progress' obtained 
(Pauli 1979, p.~358), Dirac's transformation theory did not resolve the question of the meaning of quantum mechanics.
As Darrigol (1992, p.~344) emphasises, there is no notion of state vector in Dirac's paper (the well-known 
bras and kets do not appear yet).  
C-number values, and probability distributions over c-number values, now refer to arbitrary quantities or pairs of 
quantities. As Heisenberg wrote: `there are too many c-numbers in all our 
utterances used to describe a fact
' (Pauli 1979, p.~359). Crucially, however, the energy variable and the stationary states no longer played 
a privileged role.

\subsection{Development of the `statistical view' in the report}\label{on-interference}
In the report, Born and Heisenberg appear to understand Born's collision papers (1926a,b [30]) on the one hand and 
the papers by Heisenberg (1926b [35]) and Jordan (1927a [36]) on the other broadly in the same way, as seeking to 
obtain `information .... about actual phenomena', by `direct[ing] one's attention to a {\em part} of the system' 
(p.~\pageref{nothing} of the report). And indeed, by considering coupled systems all of these papers manage to 
derive quantitative expressions for the {\em probabilities} of quantum jumps between energy eigenstates. 

Heisenberg's setting is the one chosen in the report, and since Heisenberg's treatment of interacting systems does 
not use the formalism of wave mechanics, this choice may be intended to make the point that matrix mechanics can 
indeed account for time-dependent phenomena without the aid of wave mechanics.

The form of Heisenberg's result (\ref{newyear15}) as given in the report is in terms of the expected deviation of the 
value of energy from a given initial value, for instance $n_1$:\label{Heisenberg-correction}\footnote{Note that 
$|S_{n_1m_2,a}|^2 + |S_{m_1n_2,a}|^2 =1$, because $a$ is normalised.}
  \begin{equation}
    \begin{split}
    \delta\overline{f_{n_1}} = &\overline{f(H^{(1)})}-f(W^{(1)}_{n_1})= \\
           &\{f(W^{(1)}_{n_1}) - f(W^{(1)}_{n_1})\} |S_{n_1m_2,a}|^2  +   \\
           &\{f(W^{(1)}_{m_1}) - f(W^{(1)}_{n_1})\} |S_{m_1n_2,a}|^2\ . 
    \end{split}
    \label{newyear16}
  \end{equation}
If we write $\Phi_{n_1m_1}$ for the probability of the transition 
$n_1\rightarrow m_1$ in system 1, this becomes equation (\ref{eq20}) of the report, except that
Born and Heisenberg label the matrix 
elements $S_{n_1m_2,a}$ and $S_{m_1n_2,a}$, respectively, by the transitions $n_1m_2\rightarrow n_1m_2$ and
$n_1m_2\rightarrow m_1n_2$, that is, as $S_{n_1m_2,n_1m_2}$ and $S_{n_1m_2,m_1n_2}$, omitting reference to the
state $a$.\footnote{In Heisenberg's paper, the matrix (\ref{newyear13}) corresponds to a 45-degree rotation,
so the probabilities are independent of the choice of $a$ or $b$.} 
One further difference between our description above and the one given in the report is that Born and 
Heisenberg are treating the case in which the transition $n_1\rightarrow m_1$ in system 1 
may resonate with more than one transition in system 2. In this case, the total transition probability 
$\Phi_{n_1m_1}$ is no longer equal to $|S_{n_1m_2,m_1n_2}|^2$ but to
  \begin{equation}
    \Phi_{n_1m_1}=\sum_{m_2n_2}|S_{n_1m_2,m_1n_2}|^2\ .	
    \label{newyear17}
  \end{equation}
Thus we also have equation (\ref{eq21}) of the report. 

It is only after this matrix mechanical discussion that Born and Heisenberg introduce the time-dependent 
Schr\"{o}dinger equation, as a more 
`convenient' formalism for `thinking of the system under consideration as coupled to another one and neglecting the 
reaction on the latter' (p.~\pageref{consideration}). This suggests that Born and Heisenberg may consider
the time-dependent Schr\"{o}dinger equation only as an effective description. This impression is reinforced by 
comparing with Heisenberg's book (1930b, pp.~148--50) where, after the above derivation, Heisenberg continues with 
a more general derivation of time-dependent probabilities, which he then relates to the usual time-dependent 
Schr\"{o}dinger equation. Nevertheless, Born and Heisenberg use the wave function throughout the ensuing 
discussion of probabilities, noting that this formalism `leads to a further development of the statistical 
view', by which they mean in particular the idea of interference of probabilities.\footnote{On these matters cf.\ 
also section~\ref{Time-in-quantum-theory}.} 

First of all Born and Heisenberg relate the wave function to probabilities. They take the time-dependent 
transformation matrix $S(t)$ given by the unitary evolution. For the coefficients of the wave function in 
the energy basis one has
(equation (\ref{eq25}) in the report):
  \begin{equation}
    c_n(t)=\sum_mS_{nm}(t)c_m(0)\ .
    \label{newyear18}
  \end{equation}
If now all $c_m(0)$ except one (say, $c_k(0)$) are zero, from the assumption that a system is always in a stationary 
state it is natural to conclude that the $|S_{nk}(t)|^2$ are the probabilities for transitions to the respective 
energy states (`transition probabilties'), and the $|c_n(t)|^2$ are the resulting probabilities for the stationary 
states (`state probabilities'). In support of this interpretation (which is the same as in Born's collision papers), 
the report quotes in particular Born's paper on 
the adiabatic principle (1926c [34]), that is, Born's proof that in the adiabatic case the transition probabilities 
between different states tend to zero, in accordance with Ehrenfest's (1917) principle.\footnote{Ehrenfest had the idea
that since quantised variables cannot change by arbitrarily small amounts, they should remain constant under adiabatic 
perturbations. This led him to formulate the principle stating that the classical variables to be quantised are the 
adiabatic invariants of the system. Cf.\ Born (1969, pp.~113~ff.).}

Then Born and Heisenberg come to discussing interference.\footnote{Born and Heisenberg give credit to Pauli for the notion 
of interference of probabilities (p.~\pageref{Pauli}). Note that Pauli contributed significantly to the development
of the `statistical view', albeit mainly in correspondence and discussion. Contrary to what is commonly assumed, the 
idea of a probability density for position is not contained in Born's collision papers, but appears in fact in Pauli's 
letter to Heisenberg of 19 October 1926 (Pauli 1979, p.~340--9), together with the idea of the corresponding momentum 
density, and in print in a footnote of Pauli's paper on gas degeneracy and paramagnetism (1927 [44]). (See also 
Heisenberg to Pauli, 28 October 1926, in Pauli 1979, pp.~340--52.) Jordan (1927b [39]), in his second paper 
on the transformation theory, even gives credit to Pauli for the introduction of arbitrary transition probabilities 
and amplitudes.}\label{Paulifootnote} This is also the first time in the presentation of the 
physical interpretation of the theory that {\em measurements} enter the picture. 
Born and Heisenberg note that if $c_k$ is not the only non-zero coefficient 
at $t=0$, then (\ref{newyear18}) does not imply
  \begin{equation}
    |c_n(t)|^2=\sum_m|S_{nm}(t)|^2|c_m(0)|^2\ ,
    \label{newyear18a}
  \end{equation}
but that instead one has
  \begin{equation}
    |c_n(t)|^2=|\sum_mS_{nm}(t)c_m(0)|^2\ .
    \label{newyear18b}
  \end{equation}
The passage immediately following this is both remarkable and, in our opinion, 
very significant (p.~\pageref{Born-contradiction}):\footnote{For further discussion, see 
section~\ref{Interference-in-Born-and-Heisenberg}.}

  \begin{quotation}
    it should be noted that this `interference' does not represent a contradiction with the rules 
    of the probability calculus, that is, with the assumption that the $|S_{nk}|^2$ 
    are quite usual probabilities. In fact, the composition rule [(\ref{newyear18a})] follows from the 
    concept of probability for the problem treated here when and only when the relative number, 
    that is, the probability $|c_{n}|^2$ of the atoms in the state $n$, has been {\em established} 
    beforehand {\em experimentally}. In this case the phases $\gamma_n$  are unknown in principle, 
    so that [(\ref{newyear18b})] then naturally goes over to [(\ref{newyear18a})] .... .
  \end{quotation}
(The passage ends with a reference to Heisenberg's uncertainty paper.) How do Born and Heisenberg 
propose to resolve this apparent contradiction? 

It would make sense to say that the $|S_{nk}(t)|^2$ cannot be taken in general as probabilities for quantum 
jumps, because the derivation of $|S_{nk}(t)|^2$ as a transition probability works only in a special case
(that is, presumably, if the energy at $t=0$ has in fact been measured). On this reading, there might 
conceivably exist some quite different transition probabilities, which lie outside the scope of quantum mechanics
and are presumbly of no practical value (like a deterministic completion in the case of collision 
processes).\footnote{That such probabilities can be defined (albeit non-uniquely), leading to well-defined 
stochastic processes for the quantum jumps, was shown explicitly by Bell (1984).}  

However, this reading does not seem to fit what Born and Heisenberg actually say. Their suggestion seems to be that 
the $|S_{nk}(t)|^2$ are indeed always transition probabilities, but that the $|c_{n}|^2$ are {\em not} always state 
probabilities: the $|c_{n}|^2$ will be state probabilities if and only if the energies have been measured
(non-selectively). This seems analogous to Heisenberg's (1927 [46], p.~197) idea in the uncertainty paper
that the `law of causality' is inapplicable because it is impossible in principle to know the present
with sufficient accuracy (i.e.\ the antecedent of the law of causality fails).\footnote{It appears not to be
well known that the last 20 pages of the original typescript of Heisenberg's uncertainty paper are contained in AHQP,
miscatalogued as an 'incomplete and unpublished paper (pp.~12-31)' by Kramers.\endnotemark The typescript contains slight
textual variants (as compared with the published version) and manuscript corrections in what appears to be Heisenberg's 
hand, but does not include the famous addendum in proof in response to Bohr's criticism (cf.\ p.~\pageref{addendum} 
below). On Heisenberg's treatment of the `law of causality', see also Beller (1999, pp.~110--13).}\endnotetext{AHQP-28 
(H.~A.~Kramers, notes and drafts 1926--52), section 6.} 

Born and Heisenberg swiftly move on to generalising the discussion to the case of arbitrary observables,
on the basis of Dirac's and Jordan's transformation theory (Dirac 1927a [38], Jordan 1927b,c [39]). They 
introduce the interpretation of 
$|\varphi|^2$ as a position density, and consider in particular the density $|\varphi(q',W')|^2$ 
defined by the stationary state $\varphi(q',W')$ with the energy $W'$, or in modern notation,
  \begin{equation}
    |\varphi(q',W')|^2=|\varphi_{W'}(q')|^2=|\langle q' | W' \rangle|^2\ .
    \label{newyear19}
  \end{equation}
This is immediately generalised to arbitrary pairs of observables $q$ and $Q$ with values 
$q'$ and $Q'$:
  \begin{equation}
    |\varphi(q',Q')|^2=|\langle q' | Q' \rangle|^2\ .
    \label{newyear20}
  \end{equation} 
Born and Heisenberg call this a `relative state probability', reserving the term `transition probability' for the 
case of a single observable evolving in time (or depending on some external parameter), 
always with the proviso that in general one should expect interference of the corresponding `transition 
amplitudes'.

Note that the physical interpretation of this generalisation makes sense only if one takes over from the above 
the idea that probabilities such as $|\varphi|^2$ are well-defined only upon measurement, or more precisely, that 
actual frequencies upon measurement will be given by the expression $|\varphi|^2$. Born and Heisenberg's
terminology, however, is somewhat ambiguous.\footnote{In one paragraph they refer to `the probability that 
for given energy $W'$ the coordinate $q'$ {\em is} in some given element $dq'$', whereas in the next they refer 
to `the probability, given $q'$, to {\em find} the value of $Q'$ in $dQ'$' (p.~\pageref{find}; italics added).} 

A major conceptual shift appears to be taking place, which may be easy to miss. Do quantum jumps still 
occur whenever two systems interact, or do they now occur only between measurements? Indeed, are systems 
always in stationary states, as has been explicitly assumed until now, or only when the energy is measured? 
Heisenberg's uncertainty paper (on pp.~190--1), as well as the correspondence with Pauli (Heisenberg to 
Pauli, 23 February 1927, in Pauli 1979, pp.~376--82) both mention explicitly the loss of a privileged status for 
stationary states. It seems that, even though we are not explicitly told so, the picture of quantum jumps (that is, 
of probabilistic transitions between possessed values of energy) is shifting to that of probabilistic transitions 
from one measurement to the next.

The idea that frequencies are well-defined only upon measurement appears to play the same role as 
von Neumann's projection postulate. As discussed in more detail in chapter~\ref{Interference-superposition-collapse}, 
however, it is far from clear whether that is what Born and Heisenberg have in mind. A similar notion is introduced in 
Heisenberg's uncertainty paper (1927 [46], p.~186), but again in terms that are `somewhat mystical' (Pauli to 
Bohr, 17 October 1927, in Pauli 1979, p.~411). It appears explicitly in the proceedings only in the general 
discussion, in Born's main contribution (p.~\pageref{page172}) and in the intriguing exchange between Dirac and 
Heisenberg (pp.~\pageref{Dirac-determinism}~ff.).
In chapter~\ref{Interference-superposition-collapse} and section~\ref{Time-in-quantum-theory}, we shall return to 
Born and Heisenberg's view of interference and to the question of whether, according to them, the 
collapse of the wave function and the time-dependent Schr\"{o}dinger evolution are at all fundamental processes.


\subsection{Justification and overall conclusions}\label{foundmat}
The following section III (drafted by Heisenberg) presents Jordan's axiomatic formulation of quantum mechanics 
(Jordan 1927b,c [39]),\footnote{This formulation, which is how Jordan presents his transformation theory, 
was explicitly intended as a generalisation of the formalisms of matrix mechanics, wave mechanics, q-number theory 
and of Born and Wiener's original operator formalism.} and justifies the necessity of a statistical view in the context 
of Heisenberg's notion of 
uncertainty (Heisenberg 1927 [46]). Born and Heisenberg argue as follows. Even in classical mechanics, if certain 
quantities (for instance the phases of the motion) were known only with a certain imprecision, the future evolution of 
the system would be only statistically constrained. Now, the uncertainty relations prevent one from determining the values 
of all physical quantities, providing a fundamental limit of precision. In addition, quantum mechanics prescibes different 
laws for the time evolution of the statistical constraints. Imprecise initial conditions can be described by choosing 
certain `probability functions' (this is the closest Born and Heisenberg come to discussing the `reduction of the wave 
packet' as presented in the uncertainty paper), and `the quantum mechanical laws determine the change (wave-like 
propagation) of these probability functions' (p.~\pageref{laws}). Born and Heisenberg claim that discussion of 
the cases in which these laws coincide to a very good approximation with the classical evolution of a probability 
density justifies the classical treatment of $\alpha$- and $\beta$-particle trajectories in a cloud 
chamber.\footnote{Born and Heisenberg's remarks about different laws of propagation of the probabilities may
refer to the conditions under which (\ref{newyear18b}) reduces to (\ref{newyear18a}), which would arguably be an 
early example of decoherence considerations. However, the remark is too brief, and Born and Heisenberg may be merely
comparing the spreading of the quantum probabilities with that of a Liouville distribution. For a modern treatment 
of the latter comparison, see Ballentine (2003).} They thus maintain that the statistical element in the theory can be reconciled 
with macroscopic determinism.\footnote{For further discussion of these issues see sections~\ref{QM-without-collapse} 
and \ref{Further-remarks-on-Born}.}

Section III arguably addresses the dual task set in the introduction 
of ensuring that quantum mechanics is `consistent in itself' and of showing that quantum mechanics 
can be taken to `predict unambiguously the results for all experiments conceivable in 
its domain' (p.~\pageref{unambiguously}). This task appears
to be related to two conceptual {\em desiderata}, that the theory be intuitive (anschaulich) and closed
(abgeschlossen). These are touched upon briefly in the report, especially in the introduction and conclusion, 
but are important both in the debate with Schr\"{o}dinger (see section~\ref{Schr-conflict}) and in some of 
Heisenberg's later writings (in particular, Heisenberg 1948). The report appears to be the first instance 
in which Heisenberg uses the concept of a closed theory.\footnote{Compare also Scheibe (1993) on the concept of 
closed theories in Heisenberg's thought. The origin of this concept has also been traced to two earlier papers 
(Heisenberg 1926a [28] and 1926c [60]); see for instance Chevalley (1988). However, in the first paper  
there is no mention of closed theories, only of closed systems of terms (symmetric and antisymmetric), a point 
also repeated in the second paper. In the latter, Heisenberg mentions the need to introduce equations for the matrix 
variables in order to obtain a `closed theory', but this does not seem to be the same use of the term as in the 
report. We wish to thank also Gregor Schiemann for correspondence and references on this topic.} 

As defined in the report, a closed theory is one that has achieved a definitive form, and is no longer 
liable to modification, either in its mathematical formulation or in its physical meaning. This is made more
precise in later presentations (e.g.\ Heisenberg 1948), in which Heisenberg includes the applicability of the
concepts of a theory in the analysis. All closed theories possess a specific domain of application, within which 
they are and will always remain correct. Indeed, their concepts always remain part of the scientific language and 
are constitutive of our physical understanding of the world. In the report, quantum mechanics (without the inclusion 
of electrodynamics) is indeed taken to be a closed theory,\footnote{Heisenberg (1948) lists Newtonian mechanics, 
Maxwellian electrodynamics and special-relativistic physics, thermodynamics, and nonrelativistic quantum mechanics as the four main 
examples of such theories.} so that different assumptions about the physical meaning of quantum mechanics (such as Schr\"{o}dinger's 
idea of taking $|\varphi|^2$ to be a charge density\footnote{Described in more detail in section~\ref{Schr-radiation}.}), 
would lead to contradictions with experience. Thus, the report ends on a note of utmost confidence.

\newpage
\renewcommand{\enoteheading}{\section*{Archival notes}}
\addcontentsline{toc}{section}{\em Archival notes}
\theendnotes

\setcounter{endnote}{0}
\setcounter{equation}{0}

\chapter{Schr\"{o}dinger's wave mechanics}\label{SchrEss}\chaptermark{Schr\"{o}dinger's wave mechanics}

Schr\"{o}dinger's work on wave mechanics in 1926 appears to have been driven by the idea that one could give 
a purely wave-theoretical description of matter. Key elements in this picture were the idea of particles as wave
packets (section~\ref{Schr-packets}) and the possible implications for the problem of radiation
(section~\ref{Schr-radiation}). This pure wave theory, in contrast to de Broglie's work, did away with the idea of point 
particles altogether (section~\ref{Schr-deB}). The main conflict, however, was between Schr\"{o}dinger and the proponents
of quantum 
mechanics (section~\ref{Schr-conflict}), both in its form at the time of Schr\"{o}dinger's papers and in its further 
developments as sketched in the previous chapter.  

For reference, we provide a brief chronology of Schr\"{o}dinger's writings relating 
to wave mechanics up to the Solvay conference:
  \begin{itemize}
    \item
      Paper on Einstein's gas theory, submitted 15 December 1925, published 1 March 1926 (Schr\"{o}dinger 1926a).
    \item
      First paper on quantisation, submitted 27 January 1926, addendum in proof 
      28 February 1926, published 13 March 1926 (Schr\"{o}dinger 1926b).
    \item
      Second paper on quantisation, submitted 23 February 1926, published 6 April 1926 (Schr\"{o}dinger 1926c).
    \item
      Paper on the relation between wave and matrix mechanics (`equivalence paper'), submitted 
      18 March 1926, published 4 May 1926 (Schr\"{o}dinger 1926d).
    \item
      Paper on micro- and macromechanics (coherent states for the harmonic oscillator), published 9 July 1926 
      (Schr\"{o}dinger 1926e).
    \item
      Third paper on quantisation, submitted 10 May 1926, published 13 July 1926 (Schr\"{o}dinger 1926f).
    \item
      Fourth paper on quantisation, submitted 21 June 1926, published 5 September 1926 (Schr\"{o}dinger 1926g).
    \item
      Review paper in English for the {\em Physical Review}, submitted 3 September 1926, 
      published December 1926 (Schr\"{o}dinger 1926h).
    \item
      Preface to the first edition of {\em Abhandlungen zur Wellenmechanik}, dated November 1926 
      (Schr\"{o}dinger 1926i).
    \item
      Paper on the Compton effect in wave mechanics, submitted 30 November 1926, published 
      10 January 1927 (Schr\"{o}dinger 1927a).
    \item
      Paper on the energy-momentum tensor, submitted 10 December 1926, published 10 January 1927 
      (Schr\"{o}dinger 1927b).
    \item
      Paper on energy exchange in wave mechanics, submitted 10 June 1927, published 9 August 1927
      (Schr\"{o}dinger 1927c).
  \end{itemize}

We shall now discuss the above points in turn, after a brief discussion of the planning of Schr\"{o}dinger's report 
for the conference (section~\ref{Schr-invitation}) and a summary of the report itself (section~\ref{Schr-summary}).

\section{Planning of Schr\"{o}dinger's report}\label{Schr-invitation}
As reported in chapter~\ref{HistEss}, the scientific committee of the Solvay
institute met in Brussels on 1 and 2 April 1926 to plan the fifth Solvay conference.
Lorentz had asked Ehrenfest to suggest some further names of possible participants,
and it is in Ehrenfest's letter of 30 March\endnote{Ehrenfest 
to Lorentz, 30 March 1926, AHQP-LTZ-11 (in German).} that Schr\"{o}dinger's name is first
mentioned in connection with the conference.\footnote{Schr\"{o}dinger 
had already been a participant in the fourth Solvay conference, though not a speaker; 
cf.\ Moore (1989, pp.~157--8).} On this occasion, Ehrenfest suggested Schr\"{o}dinger
on the basis of a paper in which Schr\"{o}dinger proposed an expression for the broadening 
of spectral lines due to the Doppler effect, and which applied the conservation laws to 
phenomena involving single light quanta (Schr\"{o}dinger 1922).\footnote{As Schr\"{o}dinger 
points out, the calculated broadening of the spectral lines is small compared to that 
expected on the basis of the thermal agitation of the radiating gas, otherwise the effect 
could be used as a test of the light quantum hypothesis.} Evidently, neither Lorentz 
nor Ehrenfest were yet aware of Schr\"{o}dinger's work on wave mechanics.

In the meantime, Schr\"{o}dinger sent to Lorentz the proof sheets of his first two papers 
on quantisation, also on 30 March,\endnote{Schr\"{o}dinger to Lorentz, 30 March 1926, AHQP-LTZ-8 
(in German).} thus initiating his well-known correspondence with Lorentz on wave mechanics.\footnote{Most 
of this correspondence is translated in Przibram (1967). In the letter of 30 March, not included there, 
Schr\"{o}dinger suggests reading the second paper on quantisation before the first, which should be seen rather an as  
example of an application. Also, he writes that the variational principle of the first paper is given a sensible formulation 
only in the addendum in proof. Finally, he mentions the paper on Einstein's gas theory in the {\em Physikalische 
Zeitschrift} (1926a) as a kind of preparatory work. Note that Lorentz on 27 May\label{forpage91} thanks 
Schr\"{o}dinger for the proof sheets of three papers rather than the two mentioned in Schr\"{o}dinger's 
letter. This third paper is clearly the equivalence paper (1926d), and was presumably sent separately  
(cf.\ Przibram 1967, pp.~43 and 55--6).}
Accordingly, already in a report of 8 April 1926 to the administrative commission,\endnote{Verschaffelt to 
Lef\'{e}bure, 8 April 1926, IIPCS 2573 (in French).} Schr\"{o}dinger is listed as a possible substitute for 
Heisenberg for a lecture on the `adaptation of the foundations of dynamics to the quantum theory'. (It is 
unlikely, however, that the papers reached Lorentz before the meeting in Brussels.\footnote{Lorentz had 
written to Ehrenfest on 29 March from Paris, where he had another meeting, and appears to have travelled 
to Brussels directly from there.\endnotemark})\endnotetext{Lorentz to Einstein, 6 April 1926, AHQP-86 (in German).} 
In January 1927 then, as most of the other participants, Schr\"{o}dinger was invited to the fifth 
Solvay conference.\endnote{Lorentz to Schr\"{o}dinger, 21 January 1927, AHQP-41, section 9 (in German).} 

A few weeks later, Schr\"{o}dinger had the opportunity to discuss personally with Lorentz the plans for the report
`under the beautiful palms of Pasadena', as he recalls in a letter of June 1927. In the same letter, we 
find a useful sketch of the theme and focus of Schr\"{o}dinger's report; we also gather that Schr\"{o}dinger 
was wary of the potential for a confrontation in Brussels:\endnote{Schr\"{o}dinger to Lorentz, 23 June 1927, 
AHQP-LTZ-13 (original with Schr\"{o}dinger's corrections) and AHQP-41, section 9 (carbon copy) (in German).} 
  \begin{quotation}
    .... I nurtured the quiet hope you would yet return to your first plan and
    entrust only Messrs [d]e Broglie and Heisenberg with reports on the new mechanics. 
    But now you have decided otherwise and I will of course happily perform my duty.

    Yet I fear that the `matricians' (as Mr Ehrenfest used to say) will feel 
    disadvantaged. Should it come to different views, which might after all urge on the 
    committee the wish to limit the reports to two, you know, dear Professor, 
    that I shall always happily remit my charge into your hands. 
  \end{quotation}
According to Schr\"{o}dinger's sketch, the report is to stress points of principle, rather than the (by then many)
applications of the theory. First of all, one has to distinguish clearly between two wave-mechanical theories: a 
theory of waves in space and time (which however runs into difficulties especially with the many-electron problem), 
and the highly successful theory of waves in configuration space (which however is not relativistic). 
A difficulty of principle to be discussed in the context of the spacetime theory is the possibility of developing an 
interacting theory, which seems to require distinguishing between the fields generated by different 
particles, and whether this can be done in a sensible way, or perhaps be avoided.\footnote{Cf.\ p.~\pageref{indiv} 
of the report.} In the context of the configuration-space theory, the main question is how to interpret the wave 
function. Schr\"{o}dinger mentions the widespread view that the wave describes only ensembles, as well as his own 
`preferred interpretation as a real description of the individual system, which thereby becomes a kind of ``mollusc'''. 
In the letter (but not in the report), he is explicit about some of his misgivings about the ensemble view (as well 
as about the difficulties with his own preferred understanding, namely the `failure of the electrons to stay together 
and similar'). Indeed, he points out that the Schr\"{o}dinger equation is time-symmetric (if one includes complex 
conjugation), while experience teaches us that the statistical behaviour of ensembles cannot be described 
time-symmetrically. Also, insistence on a statistical interpretation leads to `mystical' calculations with amplitudes 
and thus to problems with the laws of probability.\endnote{See also Schr\"{o}dinger to Lorentz, 16 July 1927, AHQP-LTZ-13 
(in German).}

\section{Summary of the report}\label{Schr-summary}
The eventual form of Schr\"{o}dinger's report follows roughly the sketch given 
above, with an introduction, followed by three main sections, respectively on 
the configuration-space theory, on the spacetime theory and on the many-electron 
problem.

{\em Introduction.} Schr\"{o}dinger draws the distinction between the spacetime 
theory (four-di\-men\-sion\-al) and the configuration-space
theory (multi-di\-men\-sion\-al). He states that the use of configuration 
space is a mathematical way for describing what are in fact
events in space and time. However, it is the multi-dimensional 
theory that is the most successful and has proved to be a
powerful analytic tool in relation to Heisenberg and Born's 
matrix mechanics. The multi-dimensional point of view has
not been reconciled yet with the four-dimensional one.\footnote{As becomes clear in the 
discussion, Schr\"{o}dinger thinks that the multi-dimensional theory may prove
indispensable, so that one should accept the notion of a $\psi$-function 
on configuration space and try instead to understand its physical
meaning in terms of its manifestation in space and time. 
(See the discussion in section~\ref{Schr-radiation} below.)}

I. {\em Multi-dimensional theory.} Schr\"{o}dinger sketches a 
derivation of his time-independent wave equation, noting that it
reproduces or improves on the results of Bohr's quantum theory. 
He also notes that the stationary states allow one to calculate
the transition probabilities encountered in matrix mechanics. 
If one wishes to consistently develop a formalism in which there are
only discontinuous transitions, he suggests one should take seriously 
the idea that the transitions do not occur against a
continuous time background; the appearence of a continuous time parameter would be purely 
statistical, so to speak.\footnote{Cf.\ the discussion in ch.~\ref{Time-in-quantum-theory}.} 
As an alternative, he suggests interpreting the time-independent equation as arising 
from a time-dependent one from which the time variable is eliminated by assuming a stationary solution. 
He thus arrives to the description of a quantum system in terms of a time-dependent wave 
function on configuration space. He then asks what the meaning 
of this wave function is: `{\em how does the system described by it
really look like in three dimensions\/}?' (p.~\pageref{mean}, Schr\"{o}dinger's 
emphasis). He briefly mentions the view that the $\psi$-function describes an 
ensemble of systems, which Born and Heisenberg are going to discuss. 
Schr\"{o}dinger instead finds it useful (if perhaps `a bit naive') to imagine
an individual system as continuously filling the whole of space somehow 
weighted by $|\psi|^2$ (as he further clarifies in the discussion).
Schr\"{o}dinger then carefully spells out that the spatial density 
resulting from the configuration space density is not a classical charge
density, in the sense that the action on the particles by external 
fields and the interaction between the particles are already described by
the potentials in the wave equation, and that it is inconsistent to assume 
that this spatial density is also acted upon in the manner of a classical
charge density. Instead, it is possible to interpret it as a charge density (with some 
qualifications, some of which are spelled out only in the discussion) for the purpose of calculating 
the (classical) radiation field, thus yielding a partial vindication of the idea of spatial 
densities. This, however, must be an approximation, since the observation of such emitted 
radiation is itself an interaction between the emitting atom and some other
absorbing atom or molecule, to be described again by the 
appropriate potentials in the wave equation.

II. {\em Four-dimensional theory.}  Schr\"{o}dinger shows that the 
time-de\-pend\-ent wave equation for a single particle is a nonrelativistic 
approximation (with subtraction of a rest frequency) to the wave 
equation for the de~Broglie phase wave of the particle.
The latter can be made manifestly relativistic by including vector 
potentials. (In modern terminology, this is the Klein-Gordon equation.) 
If one couples the Maxwell field to it, the 
same spatial densities discussed in section~I appear as charge 
(and current) densities. However, in the application to the electron 
in the hydrogen atom, it becomes apparent that adding the self-field of the 
electron to the (external) field of the nucleus yields the wrong 
results.\footnote{Cf.\ Schr\"{o}dinger (1927b), as mentioned in section~\ref{Schr-conflict} below.} 
Thus Schr\"{o}dinger argues that if one hopes to develop a spacetime 
theory of interacting particles, it will be necessary to consider not 
just the overall field generated by the particles, but to distinguish between the 
(spatially overlapping) fields generated by each individual particle, each 
field acting only on the {\em other} particles of the system. Finally, he notes
that the Klein-Gordon equation needs to be modified in order to describe 
spin effects, and that it may be possible to do so by considering
a vectorial instead of a scalar $\psi$.

III. {\em The many-electron problem.} Schr\"{o}dinger returns to the 
multi-dimensional theory and its treatment, by approximation, 
of the many-electron atom. His interest in this specific example relates to the 
question of whether this multi-di\-men\-sion\-al system can be understood in 
spacetime terms. The treatment first neglects the interaction 
potentials between the electrons, and as a first approximation takes products
$\psi_{k_1}\ldots\psi_{k_n}$ of the single-electron wave functions 
as solutions. One then expands the solution of the full equation in terms of
the product wave functions. The time-dependent coefficients $a_{k_1\ldots k_n}(t)$ 
in this expansion can then be calculated approximately
if the interaction between the electrons is small. Schr\"{o}dinger shows 
that --- before any approximation --- the coefficients in the equations 
for the $a_{k_1\ldots k_n}(t)$ depend only on potentials calculated from 
the spatial charge densities associated with the $\psi_{k_i}$. Thus,
although the solution to the full equation is a function on configuration 
space, it is determined by purely spatial charge densities. According
to Schr\"{o}dinger, this reinforces the hope of providing a spatial interpretation 
of the wave function. A sketch of the approximation method concludes this 
section and the report.

\section{Particles as wave packets}\label{Schr-packets}
The idea of particles as wave packets is crucial to the development of Schr\"{o}dinger's
ideas and appears to provide one of the main motivations, at least initially, behind the 
idea of a description of matter purely in terms of waves.

Schr\"{o}dinger's earliest speculation about wave packets (for both material particles and light quanta)
is found in his paper on Einstein's gas statistics (1926a), in section 5, `On the possibility of representing
molecules or light quanta through interference of phase waves'. As he explains, Schr\"{o}dinger finds it uncanny
that in de Broglie's theory one should consider the phase waves of the corpuscles to be plane waves, since it is 
clear that by appropriate superposition of different plane waves one can construct a `signal', which following Debye 
(1909) and von Laue (1914, section 2) can be constrained to a small spatial volume. He then continues:
  \begin{quote}
    On the other hand, it is of course {\em not} to be achieved by the classical wave laws, that the constructed 
    `model of a light quantum' --- which by the way extends indeed for {\em many} wavelengths in every direction --- 
    also {\em permanently} stays together. Rather, it spreads itself out [zerstreut sich] over ever larger
    volumes after passing through a focal point.

    If one could avoid this last conclusion by a quantum theoretical modification of the classical wave laws, then
    a way to deliverance from the light quantum dilemma would appear to be paved [angebahnt].  (1926a, p.~101)
  \end{quote}

Wave packets are first discussed at length in the second paper on quantisation (1926c): 
after describing the optical-mechanical analogy, Schr\"{o}dinger discusses how one can construct wave 
packets (using the analogues of the optical constructions by Debye and von Laue), and then shows that 
the centroid of such a wave packet follows the classical equations of motion. Schr\"{o}dinger conjectures 
that material points are in fact described by wave packets of small dimensions. He notes also that, for 
systems moving along very small orbits, the packet will be spread out, so that the idea of the trajectory 
or of the position of the electron inside the atom loses its meaning.  
In fact, the main problem that Schr\"{o}dinger was to face with regard to wave packets turned out to be that spreading
of wave packets is a much more generic feature than he imagined at first.

As mentioned, Schr\"{o}dinger sent to Lorentz the proofs of his first two papers on quantisation (1926b,c)  
on 30 March 1926. In his reply, among many other things, Lorentz discussed explicitly the idea of wave packets, 
indeed doubting that they would stay together (Przibram 1967, 
pp.~47--8).\footnote{As 
Lorentz remarks, the alternative would be `to dissolve the electron completely .... and to replace it by a 
system of waves' (p.~48), which, however, would make it difficult to understand phenomena such as the photoelectric effect. The latter was in fact
one of the criticisms levelled at Schr\"{o}dinger's theory by Heisenberg (see below, section~\ref{Schr-conflict}).} 
Schr\"{o}dinger commented on this both in his reply of 6 June (Przibram 1967, pp.~55--66) and in an earlier letter of 
31 May to Planck (Przibram 1967, pp.~8--11). In the latter, he admits that there will always exist spread-out states, 
because of linearity, but still hopes it will be 
possible to construct packets that stay together for hydrogen orbits of high quantum number. As he notes in the reply 
to Lorentz, this would imply that there is no general identification between hydrogen eigenstates and Bohr orbits, since 
a Bohr orbit of high quantum number would be represented by a wave packet rather than a stationary wave. (For an electron 
in a hydrogen orbit of low quantum number, Schr\"{o}dinger did not envisage an orbiting packet but indeed a spread-out 
electron.) With the reply to Lorentz, Schr\"{o}dinger further sent his paper on micro- and macromechanics (1926e), in 
which he showed that for the harmonic oscillator, wave packets do stay together. He hoped that the result would generalise 
to all quasi-periodic motions (admitting that maybe there would be `dissolution' for a free electron). 
However, on 19 June Lorentz sent Schr\"{o}dinger a calculation showing that wave packets on
a high hydrogen orbit would indeed spread out (cf.\ Przibram 1967, pp.~69--71, where the details of the calculation, 
however, are omitted). 

The relevance of high hydrogen orbits 
is, of course, the role they play in Bohr's correspondence principle. The fact that wave packets 
along such orbits do not stay together was thus a blow for any hopes Schr\"{o}dinger might have 
had of explaining the transition from micro- to macromechanics along the lines of
the correspondence principle.

A further blow to the idea of wave packets must have come with Born's papers on collision theory during the
summer of 1926 (Born 1926a,b; cf.\ section~\ref{Bornsintroduction} above). In the equivalence paper, Schr\"{o}dinger had included
a remark about scattering, for which, he wrote, it is `indeed necessary to understand clearly 
the continuous transition between the macroscopic anschaulich mechanics and the micromechanics of the atom' (1926d, 
p.~753; see below, section~\ref{Schr-conflict}, for the notion of Anschaulichkeit). Given that at the time Schr\"{o}dinger thought that electrons on high quantum orbits should be described by 
wave packets, this remark may indicate that Schr\"{o}dinger also thought that scattering should involve a deflection 
of wave packets, which would move asymptotically in straight lines.\footnote{If thus was indeed 
Schr\"{o}dinger's intuition, it may seem quite remarkable. On the other hand, so is the fact that Schr\"{o}dinger does
not seem to discuss diffraction of material particles, which seems equally problematic for the idea of wave packets.}

Born's work on collisions in the summer of 1926 made essential use of wave mechanics, but suggested a very different
picture of scattering. Born explicitly understood his work as providing an alternative interpretation both to his 
earlier views on matrix mechanics and to Schr\"{o}dinger's views. 
In turn, Schr\"{o}dinger appeared to be sceptical of Born's suggestions, writing on 25 August to Wilhelm Wien:
  \begin{quotation}
    From an offprint of Born's last work in the {\em Zeitsch.\ f.\ Phys.} I know more or less how he thinks of 
    things: the {\em waves} must be strictly causally determined through field laws, the {\em wavefunctions}
    on the other hand have only the meaning of probabilities for the {\em actual} motions of light- or 
    material-particles. I believe that Born thereby overlooks that .... it would depend on the taste of the
    observer which he now wishes to regard as real, the particle or the guiding field. (Quoted in Moore 1989, p.~225)
  \end{quotation}
A few days later, Schr\"{o}dinger submitted a review paper on `undulatory mechanics' to the {\em Physical Review} 
(1926h). In it, he qualified rather strongly the idea that `material points consist of, or are nothing but, wave 
systems' (p.~1049). Indeed, he continued (pp.~1049--50): 
  \begin{quotation}
    This extreme conception may be wrong, indeed it does not offer as yet the slightest explanation of why such 
    wave-systems seem to be realized in nature as correspond to mass-points of definite mass and charge. .... a
    thorough correlation of all features of physical phenomena can probably be afforded only by a harmonious
    union of these two extremes. 
  \end{quotation}

During the general discussion at the Solvay conference, Schr\"{o}dinger summarised the situation with the 
following words (this volume, p.~\pageref{Schr-picture}):
  \begin{quotation}
    The original picture was this, that what
    moves is in reality not a point but a domain of excitation of finite
    dimensions .... . One has since found that the naive identification
    of an electron, moving on a macroscopic orbit, with a wave packet encounters
    difficulties and so cannot be accepted to the letter. The main difficulty is
    this, that with certainty the wave packet spreads in all directions when it
    strikes an obstacle, an atom for example. We know today, from the interference
    experiments with cathode rays by Davisson and Germer, that this is part of the
    truth, while on the other hand the Wilson cloud chamber experiments have shown
    that there must be something that continues to describe a well-defined
    trajectory after the collision with the obstacle. I regard the compromise
    proposed from different sides, which consists of assuming a combination of
    waves and point electrons, as simply a provisional manner of resolving the difficulty.
  \end{quotation}

The problem of the relation between micro- and macrophysics is connected of course to the linearity of
the wave equation, which appears to lead directly to highly nonclassical states (witness Schr\"{o}dinger's famous `cat' 
example, Schr\"{o}dinger 1935).\footnote{Note, on the other hand, that Schulman (1997) shows there are `classical' 
solutions to the linear equation in quite realistic models of coupling between micro- and macrosystems (measurements), 
provided one requires them to satify appropriate boundary conditions at {\em both} the initial and the final time.}  
One might ask whether Schr\"{o}dinger himself considered the idea of a nonlinear wave equation. 
In this connection, a few remarks by Schr\"{o}dinger may be worth investigating further.
One explicit, if early, reference to nonlinearity is contained in the letter of 31 May to Planck, where, 
after noticing that linearity forces the existence of non-classical states, Schr\"{o}dinger indeed speculates that the 
equations might be only approximately linear (Przibram 1967, p.~10). In the correspondence with Lorentz, the 
question of nonlinearity arises in other contexts. In the context of the problem of radiation, it appears at 
first to be necessary for combination tones to arise (pp.~49--50). 
In the context of radiation reaction, Schr\"{o}dinger writes that the exchange with Lorentz has convinced him 
of the necessity of nonlinear terms (p.~62). In print, Schr\"{o}dinger mentioned the possibility of a 
nonlinear term in order to include radiaton reaction in his fourth paper on quantisation (1926g, 
p.~130).\footnote{An explicit proposal for including radiation reaction through a nonlinear term is due to 
Fermi\label{Fermi} (1927). 
Classically, if one includes radiation reaction, one has the third-order nonrelativistic Abraham-Lorentz equation%
\begin{equation}
\mathbf{\ddot{x}}-\tau\mathbf{\dddot{x}}=\mathbf{F}/m\ ,
\end{equation}
where $\mathbf{F}$ is the externally applied force and $\tau=(2/3)e^{2}%
/mc^{3}$ ($e$ is the electron charge, $m$ is the mass).
Fermi (1927) proposed modifying the Schr\"{o}dinger equation as follows:%
\begin{equation}
i\frac{\partial\Psi}{\partial t}=\hat{H}\Psi-m\tau\Psi\mathbf{x\cdot}%
\frac{d^{3}}{dt^{3}}\int d^{3}\mathbf{x}^{\prime}\ \Psi^{\ast}\mathbf{x}%
^{\prime}\Psi \ .
\end{equation}
That is, he added an extra `potential' $-m\tau\mathbf{x\cdot}\frac{d^{3}%
}{dt^{3}}\left\langle \mathbf{x}\right\rangle $. Rederiving the Ehrenfest
theorem, one finds that $\left\langle \mathbf{x}\right\rangle $ then obeys the
above Abraham-Lorentz equation.} Finally, a 
few years later, in his second paper on entanglement (1936, pp.~451--2), Schr\"{o}dinger mentioned the 
possibility of spontaneous decay of entanglement at spatial separation, which would have meant a yet 
untested violation of the Schr\"{o}dinger equation. 

Modern collapse theories, such as those by Ghirardi, Rimini and Weber (1986) or by Pearle (1976, 1979, 1989), 
modify the Schr\"{o}dinger equation stochastically, in a way that successfully counteracts spreading with increasing 
scale of the system. Schr\"{o}dinger's strategy based on wave packets thus appears to be viable at least if one accepts 
stochastic modifications to Schr\"{o}dinger's equation (as Schr\"{o}dinger was not necessarily likely to do).

Note that while crucial, the failure of the straightforward idea of wave packets for representing the macroscopic, or
classical, regime of the theory is distinct from the question of whether Schr\"{o}dinger's wave picture could 
adequately describe what appeared to be other examples or clear indications of particulate or `discontinuous' behaviour, 
such as encountered in the photoelectric effect. This question was particularly important in the dialogue with 
Heisenberg and Bohr (section~\ref{Schr-conflict}). We shall also see that Schr\"{o}dinger continued to explore how 
continuous waves might provide descriptions of apparently particulate or discontinuous quantum phenomena,
such as the Compton effect, quantum jumps, blackbody radiation and even the photoelectric effect 
(section~\ref{Schr-Schr-development}). 


\section{The problem of radiation}\label{Schr-radiation}
In Bohr's theory of the atom, the frequency of emitted light 
corresponded not to the frequency of oscillation of an electron on a Bohr orbit, but to the term
difference between two Bohr orbits. No known mechanism could explain the difference between the frequency of 
oscillation and the frequency of emission. Bohr's theory simply postulated quantum jumps $E_i\rightarrow E_j$ 
between the stationary states of energy $E_i$ and $E_j$, accompanied by emission (or absorption) of light of 
the corresponding frequency $\nu_{ij}=\frac{E_i-E_j}{h}$. In this respect, the Bohr-Kramers-Slater (BKS) theory 
had the advantage of postulating a collection of virtual oscillators with the observed frequencies. This was also, 
in a sense, that aspect of the BKS theory that survived into Heisenberg's matrix mechanics.\footnote{See above 
section~\ref{scientific} (p.~\pageref{forpage55}) and chapter~\ref{BornEss}, especially section~\ref{beforematrix}.}

The idea that wave mechanics could provide a continuous description of the radiation 
process (as opposed to the picture of quantum jumps) appears to have been also one of the main bones of contention 
between Schr\"{o}dinger and the Co\-pen\-ha\-gen-G\"{o}t\-ting\-en physicists. 
As Schr\"{o}dinger wrote to Lorentz on 6 June 1926:
  \begin{quotation}
    The frequency discrepancy in the Bohr model, on the other hand, seems to me, (and has indeed seemed to me
    since 1914), to be something so {\em monstrous}, that I should like to characterize the excitation of light
    in this way as really almost {\em inconceivable}. (Przibram 1967, p.~61)\footnote{The emphasis here is strong, 
    but cf.\ the context of this passage in Schr\"{o}dinger's letter.}
  \end{quotation}
The reaction by Schr\"{o}dinger to the BKS theory instead was quite enthusiastic. In part, Schr\"{o}dinger
was well predisposed towards the possibility that energy and momentum conservation be only statistically 
valid.\footnote{Cf.\ Schr\"{o}dinger to Pauli, 8 November 1922, in Pauli (1979, pp.~69--71). The idea of 
abandoning exact conservation laws seems to be derived from Schr\"{o}dinger's teacher Exner ({\em ibid.}, and 
Moore 1989, pp.~152--4). Schr\"{o}dinger publicly stated this view and allegiance in his inaugural lecture 
at the University of Z\"{u}rich (9 December 1922), published as Schr\"{o}dinger (1929a). 
} In part (as argued by de Regt, 1997), this was
precisely because the BKS theory provided a mechanism for radiation, in fact a very anschaulich 
mechanism, albeit `virtual'.
Schr\"{o}dinger published a paper on the BKS theory, containing an estimate of its energy fluctuations
(1924b).\footnote{Cf.\ Darrigol (1992, pp.~247--8) on the problem of the indefinite growth of the energy 
fluctuations with time (as raised in particular by Einstein). According to Schr\"{o}dinger, an isolated system 
would behave in this way, but the problem would disappear for a system coupled to an infinite thermal bath.}

In 1926, with the development of wave mechanics, Schr\"{o}dinger saw a new possibility of conceiving
a mechanism for radiation: the superposition of two waves would involve {\em two} frequencies, 
and emitted radiation could be understood as some kind of `difference tone'. In his first paper on quantisation (1926b), 
Schr\"{o}dinger states that this picture would be `much more pleasing [um vieles sympathischer]' than the one 
of quantum jumps (p.~375), but the idea is rather sketchy: Schr\"{o}dinger speculates that the energies of the 
different eigenstates all share a large constant term, and that if the {\em square} of the frequency 
is proportional to $mc^2+E$, then the frequency differences (and therefore the beat frequencies) are
approximately given by the hydrogen term differences. The second paper (1926c) refers to radiation only in passing.

Commenting on these papers in his letter to Schr\"{o}dinger 
of 27 May, Lorentz pointed out that while beats would arise if the time-dependent 
wave equation (which Schr\"{o}dinger did not have yet) were linear, they would still not produce radiation by any known 
mechanism. Combination tones would arise if the wave equation were nonlinear (Przibram 1967, pp.~49--50).   
As becomes clear in the following letters between Schr\"{o}dinger and Lorentz,
once the charge density of a particle is associated with a quadratic function of $\psi$,
such as $|\psi|^2$, `difference tones' in the oscillating charge density arise regardless of
whether the wave equation is linear or nonlinear. 

This idea is still the basis of today's semiclassical radiation 
theory (often used in quantum optics), that is, the determination of classical electromagnetic radiation from the 
current associated with a charge density proportional to $|\psi|^2$ (for a non-stationary $\psi$).\footnote{This method
is touched on previously. Pauli uses it (for the case of one particle) in 
calculating the scattered radiation in the Compton effect during the discussion 
after Compton's talk (p.~\pageref{Pauli-classical}).} Schr\"{o}dinger arrived at this result through his 
work connecting wave mechanics and matrix mechanics (his `equivalence paper', 1926d).
In fact, Schr\"{o}dinger showed how to express the elements of Heisenberg's matrices wave mechanically, in
particular the elements of the dipole moment matrix, which (by correspondence arguments) were interpreted as proportional 
to the radiation intensities. Schr\"{o}dinger 
now suggested it might be possible `to give an extraordinarily anschaulich interpretation of the intensity and
polarisation of radiation' (1926d, p.~755), by defining an appropriate charge density in terms of $\psi$. 
In this paper, he suggested as yet, for a single electron, to use (the real part of) the quadratic function 
$\psi\frac{\partial\psi^*}{\partial t}$.

The third paper on quantisation (1926f) is concerned with perturbation theory and its application to the Stark effect, 
but Schr\"{o}dinger notes in an addendum in proof (footnote on p.~476) that the correct charge
density is given by $|\psi|^2$. In the letter to Lorentz of 6 June 
1926, he explains in detail how this gives rise to a sensible notion of charge density also for several particles, 
each contribution being obtained by integrating over the other particles (Przibram 1967, p.~56). This idea is then 
used and discussed in print (as a `{\em heuristic hypothesis\/}') in the fourth paper on quantisation (1926g, p.~118, 
Schr\"{o}dinger's italics), where the wave function is also explicitly interpreted in terms of a superposition of 
all classical configurations of a system weighted by $|\psi|^2$, and the time constancy of $\int|\psi|^2$ is
derived (pp.~135--6).\footnote{The latter question had also been raised by Lorentz (Przibram 1967, p.~71) and
answered in Schr\"{o}dinger's next letter, not included in Przibram's collection.\endnotemark}\endnotetext{Schr\"{o}dinger to 
Lorentz, 23 June 1926, AHQP-LTZ-8 (in German).} 

This is also the picture of the wave function given by Schr\"{o}dinger in the Solvay report (pp.~\pageref{mean}~ff.).
Schr\"{o}dinger discreetly skips discussing the view of the wave function as describing only an ensemble. 
Schr\"{o}dinger's concern in interpreting the wave function is to 
understand its manifestation in spacetime. This concern also motivates Schr\"{o}dinger's discussion of 
the many-electron atom\footnote{Previously unpublished but deriving from methods used in Schr\"{o}dinger's paper
on energy exchange (1927c).}, where he stresses that the spatial charge distributions of the single (non-interacting) 
electrons already determine the wave function of the interacting electrons. 

In the report, Schr\"{o}dinger starts by rephrasing 
the question of the meaning of the wave function as that of how a system described by a 
certain (multi-particle) wave function looks like in three 
dimensions. He describes this as taking all possible configurations of the classical system simultaneously and
weighting them according to $|\psi|^2$. To this picture of the system 
as a `snapshot' (as he calls it in the report) or as a `mollusc' (as he had written to Lorentz) 
Schr\"{o}dinger then associates the corresponding electric charge density in 3-space. He is
careful to state, however, that this is not an electric charge in the usual sense. For one thing, 
the electromagnetic field does not exert forces on it. While this charge does describe the {\em sources} of the field in 
the semi-classical theory, this coupling of the field to the charges is described as `provisional', first, because of the 
problem of radiation reaction (which is not taken into account in the
Schr\"{o}dinger equation), and second, because 
within a closed system it is inconsistent to model the 
interaction between different charged particles using the semi-classical field. In particular, 
the observation of emitted radiation is in principle again
a quantum mechanical interaction, and should be described by corresponding 
potentials in the equation for the total system. 

The above picture of the wave function and the question of
regarding Schr\"{o}dinger's formal charge density as a source of
classical electromagnetic radiation evidently raised many questions. (Schr\"{o}dinger introduces the
discussion after his report by the remark: `It would seem that my description in terms of a snapshot was not very
fortunate, since it has been misunderstood', p.~\pageref{misunderstood}.) It also elicited the most discussion
following Schr\"{o}dinger's report.\footnote{The rest of the discussion includes a few technical 
questions and comments (contributions by Fowler and De Donder, Born's report on some numerical work on 
perturbation theory), and some discussion of the `three-dimensionality'
of the many-electron atom. Further discussion of the 
meaning of Schr\"{o}dinger's charge densities (and of the meaning of the Schr\"{o}dinger 
wave in the context of the transformation theory of matrix mechanics) took place 
in the general discussion, especially in contributions by Dirac and by Kramers 
(pp.~\pageref{Dirac-Schr} and \pageref{Kramers-Schr}).}

A difficulty that is spelled out only in the discussion (contributions by Bohr 
and by Schr\"{o}dinger, p.~\pageref{Bohr-Schr}), is that using Schr\"{o}dinger's dipole moment 
(equation (\ref{Sch13}) in Schr\"{o}dinger's talk) to calculate the radiation 
does not directly yield the correct intensities. As pointed out in the talk, if one evaluates 
the dipole moment for a superposition $\sum_k c_k\psi_k$, one obtains terms
containing the integrals $-e\int q\psi_k\psi_l^*d\tau$. These are the matrix
elements of the dipole moment matrix in matrix mechanics, and they can be used
within certain limits to calculate the emitted radiation (in particular, as
Schr\"{o}dinger points out, vanishing of the integral implies vanishing of the
corresponding spectral line). However, these integrals appear with
the coefficients $c_kc_l^*$, whereas both according to Bohr's old quantum
theory and to experiment, the intensity of radiation should not depend on the
coefficient of the `final state', say $\psi_l$. Thus, the use of (\ref{Sch13}) as
a classical dipole moment in the calculation of emitted radiation does not in
general yield the correct intensities. 

Bohr also drew attention to the fact that
by the time of the Solvay conference, Dirac (1927b,c) had already published his 
treatment of the interaction of the (Schr\"{o}dinger) electron with the 
quantised electromagnetic field. Schr\"{o}dinger replied that he was aware of 
Dirac's work, but had the same misgivings with q-numbers as he had with matrices: 
the lack of `physical meaning', without
which he thought the further development of a relativistic theory would be 
difficult.\footnote{Note that in his treatment of the emission of radiation in 
{\em The Physical Principles of the Quantum Theory} (Heisenberg 1930b, pp.~82--4), Heisenberg  
seems to follow and expand on the discussion of Schr\"{o}dinger's 
report. Indeed, Heisenberg
first describes two related methods for calculating the radiation, based 
respectively on calculating the matrix element of the dipole moment of the
atom (justified via the correspondence principle), and on calculating the
dipole moment of Schr\"{o}dinger's `virtual charges' (as he calls them). He 
then explains precisely the difficulty with the latter discussed here by
Bohr and Schr\"{o}dinger, and goes on to sketch (a variant of)
Dirac's treatment of the problem.}

If the interpretation of the wave function as a charge density raises problems of principle (as Born and Heisenberg 
also stress in their report, p.~\pageref{contradictions}), what is the point of 
suggesting such an interpretation? One possible way of understanding Schr\"{o}dinger's intentions is to say that he
is proposing an anschaulich image of the wave function in terms of the spatial density it defines, {\em without} in 
general specifying the dynamical role played by this density. Under certain circumstances, then, this spatial density 
takes on the dynamical role of a charge density, in particular as a source of radiation. This point of view does not 
resolve the other problems connected
with the wave conception of matter (spreading of the wave packet, Schr\"{o}dinger's cat, micro-macro question), 
but it offers a platform from which to work towards their possible resolution.  
This of course may be an overinterpretation of Schr\"{o}dinger's position, but it fits approximately with later 
developments that take Schr\"{o}dinger's wave conception seriously, e.g.\ Bell's (1990) 
idea of $|\psi|^2$ as `density of stuff' (in configuration space).

\section{Schr\"{o}dinger and de~Broglie}\label{Schr-deB}
Thus opens Schr\"{o}dinger's review paper on wave mechanics for the {\em Physical Review} (1926h, p.~1049, 
references omitted):
  \begin{quote}
    The theory which is reported in the following pages is based on the very interesting and fundamental
    researches of L.~de~Broglie on what he called `phase waves' (`ondes de phase') and thought to be
    associated with the motion of material points, especially with the motion of an electron or 
    [photon].\footnote{The text here reads `proton', which is very likely a misprint, since de~Broglie's work
    indeed focussed on electrons and photons.}
    The point of view taken here, which was first published in a series of German papers, is rather that
    material points consist of, or are nothing but, wave-systems.
  \end{quote}
This passage both illustrates the well-known fact that Schr\"{o}dinger arrived at his wave 
mechanics by developing further the ideas of de Broglie --- starting with his paper on gas statistics (Schr\"{o}dinger 
1926a) --- and emphasises the main conceptual difference between de Broglie's and Schr\"{o}dinger's approaches.

While some details of the relation between Schr\"{o}dinger's and de~Broglie's work are discussed in 
section~\ref{towards-a-complete}, in this section
we wish to raise the question of why Schr\"{o}dinger should have developed such a different picture of wave 
mechanics. Indeed, although Schr\"{o}dinger quotes de~Broglie as the rediscoverer of Hamilton's optical-mechanical
analogy (1926h, footnote 3 on p.~1052), the two authors apply the analogy in opposite directions: 
de~Broglie treats even the photon as a material particle with a trajectory, while Schr\"{o}dinger treats 
even the electron as a pure wave.\footnote{As is clear from de Broglie's thesis and as mentioned by de Broglie 
himself in his report (p.~\pageref{always-trajectories}), de Broglie had always 
assumed the picture of trajectories. The above quotation in any case makes it clear that this was Schr\"{o}dinger's 
own reading of de Broglie. Already as early as January 1926, Schr\"{o}dinger writes in his gas theory paper about
`the de~Broglie-Einstein undulatory theory of corpuscles in motion, according to which the latter are no more than 
a kind of ``foam crest'' [``Schammkaum''] on the wave radiation that constitutes the world background [Weltgrund]'
(1926a, p.~95). Other indications that this was well-known are Lorentz's comments to Schr\"{o}dinger
on the construction of electron orbits in his letter of 19 June 1926 coming `close to de Broglie's arguments'
(Przibram 1967, p.~74) and Pauli's reference to Einstein's and de Broglie's `moving point masses' in his letter 
to Jordan of 12 April 1926 (Pauli 1979, p.~316).}

As Schr\"{o}dinger writes in his first paper on quantisation, it was in particular `reflection on the spatial
distribution' of de Broglie's phase waves that gave the impulse for the development of his own theory of wave
mechanics (1926b, p.~372). We gain some insight as to what this refers to from a letter from Schr\"{o}dinger 
to Land\'{e} of 16 November 1925:
  \begin{quotation}
    I have tried in vain to make for myself a picture of the phase wave of the electron in the Kepler orbit.
    Closely neighbouring Kepler ellipses are considered as `rays'. This, however, gives horrible
    `caustics' or the like for the wave fronts. (Quoted in Moore 1989, p.~192)
  \end{quotation}
Thus, one reason for Schr\"{o}dinger to abandon the idea of trajectories in favour of the pure wave theory
might have been that well-behaved trajectories seemed to be incompatible with well-behaved 
waves. And indeed, in his presentations of the 
optical-mechanical analogy, Schr\"{o}dinger states that outside of the geometric limit the notion of 
`ray' becomes meaningless (1926b, 
pp.~495 and 507, 1926h, pp.~1052--3). This also seems to be at issue 
in the exchange between Schr\"{o}dinger and Lorentz during the 
discussion of de Broglie's report, where Schr\"{o}dinger points out that in cases of degeneracy, any arbitrary 
linear combination of solutions is allowed, and de Broglie's theory would predict very complicated orbits 
(p.~\pageref{forpage89}).
 
There are other aspects that could conceivably provide further reasons for Schr\"{o}dinger's definite abandoning 
of the trajectories. One possibility is that Schr\"{o}dinger picked up from de Broglie specifically the idea of 
particles as singularities of the wave,\footnote{Cf.\ Schr\"{o}dinger (1926a, p.~99): `The universal radiation,
as ``signals'' or perhaps singularities of which the particles are meant to occur, is thus something quite essentially 
more complicated than for instance the wave radiation of Maxwell's theory .... '} and was happy to relax 
it to the idea of wave packets. Indeed, as we have seen above, Schr\"{o}dinger placed great emphasis
on the notion of a wave packet, and if it had been an adequate notion, there would have been no need for a separate
notion of a corpuscle in order to explain the particulate aspects of matter.

We have also seen that Schr\"{o}dinger was acutely aware of the radiation problem, namely of the
discrepancy between the orbital frequency of the electron and the frequency of the emitted radiation. 
While de~Broglie in his thesis was able to derive the Bohr orbits from wave considerations,  
this would in no way seem to alleviate the problem: the frequency of revolution of the electron in de Broglie's 
theory was the same as in the Bohr model, and would lead to the wrong frequency of radiation if the electron
was treated as a classical source. (In the case of degeneracy noted above, the situation would be even 
more complicated.)\footnote{Note, however, that de Broglie (1924c) had discussed the solution of this problem 
in the correspondence limit (i.e. for the case of high quantum numbers).}

Finally, Schr\"{o}dinger appeared to be critical of proposals combining waves and particles, for instance
as appeared to be done by Born in his collision papers (see section~\ref{Schr-packets} above). Such misgivings 
could easily have applied also to de Broglie's theory.

\section{The conflict with matrix mechanics}\label{Schr-conflict}
In the early discussions on the meaning of quantum theory, the notion of `Anschaulichkeit' resurfaces time 
and again. The verb `anschauen' means `to look at', and `anschaulich', which means `clear', `vivid' or `intuitive', has 
visual connotations that the English word `intuitive' lacks. Anschaulichkeit of a physical theory is thus a quality of 
ready comprehensibility that may (but need not\footnote{Cf.\ below Heisenberg's use of the term in the uncertainty paper.}) 
include a strong component of literal picturability. For Schr\"{o}dinger, at any rate, it seems that the
possibility of grasping a theory through some kind of spatio-temporal intuition was a key component of physical
understanding.\footnote{A good discussion of the role the notion of Anschaulichkeit played for Schr\"{o}dinger is given by de Regt 
(1997, 2001), who argues that Schr\"{o}dinger's requirement of Anschaulichkeit is 
derived indirectly from Boltzmann, and is essentially a methodological requirement (as opposed to being a commitment 
to realism --- cf.\ also the introduction by Bitbol in Schr\"{o}dinger 1995, p.~4 fn.~10).} 

It is clear, however, that the conflict between wave mechanics and matrix or quantum mechanics was not, or not only, 
a philosophical issue, say about the validity of the concepts of spacetime and causality,\footnote{Note that Kant's 
conception of space and time is formulated in terms of what he calls the `Anschauungsformen' (the `forms of intuition'), 
so that the discussion has indeed strong philosophical overtones.} and that, in the minds of the parties involved, 
these issues were connected with specifically scientific questions.\footnote{This point has been recently argued 
also by Perovic (2006).} 

The list of these questions is extensive. In his `equivalence' paper (1926d), 
Schr\"{o}dinger states that mathematical equivalence is not the same as physical equivalence, in the sense that
two theories can offer quite different possibilities for generalisation and further development.\footnote{Similarly, 
in the discussion after his own report, Schr\"{o}dinger insists that finding a physical interpretation of the theory 
is `indispensable for the further development of the theory' (p.~\pageref{indispensable}).} He thinks of 
two problems in particular: first, the problem of scattering (as mentioned in section~\ref{Schr-packets} above), 
second, the problem of radiation (discussed in section~\ref{Schr-radiation}), in connection with which he then 
describes the idea of the vibrating charge density. The radiation problem in turn links to further issues at the heart
of the debate with Heisenberg and Bohr in particular, on whether there are quantum jumps or whether the process of 
radiation and the atomic transitions can be described as continuous processes in space and time. In a sense, neither 
the issue of scattering nor that of radiation resolved the debate in favour of either theory: both theories were 
modified or reinterpreted in the course of these developments, even though the result (statistical interpretation, 
Copenhagen interpretation) was not to Schr\"{o}dinger's taste.

We shall now follow how the conflict evolved, from the beginnings to the time of the Solvay conference, since both 
sides developed considerably during the crucial period between Schr\"{o}dinger's first papers and the conference.

\subsection{Early days}
At the time of Schr\"{o}dinger's first two papers on quantisation, as we have seen in the previous chapter, 
matrix mechanics was a theory that rejected the notion of electron orbits, indeed the very possibility of 
a spacetime description, substituting the classical kinematical quantities with matrix 
quantities; but it kept the postulate of stationary states and of quantum jumps between these states. 
Matrix mechanics did not describe the stationary states individually, only collectively, and allowed one to 
calculate only transition probabilities for the jumps. 

Schr\"{o}dinger's first paper on quantisation (1926b) contains only a few comments on the possible continuous picture of
atomic transitions, as opposed to quantum jumps. (But of course, at this stage, he has very little to say
about the problem of radiation, in particular nothing about intensities.) The first time he comments 
explicitly on the differences between wave mechanics and quantum mechanics is in his second paper.
There he writes that wave mechanics offers a way to interpret 
  \begin{quotation}
    the conviction, more and more coming to the fore today, that {\em first}: one should deny real significance to the 
    {\em phase} of the electron motions in the atom; {\em second}: that one may not even claim that the electron at a 
    certain 
    time is located {\em on one particular} of the quantum trajectories distinguished by the quantum conditions; 
    {\em third}: the true laws of quantum mechanics consist not in definite prescriptions for the {\em individual 
    trajectory}, rather these true laws relate through equations the elements of the whole manifold of trajectories 
    of a system through, so that apparently a certain interaction between the different trajectories obtains. (1926c, pp.~508)
  \end{quotation}
As Schr\"{o}dinger proceeds to say, these claims are in contradiction with the ideas of electron position and electron 
orbit, but should not be taken as forcing a complete surrender of spatio-temporal ideas. At the time, the mathematical
relation between wave mechanics and matrix mechanics was not yet clarified, but Schr\"{o}dinger hopes there would 
be a well-defined mathematical relation between the two, which could then complement each other.
According to Schr\"{o}dinger, Heisenberg's theory yields the line intensities, his own offers the possibility of 
bridging the micro-macro gap.\footnote{Schr\"{o}dinger's collected works (1984) reproduce this and other papers from 
Erwin and Anny Schr\"{o}dinger's own copy of the second edition of {\em Abhandlungen zur Wellenmechanik} (Schr\"{o}dinger 
1928). In this copy of the book, the passage about the micro-macro bridge is underlined.} 
Personally, he finds the conception of emitted frequencies as `beats' particularly attractive and believes it will 
provide an anschaulich understanding of the intensity formulas (1926c, pp.~513--14). 

Within weeks, the relation between the two theories was clarified independently by Schr\"{o}dinger (1926d), by
Eckart (1926) and by Pauli, in his remarkable letter to Jordan of 12 April 1926 (Pauli 1979, 
pp.~315--20). This is one of the first documented reactions to Schr\"{o}dinger's new work from a physicist of the 
Copenhagen-G\"{o}ttingen school.\footnote{Ehrenfest informed Lorentz of the `equivalence' result (and of Klein's (1926) 
theory) on 5 May 1926, when Kramers reported them in the colloquium at Leiden. Ehrenfest also mentioned the relation 
of this result to Lanczos's (1927) work.\endnotemark As noted above (footnote on p.~\pageref{forpage91}), by 27 May 
Lorentz had received a copy of the paper directly from Schr\"{o}dinger.}\endnotetext{Ehrenfest to Lorentz, 5 May 1926, 
AHQP-EHR-23 (in German).} 
In it, Pauli emphasises in fact that in Schr\"{o}dinger's theory there are no 
electron orbits, since trajectories are a concept belonging to the geometric limit of the theory. Pauli seems to say 
that insofar as Schr\"{o}dinger provides a description of individual stationary states, his theory
is not in conflict with matrix mechanics, since it also does not contain the concept of electron 
orbits.\footnote{Pauli knew very early about Schr\"{o}dinger's paper, having been informed by Sommerfeld, to 
whom at Schr\"{o}dinger's request a copy of the paper had been forwarded by Wien, the editor of {\em Annalen der Physik} 
(cf.\ Pauli 1979, pp.~278 and 293). Pauli's analysis in the letter of 12 April appears to be based only on 
Schr\"{o}dinger's first paper on quantisation, although at least the official publication date of the second paper was 
6 April 1926. See also the letters between Pauli and Schr\"{o}dinger reproduced in Pauli 
(1979).}


The tone of Schr\"{o}dinger's remarks and of the comments they elicit changes with the equivalence paper.
In a much-quoted footnote, Schr\"{o}dinger says he had known of Heisenberg's theory but was `scared away, not to say
repelled', by the complicated algebraic methods and the lack of Anschaulichkeit (1926d, p.~735). 
On 8 June 1926, Heisenberg sends Pauli an equally notorious (but often mis-quoted) comment: 
  \begin{quote}
    The more I reflect on the physical part of Schr\"{o}dinger's theory, the more disgusting [abscheulich] I find it. 
    Imagine the rotating electron, whose charge is distributed over the whole space with its axis in a fourth and 
    fifth dimension. What Schr\"{o}dinger writes of the Anschaulichkeit of his theory ``{\em would scarcely} [{\em be\/}] 
    {\em an appropriate}...'' in other words I find it poppycock [Mist].  (Pauli 1979, 
    p.~328)
  \end{quote}
It seems clear that, according to Heisenberg, it is Schr\"{o}dinger's claim of Anschaulichkeit for his 
theory that is ludicrous, presumably partly because of the spread-out electron and partly because the waves are 
in configuration space.\footnote{Alternatively, with a `fourth and fifth dimension', Heisenberg might conceivably 
be referring to Klein's five-dimensional extension of Schr\"{o}dinger's equation (Klein 1926). As one example, 
here is how Moore (1989, p.~221) quotes the passage: `The more I think of the physical part of the Schr\"{o}dinger 
theory, the more abominable I find it. What Schr\"{o}dinger writes about {\em Anschaulichkeit} makes scarcely any 
sense, in other words I think it is bullshit [Mist]'.}

\subsection{From Munich to Copenhagen}
In the summer of 1926, Schr\"{o}dinger gave a series of talks on wave mechanics in various German universities,
in particular, on 21 and 23 July, he talked in Munich at the invitation of Sommerfeld and of Wien.\footnote{Here 
we follow mainly the account given by Heisenberg (1946).} The description of a mechanism for radiation elicited 
enthusiastic comments by Wien, but criticism from Sommerfeld and from Heisenberg. The discussion was apparently 
heated, and eventually identified a crucial experiment that would decide between the idea of the continuous mechanism 
of radiation envisaged in wave mechanics and the idea of quantum jumps.\label{forpage62} This was incoherent scattering, 
i.e.\ the Raman effect (at that time neither observed nor thus named). According to the quantum prediction (Smekal 1923, 
Kramers and Heisenberg 1925), incoherent scattering would exist also for atoms in the ground state, because 
the atom could be excited by the incident light. According to Schr\"{o}dinger instead, the effect was due to induced 
vibrations for the case in which at least two atomic frequencies were already present, and so would not occur in the 
ground state.\footnote{Note the implied inequivalence of wave and matrix mechanics (despite the recent `equivalence 
proofs'). Another such possibility of experimental inequivalence is mentioned in the discussion after Schr\"{o}dinger's 
report, with regard to the quadrupole radiation of the atom (p.~\pageref{quadrupole}). Cf.\ also Muller (1997) on the
inequivalence of the two theories. 
} Apparently, Sommerfeld and 
Heisenberg were `prepared to enter a bet for its existence, while the experimental physicists were against it and 
Schr\"{o}dinger took a more wait-and-see attitude' (Heisenberg 1946, p.~5).

In a letter to Pauli of 28 July, Heisenberg gives other specific criticisms of Schr\"{o}dinger, for throwing overboard 
`everything ``quantum theoretical'': namely photoelectric effect, Franck collisions, Stern-Gerlach effect etc.' (Pauli 
1979, p.~338). As a matter of fact, in the letter to Wien of 25 August 1926, quoted above (section~\ref{Schr-packets}), 
Schr\"{o}dinger admitted that he had great conceptual difficulties with the photoelectric effect (but see below the 
discussion of Schr\"{o}dinger, 1929b). Heisenberg 
further mentioned to Pauli that, together with Schr\"{o}dinger and Wien, he had discussed Wien's experiments on the decay 
of luminescence (Wien 1923, ch.~XX).\label{Wien} This was another point where Schr\"{o}dinger thought wave mechanics 
proved superior to matrix mechanics, and Heisenberg encouraged Pauli to calculate and publish the damping coefficients 
for the hydrogen spectrum.\footnote{Cf.\ Born and Heisenberg's report (pp.~\pageref{help} and \pageref{decay}), where spin 
is considered to be problematic for wave mechanics, and where it is explicitly stated that Dirac (1927c) provides an 
explanation for the decay experiments.}

Heisenberg mentions similar criticisms in his talk on `Quantum mechanics', given at the 89th meeting of German Scientists 
and Physicians in D\"{u}sseldorf on 23 September 1926 and published in the issue of 5 November of {\em Naturwissenschaften} 
(Heisenberg 1926c). This talk could be seen as a public response to the claims that the return to a `continuum theory' was
possible. In it, Heisenberg addresses in particular the problem of Anschaulichkeit: according to Heisenberg, the usual 
notions of space and time, and in particular their application to physics with the idea that space and matter are in 
principle continuously divisible, turn out to be mistaken, first of all due to the Unanschaulichkeit of the corpuscular 
nature of matter, then through the theoretical and experimental considerations leading to the idea of stationary states and quantum jumps (Bohr, 
Franck-Hertz, Stern-Gerlach), and finally through consideration of radiation phenomena (Planck's radiation formula, Einstein's
light quantum, the Compton effect and the Bothe-Geiger experiments). This issue, it is claimed, also relates closely to the question of the 
degree of `reality' to be ascribed to material particles or light quanta. Quantum mechanics in its 
development had thus first of all to free itself from notions of Anschaulichkeit, in order to set up a new kind of
kinematics and mechanics. In discussing the wave theory, Heisenberg considers first de Broglie and Einstein 
as having developed wave-particle dualism for matter and having suggested the possibility of interference for an ensemble [Schar]
of particles. He then explains that Schr\"{o}dinger found a differential equation for the matter waves that reproduces the 
eigenvalue problem of quantum (i.e.\ matrix) mechanics. However, according to Heisenberg, the Schr\"{o}dinger theory fails 
to provide the link with de Broglie's ideas, that is, it fails to provide the analogy with light waves in ordinary space, 
because of the need to 
consider waves in configuration space; the latter therefore have only a formal significance. Heisenberg refers to the 
claim that on the basis of Schr\"{o}dinger's theory one may be able to return to `a purely continuous description of 
the quantum theoretical phenomena', and continues (Heisenberg 1926c, p.~992): 
  \begin{quote}
    In developing consequently this point of view one leaves in fact the ground of 
    de Broglie's theory, thus of Q.M. and indeed of all quantum theory and arrives in my opinion 
    at a complete contradiction with experience (law of blackbody radiation; dispersion theory). 
    This route is thus not viable. The actual reality of the de Broglie waves lies rather in the interference
    phenomena mentioned above, which defy any interpretation on the basis of classical concepts. The
    extraordinary physical significance of Schr\"{o}dinger's results lies in the realisation that an anschaulich
    interpretation of the quantum mechanical formulas contains both typical features of a corpuscular theory
    and typical features of a wave theory.  
  \end{quote}
After discussing quantum statistics (`which in any case represents a very bizarre further limiting of the
reality of the corpuscles', p.~992), Heisenberg concludes (p.~994):
  \begin{quote}
    In our anschaulich interpretation of the physical processes and mathematical formulas there is a dualism between 
    wave theory and corpuscular theory such that many phenomena are described most naturally by a wave
    theory of light as well as of matter, in particular interference and diffraction phenomena, while other phenomena
    in turn can be interpreted only on the basis of the corpuscular theory. .... The contradictions of the 
    anschaulich interpretations of different phenomena contained in the current scheme are completely unsatisfactory.
    For a contradiction-free anschaulich interpretation of the experiments, which in themselves are indeed
    contradiction-free, some essential trait in our picture of the structure of matter is currently still missing. 
  \end{quote}

After the summer, Schr\"{o}dinger visited Copenhagen for an intense round of discussions. According to Heisenberg's 
reconstruction in {\em Der Teil und das Ganze} (Heisenberg 1969), Schr\"{o}dinger argued with
Bohr precisely about the necessity of finding a mechanism for radiation, while Bohr insisted that quantum jumps were
necessary for the derivation of Planck's radiation law, as well as being directly observable in experiments. 
Heisenberg (1946, p.~6) states that the discussion ended with the recognition `that an interpretation of wave mechanics
without quantum jumps was impossible and that the mentioned crucial experiment [i.e.\ the Raman effect] in any case would 
turn out in favour of the quantum jumps'.\footnote{The well-known quotation `If this damned quantum jumping is indeed 
to stay, then I regret having worked on this subject at all' is reported both in Heisenberg (1946) and Heisenberg (1969).} 

Schr\"{o}dinger in turn admitted his difficulties. In a letter to Wien of 21 October 1926 (quoted in Pauli 1979, p.~339),
he wrote: `It is quite certain that the position of anschaulich images, which de Broglie and I take, has not nearly
been developed far enough to account even just for the most important facts. And it is downright probable that here and 
there a wrong path has been taken that needs to be abandoned'. And, commenting in his preface (dated November 1926) to 
the first edition of {\em Abhandlungen zur Wellenmechanik} (1926i) on the fact that the papers were being reprinted 
unchanged, he invoked `the impossibility at the current stage of giving an essentially more satisfactory or even 
definitive new presentation'. 

In the meantime, both Bohr and Heisenberg were working to find their own satisfactory interpretation of the 
theory.\footnote{See also section~\ref{stateprob} above.}  
In particular as regards Heisenberg, the correspondence with Pauli is very telling.\footnote{As noted in 
Pauli (1979, p.~339 and fn.~3 on p.~340), however, most of Pauli's letters to Heisenberg from this period appear to have 
been destroyed in the war.} First of all one finds the discussion of Born's collision papers (Born 1926a,b), which led
to Heisenberg's paper on fluctuation phenomena (Heisenberg 1926b). As already mentioned in section~\ref{transf-theory}, 
Heisenberg then reports to Pauli about Dirac's transformation theory (Dirac 1927a), which, however, is seen only as an 
extraordinary formal development. The next letters from Heisenberg to Pauli are from February 1927 and include on 23 
February the sketch of the ideas of the uncertainty paper (Pauli 1979, pp.~376--81). This, as was Bohr's 
simultaneous development of the idea of complementarity, was meant to provide the anschaulich picture that was still 
missing. Indeed, in the uncertainty paper, Heisenberg  (1927, p.~172) formulates Anschaulichkeit as the possibility of 
arriving at qualitative predictions in simple cases together with formal consistency. With the formulation and 
application of the uncertainty relations, he was then satisfied to have found such an interpretation.

In a sense, however, the uncertainty paper was also the end of the original notion of `quantum jumps' as transitions 
between stationary states of a system. Indeed, as had to be the case given the generalisation of transition probabilities 
to arbitrary pairs of observables (and as was implied in the Heisenberg-Pauli correspondence), the privileged role of 
stationary states had to give way. It is somewhat ironic that Heisenberg was to give up quantum jumps only a few months 
after the discussions with Schr\"{o}dinger in Copenhagen; nevertheless, quantum mechanics thus wedded to probabilistic 
transitions between measurements was just as discontinuous as the picture of quantum jumps between stationary states 
that it replaced, and, for Schr\"{o}dinger, it was equally unsatisfactory.

Note that, for Bohr at least, wave aspects played a crucial role in the resulting `Copenhagen' interpretation. 
Yet, just as in the case of Born's (1926a,b) use of wave mechanics in the discussion of collisions 
(see section~\ref{resonance}), Heisenberg appears to have been convinced that the apparent wave aspects could be 
interpreted entirely in matrix terms. This difference of opinion was reflected in the sometimes tense discussions 
between Heisenberg and Bohr at the time, in particular on the topic of Heisenberg's treatment of the $\gamma$-ray 
microscope in the uncertainty paper and the corresponding addendum in proof.\label{addendum}\footnote{For more on 
this issue, see Beller (1999, pp.~71--4 and 138--41) and Camilleri (2006). For Heisenberg's way out of the difficulty, 
see Heisenberg (1929, pp.~494--5).}

\subsection{Continuity and discontinuity}\label{Schr-Schr-development}
Between late 1926 and the time of the Solvay conference, Schr\"{o}dinger continued to work along lines that brought
out the attractive features --- and sometimes the limitations --- of the wave picture.
In late 1926 and early 1927, Schr\"{o}dinger focussed on the `four-dimensional' theory, with his papers
on the Compton effect (1927a), and on the energy-momentum tensor (1927b), while he returned to the `many-dimensional'
theory shortly before the Solvay conference, with his work on energy exchange (1927c) and the treatment of the
many-electron atom presented in his report. 

The Compton effect is of course a paradigmatic example of a `discontinuous' phenomenon, but Schr\"{o}dinger (1927a) gives 
it a wave mechanical treatment, by analogy with the classical case of reflection of a light wave when it encounters a sound 
wave (Brillouin 1922).\footnote{Schr\"{o}dinger had given a derivation of Brillouin's result by assuming that
energy and momentum were exchanged in the form of quanta (1924a). In the paper on the Compton effect he now comments 
on how, in a sense, he is reversing his own earlier reasoning.} As he remarks in the paper, and as 
remarked in the discussion after Compton's report, this treatment relies on the consideration of stationary waves and 
does not directly describe an individual Compton collision.\footnote{See in particular the remarks by Pauli 
(p.~\pageref{Pauli-Compton}) and the discussion following Schr\"{o}dinger's contribution (p.~\pageref{Schr-Compton}). 
Cf.\ also the closing paragraph of Pauli's letter to Schr\"{o}dinger of 12 December 1927 (Pauli 1979, p.~366).}

The paper on the energy-momentum tensor (1927b), following Gordon (1926), takes the Lagrangian approach to deriving
the Klein-Gordon equation, and varies also the electromagnetic potentials, thus deriving the Maxwell equations as well. 
Schr\"{o}dinger then considers in particular the energy-momentum tensor of the 
combined Maxwell and Klein-Gordon fields. As he remarks (Schr\"{o}dinger 1928, p.~x), this is a beautiful formal 
development of the theory, but it heightens starkly the difficulties with the four-dimensional view, since one 
cannot insert the electromagnetic potentials thus obtained back into the Klein-Gordon equation. For instance, 
including the self-field of the electron in the treatment of the hydrogen atom would yield the wrong results, as 
Schr\"{o}dinger also mentions in section II of his report.
In this connection, Schr\"{o}dinger states that: `The exchange of energy and momentum between the electromagnetic 
field and ``matter'' does {\em not} in reality take place in a continuous way, as the [given] fieldlike expression
suggests' (1927b, p.~271).\footnote{See again the exchange of letters between Pauli and Schr\"{o}dinger in December 1926 (Pauli 1979, 
pp.~364--8).} According to Heisenberg (1929), the further development of the four-dimensional theory was indeed purely
formal, but provided a basis for the later development of quantum field theory.\label{backreaction}\footnote{Heisenberg (1930b) 
incorporates into his view of quantum theory both the multi-dimensional theory and the four-dimensional theory of 
Schr\"{o}dinger's report. The latter is interpreted as a classical wave theory of matter that forms the background for 
a second quantisation, and includes the backreaction of the self-field via inclusion in the potential (cf.\ also
above, fn. on p.~\pageref{Fermi}).}

The paper on energy exchange (1927c), Schr\"{o}dinger's last paper before the Solvay conference, marks an attempt
to meet the criticism that wave mechanics cannot account for crucial phenomena involving `quantum jumps'. 
Following Dirac (1926c), Schr\"{o}dinger sketches the method of the variation of constants, which he is to use also 
in the discussion of the many-electron atom in his Solvay report. He then 
applies it to the system of two atoms in resonance discussed by Heisenberg (1926b) (and by Jordan (1927a)). As discussed in
section~\ref{stateprob}, Heisenberg uses this example to show how in matrix mechanics one can indeed describe change 
starting from first principles, in particular how one can determine the probabilities for quantum jumps. Now 
Schr\"{o}dinger turns the tables around and argues that the treatment of atoms in resonance using wave mechanics shows 
how one can eliminate quantum jumps from the picture. He argues that the two atoms exchange energies {\em as if\/} they 
were exchanging definite quanta. Indeed, he goes further and suggests that the idea of quantised energy itself should 
be reinterpreted in terms of wave frequency,\footnote{Schr\"{o}dinger had expressed the idea of energy as frequency 
already in his letter to Wien of 25 August 1926: `What we call the energy of an individual 
electron is its frequency. Basically it does not move with a certain speed because it has received a certain ``shove'',
but because a dispersion law holds for the waves of which it consists, as a consequence of which a wave packet of this
frequency has exactly this speed of propagation' (as quoted in Moore 1989, p.~225).} and that resonance phenomena are 
indeed the key to the `quantum postulates'. Schr\"{o}dinger then proceeds to formulate statistical considerations about 
the distributions of the amplitudes of the two systems in resonance, leading to the idea of the squares of the amplitudes 
as measures of the strength of excitation of an eigenvalue. He then returns to resonance considerations in the case of 
a system coupled to a heat bath, which he considers would suffice in principle for the derivation of Planck's radiation 
formula and of all the results of the `old quantum statistics'.

In the aftermath of the Solvay conference, although in places one finds Schr\"{o}dinger at least temporarily 
espousing views much closer to those of the Copenhagen-G\"{o}ttingen school (cf.\ Moore 1989, pp.~250--51), 
Schr\"{o}dinger continued to explore the possibilities of the wave picture.\footnote{Cf.\ also Bitbol's introduction 
in Schr\"{o}dinger (1995), p.~5.} The following is a telling example. In a short paper in {\em Naturwissenschaften}, 
Schr\"{o}dinger (1929b) proposes to illustrate `how the quantum theory in its newest phase again makes use of 
continuous spacetime functions, indeed of properties of their {\em form}, to describe the state and behaviour of a 
system ....' (p.~487). Schr\"{o}dinger quotes the example of how photochemical and photoelectric phenomena depend 
on the form of an impinging wave (i.e. on its Fourier decomposition), rather than on local properties of the
wave at the point where it impinges on the relevant system. He argues that a continuous picture can be retained,
but introduces the idea (which, as he remarks, generalises without difficulty to the case of many-particle wave 
functions) that the crucial properties of wave functions are in fact properties pertaining to the entire wave. 
Schr\"{o}dinger may thus have been the first to introduce the idea of nonlocalisable properties, as part of 
the price to pay in order to pursue a wave picture of matter.

\newpage

\renewcommand{\enoteheading}{\section*{Archival notes}}
\addcontentsline{toc}{section}{\em Archival notes}
\theendnotes

\part{\\ Quantum foundations and the 1927 Solvay conference}

\setcounter{endnote}{0}
\setcounter{equation}{0}

\chapter{Quantum theory and the measurement problem}\label{QTMeasProb}\chaptermark{Quantum 
theory and the measurement problem}

\section{What is quantum theory?}\label{what-is-QT}

For much of the twentieth century, it was widely believed that the
interpretation of quantum theory had been essentially settled by Bohr and
Heisenberg in 1927. But not only were the `dissenters' of 1927 --- in
particular de Broglie, Einstein, and Schr\"{o}dinger --- unconvinced at the
time: similar dissenting points of view are not uncommon even today. What
Popper called `the schism in physics' (Popper 1982) never really healed. Soon
after 1927 it became standard to assert that matters of interpretation had
been dealt with, but the sense of puzzlement and paradox surrounding quantum
theory never disappeared.

As the century wore on, many of the concerns and alternative viewpoints
expressed in 1927 slowly but surely revived. In 1952, Bohm revived
and extended de Broglie's theory (Bohm 1952a,b), and in 1993 the de Broglie-Bohm theory
finally received textbook treatment as an alternative formulation of quantum
theory (Bohm and Hiley 1993; Holland 1993). In 1957, Everett (1957) revived
Schr\"{o}dinger's view that the wave function, and the wave function alone, is
real (albeit in a very novel sense), and the resulting `Everett' or
`many-worlds' interpretation (DeWitt and Graham 1973) gradually won widespread
support, especially among physicists interested in quantum gravity and quantum
cosmology. Theories even closer to Schr\"{o}dinger's ideas --- collapse
theories, with macroscopic objects regarded as wave packets whose spreading is
prevented by stochastic collapse --- were developed from the 1970s onwards
(Pearle 1976, 1979; Ghirardi, Rimini and Weber 1986). As for Einstein's
concerns in 1927 about the nonlocality of quantum theory (see chapter~\ref{locality-and-incompleteness}), 
re-expressed in the famous EPR paper of 1935, matters came to a head in 1964 with the publication
of Bell's theorem (Bell 1964). In the closing decades of the twentieth
century, after many stringent experimental tests showed that Bell's inequality
was violated by entangled quantum states, nonlocality came to be widely
regarded as a central fact of the quantum world.

Other concerns, voiced by Schr\"{o}dinger just a few years after the fifth
Solvay conference, also eventually played a central role after decades of
obscurity. Schr\"{o}dinger's `cat paradox' of 1935 came to dominate
discussions about the meaning of quantum theory. And the peculiar
`entanglement' that Schr\"{o}dinger had highlighted as a key difference
between classical and quantum physics (Schr\"{o}dinger 1935) eventually found
its place as a central concept in quantum information theory: as well as being a
matter of `philosophical' concern, entanglement came to be seen as a
physical resource to be exploited for technological purposes, and as a central
feature of quantum physics that had been strangely under-appreciated for most
of the twentieth century.

The interpretation of quantum theory is probably as controversial now as it
ever has been. Many workers now recognise that standard quantum theory ---
centred as it is around the notion of `measurement' --- requires a classical
background (containing macroscopic measuring devices), which can never be
sharply defined, and which in principle does not even exist. Even so, the
operational approach to the interpretation of quantum physics is still being pursued by some, in
terms of new axioms that constrain the structure of quantum theory (Hardy
2001, 2002; Clifton, Bub and Halvorson 2003). On the other hand, those who do
regard the background problem as crucial tend to assert that everything in the
universe --- microscopic systems, macroscopic equipment, and even human
experimenters --- should in principle be described in a unified manner, and
that `measurement' processes must be regarded as physical processes like any
other. Approaches of this type include: the Everett interpretation (Everett
1957), which is being subjected to increasing scrutiny at a foundational level
(Saunders 1995, 1998; Deutsch 1999; Wallace 2003a,b); the pilot-wave theory of
de Broglie and Bohm (de Broglie 1928; Bohm 1952a,b; Bohm and Hiley 1993;
Holland 1993), which is being pursued and developed more than ever before
(Cushing, Fine and Goldstein 1996; Pearle and Valentini 2006; Valentini 2007);
collapse models (Pearle 1976, 1979, 1989; Ghirardi, Rimini and Weber 1986),
which are being subjected to ever more stringent experimental tests (Pearle
and Valentini 2006); and theories of `consistent' or `decoherent' histories
(Griffiths 1984, 2002; Gell-Mann and Hartle 1990; Hartle 1995; Omn\`{e}s 1992, 1994).

Today, it is simply untenable to regard the views of Bohr and Heisenberg
(which in any case differed considerably from each other) as in any sense
standard or canonical. The meaning of quantum theory is today an open
question, arguably as much as it was in October 1927.

\section{The measurement problem today}\label{mptoday}

\epigraph{The problem of measurement and the observer is the problem of where the
measurement begins and ends, and where the observer begins and ends. .... I
think, that --- when you analyse this language that the physicists have fallen
into, that physics is about the results of observations --- you find that on
analysis it evaporates, and nothing very clear is being said.}{J.~S.~Bell (1986, p. 48)}

\noindent The recurring puzzlement over the meaning of quantum theory often centres
around a group of related conceptual questions that usually come under the
general heading of the `measurement problem'.

\subsection{A fundamental ambiguity}\label{fundamb}

As normally presented in textbooks, quantum theory describes experiments in a
way that is certainly practically successful, but seemingly fundamentally
ill-defined. For it is usually implicitly or explicitly assumed that there is
a clear boundary between microscopic quantum systems and macroscopic classical
apparatus, or that there is a clear dividing line between `microscopic
indefiniteness' and the definite states of our classical macroscopic realm.
Yet, such distinctions defy sharp and precise formulation.

That quantum theory is therefore fundamentally ambiguous was argued with
particular clarity by Bell. For example (Bell 1986, p. 54):
\begin{quotation}
The formulations of quantum mechanics that you find in the books involve
dividing the world into an observer and an observed, and you are not told
where that division comes --- on which side of my spectacles it comes, for
example --- or at which end of my optic nerve.
\end{quotation}
The problem being pointed to here is the lack of a precise boundary
between the quantum system and the rest of the world (including the apparatus
and the experimenter).

A closely-related aspect of the `measurement problem' is the need to explain
what happens to the definite states of the everyday macroscopic domain as one
goes to smaller scales. Where does macroscopic definiteness give way to
microscopic indefiniteness? Does the transition occur somewhere between pollen
grains and macromolecules, and if so, where? On which side of the line is a virus?

Nor can quantum `indefiniteness' or `fuzziness' be easily confined to the
atomic level. For macroscopic objects are made of atoms, and so inevitably one
is led to doubt whether rocks, trees, or even the Moon, have definite
macroscopic states, especially when observers are not present. And this in the
face of remarkable developments in twentieth-century astrophysics and
cosmology, which have traced the origins of stars, galaxies, helium and the
other elements, to times long before human observers existed.

The notion of a `real state of affairs' is familiar from everyday experience:
for example, the location and number of macroscopic bodies in a laboratory.
Science has shown that there is more to the real state of things than is
immediately obvious (for example, the electromagnetic field). Further, it has
been shown that the character of the real state of things changes with scale:
on large scales we find planets, stars and galaxies, while on small scales we
find pollen grains, viruses, molecules, and atoms. Nevertheless, at least
outside of the quantum domain, the notion of `real state' remains. The
ambiguity emphasised by Bell consists of the lack of a sharp boundary between
the `classical' domain, in which `real state' is a valid concept, and the
`quantum domain', in which `real state' is not a valid concept.

Despite decades of effort, this ambiguity remains unresolved within standard
textbook quantum theory, and many critics have been led to argue that the
notion of real state should be extended, in some appropriate way, into the
quantum domain. Thus, for example, Bell (1987, pp. 29--30):
  \begin{quotation}
  Theoretical physicists live in a classical world, looking out into a
  quantum-mechanical world. The latter we describe only subjectively, in terms
  of procedures and results in our classical domain. This subjective description
  is effected by means of quantum-mechanical state functions $\psi$ .... . The
  classical world of course is described quite directly --- `as it is'. .... Now
  nobody knows just where the boundary between the classical and quantum domain
  is situated. .... A possibility is that we find exactly where the boundary
  lies. More plausible to me is that we will find that there is no boundary. It
  is hard for me to envisage intelligible discourse about a world with no
  classical part --- no base of given events .... to be correlated. On the other
  hand, it is easy to imagine that the classical domain could be extended to
  cover the whole.
  \end{quotation}

While Bell goes on to argue in favour of adding extra (`hidden') parameters to
the quantum formalism, for our purposes the key point being made here is the
need to extend the notion of real state into the microscopic domain. This
might indeed be achieved by introducing hidden variables, or by other means
(for example, the Everett approach). Whatever form the theory may take, the
real macroscopic states considered in the rest of science should be part of a
unified description of microscopic and macroscopic phenomena --- what Bell
(1987, p. 30) called `a homogeneous account of the world'.

There are in fact, as we have mentioned, several well-developed proposals for
such a unified or homogeneous (or `realist') account of the world: the
pilot-wave theory of de Broglie (1928) and Bohm (1952a,b); theories of
dynamical wave-function collapse (Pearle 1976, 1979, 1989; Ghirardi, Rimini
and Weber 1986); and the many-worlds interpretation of Everett (1957). The
available proposals that have broad scope assume that the wave function is a
real object that is part of the structure of an individual system. At the time
of writing, it is not known if realist theories may be constructed without
this feature.\footnote{For example, acording to the stochastic
hidden-variables theory of F\'{e}nyes (1952) and Nelson (1966), the wave
function merely provides an emergent description of probabilities. However,
despite appearances, it seems that for technical reasons this theory is flawed
and does not really reproduce quantum theory: the Schr\"{o}dinger equation is
obtained only for exceptional (nodeless) wave functions (Wallstrom 1994;
Pearle and Valentini 2006).}

\subsection{Measurement as a physical process: quantum theory `without
observers'}\label{measphysproc}

Another closely-related aspect of the measurement problem is the question of
how quantum theory may be applied to the process of measurement itself. For it
seems inescapable that it should be possible (in principle) to treat apparatus
and observers as physical systems, and to discuss the process of measurement
in purely quantum-theoretical terms. However, attempts to do so are
notoriously controversial and apt to result in paradox and confusion.

For example, in the paradox of `Wigner's friend' (Wigner 1961), an
experimenter A (Wigner) possesses a box containing an experimenter B (his friend)
and a microscopic system S. Suppose S is initially in, for example, a superposition of energy
states%
\[
\left\vert \psi_{0}\right\rangle =\frac{1}{\sqrt{2}}\left(  \left\vert
E_{1}\right\rangle +\left\vert E_{2}\right\rangle \right)
\]
and that the whole box is initially in a state $\left\vert \Psi_{0}%
\right\rangle =\left\vert B_{0}\right\rangle \otimes\left\vert \psi
_{0}\right\rangle $ (idealising B as initially in a pure state $\left\vert
B_{0}\right\rangle $). Let B perform an ideal energy measurement on S. If A
does not carry out any measurement, then from the point of view of A the
quantum state of the whole box evolves continuously (according to the
Schr\"{o}dinger equation) into a superposition of states%
\begin{equation}
\left\vert \Psi(t)\right\rangle =\frac{1}{\sqrt{2}}\left(  \left\vert
B_{1}\right\rangle \otimes\left\vert E_{1}\right\rangle +\left\vert
B_{2}\right\rangle \otimes\left\vert E_{2}\right\rangle \right)  \label{Wig}%
\end{equation}
(where $\left\vert B_{i}\right\rangle $ is a state such that B has found the
energy value $E_{i}$).

Now, if A wished to, could he (in principle) at later times, by appropriate
experiments on the whole box, observe interference effects involving both
branches of the superposition in (\ref{Wig})? If so, could this be consistent
with the point of view of B, according to which the energy measurement had a
definite result?

One may well question whether the above scenario is realistic, even in
principle (given the resources in our universe). For example, one might
question whether a box containing a human observer could ever be sufficiently
isolated for environmental decoherence to be negligible. However, if the above
`experimenter B' were replaced by an automatic device or machine, the scenario
may indeed become realistic, depending on the possibility of isolating the box
to sufficient accuracy.\footnote{It is perhaps worth remarking that, even if
decoherence has a role to play here, one has to realise what
the problem is in order to understand whether and how decoherence might
contribute to a solution (cf.\ Bacciagaluppi 2005).}

Most scientists agree that macroscopic equipment is subject to the laws of
physics, just like any other system, and that it should be possible to
describe the operation of such equipment purely in terms of the most
fundamental theory available. There is somewhat less consensus over the status
of human experimenters as physical systems. Some physicists have suggested, in
the context of quantum physics, that human beings cannot be treated as just
another physical system, and that human consciousness plays a special role.
For example, Wigner (1961) concluded from his paradox that `the being with a
consciousness must have a different role in quantum mechanics than the
inanimate measuring device', and that for a system containing a conscious
observer `the quantum mechanical equations of motion cannot be linear'.

Wigner's conclusion, that living beings violate quantum laws, seems
increasingly incredible given the impressive progress made in human biology
and neuroscience, in which the human organism --- including the brain --- is
treated as (ultimately) a complex electro-chemical system. There is no
evidence that human beings are able to violate, for example, the laws of
gravity, or of thermodynamics, or basic principles of chemistry, and the
conclusion that human beings in particular should be outside the domain of
quantum laws seems difficult to accept. An alternative conclusion, of course,
is that something is missing from orthodox quantum theory.

Assuming, then, that human experimenters and their equipment may in principle
be regarded as physical systems subject to the usual laws, their interaction
with microsystems ought to be analysable, and the process of measurement ought
to be treatable as a physical process like any other. One can then ask if,
over an ensemble of similar experiments, it would be possible in principle for
the external experimenter A to observe (at the statistical level) interference
effects associated with both terms in (\ref{Wig}). To deny this possibility
would be to claim (with Wigner) that a box containing a human being violates
the laws of quantum theory. To accept the possibility would seem to imply
that, at time $t$ before experimenter A makes a measurement, there was (at
least according to A) no matter of fact about the result of B's observation,
notwithstanding the explicit supposition that B had indeed carried out an
observation by time $t$.

It is sometimes said that Wigner's paradox may be evaded by noting that, if
the external experimenter A actually performs an experiment on the whole box
that reveals interference between the two branches of (\ref{Wig}), then this
operation will destroy the memory the internal experimenter B had of obtaining
a particular result, so that there is no contradiction. But this misses the
point. For while it is true that B will then not have any \textit{memory} of
having obtained a particular experimental result, the contradiction remains
with there being a purported \textit{matter of fact} (at time $t$) as to the
result of B's observation, regardless of whether or not B has subsequently
forgotten it. (Note that in this discussion we are talking about matters of
fact, not for microsystems, but for macroscopic experimental results.)

Let us examine the reasoning behind Wigner's paradox more closely. We take it
that experimenter B agreed beforehand to enter the box and perform an energy
measurement on the microscopic system S; and that it was further agreed that
after sufficient time had elapsed for B to perform the measurement, A would
decide whether or not to carry out an experiment showing interference between
the two branches of (\ref{Wig}). Considering an ensemble of similar
experiments, the paradox consists of a contradiction between the following
statements concerning the physical state of B just before A decides what to do:

\
 
\noindent (I) There is no definite state of B, because A can if he wishes perform
  measurements showing the presence of interference between different states of B.

\noindent (II) There is a definite state of B, because B is a human experimenter like
  any other, and because instead of testing for interference A can simply ask B
  what he saw.
 
\
 
The argument in (I) is the familiar one from standard quantum theory,
applied to the unusual case of a box containing an experimenter. The argument
in (II) is unusual: it requires comment and elaboration.

Because B is a human experimenter like any other, we are driven to consider
the theoretical possibility that, in the distant future, some
`super-experimenter' could decide to perform an interference experiment on a
`box' containing \textit{us} and our equipment. What would happen to our
current (macroscopically-recorded) experimental facts --- concerning for
example the outcome of a spin measurement performed in the laboratory? To be
sure, our records of these experiments could one day be erased, but it would
be illogical to suppose that the \textit{fact} of these experiments having
been carried out (with definite results) could ever be changed. Unless we
accept (II), we are in danger of encountering the paradox that facts about
what we have done in the laboratory today might later turn out 
\textit{not} to be facts.

Further support for (II) comes from Wigner's original argument, which centred
around the assumed reality of other minds. (Wigner did not regard solipsism as
worthy of serious consideration.) From this assumption Wigner inferred that,
whatever the circumstances, if one asks a `friend' what he saw, the answer
given by the friend must have been, as Wigner put it, `already decided in his
mind, before I asked him'. But then, if A decides not to perform an
interference experiment on the box, and simply asks B what he saw, A is
obliged to take B's answer as indicative of the state of B's mind before A
asked the question --- indeed, before A decided on whether or not to perform
an interference experiment. For Wigner, a superposition of the form
(\ref{Wig}) is unacceptable for a system containing a human experimenter or
`friend', because it implies that the friend `was in a state of suspended
animation before he answered my question'.

As Wigner presented it, the argument is based on the assumption that a
conscious being will always have a definite state of consciousness. In
orthodox quantum theory, of course, one might dismiss as `meaningless' the
question of whether the friend's consciousness contained one impression or the
other ($B_{1}$ or $B_{2}$) before he was asked. However, as Wigner put it, `to
deny the existence of the consciousness of a friend to this extent is surely
an unnatural attitude, approaching solipsism'.

Finally, on the topic of Wigner's paradox, it is important to emphasise the
distinction between `matters of fact' on the one hand, and `memories' (true or
false) on the other --- a distinction that is comparable to the distinction
between facts and opinions, or between truth and belief, or between
reality and appearance, distinctions that form part and parcel of the
scientific method. Thus, again, while an experimenter's memory of having
obtained a certain result might be erased in the future, the \textit{fact}
that he once obtained a certain result will necessarily remain a fact: to
assert otherwise would be a logical contradiction.

A further, more subtle motivation for treating measurement as a physical
process comes from considering the very nature of `measurement'. As is well
known to philosophers, and to experimental physicists, the process of
measurement is `theory-laden'. That is, in order to know how to carry out a
measurement correctly, or how to design a specific measuring apparatus
correctly, some prior body of theory is required: in particular, one needs
some understanding of how the equipment functions, and how it interacts with
the system being examined. For this reason, it is difficult to see how the
process of quantum measurement can be properly understood, without some prior
body of theory that describes the equipment itself and its interaction with
the `system'. And since the equipment usually belongs to the definite
macroscopic realm, and the `system' often does not, a proper understanding
seems to require a `homogeneous account of the world' as discussed above, that
is, a theory in which an objective account is provided not only of the
macroscopic apparatus, but also of the microsystem and its interaction with
the apparatus.

A common conclusion, then, is that a coherent account of quantum measurement
requires that quantum theory be somehow extended from a theory of microsystems
to a universal physical theory with an unbounded domain of application, with
our everyday macroscopic realism being somehow extended to the microscopic
level. Given such a well-defined and universal physical theory, whose subject
matter consists of the real states of the world as a whole, it would be
possible in principle to use the theory to analyse the process of measurement
as a physical process like any other (just as, for example, classical
electrodynamics may be used to analyse the process --- involving forces
exerted by magnetic fields --- by which an ammeter measures an electric current).

Thus, for example, pilot-wave theory, or the Everett interpretation, or
collapse models, may be applied to situations where a quantum measurement is
taking place. If the theory provides an unambiguous account of objective
processes in general, it will provide an unambiguous account of the quantum
measurement process in particular. The result is a quantum theory `without
observers', in the sense that observers are physical systems obeying the same
laws as all other systems, and do not have to be added to the theory as
extra-physical elements.

Conclusions as to what is actually happening during quantum measurements will,
of course, depend on the details of the theory. For example, consider again
Wigner's scenario above. In the Everett interpretation, B's observation within
the box has two results, and there is no contradiction if the external
experimenter A subsequently observes interference between them. In de
Broglie-Bohm theory, B's observation has only one result selected by the
actual configuration, but even so the empty wave packet still exists in
configuration space, and can in principle re-overlap with (and hence interfere
with) the occupied packet if appropriate Hamiltonians are applied.

\subsection{Quantum cosmology}

Further closely-related questions, again broadly under the heading of the
`measurement problem', concern the description of the distant past before
human beings and other life forms evolved on Earth, and indeed the description
of the universe as a whole in epochs before life existed.

While the basic theoretical foundations of big-bang cosmology had already been
laid by 1927 (through the work of Friedmann and Lema\^{\i}tre), at that time
any suggestion of the need to provide a quantum-theoretical account of the
early universe could easily have been dismissed as being of no practical or
experimental import. By the 1980s, however, with the development of
inflationary cosmology (Guth 1981), the theoretical question became a
practical one, with observational implications.

According to our current understanding, the small non-uniformities of
temperature observed in the cosmic microwave background originated from
classical density perturbations in the early (and approximately homogenous)
universe (Padmanabhan 1993). And according to inflationary theory, those early
classical density perturbations originated from quantum fluctuations at even
earlier times (Liddle and Lyth 2000). Here we have an example of a
cosmological theory in which a `quantum-to-classical transition' occurred long
before life (or even galaxies) developed, and whose details have left an
imprint on the sky that can be measured today. This is the measurement problem
on a cosmic scale (Kiefer, Polarski and Starobinsky 1998; Perez, Sahlmann and
Sudarsky 2006; Valentini 2006).

But the tension between `Copenhagen' quantum theory and the requirements of
cosmology was felt long before cosmology matured as an experimental science.
Thus, for example, in his pioneering work in the 1960s on quantum gravity,
when it came to applying the theory to a closed universe DeWitt wrote (DeWitt
1967, p. 1131):
  \begin{quotation}
    The Copenhagen view depends on the assumed \textit{a priori} existence of a
    classical level to which all questions of observation may ultimately be
    referred. Here, however, the whole universe is the object of inspection; there
    is no classical vantage point, and hence the interpretation question must be
    re-argued from the beginning.
  \end{quotation}
DeWitt went on to argue (pp. 1140--2) that, in the absence of a
classical level, the Everett interpretation should be adopted. According to
DeWitt (p. 1141):
  \begin{quotation}
    Everett's view of the world is a very natural one to adopt in the quantum
    theory of gravity, where one is accustomed to speak without embarrassment of
    the `wave function of the universe'.
  \end{quotation}
While DeWitt expresses a preference for the Everett interpretation, for
our purposes the central point being made is that, if the whole universe is
treated as a quantum object, with no definite (classical) background `to which
all questions of observation may ultimately be referred', then the physics
becomes unintelligible unless some form of real state (or ontology) is
ascribed to the quantum object. The Everett interpretation provides one way,
among others, to do this.\footnote{Everett's original formulation was of
course open to a number of criticisms, in particular concerning the notion of
`world' and the idea of probability. Such criticisms are arguably being met
only through more recent developments (Saunders 1995, 1998; Deutsch 1999;
Wallace 2003a,b).}

Everett himself, in 1957, had already cited the quantum theory of cosmology as
one of his main motivations for going beyond what he called the `conventional
or \textquotedblleft external observation\textquotedblright\ formulation of
quantum mechanics' (Everett 1957, p. 454). Everett's general motivation was
the need to describe the quantum physics internal to an isolated system, in
particular one containing observers. A closed universe was a special case of
such a system, and one that arguably would have to be considered as the
science of cosmology progressed (as has indeed proved to be the case). Thus
Everett wrote (p. 455):
  \begin{quotation}
    How is one to apply the conventional formulation of quantum mechanics to the
    space-time geometry itself? The issue becomes especially acute in the case of
    a closed universe. There is no place to stand outside the system to observe
    it. There is nothing outside it to produce transitions from one state to
    another. .... No way is evident to apply the conventional formulation of
    quantum mechanics to a system that is not subject to \textit{external}
    observation. The whole interpretive scheme of that formalism rests upon the
    notion of external observation. The probabilities of the various possible
    outcomes of the observation are prescribed exclusively by Process 1
    [discontinuous wave function collapse].
  \end{quotation}

In more recent years, similar concerns have motivated the development of a
`generalised quantum mechanics' based on `consistent' or `decoherent'
histories (Griffiths 1984, 2002; Gell-Mann and Hartle 1990; Hartle 1995;
Omn\`{e}s 1992, 1994), an approach that is also supposed to provide a quantum
theory `without observers', and without a presumed classical background, so as
to be applicable to quantum cosmology.

\subsection{The measurement problem in `statistical' interpretations of
$\psi$}\label{measprobstat}

The measurement problem is often posed simply as the problem of how to
interpret a macroscopic superposition of quantum states, such as (pure) states
of Schr\"{o}dinger's cat. This way of posing the measurement problem can be
misleading, however, as it usually rests on the implicit assumption (or
suggestion) that the quantum wave function $\psi$ is a real physical object
identifiable as a complete description of an individual system. It might be
that $\psi$ is indeed a real object, but not a complete description (as in
pilot-wave theory). Or, $\psi$ might not be a real object at all. Here we
focus on the latter possibility.

The quantum wave function $\psi$ might be merely a mathematical tool for
calculating and predicting the measured frequencies of outcomes over an
ensemble of similar experiments. In which case, it would be immediately wrong
to interpret a mathematical superposition of terms in $\psi$ as somehow
corresponding to a physical superposition of real states for individual
systems, and the `measurement problem' in the limited sense just mentioned
would be a pseudo-problem. This `statistical interpretation' of quantum theory
has been championed in particular by Ballentine (1970).

However, even in the statistical interpretation, the `measurement problem' in
the more general sense remains. For quantum theory is then an incomplete
theory that refers only to ensembles, and simply does not fully describe
individual quantum systems or their relation to real, individual macroscopic
states. The statistical interpretation gives no account of what happens, for
example, when an individual electron is being measured: it talks only about
the distribution of (macroscopically-registered) measurement outcomes over an
ensemble of similar experiments. Nor does the statistical interpretation
provide any sharp delineation of the boundary between `macroscopic' objects
with an individual (non-ensemble) description and `microscopic' objects with
no such description.\footnote{There are different ways of considering a
`statistical' interpretation, depending on one's point of view concerning the
nature of probability. In any case, the `measurement problem' in the general
sense still stands.}

In the statistical interpretation, then, a solution of the measurement problem
in the general sense will require the development of a complete description of
individual systems. This was Einstein's point of view (Einstein 1949, pp. 671--2):
  \begin{quotation}
    The attempt to conceive the quantum-theoretical description as the complete
    description of the individual systems leads to unnatural theoretical
    interpretations, which become immediately unnecessary if one accepts the
    interpretation that the description refers to ensembles of systems and not to
    individual systems. .... [I]t appears unavoidable to look elsewhere for a
    complete description of the individual system .... .
  \end{quotation}

Einstein was arguably the founder of the statistical
interpretation.\footnote{Cf. Ballentine (1972).} It should be noted, however,
that while Einstein's conclusion about the nature of $\psi$ might turn out to
be correct, what seems to have been his main \textit{argument} for this
conclusion now appears to be wrong, in that it was based on what now appears
to be a false premise --- the assumption of locality. For example, in a letter
to his friend Michele Besso, dated 8 October 1952, Einstein argued that the
`quantum state' $\psi$ could not be a complete characterisation of the `real
state' of an individual system, on the following grounds:
  \begin{quotation}
    A system $S_{12}$, with known function $\psi_{12}$, is composed of subsystems
    $S_{1}$ and $S_{2}$, which at time $t$ are far away from each other. If one
    makes a `complete'\ measurement on $S_{1}$, this can be done in different ways
    .... . From the measurement result \textit{and} the $\psi$-function $\psi_{12}$, 
    one can determine .... the $\psi$-function $\psi_{2}$ of the second
    system. \textit{This will take on different forms}, according to the
    \textit{kind} of measurement applied to $S_{1}$.

    But this is in contradiction with assumption (1) [that the quantum state
    characterises the real state completely], \textit{if one excludes action at a
    distance}. Then in fact the measurement on $S_{1}$ can have no influence on
    the real state of $S_{2}$, and therefore according to (1) can have also no
    influence on the quantum state of $S_{2}$ described by $\psi_{2}$. (Einstein
    and Besso 1972, pp. 487--8, emphasis in the original)
  \end{quotation}
Einstein's argument hinges on the fact that, in a local physics, the
measurement made on $S_{1}$ can have no effect on the real state of $S_{2}%
$.\footnote{The notion of locality that Einstein uses here is, to be precise,
a combination of the principles of `separability' (that widely-separated
systems have locally-defined real states) and of `no action at a distance'.
Cf. Howard (1990).}

With the development of Bell's theorem, however, it seems to be beyond
reasonable doubt that quantum physics is \textit{not} local (if one assumes
the absence of backwards causation or of many worlds). For if locality is
assumed, one may use the EPR argument to infer determinism for the outcomes of
quantum measurements at widely-separated wings of an entangled
state.\footnote{`It is important to note that .... \textit{determinism} ....
in the EPR argument .... is not assumed but \textit{inferred} [from locality].
.... It is remarkably difficult to get this point across, that determinism is
not a \textit{presupposition} of the analysis' (Bell 1987, p. 143, italics in
the original).} Following further reasoning by Bell (1964), one may then show
that any local and deterministic completion of quantum theory cannot reproduce
quantum correlations for all measurements on entangled states. Therefore,
locality contradicts quantum theory.\footnote{See, however, Fine (1999) for a
dissenting view.} Because the premise of Einstein's argument contradicts
quantum theory, the argument cannot be used to infer anything about quantum
theory or about the nature of $\psi$. Thus, Einstein's argument does not
establish the `statistical' or `ensemble' nature of $\psi$%
.\footnote{Einstein's argument above has recently been revived by Fuchs (2002)
(who is, however, not explicit about the completeness or incompleteness of
quantum theory). Fuchs states (p. 9) that Einstein `was the first person to
say in absolutely unambiguous terms why the quantum state should be viewed as
information .... . His argument was simply that a quantum-state assignment for
a system can be forced to go one way or the other by interacting with a part
of the world that should have no causal connection with the system of
interest'. Fuchs then quotes at length (p. 10) the above letter by Einstein.
Later in the same paper, Fuchs writes (p. 39): `Recall what I viewed to be the
most powerful argument for the quantum state's subjectivity --- the
Einsteinian argument of [the above letter]. Since [for entangled systems] we
can toggle the quantum state from a distance, it must not be something sitting
over there, but rather something sitting over here: It can only be our
information about the far-away system'. Again, the premise of this Einsteinian
argument -- locality -- is nowadays no longer reasonable (as of course it was
in 1952), and so the argument cannot be used to infer the subjective or
epistemic nature of the quantum state.}

\setcounter{endnote}{0}
\setcounter{equation}{0}

\chapter{Interference, superposition, and wave packet collapse}\label{Interference-superposition-collapse}%
\chaptermark{Interference, superposition, and collapse}

\section{Probability and interference}\label{prob-intrfce}

According to Feynman (1965, chap.~1, p.~1), single-particle interference is `the
\textit{only} mystery' of quantum theory. Feynman considered an experiment in
which particles are fired, one at a time, towards a screen with two holes
labelled 1 and 2. With both holes open, the distribution $P_{12}$ of particles
at the backstop displays an oscillatory pattern of bright and dark fringes. If
$P_{1}$ is the distribution with only hole 1 open, and $P_{2}$ is the
distribution with only hole 2 open, then experimentally it is found that
$P_{12}\neq P_{1}+P_{2}$. According to the argument given by Feynman (as well
as by many other authors), this result is inexplicable by `classical' reasoning.

By his presentation of the two-slit experiment (as well as by his development of the 
path-integral formulation of quantum theory), Feynman popularised the idea
that the usual probability calculus breaks down in the presence of quantum
interference, where it is probability amplitudes (and not probabilities
themselves) that are to be added. As pointed out by Koopman (1955), and by
Ballentine (1986), this argument is mistaken: the probability distributions at
the backstop --- $P_{12}$, $P_{1}$ and $P_{2}$ --- are conditional
probabilities with three distinct conditions (both slits open, one or other
slit closed), and probability calculus does \textit{not} imply any
relationship between these. Feynman's argument notwithstanding, standard
probability calculus is perfectly consistent with the two-slit experiment.

In his influential lectures on physics, as well as asserting the breakdown of
probability calculus, Feynman claimed that no theory with particle
trajectories could explain the two-slit experiment. This claim is still found
in many textbooks.\footnote{For example, Shankar (1994) discusses the two-slit
experiment at length in his chapter 3, and claims (p.~111) that the observed
single-photon interference pattern `completely rules out the possibility that
photons move in well-defined trajectories'. Further, according to Shankar (p.~112): 
`It is now widely accepted that all particles are described by
probability amplitudes $\psi(x)$, and that the assumption that they move in
definite trajectories is ruled out by experiment'.} From a historical point of
view, it is remarkable indeed that single-particle interference came to be
widely regarded as inconsistent with any theory containing particle
trajectories: for as we have seen in chapter~\ref{deBroglieEss}, in the
case of electrons this phenomenon was in fact first predicted (by de Broglie)
on the basis of precisely such a theory.

As we shall now discuss, in his report at the fifth Solvay conference de
Broglie gave a clear and simple explanation for single-particle interference
on the basis of his pilot-wave theory; and the extensive discussions at the
conference contain no sign of any objection to the consistency of de Broglie's
position on this point.

As for Schr\"{o}dinger's theory of wave mechanics, in which particles were
supposed to be constructed out of localised wave packets, in retrospect it is
difficult to see how single-particle interference could have been accounted
for. It is then perhaps not surprising that, in Brussels in 1927, no specific
discussion of interference appears in Schr\"{o}dinger's contributions.

Born and Heisenberg, on the other hand, do discuss interference in their
report, from the point of view of their `quantum mechanics'. And, they do
consider the question of the applicability of probability calculus. The views
they present are, interestingly enough, rather different from the views
usually associated with quantum mechanics today. In particular, as we shall
see below, according to Born and Heisenberg there was (in a very specific
sense) \textit{no} conflict between quantum interference and the ordinary
probability calculus.

Interference was also considered in the general discussion, in particular by
Dirac and Heisenberg: this latter material will be discussed later, in 
section~\ref{DiracandHeisenberg}.

\subsection{Interference in de Broglie's pilot-wave theory}\label{interf-deB}

At the fifth Solvay conference, the subject of interference was addressed from
a pilot-wave perspective by de Broglie in his report. In his section 5, `The
interpretation of interference', de Broglie considered interference
experiments with light of a given frequency $\mathrm{\nu}$. For a guiding wave
$\Psi$ of phase $\phi$ and amplitude $a$, de Broglie took the photon velocity
to be given by $\mathbf{v}=-\frac{c^{2}}{h\mathrm{\nu}}\mathbf{\nabla}\phi$,
while the probability distribution was taken to be $\pi=\mathrm{const\cdot
}a^{2}$. As de Broglie had pointed out, the latter distribution is preserved
over time by the assumed motion of the photons. Therefore, the usual
interference and diffraction patterns follow immediately. To quote de Broglie
(p.~\pageref{forMinEss2}):

\begin{quotation}
the bright and dark fringes predicted by the new theory will coincide with
those predicted by the old [that is, by classical wave optics].
\end{quotation}

De Broglie also pointed out that his theory gave the correct bright and dark
fringes for photon interference experiments, regardless of whether the
experiments were performed with an intense or a very feeble source. As he put
it (p.~\pageref{forMinEss2B}):

\begin{quotation}
one can do an experiment of short duration with intense irradiation, or an
experiment of long duration with feeble irradiation .... if the light quanta
do not act on each other the statistical result must evidently be the same.
\end{quotation}
De Broglie's discussion here addresses precisely the supposed
difficulty highlighted much later by Feynman. It is noteworthy that a clear
and simple answer to what Feynman thought was `the only mystery' of quantum
mechanics was published as long ago as the 1920s.

Even so, for the rest of the twentieth century, the two-slit experiment was
widely cited as proof of the non-existence of particle trajectories in the
quantum domain. Such trajectories were thought to imply the relation
$P_{12}=P_{1}+P_{2}$, which is violated by experiment. As Feynman (1965, chap.~1, p.~6) 
put it, on the basis of this argument it should `undoubtedly' be
concluded that: `It is \textit{not} true that the electrons go \textit{either}
through hole 1 or hole 2'. Feynman also suggested that, by 1965, there had
been a long history of failures to explain interference in terms of trajectories:

\begin{quotation}
Many ideas have been concocted to try to explain the curve for $P_{12}$ [that
is, the interference pattern] in terms of individual electrons going around in
complicated ways through the holes. None of them has succeeded. (Feynman 1965,
chap.~1, p.~6)
\end{quotation}
Yet, de Broglie's construction is so simple as to be almost trivial:
the quantum probability density $\left\vert \psi\right\vert ^{2}$ for particle
position obeys a continuity equation, with a local probability current; if the
trajectories follow the flow lines of the quantum current then, by
construction, an incident distribution $\left\vert \psi_{0}\right\vert ^{2}$
of particles will necessarily evolve into a distribution $\left\vert
\psi\right\vert ^{2}$ at the backstop --- with interference or diffraction, as
the case may be, depending on the potential in which the wave $\psi$ evolves.

Not only did Feynman claim, wrongly, that no one had ever succeeded in
explaining interference in terms of trajectories; he also gave an argument to
the effect that any such explanation was impossible:

\begin{quotation}
Suppose we were to assume that inside the electron there is some kind of
machinery that determines where it is going to end up. That machine must
\textit{also} determine which hole it is going to go through on its way. But
.... what is inside the electron should not be dependent .... upon whether we
open or close one of the holes. So if an electron, before it starts, has
already made up its mind (a) which hole it is going to use, and (b) where it
is going to land, we should find $P_{1}$ for those electrons that have chosen
hole 1, $P_{2}$ for those that have chosen hole 2, \textit{and necessarily}
the sum $P_{1}+P_{2}$ for those that arrive through the two holes. There seems
to be no way around this. (Feynman 1965, chap.~1, p.~10)
\end{quotation}
Feynman's argument assumes that the motion of the electron is
unaffected by opening or closing one of the holes. This assumption is violated
in pilot-wave theory, where the form of the guiding wave behind the two-slit
screen does depend on whether or not both slits are open.

A similar assumption is made in the discussion of the two-slit experiment by
Heisenberg (1962), in chapter III of his book \textit{Physics and Philosophy}.
Heisenberg considers single photons incident on a screen with two small holes
and a photographic plate on the far side, and gives the familiar argument that
the existence of particle trajectories implies the non-interfering result
$P_{1}+P_{2}$. As Heisenberg puts it:

\begin{quotation}
If [a single photon] goes through the first hole and is scattered there, its
probability for being absorbed at a certain point of the photographic plate
cannot depend upon whether the second hole is closed or open. (Heisenberg 1962)
\end{quotation}
This assertion is denied by pilot-wave theory, which provides a simple
counterexample to Heisenberg's conclusion that `the statement that any light
quantum must have gone \textit{either} through the first \textit{or} through
the second hole is problematic and leads to contradictions'.

Finally, we note that interference was also considered by Brillouin 
(pp.~\pageref{Brill-beginning}~ff.) in the discussion following de Broglie's
report, for the case of photons reflected by a mirror. Brillouin drew a
figure (p.~\pageref{Brill-figure}), with a sketch of a photon trajectory
passing through an interference region. To our knowledge, plots of
trajectories in cases of interference did not appear again in the literature
until the pioneering numerical work by Philippidis, Dewdney and Hiley (1979).

\subsection{Interference in the `quantum mechanics' of Born and Heisenberg}\label{Interference-in-Born-and-Heisenberg}

The subject of interference was considered by Born and Heisenberg, in their
report on quantum mechanics (pp.~\pageref{BH-beginning}~f.), for the
case of an atom initially in a superposition%
\begin{equation}
\left\vert \psi(0)\right\rangle =\sum_{n}c_{n}(0)\left\vert n\right\rangle
\label{supn}%
\end{equation}
of energy states $\left\vert n\right\rangle $, with 
coefficients $c_{n}(0)=\left\vert c_{n}(0)\right\vert e^{i\gamma_{n}}$ and eigenvalues $E_{n}$. The
Schr\"{o}dinger equation implies a time evolution%
\begin{equation}
c_{n}(t)=\sum_{m}S_{nm}(t)c_{m}(0)
\end{equation}
with (in modern notation) $S_{nm}(t)=\left\langle n\right\vert
U(t,0)\left\vert m\right\rangle $, where $U(t,0)$ is the evolution operator.
In the special case where $c_{m}(0)=\delta_{mk}$ for some $k$, we have
$\left\vert c_{n}(t)\right\vert ^{2}=\left\vert S_{nk}(t)\right\vert ^{2}$,
and Born and Heisenberg interpret $\left\vert S_{nk}(t)\right\vert ^{2}$ as a
transition probability. They also draw the conclusion that `the $|c_n(t)|^2$ 
must be the state probabilities' (p.~\pageref{mustbe}).

Born and Heisenberg seem to adopt a statistical interpretation, according to
which the system is always in a definite energy state, with jump probabilities
$\left\vert S_{nk}(t)\right\vert ^{2}$ and occupation (or `state')
probabilities $\left\vert c_{n}(t)\right\vert ^{2}$ (cf. section~\ref{on-interference}). 
This is stated quite explicitly (p.~\pageref{BHviewpoint}):

\begin{quotation}
From the point of view of Bohr's theory a system can always be in only  
{\em one} quantum state. .... According to Bohr's 
principles it makes no sense to say a system is simultaneously in several states. 
The only possible interpretation seems to be statistical: 
the superposition of several eigensolutions expresses that through the perturbation the initial
state can go over to any other quantum state .... .
\end{quotation}
Note that this is quite different from present-day quantum mechanics,
in which a system described by the superposition (\ref{supn}) would not
normally be regarded as always occupying only one energy state.

At the same time, Born and Heisenberg recognise a difficulty (p.~\pageref{BH-beginning}):

\begin{quotation}
Here, however, one runs into a difficulty of principle that is of great importance, 
as soon as one starts from an initial state for which not 
all the $c_n(0)$ except one vanish.
\end{quotation}
The difficulty, of course, is that for an initial superposition the
final probability distribution is given by%
\begin{equation}
\left\vert c_{n}(t)\right\vert ^{2}=\Big\vert \sum_{m}S_{nm}(t)c_{m}%
(0)\Big\vert ^{2}\label{interf}%
\end{equation}
as opposed to%
\begin{equation}
\left\vert c_{n}(t)\right\vert ^{2}=\sum_{m}\left\vert S_{nm}(t)\right\vert
^{2}\left\vert c_{m}(0)\right\vert ^{2}\label{nointerf}%
\end{equation}
which, as Born and Heisenberg remark, `one might suppose from the usual probability calculus'. (In
standard probability calculus, of course, (\ref{nointerf}) expresses
$\left\vert c_{n}(t)\right\vert ^{2}$ as a sum over conditional transition
probabilities $\left\vert S_{nm}(t)\right\vert ^{2}$ weighted by the initial
population probabilities $\left\vert c_{m}(0)\right\vert ^{2}$.)

While Born and Heisenberg refer to (\ref{interf}) as the `theorem of the \textit{interference of
probabilities}', they make the remarkable assertion that there is in fact
\textit{no} contradiction with the usual rules of probability calculus, and
that the $\left\vert S_{nm}\right\vert ^{2}$ may still be regarded as ordinary
probabilities. Further, and equally remarkably, it is claimed that in any case
where the state probabilities $\left\vert c_{n}\right\vert ^{2}$ are
experimentally established, the presence of the unknown phases $\gamma_{n}$
makes the interfering expression (\ref{interf}) reduce to the non-interfering
expression (\ref{nointerf}) (p.~\pageref{Born-contradiction}):

\begin{quotation}
.... it should be noted that this `interference' does not represent a
contradiction with the rules of the probability calculus, that is, with the
assumption that the $\left\vert S_{nk}\right\vert ^{2}$ are quite usual
probabilities. In fact, .... [(\ref{nointerf})] follows from the concept of
probability .... when and only when the relative number, that is, the
probability $\left\vert c_{n}\right\vert ^{2}$ of the atoms in the state $n$,
has been \textit{established} beforehand \textit{experimentally}. In this case
the phases $\gamma_{n}$ are unknown in principle, so that [(\ref{interf})]
then naturally goes over to [(\ref{nointerf})].... .
\end{quotation}
Here, Born and Heisenberg refer to Heisenberg's (recently-published)
uncertainty paper, which contains a similar claim. There, Heisenberg considers
a Stern-Gerlach atomic beam passing through two successive regions of field
inhomogeneous in the direction of the beam (so as to induce transitions
between energy states without separating the beam into components). If the
input beam is in a definite energy state then the beam emerging from the first
region will be in a superposition. The probability distribution for energy
emerging from the second region will then contain interference --- as in
(\ref{interf}), where the `initial' superposition (\ref{supn}) is now the
state emerging from the first region. Heisenberg asserts that, if the energy
of an atom is actually measured between the two regions, then because of the
resulting perturbation `the ``phase'' of the atom changes by amounts that are
in principle uncontrollable' (Heisenberg 1927, pp.~183--4), and
averaging over the unknown phases in the final superposition yields a
non-interfering result.

The same example, with the same phase randomisation argument, is also given by
Heisenberg (1930b) in his book \textit{The Physical Principles of the Quantum
Theory} (chapter IV, section 2), which was based on lectures delivered at
Chicago in 1929. Heisenberg asserts (p.~60) that an energy measurement for an
atom in the intermediate region `will necessarily alter the phase of the de
Broglie wave of the atom in state $m$ by an unknown amount of order of
magnitude one', so that in applying the (analogue of the) interfering
expression (\ref{interf}) each term in the sum `must thus be multiplied by the
arbitrary factor $\exp\left(  i\varphi_{m}\right)  $ and then averaged over
all values of $\varphi_{m}$'.

From a modern perspective, this argument seems strange and unfamiliar, and
indeed quite wrong. However, the argument makes rather more sense, if one
recognises that the `quantum mechanics' described by Born and Heisenberg is
not quantum mechanics as we usually know it today. In particular, the theory
as they present it appears to contain no notion of wave packet collapse (or
state vector reduction).

The argument given by Born and Heisenberg amounts to saying, in modern
language, that if the energies of an atomic population have actually been
measured, then one will have a mixture of states of the \textit{superposed}
form (\ref{supn}), with randomly-distributed phases $\gamma_{n}$. Such a
mixture is indeed statistically equivalent to a mixture of energy states
$\left\vert n\right\rangle $ with weights $\left\vert c_{n}(0)\right\vert
^{2}$, because the density operators are the same:%
\begin{multline}
\left(  \prod\limits_{k}\frac{1}{2\pi}\int d\gamma_{k}\right)  \sum
_{n,m}\left\vert c_{n}(0)\right\vert \left\vert c_{m}(0)\right\vert
e^{i(\gamma_{n}-\gamma_{m})}\left\vert n\right\rangle \left\langle
m\right\vert =\\
\sum_{n}\left\vert c_{n}(0)\right\vert ^{2}\left\vert
n\right\rangle \left\langle n\right\vert\ .  \label{rho12}%
\end{multline}
\newline However, from a modern point of view, if one did measure the atomic
energies and find the value $E_{n}$ with frequency $p_{n}$, the resulting
total ensemble would naturally be represented by a density operator $\rho
=\sum_{n}p_{n}\left\vert n\right\rangle \left\langle n\right\vert $, and there
would seem to be no particular reason to rewrite this in terms of the
alternative decomposition on the left-hand side of (\ref{rho12}) (with
$\left\vert c_{n}(0)\right\vert ^{2}=p_{n}$ and random phases $e^{i\gamma_{n}%
}$); though of course one could if one wished to. What is more, an actual
inconsistency would appear if --- having measured the atomic energies --- one
selected a particular atom that was found to have energy $E_{m}$: a subsequent
and immediate energy measurement for this particular atom should again yield
the result $E_{m}$ with certainty, as is consistent with the usual
representation of the atom by the state $\left\vert m\right\rangle $, and this
certainty would be \textit{in}consistent with what appears to be (at least in
effect) the proposed representation of the atom by a state $\sum_{n}\left\vert
c_{n}(0)\right\vert e^{i\gamma_{n}}\left\vert n\right\rangle $ with randomised
phases $\gamma_{n}$. Indeed, for any subensemble composed of the latter
states, all the energy values present in the sum will be possible outcomes of
subsequent and immediate energy measurements.\footnote{Of course, if we do not
subdivide the atomic ensemble on the basis of the measured energies, the
\textit{total} ensemble will be a mixture with density operator $\sum
_{n}\left\vert c_{n}(0)\right\vert ^{2}\left\vert n\right\rangle \left\langle
n\right\vert $ and will indeed be indistinguishable from the proposed mixture
of states $\sum_{n}\left\vert c_{n}(0)\right\vert e^{i\gamma_{n}}\left\vert
n\right\rangle $ with randomly-distributed phases. But there is nothing to
prevent an experimenter from selecting atoms according to their measured
energies.}

This inconsistency arises, however, \textit{if} one applies the modern notion
of state vector collapse --- a notion that, upon close examination, appears to
be quite absent from the theory presented by Born and Heisenberg. Instead of
applying the usual collapse rule, Born and Heisenberg seem to interpret the
quantities $\left\vert S_{nm}\right\vert ^{2}$ as `quite usual' transition probabilities 
in all circumstances, even in the presence of interference. On this view, then, an
atom that has been found to have energy $E_{m}$ can be represented by a state
$\sum_{n}\left\vert c_{n}(0)\right\vert e^{i\gamma_{n}}\left\vert
n\right\rangle $ with randomised phases $\gamma_{n}$, and the probability of
obtaining a value $E_{n}$ in an immediately successive measurement is given
not by $\left\vert c_{n}(0)\right\vert ^{2}$ (as would follow from the usual
collapse rule) but by the transition probability $\left\vert S_{nm}\right\vert
^{2}$ --- where the latter does indeed approach $\delta_{nm}$ as the time
interval between the two energy measurements tends to zero, so that the above
contradiction does not in fact arise.

Considering now the whole atomic ensemble, if the energies of the atoms have
indeed been measured, then using the $\left\vert S_{nm}\right\vert ^{2}$ as
transition probabilities and the $\left\vert c_{m}(0)\right\vert ^{2}$ as
population probabilities, application of the probability calculus gives the
non-interfering result (\ref{nointerf}). As noted by Born and Heisenberg, on
their view exactly the \textit{same} result is obtained from the `interfering'
expression (\ref{interf}), with random phases $\gamma_{m}$ appearing in the
coefficients $c_{m}(0)=\left\vert c_{m}(0)\right\vert e^{i\gamma_{m}}$.

If instead the atomic energies have not been measured, then, according to Born
and Heisenberg, the phases $\gamma_{m}$ have not been randomised and the
expression (\ref{interf}) does show interference, in contradiction with the
non-interfering expression (\ref{nointerf}). How do Born and Heisenberg
reconcile the breakdown of (\ref{nointerf}) with their claim that the ordinary
probability calculus still holds, with the $\left\vert S_{nm}\right\vert ^{2}$
being quite ordinary probabilities? The answer seems to be that, if the
energies have not been measured, then the population probabilities $\left\vert
c_{m}(0)\right\vert ^{2}$ are in some sense ill-defined, so that the usual
probability formula (\ref{nointerf}) simply cannot be applied: `[(\ref{nointerf})] 
follows from the concept of probability .... when and only when .... the probability $\left\vert
c_{n}\right\vert ^{2}$ .... has been \textit{established} beforehand
\textit{experimentally}'.

Born and Heisenberg seem to take an `operational' view of the population
probabilities, in the sense that these are to be regarded as meaningful only
when directly measured. And the cited argument in Heisenberg's uncertainty
paper suggests that it is operationally impossible to have simultaneously
well-defined phase relations and population probabilities in the same
experiment. This impossibility was presumably regarded as comparable to the
(operational) impossibility of having simultaneously a well-defined position
and momentum for a particle. What seems to be at work here, then, is some form
of uncertainty relation (or complementarity) between population probabilities
and phases: measurement of the former makes the latter ill-defined, and vice versa. 
Interference requires definite phase
relationships, which preclude a well-defined population probability, so that
the ordinary probability calculus cannot be applied.\footnote{There is an
analogy here with Heisenberg's view of causality, expressed in his uncertainty
paper, according to which causality cannot be applied because its premiss is
generally false: `.... in the sharp formulation of the law of causality, ``If we 
know the present exactly, we can calculate the future'', it is not the consequent
that is wrong, but the antecedent. We {\em cannot} in principle get to know the present 
in all [its] determining data' 
(Heisenberg 1927, p.~197).} If
instead the population probability has actually been measured, then the phases
are indefinite and averaging over them washes out any interference.

The resulting viewpoint is certainly remarkable. According to Born and
Heisenberg, in the presence of interference, the quantities $\left\vert
S_{nm}\right\vert ^{2}$ continue to be quite ordinary (transition)
probabilities, while the quantities $\left\vert c_{m}(0)\right\vert ^{2}$
cannot be regarded as population probabilities --- rendering the formula
(\ref{nointerf}) inapplicable. On this view, ordinary probability calculus is
not violated; it is simply wrong to assert that the $\left\vert c_{m}%
(0)\right\vert ^{2}$ represent state probabilities in an interfering case.

One may well object to this point of view on the grounds that, even without
measuring the energies, for a given preparation of the state (\ref{supn}) the
coefficients $c_{m}(0)$ --- and hence the values of $\left\vert c_{m}%
(0)\right\vert ^{2}$ --- will be known (up to an overall phase). However,
presumably, Born and Heisenberg would have had to assert that while the
$\left\vert c_{m}(0)\right\vert ^{2}$ always exist as mathematical quantities,
they cannot be properly interpreted as population probabilities unless the
energies have been measured directly.

From a modern perspective, Born and Heisenberg's treatment of interference is
surprising: in modern quantum mechanics, of course, in cases where
interference occurs the quantities $\left\vert S_{nm}\right\vert ^{2}$ would
not normally be interpreted as `quite usual' transition probabilities; while in cases where interference
does not occur, the non-interfering result (\ref{nointerf}) (as applied here)
would not normally be regarded as arising from the interfering result
(\ref{interf}) through a process of phase randomisation.

\section{Macroscopic superposition: Born's discussion of the cloud 
chamber}\label{macrosup}\sectionmark{Macroscopic superposition}

Quantum theory is normally understood to allow the `superposition of distinct
physical states'. However, while `superposition' is well-defined as a
mathematical term, it is hard to make sense of when applied to physical states
--- that is, when the components in a superposition are regarded as
simultaneous physical attributes of a single system. The need to understand
such `physical superposition' seems particularly acute when it is considered
at the macroscopic level. The difficulty here is closely related to the
question of wave packet collapse: how is a mathematical superposition of
macroscopically-distinct states related to the definite macroscopic states
seen in the laboratory?

The Wilson cloud chamber, as used to observe the tracks of $\alpha$-particles,
was discussed at length by Born in the general discussion. The cloud chamber
illustrates the measurement problem rather well, and is a good example of how
microscopic superpositions can become transferred to the macroscopic domain.
It also illustrates how extending the formal quantum description to the
environment does not by itself alleviate the measurement problem (despite many
claims to the contrary, for example Zurek (1991)).\footnote{For a summary of
criticisms of environmental decoherence as a solution to the measurement
problem, see Bacciagaluppi (2005).} Remarkably, as we shall see, Born asserts
that wave packet collapse is not required to discuss the cloud chamber.

The mechanism of the cloud chamber is well known. The $\alpha$-particles pass
through a supersaturated vapour. The passage of the particles causes
ionisation, and the vapour condenses around the ions, resulting in the
formation of tiny droplets. The droplets scatter light, making the particle
tracks visible.

If the emission of an $\alpha$-particle is undirected, so that the emitted
wave function is approximately spherical, how does one account for the
approximately straight particle track revealed by the cloud chamber? In the
general discussion, Born attributes this question to Einstein, and asserts
that to answer it (p.~\pageref{page172})

\begin{quotation}
.... one must appeal to the notion of `reduction of the probability packet'
developed by Heisenberg.
\end{quotation}
This notion appears in Heisenberg's uncertainty paper, which had been
published in May of 1927. In section 3, entitled `The transition from micro-
to macromechanics', Heisenberg had described how a classical electron orbit
`comes into being' through repeated observation of the electron position,
using light of wavelength $\lambda$. According to Heisenberg, the result of
each observation can be characterised by a probability packet of width
$\lambda$, where the packet spreads freely until the next observation: `Every
determination of position reduces therefore the wave packet back to its original 
size $\lambda$' (Heisenberg 1927, p.~186).

In the case of the cloud chamber, the collapse of the wave packet is applied
repeatedly to the $\alpha$-particle alone. Upon producing visible ionisation,
the wave packet of the $\alpha$-particle collapses, and then starts to spread
again, until further visible ionisation is produced, whereupon collapse occurs
again, and so on. The probability for the resulting `trajectory' is
concentrated along straight lines, accounting for the observed track in the
cloud chamber. As Born puts it (p.~\pageref{thedescription}):

\begin{quotation}
The description of the emission by a spherical wave is valid only for as long
as one does not observe ionisation; as soon as such ionisation is shown by the
appearance of cloud droplets, in order to describe what happens afterwards one
must `reduce' the wave packet in the immediate vicinity of the drops. One thus
obtains a wave packet in the form of a ray, which corresponds to the
corpuscular character of the phenomenon.
\end{quotation}

Here, the cloud chamber itself --- the ionisation, and the formation of
droplets from the vapour --- is treated as if it were an external `classical
apparatus': only the $\alpha$-particle appears in the wave function.

\subsection{Quantum mechanics without wave packet collapse?}\label{QM-without-collapse}

Born goes on to consider if wave packet reduction can be \textit{avoided} by
treating the atoms of the cloud chamber, along with the $\alpha$-particle, as
a single system described by quantum theory, a suggestion that he attributes
to Pauli (p.~\pageref{MrPauli}):

\begin{quotation}
Mr Pauli has asked me if it is not possible to describe the process without
the reduction of wave packets, by resorting to a multi-dimensional space,
whose number of dimensions is three times the number of all the particles
present .... . This is in fact possible .... but this does not lead us further
as regards the fundamental questions.
\end{quotation}

Remarkably, Born claims that a treatment without reduction is `in fact'
possible, and goes on to illustrate how, in his opinion, this can be done. As
we shall see, Born seems to make use of a `classical' probability reduction
only, without any reduction for the configuration-space wave packet.

As for Born's reference to Pauli, around the time of the Solvay conference
Pauli believed that wave packet reduction was needed only for describing
subsystems. This is clear from a letter he wrote to Bohr, on 17 October 1927
(one week before the Solvay meeting began), in which Pauli comments on wave
packet reduction (Pauli 1979, p.~411):

\begin{quotation}
This is precisely a point that was not quite satisfactory in Heisenberg [that
is, in the uncertainty paper]; there the `reduction of the packets' seemed a
bit mystical. Now however, it is to be stressed that at first such reductions
are not necessary if one \textit{includes} in the system all means of
measurement. But in order to describe observational results theoretically at
all, one has to ask what one can say about just a \textit{part} of the whole
system. And then from the complete solution one sees immediately that, in many
cases (of course not always), leaving out the means of observation can be
formally replaced by such reductions.
\end{quotation}
Thus, at that time, Pauli thought that the reduction was a formality
associated with an effective description of subsystems alone, and that if the
apparatus were included in the system then reduction would not be needed at all.

Born, then, presents a multi-dimensional treatment, in which atoms in the
cloud chamber are described by quantum theory on the same footing as the
$\alpha$-particle. Born considers the simple case of a model cloud chamber
consisting of just two atoms in one spatial dimension. There are two cases,
one with both atoms on the same side of the origin (where the $\alpha
$-particle is emitted), and the other with the atoms on opposite sides of the
origin. The two `tubes' in Born's diagram (see his figure
) represent, for
the two cases, the time development of the total (localised) packet in
3-dimensional configuration space. The coordinate $x_{0}$ of the $\alpha
$-particle is perpendicular to the page, while $x_{1},x_{2}$ are the
coordinates of the two atoms. In case I, the initial state is localised at
$x_{0}=0$, $x_{1},x_{2}>0$; in case II it is localised at $x_{0}=0$, $x_{1}%
>0$, $x_{2}<0$. Each initial packet separates into two packets moving in
opposite directions along the $x_{0}$-axis. In the first case the two
collisions (indicated by kinks in the trajectory of the packet\footnote{Note
that the motion occurs in one spatial dimension; a kink in the
configuration-space trajectory shows that the corresponding atom has undergone
a small spatial displacement.}) take place on the same side of the origin; in
the second case they take place on opposite sides. Note that, in both cases,
both branches of the wave packet --- moving in opposite directions --- are
shown; that is, the complete (`uncollapsed') packets are shown in the figure.

Born remarks (p.~\pageref{tothereduction}) that:

\begin{quotation}
To the `reduction' of the wave packet corresponds the choice of one of the two
directions of propagation $+x_{0}$, $-x_{0}$, which one must take as soon as
it is established that one of the two points 1 and 2 is hit, that is to say,
that the trajectory of the packet has received a kink.
\end{quotation}
Here Born seems to be saying that, instead of reduction, what takes
place is a choice of direction of propagation. But propagation of what? Born
presumably does not mean a choice of direction of propagation of the wave packet, 
for that would amount to wave packet reduction, which Born at the outset has 
claimed is unnecessary in a multi-dimensional treatment. (And indeed, his figure
shows the wave packet propagating in both directions.) Instead, Born seems to be
referring to a choice in the direction of propagation of the \textit{system}
(which we would represent by a point in configuration space). The wave packet
spreads in both directions, and determines the probabilities for the different
possible motions of the system. Once the direction of motion of the system is
established, by the occurrence of collisions, an ordinary (`classical')
reduction of the \textit{probability} distribution occurs --- while the wave
packet itself is unchanged. In other words, the probabilities are updated but
the wave packet does not collapse. This, at least, appears to be Born's point
of view.

This may seem a peculiar interpretation --- ordinary probabilistic collapse
without wave packet collapse --- but it is perhaps related to the intuitive
thinking behind Born's famous collision papers of the previous year (Born
1926a,b) (papers in which probabilities for results of collisions were
identified with squares of scattering amplitudes for the wave
function\footnote{Cf. the discussion in section~\ref{Bornsintroduction}.}). Born
drew an analogy with Einstein's notion of a `ghost field' that determines
probabilities for photons (see chapter~\ref{guiding-fields-in-3-space}).
As Born put it:

\begin{quotation}
In this, I start from a remark by Einstein on the relationship between the
wave field and light quanta; he said, for instance, that the waves are there
only to show the corpuscular light quanta the way, and in this sense he talked
of a `ghost field'. This determines the probability for a light quantum, the
carrier of energy and momentum, to take a particular path; the field itself,
however, possesses no energy and no momentum. .... Given the perfect analogy
between the light quantum and the electron .... one will think of formulating
the laws of motion of electrons in a similar way. And here it is natural to
consider the de Broglie-Schr\"{o}dinger waves as the `ghost field', or better,
`guiding field' .... [which] propagates according to the Schr\"{o}dinger
equation. Momentum and energy, however, are transferred as if corpuscles
(electrons) were actually flying around. The trajectories of these corpuscles
are determined only insofar as they are constrained by the conservation of
energy and momentum; furthermore, only a probability for taking a certain path
is determined by the distribution of values of the function $\psi$. (Born
1926b, pp. 803--4)
\end{quotation}

Here, Born seems to be suggesting that there are stochastic trajectories for
electrons, with probabilities for paths determined by the wave function $\psi
$. It is not clear, though, whether the $\psi$ field is to be regarded as a
physical field associated with \textit{individual} systems (as the
electromagnetic field usually is), or whether it is to be regarded merely as
relating to an ensemble. If the former, then it might make sense to apply
collapse to the probabilities without applying collapse to $\psi$ itself. For
example, this could happen in a stochastic version of de Broglie-Bohm theory:
$\psi$ could be a physical field evolving at all times by the Schr\"{o}dinger
equation (hence never collapsing), and instead of generating deterministic
particle trajectories (as in standard de Broglie-Bohm theory) $\psi$ could
generate probabilistic motions only. Further evidence that Born was indeed
thinking along such lines comes from an unpublished manuscript by Born and
Jordan, written in 1925, in which they propose a stochastic theory of photon
trajectories with probabilities determined by the electromagnetic field (as
originally envisaged by Slater) --- see Darrigol (1992, p.~253) and 
section~\ref{BornBohr}.\footnote{In this
connection it is interesting to note the following passage from Born's book
\textit{Atomic Physics} (Born 1969), which was first published in German in
1933: `A mechanical process is therefore accompanied by a wave process, the
guiding wave, described by Schr\"{o}dinger's equation, the significance of
which is that it gives the probability of a definite course of the mechanical
process'. Born's reference to the wave function as `the guiding wave' shows
the lingering influence of the ideas that had inspired him in 1926.}

Thus, in his discussion of the cloud chamber, when Born spoke of `the choice
of one of the two directions of propagation', he may indeed have been
referring to the possible directions of propagation of the system
configuration, with the $\psi$ field remaining in a superposition. On this
reading, wave packet reduction for the $\alpha$-particle would be only an
effective description, which properly corresponds to the branching of the
total wave function together with a random choice of trajectory (in
multi-dimensional configuration space). Unfortunately, however, Born's
intentions are not entirely clear: whether this really is what Born had in
mind, in 1926 or 1927, is difficult to say. Certainly, in the general
discussion at the fifth Solvay conference, Born did maintain that $\psi$ does
not really collapse (in this multi-dimensional treatment), so it is difficult
to see how he could have thought that $\psi$ gave merely a probability
distribution over an ensemble.

The claim that the wave function $\psi$ does not collapse is also found, in
effect, in section II of the report that Born and Heisenberg gave on quantum
mechanics. As we saw in section~\ref{Interference-in-Born-and-Heisenberg}, in
their discussion of interference, an energy measurement is taken to induce a
randomisation of the phases appearing in a superposition of energy states,
instead of the usual collapse to an energy eigenstate. In their example, after
an energy measurement all of the components of the superposition are still
present, and the phase relations between them are randomised. In Born's
example of the cloud chamber, it seems that, here too, all of the components
of the wave function are still present at the end of a measurement. Nothing is
said, however, about the relative phases of the components, and we do not know
whether or not Born had in mind a similar phase randomisation in this case also.

\sectionmark{Dirac and Heisenberg}
\section{Dirac and Heisenberg: interference, state reduction, and delayed
choice}\sectionmark{Dirac and Heisenberg}\label{DiracandHeisenberg}

Another striking feature of quantum theory, as normally understood, is
`interference between alternative histories'. Like superposition, interference
is mathematically well-defined but its physical meaning is ambiguous, and it
has long been considered one of the main mysteries of quantum theory (as we
saw in section~\ref{prob-intrfce}). The question of \textit{when}
interference can or cannot take place is intimately bound up with the
measurement problem, in particular, with the question of when or how definite
outcomes emerge from quantum experiments, and with the question of the
boundary between the quantum and classical domains.

\begin{figure}
    \centering
    \resizebox{\textwidth}{!}{\includegraphics[0mm,0mm][220.30mm,150.98mm]{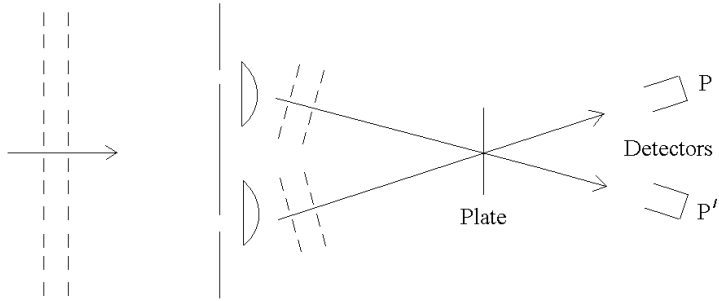}}

\caption{Delayed-choice double-slit experiment.}%
\label{JAW}%
  \end{figure}

In modern times, one of the most puzzling aspects of interference was
emphasised by Wheeler (1978). In his `delayed-choice' experiment, it appears
that the existence or non-existence of interfering histories in the past is
determined by an experimental choice made in the present. One version of
Wheeler's experiment --- a `delayed-choice double-slit experiment' with single
photons --- is shown in Fig. \ref{JAW}. A single photon is incident on a
screen with two slits. The waves emerging from the slits are focussed (by
off-centred lenses) so as to cross each other as shown. The insertion of a
photographic plate in the interference region would seem (from the
interference pattern) to imply the past existence of interfering trajectories
passing through both slits. On the other hand, if no such plate is inserted,
then a detection at P or P%
\'{}
seems to imply that the particle passed through the bottom or top slit
respectively.\footnote{This inference is commonly made, usually without
explicit justification. Some authors appeal to conservation of momentum for a
free particle, but it is not clear how such an argument could be made precise
--- after all there are no particle trajectories in standard quantum theory.
In de Broglie-Bohm theory, the inference is actually wrong: particles detected
at P or P%
\'{}
come from the top or bottom slits respectively (Bell 1980; Bell 1987, chap.~14).} 
Since the plate could have been inserted long after the photon completed
most of its journey (or journeys), it appears that an experimental choice now
can affect whether or not there was a definite photon path in the past.%

A similar point arose in 1927 in the general discussion. Dirac expounded his
view that quantum outcomes occur when nature makes a choice. Heisenberg
replied that this could not be, because of the possibility of observing
interference later on by choosing an appropriate experimental arrangement,
leading Heisenberg to conclude that outcomes occur when a choice is made (or
brought about) not by nature but by the \textit{observer}. Heisenberg's view
here bears some resemblance to Wheeler's. Dirac, in contrast, seems to say on
the one hand that stochastic collapse of the wave packet occurs for
microscopic systems, while on the other hand that if the experiment is chosen
so as to allow interference then such collapse is postponed.

Here is how Dirac expresses it (p.~\pageref{Dirac-state}):

\begin{quotation}
According to quantum mechanics the state of the
world at any time is describable by a wave function $\psi$, which normally
varies according to a causal law, so that its initial value determines its
value at any later time. It may however happen that at a certain time $t_{1}$,
$\psi$ can be expanded in the form%
\[
\psi=\sum_{n}c_{n}\psi_{n}\ ,
\]
where the $\psi_{n}$'s are wave functions of such a nature that they cannot
interfere with one another at any time subsequent to $t_{1}$. If such is the
case, then the world at times later than $t_{1}$ will be described not by
$\psi$ but by one of the $\psi_{n}$'s. The particular $\psi_{n}$ that it shall
be must be regarded as chosen by nature.
\end{quotation}

Note first of all that Dirac regards $\psi$ as describing the state `of the
world' --- presumably the whole world. Then, in circumstances where $\psi$ may
be expanded in terms of non-interfering states $\psi_{n}$, the world is
subsequently described by one of the $\psi_{n}$ (the choice being made by
nature, the probability for $\psi_{n}$ being $\left\vert c_{n}\right\vert
^{2}$). Dirac does not elaborate on precisely when or why a decomposition into
non-interfering states should exist, nor does he address the question of
whether such a decomposition is likely to be unique. Such questions, of
course, go to the heart of the measurement problem, and are lively topics of
current research.

It is interesting that, in Dirac's view (apparently), there are circumstances
in which interference is completely and irreversibly destroyed. For him, the
particular $\psi_{n}$ results from (p.~\pageref{Dirac-choice}):

\begin{quotation}
an irrevocable choice of nature, which must affect the whole of the future
course of events.
\end{quotation}

It seems that according to Dirac, once nature makes a choice of one branch,
interference with the other branches is impossible for the whole of the
future. A definite collapse has occurred, after which interference between the
alternative outcomes is no longer possible, even in principle. This view
clearly violates the Schr\"{o}dinger equation as applied to the whole world:
as Dirac states, the wave function $\psi$ of the world `normally' evolves
according to a causal law, but not always.

\begin{figure}
    \centering
     \resizebox{\textwidth}{!}{\includegraphics[0mm,0mm][220.30mm,270.98mm]{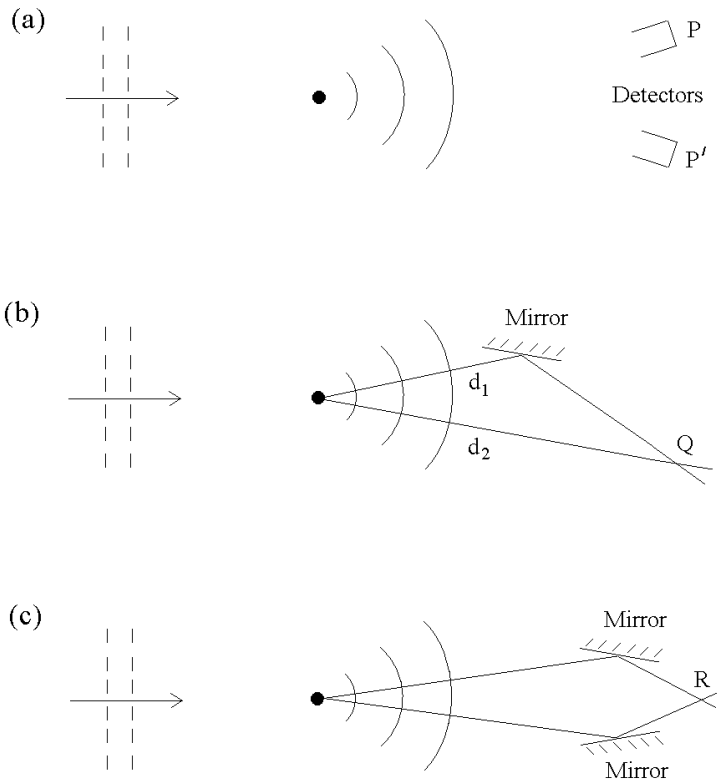}}
\caption{Reconstruction of scattering scenarios discussed by Dirac (Figs. (a),
(b)) and Heisenberg (Fig. (c)).}%
\label{DirHeis}%
  \end{figure}

But Dirac goes further, and recognises that there are circumstances where the
choice made by nature cannot have occurred at the point where it might have
been expected. Dirac considers the specific example of the scattering of an
electron. He first notes that, after the scattering, one must take the wave
function to be not the whole scattered wave but a packet moving in a specific
direction (that is, one of the $\psi_{n}$). He claims (p.~\pageref{fromresults}) 
that one could infer that nature had chosen this
specific direction:

\begin{quotation}
From the results of an experiment, by tracing back a chain of causally
connected events one could determine in which direction the electron was
scattered and one would thus infer that nature had chosen this direction.
\end{quotation}
This is illustrated in Fig.~\ref{DirHeis}(a). If the electron is
detected at P, for example, one may arguably infer a corresponding choice of
direction at the time of scattering. (Note that Figs.~\ref{DirHeis}(a)--(c)
are ours.)%

On the other hand, Dirac goes on to make the following observation (p.~\pageref{ifnow}):

\begin{quotation}
If, now, one arranged a mirror to reflect the electron wave scattered in one
direction $d_{1}$ so as to make it interfere with the electron wave scattered
in another direction $d_{2}$, one would not be able to distinguish between the
case when the electron is scattered in the direction $d_{2}$ and when it is
scattered in the direction $d_{1}$ and reflected back into $d_{2}$. One would
then not be able to trace back the chain of causal events so far, and one
would not be able to say that nature had chosen a direction as soon as the
collision occurred, but only [that] at a later time nature chose where the
electron should appear.
\end{quotation}

Dirac's modified scenario is sketched in Fig.~\ref{DirHeis}(b). The presence
of the mirror leads to interference, at Q, between parts of the electron wave
scattered in different directions $d_{1}$, $d_{2}$. And Dirac's interpretation
is that this interference is intimately related to the fact that an
experimenter observing the outgoing electron in a direction $d_{2}$ `would not
be able to distinguish between the case when the electron is scattered in the
direction $d_{2}$ and when it is scattered in the direction $d_{1}$ and
reflected back into $d_{2}$'. The experimenter would not be able to `trace
back the chain of causal events' to the point where he could say that `nature
had chosen a direction as soon as the collision occurred'. For Dirac, in this
case, nature did \textit{not} make a choice at the time of the collision, and
only later nature `chose where the electron should appear'.

What Dirac describes here is precisely the viewpoint popularised by Feynman in
his famous lectures (Feynman 1965), according to which if a process occurs by
different routes that are subsequently indistinguishable (in the sense that
afterwards an experimenter is in principle unable to tell which route was
taken) then the probability \textit{amplitudes} for the different routes are
to be added; whereas if the different routes are subsequently distinguishable
in principle, then the \textit{probabilities} are to be added.\footnote{Note
that Feynman's path-integral formulation of quantum theory --- developed in
his PhD thesis and elsewhere (Feynman 1942, 1948) --- was anticipated by Dirac
(1933). Feynman's thesis and Dirac's paper are reprinted in Brown (2005).}

Dirac's presentation of the scattering experiment with the mirror ends with
the statement (p.~\pageref{postpone}):

\begin{quotation}
The interference between the $\psi_{n}$'s compels nature to postpone her choice.
\end{quotation}
In his manuscript, a cancelled version of the sentence begins with
`Thus a \textit{possibility} of interference between .... ', while another
cancelled version begins as `Thus the \textit{existence} of .... ' (italics
added). Possibly, Dirac hesitated here because he saw that the mirror could be
added by the experimenter after the scattering had taken place, leading to
difficulties with his view that without the mirror nature makes a choice at
the time of scattering. For if, in the absence of the mirror, nature indeed
makes a choice at the time of scattering, how could this choice be undone by
subsequent addition of the mirror? Whether Dirac really foresaw this
difficulty is hard to say. In any case, precisely this point was made by
Heisenberg, and Dirac's hesitation here certainly reflects a deep difficulty
that lies at the heart of the measurement problem.

Heisenberg makes his point with disarming simplicity (p.~\pageref{Heisenberg-determinism}):

\begin{quotation}
I do not agree with Mr Dirac when he says that, in the described experiment,
nature makes a choice. Even if you place yourself very far away from your
scattering material, and if you measure after a very long time, you are able
to obtain interference by taking two mirrors. If nature had made a choice, it
would be difficult to imagine how the interference is produced.
\end{quotation}

What Heisenberg had in mind seems to have been something like the set-up shown
in Fig.~\ref{DirHeis}(c), where a pair of mirrors is placed far away from the
scattering region, causing different parts of the scattered wave to re-overlap
and interfere at R. According to Dirac's account, in the absence of any
mirrors (Fig.~\ref{DirHeis}(a)), upon detection of the particle one might say
that nature chose a specific direction at the time of scattering. Heisenberg
points out that, by placing mirrors far away (and removing the detectors at P
and P%
\'{}%
), interference may be observed a long time after the scattering took place.

While Heisenberg does not mention it explicitly, in this example the choice
between `which-way information' on the one hand, or interfering paths on the
other, may be made long after the particle has completed most of its journey
(or journeys), just as in Wheeler's delayed-choice experiment. Dirac's set-up
with no mirrors at all provides which-way information, since detection of the
particle at a point far away may be interpreted as providing information on
the direction chosen at the time of scattering (Fig.~\ref{DirHeis}(a)).
Heisenberg's modification, with the two mirrors, demonstrates interference
between alternative paths starting from the scattering region (Fig.~\ref{DirHeis}(c)). 
Unlike Wheeler, however, Heisenberg does not explicitly
emphasise that the choice of whether or not to add the mirrors could be made
at the last moment, long after the scattering takes place. On the other hand,
Heisenberg does emphasise that the measurement with the mirrors could be done
`very far away' and `after a very long time', and notes the contradiction with
nature having made a choice at the time of scattering. Thus, Heisenberg's
remarks arguably contain the essence of Wheeler's delayed-choice experiment.

Heisenberg goes on to say (p.~\pageref{Heisenberg-determinism2}) that, instead of nature
making a choice,

\begin{quotation}
I should rather say, as I did in my last paper, that the \textit{observer
himself} makes the choice, because it is only at the moment where the
observation is made that the `choice' has become a physical reality and that
the phase relationship in the waves, the power of interference, is destroyed.
\end{quotation}
From the chronology of Heisenberg's publications, here he must be
referring to his uncertainty paper (published in May 1927), in which he writes
that `all perceiving is a
choice from a plenitude of possibilities' (Heisenberg 1927, p.~197).
Heisenberg's statement above
that the observer `makes' the choice seems to be meant in the sense of the
observer `bringing about' the choice. Thus it would seem that, for Heisenberg,
a definite outcome occurs --- and there is no longer any possibility of
interference --- only when an experimenter makes an observation. Similar views
have been expressed by Wheeler (1986).

One may, however, object to this viewpoint, on the grounds that there is no
reason in principle why a more advanced being could not observe interference
between the alternative states of the detector registering interference, or,
between the alternative states of the human observer watching the detector.
(Cf. the discussion of Wigner's paradox in section~\ref{measphysproc}.) 
After all, the detector is certainly just another
physical system, built out of atoms. And as far as we can tell, human
observers can likewise be treated as physical systems built out of atoms. To
say that `the power of interference' is `destroyed' when and only when a human
observer intervenes is to make a remarkable assertion to the effect that human
beings, unlike any other physical systems, have special properties by virtue
of which they cannot be treated by ordinary physical laws but generate
deviations from those laws. As we have already mentioned, there is no evidence
that human beings are able to violate, for example, the laws of gravity or of
thermodynamics, and it would be remarkable if they were indeed able to violate
the laws of quantum physics.

It is interesting to note that, while for Heisenberg the human observer seems
to play a crucial role at the end of a quantum experiment, for Dirac the human
observer --- and his `freewill' --- seems to play a crucial role at the
beginning, in the preparation stage. For as Dirac puts it (p.~\pageref{Dirac-freewill}, 
Dirac's italics): `The disturbances that an
experimenter applies to a system to observe it are directly under his control,
and are acts of freewill by him. \emph{It is only the numbers that describe
these acts of freewill that can be taken as initial numbers for a calculation
in the quantum theory}'. (Cf. the discussion about determinism in section~\ref{det-prob}.)

Returning to Dirac's view of quantum outcomes, Heisenberg's objection
certainly causes a difficulty. If a choice --- or collapse to a particular
$\psi_{n}$ --- really does occur around the time of scattering, then a
`delayed interference experiment' of the form described by Heisenberg should
show \textit{no} interference, and Dirac's view would amount to a violation of
the quantum formalism along the lines of dynamical models of wave function
collapse (Pearle 1976, 1979, 1989; Ghirardi, Rimini and Weber 1986). And
Dirac's caveats concerning the possibility of tracing back a chain of causal
events do not lead to a really satisfactory position either. As in Feynman's
view that interference occurs only for paths that are subsequently
indistinguishable, the question is begged as to the precise definition of
subsequently distinguishable or subsequently indistinguishable paths: for in a
delayed-choice set-up, it appears to be at the later whim of the experimenter
to decide whether certain paths taken in the past are subsequently
distinguishable or not. This procedure correctly predicts the experimental
results (or statistics thereof), but it has the peculiar consequence that
whether or not there is a matter of fact about the past depends on what the
experimenter does in the present.

Finally, as discussed in section~\ref{interf-deB}, interference was 
considered from a pilot-wave perspective by de Broglie
in his report and by Brillouin in the discussion that followed. De Broglie
did not comment, however, on the exchange between Dirac and Heisenberg. From a
modern point of view it is clear that, in his theory, the particle trajectory
does take one particular route after a scattering process, while at the same
time there are portions of the scattered wave travelling along the alternative
routes. An `empty' part of the wave can be subsequently reflected by a mirror,
and if the reflected wave later reoverlaps with the part of the wave carrying
the particle, then in the interference zone the particle is indeed affected by
both components. Similarly, de Broglie's theory provides a straightforward
account of Wheeler's delayed-choice experiment, without present actions
influencing the past in any way (Bell 1980; Bell 1987, chap. 14; Bohm, Dewdney
and Hiley 1985).

\sectionmark{Further remarks on Born and Heisenberg}
\section{Further remarks on Born and Heisenberg's quantum 
mechanics}\label{Further-remarks-on-Born}\sectionmark{Further remarks on Born and Heisenberg}

As we saw in section~\ref{Interference-in-Born-and-Heisenberg}, Born and
Heisenberg's report contains some remarkable comments about the nature of
interference (in section II, `Physical interpretation'). These comments are
perhaps related to a conceptual transition that seems to occur at around this
point in their presentation. In the earlier part of their section II, Born and
Heisenberg describe a theory in which probabilistic transitions occur between
possessed values of energy; while later in the same section, in their
discussion of arbitrary observables, they emphasise probabilistic transitions
from one measurement to the next (still noting the presence of interference).
Earlier in that section they explicitly assert that a system always occupies a
definite energy state at any one time, while in the later treatment of
arbitrary observables nothing is said about whether a system always possesses
definite values or not. This is perhaps not surprising, given that the
discussion of interference (for the case of energy measurements) made it clear
that taking the quantities $\left\vert c_{n}(t)\right\vert ^{2}=\left\vert
\langle n\left\vert \psi(t)\right\rangle \right\vert ^{2}$ to be population
probabilities for energies $E_{n}$ led to a difficulty in the presence of
interference. As we saw in section~\ref{Interference-in-Born-and-Heisenberg},
Born and Heisenberg resolved the difficulty by asserting that unmeasured
population probabilities are somehow not applicable or meaningful. This does
not seem consistent with the view they expressed earlier, that atoms always
have definite energy states even when these are not measured. How could an
ensemble of atoms have definite energy states, without the energy distribution
being meaningful?

Consideration of interference, then, was likely to force a shift away from the
view that atoms are always in definite stationary states. Later, in his book
of 1930, Heisenberg did in fact explicitly deny that atoms are always in such
states. Considering again the example from his uncertainty paper, of atoms
passing through two successive regions of inhomogeneous field (see 
section~\ref{Interference-in-Born-and-Heisenberg}), Heisenberg notes that if the
energies are not actually measured in the intermediate region, then, because
of the resulting `interference of probabilities',

\begin{quotation}
it is not reasonable to speak of the atom as having been in a stationary state
between $F_{1}$ and $F_{2}$ [that is, in the intermediate region]. (Heisenberg
1930b, p.~61)
\end{quotation}

As we have already noted, again in section~\ref{Interference-in-Born-and-Heisenberg}, 
Born and Heisenberg's discussion of interference seems to dispense
with the standard collapse postulate for quantum states. Upon performing an
energy measurement, instead of the usual collapse to a single energy
eigenstate $\left\vert n\right\rangle $, Born and Heisenberg in effect replace
the superposition $\sum_{n}\left\vert c_{n}(0)\right\vert e^{i\gamma_{n}%
}\left\vert n\right\rangle $ by a similar expression with randomised phases
$\gamma_{n}$. And the justification given for this appears to be some form of
uncertainty relation or complementarity between population probabilities and
phases: if the former have been measured, then the latter are ill-defined, 
and vice versa. On this view, the definite phase
relationships associated with interference preclude the possibility of
speaking of a well-defined population probability, so that the usual formulas
of probability calculus cannot be properly applied; on the other hand, if the
population probability has been measured experimentally, then the phases are
ill-defined, and averaging over the random phases destroys interference.

One crucial point is not entirely clear, however. Was the phase randomisation
thought to occur only upon measurement of energy, or upon measurement of any
arbitrary observable?

The phase randomisation explicitly appealed to by Born and Heisenberg takes
the following form: for a quantum state%
\begin{equation}
\left\vert \psi(t)\right\rangle =\sum_{E}\left\vert E\right\rangle \langle
E\left\vert \psi(t)\right\rangle =\sum_{E}\left\vert E\right\rangle \langle
E\left\vert \psi(0)\right\rangle e^{-iEt}%
\end{equation}
an energy measurement induces a random change in each phase factor
$e^{-iEt}\rightarrow e^{-iEt}e^{i\gamma(E)}$, where each $\gamma(E)$ is random
on the unit circle. This procedure might be generalised to, for example,
measurements of position, as follows: for a state%
\begin{equation}
\left\vert \psi\right\rangle \propto\sum_{\mathbf{x}}\left\vert \mathbf{x}%
\right\rangle \langle\mathbf{x}\left\vert \psi\right\rangle \propto
\sum_{\mathbf{x}}\left(  \sum_{\mathbf{p}}\left\vert \mathbf{p}\right\rangle
e^{-i\mathbf{p}\cdot\mathbf{x}}\right)  \langle\mathbf{x}\left\vert
\psi\right\rangle
\end{equation}
(writing as if $\mathbf{x}$ and $\mathbf{p}$ were discrete, for simplicity)
one might suppose that a position measurement induces a random change
$e^{-i\mathbf{p}\cdot\mathbf{x}}\rightarrow e^{-i\mathbf{p}\cdot\mathbf{x}%
}e^{i\gamma(\mathbf{x})}$, resulting in a state%
\begin{equation}
\sum_{\mathbf{x}}\left\vert \mathbf{x}\right\rangle \langle\mathbf{x}%
\left\vert \psi\right\rangle e^{i\gamma(\mathbf{x})}%
\end{equation}
with random relative phases. Averaging over the random phases $\gamma
(\mathbf{x})$ would then destroy interference between different positions,
just as in the case of energy measurements.

However, we have found no clear evidence that Born or Heisenberg considered
any such generalisation. It then seems possible that the phase randomisation
argument for the suppression of interference was to be applied to the case of
energy measurements only. On the other hand, there is a suggestive remark by
Heisenberg in the general discussion (p.~\pageref{Heisenberg-determinism2}), quoted in
the last section. When expressing his view that definite outcomes occur only
when an experimenter makes an observation, Heisenberg refers to an example
where position measurements are made at the end of a scattering process (see
Fig.~\ref{DirHeis}(c)), and he states that it is only when the observation is
made that `the phase relationship in the waves, the power of interference, is
destroyed'. This might be read as suggesting that the waves continue to exist,
but may or may not have the capacity to interfere --- depending on whether or
not the phase relations have been randomised by the position measurement. If
Heisenberg did take such a view, his use of wave packet reduction for position
measurements in the uncertainty paper would have to be interpreted as some
sort of effective description.

Even in the case of energy measurements, the status of the phase randomisation
argument is not clear. After all, Born and Heisenberg assert that the
time-dependent Schr\"{o}dinger equation itself (which they use to discuss
interference) is only phenomenological, and applicable to subsystems only.
Fundamentally, they have a time-independent theory for a closed system.
Presumably, the phase randomisation for energy measurements was also seen as
phenomenological only, with the measured system being treated as a subsystem.

As we saw in section~\ref{DiracandHeisenberg}, in the general
discussion Dirac describes what is recognisably the process of wave packet
reduction. Born and Heisenberg, in contrast, seem to speak only of the
ordinary reduction (or conditionalisation) of `probability functions' 
as it appears in standard probability
calculus. As they put it, near the end of section III of their report 
(p.~\pageref{BHprobfunc}):

\begin{quotation}
the result of each measurement can be
expressed by the choice of appropriate initial values for probability
functions .... . Each new experiment replaces the probability functions valid
until now with new ones, which correspond to the result of the observation
.... .
\end{quotation}
On the other hand, Born and Heisenberg's contributions at the Solvay
conference do not seem sufficiently clear or complete to warrant definite
conclusions as to what they believed concerning the precise relationship
between probabilities and the wave function; sometimes it is unclear whether
or not they mean to draw a distinction between probability distributions on
the one hand and wave functions on the other.

It may be that, in October 1927, Born and Heisenberg had in some respects not
yet reached a definitive point of view, perhaps partly because of the
different perspectives that Born and Heisenberg each brought to the subject.
Born's recent thinking (in 1926) had been influenced by Einstein's idea of a
guiding field, while Heisenberg's recent thinking (in his uncertainty paper)
had been influenced by the operational approach to physics.

Concerning the question of wave packet collapse, it should also be remembered
that Pauli seems to have played an important role in Born and Heisenberg's
thinking at the time. In particular, as we saw in Born's discussion of the
cloud chamber (section~\ref{QM-without-collapse}), Pauli had
been critical of Heisenberg's use of the reduction of the wave packet in the
uncertainty paper (in a discussion of classical electron orbits), and Born ---
who by his own account was following Pauli's suggestion --- tried to show that
such reduction was unnecessary.

\setcounter{endnote}{0}
\setcounter{equation}{0}

\chapter{Locality and incompleteness}\label{locality-and-incompleteness}\chaptermark{Locality and incompleteness}

\section{Einstein's 1927 argument for incompleteness}\label{1927-incompleteness}

A huge literature arose out of the famous `EPR' paper by Einstein, Podolsky
and Rosen (1935), entitled `Can quantum-mechanical description of physical
reality be considered complete?'. The EPR paper argued, on the basis of (among
other things) the absence of action at a distance, that quantum theory must be
incomplete.\footnote{Note that, as pointed out by Fine (1986) and discussed
further by Howard (1990), the logical structure of the EPR paper (which was
actually written by Podolsky) is more complicated and less direct than
Einstein had intended.} It is less well-known that a much simpler argument,
leading to the same conclusion, was presented by Einstein eight years earlier
in the general discussion at the fifth Solvay conference (pp.~\pageref{Einstein-disc}~ff.).

Einstein compares and contrasts two views about the nature of the wave
function $\psi$, for the specific case of a single electron. According to view
I, $\psi$ represents an ensemble (or `cloud') of electrons; while according to
view II, $\psi$ is a complete description of an individual electron. Einstein
argues that view II is incompatible with locality, and that to avoid this, in
addition to $\psi$ there should exist a localised particle (along the lines of
de Broglie's theory). Thus, according to this reasoning, if one assumes
locality, then quantum theory (as normally understood today) is incomplete.

The conclusion of Einstein's argument in 1927 is the same as that of EPR in
1935, even if the form of the argument is rather different. Einstein considers
electrons striking a screen with a small hole that diffracts the electron
wave, which on the far side of the screen spreads out uniformly in all
directions and strikes a photographic film in the shape of a hemisphere with
large radius (see Einstein's figure
). Einstein's argument against view II is then as follows:

\begin{quotation}
If $\left\vert \psi\right\vert ^{2}$ were simply regarded as the probability
that at a certain point a given particle is found at a given time, it could
happen that \textit{the same} elementary process produces an action \textit{in
two or several} places on the screen. But the interpretation, according to
which $\left\vert \psi\right\vert ^{2}$ expresses the probability that
\textit{this} particle is found at a given point, assumes an entirely peculiar
mechanism of action at a distance, which prevents the wave continuously
distributed in space from producing an action in \textit{two} places on the screen.
\end{quotation}

The key point here is that, if there is no action at a distance, and if the
extended field $\psi$ is indeed a complete description of the physical
situation, then if the electron is detected at a point $P$ on the film, it
could happen that the electron is also detected at another point $Q$, or
indeed at any point where $\left\vert \psi\right\vert ^{2}$ is non-zero. Upon
detection at $P$, it appears that a `mechanism of action at a distance'
prevents detection elsewhere.

Einstein's argument is so concise that its point is easily missed, and one
might well dismiss it as arising from an elementary confusion about the nature
of probability. (Indeed, Bohr comments that he does not `understand what
precisely is the point' Einstein is making.) For example, it might be thought
that, since we are talking about a probability distribution for just one
particle, it is a matter of pure logic that only one detection can
occur.\footnote{Cf.\ Shimony (2005).} But this would be to beg the question
concerning the nature of $\psi$. Einstein's wording above attempts to convey a
distinction between probability for a `given' particle (leading to the
possibility of multiple detections) and probability for `\textit{this}'
particle (leading to single detection only). The wording is not such as to
convey the distinction very clearly, perhaps indicating an inadequate
translation of Einstein's German into French.\footnote{Unfortunately, the full
German text of Einstein's contribution to the general discussion seems to have
been lost; the Einstein archives contain only a fragment, consisting of just
the first four paragraphs (AEA 16-617.00).} But from the context, the words `probability that
\textit{this} particle is found' are clearly being used to express the
assumption that in this case $\psi$ indeed expresses the probability for just
\textit{one} particle detection.

\begin{figure}
    \centering
     \resizebox{\textwidth}{!}{\includegraphics[0mm,0mm][200.30mm,160.98mm]{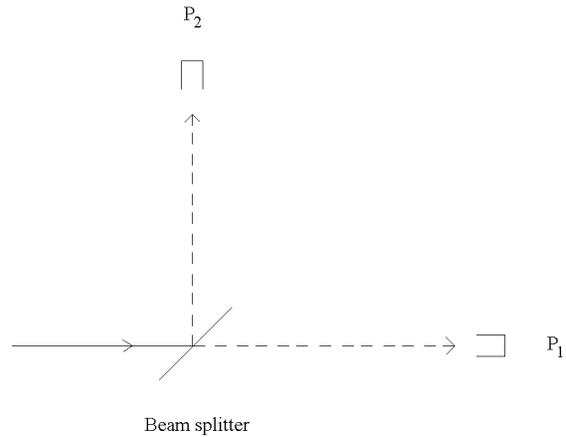}}
     \caption{Hardy's simplified version of Einstein's argument.}%
     \label{SolHardy}%
  \end{figure}

As shown by Hardy (1995), Einstein's argument may be readily put into the same
rigorous form as the later EPR argument. (See Norsen (2005) for a careful and
extensive discussion.) Hardy simplifies Einstein's example, and considers a
single particle incident on a beam splitter (Fig.~\ref{SolHardy}), so that
there are only two points $P_{1}$, $P_{2}$ at which the particle might be
detected. One may then adopt the following sufficient condition, given by EPR,
for the existence of an element of reality:

\begin{quotation}
If, without in any way disturbing a system, we can predict with certainty
(i.e., with probability equal to unity) the value of a physical quantity, then
there exists an element of physical reality corresponding to this physical
quantity. (Einstein, Podolsky and Rosen 1935, p.~777)
\end{quotation}
Now, if a detector is placed at $P_{1}$, either it will fire or it will
not. In either case, from the state of the detector at $P_{1}$ one could
deduce with certainty whether or not a detector placed at $P_{2}$ would fire.
Such deductions could be made for any individual run of the experiment. Even
though the outcome at $P_{1}$ cannot be predicted in advance, in each case the
outcome allows us to infer the existence of a definite element of reality at
$P_{2}$. If locality holds, an element of reality at $P_{2}$ cannot be
affected by the presence or absence of a detector at $P_{1}$. Therefore, even
if no detector is placed at $P_{1}$, there must still be an element of reality
at $P_{2}$ corresponding to detection \textit{or} no detection at $P_{2}$.
Since $\psi$ is a superposition of detection \textit{and} no detection at
$P_{2}$, $\psi$ contains nothing corresponding to the deduced element of
reality at $P_{2}$. Therefore, $\psi$ is not a complete description of a
single particle.\footnote{Note that in this argument the incompleteness of
quantum theory is \textit{inferred} from the assumption of locality. Cf. Bell
(1987, p.~143).}%

Thus, `the essential points in the EPR argument had already been made by
Einstein some eight years earlier at the fifth Solvay conference' (Hardy 1995,
p.~600).

Einstein concludes that:

\begin{quotation}
In my opinion, one can remove this objection only in the following way, that
one does not describe the process solely by the Schr\"{o}dinger wave, but that
at the same time one localises the particle during the propagation. I think
that Mr de Broglie is right to search in this direction. If one works solely
with the Schr\"{o}dinger waves, interpretation II of $\left\vert
\psi\right\vert ^{2}$ implies to my mind a contradiction with the postulate of relativity.
\end{quotation}
In other words, for Einstein, action at a distance can be avoided only
by admitting that the wave function is incomplete.

According to Einstein's argument, quantum theory is either nonlocal or
incomplete. For the rest of his life, Einstein continued to believe that
locality was a fundamental principle of physics, and so he adhered to the view
that quantum theory must be incomplete. However, further reasoning by Bell
(1964) showed that any completion of quantum theory would still require
nonlocality, in order to reproduce the details of quantum correlations for
entangled states (assuming the absence of backwards causation or of many
worlds\footnote{Bell's argument assumes that there is no common cause between
the hidden variables (defined at the time of preparation) and the settings of
the measuring apparatus. It also assumes that there is no backwards causation,
so that the hidden variables are not affected by the future outcomes or
apparatus settings. Further, the derivation of the Bell inequalities assumes
that a quantum measurement has only one outcome, and therefore does not apply
in the many-worlds interpretation.}). It then appears that, whether complete
or incomplete, quantum theory is necessarily nonlocal, a conclusion that would
surely have been deeply disturbing to Einstein.

It is ironic that Einstein's (and EPR's) argument started out by holding
steadfast to locality and deducing that quantum theory is incomplete. But then
the argument, as carried further by Bell, led to a contradiction between
locality and quantum correlations, so that in the end one fails to establish
incompleteness and instead establishes nonlocality (with completeness or
incompleteness remaining an open question).

\section{A precursor: Einstein at Salzburg in 1909}\label{Einstein-at-Salzburg}

In September 1909, at a meeting in Salzburg, Einstein gave a lecture entitled
`On the development of our views concerning the nature and constitution of
radiation' (Einstein 1909). Einstein summarised what he saw as evidence for
the dual nature of radiation: he held that light had both particle and wave
aspects, and argued that classical electromagnetic theory would have to be
abandoned. It seems to have gone unnoticed that one of Einstein's arguments at
Salzburg was essentially the same as the argument he presented at the 1927
Solvay conference (though applied to light quanta instead of to electrons).

Einstein began his lecture by noting that the phenomena of interference and
diffraction make it plain that, at least in some respects, light behaves like
a wave. He then went on to describe how, in other respects, light behaves as
if it consisted of particles. In experiments involving the photoelectric
effect, it had been found that the velocity of the photoelectrons was
independent of the radiation intensity. According to Einstein, this was more
consistent with `Newton's emission theory of light' than with the wave theory.
Einstein also discussed pressure fluctuations in blackbody radiation, and
showed that these contained two terms, which could be naturally identified as
contributions from particle-like and wave-like aspects of the radiation.%

  \begin{figure}
    \centering
     \resizebox{\textwidth}{!}{\includegraphics[0mm,0mm][200.30mm,140.98mm]{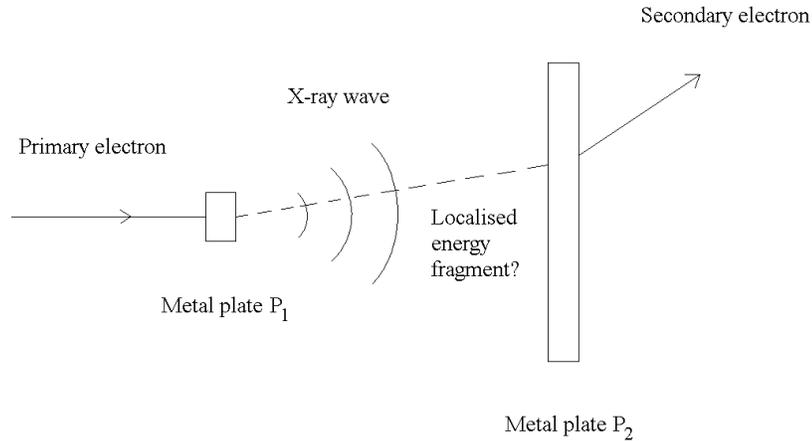}}
     \caption{Figure based on Einstein's 1909 argument for the existence of localised light quanta.
     Assuming the principle of local action, the delocalised X-ray wave can produce
     an electron (of energy comparable to that of the primary electron) in a small region of 
     the second plate only if, in addition to the
     wave, there is a localised energy fragment propagating in space from
     $\mathrm{P}_{1}$ to $\mathrm{P}_{2}$.}%
     \label{SolEin1909}%
  \end{figure} 

Of special interest here is another argument Einstein gave for the existence
of localised light quanta. Einstein considered a beam of electrons (`primary
cathode rays') incident upon a metal plate $\mathrm{P}_{1}$ and producing
X-rays (see Fig.~\ref{SolEin1909}). These X-rays, in turn, strike a second
metal plate $\mathrm{P}_{2}$ leading to the production of electrons
(`secondary cathode rays') from $\mathrm{P}_{2}$. Experimentally, it had been
found that the velocity of the secondary electrons had the same order of
magnitude as the velocity of the primary electrons. Further, the available
evidence suggested that the velocity of the secondary electrons did not depend
at all on the distance between the plates $\mathrm{P}_{1}$ and $\mathrm{P}%
_{2}$, or on the intensity of the primary electron beam, but only on the
velocity of the primary electrons. Assuming this to be strictly true, Einstein
then asked what would happen if the primary intensity were so small, or the
area of the plate $\mathrm{P}_{1}$ so small, that one could consider just one
electron striking the plate, as in Fig.~\ref{SolEin1909}. According to Einstein,

\begin{quotation}
we will have to assume that on $\mathrm{P}_{2}$ (as a result of the impinging
of the above electron on $\mathrm{P}_{1}$) either nothing is being produced or
that a secondary emission of an electron occurs on it with a velocity of the
same order of magnitude as of the electron impinging on $\mathrm{P}_{1}$. In
other words, the elementary radiation process seems to proceed such that it
does not, as the wave theory would require, distribute and scatter the energy
of the primary electron in a spherical wave propagating in all directions.
Rather, it seems that at least a large part of this energy is available at
some location of $\mathrm{P}_{2}$ or somewhere else. (Einstein 1909, English
translation, p.~388)
\end{quotation}

Einstein's argument, then, is that according to the wave theory the point of
emission of the X-ray from the first plate must be the source of waves
spreading out in space, waves whose amplitude will spread over the region
occupied by the second plate. And yet, in the second plate, \textit{all} the
energy of the X-ray becomes concentrated in the vicinity of a single point,
leading to the production of an electron with velocity comparable to that of
the primary electron.\footnote{Cf.\ Compton's report, p.~\pageref{Comp53}: 
`It is clearly impossible that all the energy of an X-ray pulse which has
spread out in a spherical wave should spend itself on this [small region]'.}
Einstein concluded from this that, in addition to the wave spreading from the
point of emission, there seems also to be a localised energy fragment
propagating from the point of emission of the X-ray wave to the point of
production of the secondary electron. As Einstein put it: `the elementary
process of radiation seems to be \textit{directed}'.

Now, Einstein's argument of 1909 implicitly assumes a principle of local
action, similar to that explicitly assumed in his published critique of
quantum theory at the 1927 Solvay conference. Because the distance between the
plates $\mathrm{P}_{1}$ and $\mathrm{P}_{2}$ can be arbitrarily large, the
wave impinging on $\mathrm{P}_{2}$ can be spread over an arbitrarily large
area. The production of an electron in a highly localised region of
$\mathrm{P}_{2}$ can then be accounted for only if, in addition to the
delocalised wave, there is a localised energy fragment propagating through
space --- for otherwise, there would have to be some mechanism by means of
which energy spread out over arbitrarily large regions of space suddenly
becomes concentrated in the neighbourhood of a single point.

It should be quite clear, then, that Einstein's 1927 argument for the
existence of localised electrons (accompanying de Broglie-Schr\"{o}dinger
waves) was identical in form to one of his 1909 arguments for the existence of
localised light quanta (accompanying electromagnetic waves). In his 1927
argument, the small hole in the screen (see his figure) acts as a source for
an electron wave, which spreads over the area of the photographic film ---
just as, in the 1909 argument, the point where the primary electron strikes
the first plate acts as a source for an X-ray wave, which spreads over the
area of the second plate. Both arguments depend crucially on the assumption
(implicit in 1909, explicit in 1927) that there is no action at a distance.

As we shall discuss further in chapter~\ref{guiding-fields-in-3-space},
by 1927 Einstein had already spent over twenty years trying to reconcile
localised energy quanta --- which he had postulated in 1905 --- with the wave
aspect of radiation. And for much of that time, he had been more or less alone
in his belief in the existence of such quanta. It is then perhaps not so
surprising that at the 1927 Solvay meeting Einstein was able to raise such a
penetrating critique of the view that the wave function is a complete
description of a single electron: from his long and largely solitary
experience pondering the wave-particle duality of light, Einstein could
immediately see that, in the analogous case of electron waves, the principle
of local action entailed the existence of localised particles moving through
space, in addition to the wave function.

The meeting in Salzburg took place four years before Bohr published his model
of the atom. After 1913, one might have simply rephrased Einstein's argument
in terms of atomic transitions. Consider an atom A that makes a transition
from an initial stationary state with energy $E_{i}$ to a final stationary
state with lower energy $E_{f}$. At a later time, the energy $E_{i}-E_{f}$
lost by atom A may be \textit{wholly} absorbed by an arbitrarily distant atom
B, if there exists an appropriate transition from the initial state of B to a
final state corresponding to an energy increase $E_{i}-E_{f}$. This process
may seem unmysterious, if one imagines atom A emitting a photon, or `localised
energy quantum', which somehow propagates through space from A to B. However,
if one tries to make do without the photon concept, and represents the
electromagnetic field in terms of (classical) waves only --- which spread out
in all directions from A --- then it is hard to understand how the energy lost
by A may be wholly transferred to B: instead, one would expect the energy to
spread out in space like the waves themselves, so that the energy density
becomes diluted.

We have laboured this point because the power of Einstein's simple argument
seems to have been generally missed, not only in 1909, but also in 1927, and
for decades afterwards. Indeed, it appears that Einstein's point did not start
to become widely appreciated until the late twentieth century (see, again,
Norsen (2005)).

In retrospect, it seems quite puzzling that Einstein's simple argument should
have taken so long to be understood. A perhaps related puzzle, emphasised by
Pais (1982, pp.~382--6), is why Einstein's light-quantum hypothesis itself
should have been largely ignored by so many physicists until the advent of the
Compton effect in 1923. Even after Millikan's experimental confirmation of
Einstein's photoelectric equation in 1916, `almost no one but Einstein himself
would have anything to do with light-quanta' (Pais 1982, p.~386).\footnote{Though 
according to Brillouin's recollections of 1962, the
situation was rather different in France, where Einstein's light quantum was
accepted (by Langevin, Perrin and Marie Curie) much earlier than it was
elsewhere (Mehra and Rechenberg 1982a, p.~580).}

We do not wish to suggest, of course, that Einstein's locality argument should
today be regarded as establishing the existence of localised photons: for the
implicit premise of Einstein's argument --- the principle of locality ---
today seems to be ruled out by Bell's theorem. Our point, rather, is that
prior to Bell (and certainly in 1909) Einstein's arguments were indeed
compelling and should have been taken more seriously.

\section{More on nonlocality and relativity}\label{morelocrel}

At the end of his long contribution to the general discussion (in which he
argued for the incompleteness of quantum theory), Einstein objected to the
multi-dimensional representation in configuration space, on the grounds that
(p.~\pageref{featureofforces})

\begin{quotation}
.... the feature of forces of acting only at small \textit{spatial} distances
finds a less natural expression in configuration space than in the space of
three or four dimensions.
\end{quotation}
As Einstein himself stated, he was here adding another argument against
what he called view II (the view that $\psi$ is a complete description of an
individual system), a view that he claimed is `essentially tied to a
multi-dimensional representation (configuration space)'.

Einstein's point seems to be that, if physics is fundamentally grounded in
configuration space, there will be no reason to expect physics to be
characterised by \textit{local} action. This objection should be seen in the
context of Einstein's concerns, in the period 1926--27, over the
nonseparability of Schr\"{o}dinger's wave mechanics for many-body systems
(Howard 1990, pp.~83--91; cf.\ section~\ref{historicalrevision}).

A certain form of classical locality survives, of course, in modern quantum
theory and quantum field theory, in the structure of the Hamiltonian or
Lagrangian, a structure that ensures the absence of controllable nonlocal
signals at the statistical level. But even so, we understand today that, in
fact, quantum physics is characterised by nonlocality. And the nonlocality may
indeed be traced to the fact that, unlike classical theory, quantum theory is
not grounded in ordinary three-dimensional space.

The setting for standard quantum theory is Hilbert space, whose tensor-product
structure allows for entanglement and associated nonlocal effects. In the
pilot-wave formulation of quantum theory, the setting is configuration space
(in which the pilot wave propagates), and in general the motions of
spatially-separated particles are nonlocally connected. Further, Bell's
theorem shows that, if we leave aside backwards causation or many worlds, then
quantum theory is in some sense nonlocal under any interpretation or
formulation. As Ballentine once pointed out, while discussing the significance
of Bell's theorem:

\begin{quotation}
Perhaps what is needed is not an explanation of nonlocality, but an
explanation of locality. Why, if locality is not true, does it work so well in
so many different contexts? (Ballentine 1987, pp.~786--7)
\end{quotation}

Einstein's fear, that there would be difficulties with locality in quantum
physics, has certainly been borne out by subsequent developments. In standard
quantum theory, there appears to be a peaceful but uneasy `coexistence' with
relativity. While from a pilot-wave (or more generally, from a deterministic
hidden-variables) point of view, statistical locality appears as an accidental
feature of the `quantum equilibrium' state (Valentini 1991b, 2002a).

In his main contribution to the general discussion Dirac (p.~\pageref{abandonment}) 
also notes that `the general theory of the wave
function in many-dimensional space necessarily involves the abandonment of
relativity', but he suggests that this problem might be solved by `quantising
3-dimensional waves' (that is, by what we would now call quantum field
theory). And de Broglie in his report, when considering the pilot-wave
dynamics of many-body systems, notes that unlike in the case of a single
particle `it does not appear easy to find a wave $\Psi$ that would define the
motion of the system taking Relativity into account' (p.~\pageref{relativityaccount}), 
a difficulty that has persisted in pilot-wave theory right up to the
present day (see, for example, Berndl \textit{et al}. (1996)).

In 1927, then, there was a fairly broad recognition that the fundamental use
of configuration space did not bode well for consistency with relativity.

\setcounter{endnote}{0}
\setcounter{equation}{0}

\chapter{Time, determinism, and the spacetime framework}\label{Time-determinism-spacetime}%
\chaptermark{Time, determinism, and spacetime framework}

\section{Time in quantum theory}\label{Time-in-quantum-theory}

By 1920, the spectacular confirmation of general relativity, during the solar
eclipse of 1919, had made Einstein a household name. Not only did relativity
theory (both special and general) upset the long-received Newtonian ideas of
space and time, it also stimulated a widespread `operationalist' attitude to
physical theories. Physical quantities came to be seen as inextricably
interwoven with our means of measuring them, in the sense that any limits on
our means of measurement were taken to imply limits on the definability, or
`meaningfulness', of the physical quantities themselves. In particular,
Einstein's relativity paper of 1905 --- with its operational analysis of
simultaneity --- came to be widely regarded as a model for the new
operationalist approach to physics.

Not surprisingly, then, as the puzzles continued to emerge from atomic
experiments, in the 1920s a number of workers suggested that the concepts of
space and time would require still further revision in the atomic domain.
Thus, Campbell (1921, 1926) suggested that the puzzles in atomic physics could
be removed if the concept of time was given a purely statistical significance:
`time, like temperature, is a purely statistical conception, having no meaning
except as applied to statistical aggregates' (quoted in Beller 1999, p.~97).
In an operationalist vein, Campbell considered `clocks' based on (random)
radioactive decays. He suggested that it might be possible to construct a
theory that did not involve time at all, and in which `all the experiments on
which the prevailing temporal conceptions are based can be described in terms
of statistics' (quoted in Beller 1999, p.~98). On the other hand, Senftleben (1923)
asserted that Planck's constant $h$ set limits to the definability of the
concepts of space and time, and concluded that spacetime must be
discontinuous. According to Beller (1999, pp.~96--101), both Campbell and
Senftleben had a significant influence on Heisenberg in his formulation of the
uncertainty principle.\footnote{For Campbell's influence on Bohr's formulation
of complementarity, see Beller (1999, pp.~135--7) and also Mehra and
Rechenberg (2000, pp.~189--90).} Certainly, in a letter to Pauli of 28 October
1926, Heisenberg expresses views very similar to Campbell's (Pauli 1979, p.~350):

\begin{quotation}
I have for all that a hope in a later solution of more or less the following
kind (but one should not say something like this aloud): that space and time
are really only statistical concepts, such as, say, temperature, pressure etc.
in a gas. I mean that spatial and temporal concepts are meaningless for
\textit{one} corpuscle and that they make more and more sense the more
particles are present. I often try to get further in this direction, but until
now it will not work.
\end{quotation}

Be that as it may, at the 1927 Solvay conference, in their report on quantum
mechanics, Born and Heisenberg seem to express the remarkable view that
temporal changes do not occur at all for closed systems, and that the
time-dependent Schr\"{o}dinger equation emerges only as an effective and
approximate description for subsystems. How these views related to Campbell's,
or indeed if they did at all, is not clear.

In their section II, `Physical interpretation', Born and Heisenberg begin with
the following statement (p.~\pageref{noticeable}):

\begin{quotation}
The most noticeable defect of the original matrix mechanics consists in 
the fact that at first it appears to give information not about actual phenomena, 
but rather only about possible states and processes. .... it says nothing about 
when a given state is present .... matrix mechanics deals only with closed 
periodic systems, and in these there are indeed no changes. In order to have 
true processes .... one must direct one's attention to a {\em part} of the 
system .... .
\end{quotation}

From a modern point of view, the original matrix mechanics did not contain the
notion of a general state $\left\vert \Psi\right\rangle $ for a system (not
even a static, Heisenberg-picture state). The only states that appeared in the
theory were the stationary states $\left\vert E_{i}\right\rangle $ (cf.\ 
chapter~\ref{BornEss}). Even as regards stationary states, there seems
to have been no notion of \textit{initial} state (`it says nothing about when a given state is present').
Instead, the matrices provided a collective representation of all the energy
eigenstates of a closed system with Hamiltonian $H$. In modern notation, the
matrices consisted of matrix elements of (Heisenberg-picture) observables
$\Omega(t)=e^{(i/\hbar)Ht}\Omega(0)e^{-(i/\hbar)Ht}$ in the energy basis:%
\begin{equation}
\left\langle E_{i}\right\vert \Omega(t)\left\vert E_{j}\right\rangle
=\left\langle E_{i}\right\vert \Omega(0)\left\vert E_{j}\right\rangle
e^{(i/\hbar)(E_{i}-E_{j})t}\ .
\end{equation}
As noted in chapter~\ref{BornEss}, the formal mathematics of matrix
mechanics then seems to represent an atomic system somewhat in the manner of
the Bohr-Kramers-Slater (BKS) theory (cf.\ chapter~\ref{guiding-fields-in-3-space}), 
with each matrix element corresponding to a virtual oscillator of
frequency $\nu_{ij}=(E_{i}-E_{j})/h$.

The matrix formalism, without a notion of initial state, amounts to a static
description.\footnote{Even if one added a notion of initial state, because the
only allowed states are energy eigenstates the description of a closed system
would still be static.} However, Born and Heisenberg add an intuitive physical
picture to the formalism, to the effect that a subsystem of a larger (closed)
system is in fact in one stationary state at any one time and performs random,
indeterministic `quantum jumps' between such states (cf.\ chapter~\ref{BornEss}).

Born and Heisenberg then go on to say that `[t]he clumsiness of the matrix theory in the description of 
processes developing in time can be avoided' (p.~\pageref{clumsiness}) by
introducing what we would now call the time-dependent Schr\"{o}dinger
equation. Here, it might appear that their view is that the mentioned 
`defect of the original matrix mechanics' is removed by generalisation to a time-dependent theory.
However, they add (p.~\pageref{consideration}) that:

\begin{quotation}
Essentially, the introduction of time as a numerical variable reduces to
thinking of the system under consideration as coupled to another one and
neglecting the reaction on the latter. But this formalism is very convenient
.... .
\end{quotation}
These words give, instead, the impression that the time-dependent
theory is regarded as only emergent in some approximation; the time-dependent
Schr\"{o}dinger equation seems to have no fundamental status.

Even so, this `convenient' formalism `leads to a
further development of the statistical view'. They include a time-dependent
external perturbation in the (time-dependent) Schr\"{o}dinger equation, and
show how to calculate the time development of any initial wave function. Born
and Heisenberg then argue that, following Bohr's original (1913) theory of
stationary states, a system can be in only one energy eigenstate at any one
time, leading to the interpretation of a superposition $\left\vert
\Psi\right\rangle =\sum_{n}c_{n}(t)\left\vert E_{i}\right\rangle $ as a
statistical mixture, with state probabilities $\left\vert c_{n}\right\vert
^{2}$. The time evolution of the wave function then describes transition
probabilities from initial to final stationary states. This might seem clear
enough, but a difficulty is then raised concerning the interpretation of a
case where the initial wave function is already a superposition, resulting in
`interference of probabilities' at later times (see section~\ref{Interference-in-Born-and-Heisenberg}).

Let us now consider what Schr\"{o}dinger had to say, in his report on wave
mechanics, concerning time in quantum theory. (De Broglie's report does not
contain any special remarks on this subject.) In his report Schr\"{o}dinger
first presents (or derives from a variational principle) what we would now
call the time-independent Schr\"{o}dinger equation for a nonrelativistic
many-body system with coordinates $q_{1}, q_{2}, ... , q_{n}$. After
noting that the eigenfunctions $\psi_{k}$, with eigenvalues $E_{k}$, may be
identified with Bohr's stationary states, Schr\"{o}dinger addresses the
question of time. He first points out (p.~\pageref{Schr-content}) that the
time-independent theory might be regarded as sufficient, providing as it does
a description of stationary states, together with expressions for jump
probabilities between them:
  \begin{quotation}
    One can take the view that one should 
    be content in principle with what has been said so far .... . The single stationary states of Bohr's theory 
    would then in a way be described by the eigenfunctions $\psi_k$, which 
    {\em do not contain time at all}. One .... can form from them .... quantities that can be aptly taken to be 
    {\em jump probabilities} between the single stationary states.
  \end{quotation}
Here, the jump probabilities are to be obtained from matrix elements
such as (in modern notation)%
\begin{equation}
\left\langle k\right\vert Q_{i}\left\vert k^{\prime}\right\rangle =\int
dq\ q_{i}\psi_{k}^{\ast}\psi_{k^{\prime}}%
\end{equation}
which can all be calculated from the eigenfunctions $\psi_{k}$.

Schr\"{o}dinger suggests further that interacting systems could be treated in
the same way, by regarding them as one single system.

Schr\"{o}dinger then goes on to discuss this point of view and its relation
with the ideas of Campbell (p.~\pageref{Sch-Campbell}):

\begin{quotation}
On this view  the {\em time variable} would play absolutely no role
in an isolated system --- a possibility to which N. Campbell .... has recently pointed.  
Limiting our attention to an isolated system, we would not perceive the passage of time in it any more
than we can notice its possible progress in space .... . What we would notice would be
merely a sequence of discontinuous transitions, so to speak a cinematic image, 
but without the possibility of
comparing the time intervals between the transitions.
\end{quotation}
According to these ideas, then, time does not exist at the level of
isolated atomic systems, and our usual (macroscopically-defined) time emerges
only from the statistics of large numbers of transitions between stationary
states. As Schr\"{o}dinger puts it:

\begin{quotation}
Only secondarily, and in fact with increasing precision 
the more extended the system, would a {\em statistical} definition of time result from 
{\em counting} the transitions taking place (Campbell's `radioactive clock').
Of course then one cannot understand the jump probability in the usual way 
as the probability of a transition calculated relative to unit time.
Rather, a {\em single} jump probability is then utterly meaningless; only
with {\em two} possibilities for jumps, the probability that the one 
may happen {\em before} the other is equal to {\em its} jump probability 
divided by the sum of the two.
\end{quotation}
Schr\"{o}dinger claims that this is the only consistent view in a
theory with quantum jumps, asserting that `[e]ither all changes in nature 
are discontinuous or not a single one'.

Having sketched a timeless view of isolated systems with discrete quantum
jumps, Schr\"{o}dinger states that such a discrete viewpoint `still poses great difficulties', and he
goes on to develop his own theory of time-dependent quantum states, in which
(continuous) time evolution does play a fundamental role even at the level of
a single atomic system. Here, a general time-dependent wave function
$\psi(q,t)$ --- a solution of the time-dependent Schr\"{o}dinger equation,
with arbitrary initial conditions --- is regarded as the description of the
continuous time development of a single isolated system.

From a contemporary perspective, it is clear that quantum theory as we know it
today is rather less radical than some expected it to be in the 1920s,
especially concerning the concepts of space and time. Both nonrelativistic
quantum mechanics, and relativistic quantum field theory, take place on a
classical spacetime background; time and space are continuous and
well-defined, even for closed systems. The evolution operator $U(t,t_{0})$
provides a continuous time evolution $\left\vert \Psi(t)\right\rangle
=U(t,t_{0})\left\vert \Psi(t_{0})\right\rangle $ for any initial quantum
state, with respect to an `external' time parameter $t$ (even in quantum field
theory, in a given inertial frame). While Schr\"{o}dinger's and de Broglie's
interpretations of the wave function did not gain widespread acceptance, their
view of time in quantum theory coincides with the one generally accepted
today.\footnote{We are of course referring here to the standard theories as
presented in textbooks. The literature contains a number of proposals, along
operational lines, calling for a `quantum spacetime' that incorporates
quantum-theoretical limits on the construction of rods and clocks. A
statistical approach to causal structure, somewhat reminiscent of Campbell's
statistical view of time, has recently been proposed by Hardy (2005).}

In contrast, the views on time expressed by Born and Heisenberg are somewhat
reminiscent of views put forward by some later workers in the context of
canonical quantum gravity. There, the wave functional $\Psi\lbrack
^{(3)}\mathcal{G}]$ on the space of 3-geometries $^{(3)}\mathcal{G}$ contains
no explicit time parameter, and obeys a `timeless' Schr\"{o}dinger equation
$\mathcal{H}\Psi=0$ (the Wheeler-DeWitt equation, with Hamiltonian density
operator $\mathcal{H}$). It is claimed that `time' emerges only
phenomenologically, through the analysis of interaction with quantum clocks,
or by the extraction of an effective time variable from the 3-metric (the
radius of an expanding universe being a popular choice) (DeWitt 1967).
However, closer analysis reveals a series of difficulties with such proposals:
for example, it is difficult to ensure the emergence of a well-behaved time
parameter $t$ such that only one physical state is associated with each value
of $t$. (See, for example, Unruh and Wald (1989).) Despite some 50 years of
effort, including the technical progress made in recent years using `loop'
variables to solve the equations (Rovelli 2004), the `problem of time' in
canonical quantum gravity remains unresolved.\footnote{Barbour (1994a,b) has
proposed a timeless formulation of classical and quantum physics. As applied
to quantum gravity, the viability of Barbour's scheme seems to depend on
unproven properties of the Wheeler-DeWitt equation.}

\section{Determinism and probability}\label{det-prob}

In the published text, the first section of the general discussion bears the
title `Causality, determinism, probability'. Lorentz's opening remarks are
mainly concerned with the importance of having a clear and definite picture of
physical processes (see section~\ref{visualisability}). He ends by
addressing the question of determinism (p.~\pageref{attheend}):

\begin{quotation}
.... I think that this notion of probability should be placed at the end, and
as a conclusion, of theoretical considerations, and not as an \textit{a
priori} axiom, though I may well admit that this indeterminacy corresponds to
experimental possibilities. I would always be able to keep my deterministic
faith for the fundamental phenomena .... .
\end{quotation}

Lorentz seems to demand that the fundamental phenomena be deterministic, and
that indeterminism should be merely emergent or effective. Probabilities
should not be axiomatic, and some theoretical explanation is needed for the
experimental limitations encountered in practice. This view would nowadays be
usually associated with deterministic hidden-variables theories, such as de
Broglie's pilot-wave dynamics (though it might also be associated with the
many-worlds interpretation of Everett).

De Broglie's basic equations --- the guidance equation and Schr\"{o}dinger
equation --- are certainly deterministic. De Broglie in his report, and
Brillouin in the subsequent discussion, give examples of how these equations
determine the trajectories (during interference and diffraction, atomic
transitions, and elastic scattering). As regards probabilities, de Broglie
pointed out that if an ensemble of systems with initial wave function
$\Psi_{0}$ begins with a Born-rule distribution $P_{0}=\left\vert \Psi
_{0}\right\vert ^{2}$, then as $\Psi$ evolves, the system dynamics will
maintain the distribution $P=\left\vert \Psi\right\vert ^{2}$ at later times.
However, nothing was said about how the initial distribution might arise in
the first place. Subsequent work has shown that, in de Broglie's theory, the
Born-rule distribution can arise from the complex evolution generated by the
dynamics itself, much as thermal distributions arise in classical dynamics,
thereby providing an example of the kind of theoretical explanation that
Lorentz wished for (Bohm 1953; Valentini 1991a, 1992, 2001; Valentini and
Westman 2005).\footnote{Bohm (1953) considered the particular case of an
ensemble of two-level molecules and argued that external perturbations would
drive it to quantum equilibrium. A general argument for relaxation was not
given, however, and soon afterwards Bohm and Vigier (1954) modified the
dynamics by adding random fluctuations that drive any system to equilibrium.
This move to a stochastic theory seems unnecessary: a general \textit{H}%
-theorem argument, analogous to the classical coarse-graining \textit{H}%
-theorem, has been given (Valentini 1991a, 1992, 2001), and numerical
simulations show a very efficient relaxation --- with an exponential decay of
the coarse-grained $H$-function --- on the basis of the purely deterministic
de Broglie-Bohm theory (Valentini and Westman 2005).} On this view, the
initial ensemble considered by de Broglie corresponds to a special
`equilibrium' state analogous to thermal equilibrium in classical physics.

Lorentz goes on to say (p.~\pageref{deepermind}):

\begin{quotation}
Could a deeper mind not be aware of the motions of these electrons? Could one
not keep determinism by making it an object of belief? Must one necessarily
elevate indeterminism to a principle?
\end{quotation}
Here, again, we now know that de Broglie's theory provides an example
of what Lorentz seems to have had in mind. For in principle, the theory allows
the existence of `nonequilibrium' distributions $P\neq\left\vert
\Psi\right\vert ^{2}$ (Valentini 1991b, 1992), just as classical physics
allows the existence of non-thermal distributions (not uniformly distributed
on the energy surface in phase space). Such distributions violate many of the
standard quantum constraints; in particular, an experimenter possessing
particles with a distribution $P$ much narrower than $\left\vert
\Psi\right\vert ^{2}$ would be able to use those particles to perform
measurements on ordinary systems more accurate than normally allowed by the
uncertainty principle; an experimenter possessing such `nonequilibrium
particles' would in fact be able to use them to observe the (normally invisible)
details of the trajectories of ordinary particles (Valentini 2002b; Pearle and
Valentini 2006). From this point of view, there is indeed no need to `elevate
indeterminism to a principle': for the current experimental limitations
(embodied in the uncertainty principle) are not built into the laws of
physics; rather, they are merely contingent features of the equilibrium state
$P=\left\vert \Psi\right\vert ^{2}$.

In the general discussion, as we saw in section~\ref{DiracandHeisenberg}, 
Dirac expressed the view that quantum outcomes occur when
nature makes a choice, a view countered by Heisenberg who claimed that the
`choice' is in some sense really made by the observer. As Lorentz noted at the
end of this exchange, the view that nature makes a choice amounts to a
fundamental indeterminism, while at the same time, Dirac and Heisenberg had
radically different views about the meaning of this indeterminism.

Dirac also gave an argument for why quantum theory had to be indeterministic.
In his view, the indeterminism was necessary because of the inevitable
disturbance involved in setting up an initial quantum state (pp.~\pageref{Dirac-determinism} f.):

\begin{quotation}
I should now like to express my views on determinism and the nature of the
numbers appearing in the calculations of the quantum theory .... . In the
classical theory one starts from certain numbers describing completely the
initial state of the system, and deduces other numbers that describe
completely the final state. This deterministic theory applies only to an
isolated system.

But, as Professor Bohr has pointed out, an isolated system is by definition
unobservable. One can observe the system only by disturbing it and observing
its reaction to the disturbance. Now since physics is concerned only with
observable quantities the deterministic classical theory is untenable.
\end{quotation}
Dirac's argument seems unsatisfactory. First of all, as a general
philosophical point, the claim that `physics is concerned only with observable
quantities' is not realistic. As is well known to philosophers of science as
well as to experimentalists, observation is `theory-laden': in order to carry
out observations (or measurements) some body of theory is required in order to
know how to carry out a correct observation (for example, some knowledge is
required of how the system being measured interacts with the apparatus, in
order to design a correctly functioning apparatus). Thus, some body of theory
is necessarily conceptually prior to observation, and it is logically
impossible to base physical theory on `observables' only. More specifically,
Dirac claims that classical determinism is untenable because of the
disturbance involved in observing a system: but there are many cases in
classical physics where experimenters can use their knowledge of the
interactions involved to compensate for the disturbance caused by the
measurement. Disturbance \textit{per se} cannot be a reason for indeterminism.

It may well be that Dirac had in mind the kind of `irreducible' or
`uncontrollable' disturbance that textbooks commonly associate with the
uncertainty principle. However, it is interesting that, in fact, Dirac goes on
to say that the disturbances applied by an experimenter \textit{are} under his
control (p.~\pageref{inthequantumtheory}, Dirac's italics):

\begin{quotation}
In the quantum theory one also begins with certain numbers and deduces others
from them. .... The disturbances that an experimenter applies to a system to
observe it are directly under his control, and are acts of freewill by him.
\emph{It is only the numbers that describe these acts of freewill that can be
taken as initial numbers for a calculation in the quantum theory}. Other
numbers describing the initial state of the system are inherently
unobservable, and do not appear in the quantum theoretical treatment.
\end{quotation}
The `disturbances' refer to the experimental operations that the
experimenter chooses to apply to the system, and indeed these are normally
regarded as freely controlled by the experimenter (at least in some effective
sense). The sense in which the word `disturbance' is being used here is quite
different from the textbook sense of uncontrollable disturbance associated
with quantum uncertainty. Dirac seems to regard quantum numbers, or
eigenvalues, as representing the extent to which an experimenter can
controllably manipulate a system. Macroscopic operations are under our
control, and through these we can prepare an initial state specified by
particular quantum numbers. For Dirac, these initial numbers represent `acts
of freewill' in the form of laboratory operations. (The final remark about
`other numbers' that are unobservable, and that do not appear in quantum
theory, is intriguing, and might be taken as suggesting that there are other
degrees of freedom that cannot be controlled by us.)

A view quite different from Dirac's is expressed by Born at the very end of
the general discussion. According to Born, the constraints on the preparation
of an initial quantum state are \textit{not} what distinguishes quantum from
classical mechanics, for in classical physics too (p.~\pageref{Bornprecision})

\begin{quotation}
.... the precision with which the future location of a particle can be
predicted depends on the accuracy of the measurement of the initial location.
\end{quotation}
For Born, the difference rather lies in the law of propagation of
probability packets:

\begin{quotation}
It is then not in this that the manner of description of quantum mechanics, by
wave packets, is different from classical mechanics. It is different because
the laws of propagation of packets are slightly different in the two cases.
\end{quotation}

\section{Visualisability and the spacetime framework}\label{visualisability}

With hindsight, from a contemporary perspective, perhaps the most
characteristic feature of quantum physics is the apparent absence of
visualisable processes taking place within a spacetime framework. From Bell's
theorem, it appears that any attempt to provide a complete description of
quantum systems (within a single world) will require some form of nonlocality,
leading to difficulties with relativistic spacetime. At the time of writing,
we possess only one hidden-variables theory of broad scope --- the pilot-wave
theory of de Broglie and Bohm --- and in this theory there is a field on
\textit{configuration} space (not 3-space) that affects the motion of quantum
systems.\footnote{As already mentioned in section~\ref{fundamb}, attempts 
to construct hidden-variables theories without an
ontological wave function, along the lines pioneered by F\'{e}nyes (1952) and
Nelson (1966), seem to fail (Wallstrom 1994; Pearle and Valentini 2006).}
Arguably, then, pilot-wave theory does not really fit into a spacetime
framework: the physics is grounded in configuration space, and the
interactions encoded in the pilot wave take place outside of 3-space. Instead
of a nonlocal hidden-variables theory, one might prefer to have a complete
account of quantum behaviour in terms of many worlds: there too, one leaves
behind ordinary spacetime as a basic framework for physics, since the totality
of what is real cannot be mapped onto a single spacetime geometry. Generally
speaking, whatever one's view of quantum theory today, the usual spacetime
framework seems too restrictive, and unable to accomodate (at least in a
natural way) the phenomena associated with quantum superposition and
entanglement.\footnote{A possible exception here is some version of quantum
theory with dynamical wave function collapse.}

We saw in section~\ref{Time-in-quantum-theory} that the quantum or matrix
mechanics of Born and Heisenberg certainly did not provide an account of
physical systems in a spacetime framework. In contrast, the initial practical
success of Schr\"{o}dinger's wave mechanics in 1926 had led some workers to
think that an understanding in terms of (wave processes in) space and time
might be possible after all. The resulting tension between Schr\"{o}dinger on
the one hand, and Bohr, Heisenberg and Pauli on the other (in the year
preceding the Solvay meeting) has been described at length in section~\ref{Schr-conflict}, 
where we saw that for Schr\"{o}dinger the notion of
`Anschaulichkeit' --- in the sense of visualisability in a spacetime framework
--- played a key role.

The clash between quantum physics and the spacetime framework was a central
theme of the fifth Solvay conference. There was a notable tension between
those participants who still hoped for a spacetime-based theory and those who
insisted that no such theory was possible. These differences are especially
apparent in the general discussion where, as we have already discussed in
section~\ref{morelocrel}, difficulties were
raised concerning locality and relativity.

Lorentz, in his opening remarks at the first session of the general discussion, seems to set the tone for
one side of the debate, by speaking in favour of space and time as a basic
framework for physics (p.~\pageref{Lorentzwish}):

\begin{quotation}
We wish to make a representation of the phenomena, to form an image of them in
our minds. Until now, we have always wanted to form these images by means of
the ordinary notions of time and space. These notions are perhaps innate; in
any case, they have developed from our personal experience, by our daily
observations. For me, these notions are clear and I confess that I should be
unable to imagine physics without these notions. The image that I wish to form
of phenomena must be absolutely sharp and definite, and it seems to me that we
can form such an image only in the framework of space and time.
\end{quotation}
Lorentz's committment to processes taking place in space and time was
shared by de Broglie and Schr\"{o}dinger, even though both men had found
themselves unable to avoid working in terms of configuration space. As we saw
in section~\ref{deB-1927-Solvay-report}, in his report de
Broglie presented his pilot-wave dynamics, with a guiding field in
configuration space, only as a makeshift: he hoped that his pilot-wave
dynamics would turn out to be an effective theory only, and that underlying it
would be a theory of wave fields in 3-space with singularities representing
particle motion (the double-solution theory). Schr\"{o}dinger, too, despite
working with a many-body wave equation in configuration space, hoped that the
physical content of his theory could be ultimately interpreted in terms of
processes taking place in 3-space (see chapter~\ref{SchrEss}).

In contrast, some of the other participants, in particular Bohr and Pauli,
welcomed --- indeed insisted upon --- the break with the spacetime framework.
As Bohr put it in the general discussion, after Einstein's remarks on locality
and completeness (p.~\pageref{Bohrfoundation}):

\begin{quotation}
The whole foundation for [a] causal spacetime description is taken away by quantum
theory, for it is based on [the] assumption of observations without interference.
\end{quotation}
In support of Bohr's contention, Pauli (pp.~\pageref{Pauliwholly}~f.) 
provided an intriguing argument to the effect
that interactions between particles cannot be understood in a spacetime
framework. Specifically, Pauli based his argument on the quantum-theoretical
account of the long-range interactions known as van der Waals forces.

Pauli begins his argument by saying (in agreement with Bohr) that the use of
multi-dimensional configuration space is

\begin{quotation}
only a technical means of formulating mathematically the laws of mutual action
between several particles, actions which certainly do not allow themselves to
be described simply, in the ordinary way, in space and time.
\end{quotation}
Here, on the one hand, configuration space has only mathematical
significance. But on the other hand, as Einstein feared (see section~\ref{morelocrel}), 
there is according to Pauli
no explicit account of local action in space and time. Pauli adds that the
multi-dimensional method might one day be replaced with what we would now call
quantum field theory. He then goes on to say that, in any case, in accordance
with Bohr's point of view, no matter what `technical means' are used to
describe `the mutual actions of several particles', such actions `cannot be
described in the ordinary manner in space and time' (p.~\pageref{ordinary}).

Pauli then illustrates his point with an example. He considers two
widely-separated hydrogen atoms, each in their ground state, and asks what
their `energy of mutual action' might be. According to the usual description
in space and time, says Pauli, for large separations there should be no mutual
action at all. And yet (p.~\pageref{whenonetreats}),

\begin{quotation}
when one treats the same question by the multi-dimensional method, the result
is quite different, and in accordance with experiment.
\end{quotation}

What Pauli is referring to here is the problem of accounting for van der Waals
forces between atoms and molecules. Classically, molecules with dipole moments
tend to align, resulting in a mean interaction energy $\propto1/R^{6}$ (where
$R$ is the distance between two molecules). However, many molecules exhibiting
van der Waals forces have zero dipole moment; and while the classical
orientation effect becomes negligible at high temperatures, van der Waals
forces do not. The problem of explaining van der Waals forces was finally
solved by quantum theory, beginning with the work of Wang in
1927.\footnote{For a detailed review, see Margenau (1939).} (Note that the
effect here is quite distinct from that of exchange forces resulting from the
Pauli exclusion principle. The latter forces are important only when the
relevant electronic wave functions have significant overlap, whereas van der
Waals forces occur between neutral atoms even when they are so far apart that
their charge clouds have negligible overlap.)

A standard textbook calculation of the van der Waals force, between two
hydrogen atoms (1 and 2) separated by a displacement $\mathbf{R}$, proceeds as
follows.\footnote{See, for example, Schiff (1955, pp.~176--80).} The
unperturbed energy eigenstate is the product $\Psi=\psi_{0}^{1}\psi_{0}^{2}$
of the two ground states. If $R=\left\vert\mathbf{R}\right\vert$ is much larger than the Bohr radius, the
classical electrostatic potential between the atoms is%
\begin{equation}
V=\frac{e^{2}}{R^{3}}\left(  \mathbf{r}_{1}\cdot\mathbf{r}_{2}-3\frac
{(\mathbf{r}_{1}\cdot\mathbf{R})(\mathbf{r}_{2}\cdot\mathbf{R})}{R^{2}%
}\right)  \ ,
\end{equation}
where $\mathbf{r}_{1}$, $\mathbf{r}_{2}$ are respectively the positions of
electrons 1, 2 relative to their nuclei. In the state $\Psi=\psi_{0}^{1}%
\psi_{0}^{2}$ the mean values of $\mathbf{r}_{1}$, $\mathbf{r}_{2}$ both
vanish, so that the expectation value of $V$ vanishes. However, the presence
of $V$ perturbs the ground state of the system, to a new and entangled state
$\Psi^{\prime}$, satisfying $(H_{0}+V)\Psi^{\prime}=(2E_{0}+\Delta
E)\Psi^{\prime}$, where $H_{0}$ is the unperturbed Hamiltonian, $E_{0}$ is the
ground-state energy of hydrogen, and $\Delta E$ is the energy perturbation
from which one deduces the van der Waals force. Standard perturbation methods
show that $\Delta E$ has a leading term proportional to $1/R^{6}$, accounting
for the (attractive) van der Waals force.

Pauli regarded this result --- which had been obtained by Wang from the
Schr\"{o}dinger equation in configuration space --- as evidence that in
quantum theory there are interactions that cannot be described in terms of
space and time. The result may be roughly understood classically, as Pauli
points out, by imagining that in each atom there is an oscillating dipole
moment that can induce (and so interact with) a dipole moment in the other
atom. But such understanding is only heuristic. In a proper treatment, using
`multi-dimensional wave mechanics', the correct result is obtained by methods
that, according to Pauli, cannot be understood in a spacetime
framework.\footnote{According to Ehrenfest, Bohr also gave an argument (in a
conversation with Einstein) against a spacetime description when treating
many-particle problems. See section~\ref{day-by-day}.}

Pauli's point seems to have been that, because the perturbed wave function
$\Psi^{\prime}$ is entangled, multi-dimensional configuration space plays a
crucial role in bringing about the correct result, which therefore cannot be
properly understood in terms of 3-space alone. If this argument seems
unwarranted, it ought to be remembered that, before Wang's derivation in 1927,
there had been a long history of failed attempts to explain van der Waals
forces classically (Margenau 1939).

Disagreements over both the usefulness and the tenability of the spacetime
framework are also apparent elsewhere in the general discussion. For example,
Kramers (p.~\pageref{advantage}) asks: `What advantage do you see in
giving a precise value to the velocity $v$ of the photons?' To this de Broglie
replies (in a spirit similar to that of Lorentz above):

\begin{quotation}
This allows one to imagine the trajectory followed by the photons and to
specify the meaning of these entities; one can thus consider the photon as a
material point having a position and a velocity.
\end{quotation}
Kramers was unconvinced:

\begin{quotation}
I do not very well see, for my part, the advantage that there is, for the
description of experiments, in making a picture where the photons travel along
well-defined trajectories.
\end{quotation}
In this exchange, Kramers had suggested that de Broglie's theory could
not explain radiation pressure from a single photon, thereby questioning the
tenability (and not just the usefulness) of the spacetime framework for the
description of elementary interactions. (Cf. section~\ref{recoil}.)

Later in the general discussion, another argument against de Broglie's theory 
is provided by Pauli. As we shall discuss at length in section~\ref{elastic-inelastic}, 
Pauli claims on the basis of an example that pilot-wave
theory cannot account for the discrete energy exchange taking place in
inelastic collisions. It is interesting to note
that, according to Pauli, the root of the difficulty lies in the attempt to
construct a deterministic particle dynamics in a spacetime framework (p.~\pageref{Paulidifficulty}):

\begin{quotation}
.... this difficulty .... is due directly to the condition assumed by Mr de
Broglie, that in the individual collision process the behaviour of the
particles should be completely determined and may at the same time be
described completely by ordinary kinematics in spacetime.
\end{quotation}

\setcounter{endnote}{0}
\setcounter{equation}{0}

\chapter{Guiding fields in 3-space}\label{guiding-fields-in-3-space}\chaptermark{Guiding fields in 3-space}

In this chapter we address proposals (by Einstein, and by Bohr, Kramers and
Slater) according to which quantum events are influenced by `guiding fields'
in 3-space. These ideas led to a predicted violation of energy-momentum
conservation for single events, in contradiction with experiment. The
contradiction was resolved only by the introduction of guiding fields in
configuration space. All this took place before the fifth Solvay conference,
but nevertheless forms an important background to some of the discussions that
took place there.

\section{Einstein's early attempts to formulate a dynamical theory of light
quanta}\label{Eearlyattempts}\sectionmark{Einstein's early attempts}

Since the publication of his light-quantum hypothesis in 1905, Einstein had
been engaged in a solitary struggle to construct a detailed theory of light
quanta, and to understand the relationship between the quanta on the one hand
and the electromagnetic field on the other.\footnote{In 1900 Planck had, of
course, effectively introduced a quantisation in the interaction between
radiation and matter; but it was Einstein in 1905 who first proposed that
radiation itself (even in free space) consisted, at least in part, of
spatially localised energy quanta. In 1918 Einstein wrote to his friend Besso:
`I do not doubt anymore the \textit{reality} of radiation quanta, although I
still stand quite alone in this conviction' (original italics, as quoted in
Pais (1982, p.~411)).} Einstein's efforts in this direction were never
published. We know of them indirectly: they are mentioned in letters, and they
are alluded to in Einstein's 1909 lecture in Salzburg. Einstein's published
papers on light quanta continued for the most part in the same vein as his
1905 paper: using the theory of fluctuations to make deductions about the
nature of radiation, without giving details of a substantial theory. Einstein
was essentially alone in his dualistic view of light, in which localised
energy fragments coexisted with extended waves, until the work of de Broglie
in 1923 --- which extended the dualism to all particles, and made considerable
progress towards a real theory (see chapter~\ref{deBroglieEss}).

A glimpse of Einstein's attempts to formulate a dynamical theory of light
quanta may be obtained from a close reading of his 1909 Salzburg lecture.
There, as we saw in section~\ref{Einstein-at-Salzburg},
Einstein marshalled evidence that light waves contain localised energy
fragments. He suggested that the electromagnetic field is associated with
singular points at which the energy is localised, and he offered the following
remarkable (if heuristic) picture:

\begin{quotation}
I more or less imagine each such singular point as being surrounded by a field
of force which has essentially the character of a plane wave and whose
amplitude decreases with the distance from the singular point. If many such
singularities are present at separations that are small compared with the
dimensions of the field of force of a singular point, then such fields of
force will superpose, and their totality will yield an undulatory field of
force that may differ only slightly from an undulatory field as defined by the
current electromagnetic theory of light. (Einstein 1909, English translation,
p.~394)
\end{quotation}
Here, each light quantum is supposed to have an extended field
associated with it, and large numbers of quanta with their associated fields
are supposed to yield (to a good approximation) the electromagnetic field of
Maxwell's theory. In other words, the electromagnetic field as we know it
emerges from the collective behaviour of large numbers of underlying fields
associated with individual quanta.

A similar view is expressed in a letter from Einstein to Lorentz written a few
months earlier, in May 1909:

\begin{quotation}
I conceive of the light quantum as a point that is surrounded by a greatly
extended vector field, that somehow diminishes with distance. Whether or not
when several light quanta are present with mutually overlapping fields one
must imagine a simple superposition of the vector fields, that I cannot say.
In any case, for the determination of events, one must have equations of
motion for the singular points in addition to the differential equations for
the vector field. (Quoted in Howard 1990, p.~75)
\end{quotation}
From the last sentence, it is clear that Einstein's conception was
supposed to be deterministic.

Einstein's view in 1909, then, is remarkably reminiscent of de Broglie's
pilot-wave theory as well as of his theory of the `double solution' (from
which de Broglie hoped pilot-wave theory would emerge, see chapter~\ref{deBroglieEss}). 
Einstein seems to have thought of each individual light
quantum as being accompanied by some kind of field in 3-space that affects the
motion of the quantum.

Fascinating reactions to Einstein's ideas appear in the recorded discussion
that took place after Einstein's lecture. Stark pointed out a phenomenon that
seemed to speak in favour of localised energy quanta in free space: `even at
great distances, up to 10 m, electromagnetic radiation that has left an X-ray
tube for the surrounding space can still achieve concentrated action on a
single electron' (Einstein 1909, English translation, p.~397). Stark's point
here is, again, Einstein's locality argument: as we noted at the end of
section~\ref{Einstein-at-Salzburg}, in retrospect it
seems puzzling that this simple and compelling argument was not widely
understood much earlier, but here Stark clearly appreciates it. However ---
and this is probably why the argument did not gain currency --- doubts were
raised as to how such a theory could explain interference. Planck spoke as follows:

\begin{quotation}
Stark brought up something in favor of the quantum theory, and I wish to bring
up something against it; I have in mind the interferences at the enormous
phase differences of hundreds of thousands of wavelengths. When a quantum
interferes with itself, it would have to have an extension of hundreds of
thousands of wavelengths. This is also a certain difficulty. (Einstein 1909,
English translation, p.~397)
\end{quotation}
To this objection, Einstein gives a most interesting reply:

\begin{quotation}
I picture a quantum as a singularity surrounded by a large vector field. By
using a large number of quanta one can construct a vector field that does not
differ much from the kind of vector field we assume to be involved in
radiations. I can well imagine that when rays impinge upon a boundary surface,
a separation of the quanta takes place, due to interaction at the boundary
surface, possibly according to the phase of the resulting field at which the
quanta reach the interface. .... I do not see any fundamental difficulty in
the interference phenomena. (Einstein 1909, English translation, p.~398)
\end{quotation}
Again, the similarity to de Broglie's later ideas is striking. Einstein
seems to think that the associated waves can affect the motions of the quanta,
in such a way as to account for interference.

It should be noted, though, that in this exchange it is somewhat unclear 
whether the subject is interference for single photons or for many photons.
Einstein talks about interference in terms of the collective behaviour of many
quanta, rather than in terms of one quantum at a time. While single-photon
interference with very feeble light was observed by Taylor (1909) in the same
year, Stark at least seems not to know of Taylor's results, for at this point
(just before Einstein's reply) he interjects that `the experiments to which
Mr~Planck alluded involve very dense radiation .... . With radiation of very
low density, the interference phenomena would most likely be different'
(Einstein 1909, English translation, p.~397). On the other hand, Planck may
well have thought of the light-quantum hypothesis as implying that light
quanta would move independently, like the molecules of an ideal gas (in which
case even for intense radiation one could consider the motion of each quantum
independently). Einstein countered precisely such a view at the beginning of
his reply to Planck, where he states that `it must not be assumed that
radiations consist of non-interacting quanta; this would make it impossible to
explain the phenomena of interference'. This might be read as implying that
interactions among different quanta are essential, and that interference would
not occur with one photon at a time. However, Einstein may simply have meant
that the light quanta cannot be thought of as free particles: they must be
accompanied by a wave as well, in order to explain interference. (This last
reading fits with Einstein's discussion, in the lecture, of thermal
fluctuations in radiation: these cannot be obtained from a gas of free and
independent particles alone; a wave-like component is also needed.)

While there is some uncertainty over the details of Einstein's proposal, in
retrospect, given our present understanding of how de Broglie's pilot-wave
theory provides a straightforward explanation of particle interference (see
section~\ref{interf-deB}), Einstein's reply to Planck seems
very reasonable. However, it appears that yet another of Einstein's arguments
was not appreciated by his contemporaries. Seven years later, after having
verified Einstein's photoelectric equation experimentally, Millikan
nevertheless completely rejected Einstein's light-quantum hypothesis, which he
called `reckless .... because it flies in the face of the thoroughly
established facts of interference' (Millikan 1916, p.~355).

\section{The failure of energy-momentum conservation}\label{failure-energy-momentum}

It appears that Einstein was still thinking along similar lines in the 1920s,
though again without publishing any detailed theory. As we saw at the end of
section~\ref{QM-without-collapse}, in his
collision papers of 1926 Born notes the analogy between his own work and
Einstein's ideas: `I start from a remark by Einstein ....  ; he said .... that the waves are
there only to show the corpuscular light quanta the way, and in this sense he
talked of a \textquotedblleft ghost field\textquotedblright' (Born 1926b, pp.~803--4). 
Born gives no reference to any published paper of Einstein's, however.

How Einstein's thinking at this time compared with that in 1909 is hard to
say. Certainly, in 1909 he thought of the electromagnetic field as being built
up from the collective behaviour of large numbers of vector fields associated
with individual quanta. In such a scenario, it seems plausible that in the
right circumstances the intensity of the emergent electromagnetic field could
act as, in effect, a probability field. (Unlike Born, Einstein would have
regarded a purely probabilistic description as a makeshift only.)

Just one year before Born's collision papers, in 1925, Einstein gave a
colloquium in Berlin where he indeed discussed the idea that every particle
(including electrons, following de Broglie) was accompanied by a
`F\"{u}hrungsfeld' or guiding field (Pais 1982, p.~441; Howard 1990, p.~72).
According to Wigner, who was present at the colloquium:

\begin{quotation}
Yet Einstein, though in a way he was fond of it, never published it. He
realized that it is in conflict with the conservation principles: at a
collision of a light quantum and an electron for instance, both would follow a
guiding field. But these guiding fields give only the probabilities of the
directions in which the two components, the light quantum and the electron,
will proceed. Since they follow their directions independently, .... the
momentum and the energy conservation laws would be obeyed only statistically
.... . This Einstein could not accept and hence never took his idea of the
guiding field quite seriously. (Wigner 1980, p.~463)
\end{quotation}

In the early 1920s, then, Einstein was still thinking along lines that are
reminiscent of de Broglie's work, but he never published these ideas because
they conflicted with the conservation laws for individual events. The
difficulty Einstein faced was overcome only by the introduction (through the
work of de Broglie, Schr\"{o}dinger, and also Born) of a guiding field in
configuration space --- a \textit{single} (and generally entangled) field that
determined probabilities for all particles collectively. In Einstein's
approach, where each particle had its own guiding field, the possibility of
entanglement was precluded, and the correlations were not strong enough to
guarantee energy-momentum conservation for single events.

As a simplified model of what Einstein seems to have had in mind, consider two
particles 1 and 2 moving towards each other in one dimension, with equal and
opposite momenta $p$ and $-p$ respectively. Schematically, let us represent
this with an initial wave function $\psi(x_{1},x_{2},0)=\psi_{1}^{+}%
(x_{1},0)\psi_{2}^{-}(x_{2},0)$, where $\psi_{1}^{+}$ and $\psi_{2}^{-}$ are
broad packets (approximating plane waves) moving along $+x$ and $-x$
respectively. Let the packets meet in a region centred around the origin,
where we imagine that an elastic collision takes place with probability
$\frac{1}{2}$. At large times, we assume that $\psi(x_{1},x_{2},t)$ takes the
schematic and entangled form%
\begin{equation}
\psi(x_{1},x_{2},t)=\frac{1}{\sqrt{2}}\left(  \psi_{1}^{+}(x_{1},t)\psi
_{2}^{-}(x_{2},t)+\psi_{1}^{-}(x_{1},t)\psi_{2}^{+}(x_{2},t)\right)  \ ,
\label{ent}%
\end{equation}
with the first branch corresponding to the particles having moved freely past
each other and the second corresponding to an elastic collision reversing
their motions. The initial state has zero total momentum. In the final state,
both possible outcomes correspond to zero total momentum. Thus, in quantum
theory, whatever the outcome of an individual run of the experiment, momentum
is always conserved. Now imagine if, instead of using a single wave function
$\psi(x_{1},x_{2},t)$ in configuration space, we made use of two 3-space waves
$\psi_{1}(x,t)$, $\psi_{2}(x,t)$ (one for each particle). It would then seem
natural that, during the collision, the initial 3-space wave $\psi
_{1}(x,0)=\psi_{1}^{+}(x,0)$ for particle 1 would evolve into%
\begin{equation}
\psi_{1}(x,t)=\psi_{1}^{+}(x,t)+\psi_{1}^{-}(x,t)\ ,
\end{equation}
while the initial 3-space wave $\psi_{2}(x,0)=\psi_{2}^{-}(x,0)$ for particle
2 would evolve into%
\begin{equation}
\psi_{2}(x,t)=\psi_{2}^{-}(x,t)+\psi_{2}^{+}(x,t)\ .
\end{equation}
If the amplitude of each 3-space wave determined the probabilities for the
respective particles, there would then be \textit{four} (equiprobable)
possible outcomes for the scattering experiment, the two stated above,
together with one in which both particles move to the right and one in which
both particles move to the left. These possibilities would correspond, in
effect, to a final (configuration-space) wave function of the separable form%
\begin{equation}
\left(  \psi_{1}^{+}+\psi_{1}^{-}\right)  \left(  \psi_{2}^{-}+\psi_{2}%
^{+}\right)  =\psi_{1}^{+}\psi_{2}^{-}+\psi_{1}^{+}\psi_{2}^{+}+\psi_{1}%
^{-}\psi_{2}^{-}+\psi_{1}^{-}\psi_{2}^{+}\ . \label{sep}%
\end{equation}
The final total momenta for the four (equiprobable) possible outcomes are $0$,
$+2p$, $-2p$, $0$. The total momentum would not be generally conserved for
individual outcomes, but conservation would hold on average. Note the
fundamental difference between (\ref{ent}) and (\ref{sep}): the former is
entangled, the latter is not.

Einstein considered such a failure of the conservation laws reason enough to
reject the idea. A theory bearing some resemblance to this scheme was,
however, proposed and published by Bohr, Kramers and Slater (1924a,b).

In the Bohr-Kramers-Slater (BKS) theory, there are no photons. Associated with
an atom in a stationary state is a `virtual radiation field' containing all
those frequencies corresponding to transitions to and from other stationary
states. The virtual field determines the transition probabilities for the atom
itself, and also contributes to the transition probabilities for other,
distant atoms. However, the resulting correlations between widely-separated
atoms are not strong enough to yield energy-momentum conservation for
individual events (cf. Compton's account of the BKS theory in his report, 
p.~\pageref{ComptonBKS}).\footnote{For a detailed account of the BKS theory,
see Darrigol (1992, chap. 9).}

To illustrate this, consider again (in modern language) an atom A emitting a
photon of energy $E_{i}-E_{f}$ that is subsequently absorbed by a distant atom
B. In the BKS theory, the transition at B is not directly caused by the
transition at A (as it would be in a simple `semiclassical' picture of the
emission, propagation, and subsequent absorption of a localised light
quantum). Rather, the virtual radiation field of A contains terms
corresponding to transitions of A, and this field contributes to the
transition probabilities at B. Yet, if B actually undergoes a transition
corresponding to an energy increase $E_{i}-E_{f}$, atom A is \textit{not}
constrained to make a transition corresponding to an energy decrease
$E_{i}-E_{f}$. The connection between the atoms is merely statistical, and
energy conservation holds only on average, not for individual
processes.\footnote{Note that Slater's original theory did contain photons,
whose motions were guided (statistically) by the electromagnetic field. The
photons were removed at the instigation of Bohr and Kramers (Mehra and
Rechenberg 1982a, pp.~543--7).}

Einstein objected to the BKS theory --- in a colloquium, and in private
letters and conversations (Mehra and Rechenberg 1982a, pp.~553--4; Pais 1982,
p.~420; Howard 1990, pp.~71--4) --- partly for the same reason he had not
published his proposals: that, as he believed, energy-momentum conservation
would hold even in the case of elementary interactions between
widely-separated systems.\footnote{In a letter to Ehrenfest dated 31 May 1924,
Einstein wrote: `This idea [the BKS theory] is an old acquaintance of mine,
but one whom I do not regard as a respectable fellow' (Howard 1990, p.~72).
Einstein listed five criticisms, including the violation of the conservation
laws, and a difficulty with thermodynamics (Mehra and Rechenberg 1982a, p.~553).} 
Einstein's expectation was subsequently confirmed by the Bothe-Geiger
and Compton-Simon experiments (Bothe and Geiger 1925b, Compton and Simon 1925).

In the Bothe-Geiger experiment, by means of counter coincidences, Compton
scattering (that is, relativistic electron-photon scattering) was studied for
individual events, to see if the outgoing scattered photon and the outgoing
recoil electron were produced simultaneously. Strict temporal coincidences
were expected on the basis of the light-quantum hypothesis, but not on the
basis of the BKS theory. Such coincidences were in fact observed by Bothe and
Geiger.\footnote{See Compton's account of `Bothe and Geiger's coincidence
experiments', pp.~\pageref{forpage13}~f., and also Mehra and Rechenberg
(1982a, pp.~609--12).} In the Compton-Simon experiment, Compton scattering was
again studied, with the aim of verifying the conservation laws for individual
events. Such conservation was in fact observed by Compton and
Simon.\footnote{See Compton's account of `Directional emission of scattered
X-rays', pp.~\pageref{Compton-Simon}~f., and also Mehra and Rechenberg
(1982a, p.~612).} As a result of these experiments, it was widely concluded
that the BKS theory was wrong.\footnote{Cf. Bohr's remarks in the discussion
of Compton's report, p.~\pageref{forpage170}.}

It appears that, even at the time of the fifth Solvay conference, Einstein was still thinking to some extent in terms of 
guiding fields in 3-space. Evidence for this comes from an otherwise incomprehensible remark Einstein made during his long 
contribution to the general discussion. There, Einstein compared two interpretations of the wave function $\psi$ for a 
single electron. On Einstein's view I, $\psi$ represents an ensemble --- or `cloud' --- of electrons, while on his view II 
$\psi$ is a complete description of an individual electron. As we have discussed at length in 
section~\ref{1927-incompleteness}, Einstein argued that interpretation II is inconsistent with locality. Now, Einstein 
also made the following remark concerning the conservation of energy and momentum according to interpretations I and II 
(p.~\pageref{secondconception}):
  \begin{quote}
    The second conception goes further than the first .... . It is only by virtue of II that the theory contains 
    the consequence that the conservation laws are valid for the elementary process; it is only from II that the 
    theory can derive the result of the experiment of Geiger and Bothe .... .
  \end{quote}
As currently understood, of course, a purely `statistical' interpretation of the wave function (cf.\ 
section~\ref{measprobstat}) would yield correct predictions, in agreement with the conservation laws 
for elementary processes (such as scattering). Why, then, did Einstein assert that interpretation I would conflict 
with such elementary conservation? It must surely be that, for whatever reason, he was thinking of interpretation 
I as tied specifically to wave functions in 3-space, resulting in a failure of the conservation laws for single 
events as discussed above. In contrast, specifically regarding interpretation II, Einstein explicitly asserted 
(p.~\pageref{essentiallytied}) that it is `essentially tied to a multi-dimensional representation (configuration space)'.

The conflict between Einstein's ideas about guiding fields in 3-space and 
energy-momentum conservation was, as we have mentioned, resolved only by
the introduction of a guiding field in configuration space --- with the
associated entanglement and nonseparability that Einstein was to find so
objectionable. Einstein's early worries about nonseparability (long before the
EPR paper) have been extensively documented by Howard (1990). Defining
`separability' as the idea that `spatio-temporally separated systems possess
well-defined real states, such that the joint state of the composite system is
wholly determined by these two separate states' (Howard 1990, p.~64), Howard
highlights the fundamental difficulty Einstein faced in having to choose, in
effect, between separability and energy-momentum conservation:

\begin{quotation}
But as long as the `guiding'\ or `virtual fields'\ [determining the
probabilities for particle motions, or for atomic transitions] are assigned
separately, one to each particle or atom, one cannot arrange \textit{both} for
the merely probabilistic behavior of individual systems \textit{and} for
correlations between interacting systems sufficient to secure strict
energy-momentum conservation in all individual events. As it turned out, it
was only Schr\"{o}dinger's relocation of the wave fields from physical space
to configuration space that made possible the assignment of joint wave fields
that could give the strong correlations needed to secure strict conservation.
(Howard 1990, p.~73)
\end{quotation}

Here, then, is a remarkable historical and physical connection between
nonseparability or entanglement on the one hand, and energy-momentum
conservation on the other. The introduction of probability waves in
configuration space, with generic entangled states, finally made it possible
to secure energy-momentum conservation for individual emission and scattering
events, at the price of introducing a fundamental nonseparability into physics.

\setcounter{endnote}{0}
\setcounter{equation}{0}

\chapter{Scattering and measurement in de Broglie's pilot-wave theory}\label{meas-in-pwt}%
\chaptermark{Scattering and measurement in de Broglie's theory}

At the fifth Solvay conference, some questions that are closely related to the
quantum measurement problem (as we would now call it) were addressed in the
context of pilot-wave theory, in both the discussion following de Broglie's
report and in the general discussion. Most of these questions concerned the
treatment of scattering (elastic and inelastic); they were raised by Born and
Pauli, and replies were given by Brillouin and de Broglie. Of special interest
is the famous --- and widely misunderstood --- objection by Pauli concerning
inelastic scattering. Another question closely related to the measurement
problem was raised by Kramers, concerning the recoil of a single photon on a mirror.

In this chapter, we shall first outline the pilot-wave theory of scattering,
as currently understood, and examine the extensive discussions of scattering
--- in the context of de Broglie's theory --- that took place at the conference.

We shall see that de Broglie and Brillouin correctly answered the query raised
by Born concerning elastic scattering. Further, we shall see that Pauli's
objection concerning the inelastic case was both more subtle and more confused
than is generally thought; in particular, Pauli presented his example in terms
of a misleading optical analogy (that was originally given by Fermi in a more 
restricted context). Contrary to a
widespread view, de Broglie's reply to Pauli \textit{did} contain the
essential points required for a proper treatment of inelastic scattering; at
the same time, Fermi's misleading analogy confused matters, and neither de
Broglie in 1927 nor Bohm in 1952 saw what the true fault with Pauli's example
was. (As we shall also see, a proper pilot-wave treatment of Pauli's example
was not given until 1956, by de Broglie.)

We shall also outline the pilot-wave theory of quantum measurement, again as
currently understood, and we shall use this as a context in which to examine
the question raised by Kramers, of the recoil of a single photon on a mirror,
which de Broglie was unable to answer.

\section{Scattering in pilot-wave theory}\label{scat-in-pwt}

Let us first consider elastic scattering by a fixed potential associated with
some scattering centre or region. An incident particle may (for a pure quantum
state) be represented by a freely-evolving wave packet $\psi_{\mathrm{inc}%
}(\mathbf{x},t)$ that is spatially finite (that is, limited both
longitudinally and laterally), and that has mean momentum $\hslash\mathbf{k}$.
During the scattering, the wave function evolves into $\psi=\psi
_{\mathrm{inc}}+\psi_{\mathrm{sc}}$, where $\psi_{\mathrm{sc}}$ is the
scattered wave. At large distances from the scattering region, and off the
axis (through the scattering centre) parallel to the incident wave vector
$\mathbf{k}$, only the scattered wave $\psi_{\mathrm{sc}}$ contributes to the
particle current density $\mathbf{j}$ --- where $\mathbf{j}$ is often used in
textbook derivations of the scattering cross section.

As is well known, the mathematics of a fully time-dependent calculation of the
scattering of a finite packet may be simplified by resorting to a
time-independent treatment in which $\psi_{\mathrm{inc}}$ is taken to be an
infinitely-extended plane wave $e^{i\mathbf{k}\cdot\mathbf{x}}$. At large
distances from the scattering region the wave function $\psi=e^{i\mathbf{k}%
\cdot\mathbf{x}}+\psi_{\mathrm{sc}}$ (a time-independent eigenfunction of the
total Hamiltonian) has the asymptotic form%
\begin{equation}
\psi=e^{ikz}+f(\theta,\phi)\frac{e^{ikr}}{r}%
\end{equation}
(taking the $z$-axis parallel to $\mathbf{k}$, using spherical polar
coordinates centred on the scattering region, and ignoring overall
normalisation). The scattering amplitude $f$ gives the differential cross
section $d\sigma/d\Omega=$ $\left\vert f(\theta,\phi)\right\vert ^{2}$.

In the standard textbook derivation of $d\sigma/d\Omega$, the current density
$\mathbf{j}$ is used to calculate the rate of probability flow into an element
of solid angle $d\Omega$, where $\mathbf{j}$ is taken to be the current
associated with $\psi_{\mathrm{sc}}$ only, even though $\psi_{\mathrm{sc}}$
overlaps with $\psi_{\mathrm{inc}}=e^{i\mathbf{k}\cdot\mathbf{x}}$. This is
justified because the plane wave $e^{i\mathbf{k}\cdot\mathbf{x}}$ is, of
course, merely an abstraction used for mathematical convenience: a real
incident wave will be spatially limited, and will not overlap with the
scattered wave at the location of the particle detector (which is assumed to
be located off the axis of incidence, so as
not to be bathed in the incident beam).

The above standard discussion of scattering may readily be recast in
pilot-wave terms, where%
\begin{equation}
\mathbf{v}=\frac{\mathbf{j}}{\left\vert \psi\right\vert ^{2}}=\frac{\hslash
}{m}\operatorname{Im}\frac{\mathbf{\nabla}\psi}{\psi}=\frac{\mathbf{\nabla}%
S}{m}%
\end{equation}
(with $\psi=\left\vert \psi\right\vert e^{(i/\hslash)S}$) is interpreted as
the actual velocity field of an ensemble of particles with positions
distributed according to $\left\vert \psi\right\vert ^{2}$. The differential
cross section $d\sigma/d\Omega$ measures the fraction of incident particles
whose actual trajectories end (asymptotically) in the element of solid angle
$d\Omega$.

That a real incident packet is always spatially finite is, of course, an
elementary point known to every student of wave optics. This (often implicit)
assumption is essential to introductory textbook treatments of the scattering
of light, whether by a Hertzian dipole or by a diffraction grating. If the
incident wave were a literally infinite plane wave, then the scattered wave
would of course overlap with the incident wave everywhere, and no matter where
a detector was placed it would be affected by the incident wave as well as by
the scattered wave.

Certainly, the participants at the fifth Solvay conference --- many of whom
had extensive laboratory experience --- were aware of this simple point. We
emphasise this because, as we shall see, the finiteness of incident wave
packets played a central role in the discussions that took place regarding
scattering in de Broglie's theory.

Let us now consider inelastic scattering: specifically, the scattering of an
electron by a hydrogen atom initially in the ground state. The atom can become
excited by the collision, in which case the outgoing electron will have lost a
corresponding amount of energy. Let the scattering electron have position
$\mathbf{x}_{s}$ and the atomic electron have position $\mathbf{x}_{a}$. In a
time-dependent description, the total wave function evolves into%
\begin{multline}
\Psi(\mathbf{x}_{s},\mathbf{x}_{a},t)=\\
\phi_{0}(\mathbf{x}_{a})e^{-iE_{0}%
t/\hslash}\psi_{\mathrm{inc}}(\mathbf{x}_{s},t)+\sum_{n}\phi_{n}%
(\mathbf{x}_{a})e^{-iE_{n}t/\hslash}\psi_{n}(\mathbf{x}_{s},t)\ .
\label{Psiscatt}%
\end{multline}
Here, the first term is an initial product state, where $\phi_{0}$ is the
ground-state wave function of hydrogen with ground-state energy $E_{0}$ and
$\psi_{\mathrm{inc}}$ is a (finite) incident packet. The scattering terms have
components as shown, where the $\phi_{n}$ are the $n$th excited states of
hydrogen and the $\psi_{n}$ are outgoing wave packets. It may be shown by
standard techniques that, asymptotically, the $n$th outgoing packet $\psi_{n}$
is centred on a radius $r_{n}=(\hslash k_{n}/m)t$ from the scattering region,
where $k_{n}$ is the outgoing wave number fixed by energy conservation.

Because the outgoing (asymptotic) packets $\psi_{n}$ expand with different
speeds, they eventually become widely separated in space. The actual scattered
electron with position $\mathbf{x}_{s}(t)$ can occupy only one of these
non-overlapping packets, say $\psi_{i}$, and its velocity will then be
determined by $\psi_{i}$ alone. Further, the motion of the atomic electron
will be determined by the corresponding $\phi_{i}$ alone, and after the
scattering the atom will be (in effect) in an energy eigenstate $\phi_{i}$.

We are using two well-known properties of pilot-wave dynamics: (a) If a wave
function $\Psi=\left\vert \Psi\right\vert e^{(i/\hslash)S}$ is a superposition
of terms $\Psi_{1}+\Psi_{2}+$ .... having no overlap in configuration space,
then the phase gradient $\nabla S$ at the occupied point of configuration
space (which gives the velocity of the actual configuration) reduces to
$\nabla S_{i}$, where $\Psi_{i}=\left\vert \Psi_{i}\right\vert e^{(i/\hslash
)S_{i}}$ is the occupied packet. (b) If the occupied packet $\Psi_{i}$ is a
product over certain configuration components, then the velocities of those
components are determined by the associated factors in the product.

Applying (a) and (b) to the case discussed here, once the $\psi_{n}$ have
separated the total wave function $\Psi$ becomes a sum of non-overlapping
packets, where only one packet $\Psi_{i}=\phi_{i}(\mathbf{x}_{a}%
)e^{-iE_{i}t/\hslash}\psi_{i}(\mathbf{x}_{s},t)$ can contain the actual
configuration $(\mathbf{x}_{a},\mathbf{x}_{s})$. The velocity of the scattered
electron is then given by $\mathbf{\dot{x}}_{s}=(\hslash/m)\operatorname{Im}%
(\nabla\psi_{i}/\psi_{i})$, while the velocity of the atomic electron is given
by $\mathbf{\dot{x}}_{a}=(\hslash/m)\operatorname{Im}(\nabla\phi_{i}/\phi
_{i})$. Thus, there takes place an effective `collapse of the wave packet' to
the state $\phi_{i}\psi_{i}$.

It is straightforward to show that, if the initial ensemble of $\mathbf{x}%
_{s}$, $\mathbf{x}_{a}$ has distribution $\left\vert \Psi\right\vert ^{2}$,
the probability for ending in the $i$th packet --- that is, the probability
for the atom to end in the state $\phi_{i}$ --- will be given by $\int
d^{3}\mathbf{x}_{s}\ \left\vert \psi_{i}(\mathbf{x}_{s},t)\right\vert ^{2}$,
in accordance with the usual quantum result.

Further, the effective `collapse' to the state $\phi_{i}\psi_{i}$ is for all
practical purposes irreversible. As argued by Bohm (1952a, p.~178), the
scattered particle will subsequently interact with many other degrees of
freedom --- making it very unlikely that distinct states $\phi_{i}\psi_{i}$,
$\phi_{j}\psi_{j}$ ($i\neq j$) will interfere at later times, as this would
require the associated branches of the total wave function to overlap with
respect to \textit{every} degree of freedom involved.

Again, for mathematical convenience, one often
considers the limit in which the incident wave $\psi_{\mathrm{inc}}$ is
unlimited (a plane wave). This makes the calculation of $\Psi$ easier. In this
limit, the outgoing wave packets $\psi_{n}$ become unlimited too, and overlap
with each other everywhere: \textit{all} the terms in (\ref{Psiscatt}) then
overlap in every region of space. If one naively assumed that this limit
corresponded to a real situation, the outgoing electron would never reach a
constant velocity because it would be guided by a superposition of overlapping
terms, rather than by a single term in (\ref{Psiscatt}). Similarly, after the
scattering, the atomic electron would not be guided by a single eigenfunction
$\phi_{i}$, and the atom would not finish in a definite energy state. In any
real situation, of course, $\psi_{\mathrm{inc}}$ will be limited in space and
time, and at large times the outgoing packets $\psi_{n}$ will separate: the
trajectory of the scattered electron will be guided by only one of the
$\psi_{n}$ and (in regions outside the path of the incident beam) will
\textit{not} be affected by $\psi_{\mathrm{inc}}$. (Note
that longitudinal finiteness of the incident wave $\psi_{\mathrm{inc}}$ leads
to a separation of the outgoing waves $\psi_{n}$ from each other, while
lateral finiteness of $\psi_{\mathrm{inc}}$ ensures that $\psi_{\mathrm{inc}}$
does not overlap with the $\psi_{n}$ in regions off the axis of incidence.)

As we shall see, apart from the practical irreversibility of the effective
collapse process, the above `pilot-wave theory of scattering' seems to have
been more or less understood by de Broglie (and perhaps also by Brillouin) in
October 1927. A detailed treatment of scattering was given by Bohm in his
first paper on de Broglie-Bohm theory (Bohm 1952a).

In Bohm's second paper, however, appendix B gives a misleading account of the
de Broglie-Pauli encounter at the fifth Solvay conference (Bohm 1952b, pp.~191--2), 
and this seems to be the source of the widespread misunderstandings
concerning this encounter. Citing the proceedings of the fifth Solvay
conference, Bohm wrote the following:

\begin{quotation}
De Broglie's suggestions met strong objections on the part of Pauli, in
connection with the problem of inelastic scattering of a particle by a rigid
rotator. Since this problem is conceptually equivalent to that of inelastic
scattering of a particle by a hydrogen atom, which we have already treated
.... , we shall discuss the objections raised by Pauli in terms of the latter example.
\end{quotation}
Bohm then describes `Pauli's argument': taking the incoming particle to
have a plane wave function, all the terms in (\ref{Psiscatt}) overlap, so that
`neither atom nor the outgoing particle ever seem to approach a stationary
energy', contrary to what is observed experimentally. According to Bohm:

\begin{quotation}
Pauli therefore concluded that the interpretation proposed by de Broglie was
untenable. De Broglie seems to have agreed with the conclusion, since he
subsequently gave up his suggested interpretation.
\end{quotation}
Bohm then gives what he regards as his own, original answer to Pauli's objection:

\begin{quotation}
.... as is well known, the use of an incident plane wave of infinite extent is
an excessive abstraction, not realizable in practice. Actually, both the
incident and outgoing parts of the $\psi$-field will always take the form of
bounded packets. Moreover, .... all packets corresponding to different values
of $n$ will ultimately obtain classically describable separations. The
outgoing particle must enter one of these packets, .... leaving the hydrogen
atom in a definite but correlated stationary state.
\end{quotation}
By way of conclusion, Bohm writes:

\begin{quotation}
Thus, Pauli's objection is seen to be based on the use of the excessively
abstract model of an infinite plane wave.
\end{quotation}

At this point, in the light of what we have said above about the use of plane
waves in elementary wave optics and in scattering theory, it is natural to ask
how a physicist of Pauli's abilities could have made the glaring mistake that
Bohm claims he made. In fact, as we shall see in the next section, Pauli's
objection was \textit{not} based on a failure to appreciate the importance of
the finiteness of initial wave packets. On the contrary, Pauli's objection
shows some understanding of the crucial role played by limited packets in
pilot-wave theory. What really happened is that Pauli's objection involved a
peculiar and misleading analogy with optics, according to which the incident
packet appeared to be \textit{necessarily} unlimited, in conditions such as to
prevent the required separation into non-overlapping components.

\sectionmark{Elastic and inelastic scattering}
\section{Elastic and inelastic scattering: Born and Brillouin, Pauli and de
Broglie}\label{elastic-inelastic}\sectionmark{Elastic and inelastic scattering}

In the discussion following de Broglie's report, Born suggests (p.~\pageref{Bornobjection}) 
that de Broglie's guidance equation will fail for an
\textit{elastic} collision between an electron and an atom. Specifically, Born
asks if the electron speed will be the same before and after the collision, to
which de Broglie simply replies that it will.

Later in the same discussion, Brillouin (pp.~\pageref{Brill-beginning}~ff.) 
gives an extensive and detailed presentation, explaining how de Broglie's
theory accounts for the elastic scattering of a photon from a mirror.
Brillouin is quite explicit about the role played by the finite extension of
the incident packet. In his Fig.~2
, Brillouin shows an incident photon trajectory (at angle $\theta$ to the
normal) guided by an incoming and \textit{laterally-limited} packet. The
packet is reflected by the mirror, producing an outgoing packet that is again
laterally-limited. Near the mirror there is an interference zone, where the
incoming and outgoing packets overlap. As Brillouin puts it:

\begin{quotation}
Let us draw a diagram for the case of a limited beam of light falling on a
plane mirror; the interference is produced in the region of overlap of two beams.
\end{quotation}
Brillouin sketches the photon trajectory, which curves away from the
mirror as it enters the interference zone, moves approximately parallel to the
mirror while in the interference zone, and then moves away again, eventually
settling into a rectilinear motion guided by the outgoing packet
(Brillouin's Fig.~2
). Brillouin describes the trajectory thus:

\begin{quotation}
.... at first a rectilinear path in the incident beam, then a bending at the
edge of the interference zone, then a rectilinear path parallel to the mirror,
with the photon travelling in a bright fringe and avoiding the dark
interference fringes; then, when it comes out, the photon retreats following
the direction of the reflected light beam.
\end{quotation}
Here we have a clear description of an incident photon, guided by a
finite packet and moving uniformly towards the mirror, with the packet then
undergoing interference and scattering, while the photon is eventually carried
away --- again with a uniform motion --- by a finite outgoing packet. (The
ingoing and outgoing motions of the photon are strictly uniform, of course,
only in the limit where the guiding packets become infinitely broad.)

Despite his description in terms of finite packets, however, in order to
calculate the precise motion of the photon in the interference zone Brillouin
uses the standard device of treating the incoming packet as an infinite plane
wave. In this (abstract) approximation, he shows that the photon moves
parallel to the mirror with a speed $v=c\sin\theta$. Because the incoming
packet is in reality limited, a photon motion parallel to the mirror is
(approximately) realised only in the interference zone, as Brillouin sketches
in his accompanying Fig.~2.

While Brillouin's figure shows a laterally-limited packet, it is
clear from subsequent discussion that the incident packet was implicitly
regarded as limited {\em longitudinally} as well (as of course it must be in any
realistic situation). For in the general discussion the question of photon
reflection by a mirror was raised again, and Einstein asked (p.~\pageref{duringreflection}) 
what happens in de Broglie's theory in the case of
normal incidence ($\theta=0$), for which the formula $v=c\sin\theta$ predicts
that the photons will have zero speed. Piccard responded to Einstein's query,
and pointed out that, indeed, the photons are stationary in the limiting case
of normal incidence. The meaning of this exchange between Einstein and Piccard
is clear: if a longitudinally-limited packet were incident normally on the
mirror, the incident packet would carry the photon towards the mirror; in the
region where the incident and reflected packets overlap, the photon would be
at rest; once the packet has been reflected, the photon will be carried away
by the outgoing packet. The discussion of the case of normal incidence
implicitly assumes that the incident packet is longitudinally-limited.

Thus, Brillouin's example of photon reflection by a mirror illustrates the
point that the use of plane waves was for calculational convenience only, and
that, when it came to the discussion of real physical examples, it was clear
to all (and hardly worth mentioning explicitly) that incident waves were in
reality limited in extent (in all directions). Brillouin's mathematical use of
plane waves parallels their use in the general theory of scattering sketched
in section~\ref{scat-in-pwt}.

Elastic scattering is also discussed in de Broglie's report, for the
particular case of electrons incident on a fixed, periodic potential --- the
potential generated by a crystal lattice. This case is especially interesting
in the present context, because it involves the separation of the scattered
wave into non-overlapping packets --- the interference maxima of different
orders, well-known from the theory of X-ray diffraction --- with the particle
entering just one of these packets (a point that is relevant to a proper
understanding of the de Broglie-Pauli encounter). Here is how de Broglie
describes the diffraction of an electron wave by a crystal lattice (p.~\pageref{willpropagate}):

\begin{quotation}
.... the wave $\Psi$ will propagate following the general equation, in which
one has to insert the potentials created by the atoms of the crystal
considered as centres of force. One does not know the exact expression for
these potentials but, because of the regular distribution of atoms in the
crystal, one easily realises that the scattered amplitude will show maxima in
the directions predicted by Mr von Laue's theory. Because of the role of pilot
wave played by the wave $\Psi$, one must then observe a selective scattering
of the electrons in these directions.
\end{quotation}
Again, as in Brillouin's discussion, there is no need to mention
explicitly the obvious point that the incident wave $\Psi$ will be spatially limited.

Let us now turn to the inelastic case. This was discussed by de Broglie in his
report, in particular in the final part, which includes a review of recent
experiments involving the inelastic scattering of electrons by atoms of
helium. De Broglie noted that, according to Born's calculations (using his
statistical interpretation of the wave function), the differential cross
section should show maxima as a function of the scattering angle (p.~\pageref{hasstudied}):

\begin{quotation}
.... Mr Born has studied .... the collision of a narrow beam of electrons with
an atom. According to him, the curve giving the number of electrons that have
suffered an inelastic collision as a function of the scattering angle must
show maxima and minima ....~.
\end{quotation}
As de Broglie then discussed in detail, such maxima had been observed
experimentally by Dymond (though the results were only in qualitative
agreement with the predictions). Having summarised Dymond's results, de
Broglie commented (p.~\pageref{aboveresults}):

\begin{quotation}
The above results must very probably be interpreted with the aid of the new
Mechanics and are to be related to Mr Born's predictions.
\end{quotation}

We are now ready for a close examination of Pauli's objection in the general
discussion, according to which there is a difficulty with de Broglie's theory
in the case of inelastic collisions. That there might be such a difficulty was
in fact already suggested by Pauli a few months earlier, in a letter to Bohr
dated 6 August 1927, already quoted in chapter~\ref{deBroglieEss}. As well as
noting the exceptional quality of de Broglie's `Structure' paper (de Broglie
1927b), in his letter Pauli states that he is suspicious of de Broglie's
trajectories and cannot see how the theory could account for the discrete
energy exchange seen in individual inelastic collisions between electrons and
atoms (Pauli 1979, pp.~404--5). (Such discrete exchange had been observed, of
course, in the Franck-Hertz experiment.) This is essentially the objection
that Pauli raises less than three months later in Brussels.

In the general discussion (p.~\pageref{Pauli-deB-beginning}), Pauli begins
by stating his belief that de Broglie's theory works for elastic collisions:

\begin{quotation}
It seems to me that, concerning the statistical results of scattering
experiments, the conception of Mr de Broglie is in full agreement with Born's
theory in the case of elastic collisions .... .
\end{quotation}
This preliminary comment by Pauli is significant. For if, as Bohm
asserted in 1952, Pauli's objection was based on `the use of the excessively
abstract model of an infinite plane wave' (Bohm 1952b, p.~192), then Pauli
would have regarded de Broglie's theory as problematic \textit{even in the
elastic case}. For in an elastic collision, if the incident wave
$\psi_{\mathrm{inc}}$ is infinitely extended (a plane wave), then any part of
the outgoing region will be bathed in the incident wave: the scattered
particle will inevitably be affected by both parts of the superposition
$\psi_{\mathrm{inc}}+\psi_{\mathrm{sc}}$, and will never settle down to a
constant speed. Since Pauli agreed that the outgoing speed \textit{would} be
constant in the elastic case --- as de Broglie had asserted (in reply to Born)
in the discussion following his report --- Pauli presumably understood that
the finite incident packet would not affect the scattered particle.

Pauli goes on to claim that de Broglie's theory will not work for inelastic
collisions, in particular for the example of scattering by a rotator. To
understand Pauli's point, it is important to distinguish between the real
physical situation being discussed, and the optical analogy used by Pauli ---
an analogy that had been introduced by Fermi as a convenient (and as we shall
see limited) means to solve the scattering problem. Pauli's objection is
framed in terms of Fermi's analogy, and therein lies the confusion.

The real physical set-up consists of an electron moving in the $(x,y)$-plane
and colliding with a rotator. The latter is a model scattering centre with one
rotational degree of freedom represented by an angle $\varphi$%
.\footnote{Classically, a rotator might consist of a rigid body free to rotate
about a fixed axis. The quantum rotator was known to have a discrete spectrum
of quantised energy levels (corresponding to quantised states of angular
momentum), and was sometimes considered as a useful and simple model of a
quantum system.} Pauli took the initial wave function to be%
\begin{equation}
\psi_{0}(x,y,\varphi)\propto e^{(i/\hslash)(p_{x}x+p_{y}y+p_{\varphi}\varphi
)}\ ,
\end{equation}
with $p_{\varphi}$ restricted to $p_{\varphi}=m\hslash$ ($m=0,1,2,...$). The
inelastic scattering of an electron by a rotator had been treated by Fermi
(1926) using an analogy with optics, according to which the (time-independent) scattering of an
electron in two spatial dimensions by a rotator is mathematically equivalent
to the (time-independent) scattering of a (scalar) light wave $\psi(x,y,\varphi)$ in three spatial dimensions by
an infinite diffraction grating, with $\varphi$ interpreted as a third spatial
coordinate ranging over the whole real line $(-\infty,+\infty)$. The infinite
`grating' constitutes a periodic potential, arising mathematically from the
periodicity associated with the original variable $\varphi$. Similarly, the function $\psi(x,y,\varphi)$
is necessarily unlimited along the $\varphi$-axis. By construction, then, both
the incident wave and the grating are unlimited along the $\varphi$-axis.
Fermi's analogy is useful, because the different spectral orders for
diffracted beams emerging from the grating correspond to the possible final
(post-scattering) energy states of the rotator. However, as we shall see,
Fermi's analogy has only a very limited validity.

Pauli, then, presents his objection in terms of Fermi's optical analogy. He
says (p.~\pageref{essentialpoint}):

\begin{quotation}
It is, however, an essential point that, in the case where the rotator is in a
stationary state before the collision, the incident wave is unlimited in the
direction of the [$\varphi$-]axis. For this reason, the different spectral
orders of the grating will always be superposed at each point of configuration
space. If we then calculate, according to the precepts of Mr de Broglie, the
angular velocity of the rotator after the collision, we must find that this
velocity is not constant.
\end{quotation}
In Fermi's three-dimensional analogy, for an incident beam unlimited
along $\varphi$ (as well as along $x$ and $y$) the scattered waves will indeed
overlap everywhere. The final configuration will then be guided by a
superposition of all the final energy states, and the final velocity of the
configuration will not be constant. It then appears (according to Pauli) that
the final angular velocity of the rotator will not be constant, contrary to
what is expected for a stationary state, and that there will be no definite
outcome for the scattering experiment (that is, no definite final energy state
for the rotator).%

  \begin{figure}
    \centering
    \resizebox{\textwidth}{!}{\includegraphics[0mm,0mm][220.30mm,160.98mm]{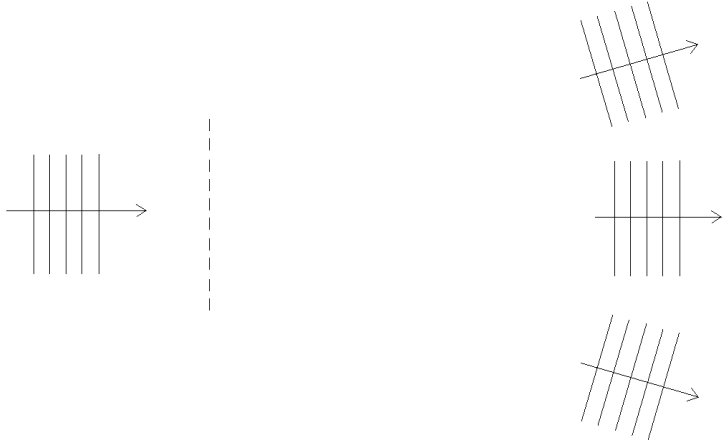}}
    \caption{Scattering of a laterally-limited wave by a finite diffraction
    grating, showing the separation of the first-order beams from the zeroth-order
    beam.}
    \label{deB-PauliA}
  \end{figure} 

\begin{figure}
    \centering
     \resizebox{\textwidth}{!}{\includegraphics[0mm,0mm][220.30mm,170.98mm]{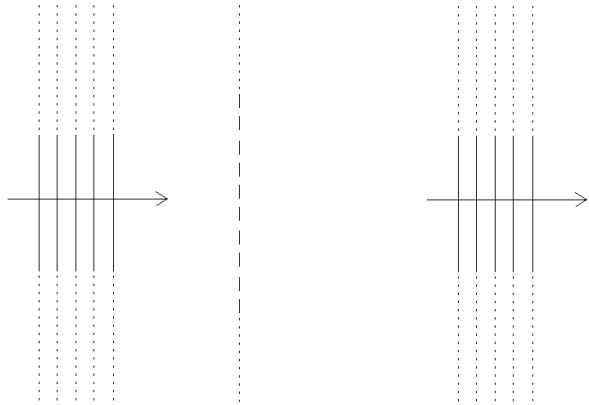}}
    \caption{Scattering of a laterally-unlimited wave by an infinite diffraction
    grating.}
    \label{deB-PauliB}
  \end{figure}

In the usual discussion of diffraction gratings in optics, it is of course
assumed that both the grating and the incident beam are laterally limited, so
that the emerging beams separate as shown in Fig.~\ref{deB-PauliA} (where only
the zeroth-order and first-order beams are drawn). But in Fermi's analogy,
there can be no such lateral limitation and no such separation. It might be
thought that, in the case of no lateral limitation, separation of an optical
beam would nevertheless take place if the incident wave were longitudinally
limited. However, as shown in Fig.~\ref{deB-PauliB}, symmetry dictates that
there will be no beams beyond the zeroth order.\footnote{Geometrically, the
presence of diverging higher-order beams (as in Fig.~\ref{deB-PauliA}) would
define a preferred central point on the grating. Roughly speaking, if one
considers Fig.~\ref{deB-PauliA} in the limit of an infinite grating (and of a
laterally unlimited incident beam), the higher-order emerging beams are
`pushed off to infinity', resulting in Fig.~\ref{deB-PauliB}.} Thus, if one
accepts Fermi's analogy with the scattering of light by an infinite grating,
even if one takes a longitudinally-limited incident light wave, there will
still be no separation and the difficulty remains . The only way to obtain a
separation of the scattered beams is through a lateral localisation along
$\varphi$ --- and according to Pauli's argument this is impossible.%

According to Fermi's analogy, then, it is inescapable that (for an initial
stationary state) the incident wave $\psi(x,y,\varphi)$ is unlimited on the
$\varphi$-axis, and one cannot avoid the conclusion that after the scattering
the rotator need not be in a definite energy eigenstate. There seems to be no
way out of Pauli's difficulty, and de Broglie's pilot-wave theory appears to
be untenable.

Today, one might answer Pauli's objection by considering the measuring
apparatus used to detect the outgoing scattered particle (or, used to measure
the energy of the atom). By including the degrees of freedom of the apparatus
in the quantum description, one could obtain a separation of the total wave
function into non-overlapping branches, resulting in a definite quantum
outcome (cf.\ section~\ref{qmeasinpwt}). However, in 1927 the
measuring apparatus was not normally considered to be part of the quantum
system; and even today, in the pilot-wave theory of scattering (sketched in
section~\ref{scat-in-pwt}), a definite outcome is generally
guaranteed by the separation of packets for the scattered particle. As we
shall now show, the usual pilot-wave theory of scattering in fact suffices in
Pauli's example too.

To see how Pauli's objection would normally be met today, note first that in
Pauli's example the real physical situation consists of an electron moving in
two spatial dimensions $x,y$ and colliding with a rotator whose angular
coordinate $\varphi$ ranges over the unit circle (from $0$ to $2\pi$). In de
Broglie's dynamics, the wave function $\psi(x,y,\varphi,t)$ yields velocities
$\dot{x},\dot{y}$ for the electron and $\dot{\varphi}$ for the rotator, where
these velocities are given by the quantum current divided by $\left\vert
\psi\right\vert ^{2}$. As we saw in the case of inelastic scattering by an
atom, the final wave function will take the form%
\begin{multline}
\psi(x,y,\varphi,t)=\\
\phi_{0}(\varphi)e^{-iE_{0}t/\hslash}\psi_{\mathrm{inc}%
}(x,y,t)+\sum_{n}\phi_{n}(\varphi)e^{-iE_{n}t/\hslash}\psi_{n}(x,y,t)\ ,
\label{P}%
\end{multline}
where now the $\phi_{n}$ are stationary states for the rotator. Once again,
for finite (localised) $\psi_{\mathrm{inc}}$, at large times the outgoing wave
packets $\psi_{n}$ will be centred on a radius $r_{n}=(\hslash k_{n}/m)t$ from
the scattering region, where again $k_{n}$ is the outgoing wave number fixed
by energy conservation. As before, because the $\psi_{n}$ expand with
different speeds they eventually become widely separated in space. The total
(electron-plus-rotator) configuration $(x,y,\varphi)$ will then occupy only
\textit{one} branch $\phi_{i}(\varphi)e^{-iE_{i}t/\hslash}\psi_{i}(x,y,t)$ of
the outgoing wave function. The scattered electron will then be guided by
$\psi_{i}$ only, and the speed of the electron will be constant. Further, the
motion of the rotator will be determined by the stationary state $\phi_{i}$,
and will also be uniform. As long as the incident wave $\psi_{\mathrm{inc}}$
is localised in $x$ and $y$, the final wave function separates in
configuration space and the scattering process has a definite outcome. Pauli's
objection therefore has a straightforward answer.

On Fermi's analogy, however, Pauli's example \textit{seems} to be equivalent
to the scattering of a laterally-unlimited light wave in three-dimensional
space by an infinite grating, for which no separation of beams can take place.
Since, for the original system, we have seen that the final state does
separate, it is clear that Fermi's analogy must be mistaken in some way.

To see what is wrong with Fermi's analogy, one must examine Fermi's original
paper. There, Fermi introduces the coordinates%
\begin{equation}
\xi=\sqrt{m}x\ ,\ \ \ \ \eta=\sqrt{m}y\ ,\ \ \ \ \zeta=\sqrt{J}\varphi\ ,
\end{equation}
and writes the time-independent Schr\"{o}dinger equation --- for the combined
rotator-plus-electron system of total energy $E$ --- in the form%
\begin{equation}
\frac{\partial^{2}\psi}{\partial\xi^{2}}+\frac{\partial^{2}\psi}{\partial
\eta^{2}}+\frac{\partial^{2}\psi}{\partial\zeta^{2}}+\frac{2}{\hslash^{2}%
}(E-V)\psi=0\ , \label{Fermi1}%
\end{equation}
where the potential energy $V$ is a periodic function of $\zeta$ with period
$2\pi\sqrt{J}$. Fermi then considers an optical analogue of the wave equation
(\ref{Fermi1}). As Fermi puts it (Fermi 1926, p.~400):

\begin{quotation}
In order to see the solution of [(\ref{Fermi1})], we consider
the optical analogy for the wave equation [(\ref{Fermi1})]. In the regions far
from the $\zeta$-axis, where $V$ vanishes, [(\ref{Fermi1})] is the wave equation
in an optically homogeneous medium; in the neighbourhood of the $\zeta$-axis,
the medium has an anomaly in the refractive index, which depends periodically
on $\zeta$. Optically this is nothing more than a linear grating of period
$2\pi\sqrt{J}$.
\end{quotation}
Fermi then goes on to consider a plane wave striking the grating, and
relates the outgoing beams of different spectral orders to different types of
collisions between the electron and the rotator.

Now, if we include the time dependence $\psi\propto e^{-(i/\hslash)Et}$, we
may write%
\begin{equation}
\frac{2}{\hslash^{2}}E\psi=-\frac{1}{c^{2}}\frac{\partial^{2}\psi}{\partial
t^{2}}%
\end{equation}
with $c\equiv\sqrt{E/2}$, and (\ref{Fermi1}) indeed coincides with the wave
equation of scalar optics for the case of a given frequency. However --- and
here is where Fermi's analogy breaks down --- because the `speed of light' $c$
depends on the energy $E$ (or frequency $E/h$), the time evolution of the
system in regions far from the $\zeta$-axis is in general \textit{not}
equivalent to wave propagation in `an optically homogeneous medium'. The
analogy holds only for one energy $E$ at a time (which is all Fermi needed to
consider). In a realistic case where the (finite) incident electron wave is a
sum over different momenta --- and as we have said, it was understood that in
any realistic case the incident wave would indeed be finite and therefore
equal to such a sum --- the problem of scattering by the rotator
\textit{cannot} be made equivalent to the scattering of light by a grating in
an otherwise optically homogeneous medium: on the contrary, the `speed of
light' $c$ would have to vary as the square root of the frequency.

Pauli's presentation does not mention that Fermi's optical analogy is valid
only for a single frequency. Since, as we have seen, the finiteness of
realistic incident packets was implicit in all these discussions, the
impression was probably given that Fermi's analogy holds generally. A finite 
(in $x$ and $y$) electron wave incident on the rotator would then be
expected to translate into a longitudinally finite but laterally infinite
(along $\zeta$) light wave incident on an infinite grating, with a resulting
lack of separation in the final state. Since, however, the quantity $c$ is
frequency-dependent, in general the analogy with light is invalid and the
conclusion unwarranted.

\begin{figure}
    \centering
     \resizebox{\textwidth}{!}{\includegraphics[0mm,0mm][220.30mm,170.98mm]{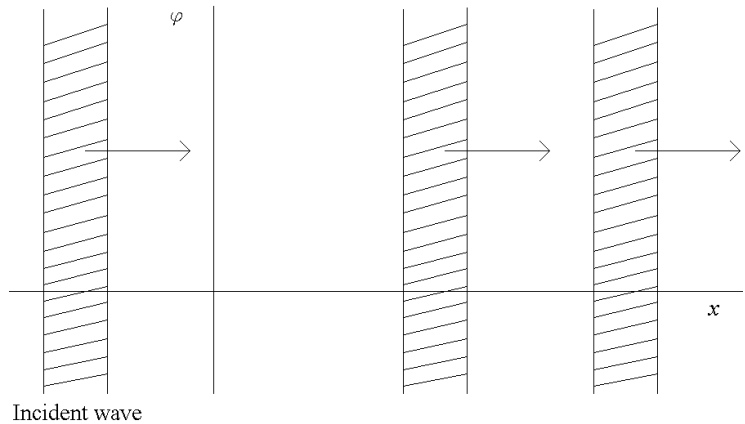}}
    \caption{True evolution of the electron-rotator wave function in Pauli's
    example. Adapted from de Broglie (1956, p.~176).}
    \label{deB-PauliC}
  \end{figure}

What really happens, then, in the two-dimensional scattering of an electron by
a rotator, if one interprets the angular coordinate $\varphi$ as a third
spatial axis \`{a} la Fermi? The answer is found in a detailed discussion of
Pauli's objection given by de Broglie, in a book published in 1956, whose
chapter 14 bears the title `Mr Pauli's objection to the pilot-wave theory' (de
Broglie 1956). There, de Broglie gives what is in fact the first proper
analysis of Pauli's example in terms of pilot-wave theory (with one spatial
dimension suppressed for simplicity). The result of de Broglie's analysis can
be easily seen by reconsidering the expression (\ref{P}), and allowing
$\varphi$ to range over the real line, with $\psi(x,y,\varphi,t)$ regarded as
a periodic function of $\varphi$ with period $2\pi$. Because $\psi$ separates
(as we have seen) into packets that are non-overlapping with respect to $x$
and $y$, the time evolution of $\psi$ will be as sketched in Fig.~\ref{deB-PauliC} 
(where we suppress $y$) --- a figure that we have adapted
from de Broglie's book (p.~176). The figure shows a wave moving along the
$x$-axis towards the rotator at $x=0$. The wave is longitudinally limited
(finite along $x$) and laterally unlimited (infinite along $\varphi$), and
separates as shown into similar packets moving at different speeds after the
scattering. (Only two of the final packets are drawn.)

The different speeds of the packets after the collision correspond, of course,
to the different possible kinetic energies of the electron after the
collision. Note that the crucial separation into packets moving at different
speeds would be ruled out if one were to mistakenly accept Fermi's analogy
with light scattering `in an optically homogeneous medium' --- for the
outgoing packets would then all have to move with the same speed (that of light).

A complete reply to Pauli's original objection should then make two points:
(1) Because of the frequency-dependent `speed of light' $c\equiv\sqrt{E/2}$,
Fermi's analogy with optics is of very limited validity and cannot be applied
to a real case with a finite incident wave. (2) For a finite incident electron
wave, a separation of packets does in fact take place with respect to the
spatial coordinates of the electron.

Let us now examine how de Broglie replied to Pauli in October 1927, in the
general discussion. As will become clear, de Broglie understood the general
separation mechanism required to yield a definite outcome, but he was misled
by the (generally false) optical analogy and phrased his answer in terms of it.

De Broglie replies by first pointing out the importance of having a
laterally-limited incident wave, to avoid overlap among the diffracted beams,
and to avoid overlap between these and the incident wave (p.~\pageref{pointedout}):

\begin{quotation}
The difficulty pointed out by Mr Pauli has an analogue in classical optics.
One can speak of the beam diffracted by a grating in a given direction only if
the grating and the incident wave are laterally limited, because otherwise all
the diffracted beams will overlap and be bathed in the incident wave. In
Fermi's problem, one must also assume the wave $\psi$ to be limited laterally
in configuration space.
\end{quotation}

De Broglie then notes that, if one can assume $\psi$ to be limited laterally,
then the system velocity will become constant once the diffracted waves have
separated (from each other and from the incident beam):

\begin{quotation}
.... the velocity of the representative point of the system will have a
constant value, and will correspond to a stationary state of the rotator, as
soon as the waves diffracted by the $\varphi$-axis will have separated from
the incident beam.
\end{quotation}
What de Broglie is describing here is precisely the separation into
non-overlapping packets in configuration space, which the pilot wave must
undergo in order for the scattering experiment to have a definite outcome ---
just as we have discussed in section~\ref{scat-in-pwt} above.

However, in Pauli's example, it appeared that the incident $\psi$ could not be
limited laterally. On this point, de Broglie claims that Fermi's configuration
space is artificial, having been formed `by rolling out along a line the
cyclic variable $\varphi$'. But as de Broglie's treatment of 1956 shows,
extending the range of $\varphi$ from the unit circle to the real line does
not really make any difference: it simply distributes copies of the
configuration space along the $\varphi$-axis (see Fig.~\ref{deB-PauliC}).

Pauli's presentation did not mention the frequency-dependent speed
$c\equiv\sqrt{E/2}$, and in his reply de Broglie did not mention it either.
Indeed, it seems that neither de Broglie in 1927 nor Bohm in 1952 noticed this
misleading aspect of Pauli's objection.

Note that finiteness of the incident wave with respect to $x$ and $y$ had not
been questioned by anyone. As we have seen, the finite spatial extension of
realistic wave packets was implicitly assumed by all. Pauli had raised the
impossibility of finiteness \textit{specifically} with respect to the
$\varphi$-axis, and de Broglie responded on this specific point alone. While
de Broglie was misled by Fermi's analogy, even so his remarks contain the key
point, later developed in detail by Bohm --- that finiteness of the initial
packet will ensure a separation into non-overlapping final packets in
configuration space. As de Broglie himself put it in his book of 1956, in
reference to his 1927 reply to Pauli: `Thus I had indeed realised .... that 
the answer to Mr Pauli's objection had to rest on
the fact that the wave trains are always limited, an idea that has been
taken up again by Mr Bohm in his recent papers' (de
Broglie 1956, p.~176).

Finally, we point out that --- leaving aside the misleading nature of Fermi's
optical analogy --- de Broglie's audience may well have understood his
description of wave packets separating in configuration space, for Born had
already described something very similar for the Wilson cloud chamber, earlier
in the general discussion (pp.~\pageref{Bornbeginning}~ff.). As we have seen 
in section~\ref{macrosup}, Born discussed the formation of a track in a cloud chamber
in terms of a branching of the total wave function in a multi-dimensional
configuration space (formed by all the particles involved). In particular, for
the simple case of an $\alpha$-particle interacting with two atoms in one
dimension, Born described an initially localised packet that separates into
two non-overlapping branches in (three-dimensional) configuration space --- as
sketched in Born's figure
. By the time de
Broglie came to reply to Pauli, then, the audience was already familiar with
the idea of a wave function evolving in configuration space and developing
non-overlapping branches. Thus, it seems more likely than not that this aspect
of de Broglie's reply to Pauli would have been understood.\footnote{It may
also be worth noting that, in his report, when presenting the pilot-wave
theory of many-body systems, de Broglie explicitly asserts (p.~\pageref{fullyaccords}) 
that his probability formula in configuration space
`fully accords .... with the results obtained by Mr Born for the collision of
an electron and an atom, and by Mr Fermi for the collision of an electron and
a rotator'. Because the printed version of de
Broglie's lecture was already complete by the time the conference took place,
this remark could not have been added after de Broglie's clash with Pauli. 
(De Broglie wrote to Lorentz on 11 October 1927 --- AHQP-LTZ-11, in French --- that he had
received the proofs of his report from Gauthier-Villars and corrected them.)}

\section{Quantum measurement in pilot-wave theory}\label{qmeasinpwt}

The general theory of quantum measurement in pilot-wave theory was first
developed by Bohm in his second paper on the theory (Bohm 1952b). Bohm
understood that the degrees of freedom associated with the measurement
apparatus were simply extra coordinates that should be included in the total
system (as in, for example, Born's 1927 discussion of the cloud chamber).
Thus, denoting by $x$ the coordinates of the `system' and $y$ the coordinates
of the `apparatus', the dynamics generates a trajectory $(x(t),y(t))$ for the
total configuration, guided by a wave function $\Psi(x,y,t)$ (where again the
velocity field $(\dot{x},\dot{y})$ is given by the quantum current of $\Psi$
divided by $\left\vert \Psi\right\vert ^{2}$). Bohm showed how, in the
circumstances corresponding to a quantum measurement, $\Psi$ separates into
non-overlapping branches and the system coordinate $x(t)$ is eventually guided
by an effectively `reduced' packet for the system.

As a simple example, let the system have an initial wave function%
\begin{equation}
\psi_{0}(x)=\sum_{n}c_{n}\phi_{n}(x)\ ,
\end{equation}
where the $\phi_{n}$ are eigenfunctions of some Hermitian operator $Q$ with
eigenvalues $q_{n}$. Suppose an experiment is performed that in quantum theory
would be called `a measurement of the observable $Q$'. This might be done by
coupling the system to a `pointer' with coordinate $y$ and initial wave
function $g_{0}(y)$, where $g_{0}$ is narrowly peaked around $y=0$. An
appropriate coupling might be described by an interaction Hamiltonian
$H=aQP_{y}$, where $a$ is a coupling constant and $P_{y}$ is the momentum
operator conjugate to $y$. If we neglect the rest of the Hamiltonian, for an
initial product wave function $\Psi_{0}(x,y)=\psi_{0}(x)g_{0}(y)$ the
Schr\"{o}dinger equation%
\begin{equation}
i\hslash\frac{\partial\Psi}{\partial t}=aQ\left(  -i\hslash\frac{\partial
}{\partial y}\right)  \Psi
\end{equation}
has the solution%
\begin{equation}
\Psi(x,y,t)=\sum_{n}c_{n}\phi_{n}(x)g_{0}(y-aq_{n}t)\ .
\end{equation}
Because $g_{0}$ is localised, $\Psi$ evolves into a superposition of terms
that \textit{separate} with respect to $y$ (in the sense of having negligible
overlap with respect to $y$). As we saw in the case of scattering, the final
configuration $(x,y)$ can be in only one `branch' of the superposition, say
$\phi_{i}(x)g_{0}(y-aq_{i}t)$, which will guide $(x(t),y(t))$ thereafter. And
because the active branch is a product in $x$ and $y$, the velocity of $x$
will be determined by $\phi_{i}(x)$ alone. In effect, at the end of the
quantum measurement, the system is guided by a `reduced' wave function
$\phi_{i}(x)$. Thus, the pointer plays the same role in the pilot-wave theory
of measurement as the scattered particle does in the pilot-wave theory of scattering.

It is also readily shown that, if the initial ensemble of $x,y$ has a
distribution $\left\vert \Psi_{0}(x,y)\right\vert ^{2}$, then the probability
of $x,y$ ending in the branch $\phi_{i}(x)g_{0}(y-aq_{i}t)$ is given by
$\left\vert c_{i}\right\vert ^{2}$, in agreement with the standard Born rule.

Finally, again as we saw in the case of scattering, the effective `collapse'
to the state $\phi_{i}(x)g_{0}(y-aq_{i}t)$ is for all practical purposes
irreversible. As noted by Bohm (1952b, p.~182), the apparatus coordinate $y$
will subsequently interact with many other degrees of freedom, making it very
unlikely that distinct states $\phi_{i}(x)g_{0}(y-aq_{i}t)$, $\phi_{j}%
(x)g_{0}(y-aq_{j}t)$ ($i\neq j$) will interfere later on, because the
associated branches of the total wave function would have to overlap with
respect to \textit{every} degree of freedom involved.

\section{Recoil of a single photon: Kramers and de Broglie}\label{recoil}\sectionmark{Recoil of a single photon}

We have sketched the pilot-wave theory of quantum measurement to emphasise
that, when one leaves the limited domain of particle scattering (by atoms, or
by fixed obstacles such as mirrors or diffracting screens), it can become
essential to describe the apparatus itself in terms of pilot-wave dynamics. In
some situations, the coordinates of the apparatus \textit{must} be included in
the total wave function (for the `supersystem' consisting of system plus
apparatus), to ensure that the total wave function separates into
non-overlapping branches in configuration space.

As we shall see in chapter~\ref{Pilot-wave-in-retrospect}, it seems that this point
was not properly appreciated by de Broglie in 1927. Indeed, most theoreticians
at the time simply applied quantum theory (in whatever form they preferred) to
microscopic systems only. Macroscopic apparatus was usually treated as a given
classical background. However, Born's treatment of the cloud chamber in the
general discussion shows that the key insight was already known: the apparatus
is made of atoms, and should ultimately be included in the wave function,
which will develop a branching structure as the measurement proceeds. All that
was needed, within pilot-wave theory, was to carry through the details
properly for a general quantum measurement --- something that de Broglie did
not see in 1927 and that Bohm did see in 1952.

Now, the need to include macroscopic equipment in the wave function is
relevant to a problem raised by Kramers in the general discussion, concerning
the recoil of a mirror due to the reflection of a \textit{single} photon. The
discussion had returned to the question of how de Broglie's theory accounts
for radiation pressure on a mirror (a subject that had been discussed at the
end of de Broglie's lecture). As Kramers put it (p.~\pageref{Kramersquestion}):

\begin{quotation}
But how is radiation pressure exerted in the case where it is so weak that
there is only one photon in the interference zone? .... And if there is only
one photon, how can one account for the sudden change of momentum suffered by
the reflecting object?
\end{quotation}
Neither de Broglie nor Brillouin were able to give an answer. De
Broglie claimed that pilot-wave theory in its current form was able to give
only the mean pressure exerted by an ensemble (or `cloud') of photons.

What was missing from de Broglie's understanding was that, in such a case, the
position of the mirror would have to be treated by pilot-wave dynamics and
included in the wave function. Schematically, let $x_{p}$ be the position of
the photon (on an axis normal to the surface of the mirror) and let $x_{m}$ be
the position of the reflecting surface. Let us treat the mirror as a very
massive but free body, with initial wave function $\phi(x_{m},0)$ localised
around $x_{m}=0$ (at $t=0$). The incident photon initially has a localised
wave function $\psi_{\mathrm{inc}}(x_{p},0)$ directed towards the mirror, with
mean momentum $\hslash k$. Roughly, if the photon packet strikes the mirror at
time $t_{0}$, then the initial total wave function $\Psi(x_{p},x_{m}%
,0)=\psi_{\mathrm{inc}}(x_{p},0)\phi(x_{m},0)$ will evolve into a wave
function of the schematic form%
\begin{equation}
\Psi(x_{p},x_{m},t)\sim\psi_{\mathrm{ref}}(x_{p},t)\phi(x_{m}-(\Delta
p/M)(t-t_{0}),0)e^{(i/\hslash)(\Delta p)x_{m}}\ ,
\end{equation}
where $\psi_{\mathrm{ref}}$ is a reflected packet directed away from the
mirror, $\Delta p=2\hslash k$ is the momentum transferred to the mirror, $M$
is the mass of the mirror, and $\phi(x_{m}-(\Delta p/M)(t-t_{0}),0)$ is a
packet (whose spreading we ignore) moving to the right with speed $\Delta
p/M$. The actual coordinates $x_{p},x_{m}$ will be guided by $\Psi$ in
accordance with de Broglie's equation, and the position $x_{m}(t)$ of the
mirror will follow the moving packet. The recoil of the mirror can therefore
be accounted for.

Thus, while de Broglie had the complete pilot-wave dynamics of a many-body
system, he seems not to have understood that it is sometimes necessary to
include the coordinates of macroscopic equipment in the pilot-wave
description. Otherwise, he might have been able to answer Kramers.

On the other hand, in ordinary quantum theory too, a proper explanation for
the recoil of the mirror would also have to treat the mirror as part of the
quantum system. If the mirror is regarded as a classical object, then quantum
theory would strictly speaking be as powerless as pilot-wave theory. Perhaps
this is why Brillouin made the following remark (p.~\pageref{notheory}):

\begin{quotation}
No theory currently gives the answer to Mr Kramers' question.
\end{quotation}

\setcounter{endnote}{0}
\setcounter{equation}{0}

\chapter{Pilot-wave theory in retrospect}\label{Pilot-wave-in-retrospect}\chaptermark{Pilot-wave theory in retrospect}

As we discussed in section~\ref{deB-1927-Solvay-report}, in his
Solvay lecture of 1927 de Broglie presented the pilot-wave dynamics of a
nonrelativistic many-body system, and outlined some simple applications of his
`new dynamics of quanta' (to interference, diffraction, and atomic
transitions). Further, as we saw in section~\ref{elastic-inelastic}, contrary 
to a widespread misunderstanding, in the general discussion
de Broglie's reply to Pauli's objection contained the essential points needed
to treat inelastic scattering (even if Fermi's misleading optical analogy
confused matters): in particular, de Broglie correctly indicated how definite
quantum outcomes in scattering processes arise from a separation of wave
packets in configuration space. We also saw in section~\ref{recoil} 
that de Broglie was unable to reply to a query from Kramers
concerning the recoil of a single photon on a mirror: to do so, he would have
had to introduce a joint wave function for the photon and the mirror.

De Broglie's theory was revived by Bohm 25 years later (Bohm 1952a,b) (though
with the dynamics written in terms of a law for acceleration instead of a law
for velocity). Bohm's truly new and very important contribution was a
pilot-wave account of the general quantum theory of measurement, with
macroscopic equipment (pointers, etc.) treated as part of the quantum system.
In effect, in 1952 Bohm provided a detailed derivation of quantum
phenomenology from de Broglie's dynamics of 1927 (albeit with the dynamical
equations written differently).

Despite this success, until about the late 1990s most physicists still
believed that hidden-variables theories such as de Broglie's could not
possibly reproduce the predictions of quantum theory (even for simple cases
such as the two-slit experiment, cf. section~\ref{prob-intrfce}).
Or, they believed that such theories had been disproved by experiments testing
EPR-type correlations and demonstrating violations of Bell's inequality. As
just one striking example of the latter belief, in the early 1990s James T.~Cushing, 
then a professor of physics and of philosophy at the University of
Notre Dame, submitted a research proposal to the US\ National Science
Foundation `for theoretical work to be done, within the framework of Bohm's
version of quantum theory, on some foundational questions in quantum
mechanics' (Cushing 1996, p.~6), and received the following evaluation:

\begin{quotation}
The subject under consideration, the rival Copenhagen and causal (Bohm)
interpretations of the quantum theory, has been discussed for many years and
in the opinion of several members of the Physics Division of the NSF, the
situation has been settled. The causal interpretation is inconsistent with
experiments which test Bell's inequalities. Consequently .... funding .... a
research program in this area would be unwise. (Cushing 1996, p.~6)
\end{quotation}
This is ironic because, as even a superficial reading of Bell's
original papers shows, the nonlocal theory of de Broglie and Bohm was a
primary motivation for Bell's work on his famous inequalities (Bell 1964,
1966). Bell knew that pilot-wave theory was empirically equivalent to quantum
theory, and wanted to find out if the nonlocality was a peculiarity of this
particular model, or if it was a general feature of all hidden-variables
theories. Bell's conclusion was that local theories have to satisfy his
inequality, which is inconsistent with EPR-type correlations, and that
therefore any viable theory must be nonlocal --- like pilot-wave theory. There
was never any question, in experimental tests of Bell's inequality, of testing
the nonlocal theory of de Broglie and Bohm; rather, it was the class of local
theories that was being tested.\footnote{See chapter 9 of Cushing (1994) for
an extensive discussion of the generally hostile reactions to and
misrepresentations of Bohm's 1952 papers.}

Despite this and other misunderstandings, in recent years the pilot-wave
theory of de Broglie and Bohm, with particle trajectories guided by a
physically-real wave function, has gained wide acceptance as an alternative
(though little used) formulation of quantum theory. While it is still
occasionally asserted that any theory with trajectories (or other hidden
variables) must disagree with experiment, such erroneous claims have become
much less frequent.\footnote{For example, in his book \textit{The Elegant
Universe}, Greene (2000) asserts that not only are quantum particle
trajectories unmeasurable (because of the uncertainty principle), their very
existence is ruled out by experiments testing the Bell inequalities: `....
theoretical progress spearheaded by the late Irish physicist John Bell and the
experimental results of Alain Aspect and his collaborators have shown
convincingly that .... [e]lectrons --- and everything else for that matter ---
cannot be described as simultaneously being at such-and-such location
\textit{and} having such-and-such speed' (p.~114). This is corrected a few
years later in \textit{The Fabric of the Cosmos}, where, commenting on `Bohm's
approach' Greene (2005) writes that it `does not fall afoul of Bell's results
because .... possessing definite properties forbidden by quantum uncertainty
is \textit{not} ruled out; only locality is ruled out .... ' (p.~206, original
italics).} This change in attitude seems to have been largely a result of the
publication in 1987 of Bell's influential book on the foundations of quantum
theory (Bell 1987), several chapters of which consisted of pedagogical
explanations of pilot-wave theory as an objective and deterministic account of
quantum phenomena. Leaving aside the question of whether or not pilot-wave
theory (or de Broglie-Bohm theory) is closer to the truth about the quantum
world than other formulations, it is a remarkable fact that it took about
three-quarters of a century for the theory to become widely accepted as an
internally consistent alternative.

Even today, however, there are widespread misconceptions not only about the
physics of pilot-wave theory, but also about its history. It is generally
recognised that de Broglie worked along pilot-wave lines in the 1920s, that
Bohm developed and extended the theory in 1952, and that Bell publicised the
theory in his book in 1987. But the full extent of de Broglie's contributions
in the 1920s is usually not recognised.

A careful examination of the proceedings of the fifth Solvay conference
changes our perception of pilot-wave theory, both as a physical theory and as
a part of the history of quantum physics.

\section{Historical misconceptions}\label{hist-mis}

Many of the widespread misconceptions about the history of pilot-wave theory
are conveniently summarised in the following extract from the book by Bohm and
Hiley (1993, pp.~38--9):

\begin{quotation}
The idea of a `pilot wave' that guides the movement of the electron was first
suggested by de Broglie in 1927, but only in connection with the one-body
system. De Broglie presented this idea at the 1927 Solvay Congress where it
was strongly criticised by Pauli. His most important criticism was that, in a
two-body scattering process, the model could not be applied coherently. In
consequence de Broglie abandoned his suggestion. The idea of a pilot wave was
proposed again in 1952 by Bohm in which an interpretation for the many-body
system was given. This latter made it possible to answer Pauli's criticism
.... .\footnote{Similar historical misconceptions appear in Cushing (1994, 
pp.~118--21, 149).}
\end{quotation}
As we have by now repeatedly emphasised, in his 1927 Solvay report (pp.~\pageref{deBsection7}~f.) 
de Broglie did in fact present pilot-wave theory
in configuration space for a many-body system, not just the one-body theory in
3-space; and further, as we saw in section~\ref{elastic-inelastic}, 
de Broglie's reply to Pauli's criticism contained the essential
ideas needed for a proper rebuttal.\footnote{To our knowledge, Bonk (1994) is
the only other author to have noticed that de Broglie's reply to Pauli was
indeed along the right lines.} As we shall see below, the claim that de
Broglie abandoned his theory because of Pauli's criticism is also not true.

Contrary to widespread belief, then, the many-body theory with a guiding wave
in configuration space is originally due to de Broglie and not Bohm; and in
1927, de Broglie did understand the essentials of the pilot-wave theory of
scattering. Thus the main content of Bohm's first paper of 1952 (Bohm 1952a)
--- which presents the dynamics (though in terms of acceleration), with
applications to scattering --- was already known to de Broglie in 1927.

Note that the theory was regarded as only provisional, by both de Broglie in
1927 and by Bohm in 1952. In particular, as we saw in sections~\ref{sigStruc} 
and \ref{deB-1927-Solvay-report}, de Broglie regarded the introduction of a pilot wave in
configuration space as a provisional measure. Bohm, on the other hand,
suggested that the basic principles of the theory would break down at nuclear
distances of order $10^{-13}$ cm (Bohm 1952a, pp.~178--9).

As we have said, in contrast with de Broglie's presentation of 1927, Bohm's
dynamics of 1952 was based on acceleration, not velocity. For de Broglie, the
basic law of motion for particles with masses $m_{i}$ and wave function
$\Psi=\left\vert \Psi\right\vert e^{(i/\hbar)S}$ was the guidance equation%
\begin{equation}
m_{i}\frac{d\mathbf{x}_{i}}{dt}=\mathbf{\nabla}_{i}S\ , \label{ge}%
\end{equation}
whereas for Bohm, the basic law of motion was the Newtonian equation%
\begin{equation}
\frac{d\mathbf{p}_{i}}{dt}=m_{i}\frac{d^{2}\mathbf{x}_{i}}{dt^{2}%
}=-\mathbf{\nabla}_{i}(V+Q)\ , \label{nl}%
\end{equation}
where%
\begin{equation}
Q\equiv-\sum_{i}\frac{\hslash^{2}}{2m_{i}}\frac{\nabla_{i}^{2}\left\vert
\Psi\right\vert }{\left\vert \Psi\right\vert } \label{Q}%
\end{equation}
is the `quantum potential'.

Taking the time derivative of (\ref{ge}) and using the Schr\"{o}dinger
equation yields precisely (\ref{nl}). For Bohm, however, (\ref{ge}) was not a
law of motion but rather a constraint $\mathbf{p}_{i}=\mathbf{\nabla}_{i}S$ to
be imposed on the \textit{initial} momenta (Bohm 1952a, p.~170). This initial
constraint happens to be preserved in time by (\ref{nl}), which Bohm regarded
as the true law of motion. Indeed, Bohm suggested that the initial constraint
$\mathbf{p}_{i}=\mathbf{\nabla}_{i}S$ could be relaxed, leading to corrections
to quantum theory:

\begin{quotation}
.... this restriction is not inherent in the conceptual structure .... it is
quite consistent in our interpretation to contemplate modifications in the
theory, which permit an arbitrary relation between $\mathbf{p}$ and
$\mathbf{\nabla}S(\mathbf{x})$. (Bohm 1952a, pp.~170--71)
\end{quotation}
For de Broglie, in contrast, there was never any question of relaxing
(\ref{ge}): he regarded (\ref{ge}) as the basic law of motion for a new form
of particle dynamics, and indeed for him (\ref{ge}) embodied --- as we saw in
chapter~\ref{deBroglieEss} --- the unification of the principles of
Maupertuis and Fermat, a unification that he regarded as the guiding principle
of his new dynamics. (De Broglie did mention in passing, however, the
alternative formulation in terms of acceleration, both in his `Structure'
paper (cf.\ section~\ref{Structure}) and in his Solvay report
(p.~\pageref{anotherobservation}).)

Some authors seem to believe that the `recasting' of Bohm's second-order
dynamics into first-order form was due to Bell (1987). But in fact, Bell's
(pedagogical) presentation of the theory --- based on the guidance equation
for velocity, and ignoring the notion of quantum potential --- was identical
to de Broglie's original presentation. De Broglie's first-order dynamics of
1927 is sometimes referred to as `Bohmian mechanics'. As already noted in
chapter~\ref{deBroglieEss}, this is a misnomer: firstly because of de
Broglie's priority, and secondly because Bohm's mechanics of 1952 was actually
second-order in time.

Another common historical misconception concerns the reception of de Broglie's
theory at the Solvay conference. It is usually said that de Broglie's ideas
attracted hardly any attention. It is difficult to understand how such an
impression originated, for even a cursory perusal of the proceedings reveals
that de Broglie's theory was extensively discussed, both after de Broglie's
lecture and during the general discussion. Nevertheless, in his classic
account of the historical development of quantum theory, Jammer asserts that
when de Broglie presented his theory:

\begin{quotation}
It was immediately clear that nobody accepted his ideas .... . In fact, with
the exception of some remarks by Pauli .... de Broglie's causal interpretation
was not even further discussed at the meeting. Only Einstein once referred to
it \textit{en passant}. (Jammer 1966, p.~357)\footnote{Jammer adds footnoted
references to pp. 280 and 256 respectively of the original proceedings, where Pauli's
objection (involving Fermi's treatment of the rotator) appears, and where, in
his main contribution to the general discussion, Einstein comments that in his
opinion de Broglie `is right to search in this direction' (p.~\pageref{searchdirection}).}
\end{quotation}
It appears to have escaped Jammer's attention that the general
discussion contains extensive and varied comments on many aspects of de
Broglie's theory (including a query by Einstein about the speed of photons),
as does the discussion after de Broglie's lecture, and that support for de
Broglie's ideas was expressed by Brillouin and by Einstein.

In a later historical study, again, Jammer (1974, pp.~110--11) writes --- in
reference to the fifth Solvay conference --- that de Broglie's theory `was
hardly discussed at all', and that `the only serious reaction came from
Pauli'. In the same study, Jammer quotes extensively from de Broglie's report
and from the general discussion, apparently without noticing the extensive
discussions of de Broglie's theory that appear both after de Broglie's report
and in the general discussion.

But Jammer is by no means the only historian to have given short shrift to de
Broglie's major presence at the 1927 Solvay conference.

In his book \textit{The Solvay Conferences on Physics}, Mehra (1975, p.~xvi)
quotes de Broglie himself as saying, with reference to his presentation of
pilot-wave theory in 1927, that `it received hardly any attention'. But these
words were written by de Broglie some 46 years later (de Broglie 1974), and de
Broglie's recollection (or misrecollection) after nearly half a century is
belied by the content of the published proceedings.\footnote{In fact, several
commentators have drawn erroneous conclusions about de Broglie, by relying on
mistaken `recollections' written by de Broglie himself decades later.}

In volume 6 of their monumental \textit{The Historical Development of Quan\-tum
Theory}, Mehra and Rechenberg (2000, pp.~246--50) devote several pages to the
published proceedings of the 1927 Solvay conference, focussing on the general
discussion --- mainly on the comments by Einstein, Dirac and Heisenberg --- as
well as on the unpublished comments by Bohr. The rest of the general
discussion is summarised in a single sentence (p.~250):

\begin{quotation}
After the Einstein-Pauli-Dirac-Heisenberg exchange, the general discussion
turned to more technical problems connected with the description of photons
and electrons in quantum mechanics, as well as with the details of de
Broglie's recent ideas.
\end{quotation}
It is added that `though these points possess some intrinsic interest,
they do not throw much light on the interpretation debate'.

As we shall see in section~\ref{standardhistory}, it
would appear that, for Mehra and Rechenberg, as indeed for most commentators,
the `interpretation debate' centred mainly around private or semi-private
discussions between Bohr and Einstein, and that the many pages of published
discussions were of comparatively little interest. Thus the remarkable
downplaying of the discussion of de Broglie's theory, as well as of other
ideas, is coupled with a strong tendency --- on the part of many authors ---
to portray (incorrectly) the 1927 conference as focussed primarily on the
confrontation between Bohr and Einstein.

We have already noted the widespread historical misconceptions concerning the
de Broglie-Pauli encounter in the general discussion. A related misconception
concerns de Broglie's thinking in the immediate aftermath of the Solvay
conference. It is often asserted that, soon after the conference, de Broglie
abandoned his theory primarily because of Pauli's criticism. This is not
correct. In his book \textit{An Introduction to the Study of Wave Mechanics}
(de Broglie 1930), which was published just three years after the Solvay
meeting, de Broglie gives three main reasons for why he considers his
pilot-wave theory to be unsatisfactory. First, de Broglie considers (p.~120) a
particle incident on an imperfectly reflecting mirror, and notes that if the
particle is found in the transmitted beam then the reflected part of the wave
must disappear (this being `a necessary consequence of the interference
principle'). De Broglie concludes that `the wave is not a physical phenomenon
in the old sense of the word. It is of the nature of a symbolic representation
of a probability .... '. Here, de Broglie did not understand how pilot-wave
theory accounts for the effective (and practically irreversible) collapse of
the wave packet, by means of a separation into non-overlapping branches
involving many degrees of freedom (cf.\ sections~{scat-in-pwt}
and \ref{qmeasinpwt}). Second, de Broglie notes (pp.~121, 133)
that a particle in free space guided by a superposition of plane waves would
have a rapidly-varying velocity and energy, and he cannot see how this could
be consistent with the outcomes of quantum energy measurements, which would
coincide strictly with the energy eigenvalues present in the superposition. To
solve this second problem, de Broglie would have had to apply pilot-wave
dynamics to the process of quantum measurement itself --- including the
apparatus in the wave function if necessary --- as done much later by Bohm
(see section\ref{qmeasinpwt}~). This question of energy
measurement bears some similarity to that raised by Pauli, and perhaps Pauli's
query set de Broglie thinking about this problem. But even so, de Broglie in
effect gave the essence of a correct reply to Pauli's query, and the problem
of energy measurement was posed by de Broglie himself. Third, in applying
pilot-wave theory to photons, de Broglie finds (p.~132) that in some
circumstances (specifically, in the interference zone close to an imperfectly
reflecting mirror) the photon trajectories have superluminal speeds, which he
considers unacceptable. De Broglie's book does not mention Pauli's criticism.

It is also often claimed that, when de Broglie abandoned his pilot-wave theory
(soon after the Solvay conference), he quickly adopted the views of Bohr and
Heisenberg. Thus, for example, Cushing (1994, p.~121) writes: `By early 1928
he [de Broglie] had decided to adopt the views of Bohr and
Heisenberg'.\footnote{As evidence for this, Cushing cites later recollections
by de Broglie in his book \textit{Physics and Microphysics} (de Broglie 1955),
which was originally published in French in 1947. Again, de Broglie's
recollections decades later do not seem a reliable guide to what actually
happened circa 1927.} But de Broglie's book of 1930, in which the above
difficulties with pilot-wave theory are described, contains a `General
introduction' that is `the reproduction of a communication made by the author
at the meeting of the British Association for the Advancement of Science held
in Glasgow in September, 1928' (de Broglie 1930, p.~1). This introduction
therefore gives an overview of de Broglie's thinking in late 1928, almost a
year after the fifth Solvay conference. While de Broglie makes it clear (p.~7)
that in his view it is `not possible to regard the theory of the pilot-wave as
satisfactory', and states that the `point of view developed by Heisenberg and
Bohr .... appears to contain a large body of truth', his concluding paragraph
shows that he was still not satisfied:

\begin{quotation}
To sum up, the physical interpretation of the new mechanics remains an
extremely difficult question .... the dualism of waves and particles must be
admitted .... . Unfortunately the profound nature of the two members in this
duality and the precise relation existing between them still remain a mystery.
(de Broglie 1930, p.~10)
\end{quotation}
At that time, de Broglie seems to have accepted the formalism of
quantum theory, and the statistical interpretation of the wave function, but
he still thought that an adequate physical understanding of wave-particle
duality had yet to be reached. Yet another year later, in December 1929,
doubts about the correct interpretation of quantum theory could still be
discerned in de Broglie's Nobel Lecture (de Broglie 1999):

\begin{quotation}
Is it even still possible to assume that at each moment the corpuscle occupies
a well-defined position in the wave and that the wave in its propagation
carries the corpuscle along in the same way as a wave would carry along a
cork? These are difficult questions and to discuss them would take us too far
and even to the confines of philosophy. All that I shall say about them here
is that nowadays the tendency in general is to assume that it is not
constantly possible to assign to the corpuscle a well-defined position in the wave.
\end{quotation}

This brings us to another historical misconception concerning de Broglie's
work. Nowadays, de Broglie-Bohm theory is often presented as a `completion' of
quantum theory.\footnote{An often-cited motivation for introducing the
trajectories is, of course, to solve the measurement problem.} Critics
sometimes view this as an arbitrary addition to or amendment of the quantum
formalism, the trajectories being viewed as an additional `baggage' being
appended to an already given formalism. Regardless of the truth or otherwise
of pilot-wave theory as a physical theory, such a view certainly does not do
justice to the historical facts. For the elements of pilot-wave theory ---
waves guiding particles via de Broglie's velocity formula --- were already in
place in de Broglie's thesis of 1924, before either matrix or wave mechanics
existed. And it was by following the lead of de Broglie's thesis that
Schr\"{o}dinger developed the wave equation for de Broglie's matter waves.
While Schr\"{o}dinger dropped the trajectories and considered only the waves,
nevertheless, historically speaking the wave function $\psi$ and the
Schr\"{o}dinger equation both grew out of de Broglie's phase
waves.\footnote{See also the discussion in section~\ref{Schr-deB}.}

The pilot-wave theory of 1927 was the culmination of de Broglie's independent
work from 1923, with a major input from Schr\"{o}dinger in 1926. There is no
sense in which de Broglie's trajectories were ever `added to' some
pre-existing theory. And when Bohm revived the theory in 1952, while it may
have seemed to Bohm's contemporaries (and indeed to Bohm himself) that he was
adding something to quantum theory, from a historical point of view Bohm was
simply reinstating what had been there from the beginning.

The failure to acknowledge the priority of de Broglie's thinking from 1923 is
visible even in the discussions of 1927. In the discussion following de
Broglie's lecture, Pauli (p.~\pageref{smallremark}) presents what he claims
is the central idea of de Broglie's theory:

\begin{quotation}
I should like to make a small remark on what seems to me to be the
mathematical basis of Mr de Broglie's viewpoint concerning particles in motion
on definite trajectories. His conception is based on the principle of
conservation of charge .... if in a field theory there exists a conservation
principle .... it is always formally possible to introduce a velocity vector
... and to imagine furthermore corpuscles that move following the current
lines of this vector.
\end{quotation}
Pauli's assertion that de Broglie's theory is based on the conservation
of charge makes sense only for one particle: for a many-body system, de
Broglie's velocity field is associated with conservation of probability in
configuration space, whereas conservation of charge is always tied to 3-space.
Still, the point remains that in any theory with a locally conserved
probability current, it is indeed possible to introduce particle trajectories
following the flow lines of that current. This way of presenting de Broglie's
theory then makes the trajectories look like an addendum to a pre-existing
structure: given the Schr\"{o}dinger equation with its locally conserved
current, one can add trajectories if one wishes. But to present the theory in
this way is a major distortion of the historical facts and priorities. The
essence of de Broglie's dynamics came before Schr\"{o}dinger's work, not
after. Further, de Broglie obtained his velocity law not from the
Schr\"{o}dinger current (which was unknown in 1923 or 1924) but from his
postulated relation between the principles of Maupertuis and Fermat. And
finally, while de Broglie's trajectory equation did not in fact owe anything
to the Schr\"{o}dinger equation, again, the latter equation arose out of
considerations (of the optical-mechanical analogy) that had been initiated by
de Broglie. It seems rather clear that the historical priority of de Broglie's
work was being downplayed by Pauli's remarks, as it has been more or less ever since.

A related historical misconception concerns the status of the Schr\"{o}dinger
equation in pilot-wave theory. It is sometimes argued that this equation has
no natural place in the theory, and that therefore the theory is artificial.
For example, commenting on what he calls `Bohm's theory', Polkinghorne (2002,
pp.~55, 89) writes:

\begin{quotation}
There is an air of contrivance about it that makes it unappealing. For
example, the hidden wave has to satisfy a wave equation. Where does this
equation come from? The frank answer is out of the air or, more accurately,
out of the mind of Schr\"{o}dinger. To get the right results, Bohm's wave
equation must be the Schr\"{o}dinger equation, but this does not follow from
any internal logic of the theory and it is simply an \textit{ad hoc} strategy
designed to produce empirically acceptable answers. .... It is on these
grounds that most physicists find the greatest difficulty with Bohmian ideas
.... the \textit{ad hoc} but necessary appropriation of the Schr\"{o}dinger
equation as the equation for the Bohmian wave has an unattractively
opportunist air to it.
\end{quotation}
Polkinghorne's comments are a fair criticism of Bohm's 1952
reformulation of de Broglie's theory. For as we have seen, Bohm based his
presentation on the Newtonian equation of motion (\ref{nl}) for acceleration,
with a `quantum potential' $Q$ determined by the wave function $\Psi$ through
(\ref{Q}): according to Bohm, $\Psi$ generates a `quantum force'
$-\mathbf{\nabla}_{i}Q$, which accounts for quantum effects. From this
Newtonian standpoint, the wave equation for $\Psi$ does indeed have nothing to
do with the internal logic of the theory: it is then fair to say that, in
Bohm's formulation, the Schr\"{o}dinger equation is `appropriated' for a
purpose quite foreign to the origins of that equation. However, Polkinghorne's
critique does not apply to pilot-wave theory in its original de Broglian
formulation, as a new form of dynamics in which particle velocities are
determined by guiding waves (rather than particle accelerations being
determined by Newtonian forces). For as a matter of historical fact, the
Schr\"{o}dinger equation \textit{did} follow from the internal logic of de
Broglie's theory. After all, Schr\"{o}dinger set out in the first place to
find the general wave equation for de Broglie's waves; and his derivation of
that equation owed much to the optical-mechanical analogy, which was a key
component of de Broglie's approach to dynamics.\footnote{Nowadays, it is
common in textbooks to motivate the free-particle Schr\"{o}dinger equation as
the simplest equation satisfied by a plane de Broglie wave $e^{i(\mathbf{k}%
\cdot\mathbf{x}-\omega t)}$ with the nonrelativistic dispersion relation
$\hslash\omega=(\hslash k)^{2}/2m$. This `derivation' is just as natural in
pilot-wave theory as it is in standard quantum theory.} It cannot be said that
de Broglie `appropriated' the Schr\"{o}dinger equation for a purpose foreign
to its origins, when the original purpose of the Schr\"{o}dinger equation was
in fact to describe de Broglie's waves.

\section{Why was de Broglie's theory rejected?}\label{why-rejected}

One might ask why de Broglie's theory did not gain widespread support soon
after 1927. This question has been considered by Bonk (1994), who applies
Bayesian reasoning to some of the discussions at the fifth Solvay conference,
in an attempt to understand the rapid acceptance of the `Copenhagen'
interpretation. It has also been suggested that were it not for certain
historical accidents, de Broglie's theory might have triumphed in 1927 and
emerged as the dominant interpretation of quantum theory (Cushing 1994,
chapter~10). It is difficult to evaluate how realistic Cushing's `alternative
historical scenario' might have been. Here, we shall simply highlight two
points that are usually overlooked, and which are relevant to any evaluation
of why de Broglie's theory did not carry the day.

Our first point is that, because de Broglie did in fact give a reply to
Pauli's criticism that contained the essence of a correct rebuttal --- and
because in contrast de Broglie completely failed to reply to the difficulty
raised by Kramers --- the question of whether or not de Broglie made a
convincing case at the fifth Solvay conference (and if not, why not) should be reconsidered.

Our second point is that there was a good technical reason for why `standard'
quantum theory had an advantage over pilot-wave theory in 1927. By the use of
a simple `collapse postulate' for microsystems, it was generally possible to
account for quantum measurement outcomes without having to treat the
measurement process (including the apparatus) quantum-mechanically. In
contrast, it was easy to find examples where in pilot-wave theory it was
essential to use the theory itself to analyse the measurement process: as we
saw in the last section, in his book of 1930 de Broglie could not see how, for
a particle guided by a superposition of energy eigenfunctions, an energy
measurement would give one of the results expected from quantum theory.
Agreement with quantum theory requires an analysis of the measurement process
in terms of pilot-wave dynamics, with the apparatus included as part of the
system, as shown by Bohm (1952b). In contrast, in ordinary quantum theory it
usually suffices in practice simply to apply a collapse rule to the
microscopic system alone. Such a collapse rule is of course merely pragmatic,
and defies precise formulation (there being no sharp boundary between
`microscopic' and `macroscopic', cf. section~\ref{mptoday}); yet, in 
ordinary laboratory situations, it yields predictions that
may be compared with experiment.

These two points should be taken into account in any full evaluation of why de
Broglie's theory was rejected in 1927 and shortly thereafter.

Further relevant material, that seems to have never been considered before,
consists of comments by Heisenberg on the possibility of a deterministic
pilot-wave interpretation. These comments do not appear in the proceedings of
the fifth Solvay conference, nor are they directed at de Broglie's theory.
Rather, they appear in a letter Heisenberg wrote to Einstein a few months
earlier, on 10 June 1927, and they concern \textit{Einstein's} version of
pilot-wave theory. Heisenberg's remarks could just as well have been directed
at de Broglie's theory, however. Both Einstein's theory and Heisenberg's
comments thereon are discussed in the next section.

\sectionmark{Einstein's alternative pilot-wave theory}
\section{Einstein's alternative pilot-wave theory (May 1927)}%
\label{EinsteinHV}\sectionmark{Einstein's alternative pilot-wave theory}

As we saw in section~\ref{Structure}, de Broglie first
arrived at pilot-wave theory in a paper published in \textit{Journal de
Physique} in May 1927 (de Broglie 1927b). In the same month, Einstein proposed
what in retrospect appears to be an alternative version of pilot-wave theory,
with particle trajectories determined by the many-body wave function but in a
manner different from that of de Broglie's theory. This new theory was
described in a paper entitled `Does Schr\"{o}dinger's wave mechanics determine
the motion of a system completely or only in the sense of statistics?', which
was presented on 5 May 1927 at a meeting of the Prussian Academy of Sciences.
On the same day Einstein wrote to Ehrenfest that `.... in a completely
unambiguous way, one can associate definite movements with the solutions [of
the Schr\"{o}dinger equation]' (Howard 1990, p.~89). However, on 21 May,
before the paper appeared in print, Einstein withdrew it from publication
(Kirsten and Treder 1979, p.~135; Pais 1982, p.~444). The paper remained
unpublished, but its contents are nevertheless known from a manuscript version
in the Einstein archive --- see Howard (1990, pp.~89--90) and Belousek 
(1996).\footnote{Archive reference: AEA 2-100.00 (in German); currently available on-line at 
http://www.alberteinstein.info/db/ViewDetails.do?DocumentID=34338 .}

Einstein's unpublished version of pilot-wave theory has some relevance to his
argument for the incompleteness of quantum theory, given in the general
discussion. As we saw in section~\ref{1927-incompleteness}, according 
to Einstein's argument, locality requires `that
one does not describe the [diffraction] process solely by the Schr\"{o}dinger
wave, but that at the same time one localises the particle during the
propagation' (p.~\pageref{locprop}). Einstein added: `I think that Mr de
Broglie is right to search in this direction' --- without mentioning that he
himself had recently made an attempt in the same direction.

It quite possible that, before abandoning his version of pilot-wave theory,
Einstein had considered presenting it at the fifth Solvay conference. Indeed,
had Einstein been happy with his new theory, there is every reason to think he
would have presented it a few months later in Brussels. As discussed in
section~\ref{scientific}, Lorentz had asked Einstein to give a report on
particle statistics, and while Einstein had agreed to do so, he was reluctant.
Less than a month after withdrawing his pilot-wave paper, Einstein withdrew
his committment to speak at the Solvay conference, writing to Lorentz on 17
June: `.... I kept hoping to be able to contribute something of value in
Brussels; I have now given up that hope. .... I did not take this lightly but
tried with all my strength .... ' (quoted from Pais 1982, p.~432). It then 
seems indeed probable that, instead of (or in addition to) speaking about 
particle statistics, Einstein had hoped to present something like his version of pilot-wave theory.

Let us now describe what Einstein's proposal was. (For more detailed
presentations, see Belousek (1996) and Holland (2005).)

Einstein's starting point is the time-independent Schr\"{o}dinger
equation\footnote{The particle masses make no explicit appearance, because
they have been absorbed into the configuration-space metric $g_{\mu\nu}$.
Einstein is following Schr\"{o}dinger's usage. In his second paper on wave
mechanics, Schr\"{o}dinger (1926c) introduced a non-Euclidean metric
determined by the kinetic energy (see also Schr\"{o}dinger's report, p.~\pageref{Hertz}). 
The Laplacian $\nabla^{2}$ is then understood in the
Riemannian sense, and the Schr\"{o}dinger equation indeed takes the form
(\ref{SchEin}) with no masses appearing explicitly. Note that also in
analytical mechanics it is sometimes convenient to write the kinetic energy
$\frac{1}{2}M_{\mu\nu}\dot{q}^{\mu}\dot{q}^{\nu}$ as $\frac{1}{2}(ds/dt)^{2}$
where $ds^{2}=M_{\mu\nu}dq^{\mu}dq^{\nu}$ is a line element with metric
$M_{\mu\nu}$ (Goldstein 1980, pp.~369--70).}%
\begin{equation}
-\frac{\hslash^{2}}{2}\nabla^{2}\Psi+V\Psi=E\Psi\label{SchEin}%
\end{equation}
for a many-body system with potential energy $V$, total energy $E$, and wave
function $\Psi$ on an $n$-dimensional configuration space. Einstein considers
(\ref{SchEin}) to define a kinetic energy%
\begin{equation}
E-V=-\frac{\hslash^{2}}{2}\frac{\nabla^{2}\Psi}{\Psi}\ , \label{KE}%
\end{equation}
which may also be written as%
\begin{equation}
E-V=\frac{1}{2}(ds/dt)^{2}=\frac{1}{2}g_{\mu\nu}\dot{q}^{\mu}\dot{q}^{\nu}\ ,
\label{KE2}%
\end{equation}
where $ds^{2}=g_{\mu\nu}dq^{\mu}dq^{\nu}$ is the line element in configuration
space and $\dot{q}^{\mu}$ is the system velocity. The theory is expressed in
terms of arbitrary coordinates $q^{\mu}$ with metric $g_{\mu\nu}$. The
Laplacian is then given by $\nabla^{2}\Psi=g^{\mu\nu}\nabla_{\mu}\nabla_{\nu
}\Psi$, where $\nabla_{\mu}$ is the covariant derivative. Einstein writes
$\nabla_{\mu}\nabla_{\nu}\Psi$ as $\Psi_{\mu\nu}$ (which he calls the `tensor
of $\Psi$-curvature'), and seeks unit vectors $A^{\mu}$ that extremise
$\Psi_{\mu\nu}A^{\mu}A^{\nu}$, leading to the eigenvalue problem%
\begin{equation}
(\Psi_{\mu\nu}-\lambda g_{\mu\nu})A^{\nu}=0\ ,
\end{equation}
with $n$ real and distinct solutions $\lambda_{(\alpha)}$. At each point, the
$A_{(\alpha)}^{\mu}$ define a local orthogonal coordinate system (with
Euclidean metric at that point). In this coordinate system, both $\Psi_{\mu
\nu}$ and $g_{\mu\nu}$ are diagonal, with components $\bar{\Psi}_{\alpha\beta
}=\lambda_{(\alpha)}\delta_{\alpha\beta}$ and $\bar{g}_{\alpha\beta}%
=\delta_{\alpha\beta}$. The two expressions (\ref{KE}) and (\ref{KE2}) for the
kinetic energy then become%
\begin{equation}
E-V=-\frac{\hslash^{2}}{2}\frac{1}{\Psi}\sum_{\alpha}\bar{\Psi}_{\alpha\alpha}%
\end{equation}
and%
\begin{equation}
E-V=\frac{1}{2}\sum_{\alpha}\overline{\dot{q}}_{\alpha}^{2}%
\end{equation}
respectively. Einstein then introduces the hypothesis that these two
expressions match term by term, so that%
\begin{equation}
\overline{\dot{q}}_{\alpha}^{2}=-\frac{\hslash^{2}}{\Psi}\bar{\Psi}%
_{\alpha\alpha}\ .
\end{equation}
Using $\bar{\Psi}_{\alpha\alpha}=\lambda_{(\alpha)}$, and transforming back to the original 
coordinate system, where $\dot{q}^{\mu}
=\sum_{\alpha}\overline{\dot{q}}_{\alpha}A_{(\alpha)}^{\mu}$, Einstein obtains the final
result%
\begin{equation}
\dot{q}^{\mu}=\hslash\sum_{\alpha}\pm\sqrt{\frac{-\lambda_{(\alpha)}}{\Psi}%
}A_{(\alpha)}^{\mu}\ , \label{Einvel}%
\end{equation}
which expresses $\dot{q}^{\mu}$ in terms of $\lambda_{(\alpha)}$ and
$A_{(\alpha)}^{\mu}$, where at each point in configuration space
$\lambda_{(\alpha)}$ and $A_{(\alpha)}^{\mu}$ are determined by the local
values of the wave function $\Psi$ and its derivatives. (The ambiguity in sign
is, according to Einstein, to be expected for quasiperiodic motions.)

Thus, according to (\ref{Einvel}), the system velocity $\dot{q}^{\mu}$ is
locally determined (up to signs) by $\Psi$ and its derivatives. This is
Einstein's proposed velocity law, to be compared and contrasted with de Broglie's.

The manuscript contains an additional note `added in proof', in which Einstein
asserts that the theory he has just outlined is physically unacceptable,
because it predicts that for a system composed of two independent subsystems,
with an additive Hamiltonian and a product wave function, the velocities for
one subsystem generally depend on the instantaneous coordinates of the other
subsystem. However, Einstein adds that Grommer has pointed out that this
problem could be avoided by replacing $\Psi$ with $\ln\Psi$ in the
construction of the velocity field. According to Einstein: `The elaboration of
this idea should occasion no difficulty ....' (Howard 1990, p.~90). Despite
Einstein's apparent optimism that Grommer's modification would work, it has
been assumed (Howard 1990, p.~90; Cushing 1994, p.~128) that Einstein withdrew
the paper because he soon realised that Grommer's suggestion did not work. But
it has been shown by Holland (2005) that the replacement $\Psi\rightarrow
\ln\Psi$ does in fact remove the difficulty raised by Einstein. Holland shows
further, however, that there are \textit{other} difficulties with Einstein's
theory --- with or without Grommer's modification. It is not known if Einstein
recognised these other difficulties, but if he did, they would certainly have
been convincing grounds on which to abandon the scheme altogether.

The difficulties raised by Holland are as follows. First, the theory applies
only to a limited range of quantum states: real stationary states with $E\geq
V$. Second, the system velocity $\dot{q}^{\mu}$ is defined only in a limited
domain of configuration space (for example, without Grommer's modification,
$\lambda_{(\alpha)}/\Psi$ must be negative for $\dot{q}^{\mu}$ to be real).
Third, even where $\dot{q}^{\mu}$ is defined, the continuity equation
$\nabla_{\mu}(\left\vert \Psi\right\vert ^{2}\dot{q}^{\mu})=0$ is generally
not satisfied (where $\left\vert \Psi\right\vert ^{2}$ is time-independent for
a stationary state), so that the velocity field $\dot{q}^{\mu}$ does not map
an initial Born-rule distribution into a final Born-rule distribution. This
last feature removes any realistic hope that the theory could reproduce the
predictions of standard quantum theory (Holland 2005).\footnote{As Holland
(2005) also points out, from a modern point of view the mutual dependence of
particle motions for product states, which Einstein found so unacceptable,
need not be a real difficulty. We now know that a hidden-variables theory must
be nonlocal, and there is no reason why in some theories the underlying
nonlocality could not exist for factorisable quantum states as well as for
entangled ones. Indeed, Holland gives an example of just such a theory.}

At the end of his manuscript (before the note added in proof), Einstein
writes: `.... the assignment of completely determined motions to solutions of
Schr\"{o}dinger's differential equation is .... just as possible as is the
assignment of determined motions to solutions of the Hamilton-Jacobi
differential equation in classical mechanics' (Belousek 1996, pp.~441--2).
Given the close relationship between the Hamilton-Jacobi function and the
phase $S$ of $\Psi$ (the latter reducing to the former in the short-wavelength
limit), it is natural to ask why Einstein did not consider de Broglie's
velocity field --- proportional to the phase gradient $\nabla S$ --- which is
a straightforward generalisation of the Hamilton-Jacobi velocity formula. As
well as being much simpler than Einstein's, de Broglie's velocity field
immediately satisfies Einstein's desired separability of particle motions for
product states.\footnote{It is not known whether or not Einstein noticed the
nonlocality of de Broglie's theory for entangled wave functions.} Why did
Einstein instead propose what seems a much more complicated and unwieldy
scheme to generate particle velocities from the wave function?

Perhaps Einstein did not adopt de Broglie's velocity field simply because, for
the real wave functions Einstein considered, the phase gradient $\nabla
S=\hslash\operatorname{Im}(\nabla\Psi/\Psi)$ vanishes; though it is not clear
why Einstein thought one could restrict attention to such wave functions. As
for the seemingly peculiar construction that Einstein did adopt, it should be
noted that Einstein was (as he himself states) following Schr\"{o}dinger in
his use of a non-Euclidean metric determined by the kinetic energy
(Schr\"{o}dinger 1926c). Further details of Einstein's construction may have
been related to another idea Einstein was pursuing: that quantisation
conditions could arise from a generally-covariant and `overdetermined' field
theory that constrains initial states, with particles represented by
singularities (Pais 1982, pp.~464--8), an idea that, in retrospect, seems
somewhat reminiscent of de Broglie's double-solution theory (section~\ref{Structure}).

It would of course have been most interesting to see how physicists would have
reacted had Einstein in fact published his paper or presented it at the fifth
Solvay conference. It so happens that Heisenberg had heard about Einstein's
theory through Born and Jordan, and --- on 19 May, just two days before
Einstein withdrew the paper --- wrote to Einstein asking about it. On 10 June
1927, Heisenberg wrote to Einstein again, this time with detailed comments and
arguments against what Einstein was (or had been) proposing.\footnote{Heisenberg to Einstein, 19 May and 
10 June 1927, AEA~12-173.00 and 12-174.00 (both in German).}

At the beginning of this letter, after thanking Einstein
for his `friendly letter', Heisenberg says he would like to explain why he
believes indeterminism is necessary and not just possible. He characterises
Einstein as thinking that, while all experiments will agree with the
statistical quantum theory, nevertheless it will be possible to talk about
definite particle trajectories. Heisenberg then outlines an objection. He
considers free electrons with a constant and very low velocity --- hence large
de Broglie wavelength $\lambda$ --- striking a grating with spacing comparable
to $\lambda$. He remarks that, in Einstein's theory, the electrons will be
scattered in discrete spatial directions, and that if the initial position of
a particle were known one could calculate where the particle will hit the
grating. Heisenberg then asserts that one could set up an obstacle at that
point, so as to deflect the particle in an arbitrary direction, independently
of the rest of the grating. Heisenberg says that this could be done, if the
forces between the particle and the obstacle act only at short range (over
distances much smaller than the spacing of the grating). Heisenberg then adds
that, in actual fact, the electron will be scattered in the usual discrete
directions regardless of the obstacle. Heisenberg goes on to say that one
could escape this conclusion if one `sets the motion of the particle again in
direct relation to the behaviour of the waves'. But this means, says
Heisenberg, that the size of the particle --- or the range of its interaction
--- depends on its velocity. Heisenberg asserts that making such assumptions
actually amounts to giving up the word `particle' and leads to a loss in
understanding of why the simple potential energy $e^{2}/r$ appears in the
Schr\"{o}dinger equation or in the matrix Hamiltonian function. On the other
hand, Heisenberg agrees that: `If you use the word \textquotedblleft
particle\textquotedblright\ so liberally, I consider it as very well possible
that one can again also define particle trajectories'. But then, adds
Heisenberg, one loses the simplicity of quantum theory, according to which the
particle motion takes place classically (to the extent that one can speak
about motion in quantum theory). Heisenberg notes that Einstein seems willing
to sacrifice this simplicity for the sake of maintaining causality. He remarks
further that, in Einstein's conception, many experiments are still determined
only statistically, and `we could only console ourselves with the fact that,
while for us the principle of causality is meaningless, because of the
uncertainty relation $p_{1}q_{1}\sim h$, however the dear God knows in
addition the position of the particle and thereby keeps the validity of the
causal law'. Heisenberg adds that he finds it unattractive to try to describe
more than just the `connection between experiments' [Zusammenhang der Experimente].

It is likely that Heisenberg had similar views of de Broglie's theory, and
that if he had commented on de Broglie's theory at the fifth Solvay conference
he would have said things similar to the above.

Heisenberg's objection concerning the electron and the grating seems to be
based on inappropriate reasoning taken from classical physics (a common
feature of objections to pilot-wave theory even today), and Heisenberg agrees
that if the motion of the particle is strictly tied to that of the waves, then
it should be possible to obtain consistency with observation (as we now know
is indeed the case for de Broglie's theory). Even so, Heisenberg seems to
think that such a highly nonclassical particle dynamics would lack the
simplicity and intelligibility of certain classical ideas that quantum theory
preserves. Finally, Heisenberg is unhappy with a theory containing
unobservable causal connections. Similar objections to pilot-wave theory are
considered in the next section.

\section{Objections: in 1927 and today}\label{objections}

It is interesting to observe that many of the objections to pilot-wave theory
that are commonly heard today were already voiced in 1927.

Regarding the existence or non-existence of de Broglie's trajectories,
Brillouin --- in the discussion following de Broglie's lecture (p.~\pageref{candoubt}) 
--- could just as well have been replying to a
present-day critic of de Broglie-Bohm theory when he said that:

\begin{quotation}
Mr Born can doubt the real existence of the trajectories calculated by L. de
Broglie, and assert that one will never be able to observe them, but he cannot
prove to us that these trajectories do not exist. There is no contradiction
between the point of view of L. de Broglie and that of other authors .... .
\end{quotation}
Here we have a conflict between those who believe in hidden entities
because of the explanatory role they play, and those who think that what
cannot be observed in detail should play no theoretical role. Such conflicts
are not uncommon in the history of science: for example, a similar
polarisation of views occurred in the late nineteenth century regarding the
reality of atoms (which the `energeticists' regarded as metaphysical fictions).

The debate over the reality of the trajectories postulated by de Broglie has
been sharpened in recent years by the recognition that, from the perspective
of pilot-wave theory itself, our inability to observe those trajectories is
not a fundamental constraint built into the theory, but rather an accident of
initial conditions with a Born-rule probability distribution $P=\left\vert
\Psi\right\vert ^{2}$ (for an ensemble of systems with wave function $\Psi$).
The statistical noise associated with such `quantum equilibrium' distributions
sets limits to what can be measured, but for more general `nonequilibrium'
distributions $P\neq\left\vert \Psi\right\vert ^{2}$, the uncertainty
principle is violated and observation of the trajectories becomes possible
(Valentini 1991b, 1992, 2002b; Pearle and Valentini 2006). Such nonequilibrium
entails, of course, a departure from the statistical predictions of quantum
theory, which are obtained in pilot-wave theory only as a special
`equilibrium' case. Arguably, the above disagreement between Born and
Brillouin might nowadays turn on the question of whether one is willing to
believe that quantum physics is merely the physics of a special statistical state.

Another objection sometimes heard today is that a velocity law different from
that assumed by de Broglie is equally possible. In the discussion following de
Broglie's lecture, Schr\"{o}dinger (p.~\pageref{SchrdeBdisc}) raised the
possibility of an alternative particle velocity defined by the momentum
density of a field. Pauli pointed out that, for a relativistic field, if the
particle velocity were obtained by dividing the momentum density by the energy
density, then the resulting trajectories would differ from those obtained by
de Broglie, who assumed a velocity defined (in the case of a single particle)
by the ratio of current density to charge density. Possibly, then as now, the
existence of alternative velocities may have been interpreted as casting doubt
on the reality of the velocities actually assumed by de Broglie (though the
true velocities would become measurable in the presence of quantum nonequilibrium).

In addition to the criticism that the trajectories cannot be observed, today
it is also often objected that the trajectories are rather strange from a
classical perspective. The peculiar nature of de Broglie's trajectories was
addressed in the discussion following de Broglie's lecture (pp.~\pageref{deBDisc-beginning}~ff.), 
and again in the general discussion (pp.~\pageref{duringreflection}, \pageref{Lordiff}~f.). It was, for
example, pointed out that the speed of an electron could be zero in a
stationary state, and that for general atomic states the orbits would be very
complicated. At the end of his discussion of photon reflection by a mirror,
Brillouin (p.~\pageref{Lewisparadox}) argued that de Broglie's
non-rectilinear photon paths (in free space) were necessary in order to avoid
a paradox posed by Lewis, in which in the presence of interference it appeared
that photons would collide with only one end of a mirror, causing it to
rotate, even though from classical electrodynamics the mean radiation pressure
on the mirror is expected to be uniform (see Brillouin's Fig.~3
).

This last example of Brillouin's recalls present-day debates involving certain
kinds of quantum measurements, in which the trajectories predicted by
pilot-wave theory have counter-intuitive features that some authors have
labelled `surreal' (Englert \textit{et al}. 1992, Aharonov and Vaidman 1996),
while other authors regard these features as perfectly understandable from
within pilot-wave theory itself (Valentini 1992, p. 24, Dewdney, Hardy and
Squires 1993, D\"{u}rr \textit{et al}. 1993). A key question here is whether
it is reasonable to expect a theory of subquantum dynamics to conform to
classical intuitions about measurement (given that it is the underlying
dynamics that should be used to analyse the measurement process).

Pilot-wave theory is sometimes seen as a return to classical physics (welcomed
by some, criticised by others). But in fact, de Broglie's velocity-based
dynamics is a new form of dynamics that is simply quite distinct from
classical theory; therefore, it is to be expected that the behaviour of the
trajectories will \textit{not} conform to classical expectations. As we saw in
detail in chapter~\ref{deBroglieEss}, de Broglie did indeed originally regard
his theory as a radical departure from the principles of classical dynamics.
It was Bohm's later revival of de Broglie's theory, in an unnatural
pseudo-Newtonian form, that led to the widespread and mistaken perception that
de Broglie-Bohm theory constituted a return to classical physics. In more
recent years, de Broglie's original pilot-wave dynamics has again become
recognised as a new form of dynamics in its own right (D\"{u}rr, Goldstein and
Zangh\`{\i} 1992, Valentini 1992).

\setcounter{endnote}{0}
\setcounter{equation}{0}

\chapter{Beyond the Bohr-Einstein debate}\label{Beyond-Bohr-Einstein}\chaptermark{Beyond the Bohr-Einstein debate}

The fifth Solvay conference is usually remembered for the clash that took
place between Bohr and Einstein, supposedly concerning in particular the
possibility of breaking the uncertainty relations. It might be assumed that
this clash took the form of an official debate that was the centrepiece of the
conference. However, no record of any such debate appears in the published
proceedings, where both Bohr and Einstein are in fact relatively silent.

The available evidence shows that in 1927 the famous exchanges between Bohr
and Einstein actually consisted of informal discussions, which took place
semi-privately (mainly over breakfast and dinner), and which were overheard by
just a few of the participants, in particular Heisenberg and Ehrenfest. The
historical sources for this consist, in fact, entirely of accounts given by
Bohr, Heisenberg and Ehrenfest. These accounts essentially ignore the
extensive formal discussions appearing in the published proceedings.

As a result of relying on these sources, the perception of the conference by
posterity has been skewed on two counts. First, at the fifth Solvay conference
there occurred much more that was memorable and important besides the
Bohr-Einstein clash. Second, as shown in detail by Howard (1990), the real
nature of Einstein's objections was in fact misunderstood by Bohr, Heisenberg
and Ehrenfest: for Einstein's main target was not the uncertainty relations,
but what he saw as the nonseparability of quantum theory.

Below we shall indicate how these misunderstandings arose, summarise what now
appear to have been Einstein's true concerns, and end by urging physicists,
philosophers and historians to reconsider what actually took place in Brussels
in October 1927, bearing in mind the deep questions that we still face
concerning the nature of quantum physics.

\section{The standard historical account}\label{standardhistory}

According to Heisenberg, the discussions that took place at the fifth Solvay
conference `contributed extraordinarily to the clarification of the physical
foundations of the quantum theory' and indeed led to `the outward completion
of the quantum theory, which now can be applied without worries as a theory
closed in itself' (Heisenberg 1929, p.~495). For Heisenberg, and perhaps for others in
the Copenhagen-G\"{o}ttingen camp, the 1927 conference seems to have played a
key role in finalising the interpretation of the theory.

However, the perception that the interpretation had been finalised proved to
be mistaken. As we have shown at length in chapter~\ref{QTMeasProb}, the 
interpretation of quantum theory is today still an open
question, and deep concerns as to its meaning have stubbornly persisted.
Further, as we have seen throughout part II of this book, many of today's
fundamental concerns were voiced (often at considerable length) at the fifth
Solvay conference. Given that these concerns are still very much alive, it has
evidently been a mistake to allow recollections of private discussions between
Bohr and Einstein to overshadow our historical memory of the rest of the conference.

Note that, as we saw in chapter~\ref{HistEss} (p.~\pageref{Mehrafootnote}), while Mehra 
(1975, p.~152) and also Mehra and Rechenberg (2000, p.~246) state that the general
discussion was a discussion following `Bohr's report', in fact Bohr did not
present a report at the conference, nor was he invited to give one. This
misunderstanding seems to have arisen because, at Bohr's request, a
translation of his Como lecture appears in the published proceedings, to
replace his remarks in the general discussion; and this has no doubt
contributed to the common view that Bohr played a central role at the
conference, when in fact it is clear from the proceedings that at the official
meetings both he and Einstein played a rather marginal role.\footnote{Since
Bohr had not been invited to give a report, one might also question the
propriety of his request that his remarks in the general discussion be
replaced by a translation of a rather lengthy paper he was already publishing
elsewhere.} Further, it seems rather clear that the standard (and unbalanced)
version of events was propagated in particular by Bohr and Heisenberg,
especially through their writings decades later.

There is very little independent evidence from the time as to what was said
between Bohr and Einstein. Thus, for example, Mehra and Rechenberg (2000, p.~251) 
note the `little evidence of the Bohr-Einstein debate in the official
conference documents', and rely on eyewitness reports by Ehrenfest, Heisenberg
and Bohr to yield `a fairly consistent historical picture of the great
epistemological debate between Bohr and Einstein' (p.~256).

One frequently-cited piece of contemporary evidence is a description of the
conference written by Ehrenfest a few days later, in a letter to his students
and associates in Leiden. This letter is cited at length by Mehra and
Rechenberg (2000, pp.~251--3). An extract reads:

\begin{quotation}
Bohr towering completely over everybody. .... step by step defeating
everybody. .... It was delightful for me to be present during the
conversations between Bohr and Einstein. .... Einstein all the time with new
examples. In a certain sense a sort of Perpetuum Mobile of the second kind to
break the UNCERTAINTY RELATION. Bohr .... constantly searching for the tools
to crush one example after the other. Einstein .... jumping out fresh every
morning. .... I am almost without reservation pro Bohr and contra Einstein.
\end{quotation}
This letter has often been taken as representative of the conference.
However, there is a marked contrast with the published proceedings (in which
Bohr and Einstein are mostly silent), a contrast which has not been taken into
account.\footnote{We have attempted to compare Ehrenfest's contemporary
account with that in letters written by other participants soon after the
conference, but have found nothing significant.}

After examining the published proceedings, Mehra and Rechenberg (2000, 
pp.~250--56) go on to consider at greater length recollections by Heisenberg and
Bohr --- written decades after the conference --- concerning the discussions
between Bohr and Einstein. With hindsight, again given that the interpretation
of quantum theory is today an open question, it would be desirable to have a
more balanced view of the conference, focussing more on the content of the
published proceedings, and rather less on these later recollections by just
two of the participants.

Here is an extract from Heisenberg's recollection, written some 40 years later
(Heisenberg 1967, p.~107):

\begin{quotation}
The discussions were soon focussed upon a duel between Einstein and Bohr~....~. 
We generally met already at breakfast in the hotel, and Einstein began to
describe an ideal experiment in which he thought the inner contradictions of
the Copenhagen interpretation were especially clearly visible. Einstein, Bohr
and I walked together from the hotel to the conference building, and I
listened to the lively discussion between those two people whose philosophical
attitudes were so different, .... at lunch time the discussions continued
between Bohr and the others from Copenhagen. Bohr had usually finished the
complete analysis of the ideal experiment by late afternoon and would show it
to Einstein at the supper table. Einstein had no good objection to this
analysis, but in his heart he was not convinced.
\end{quotation}
From this account, the Bohr-Einstein clash appears indeed to have been
a private discussion, with a few of Bohr's close associates in attendance. And
yet, there has been a marked tendency to portray this discussion as the
centrepiece of the whole conference. Thus, for example, in the preface to
Mehra's book \textit{The Solvay Conferences on Physics}, Heisenberg wrote the
following about the 1927 conference (Mehra 1975, pp.~v--vi):

\begin{quotation}
Therefore the discussions at the 1927 Solvay Conference, from the very
beginning, centred around the paradoxa of quantum theory. .... Einstein
therefore suggested special experimental arrangements for which, in his
opinion, the uncertainty relations could be evaded. But the analysis carried
out by Bohr and others during the Conference revealed errors in Einstein's
arguments. In this situation, by means of intensive discussions, the
Conference contributed directly to the clarification of the
quantum-theoretical paradoxa.
\end{quotation}
Heisenberg says nothing at all about the alternative theories of de
Broglie and Schr\"{o}dinger, or about the views of Lorentz or Dirac (for
example), or about the other extensive discussions recorded in the proceedings.

As for the text of Mehra's book on the Solvay conferences, the chapter devoted
to the fifth Solvay conference contains a summary of the general discussion,
which says nothing about the published discussions beyond providing a list of
the participants. Mehra's summary states that a debate took place between Bohr
and Einstein, and that the famous Bohr-Einstein dialogue began here in 1927.
Mehra then adds an appendix, reproducing Bohr's famous essay `Discussion with
Einstein on epistemological problems in atomic physics' (Bohr 1949), written
more than 20 years after the conference took place. Once again, the published
discussions are made to appear rather insignificant compared to (Bohr's
recollection of) the informal discussions between Bohr and Einstein.

The above essay by Bohr is in fact the principal and most detailed historical
source for the Bohr-Einstein debate. This essay was Bohr's contribution to the
1949 festschrift for Einstein's seventieth birthday. It is reprinted as the
very first paper in Wheeler and Zurek's (1983) influential collection
\textit{Quantum Theory and Measurement}, as well as elsewhere. It gives a
detailed account of Bohr's discussions with Einstein at the fifth Solvay
conference (as well as at the sixth Solvay conference of 1930). According to
Bohr, the discussions in 1927 centred around, among other things, a version of
the double-slit experiment, in which according to Einstein it was possible to
observe interference while at the same time deducing which path the particle
had taken, a claim conclusively refuted by Bohr.

Were Bohr's recollections accurate? Jammer certainly thought so:

\begin{quotation}
Bohr's masterly report of his discussions with Einstein on this issue, though
written more than 20 years after they had taken place, is undoubtedly a
reliable source for the history of this episode. (Jammer 1974, p.~120)
\end{quotation}
Though as Jammer himself adds (p.~120): `It is, however, most
deplorable that additional documentary material on the Bohr-Einstein debate is
extremely scanty'.

We now know that, as we shall now discuss, Bohr's recollection of his
discussions with Einstein did not properly capture Einstein's true intentions,
essentially because, at the time, no one understood what Einstein's principal
concern was: the nonseparability of quantum theory.

\section{Towards a historical revision}\label{historicalrevision}

Separability --- the requirement that the joint state of a composite of
spatially separated systems should be determined by the states of the
component parts --- was a condition basic to Einstein's field-theoretic view
of physics (as indeed was the absence of action at a distance). As already
mentioned in section~\ref{failure-energy-momentum}, Howard
(1990) has shown in great detail how Einstein's concerns about the failure of
separability in quantum theory date back to long before the famous EPR paper
of 1935.\footnote{See also Fine (1986, chapter~3).} There is no doubt that, by
1909, Einstein understood that if light quanta were treated like the spatially
independent molecules of an ideal gas, then the resulting fluctuations were
inconsistent with Planck's formula for blackbody radiation; and certainly, in
1925, Einstein was concerned that Bose-Einstein statistics entailed a
mysterious interdependence of photons. Also in 1925, as we discussed in
section~\ref{failure-energy-momentum}, Einstein's theory of
guiding fields in 3-space --- in which spatially separated systems each had
their own guiding wave, in accordance with Einstein's separability criterion
--- conflicted with energy-momentum conservation for single events. Howard
(pp.~83--91) argues further that, in spring 1927, Einstein must have realised
that Schr\"{o}dinger's wave mechanics in configuration space violated
separability, because a general solution to the Schr\"{o}dinger equation for a
composite system could \textit{not} be written as a product over the
components.\footnote{Howard's argument here is somewhat circumstantial,
appealing in part to the difficulty with separability that Einstein had with
his own hidden-variables amendment of wave mechanics (cf. section~\ref{EinsteinHV}).} 
In other words, Einstein objected
to what we would now call entanglement, and concluded that the wave function
in configuration space could not represent anything physical.

Einstein's concerns about separability continued up to and beyond the fifth
Solvay conference. While it seems that Einstein did have some early doubts
about the validity of the uncertainty relations, Howard's reconstruction shows
that Einstein's main concern lay elsewhere. The primary aim of the famous
thought experiments that Einstein discussed with Bohr, in 1927 and
subsequently, was not to defeat the uncertainty relations but to highlight the
(for him disturbing) feature of quantum theory, that spatially separated
systems cannot be treated independently. As Howard (p.~94) puts it, regarding
the 1927 Solvay conference:

\begin{quotation}
But if the uncertainty relations \textit{really} were the main sticking point
\textit{for Einstein}, why did Einstein not say so in the published version of
his remarks, or anywhere else for that matter in correspondence or in print in
the weeks and months following the Solvay meeting?
\end{quotation}

We have indeed seen in chapter~\ref{locality-and-incompleteness} that
Einstein's criticism in the general discussion concerned locality and
completeness (just like the later EPR argument), not the uncertainty
relations. Bohr, in his reply, states that he does not `understand what
precisely is the point' Einstein is making. It seems rather clear that,
indeed, in 1927 Bohr did not understand Einstein's point, and it is remarkable
that what is most often recalled about the fifth Solvay conference was in fact
largely a misunderstanding.

According to Bohr's later recollections (Bohr 1949), at the fifth Solvay
conference Einstein proposed a version of the two-slit experiment in which
measurement of the transverse recoil of a screen with a single slit would
enable one to deduce the path of a particle through a second screen with two
slits, while at the same time observing interference on the far side of the
second screen. (Consideration of this experiment must have taken place
informally, not in the official discussions.) This experiment has been
analysed in detail by Wootters and Zurek (1979), who show the crucial role
played by quantum nonseparability between the particle and the first screen.
While the evidence is somewhat sketchy in this particular instance, according
to Howard the main point that concerned Einstein in this experiment was
precisely such nonseparability.

That separability was indeed Einstein's central concern is clearer in the
later `photon-box' thought experiment he discussed with Bohr at the sixth
Solvay conference of 1930, involving weighing a box from which a photon
escapes. Again, Bohr discusses this experiment at length in his recollections,
where according to him it was yet another of Einstein's attempts to circumvent
the uncertainty relations. Specifically, according to Bohr, Einstein's
intention was to beat the energy-time uncertainty relation, by measuring both
the energy of the emitted photon (by weighing the box before and after) and
its time of emission (given by a clock controlling the shutter releasing the
photon). On Bohr's account, this attempt failed, ironically, because of the
time dilation in a gravitational field implied by Einstein's own general
theory of relativity.

However, it seems that in fact, the photon-box experiment was (like Einstein's
published objection of 1927) really a form of the later EPR argument for
incompleteness. This is shown by a letter Ehrenfest wrote to Bohr on 9 July
1931, just after Ehrenfest had visited Einstein in Berlin (Howard 1990, pp.~98--9). 
Ehrenfest reports that Einstein said he did not invent the photon-box
experiment to defeat the uncertainty relations (which he had for a long time
no longer doubted), but `for a totally different purpose'. Ehrenfest then
explains that Einstein's real intention was to construct an example in which
the choice of measurement at one location would enable an experimenter to
predict \textit{either} one \textit{or} the other of two incompatible
quantities for a system that was far away at the time of the measurement. In
the example at hand, if the escaped photon is reflected back towards the box
after having travelled a great distance, then the time of its return may be
predicted with certainty if the experimenter checks the clock reading (while
the photon is still far away); alternatively, the energy (or frequency) of the
returning photon may be predicted with certainty if, instead, the experimenter
chooses to weigh the box (again while the photon is still far away). Because
the two possible operations take place while the photon is at a great
distance, the assumptions of separability and locality imply that both the
time and energy of the returning photon are in reality determined in advance
(even if in practice an experimenter cannot carry out both predictions
simultaneously), leading to the conclusion that quantum theory is
incomplete.\footnote{The reasoning here is similar to that in Einstein's own
(and simpler) version of the EPR argument, which first appears in a letter
from Einstein to Schr\"{o}dinger of 19 June 1935, one month after the EPR
paper was published (Fine 1986, chap.~3). The argument --- which Einstein
repeated and refined between 1936 and 1949 --- runs essentially as follows. A
complete theory should associate one and only one theoretical state with each
real state of a system; in an EPR-type experiment on correlated systems,
depending on what measurement is carried out at one wing of the experiment,
quantum theory associates different wave functions with what must (assuming
locality) be the same real state at the other distant wing. Therefore, quantum
theory is incomplete. For a detailed discussion, see Howard (1990).}

The true thrust of Einstein's argument was not appreciated at the time,
perhaps because Bohr and his associates tended to identify the existence of
physical quantities with their experimental measurability: if two quantities
could not be measured simultaneously in the same experiment, they did not
exist simultaneously in the same experiment. With this attitude in mind, it
would be natural to mistake Einstein's claim of simultaneous existence for a
claim of simultaneous measurability.

We feel that Howard's reappraisal of the Bohr-Einstein debate, as well as
being of great intrinsic interest, also provides an instructive example of how
the history of quantum physics should be reconsidered in the light of our
modern understanding of quantum theory and its open problems. There was
certainly much more to the fifth Solvay conference than the Bohr-Einstein
clash, and a similar reappraisal of other crucial encounters at that time
seems overdue.

If the history of quantum theory is written on the assumption that Bohr,
Heisenberg and Born were right, and that de Broglie, Schr\"{o}dinger and
Einstein were wrong, the resulting account is likely to be unsatisfactory:
opposing views will tend to be misunderstood or underestimated, supporting
views over-emphasised, and valid alternative approaches ignored.

A reconsideration of the fifth Solvay conference certainly entails a
re-evaluation of de Broglie's pilot-wave theory as a coherent but (until very
recently) `forgotten' formulation of quantum theory. Schr\"{o}dinger's ideas,
too, seem more plausible today, in the light of modern collapse models. One
should also reconsider what Born and Heisenberg's `quantum mechanics' really
was, in particular as concerns the role of time and the collapse of the wave function.

There is no longer a definitive, widely-accepted interpretation of quan\-tum
mechanics; it is no longer clear who was right and who was wrong in October
1927. Therefore, it seems particularly important at this time to return to the
historical sources and re-evaluate them. We hope that physicists, philosophers
and historians will reconsider the significance of the fifth Solvay
conference, both for the history of physics and for current research in the
foundations of quantum theory.

\part{\\ The proceedings of the 1927 Solvay conference}


\chapter*{H.~A.~Lorentz \dag}\markboth{{\it H.~A.~Lorentz} \dag}{{\it H.~A.~Lorentz} \dag}
\addcontentsline{toc}{schapter}{H.~A.~Lorentz \dag}

\noindent Hardly a few months have gone by since the meeting of the fifth physics conference 
in Brussels, and now I must, in the name of the scientific
committee, recall here all that meant to the Solvay International Institute of Physics
he who was our chairman and the moving spirit of our meetings. The
illustrious teacher and physicist, H.~A. Lorentz, was taken away in February 1928
by a sudden illness, when we had just admired, once again, his magnificent
intellectual gifts which age was unable to diminish in the least.

Professor Lorentz, of a simple and modest demeanour, nevertheless enjoyed an
exceptional authority, thanks to the combination of rare qualities in a
harmonious whole. Theoretician with profound views --- eminent teacher in the
highest forms of instruction and tirelessly devoted to this task --- fervent
advocate of all international scientific collaboration --- he found, wherever
he went, a grateful circle of pupils, disciples and those who carried on his
work. Ernest Solvay had an unfailing appreciation of this moral and intellectual
force, and it was on this that he relied to carry through a plan that was dear
to him, that of serving Science by organising conferences composed of a
limited number of physicists, gathered together to discuss subjects where the
need for new insights is felt with particular intensity. Thus was born the Solvay
International Institute of Physics, of which Ernest Solvay followed the
beginnings with a touching concern and to which Lorentz devoted a loyal and
fruitful activity.

All those who had the honour to be his collaborators know what he was as
chairman of these conferences and of the preparatory meetings. His thorough
knowledge of physics gave him an overall view of the problems to be examined.
His clear judgement, his fair and benevolent spirit guided the scientific
committee in the choice of the assistance it was appropriate to call upon. When we
then were gathered together at a conference, one could only admire
without reservations the mastery with which he conducted the chairmanship. His
shining intellect dominated the discussion and followed it also in the details,
stimulating it or preventing it from drifting, making sure
that all opinions could be usefully expressed, bringing out the final
conclusion as far as possible. His perfect knowledge of languages allowed him
to interpret, with equal facility, the words uttered by each one. Our chairman
appeared to us, in fact, gifted with an invincible youth, in his passion for
scientific truth and in the joy he had in comparing opinions, sometimes with a
shrewd smile on his face, and even a little mischievousness when confronted
with an unforeseen aspect of the question. Respect and affection went 
to him spontaneously, creating a cordial and friendly atmosphere,
which facilitated the common work and increased its efficiency.

True creator of the theoretical edifice that explains optical and
electromagnetic phenomena by the exchange of energy between electrons
contained in matter and radiation viewed in accordance with Maxwell's theory,
Lorentz retained a devotion to this classical theory. All the more remarkable
is the flexibility of mind with which he followed the disconcerting evolution
of the quantum theory and of the new mechanics.

The impetus that he gave to the Solvay institute will be  a memory and an example
for the scientific committee. May this volume, faithful report of the
work of the recent physics conference, be a tribute to the memory of he who,
for the fifth and last time, honoured the conference by his presence and by
his guidance.

\

\hfill{\sc M.~Curie}



\chapter*{Fifth physics conference}\markboth{{\it Fifth physics conference}}{{\it Fifth physics conference}}
\addcontentsline{toc}{schapter}{Fifth physics conference}

\noindent The fifth of the physics conferences, provided for by article 10 of the
statutes of the international institute of physics founded by Ernest Solvay, 
held its sessions in Brussels on the premises of the institute from 24 to 
29 October 1927.

\ 

The following took part in the conference:

\

\noindent Mr~H.~A.~{\sc Lorentz} \dag, of Haarlem, {\em  Chairman}.

\

\noindent Mrs~P.~{\sc Curie}, of Paris; Messrs~N.~{\sc Bohr}, of Copenhagen; 
M.~{\sc Born}, of G\"{o}ttingen; W.~L.~{\sc Bragg}, of Manchester; 
L.~{\sc Brillouin}, of Paris; A.~H.~{\sc Compton}, of Chicago; 
L.-V.~{\sc de Broglie}, of Paris; P.~{\sc Debye}, of Leipzig; 
P.~A.~M.~{\sc Dirac}, of Cambridge; P.~{\sc Ehrenfest}, of Leiden; 
A.~{\sc Einstein}, of Berlin; R.~H.~{\sc Fowler}, of Cambridge; 
Ch.-E.~{\sc Guye}, of Geneva; W.~{\sc Heisenberg}, of Copenhagen; 
M.~{\sc Knudsen}, of Copenhagen; H.~A.~{\sc Kramers}, of Utrecht; 
P.~{\sc Langevin}, of Paris; W.~{\sc Pauli}, of Hamburg; M.~{\sc Planck}, 
of Berlin; O.~W.~{\sc Richardson}, of London; [E.~{\sc Schr\"{o}dinger}, of Zurich;] 
C.~T.~R.~{\sc Wilson}, of Cambridge, {\em  Members}.

\ 

\noindent Mr~J.-E.~{\sc Verschaffelt}, of Gent, fulfilled the duties of
{\em  Secretary}.

\ 

\noindent Messrs~Th.~{\sc De~Donder}, E.~{\sc Henriot} and Aug.~{\sc Piccard},
professors at the University of Brussels, attended the meetings of the
conference as guests of the scientific committee, Mr~Ed.~{\sc Herzen},
professor at the \'{E}cole des Hautes \'{E}tudes de Bruxelles, as representative of the
Solvay family.

\ 

\noindent Professor I.~{\sc Langmuir}, of Schenectady (U.~S. of America), visiting
Europe, attended the meetings as a guest.

\ 

\noindent Mr~Edm.~van~{\sc Aubel}, member of the Scientific Committee, and 
Mr~H. {\sc Deslandres}, director of the Meudon observatory, invited to
participate in the conference meetings, had been excused.

\ 

\noindent Sir~W.~H.~{\sc Bragg}, member of the scientific committee, who had handed in
his resignation before the meetings and requested
to be excused, also did not attend the sessions.

\ 

The administrative commission of the institute was composed of:

\ 

\noindent Messrs~Jules {\sc Bordet}, professor at the University of Brussels,
appointed by H.~M. the King of the Belgians; Armand {\sc Solvay}, engineer,
manager of Solvay and Co.; Maurice {\sc Bourquin}, professor at the
University of Brussels; \'{E}mile {\sc Henriot}, professor at the University of
Brussels; Ch.~{\sc Lef\'{e}bure}, engineer, appointed by the family of Mr~Ernest 
Solvay, {\em  Administrative Secretary}.

\ 

The scientific committee was composed of:

\ 

\noindent Messrs~H.~A.~{\sc Lorentz}\dag, professor at the University of Leiden,
{\em  Chairman}; M.~{\sc Knudsen}, professor at the University of
Copenhagen, {\em  Secretary}; W.~H.~{\sc Bragg}, professor at the
University of London, president of the Royal Institution; Mrs~Pierre
{\sc Curie}, professor at the Faculty of Sciences of Paris; 
Messrs~A.~{\sc Einstein},\footnote{Chosen in replacement of Mr~H.~Kamerlingh Onnes,
deceased.} professor, in Berlin; Charles-Eug.~{\sc Guye}, professor at the
University of Geneva; P.~{\sc Langevin}, professor at the Coll\`{e}ge de
France, in Paris; O.~W.~{\sc Richardson}, professor at the University of
London; Edm.~van~{\sc Aubel}, professor at the University of Gent.

\ 

\noindent Sir W. H. {\sc Bragg}, resigning member, was replaced by Mr B.
{\sc Cabrera}, professor at the University of Madrid.

\ 

\noindent To replace its late chairman, the scientific committee chose Professor 
{\sc P.~Langevin}.


\setcounter{endnote}{0}
\setcounter{equation}{0}

\chapter*{The intensity of X-ray reflection$^{\scriptsize\hbox{a}}$}\markboth{{\it W.~L.~Bragg}}{{\it The intensity of X-ray reflection}}
\addcontentsline{toc}{chapter}{The intensity of X-ray reflection ({\em W.~L.~Bragg\/})}

\begin{center}{\sc By Mr W.~L.~BRAGG}\footnotetext[1]{We follow Bragg's original English typescript, 
from the copy in the Richardson collection, AHQP-RDN, document M-0059 (indexed as `unidentified author'
in the microfilmed catalogue). Obvious typos are corrected mostly tacitly and some of the spelling has been 
harmonised with that used in the rest of the volume. Discrepancies between the original English and the
published French are endnoted (\textit{eds}.).}\end{center}

\

\begin{center}
\par
\Needspace{5\baselineskip}
{\sc 1. --- The classical treatment of x-ray diffraction 
phenomena}\addcontentsline{toc}{section}{The classical treatment of X-ray 
diffraction phenomena}
\end{center}
The earliest experiments on the diffraction of X-rays by crystals showed that
the directions in which the rays were diffracted were governed by the
classical laws of optics. Laue's original paper on the diffraction of white\endnote{Here and 
in a few other places, the French adds (or omits) inverted commas.}
radiation by a crystal, and the work which my father and I initiated on the
reflection of lines\endnote{[r\'{e}flexion des radiations des raies]} in the X-ray spectrum, were alike based on the laws of
optics which hold for the diffraction grating. The high accuracy which has
been developed by Siegbahn and others in the realm of X-ray spectroscopy is
the best evidence of the truth of these laws. Advance in accuracy has shown
the necessity of taking into account the very small refraction of X-rays by
the crystal, but this refraction is also determined by the classical laws and
provides no exception\endnote{[ne fait pr\'{e}voir aucun \'{e}cart]} to the above statement.

The first attempts at crystal analysis showed further that the strength of the
diffracted beam was related to the structure of the crystal in a way to be
expected by the optical analogy. This has been the basis of most work on the
analysis of crystal structure. When monochromatic X-rays are reflected from a
set of crystal planes, the orders of reflection are strong, weak, or absent in
a way which can be accounted for qualitatively by the arrangement of atoms\endnote{The French
edition adds `en couches' [in layers].}
parallel to these planes. In the analysis of many structures, it is not
necessary to make a strict examination of the strength of the diffracted
beams. Slight displacements of the atoms cause the intensities of the higher
orders to fluctuate so rapidly, that it is possible to fix the atomic
positions with high accuracy by using a rough estimate of the relative
intensity of the different orders.

When we attack the problem of developing an accurate quantitative theory of
intensity of diffraction, many difficulties present themselves. These
difficulties are so great, and the interpretation of the experimental results
has often been so uncertain, that it has led\endnote{Typescript: `have often been .... it has led'; 
French version: `a souvent \'{e}t\'{e}~.... elles ont conduit'.} to a natural distrust of
deductions drawn from intensity measurements. Investigators of crystal
structures have relied on qualitative methods,\endnote{[ont eu confiance dans les m\'{e}thodes quantitatives]} 
since these were in many cases quite adequate. The development of the quantitative analysis has always
interested me personally, particularly as a means of attacking the more
complicated crystalline structures, and it would seem that at the present time
the technique has reached a stage when we can rely on the results. It is my
purpose in this paper to attempt a critical survey of the present development
of the subject. It is of considerable interest because it is our most direct
way of analysing atomic and molecular structure.\label{Bragg-direct}

In any X-ray examination of a crystalline body, what we actually measure is a
series of samples\endnote{[portions]} of the coherent radiation scattered in certain definite
directions by the unit of the structure. This unit is, in general, the element
of pattern of the crystal, while in certain simple cases it may be a single atom.

In the examination of a small body by the microscope, the objective receives
the radiation scattered in different directions by the body, and the
information about its structure, which we get by viewing the final image, is
contained at an earlier stage\endnote{[sous une forme plus primitive]} in 
these scattered beams. Though the two cases
of microscopic and X-ray examination are so similar, there are certain
important differences. The scattered beams in the microscope can be combined
again to form an image, and in the formation of the image the phase
relationship between beams scattered in different directions plays an
essential part. In the X-ray problem, since we can only measure the intensity
of scattering in each direction, this phase relationship cannot be determined
experimentally, though in many cases it can be inferred.\endnote{[il soit possible de les trouver]} Further, the
microscope receives the scattered beams over a continuous range of directions,
whereas the geometry of the crystalline structure limits our examination to
certain directions of scattering. Thus we cannot form directly an image of the
crystalline unit which is being illuminated by X-rays. We can only measure
experimentally the strength of the scattered beams, and then build up an image
piece by piece from the information we have obtained.

It is important to note that in the case of X-ray examination all work is
being carried out at what is very nearly the theoretical limit of the
resolving power of our instruments. The range of wavelength which it is
convenient to use lies between 0.6 \AA\ and 1.5
\AA . This range is of sufficiently small wavelength for
work with the details of crystal structure, which is always on a scale of
several \AA ngstr\"{o}m units, but the wavelengths are
inconveniently great for an examination into atomic structure. It is
unfortunate from a practical point of view that there is no convenient steady
source of radiation between the K lines of the metal palladium, and the very
much shorter K lines of tungsten. This difficulty will no doubt be overcome,
and a technique of `ultraviolet' X-ray microscopy will be developed, but at
present all the accurate work on intensity of reflection has been done with
wavelengths in the neighbourhood of 0.7~\AA .

We may conveniently\endnote{[logiquement]} divide the process of analysis into three stages.

\

\noindent a) The experimental measurement of the intensities of the diffracted beams.

\

\noindent b) The reduction of these observations, with the aid of theoretical formulae,
to measurements of the amplitudes of the waves scattered by a single unit of
the structure, when a wave train of given amplitude falls on it.

\

\noindent c) The building up of the image, or deduction of the form of the unit, from
these measurements of scattering in different directions.

\

\begin{center}
\par
\Needspace{5\baselineskip}
{\sc 2. --- History of the use of quantitative 
methods}\addcontentsline{toc}{section}{History of the use of quantitative methods}
\end{center}
The fundamental principles of a mathematical analysis of X-ray reflection were
given in Laue's original paper [1], but the precise treatment of intensity of
reflection may be said to have been initiated by Darwin [2] with two papers in
the {\em Philosophical Magazine\/} early in 1914, in which he laid down the
basis for a complete theory of X-ray reflection based on the classical laws of
electrodynamics.\endnote{[thermodynamique]} The very fundamental and independent treatment of the whole
problem by Ewald [3], along quite different lines, has confirmed Darwin's
conclusions in all essentials. These papers established the following
important points.

\

\noindent 1. Two formulae for the intensity of X-ray reflection can be deduced, depending
on the assumptions which are made. The first of these has since come to be
known as the formula for the `ideally imperfect crystal' or `mosaic
crystal'.\footnote{I believe we owe to Ewald the happy suggestion of the word
`mosaic'.} It holds for a crystal in which the homogeneous blocks are so small
that the reduction in intensity of a ray passing through each block, and being
partly reflected by it, is wholly accounted for by the ordinary absorption
coefficient. This case is simple to treat from a mathematical point of view,
and in actual fact many crystals approach this physical condition of a perfect mosaic.

The second formula applies to reflection by an ideally perfect crystal. Here
ordinary\endnote{Word omitted in the French version.} absorption plays no part in intensity of reflection. This is perfect
over a finite range of glancing angles, all radiation being reflected within
this range. The range depends on the efficiency of the atom planes in
scattering. The second formula is entirely different from the first, and leads
to numerical results of a different order of magnitude.

\

\noindent 2. The actual intensity of reflection in the case of rocksalt is of the order
to be expected from the imperfect crystal formula.

\

\noindent  3. The observed rapid decline in intensity of the high orders is only partly
accounted for by the formula for reflection, and must be due in addition to the
spatial distribution of scattering matter in the atoms (electron distribution).

\

\noindent 4. When a crystal is so perfect that it is necessary to allow for the
interaction of the separate planes, the transmitted beam is extinguished more
rapidly than corresponds to the true absorption of the crystal (extinction).

\

\noindent 5. There exists a refractive index for both crystalline and amorphous
substances, slightly less than unity, which causes small deviations from the
law $n\lambda=2d\sin\theta$.

\

Another important factor in intensity of reflection had been already examined
theoretically by Debye [4], this being the diminution in intensity with rising
temperature due to atomic movement. Though subsequent work has put Debye's and
Darwin's formulae in modified and more convenient forms, the essential
features were all contained in these early papers.

On the experimental side,the first accurate quantitative measurements were
made by W.~H.~Bragg [5].\endnote{The French edition adds `Sir'.} The crystal 
was moved with constant angular velocity through the reflecting position, and the total amount of reflected radiation
measured. He showed that the reflection\endnote{[les donn\'{e}es obtenues par r\'{e}flexion]} from rocksalt 
for a series of faces lay
on a smooth curve when plotted against the sine of the glancing angle,
emphasising that a definite physical constant was being measured. This method
of measurement has since been widely used. The quantity $\frac{E\omega}{I}$,
where $E$ is the total energy of radiation\endnote{Here following the French edition; 
the typescript reads `total radiation'.} reflected, $\omega$ the angular
velocity of rotation, and $I$ the total radiation falling on the crystal face
per second, is independent of the experimental arrangements, and is a constant
for a given reflection from a mosaic crystal; it is generally termed the
`\textit{integrated reflection}'.\endnote{Emphasis omitted in the French edition.} It is related in a simple way to the energy
measurements from a powdered crystal, which have also been employed for
accurate quantitative work. W. H. Bragg's original measurements were
comparisons\endnote{[servirent \`{a} comparer]} of this quantity for different faces, not absolute measurements in
which the strength of an incident beam was considered.

W. H. Bragg further demonstrated the existence of the extinction effect
predicted by Darwin, by passing X-rays through a diamond crystal set for
reflection and obtaining an increased absorption. He made measurements of the
diminution in intensity of reflection\endnote{The French omits `of reflection'.} with rising temperature predicted by
Debye, and observed\endnote{[d\'{e}j\`{a} observ\'{e}e]} by Laue, and showed that the effect was of the expected
order. In the Bakerian Lecture in 1915~[6] he described measurements in the
intensity of a very perfect crystal, calcite, which seemed to show that the
intensity was proportional to the scattering power of the atomic planes and
not to the square of the power (this is to be expected from the formula for
reflection by a perfect crystal). In the same address he proposed the use of
the Fourier method of interpreting the measurements\endnote{[il proposa d'employer, pour l'interpr\'{e}tation des mesures, 
la m\'{e}thode de Fourier]} which has been recently
used with such success by Duane, Havighurst, and Compton, and which is dealt
with in the fourth section of this summary.\endnote{[rapport]} At about the same time, Debye and
Compton independently discussed the influence of electronic distribution in
the atom on the intensity of reflection.

The next step was made by Compton [7] in 1917. Darwin's formula for the mosaic
crystal was deduced by a different method, and was applied to the
interpretation of W.~H.~Bragg's results with rocksalt. Compton concluded that
the electronic distribution in the atoms was of the type to be expected from
Bohr's atomic model. Compton then published the first measurements of the
\textit{absolute} intensity of ref1exion. A monochromatic beam of X-rays was
obtained by reflection from a crystal, and this was reflected by a second
rotating crystal (rocksalt and\endnote{[ou]} calcite). The absolute value of the integrated
reflection $\frac{E\omega}{I}$ was found to be of the right order for rocksalt
when calculated by the imperfect crystal formula, but to be very low for
calcite indicating strong extinction or a wrong formula, in the second case.

In 1921 and 1922 I published with James and Bosanquet a series of measurements on rocksalt in which we
tried to obtain a high accuracy. We made absolute measurements of intensity
for the strongest reflections,\footnote{In our paper we failed to give due
acknowledgement to Compton's absolute measurements in 1917 of which we were
not aware at the time.} and compared the weaker reflections with them. Our main
contributions in these papers were a more accurate set of measurements of
integrated reflection for a large number of planes, and a method for estimating
and correcting for the effect of extinction. As Darwin showed in a paper in
1922 [9] on the theoretical interpretation of our results, we only succeeded
in correcting for extinction of the kind he termed `secondary' and not for
`primary' extinction.\footnote{Primary extinction is an excessive absorption
of the beam which is being reflected in each homogeneous block of crystal,
secondary extinction a statistical excessive absorption of the beam in the
many small blocks of a mosaic crystal.} Since then measurements by Havighurst
[10], by Harris, Bates and McInnes\endnote{[Mc Innes]} [11] and by Bearden~[12] have been made on
the reflecting power of powdered sodium chloride when extinction is absent.
Their measurements have agreed with ours very closely indeed, confirming one's
faith in intensity measurements, and showing that we were fortunate in
choosing a crystal for our examination where primary extinction was very
small. In the same papers we tried to make a careful analysis of the results
in order to find how much information about atomic structure could be
legitimately deduced from them, and we published curves showing the electron
distribution in sodium and chlorine\endnote{[potassium]} atoms.

In this discussion, I have refrained from any reference to the question of
reflection by `perfect' crystals. The formula for reflection by such crystals
was first obtained by Darwin, and has been arrived at independently by Ewald.
The reflection by such crystals has been examined amongst others by
Bergen Davis\endnote{The typescript has a spurious comma after `Bergen'.} 
and Stempel [13], and by Mark [14] and predictions of the
theory have been verified. It is not considered here, because I wish to
confine the discussion to those cases where a comparison of the intensity of
incident and reflected radiation leads to accurate quantitative estimates of
the distribution of scattering matter. This ideal can be attained with actual
crystals,\endnote{[\`{a} l'aide de cristaux]} when they are of the imperfect or mosaic type, though allowance for
extinction is sometimes difficult in the case of the stronger reflections. On
the other hand, it is far more difficult to know what one is measuring in the
case of crystals which approximate to the perfect type. It is a fortunate
circumstance that mathematical formulae can be applied most easily to the type
of imperfect crystal more common in nature.

\

\begin{center}
\par
\Needspace{5\baselineskip}
{\sc 3. --- Results of quantitative analysis}\addcontentsline{toc}{section}{Results of quantitative analysis}
\end{center} 
For the sake of conciseness, only one of the many intensity formulae will be
given here, for it illustrates the essential features of them all. Let us
suppose that the integrated reflection is being measured when X-rays fall on
the face of a rotating crystal of the mosaic type. We then have%
\[
\rho=\frac{Q}{2\mu}\frac{1+\cos^{2}2\theta}{2}\ .
\]

\noindent (a) $\mu$ is the effective absorption coefficient, which may be greater than
the normal coefficient, owing to the existence of extinction at the reflecting angle.

\

\noindent (b) The factor $\frac{1+\cos^{2}2\theta}{2}$ is the `polarisation factor',
which arises because the incident rays are assumed to be unpolarised.

\

\noindent (c)%
\[
Q=\left(  \frac{Ne^{2}}{mc^{2}}F\right)  ^{2}\frac{\lambda^{3}}{\sin2\theta
}\ ,
\]
where $e$ and $m$ are the electronic constants,\endnote{[les deux constantes \'{e}lectroniques]} $c$ the velocity of light,
$\lambda$ the wavelength used, $N$ the number of scattering units per unit
volume, and $\theta$ the glancing angle.

\

\noindent (d) $F$ is the quantity we are seeking to deduce. It represents the scattering
power of the crystal unit in the direction under consideration, measured in
terms of the scattering power of a single\endnote{Here and in some other instances, 
the French renders `single' as `simple'.} electron according to the classical
formula of J. J. Thomson. It is defined by Ewald as the `Atomfaktor'\endnote{[`facteur atomique']} when it
applies to a single atom.

\

Formulae applicable to other experimental arrangements (the powder method for
instance) are very similar, and contain the same quantity $Q$. Our
measurements of reflection thus lead to values of $Q$, and so of $F$, since all
other quantities in the formulae are known. Measurements on a given crystal
yield a series of values for $F$, and all the information that can be found
out about this crystalline or atomic structure is represented by these values.
They are the same for the same crystal whatever wavelength is employed (since
$F$ is a function of $\frac{\sin\theta}{\lambda}$), though of course with
shorter wavelength we have the advantage of measuring a much greater number of
these coefficients (increased resolving power).

At this stage the effect of the thermal agitation of the atom will be
considered as influencing the value of $F$. If we wish to make deductions
about atomic structure, the thermal agitation must be taken into account.
Allowance for it is a complicated matter, because not only do some atoms move
more than others, but also they change their relative mean positions as the
temperature alters in the more complex crystals.

This will be dealt with more fully below.

A series of examples will now be given to show that these quantitative
formulae, when tested, lead to results which indicate that the theory is on
the right lines. It is perhaps more convincing to study the results obtained
with very simple crystals, though I think that the success of the theory in
analysing highly complex structures is also very strong evidence, because we
have covered such a wide range of substances.

In the simple crystals, where the positions of the atoms are definite, we can
get the scattering power of individual atoms. The results should both indicate
the correct number of electrons in the atom, and should outline an atom of
about the right size. When $F$ is plotted against $\frac{\sin\theta}{\lambda}$
its value should tend to the number, $N$, of electrons in the atom for small
values of $\frac{\sin\theta}{\lambda}$, and should fall away as $\frac
{\sin\theta}{\lambda}$ increases, at a rate which is reasonably explained by
the spatial extension of the atom. In Fig.~1, the full lines give $F$ curves
obtained experimentally by various observers. The dotted lines are $F$ curves
calculated for the generalised atomic model of Thomas [15], of appropriate
atomic number.
  \begin{figure}
    \centering
      \resizebox{\textwidth}{!}{\includegraphics[0mm,0mm][220.30mm,185.98mm]{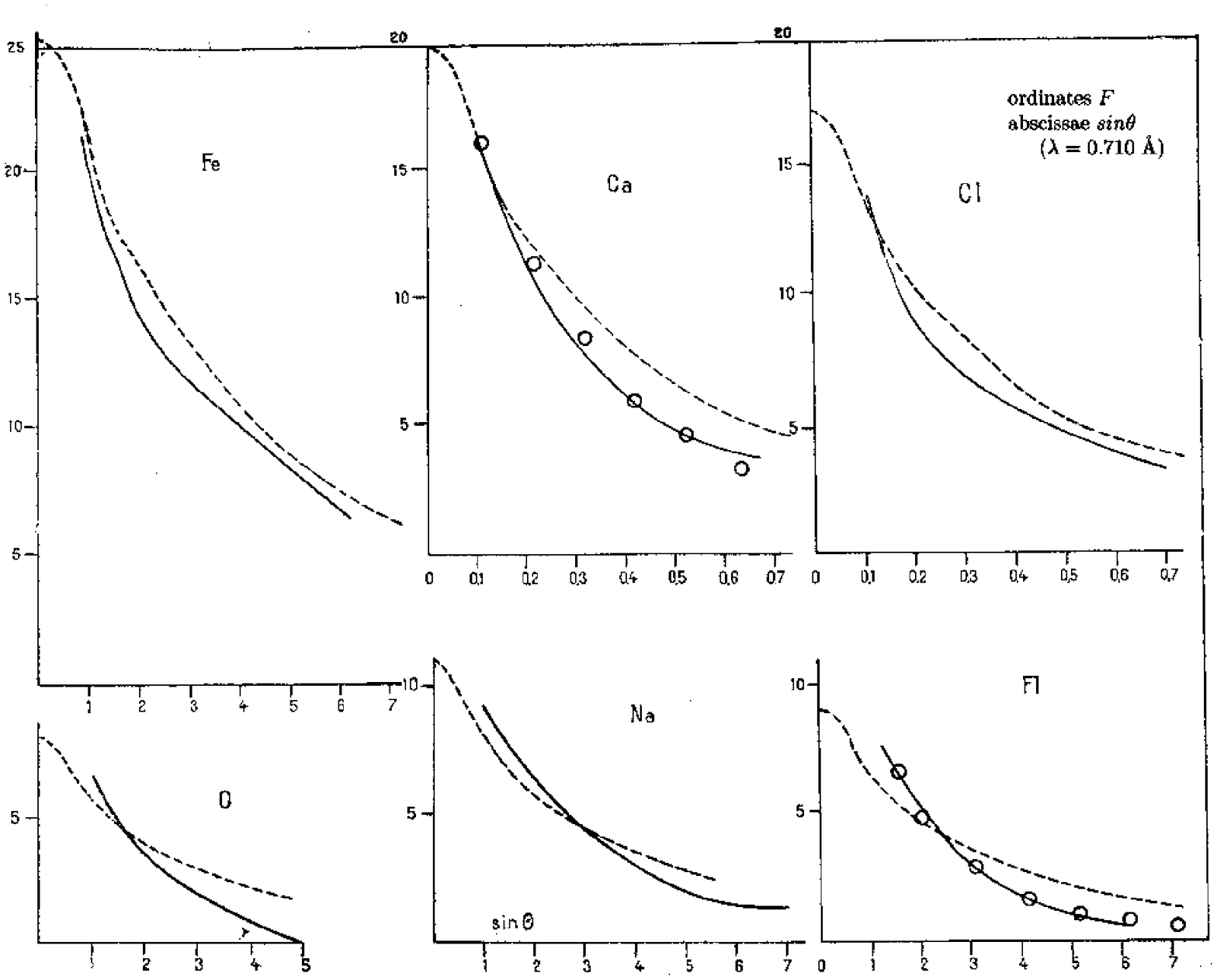}}
    \unnumberedcaption{Fig.~1.}
  \end{figure} 
The Thomas atomic model, which has been shown for comparison, is most useful
as it gives us the approximate electronic distribution in an atom of any
atomic number. Thomas calculates an ideal distribution of electrons in an atom
of high atomic number. He assumes spherical symmetry for the atom, and
supposes that `electrons are distributed uniformly in the six-dimensional
phase-space for the motion of an electron, at the rate of two for each $h^{3}$
of (six) volume'.\endnote{Not printed as a quotation in the French edition.} He thus obtains an ideal electron atmosphere around the
nucleus, the constants of which can be simply adjusted\endnote{choisies simplement} so as to be suitable
for any given nuclear charge. It is of course to be expected that the lower
the atomic weight, the more the actual distribution of scattering matter will
depart from this arrangement, and will reflect the idiosyncrasies of the 
particular atom in question. The figure will
show, however, that the actual curves are very similar to those calculated for
Thomas' models. In particular, it will be clear that they tend to maximum
values not far removed from the number of electrons in the atom in each case.
The general agreement between the observed and calculated $F$ curves must mean
that our measurements of $F$ are outlining a picture of the atom. The
agreement holds also for other atomic models than those of Thomas, which all
lead to atoms with approximately the same spatial extension and electronic
distribution, as is well known.

All these measurements of $F$ necessitate absolute values for the integrated
reflection. It is not necessary to measure these directly in each case. When
any one reflection has been measured in absolute value (by comparison of
incident and reflected radiation), other crystals may be compared with it. The
standard which has been used in every case, as far as I am aware, is the
rocksalt crystal. Absolute measurements on this have been made by Compton [7],
by Bragg, James and Bosanquet [8], and by Wasastjerna [18] which agree
satisfactorily with each other.

\

\begin{center}
\par
\Needspace{5\baselineskip}
{\sc 4. --- Interpretation of measurements of 
$F$}\addcontentsline{toc}{section}{Interpretation of measurements of $F$}
\end{center} 
In interpreting these measurements of scattering power, we may either
calculate the scattering of a proposed atomic model and compare it with the
observed $F$ curve, or we may use the observations to calculate the
distribution of scattering matter directly. The latter method is the more
attractive, and in the hands of Duane, Havighurst, and Compton it has yielded
highly interesting `images' of the atomic structure seen by X-rays. There is a
close analogy between the examination of a series of parallel planes by means
of X-rays, and the examination of a diffraction grating, by a microscope,
which is considered in Abbe's theory of
microscopic vision.\footnote{See for instance the discussion of this theory
and of A.~B.~Porter's experiments to illustrate it in Wood's \textit{Optics},
Chapter VIII.} The objective of the microscope may be considered as receiving
a limited number of orders of spectra from the grating. These spectra in their
turn build up the image viewed by the eyepiece, and the perfection of this
image depends on the number of spectra received. The strength of each spectral
order depends on the magnitude of the corresponding coefficient in that
Fourier series which represents the amplitude of the light transmitted at each
point of the grating. The extension of this well-known optical principle to
the X-ray field was suggested by W.~H.~Bragg [6] in 1915. He had formed the
conclusion\endnote{The French adds `de ses exp\'{e}riences'.} that the amplitudes 
of the scattered wave from rocksalt were inversely proportional to the square of 
the order of reflection, and he showed that\endnote{The French adds `dans ces 
conditions'.} `the periodic function which represents the density of the medium 
must therefore be of the form\endnote{This is indeed a quotation from p.~272 from
the lecture by W.~H.~Bragg. The typescript has a comma instead of the closing 
quotation mark, while the French edition omits the opening quotation mark. The typescript 
has a spurious denominator `$a$' instead of `$d$' in the second and third terms (but 
tacitly corrects another typo in the original).}%
\[
\mathrm{const}+\frac{\cos2\pi\frac{x}{d}}{1^{2}}+\frac{\cos4\pi\frac{x}{d}%
}{2^{2}}+\ ...\ +\frac{\cos2n\pi\frac{x}{d}}{n^{2}}+\ ...\ \mbox{'}
\]
and in this way built up a curve showing the periodic density of the rocksalt
grating. The method was not applied, however, to the much more accurate
measurements which are now available until recently, when Duane and Havighurst
showed how much could be done with it. Duane independently arrived at a more
general formula of the same type, giving the density of scattering matter at
any point in the whole crystal as a triple Fourier series, whose coefficients
depend on the intensity of reflection from planes of all possible indices.
Havighurst applied this principle to our measurements of rocksalt, and to
measurements which he has made on other crystals, and obtained a picture of
the relative density of scattering matter along certain lines in these
crystals. Compton made the further step of putting the formulae in a form
which gives the absolute density of electronic distribution (assuming the
scattering to be by electrons obeying the classical laws). Compton gives a
very full discussion of the whole matter in his book \textit{X-rays and
Electrons}.\endnote{[dans son livre sur `les rayons X et les \'{e}lectrons']} It is not only an extremely attractive way of making clear just
what has been achieved by the X-ray analysis, but also the most direct method
of determining the structure.

The formula for the distribution of scattering matter in parallel sheets, for
a crystal with a centre of symmetry, is given by Compton as follows%
\[
P_{z}=\frac{Z}{a}+\frac{2}{a}\sum_{1}^{\infty}F_{n}\cos\frac{2\pi nz}{a}\ .
\]
Here $z$ is measured perpendicularly to the planes which are spaced a distance
$a$ apart. $P_{z}dz$ is the amount of scattering matter between planes at
distances $z$ and $z+dz$, and $Z$ ($=\int_{0}^{a}P_{z}dz$) is the total
scattering matter of the crystal unit. This is a simplified form of Duane's
formula for a Fourier series of which the general term is%
\[
A_{n_{1}n_{2}n_{3}}\sin\left(  \frac{2\pi n_{1}x}{a_{1}}-\delta_{n_{1}%
}\right)  \sin\left(  \frac{2\pi n_{2}y}{a_{2}}-\delta_{n_{2}}\right)
\sin\left(  \frac{2\pi n_{3}z}{a_{3}}-\delta_{n_{3}}\right)  \ ,
\]
$A_{n_{1}n_{2}n_{3}}$ being proportional to the amplitude of the scattered
wave from the plane $(n_{1}n_{2}n_{3})$.

Another Fourier series, due to Compton, gives the radial distribution of
scattering matter, i.e. the values of $U_{n}$ where $U_{n}dr$ is the amount of
scattering matter between radii $r$ and $r+dr$%
\[
U_{n}=\frac{8\pi r}{a^{2}}\sum_{1}^{\infty}nF_{n}\sin\frac{2\pi nr}{a}\ ,
\]
where $a$ is chosen so that values of $F$ occur at convenient intervals on the
graph for $F$.

If we know the values of $F$ for a given atom over a sufficiently wide range,
we can build up an image of the atom either as a `sheet distribution' parallel
to a plane, or as a radial distribution of scattering matter around the
nucleus. In using these methods of analysis, however, it is very necessary to
remember that we are working right at the limit of resolving power of our
instruments, and in fact are attempting a more ambitious problem than in the
corresponding optical case. In A.~B.~Porter's experiments to test Abbe's theory, 
he viewed the image of a diffraction grating and removed
any desired group of diffracted rays by cutting them off with a screen. The
first order gives blurred lines, four or five orders give sharper lines with a
fine dark line down the centre, eight orders give two dark lines down the
centre of each bright line and so forth. These imperfect images are due to the
absences of the higher members in building up the Fourier series. In exactly
the same way we get false detail in our X-ray image, owing to ignorance of the
values of the higher members in the $F$ curve. Similarly, the fine structure
which actually exists may be glossed over, since by using a wavelength of 
0.7~\AA , we cannot hope to `resolve' details of atomic
structure on a scale of less than half this value.

The ignorance of the values of higher members of the Fourier series matters
much less in the curve of sheet distribution than in that for radial
distribution, since the latter converges far more slowly. Examples of the
Fourier method of analysis are given in the next paragraph.

As opposed to this method of building up an image from the X-ray results,\endnote{The French
adds `on peut proc\'{e}der de la fa\c{c}on inverse, c'est-\`{a} dire'.} we
may make an atomic model and test it by calculating an $F$ curve for it which
can be compared with that obtained experimentally. This is the most
satisfactory method of testing models arrived at by other lines of research,
for nothing has to be assumed about the values of the higher coefficients $F$.
It is of course again true that our test only applies to details of the
proposed model on a scale comparable with the wavelength we are using. Since
we can reflect X-rays right back from an atomic plane, we may get a resolving
power for a given wavelength with the X-ray method twice as great as the best
the microscope can yield.

It is perhaps worth mentioning the methods I used with James and Bosanquet in
our determination of the electronic distribution in sodium and chlorine in
1922. We tried to avoid extrapolations of the $F$ curve beyond the limit of
experimental investigation. We divided the atom arbitrarily into a set of
shells, with an unknown number of electrons in each shell. These unknowns were
evaluated by making the scattering due to them fit the $F$ curve over the
observed range, this being simply done by solving simultaneous linear
equations. We found we got much the same type of distribution however the
shells were chosen, and that a limit to the electronic distribution at a
radius of about 1.1 \AA\ in sodium and 1.8
\AA\ in chlorine was clearly indicated. Our distribution
corresponds in its general outline to that found by the much more direct
Fourier analysis, as the examples in paragraph 7 will show.

  \begin{figure}
    \centering
      \resizebox{\textwidth}{!}{\includegraphics[0mm,0mm][220.53mm,225.17mm]{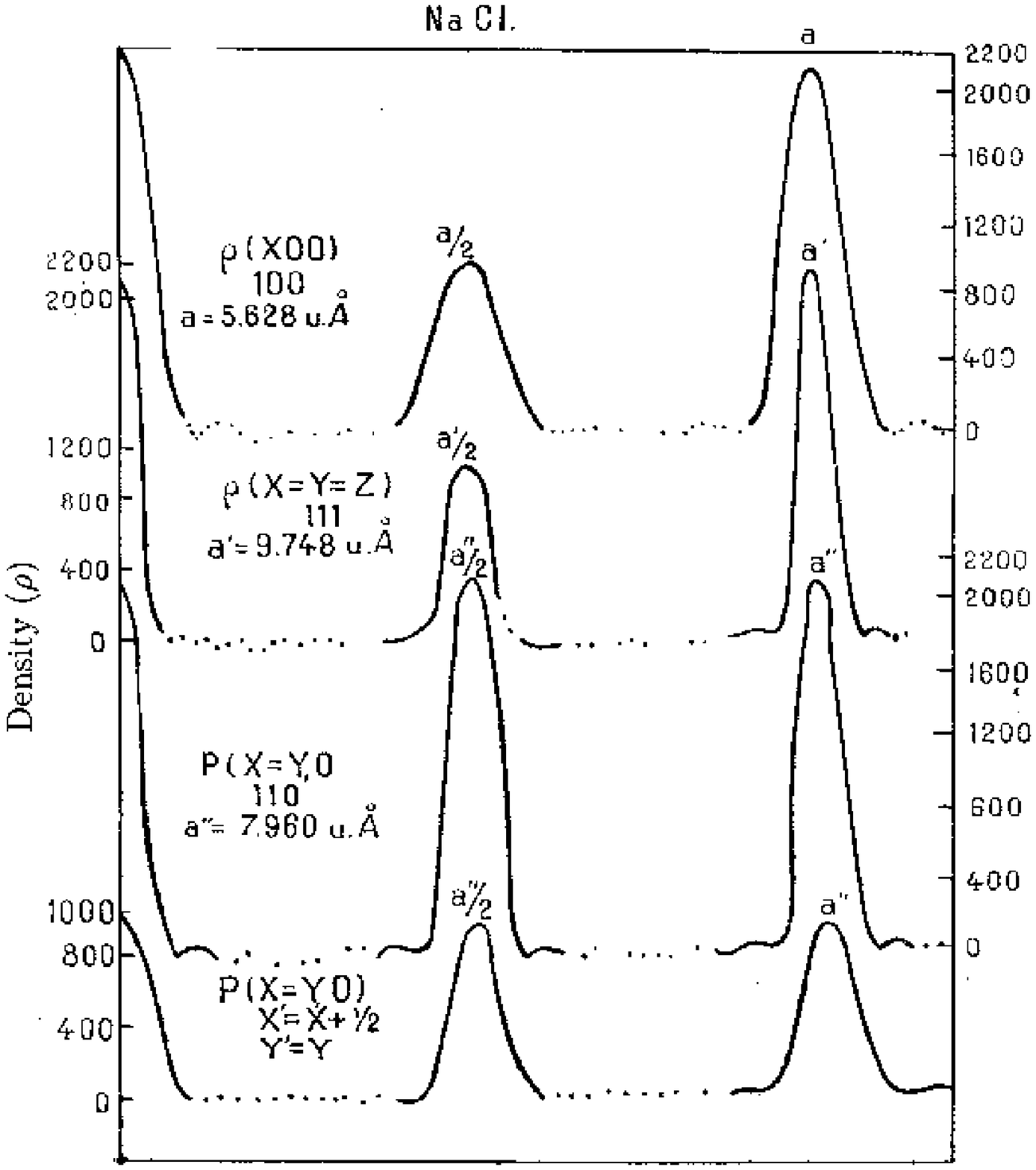}}
    \unnumberedcaption{Fig.~2.}
  \end{figure}

\begin{center}
\par
\Needspace{5\baselineskip}
{\sc 5. --- Examples of analysis}\addcontentsline{toc}{section}{Examples of analysis}
\end{center} 
We owe to Duane [20] the appreciation of the very attractive way in which the
Fourier analysis represents the results of X-ray examinations. It has the
great merit of representing, in the form of a single curve, the information
yielded by all orders of reflection from a given plane, or from the whole
crystal. It is of course only an alternative way of interpreting the results,
and the deductions we can make about atomic or molecular structures depend in
the end on the extent to which we can trust our experimental observations, and
not on the method of analysis we use. The Fourier method is so direct however,
and its significance so easy to grasp, that Duane's introduction of it marks a
great advance in technique of analysis.

I have reserved to paragraph 7 the more difficult problem of the arrangement
of scattering matter in the atoms themselves, and the examples given here are
of a simpler character. They illustrate the application of analysis to the
general problem of the distribution of scattering matter in the whole crystal,
when we are not so near the limit of resolving power. The curves in Fig.~2
represent the first application of the new method of Fourier analysis to
accurate data, carried out by Havighurst [21] in 1925. He used our
determinations of $F$ for sodium chloride, and Duane's three-dimensional
Fourier series, and calculated the density of scattering matter along a cube
edge through sodium and chlorine centres, along a cube diagonal through the
same atoms, and along two face diagonals chosen so as to pass through chlorine
atoms alone or sodium atoms alone in the crystal. The atoms show as peaks in
the density distribution.\addtocounter{endnote}{1}\endnotetext{`Beryl.' omitted in French edition.}
  \begin{figure}
    \centering
      \resizebox{\textwidth}{!}{\includegraphics[0mm,0mm][220.59mm,165.27mm]{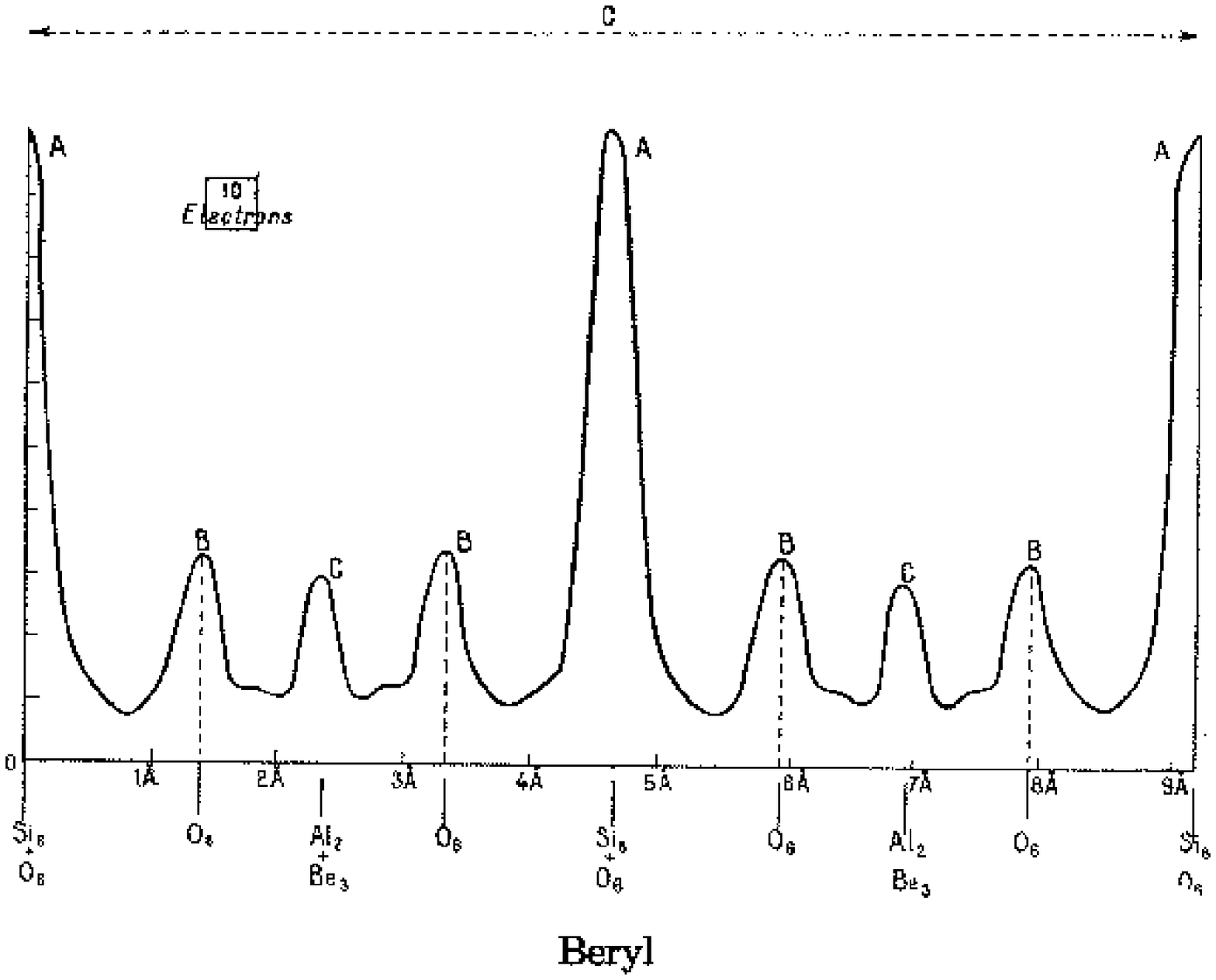}}
    \unnumberedcaption{Fig.~3a. --- Distribution of electrons in sheets parallel to 0001.$^{37}$}
  \end{figure} 
In the other examples, the formula for distribution in sheets has been applied
to some results we have obtained in our work on crystal structure at
Manchester. I have given them because I feel they are convincing evidence of
the power of quantitative measurements, and show that all methods of
interpretations lead to the same results.
  \begin{figure}
    \centering
      \resizebox{\textwidth}{!}{\includegraphics[0mm,0mm][220.30mm,85,47mm]{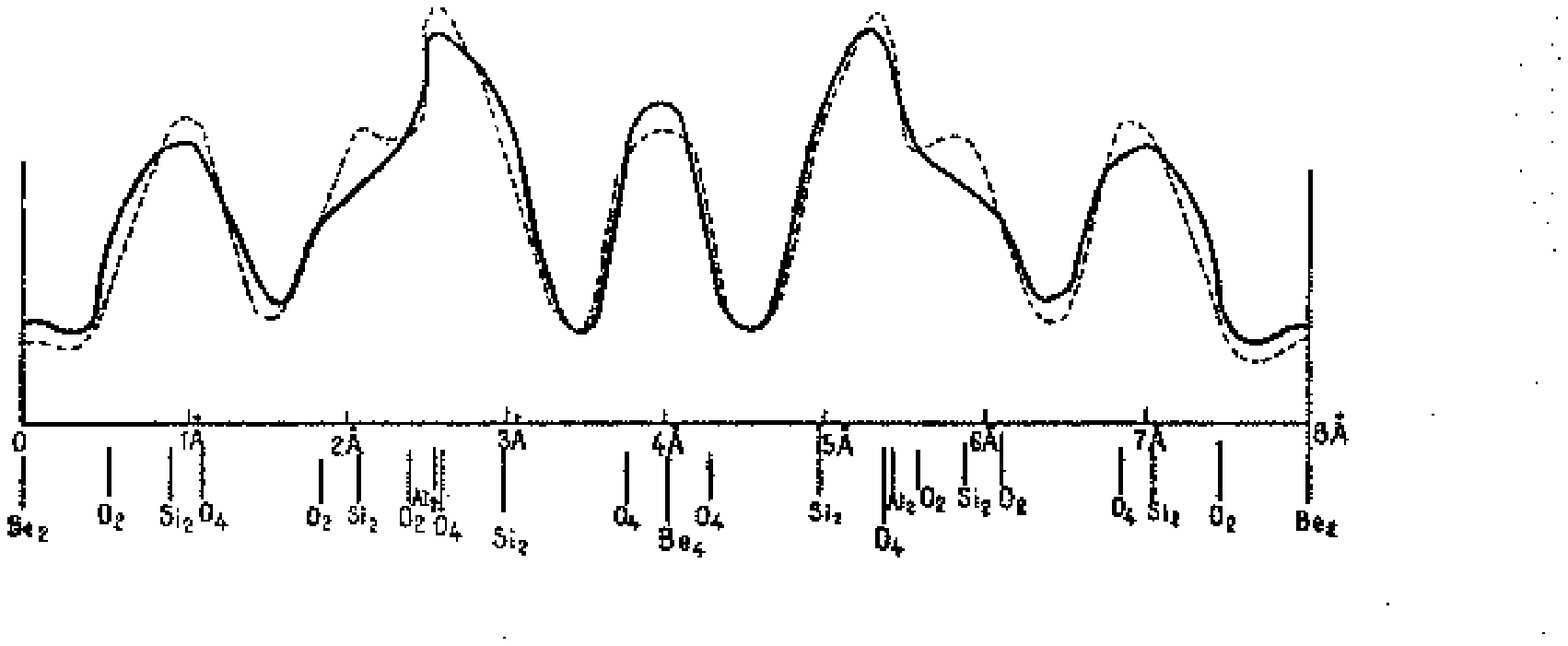}}
    \unnumberedcaption{Fig.~3b. --- Distribution of electrons in sheets parallel to $10\bar{1}0$.$^{39}$} 
  \end{figure}

Mr West and I [22] recently analysed the hexagonal crystal beryl, 
Be$_{3}$A1$_{2}$Si$_{6}$O$_{18}$,\endnote{The French edition uses superscripts throughout.}%
\addtocounter{endnote}{1}\endnotetext{The French edition omits the overbar in the caption.} 
which has a structure of some complexity, depending on
seven parameters. We obtained the atomic positions by the usual method of
analysis, using more or less known $F$ curves for the atoms in the crystal,
and moving them about till we explained the observed $F$s due to the crystal
unit. Fig.~3 shows the reinterpretations of this result by the Fourier
method. Fig.~3a gives the electron density in sheets perpendicular to the
principal axis of the crystal, which is of a very simple type. The particular
point to be noted is the correspondence between the position of the line B in
the figure and the hump of the Fourier analysis. The line B marks the position
of a group of oxygen atoms which lies between two other groups A and C fixed
by symmetry, the position of B being fixed by a parameter found by familiar
methods of crystal analysis. The hump represents the same group fixed by the
Fourier analysis, and it will be seen how closely they correspond. In Figs.~3b 
and 3c more complex sets of planes are shown. The dotted curve represents
the interpretation of our results by Fourier analysis. The full curve is got
by adding together the humps due to the separate atoms shown below, the
position of these having been obtained by our X-ray analysis and their sizes
by the aid of the curve in Fig.~3a in which the contribution of the atoms
can be separated out. The correspondence between the two shows that the older
methods and the Fourier analysis agree. It is to be noted that the crystal had
first to be analysed by the older methods, in order that the sizes of the
Fourier coefficients might be known.
  \begin{figure}
    \centering
      \resizebox{\textwidth}{!}{\includegraphics[0mm,0mm][220.40mm,190,78mm]{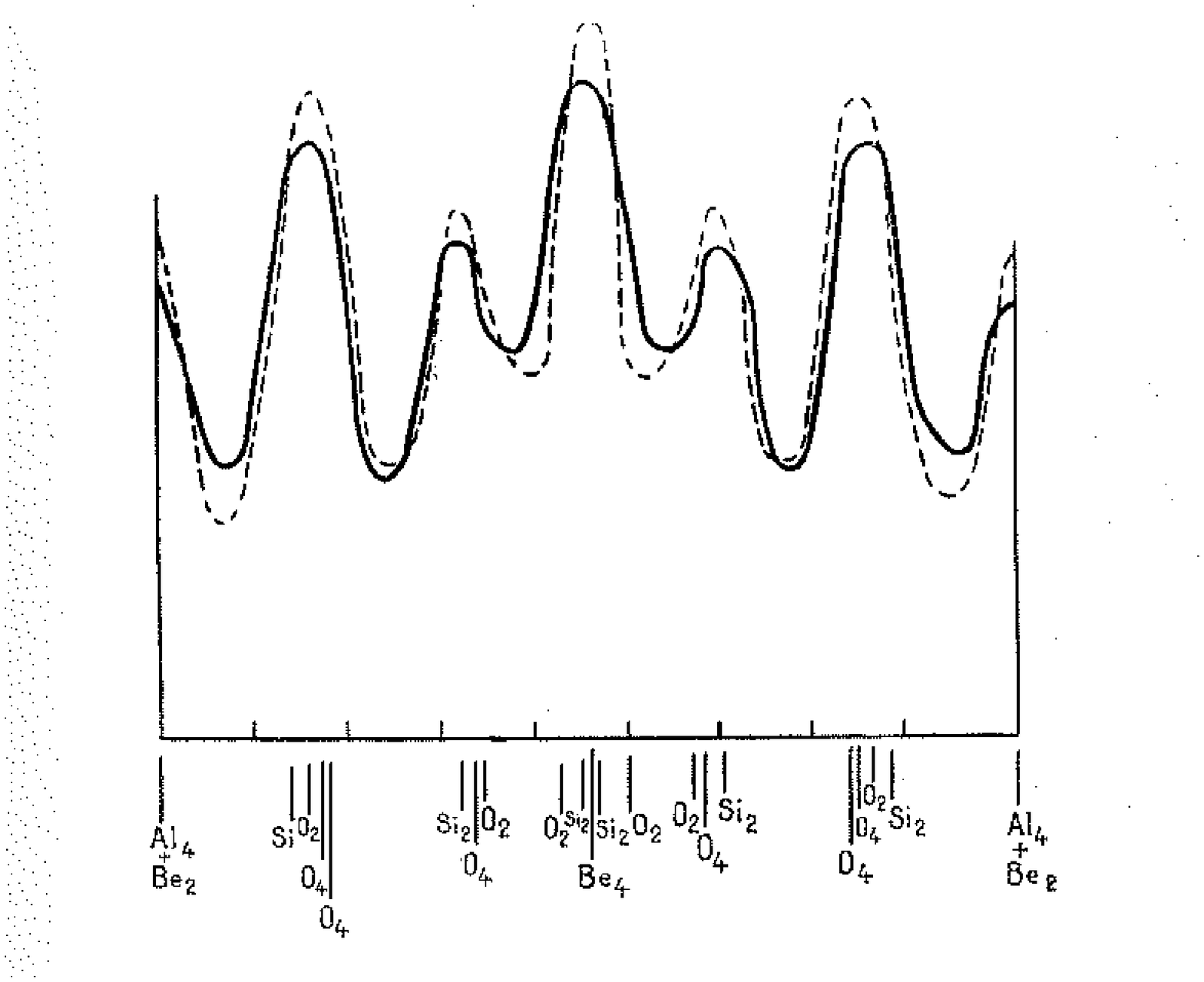}}
    \unnumberedcaption{Fig.~3c. --- Distribution of electrons in sheets parallel to $11\bar{2}0$.$^{42}$}
  \end{figure}

In Fig.~4 I have given a set of curves for the alums, recently analysed by
Professor Cork [23]. The alums are complicated cubic crystals with such
formulae as KAl(SO$_{4}$)$_{2}$.12 H$_{2}$O. Wyckoff\endnote{[Wyckhoff]} has shown that the
potassium and aluminium atoms\endnote{Word omitted in French edition.}\addtocounter{endnote}{1}\endnotetext{Again, 
the French edition omits the overbar in the caption.} occupy the same positions in the cubic cell as
the sodium atom in rocksalt. Now we can replace the potassium by ammonium,
rubidium, caesium, or thallium, and the aluminium by chromium, or other
trivalent metals. Though the positions of the other atoms in the crystals are
not yet known, they will presumably be much the same in all these crystals. If
we represent by a Fourier series the quantitative measurements of the alums,
we would expect the density of scattering matter to vary from crystal to
crystal at the points occupied by the metal atoms, but to remain constant
elsewhere. The curves show this in the most interesting\endnote{[frappante]} way.
  \begin{figure}
    \centering
      \resizebox{\textwidth}{!}{\includegraphics[0mm,0mm][220.35mm,202.71mm]{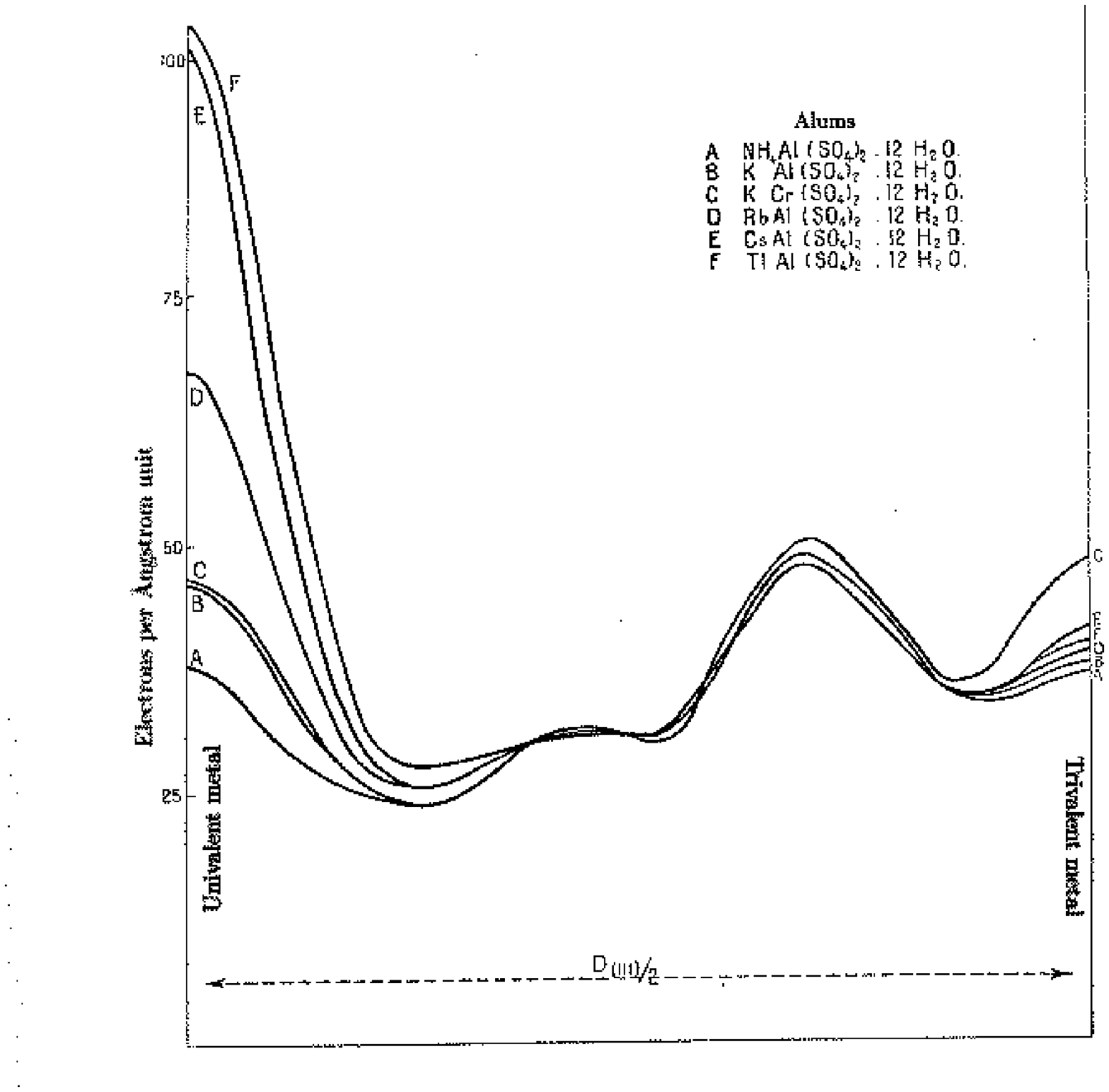}}
    \unnumberedcaption{Fig.~4.}
  \end{figure} 

The effect of heat motion on the movements of the atoms has already been
mentioned. It was first treated theoretically by Debye [4]. Recently Waller
[24] has recalculated Debye's formula, and has arrived at a modified form of
it. Debye found that the intensities of the interference maxima in a simple
crystal should be multiplied by a factor $e^{-M}$, where%
\begin{align*}
M  &  =\frac{6h^{2}}{\mu k\Theta}\frac{\varphi(x)}{x}\frac{\sin^{2}\theta
}{\lambda^{2}}\ ,\\[1ex]
x  &  =\frac{\Theta}{T}=\frac
{\mathrm{characteristic\ temperature\ of\ crystal}}%
{\mathrm{absolute\ temperature}}\ .
\end{align*}
  \begin{figure}
    \centering
      \resizebox{\textwidth}{!}{\includegraphics[0mm,0mm][220.11mm,230.10mm]{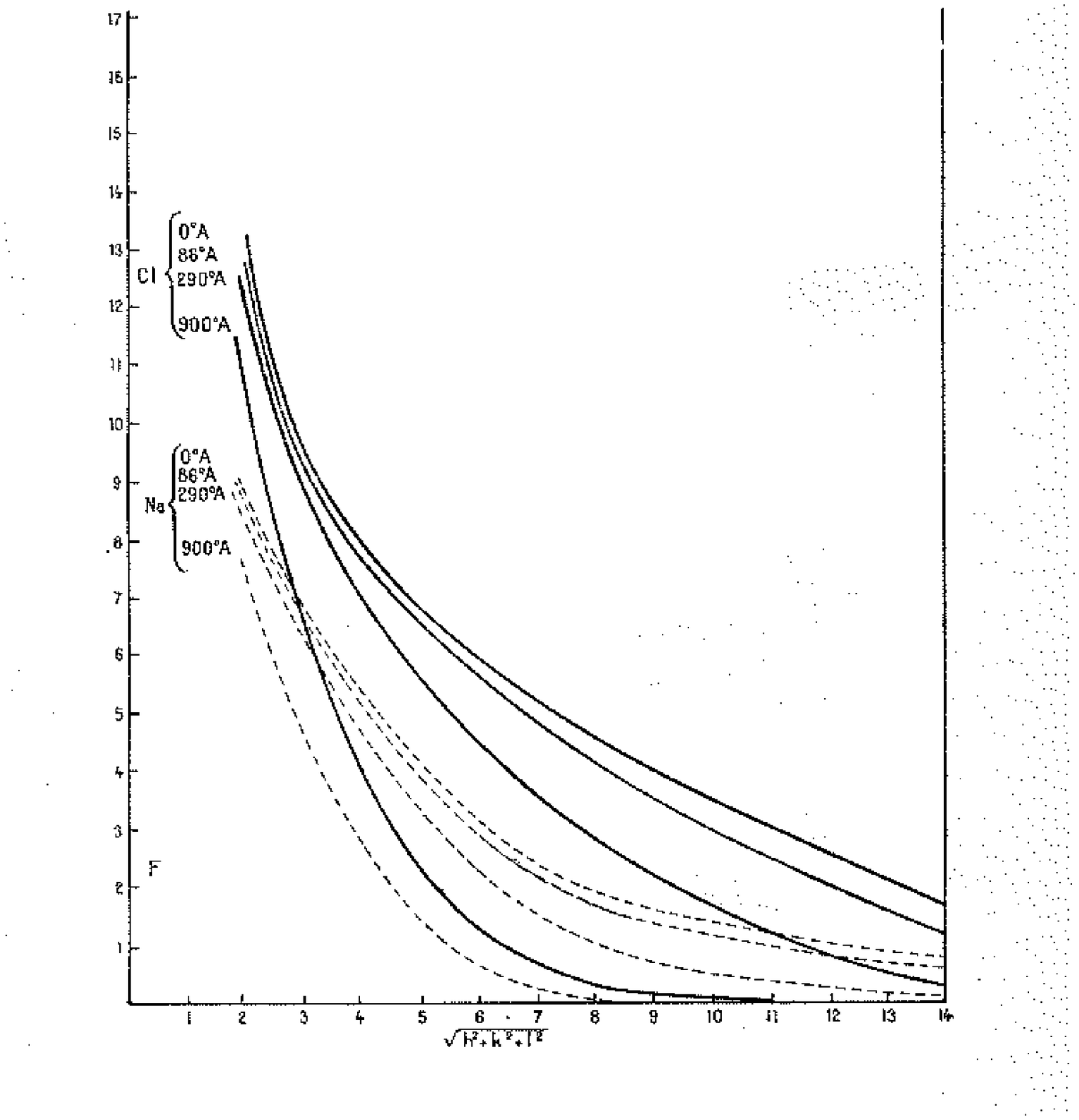}}
    \unnumberedcaption{Fig.~5.}
  \end{figure} 

Without going into further detail, it is sufficient to note that Waller's
formula differs from Debye's by making the factor $e^{-2M}$, not $e^{-M}$.
James and Miss Firth [25] have recently carried out a series of measurements
for rocksalt between the temperatures 86$%
{{}^\circ}%
$ abs. and 900$%
{{}^\circ}%
$ abs. They find that Waller's formula is very closely followed up to 500$%
{{}^\circ}%
$ abs., though at higher temperatures the decline in intensity is even more
rapid, as is perhaps to be expected owing to the crystal becoming more loosely
bound. I have given the results of the measurements in Figs. 5 and 6, both as
an example of the type of information which can be got from X-ray
measurements, and because these actual figures are of interest as a set of
careful and accurate measurements of scattering power.

Fig. 5 shows the $F$ curves for sodium and chlorine at different temperatures.
The rapid decline in intensity for the higher orders will be realised when it
is remembered that they are proportional to $F^{2}$. The curve for absolute
zero is an extrapolation from the others, following the Debye formula as
modified by Waller.
  \begin{figure}
    \centering
      \resizebox{\textwidth}{!}{\includegraphics[0mm,0mm][220.06mm,170.57mm]{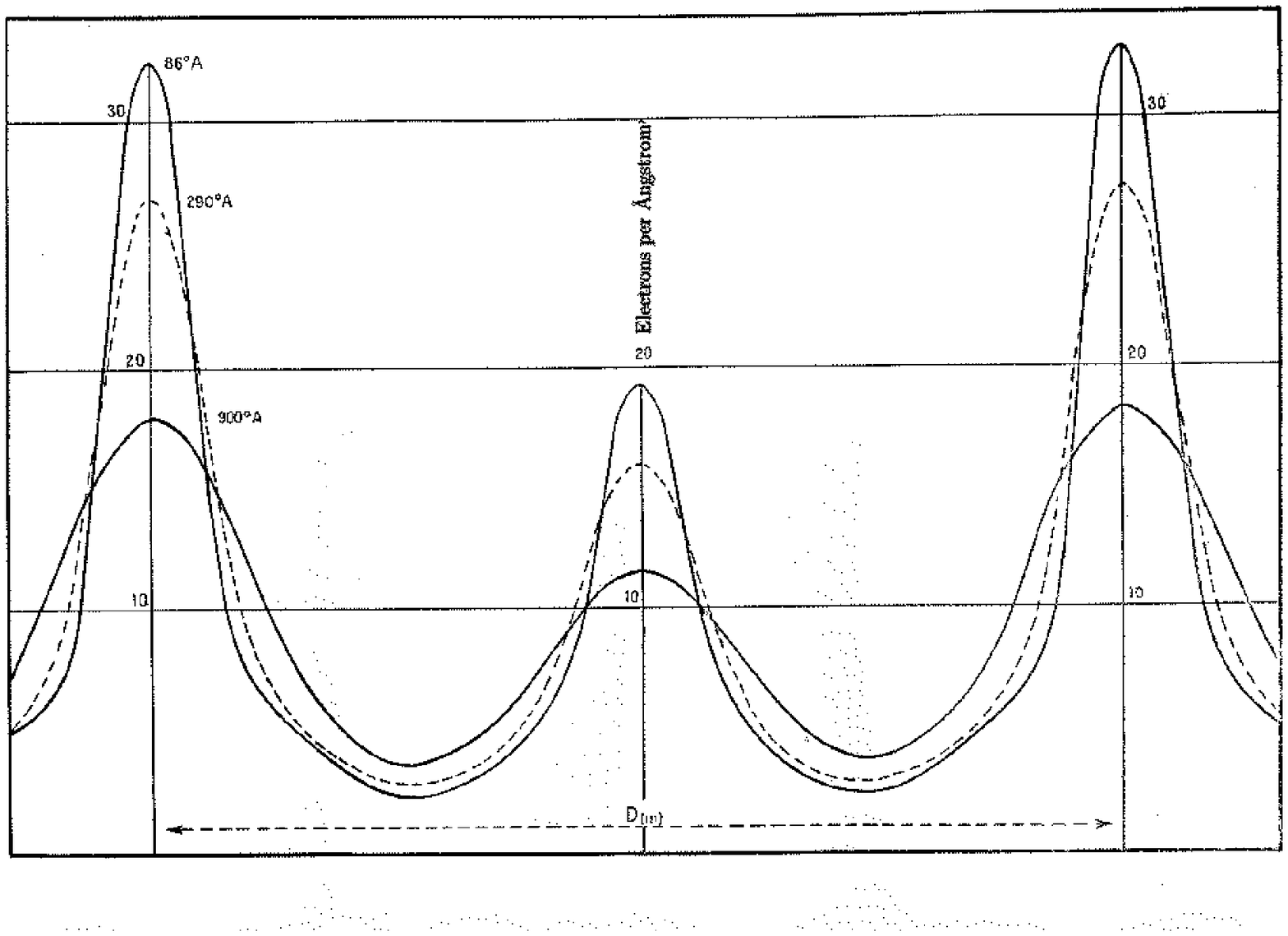}}
    \unnumberedcaption{Fig.~6.}
  \end{figure} 

In Fig. 6 the same results are interpreted by the Fourier analysis. The curve
at room temperatures for NaCl is practically identical with the interpretation
of our earlier figures by Compton, in his book \textit{X-rays and 
Electrons},\endnote{[son livre sur les rayons X et les \'{e}lectrons]}
though the figures on which it is based should be more accurate.\endnote{[bien que les figures ({\em sic})
sur lesquelles la nouvelle courbe se base soient plus exactes]} The curves
show the manner in which the sharply defined peaks due to Cl and Na at low
temperatures become diffuse owing to heat motion at the higher temperatures.
  \begin{figure}
    \centering
      \resizebox{\textwidth}{!}{\includegraphics[0mm,0mm][220.31mm,170.00mm]{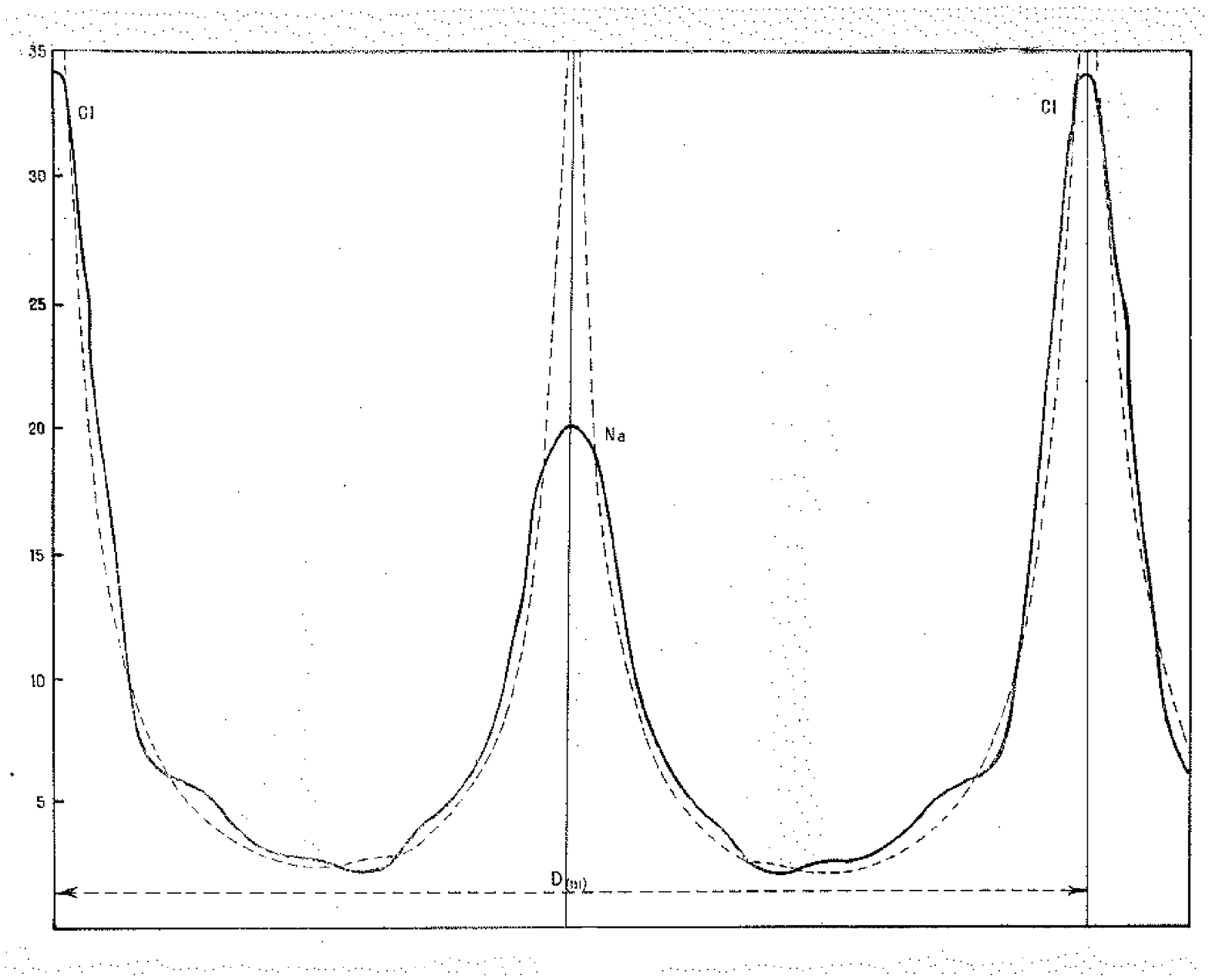}}
    \unnumberedcaption{Fig.~7.}
  \end{figure} 

Several interesting points arise in connection with this analysis. In the
first place, James and Firth\endnote{Here and in several other places, the French adds `Mlle'.} 
find that the heat factor is different for sodium
and chlorine, the sodium atoms moving with greater average amplitudes than the
chlorine atoms. This has a very interesting bearing on the crystal dynamics
which is being further investigated by Waller. To a first approximation both
atoms are affected equally by the elastic waves travelling through the
crystal, but in a further approximation it can be seen that the sodium atoms
are more loosely bound than the chlorine atoms. If an atom of either kind were
only fixed in position by the six atoms immediately surrounding it, Waller has
shown that there would be no difference between the motions of a sodium atom
between six chlorine atoms, or a chlorine atom between six sodium atoms.
However, the chlorine is more firmly pinned in position because it has in
addition twelve large chlorine neighbours, whereas the sodium atom is much
less influenced by the twelve nearest sodium atoms. Hence arises the
difference in their heat motions. It is important to find the correct method
for reducing observations to absolute zero, and this difference in heat motion
must be satisfactorily analysed before this is possible.

In the second place, the accuracy which can be attained by the experimental
measurements holds out some hope that we may be able to test directly whether
there is zero-point energy\endnote{The French reads `une \'{e}nergie au z\'{e}ro absolu (\'{e}nergie de structure)'.} 
or not. This is being investigated by James and
Waller. If a reliable atomic model is available, it would seem that the
measurements can tell whether there is vibration at absolute zero or not, for
the theoretical diminution in intensity due to the vibration is much larger
than the experimental error in measuring $F$. I feel considerable diffidence
in speaking of the question of zero-point energy, and would like to have the
advice of the mathematical physicists present.

We may calculate, either from the measured heat factor or directly from the
Fourier analyses, the average amplitude of vibration for different
temperatures. James and Firth find by both methods, for instance, that at room
temperature the mean amplitude of vibration for both atoms is 0.21
\AA , and at 900$%
{{}^\circ}%
$ abs.\ it is about 0.58 \AA . They examined the form which
the Fourier curve at 0$%
{{}^\circ}%
$ abs. assumes when it is deformed by supposing all the atoms to be in
vibration with the same mean amplitude.

It has been already remarked that the observed $F$ curves for atoms are very
similar to those calculated for the Thomas atomic model. 
The same comparison
may be made between the distributions of scattering matter. In Fig. 7 the
distribution in sheets for NaCl at absolute zero is shown as a full curve. The
dotted curve shows the horizontal distribution in sheets for atoms of atomic
number 17 and 11. In Thomas' model the density rises towards an infinite value
very close to the nucleus, and this is represented by the very sharp peaks at
the atomic centres in the dotted curve. We would not expect the observed
distribution to correspond to the actual Thomas distribution at these points.
Throughout the rest of the crystal the distribution is very similar. The
comparison is interesting, because it shows how delicate a matter it is to get
the fine detail of atomic structure from the observations. Thomas'
distribution is quite continuous and takes no account of K, L and M sets of
electrons. The slight departures of the observed curve from the smooth Thomas
curve represent the experimental evidence for the existence of all the
individual features of the atom.

\

\begin{center}
\par
\Needspace{5\baselineskip}
{\sc  6. --- The mechanism of X-ray scattering}\addcontentsline{toc}{section}{The mechanism of X-ray scattering}
\end{center} 
Before going on to discuss the application of the analysis to atomic
structure, it is necessary to consider what is being measured when a
distribution of scattering matter is deduced from the X-ray results. The
classical treatment regards the atom as containing a number of electrons, each
of which scatters radiation according to the formula of J.~J.~Thomson. Since a
vast number of atoms contribute to the reflection by a single crystal plane, we
should obtain a picture of the \textit{average electronic distribution}. The
quantity $F$ should thus tend to a maximum value, at small angles of
scattering, equal to the number of electrons in the atom, and should fall away
owing to their spatial distribution as $\frac{\sin\theta}{\lambda}$ increases.
The observed\endnote{Word missing in the French edition.} $F$ curves are of this character, as has been seen. When
interpreted as an atomic distribution, they give atoms containing the correct
number of electrons, and this seems satisfactory from the classical viewpoint.
On the other hand, the evidence of the Compton effect would appear at first
sight to cast doubt on the whole of our analysis. What we are measuring is
essentially the \textit{coherent} radiation diffracted by the crystal, whereas
the Compton effect shows that a part of the radiation which is scattered is of
different wavelength. Further, this radiation of different wavelength is
included with the coherent radiation, when the total amount of scattered
radiation is measured, and found to agree under suitable conditions with the
amount predicted by J.~J.~Thomson's formula. It would therefore seem wrong to
assume that we obtain a true picture of electronic distribution by the aid of
measurements on the coherent radiation alone.

Even before the advent of the new mechanics, Compton's original treatment of
the effect which he discovered suggested a way out of this difficulty. The
recoil electron is given an amount of energy%
\[
2\frac{h^{2}}{m}\frac{\nu^{\prime}}{\nu}\left(  \frac{\sin\theta}{\lambda
}\right)  ^{2}\ ,
\]
where $\nu$ and $\nu^{\prime}$ are the frequencies of the modified and
unmodified radiations. If the electron is ejected from the atom the radiation
is modified in wavelength, if not coherent waves are scattered. Since there is
little modified scattering at small angles, the $F$ curve will tend to a
maximum equal to the number of electrons in the atom, and any interpretation
of the curve will give an atom containing the correct number of electrons. As
$\frac{\sin\theta}{\lambda}$ increases, more and more of the scattered
radiation will be modified, and in calculating the $F$ curve this must be
taken into account. However, if $\frac{\nu^{\prime}}{\nu}$ is not far from
unity, the $F$ curve will remain a function of $\frac{\sin\theta}{\lambda}$,
since whatever criterion is applied for the scattering of modified or
unmodified radiation, it will depend on the energy imparted to the scattering
electron, which is itself a function of $\frac{\sin\theta}{\lambda}$. Our
X-ray analysis would thus give us an untrue picture of the atom, but one which
is consistently the same whatever wavelength is employed. Williams [27] and
Jauncey [28] have recalculated $F$ curves from atomic models using this
criterion, and found a better fit to the experimental curves when the Compton
effect was taken into account. (Examples of this closer approximation will be
found in the paper by Williams [27] in 1926. See also a discussion by Kallmann
and Mark [26]).\endnote{Bracket printed as a footnote in the French edition.}

  \begin{figure}
    \centering   
      \resizebox{\textwidth}{!}{\includegraphics[0mm,0mm][220.83mm,250.22mm]{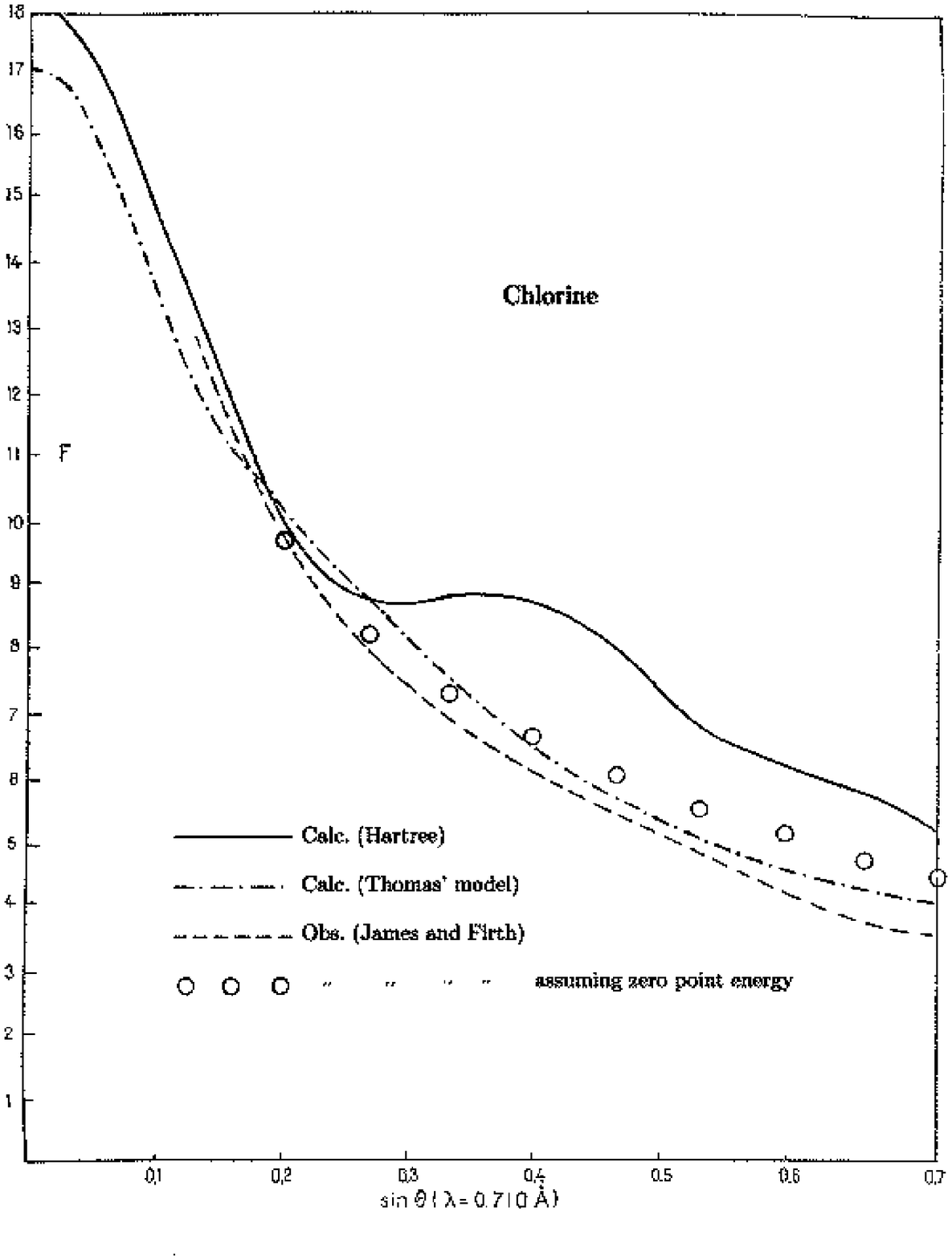}}
    \unnumberedcaption{Fig.~8.}
  \end{figure} 

The point at issue is illustrated by the curves in Fig. 8. Three $F$ curves
for chlorine are plotted in the figure. The dotted line represents the
observed $F$ curve (James and Firth). The continuous line is the $F$ curve
calculated from Hartree's [29] atomic model for chlorine. It shows a hump at a
value of $\sin\theta$ of $0.4$, which is not present in the observed curve.
This hump arises from the fact that the outer electrons in the chlorine model
give negative values for $F$ just short of this point,\endnote{[tout pr\`{e}s de ce point]} and positive values
again at the point itself. All atomic models calculated with electronic orbits
show similar irregularities which are not actually observed. When, however,
the Compton effect is taken into consideration, these outer electrons are
found to give a very small contribution to the $F$ curve at the large angles
where the humps\endnote{[irr\'{e}gularit\'{e}s]} occur, because they scatter so much modified radiation. The
allowance for the Compton effect smooths out the hump, and leads to $F$ curves
much more like those observed. The third curve shows the $F$ curve due to the
continuous Thomas distribution and is a close fit to the observed curve.

I have quoted from a note by Dr Ivar Waller, in the following tentative
summary of the interpretation which the new mechanics gives us of this
phenomenon.\footnote{Space forbids a reference to the many theoretical papers
which have contributed towards this interpretation.} In a recent letter to
\textit{Nature} [30], Waller discusses the transition for the whole range from
ordinary dispersion into Compton effect. His note only refers to scattering by
a single electron, but it can probably be extended to many-electron atoms.
Waves of continually decreasing wavelength are supposed to fall upon the atom,
and the transition is traced through the following stages.

\

\noindent a) While the wavelength of the radiation remains long compared with atomic
dimensions, the dispersion formula for optical frequencies gradually
transforms into the scattering for free electrons given by the classical J.~J.~Thomson 
formula. This formula holds approximately to\endnote{[pour]} wavelengths approaching
atomic dimensions.

\

\noindent b) At this point the scattering of coherent radiation will diminish, owing to
interference, and become more concentrated in the forward direction of the
incident light. This is the phenomenon we are studying, with X-rays, and our
$F$ curves map out the distribution of the coherent radiation where the
wavelength is of atomic dimensions.

\

\noindent c) At the same time, the scattering of incoherent radiation will become
appreciable, and approximate more and more closely in change of wavelength and
intensity distribution to the Compton effect. It will have practically merged
into the Compton effect when the momentum of a quantum of the incident light
is large compared with that corresponding to electronic motions in the atom.

\

\noindent d) Up to this point the Thomson formula holds for the total intensity of light
scattered in any direction, coherent and incoherent radiation being summed
together. It first ceases to hold, when the frequency displacement due to the
Compton effect is no longer small compared with the frequency of the incident light.

\

The point of importance for our present problem is that `the coherent part of
the radiation is to be directly calculated from that continuous distribution
of electricity which is defined by the Schr\"{o}dinger density-distribution in
the initial state of the atom'.\label{Bragg-Schr1} The classical treatment supposes each point
electron to scatter according to the J.~J.~Thomson formula in all directions.
In the new treatment, the electron is replaced by a spatial distribution of
scattering matter, and so each electron has an `$F$ curve' of its own.
It will still scatter coherent radiation in all directions, but its amount
will fall away from that given by the classical formula owing to interference
as $\frac{\sin\theta}{\lambda}$ increases, and this decline will be much more
rapid for the more diffuse outer electrons than for the concentrated inner
electrons. The total amount of radiation $T$ scattered in any direction by the
electron is given by the Thomson formula. A fraction $f^{2}T$ will be
coherent, and will be calculated by the laws of interference from the
Schr\"{o}dinger distribution, and the remainder, $(1-f^{2})T$, will be
incoherent. Thus the total coherent radiation will be $F^{2}T$ where $F$ is
calculated from the Schr\"{o}dinger distribution for the whole atom. An amount
$(N-\sum f^{2})T$ will be scattered with change of wavelength. Our
measurements of X-ray diffraction, if this be true, can be trusted to measure
the Schr\"{o}dinger continuous distribution of electricity in the crystal lattice.

A very interesting point arises in the case where characteristic absorption
frequencies of the scattering atom are of shorter wavelength than the
radiation which is being scattered. In general, this has not been so when
careful intensity measurements have been made since atoms of low atomic weight
have alone been investigated. On the classical analogy, we would expect a
reversal in phase of the scattered radiation, when an electron has a
characteristic frequency greater than that of the incident light. A
fascinating\endnote{[brillante]} experiment by Mark and Szilard [31] has shown that something very
like this takes place. They investigated the (111) and (333) reflections of
RbBr, which are extremely weak because Rb and Br oppose each other and are
nearly equal in atomic number. They found that these `forbidden' reflections
were indeed absent when the soft Cu$_{\mathrm{K}}$ or hard Ba$_{\mathrm{K}}$
radiation was used, but that Sr$_{\mathrm{K}}$ radiation was appreciably
reflected (Sr$_{\mathrm{K}\alpha}$ $\lambda$ 0.871 \AA ;\endnote{[$\lambda$ 
Sr$_{\mathrm{K}\alpha}=0.871 \mathrm{\mathring{A}}$]}
absorption edges of Rb$_{\mathrm{K}}$ and Br$_{\mathrm{K}}$, 0.814
\AA\ and 0.918 \AA ). The atoms are
differentiated because a reversal of phase in scattering by the K electrons
takes place in the one case and not in the other.

\begin{center}
\par
\Needspace{5\baselineskip}
{\sc 7. --- The analysis of atomic structure by X-ray intensity
measurements}\addcontentsline{toc}{section}{The analysis of atomic structure by X-ray intensity
measurements}
\end{center} 
It has been seen that the intensity measurements assign the correct number of
electrons to each atom in a crystal, and indicate a spatial extension of the
atoms of the right order. In attempting to make the further step of deducing
the arrangement of the electrons in the atom, the limitations of the method
begin to be very apparent.

  \begin{figure}
    \centering
      \resizebox{\textwidth}{!}{\includegraphics[0mm,0mm][220.66mm,250.29mm]{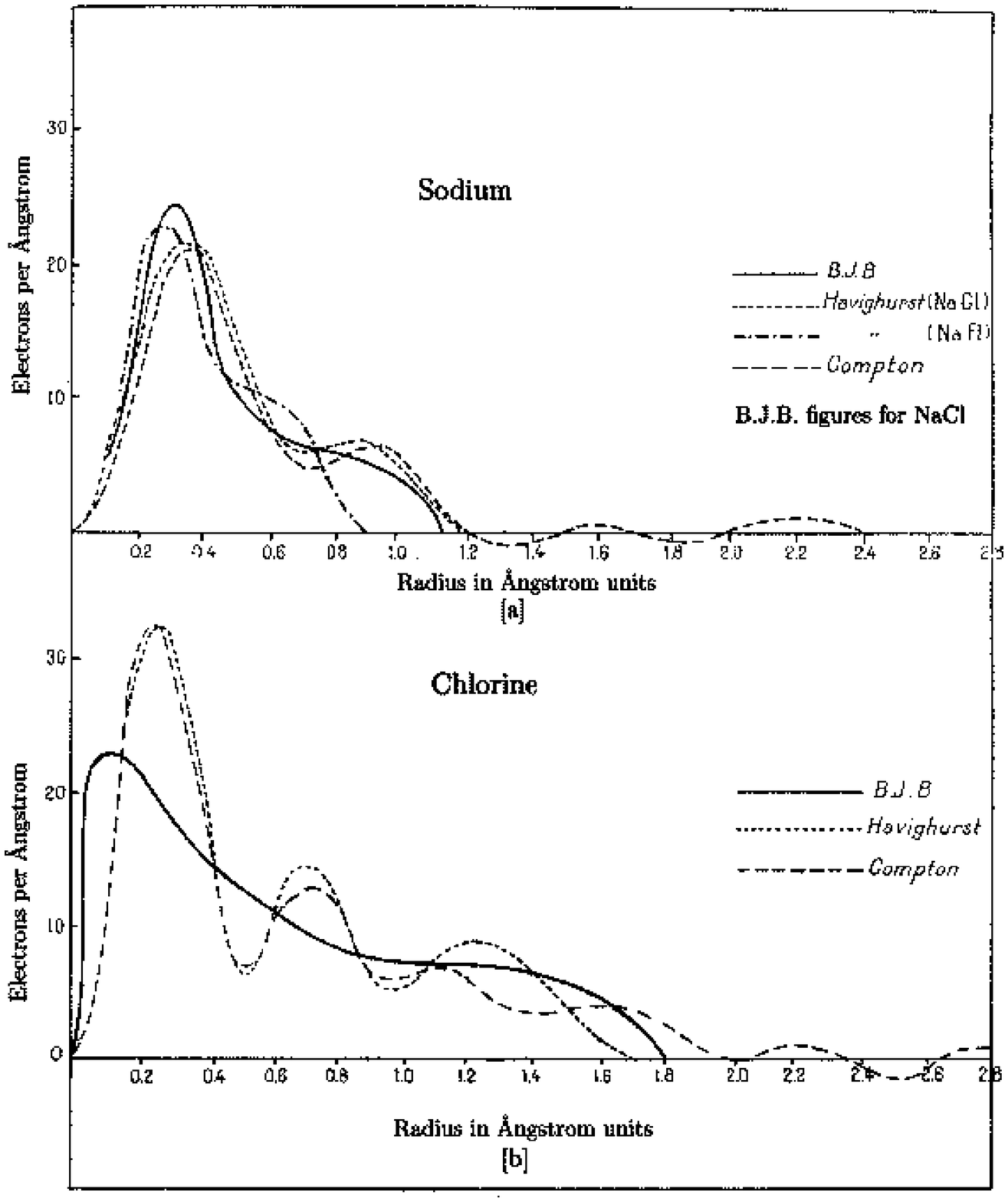}}
    \unnumberedcaption{Fig.~9.$^{56}$}
  \end{figure} 

In all cases where analysis has been attempted, the atom has been treated as
spherically symmetrical. The analysis is used to determine the amount of
scattering matter $U_{n}dr$ between radii $r$ and $r+dr$. All methods of
analysis give a distribution of the same general type. I have given, for
instance, a series of analyses of sodium and chlorine in Fig.~9. In these
figures, $U_{n}$ is plotted as ordinate against $r$ as abscissa. The total
area of the curve in each case is equal to the number of electrons in each
atom, since $\int_{0}^{\infty}U_{n}dr=N$. The full-line curves are our original interpretations of
the distribution in sodium and chlorine, based on our 1921 figures.\endnote{[faites d'apr\`{e}s 1921 figures]}%
\addtocounter{endnote}{1}\endnotetext{The French omits `B.~J.~B.\ figures for NaCl'.} 
The other curves are the interpretations of the same or closely similar sets of figures\endnote{[figures]}
by Havighurst [32] and by Compton (\textit{X-rays and Electrons}) using the
Fourier method of analysis.

In Fig. 9a are included our analysis of sodium in NaCl, two analyses by
Havighurst of sodium in NaCl and NaF obtained by using Duane's triple Fourier
series, and an analysis of our figures\endnote{Again, in the French, the false friend `figures'.} by Compton using the Fourier formula
for radial distribution. It will be seen that the general distribution of
scattering matter and the limits of the atom are approximately the same in
each case. The same holds for the chlorine curves in Fig. 9b.

The interesting point which is raised is the reality of the humps which are
shown by the Fourier analysis. We obtained similar humps in our analysis by
means of shells but doubted their reality because we found that if we smoothed
them out and recalculated the $F$ curve, it agreed with the observed curve
within the limits of experimental error. The technique of measurement has
greatly improved since then, and it would even appear from later results that
we over-estimated the possible errors of our first determinations of $F$. It
is obvious, however, that great care must still be taken in basing conclusions
on the finer details shown by any method of analysis. The formula which is
used in the Fourier analysis,%
\[
U_{n}=\frac{4\pi r}{a}\sum_{1}^{\infty}\frac{2nF_{n}}{a}\sin\frac{2\pi nr}%
{a}\ ,
\]
is one which converges very slowly, since the successive coefficients $F_{n}$
are multiplied by $n$. The observed $F$ curve must be extrapolated to a point
when $F$ is supposed to fall to zero, and the precise form of the curve reacts
very sensitively to the way in which this extrapolation is carried out.

  \begin{figure}
    \centering
      \resizebox{\textwidth}{!}{\includegraphics[0mm,0mm][220.71mm,260.10mm]{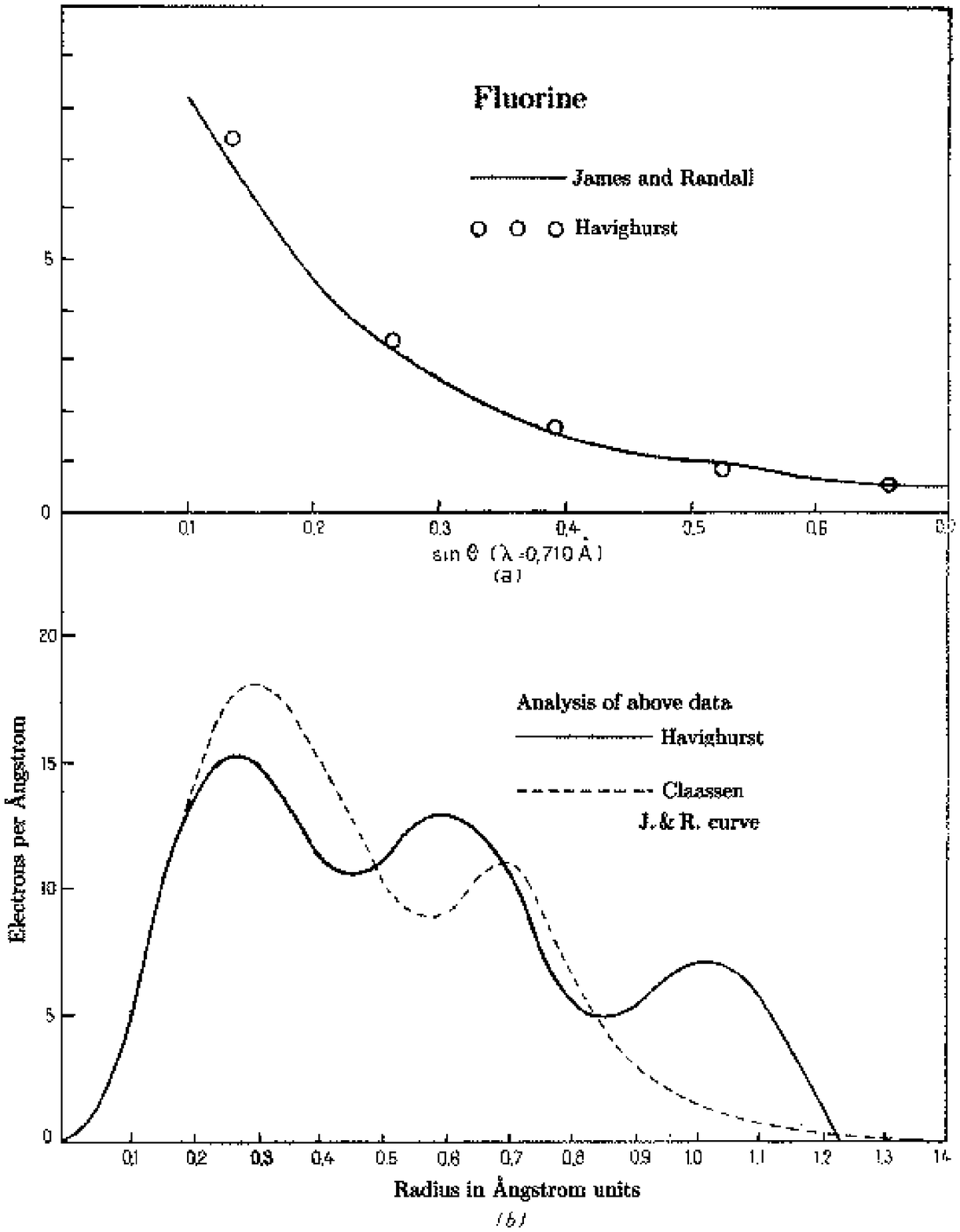}}
    \unnumberedcaption{Fig.~10.}
  \end{figure} 

The curves in Fig.~10 will illustrate the extent to which the analysis can be
considered to give us information about the actual atomic distribution. In
Fig.~10a the curve shows the $F$ values for fluorine obtained by James and Randall
[17]. The circles are points obtained\endnote{[d\'{e}duits]} by Havighurst from 
measurements on CaF, LiF, NaF;\endnote{[CaFl, LiFl, NaFl]} it will be seen that the two sets of experimental data are in very
satisfactory agreement. In Fig.~10b I have shown on the one hand
Havighurst's interpretations of the $F$ curve drawn through his points, and on
the other an analysis carried out by Claassen [16] of James and Randall's
using the Fourier method. The distributions are the same in their main
outlines, but the peaks occur in quite different places.
  \begin{figure}
    \centering
      \resizebox{\textwidth}{!}{\includegraphics[0mm,0mm][220.35mm,240.81mm]{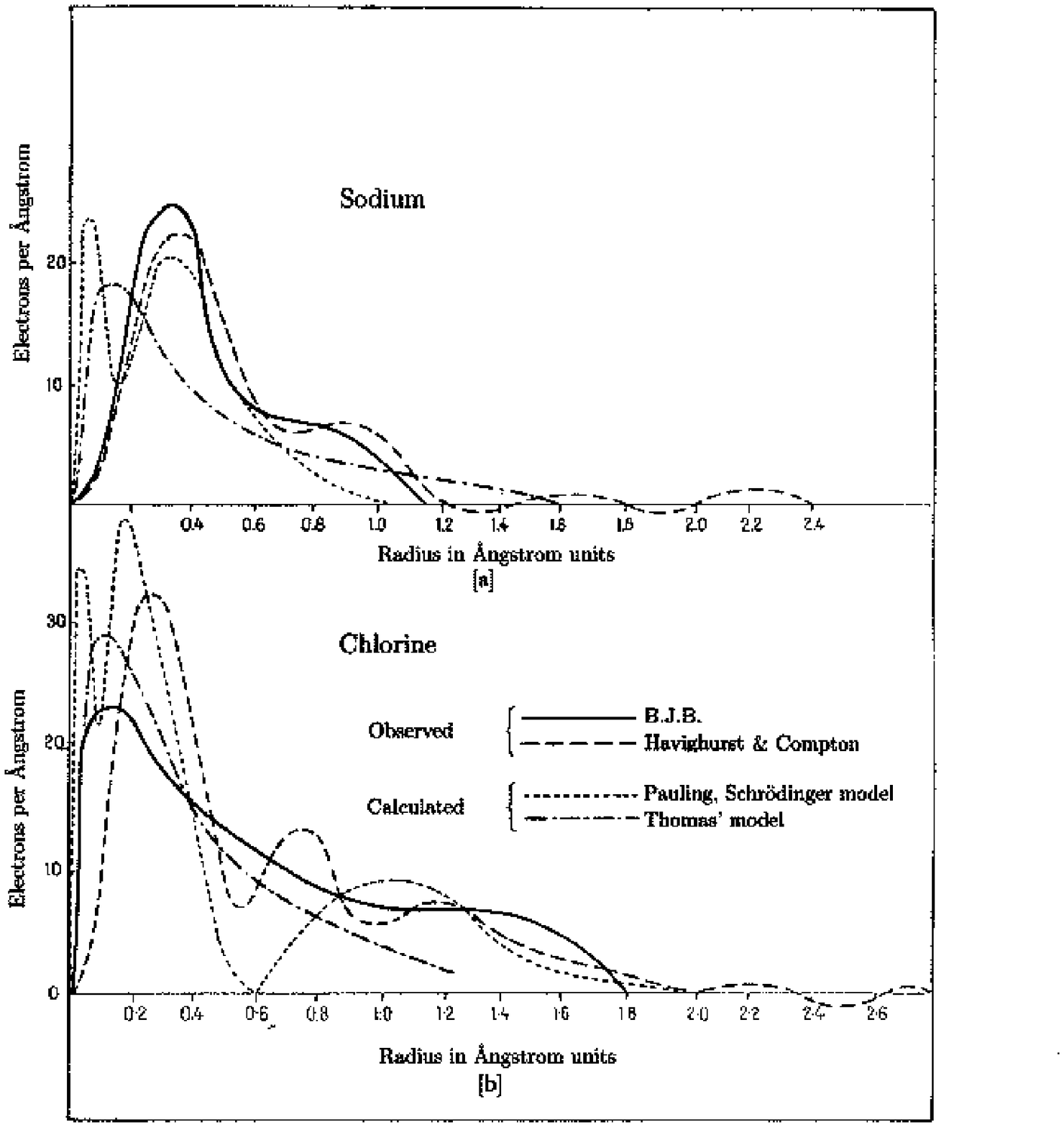}}
    \unnumberedcaption{Fig.~11.$^{62}$}
  \end{figure} 

Compton (\textit{X-rays and Electrons}, p.~167) in discussing his diagrams of
radial distribution has remarked that slight differences in the $F$ curves
lead to wide differences in details of the curves, and that too much
confidence should not be placed on these details. Havighurst [32] discusses
the significance of the analysis very fully in his paper on electron
distribution in the atoms. Our data are not yet sufficiently accurate or
extensive. Nevertheless, we are so near to attaining an accuracy of a
satisfactory order, and the results of the analysis seem to indicate so
clearly its fundamental correctness, that it appears to be well worth while to
pursue enquiry further. Work with shorter wavelengths, and at low
temperatures, when heat motion is small and a large range of $F$ values can be
measured, should yield us accurate pictures of the atomic structure itself.
Given accurate data,\endnote{[Une fois que nous disposerons de donn\'{e}es pr\'{e}cises]}%
\addtocounter{endnote}{1}\endnotetext{The French edition omits `\&\ Compton' and has 
`Mod\`{e}le de Pauling et Schr\"{o}dinger'.} 
the Fourier method of analysis provides a direct way of utilising them.

The radial distribution of scattering power outlined in this way is in general
agreement with any reasonable atomic model. We have seen, in particular, that
the $F$ curves, and therefore the radial distributions, of Thomas' model\endnote{The French 
translates as if the comma were after `of Thomas' model' rather than before.} are
in approximate accord with those actually observed. If it is true that the
scattering of coherent radiation is to be calculated in all cases by the
Schr\"{o}dinger density distribution, we should test our model against this distribution.

An interesting attempt along these lines has been recently made by Pauling
[33]. He has used certain simplifying assumptions to obtain an approximate
Schr\"{o}dinger density-distribution for many-electron atoms. I have shown in 
Fig.~11 four sets of curves. The radial electron distributions deduced by
Havighurst and by Compton are shown as one curve since they are very similar.
The figure shows also our first analysis of electron distribution. Matched
against these are plotted the generalised distribution of the Thomas model,
and the Schr\"{o}dinger density distribution calculated by Pauling.\label{Bragg-Schr2}

We have obviously not yet reached a point when we can be satisfied with the
agreement between theory and experiment, yet the success attained so far is a
distinct encouragement to further investigation.

\

\begin{center}
\par
\Needspace{5\baselineskip}
{\sc 8. --- The refraction of X-rays}\addcontentsline{toc}{section}{The refraction of X-rays}
\end{center} 
At Professor Lorentz's\endnote{[M.~Lorentz]} suggestion I have added a very brief note on the
refraction of X-rays, since the phenomenon is so intimately connected with the
question of intensity of reflection and scattering, and is another example of
the successful application of classical laws. The diffraction phenomenon dealt
with above (intensity of reflection) arises from the scattering of coherent
radiation in all directions by the atoms of a crystal. The refractive index
may be considered as being due to the scattering in the forward direction of
coherent radiation, which interferes with the primary beam. The arrangement of
the scattering matter plays no part, so that the body may be crystalline or
amorphous. The measurement of the refractive index is thus a direct measure of
the amount of coherent radiation scattered in the forward direction of the
incident beam.

\

\noindent 1. Darwin [2] appears to have been first in pointing out that theory assigns a
refractive index for X-rays differing from unity by about one part in a
million. He predicted that a very slight departure from the law of reflection%
\[
n\lambda=2d\sin\theta_{0}%
\]
would be found, the actual angle $\theta$ being given by Darwin's formula%
\[
\theta-\theta_{0}=\frac{1-\mu}{\sin\theta\cos\theta}\ .
\]
Ewald's [34] independent treatment of X-ray reflection leads to an equivalent
result, though the problem is approached along quite different lines.

As is well known, the first experimental evidence of an index of refraction
was found in a departure from the reflection laws. Stenstr\"{o}m [35] observed
differences in the apparent wavelength of soft X-rays (3~$\mathrm{\mathring
{A}}$) as measured in the different orders, which were explained by Ewald's
laws of X-ray reflection. The increased accuracy of X-ray spectroscopy has
shown that similar deviations from the simple law of reflection exist for
harder rays, though the deviations are much smaller than in the ordinary X-ray
region.\endnote{The typescript reads `much smaller in the ordinary X-ray
region', but given the context the text should be amended as shown (as also done
in the French version).} Thus the deviations have been detected for hard rays by Duane and
Patterson [36] and by Siegbahn and Hjalmar [37]. It is difficult to measure
the refractive index by means of these deviations in the ordinary way, since
they are so small, but Davis [38, 39] developed a very ingenious way of
greatly increasing the effect. A crystal is ground so that the rays reflected
by the atomic planes enter or leave a face at a very fine glancing angle, and
thus suffer a comparatively great deflection.

Compton [40] discovered the total reflection of X-rays, and measured the index
of refraction in this way. The refractive index is slightly less than unity,
hence X-rays falling at a very fine glancing angle on a plane surface of a
body are totally reflected, none of the radiation passing into the body.
Compton showed that, although the refractive index is so nearly unity, yet the
critical glancing angle is quite appreciable.

Finally, the direct effect of refraction by a prism has been observed by
Larsson, Siegbahn and Waller [41]. X-rays entered one face of a glass prism
at a very fine glancing angle, and suffered a measurable deflection. They
obtained in this way a dispersion spectrum of X-rays.

\

\noindent 2. In all cases where the frequency of the X-radiation is great compared with
any frequency characteristic of the atom, the refractive index measured by any
of these methods is in close accord\endnote{[parfaitement d'accord]} with the formula%
\[
1-\mu=\frac{ne^{2}}{2\pi m\nu^{2}}\ ,
\]
where $n$ is the number of electrons per unit volume in the body, $e$ and $m$
are the electronic constants, and $\nu$ the frequency of the incident
radiation. The formula follows directly from the classical Drude-Lorentz
theory of dispersion, in the limiting case where
the frequency of the radiation is large compared with the `free periods' of
the electrons in the atom. It can be put in the form [42]\endnote{Reference omitted in the French edition.}%
\[
1-\mu=2.71\times10^{-6}\,\frac{\rho Z}{A}\lambda^{2}\ ,
\]
where $\lambda$ is the wavelength in \AA ngstr\"{o}m units
of the incident radiation, $\rho$ the density of the substance, $Z$ and $A$
the average atomic number and atomic weight of its constituents (for all light
atoms $Z/A$ is very nearly 0.5).\endnote{[$\frac{Z}{A}$ la valeur moyenne du 
rapport du nombre atomique au poids atomique pour ses divers constituants (pour 
tous les atomes l\'{e}gers ce rapport est \`{a} peu pr\`{e}s \'{e}gal \`{a} 0.5]} 
Expressed in this form, the order of
magnitude of $1-\mu$ is easily grasped. The critical glancing angle $\theta$
for total reflection is given by%
\[
\cos\theta=\mu\ ,
\]
whence%
\[
\theta=\sqrt{\frac{ne^{2}}{\pi m\nu^{2}}}\ .
\]
Expressing $\theta$ in minutes of arc, and $\lambda$ in \AA ngstr\"{o}m units as before,%
\[
\theta=8.0\,\lambda\sqrt{\frac{\rho Z}{A}}\ .
\]
Measurements of refractive index have been made by Compton and by Doan using
the method of total reflection, by Davis, Hatley and Nardroff using reflection
in a crystal, and by Larsson, Siegbahn and Waller with a prism. A variety of
substances has been examined, and wavelengths between 0.5 and 2
\AA\ have been used. The accuracy of the experimental
determination of $1-\mu$ is of the order of one to five per cent. As long as
the critical frequencies of the atom have not been approached, the results
have agreed with the above formula within experimental error. Just as in the
measurements of intensity of reflection the $F$ curves approach a limit at
small angles equal to the number of electrons in the atom, so these
measurements of refractive index when interpreted by classical theory lead to
a very accurate numbering of the electrons in the scattering units.

\

\noindent 3. A highly interesting field is opened up by the measurements of refractive
index for wavelengths in the neighbourhood of a critical frequency of the
atom. It is a striking fact that the simple dispersion formula%
\[
\mu-1=\frac{e^{2}}{2\pi m}\sum_{1}^{n}\frac{n_{s}}{\nu_{s}^{2}-\nu^{2}}%
\]
still gives values for the refractive index agreeing with experiment in this
region, except when the critical frequency is very closely approached indeed.
Davis and von Nardroff reflected Cu$_{\mathrm{K}\alpha}$ and Cu$_{\mathrm{K}%
\beta}$ X-rays\endnote{[rayons]} from iron pyrites, and found that the refractive indices could
be reproduced by substituting constants in the formulae corresponding to two K
electrons in iron with the frequency of the K absorption edge.\endnote{Typescript: `of the K adsorption edge'; 
French version: `de la discontinuit\'{e} K'.} R.~L.~Doan [44]
has recently made a series of measurements by the total reflection method. His
accurate data support the conclusion that the Drude-Lorentz theory of
dispersion represents the facts, `not only in regions remote from the
absorption edge,\endnote{Typescript: `adsorption edge'; French: `bord d'absorption'.} but also in some instances in which the radiation approaches
the natural frequencies of certain groups of electrons'. The existence of two
K electrons\endnote{The French adds `dans la pyrite'.} is very definitely indicated. Kallmann and Mark [43] have gone
more deeply into the form of the dispersion curve in the neighbourhood of the
critical frequencies. The change in scattering power of an atom as the
frequency of the scattered radiation passes through a critical value is of
course another aspect of this anomalous dispersion; the experiment of Mark and
Szilard which showed this effect has been described above. There is ample
evidence that measurements of refractive index will in future prove to be a
most fruitful means of investigating the response of the atom to incident
radiation of frequency very near each of its own characteristic frequencies.


\newpage

\par
\Needspace{5\baselineskip}
\begin{center}

{\bf References}\addcontentsline{toc}{section}{References}\footnote{The style 
of the references has been modernised and uniformised ({\em eds.}).}\\
\hfill\\ 
\end{center}
\noindent [1] [W.~Friedrich, P.~Knipping and] M.~v.~Laue, {\em Bayr.\ Akad.\ d.\ Wiss.\ Math.\ phys.\ Kl.\ }(1912), 303.

\noindent [2] C.~G.~Darwin, {\em Phil.\ Mag.}, {\bf 27} (1914), 315, 675.

\noindent [3] P.~P.~Ewald, {\em Ann.\ d.\ Phys.}, {\bf 54} ([1917]), 519. 

\noindent [4] P.~Debye, {\em Ann.\ d.\ Phys.}, {\bf 43} (1914), 49. 

\noindent [5] W.~H.~Bragg, {\em Phil.\ Mag.}, {\bf 27} (1914), 881.

\noindent [6] W.~H.~Bragg, {\em Phil.\ Trans.\ Roy.\ Soc.\ }[{\em A}], {\bf 215} (1915), 253.

\noindent [7] A.~H.~Compton, {\em Phys.\ Rev.}, {\bf 9} (1917), 29; {\bf 10} (1917), 95.

\noindent [8] W.~L.~Bragg, R.~W.~James and C.~H.~Bosanquet, {\em Phil.\ Mag.}, {\bf  41} (1921), 309; 
{\bf 42} (1921), 1; {\bf 44} (1922), 433.

\noindent [9] C.~G.~Darwin, {\em Phil.\ Mag.}, {\bf 43} (1922), 800.

\noindent [10] R.~J.~Havighurst, {\em Phys.\ Rev.},  {\bf  28} (1926), n.~5, 869 and 882.

\noindent [11] L.~Harris, S.~J.~Bates and D.~A.~MacInnes,\endnote{French edition: 
`Mac Innes'.} {\em Phys.\ Rev.}, {\bf  28} (1926), 235.

\noindent [12] J.~A.~Bearden, {\em Phys.\ Rev.}, {\bf  27} (1926), 796; {\bf 29} (1927), 20.

\noindent [13] Bergen Davis and W.~M.~Stempel, {\em Phys.\ Rev.}, [{\bf  17}] (1921), 608.

\noindent [14] H.~Mark, {\em Naturwiss.}, {\bf 13} (1925), n.~49/50, 1042.

\noindent [15] L.~H.~Thomas, {\em Proc.\ Camb.\ Phil.\ Soc.}, {\bf  23} (1927), 542.

\noindent [16] A.~Claassen, {\em Proc.\ Phys.\ Soc.\ London}, {\bf  38} [pt] 5 (1926), 482.

\noindent [17] R.~W.~James and J.~T.~Randall, {\em Phil.\ Mag.\ }[(7)], {\bf  1} (1926), 1202.

\noindent [18] J.~A.~Wasastjerna, {\em Comm.\ Fenn.,} {\bf  2} (1925), 15.

\noindent [19] W.~L.~Bragg, C.~G.~Darwin and R.~W.~James, {\em Phil.\ Mag.\ } (7), {\bf 1} (1926), 897. 

\noindent [20] W.~Duane, {\em Proc.\ Nat.\ Acad.\ Sci.}, {\bf 11} (1925), 489.

\noindent [21] R.~J.~Havighurst, {\em Proc.\ Nat.\ Acad.\ Sci.}, {\bf 11} (1925), 502.

\noindent [22] W.~L.~Bragg and J.~West, {\em Roy.\ Soc.\ Proc.\ A}, {\bf 111} (1926), 691.

\noindent [23] [J.~M.] Cork, {\em Phil.\ Mag.\ }[(7), {\bf 4} (1927), 688].

\noindent [24] I.~Waller, {\em Upsala Univ.\ \AA rsskr.\ 1925}, [11]; {\em Ann.\ d.\ Phys.}
, {\bf 83} (1927), 153.

\noindent [25] R.~W.~James and E.~Firth, {\em Roy.\ Soc.\ Proc.\ }[{\em A}, {\bf 117} (1927), 62].

\noindent [26] [H.] Kallmann and H.~Mark, {\em Zeit.\ f.\ Phys.}, {\bf 26} (1926), [n.]~2, [120].

\noindent [27] E.~J.~Williams, {\em Phil.\ Mag.\ }[(7)], {\bf 2} (1926), 657.

\noindent [28] [G.~E.~M.] Jauncey, {\em Phys.\ Rev.}, {\bf 29} (1927), 605.

\noindent [29] D.~R.~Hartree, {\em Phil.\ Mag.}, {\bf 50} (1925), 289.

\noindent [30] I.~Waller, {\em Nature}, [{\bf 120}] (July 1927), [155].

\noindent [31] H.~Mark and L.~Szilard, {\em Zeit.\ f.\ Phys.}, {\bf 33} (1925), 688.

\noindent [32] R.~J.~Havighurst, {\em Phys.\ Rev.}, {\bf 29} (1927), 1.

\noindent [33] L.~Pauling, {\em Roy.\ Soc.\ Proc.\ A}, {\bf 114} (1927), 181.

\noindent [34] P.~P.~Ewald, {\em Phys.\ Zeitsch.}, {\bf 21} (1920), 617; {\em Zeitschr.\ f.\ Physik}, {\bf 2} (1920), 332.

\noindent [35] W.~Stenstr\"{o}m, Exper[imentelle] Unters[uchungen] d[er] R\"{o}ntgenspektra. Dissertation, Lund (1919). 

\noindent [36] [W.] Duane and [R.~A.] Patterson, {\em Phys.\ Rev.}, {\bf 16} (1920), [526].\endnote{Typescript and 
French edition both have `532'.}

\noindent [37] M.~Siegbahn, {\em Spektroskopie der R\"{o}ntgenstrahlen} [(Berlin: Springer, 1924)].

\noindent [38] [C.~C.~Hatley and Bergen Davis], {\em Phys.\ Rev.}, {\bf 23} (1924), 
290.\endnote{Both typescript and French edition give this reference as `B.~Davis and C.~C.~Hatley'. 
The typescript has `291'.}

\noindent [39] B[ergen] Davis and R.~von Nardroff, {\em Phys.\ Rev.}, {\bf 23} (1924), 291.

\noindent [40] A.~H.~Compton, {\em Phil.\ Mag.}, {\bf 45} (1923), 1121.

\noindent [41] [A.] Larsson, [M.] Siegbahn and [I.] Waller, {\em Naturwiss.}, {\bf 12} ([1924]), 1212.

\noindent [42] I.~Waller, [Theoretische Studien zur] Interferenz- und Dispersionstheorie 
der R\"{o}ntgenstrahlen. [Dissertation, Upsala (1925)].

\noindent [43] H.~Kallmann and H.~Mark, {\em Ann.\ d.\ Physik}, {\bf 82} (1927), 585.

\noindent [44] R.~L.~Doan, {\em Phil.\ Mag.\ }[(7), {\bf 4} n.]~20 (1927), [100].

\

A very complete account of work on intensity of reflection is given by Compton
in {\em X-rays and Electrons} and by Ewald in volume 24 of the {\em Handbuch der Physik}
by H.~Geiger and K.~Scheel,\endnote{Authors added in the French edition.} {\em Aufbau der 
festen Materie und seine Erforschung durch R\"{o}ntgenstrahlen}, section~18.

\newpage

\section*{Discussion of Mr Bragg's report}\markboth{{\it W.~L.~Bragg}}{{\it Discussion}}
\addcontentsline{toc}{section}{Discussion of Mr Bragg's report}

{\sc Mr Debye.}~---~To what extent can you conclude that there exists an
energy at absolute zero?\\

{\sc Mr Bragg.}~---~Waller and James have recently submitted a paper to the
Royal Society in which they discuss the relation between the influence of
temperature on the intensity of reflection (Debye effect) and the elastic
constants of a crystal. Using the experimentally determined value of the Debye
coefficient, they deduce the scattering by an atom at rest from the scattering
by the atom at the temperature of liquid air (86$%
{{}^\circ}%
$ abs.). The curve deduced for the scattering by a perfectly motionless atom
can of course take two forms, according to whether or not, in interpreting the
results of the experiment, one assumes the existence of an energy at absolute zero.

If one assumes the existence of such an energy, the curve deduced from the
experimental results agrees with that calculated by Hartree by applying
Schr\"{o}dinger's mechanics. The agreement is really very good for sodium as
well as for chlorine. On the other hand, the curve that one obtains if one
does not assume any energy at absolute zero deviates considerably from the
calculated curve by an amount that exceeds the possible experimental error.

If these experimental results\footnote{\textit{Note added 5 April}
\textit{1928}. The results to which allusion is made here have just been
published in detail by Messrs James, Waller and Hartree in a paper entitled:
`An investigation into the existence of zero-point energy in the rock-salt
lattice by an X-ray diffraction method' (\textit{Proc.\ Roy.\ Soc.\ A}, 
\textbf{118} (1928), 334).
} are confirmed by new experiments, they provide a
direct and convincing proof of the existence of an energy at absolute zero.\\

{\sc Mr Debye.}~---~Would the effect not be larger if one did the
experiments with diamond?\\

{\sc Mr Bragg.}~---~In the case of diamond, it is difficult to interpret
the results obtained using a single crystal, because the structure is very
perfect and the `extinction' is strong. One would have to work with diamond
powder. But I cannot say if it would be easy to find that there exists an
energy at absolute zero in diamond; I should consider it further.\\

{\sc Mr Fowler.}~---~\label{Hartree}Here is how Hartree calculates the atomic fields.
Starting from Thomas' atomic field, taken as a first approximation, he
calculates the Schr\"{o}dinger functions for an electron placed in this field,
then the density of charge in the atom corresponding to the Schr\"{o}dinger
functions, and then the corresponding atomic field, which will differ from
that of Thomas. By successive approximations one modifies the field until the
calculations yield the field which served as a starting point. This method
gives very good values for the levels corresponding to X-rays and to visible
light, and leads to the atom that Mr Bragg considered for comparison with experiments.\\

{\sc Mr Heisenberg.}~---~How can you say that Hartree's method gives exact
results, if it has not given any for the hydrogen atom? In the case of
hydrogen the Schr\"{o}dinger functions must be calculated with the aid of his
differential equation, in which one introduces only the electric potential due
to the nucleus. One would not obtain correct results if one added to this potential
the one coming from a charge distribution by which one had replaced the
electron. One may then obtain exact results only by taking the charge density
of all the electrons, except the one whose motion one wishes to calculate.
Hartree's method is certainly very useful and I have no objection to it, but
it is essentially an approximation.\\

{\sc Mr Fowler.}~---~I may add to what I have just said that Hartree is
always careful to leave out the field of the electron itself in each state, so
that, when he considers an L electron, for example, the central part of the
field of the whole atom is diminished by the field of an L electron, as far as
this may be considered as central. Hartree's method would then be entirely
exact for hydrogen and in fact he has shown that it is extremely close to
being exact for helium. (One finds a recent theoretical discussion of
Hartree's method, by Gaunt, in \textit{Proc.\ Cambr.\ Phil.\ Soc.},
\textbf{24} (1928), 328.
)\\

{\sc Mr Pauli.}~---~In my opinion one must not perform the calculations, as
in wave mechanics, by considering a density $\left\vert \psi(x,y,z)\right\vert
^{2}$ in three-dimensional space,\endnote{Here and in the following displayed formula,
the published version has square brackets instead of absolute bars.} but must consider a density in several
dimensions%
\[
\left\vert \psi(x_{1},\;y_{1},\;z_{1},\;...,\;x_{N},\;y_{N},\;z_{N}%
)\right\vert ^{2}\;,
\]
which depends on the $N$ particles in the atom. For sufficiently short
waves the intensity of coherent scattered radiation is then proportional
to\endnote{Arrow missing on $\overrightarrow{r_{k}}$ in the published volume.}%
\[
\int...\int\sum_{1}^{N}e^{\frac{2\pi i}{\lambda}\left(  \overrightarrow{n_{d}%
}-\overrightarrow{n_{u}},\,\overrightarrow{r_{k}}\right)  }\left\vert
\psi(x_{1},\;...,\;z_{N})\right\vert ^{2}dx_{1}\;...\;dz_{N}\;,
\]
where $\lambda$ is the wavelength of the incident radiation, $\overrightarrow
{n_{u}}$ a unit vector in the direction of propagation, and $\overrightarrow
{n_{d}}$ the corresponding unit vector for the scattered radiation; the sum
must be taken over all the particles. The result that one obtains by assuming
a three-dimensional density cannot be rigorously exact; it can only be so to a
certain degree of approximation.\\

{\sc Mr Lorentz.}~---~How have you calculated the scattering of radiation
by a charge distributed over a region comparable to the volume occupied by the atom?\\

{\sc Mr Bragg.}~---~To interpret the results of observation as produced by
an average distribution of the scattering material, we applied J.~J.~Thomson's
classical formula for the amplitude of the wave scattered by a single electron.\\

{\sc Mr Compton.}~---~If we assume that there is always a constant ratio
between the charge and mass of the electron, the result of the classical
calculation of reflection by a crystal is exactly the same, whether the
charge and mass are assumed concentrated in particles (electrons) or
distributed irregularly in the atom. The intensity of reflection is determined
by the average density of the electric charge in different parts of the atom.
That may be represented either by the probability that a point charge occupies
this region or by the volume density of an electric charge distributed in a
continuous manner through this region.\\

{\sc Mr Kramers.}~---~The use that one may make of the simple Thomas model
of the atom in the search for the laws of reflection is extremely interesting.
It would perhaps not be superfluous to investigate what result would be
obtained for the electron distribution if, instead of restricting oneself to
considering a single centre of attraction, one applied Thomas' differential
equation to an infinity of centres distributed as in a crystal grating. Has
anyone already tried to solve the problem of which Mr Bragg has just spoken,
of the calculation of the general distribution of the electronic density
around the nucleus of a heavy atom, in the case where there are many nuclei,
as in a crystal?\\

{\sc Mr Bragg.}~---~No, no one has yet attacked this problem, which I only
mentioned because it is interesting.\\

{\sc Mr Dirac.}~---~Do the scattering curves depend on the phase relations
between the oscillations of different atoms?\\

{\sc Mr Bragg.}~---~No, because the results of our experiments give only
the average scattering produced in each direction by a very large number of atoms.\\

{\sc Mr Dirac.}~---~What would happen if you had two simple oscillators
performing harmonic vibrations? Would they produce a different scattering when
in phase than when out of phase?\\

{\sc Mr Born.}~---~The correct answer to the question of scattering by
an atom is contained in the remark by Mr Pauli. Strictly speaking there is no
three-dimensional charge distribution that may describe exactly how an atom
behaves; one always has to consider the total configuration of all the
electrons in the space of $3n$ dimensions. A model in three dimensions only
ever gives a more or less crude approximation.\\

{\sc Mr Kramers} asks a question concerning the influence of the Compton
effect on the scattering.\\

{\sc Mr Bragg.}~---~I have already said something on that subject in my
report.\footnote{Cf.\ Bragg's report, section 6 (\textit{eds}).} Assuming a
model of the atom of the old type, Jauncey and Williams have used the
criterion that the wavelength is modified when the recoil of the scattering
electron is sufficient to take it entirely outside the atom. Williams was the
first to apply this criterion to scattering curves obtained with crystals. He
pointed out that while the speed of the recoil electron depends on both the
scattering angle and the wavelength, any criterion one uses is a function of
$\frac{\sin\theta}{\lambda}$, just as the interference effects depend on
$\frac{\sin\theta}{\lambda}$. This implies that the existence of the Compton
effect modifies the scattering curve such that we can always assign the same
scattering curve to no matter what type of atom, whatever the wavelength may be.\\

{\sc Mr Fowler.}~---~If I have understood properly, Mr Bragg uses
theoretical calculations by Waller that have not yet been published. When
light is scattered by an atom in accordance with the interpretation given by
Mr Waller by means of the new mechanics, the \textit{total} amount of
scattered light is given exactly by J.~J.~Thomson's classical formula (except
for very hard $\gamma$-rays). This light is composed of the coherent scattered
radiation and of the modified light (Compton scattering). In the theorem of
the reflection of X-rays only the coherent scattered light must be used, and
indeed it is; and this light is given exactly by the \textit{F} curves like
those proposed by Hartree. These \textit{F} curves for atomic scattering are
obviously given simply by the classical scattering for each electron,
diminished by interference.\\

{\sc Mr Bragg.}~---~I should like to develop Mr Fowler's remark by
recalling Waller and Wentzel's conclusions briefly sketched in my report. The
scattering by one of the electrons in an atom partly remains the same and
partly is modified. Within certain limits the total amount of scattered
radiation is given by J.~J.~Thomson's formula. A fraction $f^{2}$ of this
amount is not modified, $f$ being a coefficient smaller than 1, depending on
the interference of the spatial distribution of the charge according to
Schr\"{o}dinger and calculated according to the classical laws of optics. The
remaining fraction $1-f^{2}$ is modified.\endnote{The original text
mistakenly states that both fractions are `not modified'.}\\

{\sc Mr Lorentz.}~---~It is, without doubt, extremely noteworthy that the
total scattering, composed of two parts of quite different origin, agrees with
Thomson's formula.\\

{\sc Mr Kramers} makes two remarks:\\

\noindent 1. As Mr Bragg has pointed out the importance of there being interest in
having more experimental data concerning the refrangibility of X-rays in the
neighbourhood of the absorption limit, I should like to draw attention to
experiments performed recently by Mr Prins in the laboratory of Professor
Coster at Groningen. By means of his apparatus (the details of the experiments
and the results obtained are described in a paper published recently in
\textit{Zeitschrift f\"{u}r Physik}, \textbf{47} (1928), [479]
), Mr Prins finds
in a single test the angle of total reflection corresponding to an extended
region of frequencies. In the region of the absorption limit of the metal, he
finds an abnormal effect, which consists mainly of a strong decrease in the
angle of total reflection on the side of the absorption limit located towards
the short wavelengths. This effect is easily explained taking into account the
influence of absorption on the total reflection, without it being necessary to
enter into the question of the change in refrangibility of the X-rays. In
fact, the absorption may be described by considering the refractive index $n$
as a complex number, whose imaginary part is related in a simple manner to the
absorption coefficient. Introducing this complex value for $n$ in the
well-known formulas of Fresnel for the intensity of reflected rays, one finds
that the sharp limit of total reflection disappears, and that the manner in which
the intensity of reflected rays depends on the angle of incidence is such that
the experiment must give an `effective angle of total reflection' that is
smaller than in the case where there is no absorption and that decreases as
the absorption increases.

According to the atomic theory one would also expect to find, in the region of
the absorption limit, anomalies in the real part of the refractive index,
producing a similar though less noticeable decrease of the effective angle of
total reflection on the side of the absorption edge directed towards the large
wavelengths. Mr Prins has not yet succeeded in showing that the experiments
really demonstrate this effect.\footnote{Continuing his research Mr Prins has
established (February 1928) the existence of this effect, in agreement with
the theory.}

The theory of these anomalies in the real part of the refractive index
constitutes the subject of my second remark.

\

\noindent 2. Let us consider plane and polarised electromagnetic waves, in which the
electric force can be represented by the real part of $Ee^{2\pi i\mathrm{\nu
}t}$, striking an atom which for further simplicity we shall assume to be
isotropic. The waves make the atom behave like an oscillating dipole, giving,
by expansion in a Fourier series, a term with frequency $\mathrm{\nu}$. Let us
represent this term by the real part of $Pe^{2\pi i\mathrm{\nu}t}$, where $P$
is a complex vector having the same direction as the vector $E$ to which it
is, moreover, proportional. If we set%
\begin{equation}
\frac{P}{E}=f+ig\ , \tag{1}%
\end{equation}
where $f$ and $g$ are real functions of $\mathrm{\nu}$, the real and imaginary
parts of the refractive index of a sample of matter are related in a simple way
to the functions $f$ and $g$ of the atoms contained in the sample.

Extending the domain of values that $\mathrm{\nu}$ may take into the negative
region and defining $f$ as an even function of $\mathrm{\nu}$, $g$ as an odd
function, one easily verifies that the dispersion formulas of Lorentz's
classical theory and also those of modern quantum mechanics are equivalent to
the formula%
\begin{equation}
f(\mathrm{\nu})=\frac{1}{\pi}\dashint_{-\infty}^{+\infty}\frac{g(\mathrm{\nu
}^{\prime})}{\mathrm{\nu-\nu}^{\prime}}d\mathrm{\nu}^{\prime}\mathrm{\;,}
\tag{2}%
\end{equation}
where the sign $\dashint$ indicates the `principal' value of the integral.

This formula can easily be applied to atoms showing continuous absorption
regions and is equivalent to the formulas proposed for these cases by R.~de~Laer 
Kronig and by Mark and Kallmann. There is hardly any doubt that this
general formula may be derived from quantum mechanics, if one duly takes into
account the absorption of radiation, basing oneself on Dirac's theory, for example.

From a mathematical point of view, formula (2) gives us the means to construct
an analytic function of a complex variable $\mathrm{\nu}$ that is holomorphic
below the real axis and whose real part takes the values $g(\mathrm{\nu
}^{\prime})$ on this axis. If one considers $\mathrm{\nu}$ as a real variable,
the integral equation (2) has the solution%
\begin{equation}
g(\mathrm{\nu})=-\frac{1}{\pi}\dashint_{-\infty}^{+\infty}\frac{f(\mathrm{\nu
}^{\prime})}{\mathrm{\nu-\nu}^{\prime}}d\mathrm{\nu}^{\prime}\mathrm{\;,}
\tag{3}%
\end{equation}
which shows that the imaginary part of the refractive index depends on the
real part in nearly the same way as the real part depends on the imaginary
part. The fact that the analytic function $f$ of the complex variable
$\mathrm{\nu}$, defined by (2) for the lower half of the complex plane, has no
singularity in this half-plane, means that dispersion phenomena, when one
studies them by means of waves whose amplitude grows in an exponential manner
($\mathrm{\nu}$ complex), can never give rise to singular behaviour for the atoms.\\

{\sc Mr Compton.}~---~The measurements of refractive indices of X-rays made
by Doan agree better with the Drude-Lorentz formula than with the expression
derived by Kronig based on the quantum theory of dispersion.\\

{\sc Mr de Broglie.}~---~I should like to draw attention to recent
experiments carried out by Messrs J.~Thibaud and A.~Soltan,\footnote[1]%
{\textit{C.\ R.\ Acad.\ Sc.}, \textbf{185} (1927), 642.
} which touch on the
questions raised by Mr Bragg. In these experiments Messrs Thibaud and Soltan
measured, by the tangent grating method, the wavelength of a certain number of
X-rays in the domain 20 to 70 \AA . Some of these
wavelengths had already been determined by Mr Dauvillier using diffraction by
fatty-acid gratings. Now, comparing the results of Dauvillier with those of
Thibaud and Soltan, one notices that there is a systematic discrepancy
between them that increases with wavelength. Thus for the $\mathrm{K}_{\alpha
}$ line of boron, Thibaud and Soltan find 68 \AA , while
Dauvillier had found 73.5 \AA , that is, a difference of
5.5 \AA . This systematic discrepancy appears to be due to
the increase of the refractive index with wavelength. The index does not
actually play a role in the tangent grating method, while it distorts in a
systematic way the results obtained by crystalline diffraction when one
uses the Bragg formula. Starting from the difference between their results and
those of Mr Dauvillier, Messrs Thibaud and Soltan have calculated the value of
the refractive index of fatty acids around 70 \AA\ and
found%
\[
\delta=1-\mu=10^{-2}%
\]
thereabouts. This agrees well with a law of the form $\delta=K\lambda^{2}$;
since in the ordinary X-ray domain the wavelengths are about 100 times
smaller, $\delta$ is of order $10^{-6}$. One could object that, according to
the Drude-Lorentz law, the presence of K discontinuities of oxygen, nitrogen
and carbon between 30 and 45 \AA\ should perturb the law in
$\lambda^{2}$. But in the X-ray domain the validity of Drude's law is
doubtful, and if one uses in its place the formula proposed by Kallmann and
Mark\footnote[1]{\textit{Ann. d. Phys.}, \textbf{82} (1927), 585.} the
agreement with the experimental results is very good. Let us note finally that
the existence of an index appreciably different from 1 can contribute to
explaining why large-wavelength lines, obtained with a fatty-acid grating, are
broad and spread out.\\

{\sc Mr Lorentz}\label{forHistEss12} makes a remark concerning the refractive index of a
crystal for R\"{o}ntgen rays and the deviations from the Bragg law. It is
clear that, according to the classical theory, the index must be less than
unity, because the electrons contained in the atoms have eigenfrequencies
smaller than the frequency of the rays, which gives rise to a speed of
propagation greater than $c$. But in order to speak of this speed, one must
adopt the macroscopic point of view, abstracting away the molecular
discontinuity. Now, if one wishes to explain Laue's phenomenon in all its
details, one must consider, for example, the action of the vibrations excited
in the particles of a crystallographic layer on a particle of a neighbouring
layer. This gives rise to series that one cannot replace by integrals. It is
for this reason that I found some difficulty in the explanation of deviations
from the Bragg law.\endnote{The mixing of first and third person, here and in
a few similar instances throughout the discussions, is as in the published text.}\\

{\sc Mr Debye.}~---~Ewald has tried to do similar calculations.\\

{\sc Mr Lorentz.}~---~It is very interesting to note that with R\"{o}ntgen
rays one finds, in the vicinity of an absorption edge, phenomena similar to
those that in classical optics are produced close to an absorption band. There
is, however, a profound difference between the two cases, the absorption edge
not corresponding to a frequency that really exists in the particles.


\newpage

\renewcommand{\enoteheading}{\section*{Notes to the translation}}
\addcontentsline{toc}{section}{\it Notes to the translation}
\theendnotes

\setcounter{endnote}{0}
\setcounter{equation}{0}

\chapter*{Disagreements between experiment and the electromagnetic theory of 
radiation$^{\scriptsize\hbox{a}}$}\markboth{{\it A.~H.~Compton}}{{\it Experiment and the electromagnetic theory}}
\addcontentsline{toc}{chapter}{Disagreements between experiment and the electromagnetic theory
of radiation ({\em A.~H.~Compton\/})}
\begin{center}{\sc By Mr Arthur H.\ COMPTON}\end{center}\footnotetext[1]{An English version of this report (Compton 1928) was published 
in the {\em Journal of the Franklin Institute}. The French version appears to be essentially a translation of the English
paper with some additions. Whenever there are no discrepancies, we reproduce Compton's own English (we have 
corrected some obvious typos and harmonised some of the spelling). Interesting variants are footnoted. Other discrepancies between the 
two versions are reported in the endnotes ({\em eds.}).}

\

\begin{center}

\par
\Needspace{5\baselineskip}
{\sc Introduction}\addcontentsline{toc}{section}{Introduction}
\end{center}
Professor W.\ L.\ Bragg has just discussed a whole series of radiation phenomena in which the electromagnetic 
theory is confirmed. He has even dwelt on some of the limiting 
cases, such as the reflection of X-rays by crystals, in which the electromagnetic theory of radiation gives us, at 
least approximately, a correct interpretation of the facts, although there are reasons to doubt that its predictions 
are truly exact. I have been left the task of pleading the opposing cause to that of the electromagnetic theory of 
radiation, seen from the experimental viewpoint.\label{accuser}

I have to declare from the outset that in playing this role of the accuser I have no intention of diminishing the 
importance of the electromagnetic theory as applied to a great variety of problems.\footnote[2]{The opening has been 
translated from the French edition. The English version has the following different opening ({\em eds.}):
  \begin{quote} 
    During the last few years it has become increasingly evident that the classical electromagnetic theory of 
    radiation is incapable of accounting for certain large classes of phenomena, especially those concerned with 
    the interaction between radiation and matter. It is not that we question the wave character of light --- the 
    striking successes of this conception in explaining polarisation and interference of light can leave no doubt 
    that radiation has the characteristics of waves; but it is equally true that certain other properties of 
    radiation are not easily interpreted in terms of waves. The power of the electromagnetic theory as applied 
    to a great variety of problems of radiation is too well known to require emphasis.
  \end{quote}
} It is, however, only by acquainting ourselves with 
the real or apparent\endnote{The words `r\'{e}els ou apparents' are present only in the French version.} failures of 
this powerful theory that we can hope to develop a more complete theory of radiation which will describe the facts as 
we know them.

The more serious difficulties which present themselves in connection with the theory that radiation consists of 
electromagnetic waves, propagated through space in accord with the demands of Maxwell's equations, may be classified 
conveniently under five heads:\endnote{The English version has only four headings (starting with `(1) How are the waves 
produced?'), and accordingly omits the next section, on `The problem of the ether', and later references to the ether.}

\

\noindent (1) Is there an ether? If there are oscillations, there must be a medium in which these oscillations are produced. 
Assuming the existence of such a medium, however, one encounters great difficulties.

\

\noindent (2) How are the waves produced? The classical electrodynamics requires as a source of an electromagnetic wave an 
oscillator of the same frequency as that of the waves it radiates. Our studies of spectra,\endnote{[d'apr\`{e}s les 
r\'{e}sultats de l'\'{e}tude des spectres]} however, make it appear impossible that an atom should contain oscillators 
of the same frequencies as the emitted rays.

\

\noindent (3) The photoelectric effect. This phenomenon is wholly anomalous when viewed from the standpoint of waves.

\

\noindent (4) The scattering of X-rays, and the recoil electrons, phenomena in which we find gradually increasing departures 
from the predictions of the classical wave theory as the frequency increases.

\

\noindent (5) Experiments on individual interactions between quanta of radiation and electrons. If the results of the experiments 
of this type are reliable, they seem to show definitely that individual quanta of radiation, of energy $h\nu$, proceed 
in definite directions.

\

\noindent {\em The photon hypothesis}.\endnote{The English edition distinguishes sections and subsections more systematically 
than the French edition, and in this and other small details of layout we shall mostly follow the former.} --- In order 
to exhibit more clearly the difficulties with the classical theory of radiation, it will be helpful to keep in mind the 
suggestion that\endnote{[rappeler qu'il existe une th\'{e}orie dans laquelle]} light consists of corpuscles. We need not 
think of these two views as necessarily alternative. It may well 
be that the two conceptions are complementary. Perhaps the corpuscle is related to the wave in somewhat the same manner 
that the molecule is related to matter in bulk; or there may be a guiding wave\label{Comp46} which directs the corpuscles which carry 
the energy. In any case, the phenomena which we have just mentioned suggest the hypothesis that radiation is divisible 
into units possessing energy $h\nu$, and which proceed in definite directions with momentum $h\nu /c$. This is obviously 
similar to Newton's old conception of light corpuscles. It was revived in its present form by Professor 
Einstein,\endnote{The words `le professeur' are present only in the French edition.} it was defended under the name of the 
`Neutron Theory' by Sir William [H.] Bragg, and has been given new life by the recent discoveries associated with the 
scattering of X-rays.

In referring to this unit of radiation I shall use the name `photon', suggested recently by 
G.~N.~Lewis.\footnote{G.~N.~Lewis, {\em Nature}, [{\bf 118}], [874] (Dec.\ 18, 1926).} This word avoids any 
implication regarding the nature of the unit, as contained for example in the 
name `needle ray'. As compared with the terms `radiation quantum' and `light quant',\endnote{[`\'{e}l\'{e}ment de 
radiation' ou `quantum de lumi\`{e}re']} this name has the advantages of brevity and of avoiding any implied dependence 
upon the much more general quantum mechanics or quantum theory of atomic structure.

\

\noindent {\em Virtual radiation.}\label{ComptonBKS} --- Another conception of the nature of radiation which it will be desirable to compare with 
the experiments is Bohr, Kramers and Slater's important theory of virtual radiation.\footnote{N.~Bohr, 
H.~A.~Kramers and J.~C.~Slater, {\em Phil.\ Mag.}, {\bf 47} (1924), 785; {\em Zeits.\ f.\ Phys.}, {\bf 24} (1924), 69.} 
According to this theory, an atom in an excited 
state is continually emitting virtual radiation, to which no energy characteristics are to be ascribed. The normal atoms 
have associated with them virtual oscillators, of the frequencies corresponding to jumps of the atom to all of the 
stationary states of higher energy. The virtual radiation may be thought of as being absorbed by these virtual oscillators, 
and any atom which has a virtual oscillator absorbing this virtual radiation has a certain probability of jumping suddenly 
to the higher state of energy corresponding to the frequency of the particular virtual oscillator. On the average, if the 
radiation is completely absorbed, the number of such jumps to levels of higher energy is equal to the number of emitting 
atoms which pass from higher to lower states. But there is no direct connection between the falling of one atom from a 
higher to a lower state and a corresponding rise of a second atom from a lower to a higher state. Thus on this view the 
energy of the emitting atoms and of the absorbing atoms is only statistically conserved. 

\

\begin{center}
\par
\Needspace{5\baselineskip}
{\sc The problem of the ether}\addcontentsline{toc}{section}{The problem of the ether}\endnote{This 
section is present only in the French version.}\end{center}
The constancy of the speed of radiation of different wavelengths has long been considered as one of the most powerful 
arguments in favour of the wave theory of light. This constancy suggests that a perturbation is travelling through a 
fixed medium in space, the ether.

If experiments like those by Michelson and Morley's were to show the existence of a relative motion with respect to such 
a medium, this argument would be considerably strengthened. For then we could imagine light as having a speed determined 
with reference to a fixed axis in space. But, except for the recent and quite doubtful experiments by 
Miller,\footnote{D.~C.~Miller, {\em Nat.\ Acad.\ Sci.\ Proc.}, {\bf 11} (1925), 306.} no-one has 
ever detected such a relative motion. We thus find ourselves in the difficult position of having to imagine a medium in 
which perturbations travel with a definite speed, not with reference to a fixed system of axes, but with reference to 
each individual observer, whatever his motion. If we think of the complex properties a medium must have in order to 
transmit a perturbation in this way, we find that the medium differs so considerably from the simple ether from which 
we started that the analogy between a wave in such a medium and a pertubation travelling in an elastic medium is very 
distant. It is true that doubts have often been expressed as to the usefulness of retaining the notion of the ether. 
Nevertheless, if light is truly a wave motion, in the sense of Maxwell, there must be a medium in order to transmit this 
motion, without which the notion of wave would have no meaning. This means that, instead of being a support for the wave 
theory, the concept of the ether has become an uncomfortable burden of which the wave theory has been unable to rid 
itself.

If, on the other hand, we accept the view suggested by the theory of relativity, in which for the motion of matter or 
energy there is a limiting speed relative to the observer, it is not surprising to find a form of energy that moves at 
this limiting speed. If we abandon the idea of an ether, it is simpler to suppose that this energy moves in the form of 
corpuscles rather than waves.

\

\begin{center}
\par
\Needspace{5\baselineskip}
{\sc The emission of radiation}\label{emission}\addcontentsline{toc}{section}{The emission of radiation}\end{center}
When we trace a sound to its origin, we find it coming from an oscillator vibrating with the frequency of the sound 
itself. The same is true of electric waves, such as radio waves, where the source of the radiation is a stream of 
electrons oscillating back and forth in a wire. But when we trace a light ray or an X-ray back to its origin, we fail 
to find any oscillator which has the same frequency as the ray itself. The more complete our knowledge becomes of the 
origin of spectrum lines, the more clearly we see that if we are to assign any frequencies to the electrons within the 
atoms, these frequencies are not the frequencies of the emitted rays, but are the frequencies associated with the 
stationary states of the atom. This result cannot be reconciled with the electromagnetic theory of radiation, nor has any 
mechanism been suggested whereby radiation of one frequency can be excited by an oscillator of another frequency. The wave 
theory of radiation is thus powerless to suggest how the waves originate.

The origin of the radiation is considerably simpler when we consider it from the photon viewpoint. We find that an atom 
changes from a stationary state of one energy to a state of less energy, and associated with this change radiation is 
emitted. What is simpler than to suppose that the energy lost by the atom is radiated away as a single photon? It is on 
this view unnecessary to say anything regarding the frequency of the radiation. We are concerned only with the energy of 
the photon, its direction of emission, and its state of polarisation.

The problem of the emission of radiation takes an especially interesting form when we consider the production of the 
continuous X-ray spectrum.\footnote{The difficulty here discussed was first emphasised by D.~L.~Webster, 
{\em Phys.\ Rev.}, {\bf 13} (1919), 303.} Experiment shows 
that both the intensity and the average frequency of the X-rays emitted at angles less than 90 degrees with the cathode-ray 
stream are greater than at angles greater than 90 degrees. This is just what we should expect due to the Doppler effect if 
the X-rays are emitted by a radiator moving in the direction of the cathode rays. In order to account for the observed 
dissymmetry between the rays in the forward and backward directions, the particles emitting the radiation must be moving 
with a speed of the order of 25 per cent that of light. This means that the emitting particles must be free electrons, 
since it would require an impossibly large energy to set an atom into motion with such a speed.

But it will be recalled that the continuous X-ray spectrum has a sharp upper limit. Such a sharp limit is, however, 
possible on the wave theory only in case the rays come in trains of waves of considerable length, so that the interference 
between the waves in different parts of the train can be complete at small glancing angles of reflection from the crystal. 
This implies that the oscillator which emits the rays must vibrate back and forth with constant frequency a large number 
of times while the ray is being emitted. Such an oscillation might be imagined for an electron within an atom; but it is 
impossible for an electron moving through an irregular assemblage of atoms with a speed comparable with that of light.

Thus the Doppler effect in the primary X-rays demands that the rays shall be emitted by rapidly moving electrons, while 
the sharp limit to the continuous spectrum requires that the rays be emitted by an electron bound within an atom.

The only possible escape from this dilemma on the wave theory is to suppose that the electron is itself capable of 
internal oscillation of such a character as to emit radiation. This would, however, introduce an undesirable complexity 
into our conception of the electron, and would ascribe the continuous X-rays to an origin entirely different from that of 
other known sources of radiation.

Here again the photon theory affords a simple solution. It is a consequence of Ehrenfest's adiabatic 
principle\footnote{The adiabatic principle consists in the following. Since for a quantised quantity there should be 
no quantum jumps induced by an infinitely slowly varying external force (in this case, one that gently accelerates a 
radiator), there is an analogy between these quantities and the 
classical adiabatic invariants. Ehrenfest (1917) accordingly formulated a principle identifying the 
classical quantities to be quantised as the adiabatic invariants of a system ({\em eds.}).} that photons 
emitted by a moving radiator will show the same Doppler effect, with regard to both frequency 
and intensity, as does a beam of waves.\label{Hewlett}\footnote{Cf., e.g., A.~H.~Compton, {\em Phys.\ Rev.}, {\bf 21} (1923), 483.} 
But if we suppose that photons are radiated by the moving cathode electrons, the energy of each photon will be the energy 
lost by the electron, and the limit of the 
X-ray spectrum is necessarily reached when the energy of the photon is equal to the initial energy of the electron, i.e., 
$h\nu=eV$. In this case, if we consider the initial state as an electron approaching an atom with large kinetic energy and 
the final state as the electron leaving the atom with a smaller kinetic energy, we see that the emission of the continuous 
X-ray spectrum is the same kind of event as the emission of any other type of radiation.

\

\noindent {\em Absorption of radiation.} --- According to the photon theory, absorption occurs when a photon meets an atom and 
imparts its energy to the atom. The atom is thereby raised to a stationary state of higher energy~--- precisely the 
reverse of the emission process.

On the wave theory, absorption is necessarily a continuous process, if we admit the conservation of energy, since on no 
part of the wave front is there enough energy available to change the atom suddenly from a state of low energy to a state 
of higher energy. What evidence we have is, however, strongly against the atom having for any considerable length of time 
an energy intermediate between two stationary states; and if such intermediate states cannot exist, the gradual absorption 
of radiation is not possible. Thus the absorption of energy from waves\endnote{[\'{e}nergie ondulatoire]} is irreconcilable 
with the conception of stationary states.

We have seen that on the theory of virtual radiation the energy of the emitting atoms and of the absorbing atoms 
is only statistically conserved. There is according to this view therefore no difficulty with supposing that the 
absorbing atom suddenly jumps to a higher level of energy, even though it has not received from the radiation as 
much energy as is necessary to make the jump. It is thus possible through virtual oscillators and virtual radiation 
to reconcile the wave theory of radiation with the sudden absorption of energy, and hence to retain the idea of 
stationary states.\label{Comp50} 

\

\begin{center}
\par
\Needspace{5\baselineskip}
{\sc The photoelectric effect}\addcontentsline{toc}{section}{The photoelectric effect}\end{center}
It is well known that the photon hypothesis was introduced by Einstein to account for the photoelectric 
effect.\footnote{A.~Einstein, {\em Ann.\ d.\ Phys.}, {\bf 17} 
(1905), [132].\endnotemark}\endnotetext{The original footnote gives page `145'.} The assumption that light consists of discrete units which can be absorbed by atoms only as units, 
each giving rise to a photoelectron, accounted at once for the fact that the number of photoelectrons is proportional to 
the intensity of the light; and the assumption that the energy of the light unit is equal to $h\nu$, where $h$ is Planck's 
constant, made it possible to predict the kinetic energy with which the photoelectrons should be ejected, as expressed 
by Einstein's well-known photoelectric equation,
  \begin{equation}
    mc^2\Big(\frac{1}{\sqrt{1-\beta^2}}-1\Big)=h\nu-w_p\ .
    \label{compton1}
  \end{equation}

Seven years elapsed before experiments by Richardson and Compton\footnote{O.~W.~Richardson and K.~T.~Compton, 
{\em Phil.\ Mag.}, {\bf 24} (1912), 575.} and by Hughes\footnote{A.~L.~Hughes, {\em Phil.\ Trans.\ A}, {\bf 212} 
(1912), 205.\endnotemark}\endnotetext{The English edition has `{\bf 213}'.} 
showed that the energy of the emitted electrons was indeed proportional to the frequency less a constant,\endnote{[\`{a} 
part une constante]} and that the factor of proportionality was close to the value of $h$ calculated from Planck's 
radiation formula. Millikan's more recent precision photoelectric experiments with the alkali 
metals\footnote{R.~A.~Millikan, {\em Phys.\ Rev.}, {\bf 7} (1916), 355.} confirmed the identity of the constant 
$h$ in the photoelectric equation with that in Planck's radiation formula. De Broglie's beautiful 
experiments\footnote{M.~de~Broglie, {\em Jour.\ de Phys.}, {\bf 2} (1921), 265.} with the magnetic 
spectrograph showed that in the region of X-ray frequencies the same equation holds, if only we interpret the work 
function $w_p$ as the work required to remove the electron from the $p$th energy level of the 
atom. Thibaud has made use of this 
result\footnote{J.~Thibaud, {\em C.\ R.}, {\bf 179} (1924), 165, 1053 and 1322.} in comparing the velocities of 
the photoelectrons ejected by $\gamma$-rays from different elements, 
and has thus shown that the photoelectric equation (\ref{compton1}) holds with precision even for $\beta$-rays of the 
highest speed. Thus from light of frequency so low that it is barely able to eject photoelectrons from metals to 
$\gamma$-rays that eject photoelectrons with a speed almost as great as that of light, the photon theory expresses 
accurately the speed of the photoelectrons.

The direction in which the photoelectrons are emitted is no less instructive than is the velocity. Experiments using 
the cloud expansion method, performed\endnote{[perfectionn\'{e}e]} by 
C.~T.~R.~Wilson\label{Wilson}\footnote{C.~T.~R.~Wilson, {\em Proc.\ Roy.\ Soc.\ A}, {\bf 104} (1923), 1.} and 
others,\footnote{A.~H.~Compton, {\em Bull.\ Natl.\ Res.\ Coun.}, No.~20  (1922), 25; F.~W.~Bubb, {\em Phys.\ Rev.}, 
{\bf 23} (1924), 137; P.~Auger, {\em C.\ R.}, {\bf 178} (1924), 1535; D.~H.~Loughridge, {\em Phys.\ Rev.}, {\bf 26} 
(1925), 697; F.~Kirchner, {\em Zeits.\ f.\ Phys.}, 
{\bf 27} (1926), 385.} have shown that the most probable direction in which the photoelectron is ejected 
  \begin{figure}
    \centering
     \resizebox{\textwidth}{!}{\includegraphics[0mm,0mm][220.91mm,115.28mm]{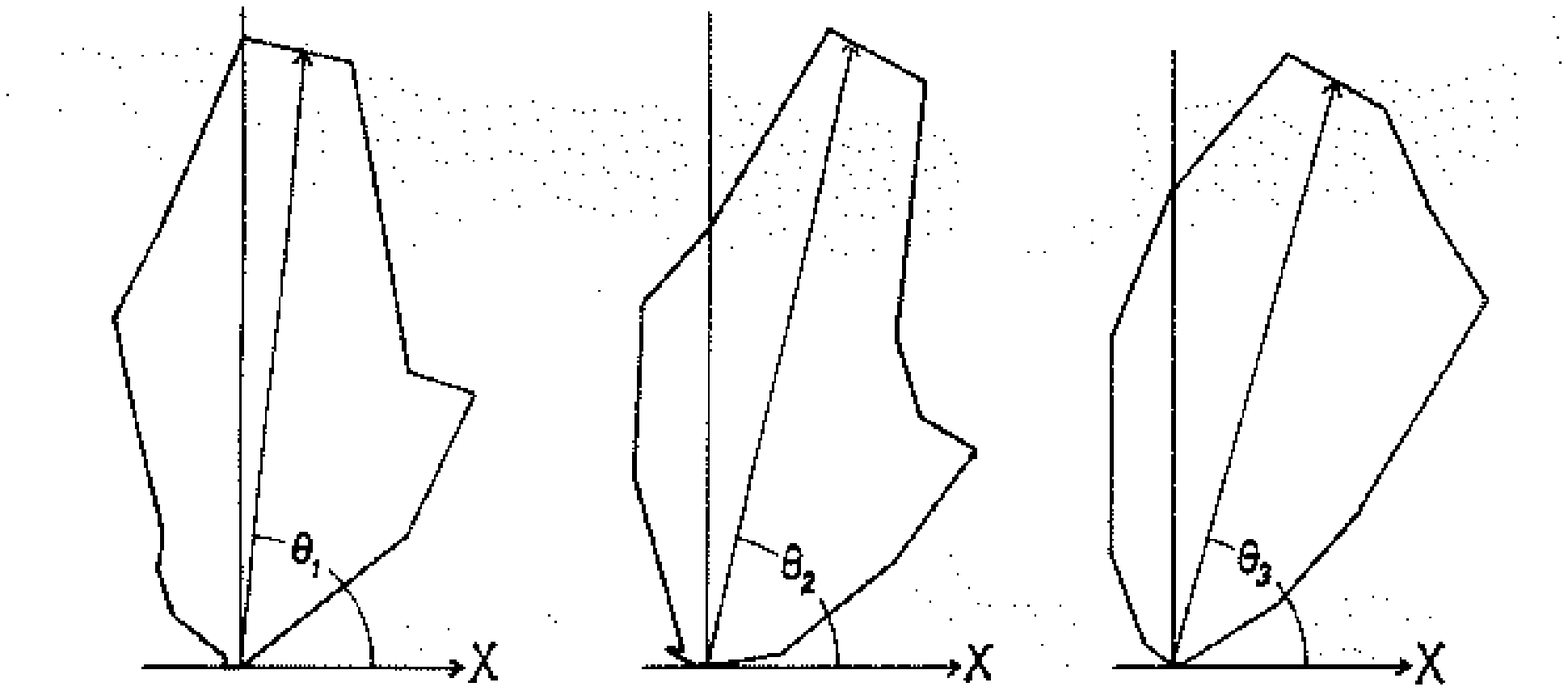}}
    \unnumberedcaption{\small Fig.~1. Longitudinal distribution of photoelectrons for X-rays of three different  
    effective wavelengths, according to Auger.}
  \end{figure} 
from an atom is nearly the direction of the electric vector of the incident wave, but with an appreciable forward 
component to its motion. There is, however, a very considerable variation in the direction of emission. For example, if we 
plot the number of photoelectrons ejected at different angles with the primary beam we find, according to Auger, 
the distribution shown in Fig.~1.
  \begin{figure}
    \centering
     \resizebox{\textwidth}{!}{\includegraphics[0mm,0mm][220.08mm,210.27mm]{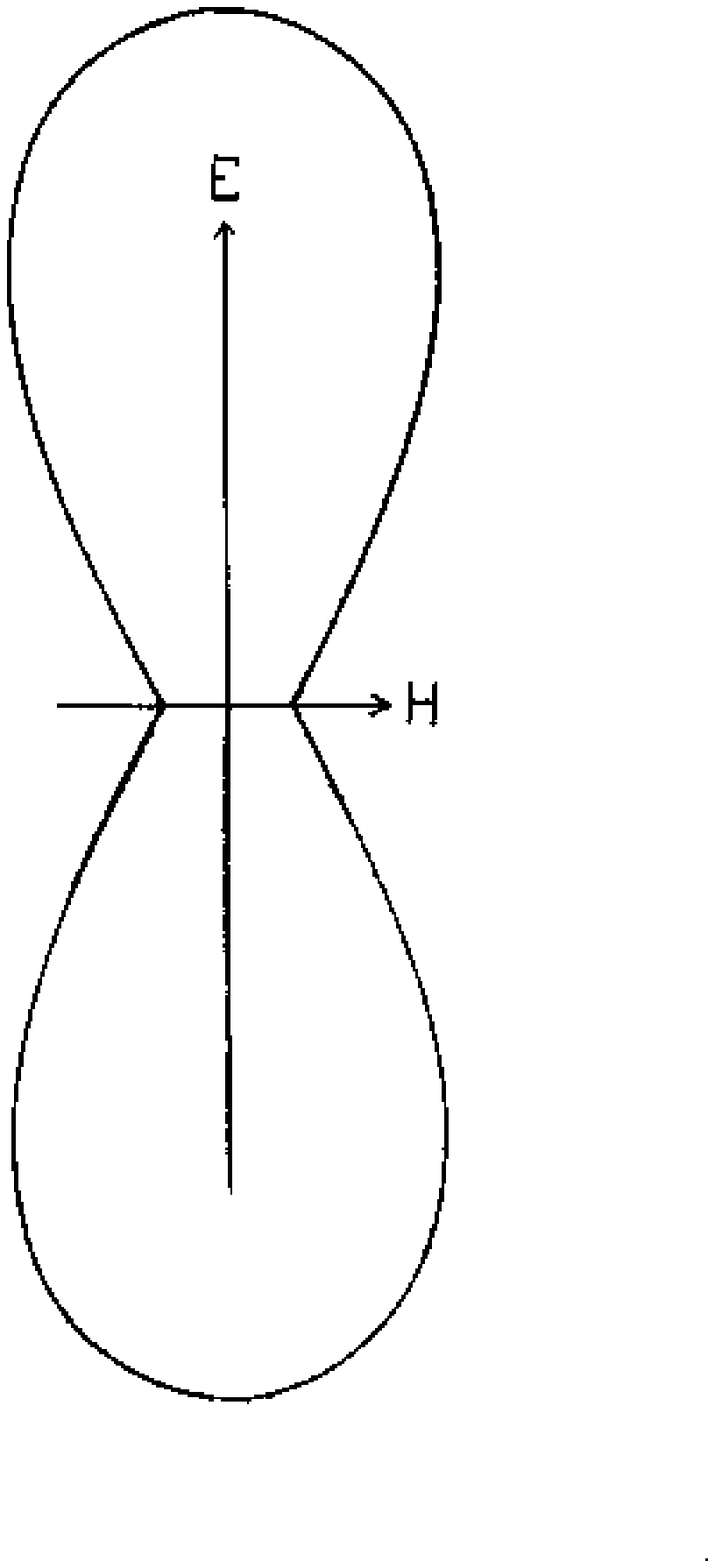}}
    \unnumberedcaption{\small Fig.~2. Lateral distribution of photoelectrons for incompletely polarised X-rays, 
    according to Bubb.}
  \end{figure}

Each of these curves, taken at a different potential, represents the distribution of about 200 photoelectron tracks. It 
will be seen that as the potential on the X-ray tube increases, the average forward component of the photoelectron's 
motion also increases.

When polarised X-rays are used, there is a strong preponderance of the photoelectrons in or near the plane including the 
electric vector of the incident rays. Thus Fig.~2
shows the distribution found by Bubb of the direction of the 
photoelectrons ejected from moist air when traversed by X-rays that have been polarised by scattering at right angles 
from a block of paraffin. Because of multiple scattering in the paraffin, the scattered rays are not completely polarised, 
and this is probably sufficient to account for the fact that some photoelectrons appear to start at right angles with the 
electric vector. This effect with X-rays is doubtless similar in character to the selective photoelectric effect 
discovered many years ago by Pohl and Pringsheim, in which the number of electrons ejected by light from the liquid 
surface of sodium-potassium alloy is greater when the electric vector is in a plane perpendicular to the surface than 
when parallel to the surface.

Recent experiments have shown that the direction in which the photoelectrons are ejected by X-rays is at least very 
nearly independent of the material from which the electrons come.\footnote{E.~A.~Owen, {\em Proc.\ Phys.\ Soc.}, 
{\bf 30} (1918), 133; Auger, Kirchner, Loughridge, {\em loc.\ cit.}\endnotemark}\endnotetext{The second part of the 
footnote is printed only in the English edition.}

\

\noindent {\em Can electromagnetic waves produce photoelectrons?} --- Before discussing the production of photoelectrons from the 
standpoint of radiation quanta, let us see what success meets the attempt\endnote{The French edition here includes the 
clause `qui a \'{e}t\'{e} faite'.} to account for them on the basis of electromagnetic waves. The fact that they are 
emitted approximately in the direction of the electric vector would suggest that the photoelectrons are ejected by the 
direct action of the electric field of the incident rays. If this were the case, however, we should expect the speed of 
the ejected electrons to be greater for greater intensity of radiation, whereas experiment shows that for the same 
wavelength intense sunlight ejects an electron no faster than does the feeble light from a star. Furthermore, the 
energy available from the electromagnetic wave is wholly inadequate. Thus in a recent experiment performed by Joffe and 
Dobronrawov,\footnote{A.~Joffe and N.~Dobronrawov, {\em Zeits.\ f.\ Phys.}, {\bf 34}, 889 (1925).} X-rays were produced by 
the impact on a target of $10^4$ to $10^5$ electrons per second. Since on the electromagnetic theory an X-ray pulse is 
of the order of $10^3$ waves in length or $10^{-16}$ seconds in duration, the X-ray pulses must have followed each other 
at widely separated intervals. It was found, however, that photoelectrons were occasionally ejected from a bismuth 
particle which subtended a solid angle not greater than $10^{-5}$. It is clearly impossible\label{clearly-impossible} that all the energy of an 
X-ray pulse which has spread out in a spherical wave should spend itself on this bismuth particle.\label{Comp53} Thus on the wave 
theory the ejection of the photoelectron, which has almost as much energy as the original cathode electron, could not 
have been accomplished by a single\endnote{Here and in several places in the following, the French edition has `simple' 
where the English one has `single'.} pulse. It cannot therefore be the direct action of the electric vector of the wave, 
taken in the usual sense,\endnote{[l'action directe du vecteur \'{e}lectrique de l'onde, prise dans le sens ordinaire]} 
which has ejected the electron.

We may assume, on the other hand, that the energy is gradually absorbed in the bismuth particle of Joffe's experiment 
until an amount $h\nu$ has accumulated, which is then spent in ejecting the photoelectron. We have already called 
attention to the fact that this gradual absorption hypothesis implies the existence of stationary states in the atom 
having infinitesimal gradations of energy, whereas the evidence is very strong that atoms cannot endure except in certain 
definitely defined stationary states. But new difficulties also arise. Why do the photoelectrons tend to start in the 
direction of the electric field of the incident wave? If we suppose that it is the gradual absorption of energy from a 
wave which liberates the electron, why does there exist a tendency for the electron to start with a large component of 
its motion in a forward direction?\endnote{[dans la direction de propagation de l'onde]} The forward impulse due to the 
radiation pressure as\endnote{[puisque]} the energy is gradually absorbed will be transferred to the atom and not left 
with [the] absorbing electron. The accumulation hypothesis is thus difficult to defend.

\

\noindent {\em Photons and photoelectrons.} --- On the photon theory it is possible to account in a simple manner for most of the 
properties of the photoelectrons. We have seen how Einstein was able to predict accurately the velocity of the 
photoelectrons, assuming only that energy is conserved when a photon acts on an electron. In order to account for the 
direction of emission we must ascribe to the photon some of the properties of an electromagnetic pulse. Bubb introduced 
the suggestion\footnote{F.~W.~Bubb, {\em Phys.\
Rev.}, {\bf 23} (1924), 137.} that we ascribe to the photon a vector property similar to the electric vector of an 
electromagnetic wave, so that when the photon traverses an atom the electrons and the nucleus receive impulses in opposite 
directions perpendicular to the direction of propagation. Associated with this electric vector, we should also expect to 
find a magnetic vector. Thus if an electron is set in motion by the electric vector of the photon at right angles to the 
direction of propagation, the magnetic vector of the photon will act on the 
moving electron in the direction of propagation. This is strictly analogous to the radiation pressure 
exerted by an electromagnetic wave on an electron which it traverses, and means that the forward momentum of the absorbed 
photon is transferred to the photoelectron.

In the simplest case, where we neglect the initial momentum of the electron in its orbital motion in the atom, the angle 
between the direction of the incident ray and the direction of ejection is found from these assumptions to be
  \begin{equation}
    \theta=\tan^{-1}\sqrt{2/\alpha}\ ,
    \label{compton2}
  \end{equation}
where $\alpha=\gamma/\lambda$, and $\gamma=h/mc=0.0242$ \AA. The quantity $\alpha$ is small compared with unity, except 
for very hard X-rays and $\gamma$-rays. Thus for light, equation (\ref{compton2}) predicts the expulsion of 
photoelectrons at nearly 90 degrees. This is in accord with the rather uncertain data which have been 
obtained with visible and ultra-violet light.\footnote{Cf.\ A.~Partsch and W.~Hallwachs, {\em Ann.\ d.\ 
Phys.}, {\bf 41} (1913), 247.}

The only really significant test of this result is in its application to X-ray photoelectrons. In Fig.~1 are drawn the 
lines $\theta_1$, $\theta_2$ and $\theta_3$ for the three curves, at the angles calculated by Auger from equation 
(\ref{compton2}). It will be seen that they fall very satisfactorily in the direction of maximum emission of the 
photoelectrons. Similar results have been obtained by other investigators.\footnote{W.~Bothe, {\em Zeits.\ f.\ 
Phys.}, {\bf 26} (1925), 59; F.~Kirchner, {\em Zeits.\ f.\ Phys.}, {\bf 27} (1926), 385.\endnotemark}\endnotetext{This 
footnote is only present in the English edition.} This may be taken as proof that a photon imparts not only its energy, 
but also its momentum to the photoelectrons.\label{Since}\footnote{The English version includes here the following 
footnote. Cf.\ also the comments by Bragg on p.~\pageref{BraggContribution} and the ensuing discussion ({\em eds.}).
  \begin{quote}
    Since this was written, experiments by [D. H.] Loughridge ({\em Phys.\ Rev.}, {\bf 30} (1927), [488]) have been 
    published which show a forward component to the photoelectron's motion which 
    seems to be greater than that predicted by equation (\ref{compton2}). Williams, in experiments as yet unpublished, 
    finds that the forward component is almost twice as great as that predicted by this theory. These results indicate 
    that the mechanism of interaction between the photon and the atom must be more complex than here postulated. The fact 
    that the forward momentum of the photoelectron is found to be of the same order of magnitude as that of the incident 
    photon, however, suggests that the momentum of the photon is acquired by the photoelectron, while an additional 
    forward impulse is imparted by the atom. Thus these more recent experiments also support the view that the 
    photoelectron acquires both the energy and the momentum of the photon. 
  \end{quote}
}

Honesty\footnote{This paragraph is present only in the French edition. The corresponding one in the English edition 
reads:
  \begin{quote}
    If the angular momentum of the atomic system from which the photoelectron is ejected is to be conserved when acted 
    upon by the radiation, the electron cannot be ejected exactly in the direction of $\theta$, but must receive an 
    impulse in a direction determined by the position of the electron in the atom at the instant it is traversed by the 
    photon.$^*$ Thus we should probably consider the electric vector of the X-ray wave as defining merely the most 
    probable direction in which the impulse should be imparted to the electron. This is doubtless the chief reason why 
    the photoelectrons are emitted over a wide range of angles instead of in a definite direction, as would be suggested 
    by the calculation just outlined. 
  \end{quote}
With the footnote: $^*$Cf.\ A.~H.~Compton,  {\em Phys.\ Rev.}, [{\bf 31}]  (1928), [59] ({\em eds.}).} 
obliges me to point out a difficulty that arises in this explanation of the motion of the photoelectrons. It is the 
failure of the attempts made to account properly for the fact that the photoelectrons are emitted over a wide range of 
angles instead of in a definite direction, as would be suggested by the calculation just outlined. The most interesting 
of these attempts is that of Bubb,\footnote{F.~W.~Bubb, {\em Phil.\ Mag.}, {\bf 49} (1925), 824.} 
who takes into account the momentum of the electron immediately before the absorption of the photon. Bubb finds a 
dispersion of the directions of emission of the photoelectrons of the correct order of magnitude, but which is larger 
when the electron issues from a heavy atom than when it issues from a light one. We have seen, however, that experiment 
has shown this dispersion of the directions of emission to be notably independent of the element from which the 
photoelectron originates.

Whatever may be the cause of the dispersion in the directions of motion of the photoelectrons,\endnote{The preceding 
clause is only present in the French edition.} it will readily be seen that if the time during which the photon exerts a 
force on the electron is comparable with the natural period of the 
electron\endnote{[la p\'{e}riode de l'\'{e}lectron dans son mouvement orbital]} in the atom, the impulse imparted to the 
electron will be transferred in part to the positive nucleus about which the electron is moving. The fact that the 
photoelectrons are ejected with a forward component equal, within the limits of experimental error, to the momentum of 
the incident photon\endnote{In the English edition this reads: `The fact that the photoelectrons receive the momentum of 
the incident photon'.} means that no appreciable part of the photon's momentum is spent on the remainder of the atom. This 
can only be the case if the time of action of the photon on the electron is short compared with the time of revolution of 
the electron in its orbit.\footnote{The English edition includes the further sentence: `Such a short duration of 
interaction is a natural consequence of the photon conception of radiation, but is quite contrary to the consequences 
of the electromagnetic theory' ({\em eds.}).}

\

\noindent {\em The photoelectric effect and virtual radiation.} --- It is to be noted that none of these properties of the 
photoelectron is inconsistent with the virtual radiation theory of Bohr, Kramers and Slater. The difficulties which 
applied to the classical wave theory do not apply here, since the energy and momentum are conserved only statistically. 
There is nothing in this theory, however, which would enable us to predict anything regarding the motion of the 
photoelectrons. The degree of success that has attended the application of the photon hypothesis to the motion of 
these electrons has come directly from the application of the conservation principles to the individual action of a 
photon on an electron. The power of these principles as applied to this case is surprising if the assumption is correct 
that they are only statistically valid.

\

\begin{center}
\par
\Needspace{5\baselineskip}
{\sc Phenomena associated with the scattering of X-rays}\addcontentsline{toc}{section}{Phenomena associated 
with the scattering of X-rays}\end{center}
As is now well known, there is a group of phenomena associated with the scattering of X-rays for which the classical wave 
theory of radiation fails to account. These phenomena may be considered under the heads of: (1) The change of wavelength 
of X-rays due to scattering, (2) the intensity of scattered X-rays, and (3) the recoil electrons.

The earliest experiments on secondary X-rays and $\gamma$-rays\endnote{[sur les rayons X secondaires et les rayons 
$\gamma$]} showed a difference in the penetrating power of the primary and the secondary rays. In the case of X-rays, 
Barkla and his collaborators\footnote{C.~[G.]~Barkla and C.~A.~Sadler, {\em Phil.\ Mag.}, {\bf 16}, 550 
(1908).\endnotemark}\endnotetext{This footnote appears only in the French edition.} showed that the secondary rays from the 
heavy elements consisted largely of fluorescent radiations characteristic of the radiator, and that it was the presence 
of these softer rays which was chiefly responsible for the greater absorption of the secondary rays. When later 
experiments\footnote{C.~A.~Sadler and P.~Mesham, {\em Phil.\ Mag.}, {\bf 24} (1912), 138; 
J.~Laub, {\em Ann.\ d.\ Phys.}, {\bf 46} (1915), 785.} showed a measurable difference in penetration 
even for light elements such as carbon, from which no fluorescent K or L radiation appears, it was natural to 
ascribe\footnote{[C.~G.] Barkla and [M.~P.] White, {\em Phil.\ Mag.}, {\bf 34} (1917), 270; J.~Laub, 
{\em Ann.\ d.\ Phys.}, {\bf 46} (1915), 785, {\em et al.}} this difference to a new type of fluorescent radiation, 
similar to the K and L types, but of shorter wavelength. Careful absorption 
measurements\footnote{E.g., [F.~K.]~Richtmyer and [K.]~Grant, {\em Phys.\ Rev.}, {\bf 15} (1920), 547. } 
failed, however, to reveal any critical absorption limit for these assumed `J' radiations similar to those corresponding 
to the K and L radiations. Moreover, direct spectroscopic 
observations\footnote{E.g., [W.] Duane and 
[T.] Shimizu, {\em Phys.\ Rev.}, {\bf 13} (1919), [289]; {\em ibid.}, 
{\bf 14} (1919), 389.} failed to reveal the 
existence of any spectrum lines\endnote{[ne 
fournirent aucune preuve de l'existence d'un spectre de raies]} under conditions for which the supposed J-rays should 
appear. It thus became evident that the softening of the secondary X-rays from the lighter elements was due to a 
different kind of process than the softening of the secondary rays from heavy elements where fluorescent X-rays are 
present.

A series of skilfully devised absorption experiments performed by J.~A.~Gray\footnote{J.~A.~Gray, {\em Phil.\ 
Mag.}, {\bf 26} (1913), 611; {\em Jour.\ Frank.\ Inst.}, [{\bf 190}], 643 (Nov.\ 1920).} showed, on the other hand, that 
both in the case of $\gamma$-rays and in that of X-rays an increase in wavelength accompanies the scattering of the rays 
of light elements.

It was at this stage that the first spectroscopic investigations of the secondary X-rays from light elements were 
made.\footnote{A.~H.~Compton, {\em Bull.\ Natl.\ Res.\ Coun.}, No.~20, [18] ([October] 1922); {\em Phys.\ Rev.}, 
{\bf 22} (1923), 409.} According to the usual electron theory of scattering it 
is obvious that the scattered rays will be of the same frequency as the forced oscillations of the electrons which emit 
them, and hence will be identical in frequency with the primary waves which set the electrons in motion. Instead of 
showing scattered rays of the same wavelength as the primary rays, however, these spectra revealed 
lines in the secondary rays corresponding to those in the 
primary beam, but with each line displaced slightly toward the longer wavelengths.

This result might have been predicted from Gray's absorption measurements; but the spectrum measurements had the advantage 
of affording a quantitative measurement of the change in wavelength, which gave a basis for its theoretical 
interpretation.

The spectroscopic experiments which have shown this change in wavelength are too well known\footnote{Cf., e.g., A.~H.~Compton, 
{\em Phys.\ Rev.}, {\bf 22} (1923), 409; P.~A.~Ross, 
{\em Proc.\ Nat.\ Acad.}, {\bf 10} (1924), 304.} to require discussion. The interpretation of the wavelength change in terms of photons being 
deflected by individual\endnote{This word is missing in the French edition.} electrons and imparting a part of their 
energy to the scattering electrons is also very familiar. For purposes of discussion, however, let us recall that when we 
consider the interaction of a single photon with a single electron the principles of the conservation of energy and 
momentum lead us\footnote{A.~H.~Compton,  
{\em Phys.\ Rev.}, [{\bf 21}] (1923), 483; P.~Debye, {\em Phys.\ Zeits.}, {\bf 24} (1923), 161.} to the result that 
the change in wavelength of the deflected photon is
  \begin{equation}
    \delta\lambda=\frac{h}{mc}(1-\cos\varphi)\ ,
    \label{compton3}
  \end{equation} 
where $\varphi$ is the angle through which the photon is deflected. The electron at the same time recoils from the photon 
at an angle of $\theta$ given by,\endnote{The English edition reads `$(1+x)$'.}
  \begin{equation}
    \cot\theta=-(1+\alpha)\tan\frac{1}{2}\varphi\ ;
    \label{compton4}
  \end{equation}
and the kinetic energy of the recoiling electron is, 
  \begin{equation}
    E_{\mbox{\scriptsize kin}}=h\nu\frac{2\alpha\cos^2\theta}{(1+\alpha)^2-\alpha^2\cos^2\theta}\ .
    \label{compton5}
  \end{equation}

The experiments show in the spectrum of the scattered rays two lines corresponding to each line of the primary ray. One 
of these lines is of precisely the same wavelength as the primary ray, and the second line, though somewhat broadened, 
has its centre of gravity displaced by the amount predicted by equation (\ref{compton3}). According to experiments by 
Kallman and Mark\footnote{H.~Kallman and H.~Mark, {\em Naturwiss.}, {\bf 13} (1925), 297. } and 
by Sharp,\footnote{H.~M.~Sharp, {\em Phys.\ Rev.}, {\bf 26} (1925), 691.} this agreement between the 
theoretical\endnote{[pr\'{e}dit]} and the observed shift is precise within a small fraction of 1 per cent. 

\

\noindent {\em The recoil electrons.} --- From the quantitative agreement between the theoretical and the observed wavelengths of 
the scattered rays, the recoil electrons predicted by the photon theory of scattering were looked for with some 
confidence.\endnote{[on eut quelque confiance dans les \'{e}lectrons de recul]} When this theory was proposed, there was no 
direct evidence for the existence of such electrons, though indirect evidence suggested that the secondary 
$\beta$-rays ejected from matter by hard $\gamma$-rays are mostly of this type. Within a few months of their prediction, however, 
C.~T.~R.~Wilson\footnote{C.~T.~R.~Wilson, {\em Proc.\ Roy.\ Soc.\ }[{\em A}], {\bf 104} (1923), 1.} and 
W.~Bothe\footnote{W.~Bothe, {\em Zeits.\ f.\ Phys.}, {\bf 16} (1923), 319.} independently announced their discovery. The recoil electrons show as short tracks, pointed in the 
direction of the primary X-ray beam, mixed among the much longer tracks due to the photoelectrons ejected by the X-rays.

Perhaps the most convincing reason for associating these short tracks with the scattered X-rays comes from a study of 
their number. Each photoelectron in a cloud photograph represents a quantum of truly absorbed X-ray energy. If the short 
tracks are due to recoil electrons, each one should represent the scattering of a photon. Thus the ratio $N_r/N_p$ of the 
number of short tracks to the number of long tracks should be the same as the ratio $\sigma/\tau$ of the scattered to the 
truly absorbed energy\endnote{The French edition uses $\rho$ instead of $\tau$ in the text, but uses $\tau$ in the 
discussion (where $\rho$ is used for matter density). The English edition uses $t$.} when the X-rays pass through air. 
The latter ratio is known from absorption measurements, and the former ratio can be determined by counting the tracks on 
the photographs. The satisfactory agreement between the two 
ratios\footnote{A.~H.~Compton and A.~W.~Simon, {\em Phys.\ Rev.}, {\bf 25} (1925), 306; J.~M.~Nuttall and 
E.~J.~Williams, {\em Manchester Memoirs}, {\bf 70} (1926), 1.} for X-rays of different wavelengths means 
that on the average there is about one quantum of energy scattered for each short track that is produced.

This result is in itself contrary to the predictions of the classical wave theory, since on this basis all the energy 
spent on a free electron (except the insignificant effect of radiation pressure) should reappear as scattered X-rays. In 
these experiments, on the contrary, 5 or 10 per cent as much energy appears in the motion of the recoil electrons as 
appears in the scattered X-rays.

That these short tracks associated with the scattered X-rays correspond to the recoil electrons predicted by the photon 
theory of scattering becomes clear from a study of their energies. The energy of the electron which produces a track can 
be calculated from the range of the track. The ranges of tracks which start in different directions have been 
studied\footnote{Compton and Simon, {\em loc.\ cit.}\endnotemark}\endnotetext{Footnote mark missing in the French edition.} 
using primary X-rays of different wavelengths, with the result that equation 
(\ref{compton5})\endnote{The French edition gives (\ref{compton4}).} has been satisfactorily verified.

In view of the fact that electrons of this type were unknown at the time the photon theory of scattering was presented, 
their existence, and the close agreement with the predictions as to their number, direction and velocity, supply strong 
evidence in favour of the fundamental hypotheses of the theory.

\

\noindent {\em Interpretation of these experiments.} --- It is impossible to account for scattered rays of altered frequency, and 
for the existence of the recoil electrons, if we assume that X-rays consist of electromagnetic waves in the usual sense. 
Yet some progress has been made on the basis of semi-classical theories. It is an interesting fact that the wavelength 
of the scattered ray according to equation 
(\ref{compton3})\endnote{The French edition gives (\ref{compton2}).} varies with the angle just as one would expect from 
a Doppler effect if the rays are scattered from an electron moving in the direction of the primary beam. Moreover, the 
velocity that must be assigned to the electron in order to give the proper magnitude to the change of wavelength is that 
which the electron would acquire by radiation pressure if it should absorb a quantum of the incident rays. Several 
writers\footnote{C.~R.~Bauer, {\em C.\ R.}, {\bf 177} (1923), 1211; C.~T.~R.~Wilson, 
{\em Proc.\ Roy.\ Soc.\ }[{\em A}], {\bf 104} (1923), 1; K.~Fosterling, {\em Phys.\ Zeits.}, {\bf 25} (1924), 313; O.~Halpern, 
{\em Zeits.\ f.\ Phys.}, {\bf 30} (1924), 153.\endnotemark}\endnotetext{This footnote is present only in the English 
edition.} have therefore assumed that an electron takes from the incident beam a whole quantum of the incident radiation, 
and then emits this energy as a spherical wave while moving 
forward\endnote{This word is missing in the French edition.} with high velocity.

This conception that the radiation occurs in spherical waves, and that the scattering electron can nevertheless acquire 
suddenly the impulses from a whole quantum of incident radiation is inconsistent with the principle of energy 
conservation. But there is the more serious experimental difficulty that this theory predicts recoil electrons all moving 
in the same direction and with the same velocity. The experiments show, on the other hand, a variety of directions and 
velocities, with the velocity and direction correlated as demanded by the photon hypothesis. Moreover, the maximum range 
of the recoil electrons, though in agreement with the predictions of the photon theory, is found to be about four times 
as great as that predicted by the semi-classical theory. 

There is nothing in these experiments, as far as we have described them, which is inconsistent with the idea of virtual 
oscillators continually scattering virtual radiation. 
In order to account for the change of wavelength on this view, Bohr, Kramers and Slater assumed that the virtual 
oscillators scatter as if moving in the direction of the primary beam, accounting for the change of wavelength as a 
Doppler effect. They then supposed that 
occasionally an electron, under the stimulation of the primary virtual rays, will suddenly 
move forward with a momentum large compared with the impulse received from the radiation pressure. Though we have seen 
that not all of the recoil electrons move directly forward, but in a variety of different directions, the theory could 
easily be extended to include the type of motion that is actually observed.

The only objection that one can raise against this virtual radiation theory in connection with the scattering phenomena 
as viewed on a large scale, is that it is difficult to see how such a theory could by itself predict the change of 
wavelength and the motion of the recoil electrons. These phenomena are directly predictable if the conservation of 
energy and momentum are assumed to apply to the individual actions of radiation on electrons; but this is precisely 
where the virtual radiation theory denies the validity of the conservation principles.

We may conclude that the photon theory predicts quantitatively and in detail the change of wavelength of the scattered 
X-rays and the characteristics of the recoil electrons. The virtual radiation theory is probably not inconsistent with 
these experiments, but is incapable of predicting the results. The classical theory, however, is altogether helpless to 
deal with these phenomena.

\

\noindent {\em The origin of the unmodified line} --- The unmodified line is probably due to X-rays which are scattered by electrons 
so firmly held within the atom that they are not ejected by the impulse from the deflected photons. This view is adequate 
to account for the major characteristics of the unmodified rays, though as yet no quantitatively satisfactory theory of 
their origin has been published.\footnote{Cf., however, G.~E.~M.~Jauncey, {\em Phys.\ Rev.}, {\bf 25} (1925), 314 and 
{\em ibid.}, 723; 
G.~Wentzel, {\em Zeits.\ f.\ Phys.}, {\bf 43} (1927), 14, 779; I.~Waller, {\em Nature}, [{\bf 120}, 155] (July 30, 
1927).\endnotemark [The footnote in the English edition continues with the sentence: `It is 
possible that the theories of the latter authors may be satisfactory, but they have not yet been stated in a form suitable 
for quantitative test' ({\em eds.}).]}\endnotetext{The French edition 
reads `J.~Waller'.} It is probable that a detailed account of these rays will involve definite assumptions regarding 
the nature and the duration of the interaction between a photon and an electron; but it is doubtful whether such 
investigations will add new evidence as to the existence of the photons themselves.

A similar situation holds regarding the intensity of the scattered X-rays. Historically it was the fact that the classical 
electromagnetic theory is unable to account for the low intensity of the scattered X-rays which called attention to the 
importance of the problem of scattering. But the solutions which have been offered by 
Breit,\footnote{G.~Breit,  {\em Phys.\ Rev.}, 
{\bf 27} (1926), 242.} Dirac\footnote{P.~A.~M.~Dirac, {\em Proc.\ Roy.\ Soc.\ A}, [{\bf 111}] (1926), [405].} and 
others\footnote{W.~Gordon,  {\em Zeits.\ f.\ Phys.},  
{\bf 40} (1926), 117; E.~Schr\"{o}dinger,  {\em Ann.\ d.\ Phys.},  
{\bf 82} (1927), 257; O.~Klein,  
{\em Zeits.\ f.\ Phys.}, {\bf 41} (1927), 407; G.~Wentzel, {\em Zeits.\ f.\ Phys.}, {\bf 43}  (1927), 
1, 779.\endnotemark}\endnotetext{The page numbers for Wentzel appear only in the French edition.} of this intensity problem 
as distinguished from that of the change of wavelength, seem to introduce no new 
concepts regarding the nature of radiation or of the scattering process. Let us therefore turn our attention to the 
experiments that have been performed on the individual process of interaction between photons and electrons.

\

\begin{center}
\par
\Needspace{5\baselineskip}
{\sc Interactions between radiation and single electrons}\addcontentsline{toc}{section}{Interactions 
between radiation and single electrons}\endnote{The English edition describes only three 
experiments, omitting the section on the composite photoelectric effect as well as references to it later.}\end{center}

The most significant of the experiments which show departures from the predictions of the classical wave theory are 
those that study the action of radiation on individual atoms or on individual electrons. Two methods have been found 
suitable for performing these experiments, Geiger's point counters, and Wilson's cloud expansion photographs.

\

\noindent (1) {\em Test for coincidences with fluorescent X-rays.} --- Bothe has performed an 
experiment\footnote{W.~Bothe,  {\em Zeits.\ f.\ Phys.}, {\bf 37} (1926), 547.} in which 
fluorescent K radiation from 
a thin copper foil is excited by a beam of incident X-rays. The emitted rays are so feeble that only about five quanta 
of energy are radiated per second. Two point counters are mounted, one on either side of the copper foil in each of which 
an average of one photoelectron is produced and recorded for about twenty quanta radiated by the foil. If we assume that 
the fluorescent radiation is emitted in quanta of energy, but proceed[s] in spherical\label{Comp61} waves in all directions, there 
should thus be about 1 chance in 20 that the recording of a photoelectron in one chamber should be simultaneous with 
the recording of a photoelectron in the other.

The experiments showed no coincidences other than those which were explicable by such sources as high-speed 
$\beta$-particles which traverse both counting chambers.

This result is in accord with the photon hypothesis,\footnote{The English edition continues: `For if a photon of 
fluorescent radiation produces a $\beta$-ray in one counting chamber it cannot traverse the second chamber. Coincidences 
should therefore not occur' ({\em eds.}).} according to which coincidences should not occur. It is, nevertheless, equally in accord 
with the virtual radiation hypothesis, if one assumes that the virtual oscillators in the copper continuously emit 
virtual fluorescent radiation, so that the photoelectrons should be observed in the counting chambers at arbitrary 
intervals.\footnote{At this point in the English version Compton is much more critical of the BKS theory ({\em eds.}):
  \begin{quote}
    According to the virtual radiation hypothesis, however, coincidences should have been observed. For on this view the 
    fluorescent K radiation is emitted by virtual oscillators associated with atoms in which there is a vacancy in the 
    K shell. That is, the copper foil can emit fluorescent K radiation only during the short interval of time following 
    the expulsion of a photoelectron from the K shell, until the shell is again occupied by another electron. This time 
    interval is so short (of the order of $10^{-15}$ sec.) as to be sensibly instantaneous on the scale of Bothe's 
    experiments. Since on this view the virtual fluorescent radiation is emitted in spherical waves, the counting chambers 
    on both sides of the foil should be simultaneously affected, and coincident pulses in the two chambers should 
    frequently occur. The results of the experiment are thus contrary to the predictions of the virtual radiation 
    hypothesis.
  \end{quote}
}

But the experiment is important in the sense that it refutes the often suggested idea  that a quantum of radiation energy 
is suddenly emitted in the form of a spherical wave when an atom passes from one stationary state to another. 

\

\noindent (2) {\em  The composite photoelectric effect.}\endnote{This section is present only in the French edition.} --- 
Wilson\footnote{C.~T.~R.~Wilson, {\em Proc.\ Roy.\ Soc.\ A}, {\bf 104} (1923), 1.} and 
Auger\footnote{P.~Auger, {\em Journ.\ d.\ Phys.}, {\bf 6} (1926), 183. } have noticed in their 
cloud expansion photographs that when X-rays eject photoelectrons from heavy atoms, it often occurs that two or more 
electrons are ejected simultaneously from the same atom. Auger has deduced from studying the ranges of these electrons 
that, when this occurs, the total energy of all the emitted electrons is no larger than that of a quantum of the incident 
radiation. When two electrons are emitted simultaneously it is usually the case that the enegy of one of them is
  \[
    E_{\mbox{\scriptsize kin}}=h\nu-h\nu_K\ ,
  \]
which according to the photon theory means that this electron is due to the absorption of an incident photon accompanied 
by the ejection of an electron from the K energy level. The second electron has in general the energy
  \[
    E_{\mbox{\scriptsize kin}}=h\nu_K-h\nu_L\ .
  \]

This electron can be explained as the result of the absorption by an L electron of the K$\alpha$-ray emitted when another 
L electron occupies the place left vacant in the K orbit by the primary photoelectron. It is established that all the 
electrons that are observed in the composite 
photoelectric effect have to be interpreted in the same way. Their interpretation according to the photon theory thus 
meets with no difficulties.

With regard to the virtual radiation theory, we can take two points of view: first, under the influence of the excitation 
produced by the primary virtual radiation, virtual fluorescent K radiation is emitted by virtual oscillators associated 
with all the atoms traversed by the primary beam. In this view, the probability that this virtual fluorescent radiation 
will cause the ejection of a photoelectron from the same atom as the one that has emitted the primary photoelectron is 
so small that such an event will almost never occur; second, we can alternatively assume that a virtual oscillator 
emitting virtual K radiation is associated only with an atom in which there is a vacant place in the K shell. In this 
case, since the virtual radiation proceeds from the atom that has emitted the primary photoelectron, we could expect 
with extremely large probability that it should excite a photoelectron from the L shell of its own atom, thus accounting 
for the composite photoelectric effect. But in this view the virtual fluorescent radiation is emitted only during a very 
short interval after the ejection of the primary photoelectron, in which case Bothe's fluorescence experiment, described 
above, should have shown some coincidences.

One sees thus that the virtual radiation hypothesis is irreconcilable both with the composite photoelectric effect and 
with the absence of coincidences in Bothe's fluorescence experiment. The photon hypothesis, instead, is in complete accord 
with both these experimental facts.

\

\noindent (3) {\em Bothe and Geiger's coincidence experiments.}\label{forpage13}\endnote{This and the next section are of course numbered (2) and 
(3) in the English edition.} --- We have seen that according to Bohr, Kramers and Slater's theory, virtual 
radiation\endnote{[rayonnement de fluorescence]} is being continually scattered by matter traversed by X-rays, but only 
occasionally is a recoil electron emitted. This is in sharp contrast with the photon theory, according to which a recoil 
electron appears every time a photon is scattered. A crucial test between the two points of view is afforded by an 
experiment devised and brilliantly performed\endnote{[une exp\'{e}rience cruciale entre les deux points de vue a 
\'{e}t\'{e} imagin\'{e}e et brillamment r\'{e}alis\'{e}e]} by Bothe and 
Geiger.\footnote{W.~Bothe and H.~Geiger, {\em Zeits.\ f.\ Phys.}, {\bf 26} (1924), 44; {\bf 32} (1925), 639.} 
X-rays were passed through hydrogen gas, and the resulting recoil electrons and scattered rays were detected by means of 
two different point counters placed on opposite sides of the column of gas. The chamber for counting the recoil electrons 
was left open, but a sheet of thin platinum prevented the recoil electrons from entering the chamber for counting the 
scattered rays. Of course not every photon entering the second counter could be noticed, for its detection depends upon 
the production of a $\beta$-ray. It was found that there were about ten recoil electrons for every scattered photon that 
recorded itself.

The impulses from the counting chambers were recorded on a moving photographic film. In observations over a total period 
of over five hours, sixty-six such coincidences were observed. Bothe and Geiger calculate that according to the statistics 
of the virtual radiation theory the chance was only 1 in $400\,000$ that so many coincidences should have occurred. This 
result therefore is in accord with the predictions of the photon theory, but is directly contrary to the statistical view 
of the scattering process.

\noindent (4) {\em Directional emission of scattered X-rays.}\label{Compton-Simon} --- Additional information regarding the nature of scattered X-rays 
has been obtained by studying the relation between the direction of ejection of the recoil electron and the direction in 
which the associated photon proceeds. According to the photon theory, we have a definite relation (equation 
(\ref{compton4})) between the angle at which the photon is scattered and the angle at which the recoil electron is 
ejected. But according to any form of spreading wave theory, including that of Bohr, Kramers and Slater, the scattered 
rays may produce effects in any direction whatever, and there should be no correlation between the directions in which 
the recoil electrons proceed and the directions in which the secondary $\beta$-rays are ejected by the scattered X-rays.

A test to see whether such a relation exists has been made,\footnote{A.~H.~Compton and A.~W.~Simon, {\em Phys.\ Rev.}, 
{\bf 26} (1925), 289.} using Wilson's cloud apparatus, in the manner shown diagrammatically in Fig.~3.
  \begin{figure}
    \centering
     \resizebox{\textwidth}{!}{\includegraphics[0mm,0mm][220.56mm,180.35mm]{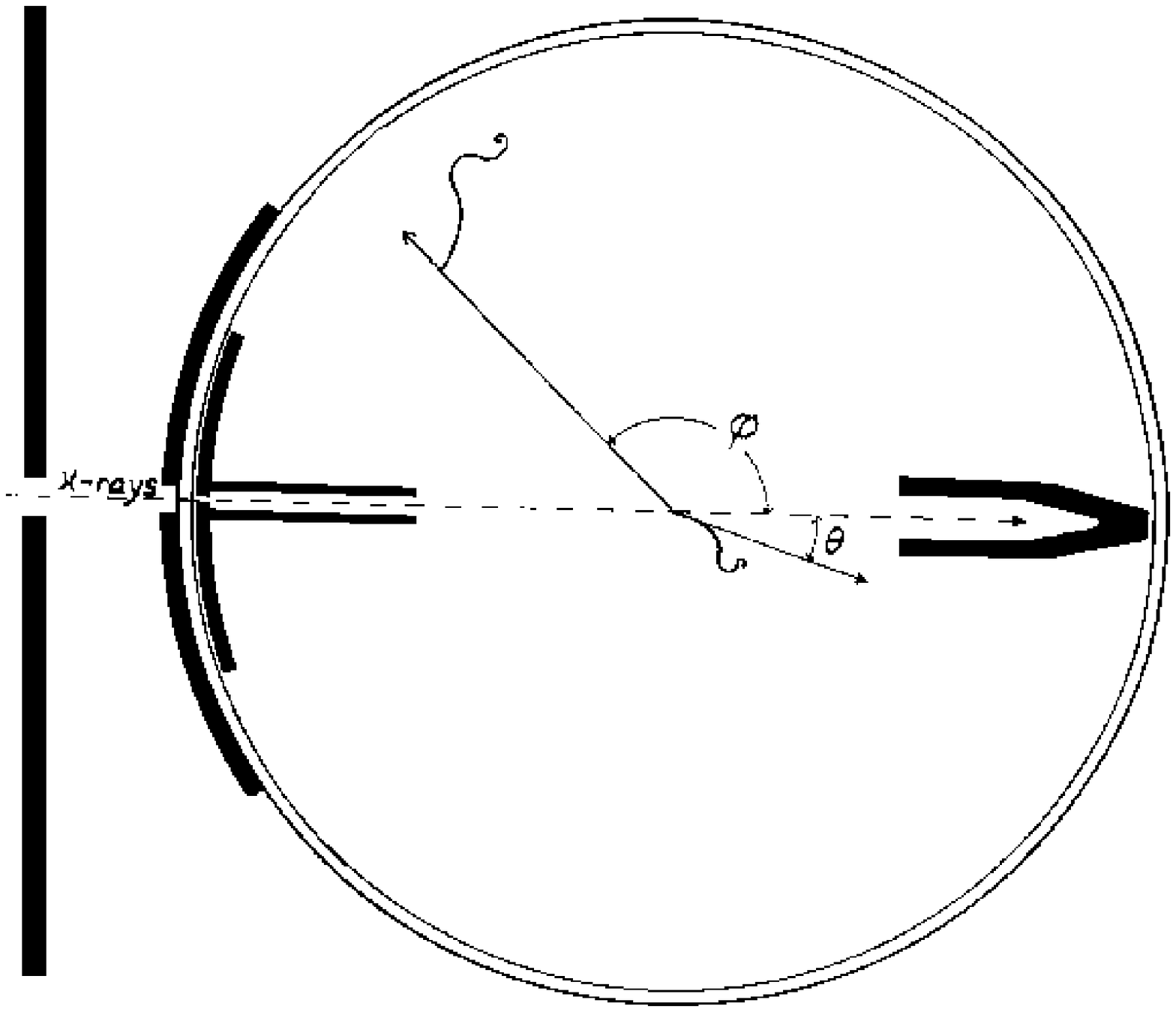}}
    \unnumberedcaption{\small Fig.~3. If the X-rays excite a recoil electron at an angle $\theta$, the photon theory 
    predicts a secondary $\beta$-particle at an angle $\varphi$.}  
  \end{figure}
Each recoil electron produces a visible track, and occasionally a secondary track is produced by 
the scattered X-ray. When but one recoil electron appears on the same plate with the track due to the scattered rays, it 
is possible to tell at once whether the angles satisfy equation (\ref{compton4}). If two or three recoil tracks 
appear,\endnote{The French edition includes also `en m\^{e}me temps'.} the measurements on each track can be 
approximately\endnote{[d'une fa\c{c}on appropri\'{e}e]} weighted.

Out of 850 plates taken in the final series of readings, thirty-eight show both recoil tracks and secondary $\beta$-ray 
tracks. On eighteen of these plates the observed angle $\varphi$\endnote{The French edition has `$\theta$'.} is within 20 
degrees of the angle calculated from the measured value of $\theta$, while the other twenty tracks are distributed at 
random angles. This ratio 18:20 is about that to be expected for the ratio of the rays scattered by the part of the air 
from which the recoil tracks could be measured to the stray rays from various sources. There is only about 1 chance in 
250 that so many secondary $\beta$-rays should have appeared at the theoretical angle.

If this experiment is reliable, it means that there is scattered X-ray energy associated with each recoil electron 
sufficient to produce a $\beta$-ray, and proceeding in a direction determined at the moment of ejection of the recoil 
electron. In other words, the scattered X-rays proceed in photons, that 
is\endnote{The words `photons, c'est-\`{a}-dire' are present only in the French edition.} in directed quanta of radiant 
energy.

This result, like that of Bothe and Geiger, is irreconcilable with Bohr, Kramers and Slater's hypothesis of the 
statistical production of recoil and photoelectrons. On the other hand, both of these experiments are in complete 
accord with the predictions of the photon theory.

\

\begin{center}
\par
\Needspace{5\baselineskip}
{\sc Reliability of experimental evidence}\addcontentsline{toc}{section}{Reliability of experimental 
evidence}\end{center}
While all of the experiments that we have considered are difficult to reconcile with the classical theory that radiation 
consists of electromagnetic waves, only those dealing with the individual scattering 
process\endnote{[au ph\'{e}nom\`{e}ne de la diffusion par les \'{e}lectrons individuels]} afford crucial tests between the 
photon theory and the statistical theory of virtual radiation. It becomes of especial importance, therefore, to consider 
the errors to which these experiments are subject.

When two point counters are set side by side, it is very easy to obtain coincidences from extraneous sources. Thus, for 
example, the apparatus must be electrically shielded so perfectly that a spark on the high-tension outfit that operates 
the X-ray tube may not produce coincident impulses in the two counters. Then there are high-speed $\alpha$- and 
$\beta$-rays, due to radium emanation in the air and other radioactive impurities, which may pass through both chambers 
and produce spurious coincidences. The method which Bothe and Geiger used to detect the coincidences, 
of\endnote{[ou]} recording on a photographic film the time of each pulse, makes it possible to estimate 
reliably\endnote{[avec certitude]} the probability that the  coincidences are due to chance. Moreover, it is possible by 
auxiliary tests to determine whether spurious coincidences are occurring --- for example, by operating the outfit as 
usual, except that the X-rays are absorbed by a sheet of lead. It is especially worthy of note that in the fluorescence 
experiment the photon theory predicted absence of coincidences, while in the scattering experiment it predicted their 
presence. It is thus difficult to see how both of these counter experiments can have been seriously affected by systematic 
errors.

In the cloud expansion experiment the effect of stray radiation is to hide the effect sought for, rather than to introduce 
a spurious effect. It is possible that due to radioactive contamination and to stray scattered X-rays $\beta$-particles 
may appear in different parts of the chamber, but it will be only a matter of chance if these $\beta$-particles appear in 
the position predicted from the direction of ejection of the recoil electrons. It was in fact only by taking great care to 
reduce such stray radiations to a minimum that the directional relations were clearly observed in the photographs. It 
would seem that the only form of consistent error that could vitiate the result of this experiment would be the 
psychological\endnote{[physiologique]} one of misjudging the angles at which the $\beta$-particles appear. It 
hardly seems possible, however, that errors in the measurement of these angles could be large enough to account for the 
strong apparent tendency for the angles to fit with the theoretical formula.

It is perhaps worth mentioning further that the initial publications of the two experiments on the individual scattering 
process were made simultaneously, which means that both sets of experimenters had independently reached a conclusion 
opposed to the statistical theory of the production of the $\beta$-rays.

Nevertheless,\endnote{This sentence is only printed in the French edition.} given the difficulty of the experiments and 
the importance of the conclusions to which they have led, it is highly desirable that both experiments should be repeated 
by physicists from  other laboratories.

\

\begin{center}
\par
\Needspace{5\baselineskip}
{\sc Summary}\addcontentsline{toc}{section}{Summary}\end{center}
The classical theory that radiation consists of electromagnetic waves propagated in all directions through 
space\endnote{The English edition omits reference to the ether and continues directly with `affords no adequate picture'.} 
is intimately connected to the idea of the ether, which is difficult to conceive. It affords no adequate picture of 
the manner in which radiation is emitted or absorbed. It is inconsistent with the experiments on the photoelectric 
effect, and is entirely helpless to account for the change of wavelength of scattered radiation or the production of 
recoil electrons.

The theory of virtual oscillators and virtual radiation which are associated statistically with sudden jumps of atomic 
energy and the emission of photoelectrons and recoil electrons, does not seem to be inconsistent with any of these 
phenomena as viewed on a macroscopic scale. This theory, 
however,\endnote{The English edition continues directly with: `seems powerless'.} retains the difficulties 
inherent in the conception of the ether and seems powerless to predict the characteristics of the photoelectrons and the 
recoil electrons. It\endnote{The English edition continues: `is also contrary to'.} is further difficult to 
reconcile with the composite photoelectric effect and is also contrary to Bothe's and Bothe and Geiger's coincidence 
experiments and to the ray track experiments relating the directions of ejection of a recoil electron and of emission 
of the associated scattered X-ray.

The photon theory avoids the difficulties associated with the conception of the ether.\endnote{In the English edition this
reads simply: `According to the photon theory, the production .... '.} The production and absorption of radiation 
is very simply connected with the modern idea of stationary states. It supplies a straightforward explanation of the major 
characteristics of the photoelectric effect, and it accounts in the simplest possible manner for the change of wavelength 
accompanying scattering and the existence of recoil electrons. Moreover, it predicts accurately the results of the 
experiments with individual radiation quanta, where the statistical theory fails.

Unless the four\endnote{In the English edition: `three'.} experiments on the individual events\endnote{[processus]} are 
subject to improbably large experimental errors, the conclusion is, I believe, unescapable that radiation consists of 
directed quanta of energy, i.e., of photons, and that energy and momentum are conserved when these photons interact with 
electrons or atoms.

Let me say again that this result does not mean that there is no truth in the concept of waves of radiation. The 
conclusion is rather that energy is not transmitted by such waves. The power of the wave concept in problems of 
interference, refraction, etc., is too well known to require emphasis. Whether the waves serve to guide the photons,\label{Comp67} 
or whether there is some other relation between photons and waves is another and a difficult question.\\


\newpage

\section*{Discussion of Mr Compton's report}\markboth{{\it A.~H.~Compton}}{{\it Discussion}}
\addcontentsline{toc}{section}{Discussion of Mr Compton's report}

{\sc Mr Lorentz.}~---~I would like to make two comments. First on the question
of the ether. Mr Compton considers it an advantage of the photon theory that it
allows us to do without the hypothesis of an ether which leads to great
difficulties. I must say that these difficulties do not seem so great to me and
that in my opinion the theory of relativity does not necessarily rule out the
concept of a universal medium. Indeed, Maxwell's 
equations are compatible with relativity, and one can well imagine
a medium for which these equations hold. One can even,
as Maxwell and other physicists have done with some success, construct a mechanical 
model of such a medium. One would have
to add only the hypothesis of the permeability of ponderable matter by the ether 
to have all that is required. Of course, in making these remarks, I should not wish to return in any way 
to these mechanical models, from which physics has
turned away for good reasons. One can be satisfied with the concept of a medium
that can pass freely through matter and to which Maxwell's equations can be
applied.

In the second place: it is quite certain that, in the phenomena of light, 
there must yet be something other than the photons. For instqncem in a diffraction experiment 
performed with very weak light, it can happen that the number of photons 
present at a given instant between the diffracting screen and the plane
on which one observes the distribution of light, is very limited. The average
number can even be smaller than one, which means that there are instants when
no photon is present in the space under consideration.

This clearly shows that the diffraction phenomena cannot
be produced by some novel action among the photons.
There must be something that guides them in their progress
and it is natural to seek this something in the electromagnetic field as 
determined by  the
classical theory. This notion of electromagnetic field, with its waves and 
vibrations would bring us back, in Mr Compton's view, to the notion of ether.\\

{\sc Mr Compton.}~---~It seems, indeed, difficult to avoid
the idea of waves in the discussion of optical phenomena. According to Maxwell's
theory the electric and magnetic properties of space lead to the idea of waves
as directly as did the elastic ether imagined by Fresnel. Why the space having
such magnetic properties should bear the name of ether is perhaps simply a 
matter of words. The fact that these properties of space immediately lead to the
wave equation with velocity $c$ is a much more solid basis for the hypothesis of
the existence of waves than the old elastic ether. That {\em something} ($E$ and
$H$) propagates like a wave with velocity $c$ seems evident. However, experiments 
of the kind we have just discussed 
show, if they are correct, that the {\em energy} of the bundle
of X-rays propagates in the form of {\em particles} and not in the form of
extended waves. So then, not even the 
electromagnetic ether appears to be satisfactory. \\

{\sc Mr Bragg.}~---~\label{BraggContribution}In his report Mr Compton has 
discussed the average momentum component of the electrons
in the direction of motion of the photon, and he has informed us of  the conclusion, 
at which several experimenters have arrived, that this forward average component is equal to
the momentum of the light quantum whose energy has been absorbed and is found
again in that of the photoelectron.

I would like to report in this
connection some results obtained by Mr Williams.\footnote{Cf.\ the relevant footnote on 
p.~\pageref{Since} ({\em eds.}).} 
Monochromatic X-rays, with wavelength lying between $0.5$ \AA\ and $0.7$ \AA,
enter a Wilson cloud chamber containing oxygen or nitrogen. 
The trajectories of the photoelectrons are observed
through a stereoscope and their initial directions are measured. Since the 
speed of the photoelectrons is exactly  known (the ionisation energy being
weak by comparison to the quantity $h\nu$), a measurement
of the initial direction is equivalent to a measurement of momentum in the
forward direction. Williams finds that the average momentum component in this
direction is in all cases markedly larger than the quantity $\frac{h\nu}{c}$ or 
$\frac{h}{\lambda}$. These results can be
summarised by a comparison with the scheme proposed by Perrin and Auger
([P.~Auger and F.~Perrin], {\em Journ.\ d.\ Phys.} [6th series, vol.~{\bf 8}] (February 1927), [93]). 
They are in perfect agreement with
the $\cos^2\theta$ law, provided one assumes that the magnetic impulse $T_m$ is equal to $1.8\frac{h\nu}{c}$ 
and not just $\frac{h\nu}{c}$ as these
authors assume. One should not attach any particular importance to this number 
$1.8$, because the range of the
examined wavelengths is too small. I mention it only to show that it is 
possible that the simple law proposed by Mr Compton might not be exact.

I would like to point out that this method of measuring the forward component of the
momentum is more precise than an attempt made to establish results about the most probable 
direction of emission.\\

{\sc Mr Wilson} says that his own observations, discussed in his Memoir of
1923\footnote{Referenced in footnote on p.~\pageref{Wilson} ({\em eds.}).} (but which 
do not pretend to be very precise) seem to show that in fact
the forward momentum component of the photoelectrons is, on average,
much larger than what one would derive from the idea that the absorbed quantum
yields all of its momentum to the expelled electron.  \\

{\sc Mr Richardson.}~---~When they are expelled by certain X-rays, 
the electrons have a momentum in the direction of propagation of the rays equal
to $1.8\frac{h\nu}{c}$. If I have understood Mr Bragg correctly,
this result is not the effect of some specific elementary process [action], but the 
average result for a great number of observations in which the
electrons were expelled in different directions. Whether or not the laws of
energy and momentum conservation apply to an elementary process, it 
is certain that they apply to the average result for a
great number of these processes. Therefore, the process [processus]
we are talking about must be governed by the equations for momentum and
energy. If for simplicity we ignore the refinements 
introduced by relativity, these equations are
  \[
    \frac{h\nu}{c}=m\overline{v}+M\overline{V}
  \]
and
  \[
    h\nu=\frac{1}{2}m\overline{v}^2+\frac{1}{2}M\overline{V}^2\ ,
  \]
where $m$ and $M$ are the masses, $v$ and $V$ the velocities of the electrons
and of the positive residue; the overbars express
that these are averages. The experiments show that the average value of $mv$ is
$1.8\frac{h\nu}{c}$ and not $\frac{h\nu}{c}$.
This means that $M\overline{V}$ is not zero, so that we cannot ignore this term
in the equation. If we consider, for instance, the photoelectric effect on
a hydrogen atom, we have to take the collision energy  of the hydrogen nucleus 
into account in the energy equation.\\

{\sc Mr Lorentz.}~---~The term $\frac{1}{2}M\overline{V}^2$ will 
however be much smaller than $\frac{1}{2}m\overline{v}^2$?\endnote{Overbars 
have been added.}\\

{\sc Mr Richardson.}~---~It is approximately its 1850th
part: that cannot always be considered negligible.\\

{\sc Mr Born} thinks that he is speaking also for several other members 
in asking Mr Compton to explain why one should expect that
the momentum imparted to the electron be equal to $\frac{h\nu}{c}$. \\

{\sc Mr Compton.}~---~When radiation of energy $h\nu$ is absorbed by an atom ---
which one surely has to assume in order to account for the kinetic energy
of the photoelectron --- the momentum imparted to the atom by this radiation is
$\frac{h\nu}{c}$. According to 
the classical electron theory, when an atom composed of a negative charge $-e$
of mass $m$ and a positive charge $+e$ of mass $M$ absorbs energy from an
electromagnetic wave, the momenta imparted to the two elementary charges 
[\'{e}lectrons] are inversely
proportional to their masses. This depends on the fact that the forward momentum is 
due to the magnetic vector, which acts with a force proportional to the velocity and 
consequently more strongly on the charge having the smaller mass.\endnote{The French 
text reads `avec moins d'intensit\'{e} sur la charge ayant la plus petite masse'.
This is evidently an error: the (transverse) velocity of the charges stems from the 
electric field, which imparts the larger velocity to the charge with the smaller mass, 
which therefore experiences the larger magnetic force.}
Effectively, the momentum is thus received by the charge
with the smaller mass. \\

{\sc Mr Debye.}~---~Is the reason why you think that the rest of the atom does
not receive any of the forward momentum purely theoretical? \\

{\sc Mr Compton.}~---~The photographs of the trajectories of the 
photoelectrons show, in accordance with Auger's prediction, that the
forward component of the momentum of the photoelectron is, on average,
the same as that of the photon. That means, clearly, that on average the
rest of the atom does not receive any momentum. \\

{\sc Mr Dirac.}~---~I have examined 
the motion of an electron placed in an arbitrary force field according 
to the classical theory, when it is subject to incident radiation, and
I have shown in a completely general way that at every instant the fraction
of the rate of change [vitesse de variation]
of the forward momentum of the electron due to the incident radiation is
equal to $\frac{1}{c}$ times the fraction of the rate of
change of the energy due to the incident radiation. The nucleus and the
other electrons of the atom produce changes of momentum and of energy that
at each instant are simply added to those produced by the incident radiation. Since the radiation
must modify the electron's orbit, it must also change the fraction of the
rate of change of the momentum and of the energy that comes from the nucleus
and the other electrons, so that it would be necessary to integrate the
motion in order to determine the total change produced by the incident
radiation in the energy and the momentum.\\

{\sc Mr Born.}~---~I would like to mention here a paper by Wentzel,\footnote{Born is presumably 
referring to Wentzel's second paper on the photoelectric effect (Wentzel 1927). Compare Mehra and 
Rechenberg (1987, pp.~835~ff.) ({\em eds.}).} which contains a rigorous treatment 
of the scattering of light by atoms according to 
quantum mechanics. In it, the author considers also the influence of the magnetic force, which 
allows him to obtain the quantum analogue of the classical light pressure. It is only in the limiting
case of very short wavelengths that one finds that the momentum of light $\frac{h\nu}{c}$  
is completely transmitted to the
electron; in the case of large wavelengths an influence of the binding forces appears. \\

{\sc Mr Ehrenfest.}~---~One can show by a very simple example
where the surplus of forward momentum, which we have
just discussed, can have its origin. Take a box whose inner walls reflect light
completely, but diffusedly, and assume that on
the bottom there is a little hole. Through the latter I shine a ray of light into the box 
which comes and goes inside the box and pushes away  its lid and bottom. The lid then has
a surplus of forward momentum.  \\

{\sc Mr Bohr.}\label{forpage170}\footnote{This discussion contribution by Bohr is reprinted and translated
also in vol.~5 of Bohr's {\em Collected Works} (Bohr 1984, pp.~207--12). ({\em eds.}).} --- With regard to 
the question of waves or photons discussed by 
Mr Compton, I would like to make a few remarks, without pre-empting the general discussion. 
The radiation experiments have indeed revealed features that are not easy to reconcile within a classical
picture. This difficulty arises particularly in the Compton effect itself. Several aspects
of this phenomenon can be described very simply with the aid of photons, but
we must not forget that the change of frequency that takes place is measured using
instruments whose functioning is interpreted
according to the wave theory. There seems to be a logical contradiction here, since
the description of the incident wave as well as that of the scattered
wave require that these waves be finitely extended [limit\'{e}es] 
in space and time, while the change in energy
and in momentum of the electron is considered as an instantaneous phenomenon at
a given point in spacetime. It is precisely because of such
difficulties that Messrs Kramers, Slater and myself were led to think that one
should completely reject the idea of the existence of photons and assume that the laws of conservation
of energy and momentum are true only in a statistical way.

The well-known experiments by Geiger and Bothe and by Compton and
Simon, however, have shown that this point of view
is not admissible and that
the conservation laws are valid for the individual processes, in accordance
with the concept of photons.
But the dilemma before which we are placed regarding the nature of light is only a typical example of the
difficulties that one encounters when one wishes to interpret the atomic
phenomena using classical concepts. The logical difficulties with a description
in space and time have since been removed in large
part by the fact that it has been realised that one encounters a similar paradox
with respect to the nature of material particles.
According  to the fundamental ideas of Mr
de Broglie, which have found such perfect
confirmation in the experiments of Davisson and Germer, the concept of
waves is as indispensable in the interpretation of the properties of
material particles as in the case of light. We know thereby that it is 
equally necessary to attribute to the wave field a finite extension
in space and in time, if one wishes to define the energy and the momentum
of the electron, just as one has to assume a similar finite extension in the case 
of the light quantum in order to be able to talk about frequency and wavelength.

Therefore, in the case of the scattering process, in order to
describe the two changes affecting the electron and the light we must
work with four wave fields (two for the electron, before and after
the phenomenon, and two for the quantum of light, incident and
scattered), finite in extension, which meet 
in the same region of spacetime.\footnote{Compare also the discussion 
contributions below, by Pauli, Schr\"{o}dinger and others ({\em eds.}).} 
In such a representation all possibility 
of incompatibility with a description in space and time disappears. I
hope the general discussion will give me the opportunity to enter more
deeply into the details of this question, which is intimately tied to
the general problem of quantum theory.\\

{\sc Mr Brillouin.}~---~I have had the opportunity to discuss Mr Compton's
report with Mr Auger,\footnote{As noted in section~\ref{details}, this and other reports had been circulated among the 
participants before the conference ({\em eds.}).} and wish to make a few comments on this topic. A
purely corpuscular description of radiation is not sufficient to understand the
peculiarities of the phenomena; to assume that energy is transported by photons $h\nu$ is
not enough to account for all the effects of radiation.
It is essential to complete our information
by giving the direction of the electric field; we cannot do without this field, whose role 
in the wave description is well known.

I shall recall in this context a simple argument, recently given by Auger
and F.~Perrin, and which illustrates clearly this remark. Let us
consider the emission of electrons by an atom subject to radiation, and let us examine the distribution of the 
directions of emission. This distribution has usually been observed in a
plane containing the light ray and the direction of
the electric field (the incident radiation is assumed to be polarised); let
$\varphi$ be the angle formed by the direction of emission of the 
photoelectrons and the electric field $h$; as long as the incident radiation is not
too hard, the distribution of the photoelectrons is symmetric around the
electric field; one can then show that the probability law necessarily takes
the form $A\cos^2\varphi$. Indeed, instead
of observing the distribution in the plane of incidence (Fig.~1),
  \begin{figure}
    \centering
      \resizebox{\textwidth}{!}{\includegraphics[0mm,0mm][220.58mm,70.79mm]{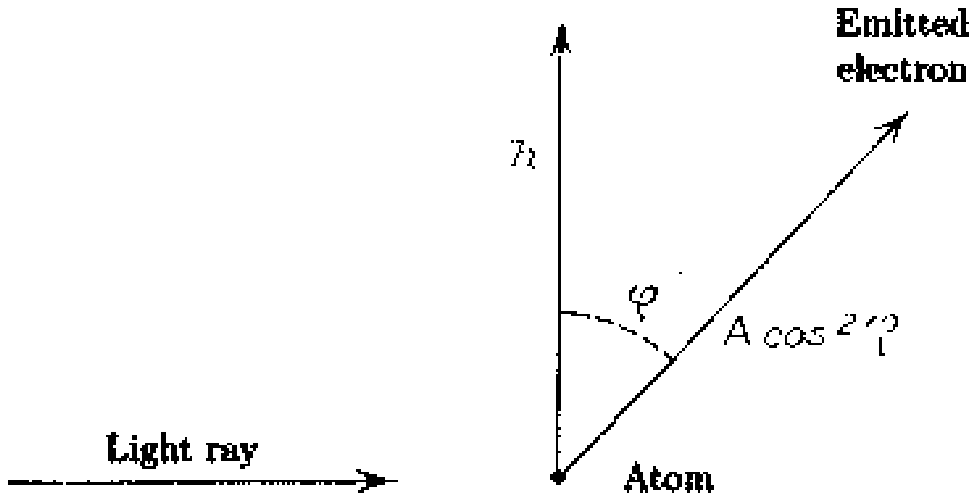}}
    \unnumberedcaption{\small Fig.~1.}
  \end{figure} 
let us examine it in the plane of the wave; the same distribution law will still be valid; 
and it is the only one that would allow us to obtain, through the superposition of two
waves polarised at right angles, an entirely symmetric distribution
  \[
    A\cos^2\varphi + A\sin^2\varphi = A\ .
  \]
Now, from the point of view of waves, one must necessarily obtain
this result, a beam [rayonnement] of natural light having no
privileged direction in the plane of the wave. These symmetry 
considerations, which any theory of radiation must respect, provide a substantial 
difficulty for the structural theories of the
photon (Bubb's quantum vector, for instance).

Summing up, the discontinuity of the radiation manifests itself just in the most
elementary way, through the laws of conservation of energy and momentum, but
the detailed analysis of the phenomena is interpreted more naturally from
the continuous point of view. For the problem of emission of the 
photoelectrons, a complete theory has been given by Wentzel, by means of
wave mechanics.\footnote{This is presumably Wentzel's treatment from his first paper on the 
photoelectric effect (Wentzel 1926). Compare again Mehra and 
Rechenberg (1987, pp.~835~ff.) ({\em eds.}).} He finds 
the $A\cos^2\varphi$ law of F.~Perrin and Auger for radiation of low penetration; when the 
radiation is harder, Wentzel obtains a more
complex law, in which the electrons tend to be emitted in larger numbers in
the forward direction. His theory, however, seems incomplete with regard to this point since, if
I am not mistaken, he has assumed the
immobility of the atomic nucleus; now, nothing tells us {\em a priori} how
the momentum $\frac{h\nu}{c}$ of the photon is going to be 
distributed between the nucleus and the emitted electron.\\

{\sc Mr Lorentz.}~---~Allow me to point out that
according to the old electron theory, when one has a nucleus and
an electron on which a beam of polarised light falls, the initial angular
momentum of the system is always conserved. The angular momentum
imparted to the electron-nucleus system will be provided at the expense
of the angular momentum of the radiation field.\\

{\sc Mr Compton.}~---~The conception of the photon differs from the classical 
theory in that, when a photoelectron is emitted,
the photon is completely absorbed and no radiation field is left. The
motion of the photoelectron must thus be such that the final angular
momentum of the electron-nucleus system will be the same as the
initial momentum of the photon-electron-nucleus system.
This condition restricts the possible trajectories of the emitted photoelectron.\\

{\sc Mr Kramers.}~---~In order to interpret his experiments, Mr Compton needs
to know how the absorption $\mu$ is divided between a 
component $\tau $, due to the `true' absorption, and a
component $\sigma$ due to the scattering. We do not know with certainty that, if $\mu$
can be written in the form $C\lambda^{a}+D$, the constant $D$ truly represents
the scattering for large wavelengths,
where $C\lambda^{a}$ is no longer small compared to $D$. In general, thus,
specific measurements of $\sigma$ are necessary.
Did you have sufficient information regarding the values of $\sigma$ and $\tau$
in your experiments? \\

{\sc Mr Compton.}~---~The most important case in which it is necessary to
distinguish between the true absorption $\tau$ and the absorption $\sigma$
due to the scattering, is that of carbon. For this case, Hewlett\footnote{
[C.~W.~Hewlett, {\em Phys.\ Rev.}, {\bf 17} (1921), 284.]
} has measured $\sigma$
directly for the wavelength $0.71$ \AA\ and the total
absorption $\mu$ over a large range of wavelengths. The difference
between $\mu$ and $\sigma$ for the wavelength $0.71$ \AA\ corresponds to 
$\tau$ for this wavelength.
According to Owen's formula this $\tau$ is proportional to $\lambda^3$;
we can thus calculate $\tau$ for all wavelengths. The difference between
this value of $\tau$ and the measured value of $\mu$ corresponds to
the value of $\sigma$ for the wavelengths considered. Since  $\tau$ is relatively small in the case
of carbon, especially for small wavelengths, this procedure  yields
a value for $\sigma$ that cannot be very imprecise.\\

{\sc Mr Bragg.}~---~When one consults the original literature on this subject, 
one is struck by how much the X-ray absorption measurements
leave to be desired, both with regard to precision
as well as with regard to the extent of the scale of wavelengths for which they
have been performed.\\

{\sc Mr Pauli.}~---~How large is the broadening of the
modified rays?\\

{\sc Mr Compton.}~---~The experiments have shown clearly that the modified 
ray is broader than the unmodified ray. In the typical case of the ray
$\lambda\ 0.7$ \AA\ scattered by carbon, 
the broadening is of
order $0.005$ angstr\"{o}m. Unfortunately, the experiments concerning 
this point are far from being satisfactory, and 
this number should be considered only as a rough approximation.\\

{\sc Mr Pauli.}~---~The broadening of the modified ray can be interpreted 
theoretically in two ways, which to tell the truth reduce to the same according to quantum mechanics. 
First, the electron, in a given stationary state of the atom, has a certain velocity
distribution with regard to magnitude and direction. That gives rise to a broadening
of the frequency of the scattered rays through the Doppler effect, a broadening
whose order of magnitude is $\frac{\Delta\lambda}{\lambda}=\frac{v}{c}$,
where $v$ denotes the average velocity of the electron in the atom.

In order to convey\label{Pauli-Compton} the second means of explanation, I would like to sketch briefly 
the meaning of the Compton effect in wave mechanics.\footnote{For 
a modern discussion, see Bj\"{o}rken and Drell (1964, Chapter 9) ({\em eds.}).} 
This meaning is based first of all on the wave equation
  \[
    \sum_\alpha \frac{\partial^2\psi}{\partial x_\alpha^2}
   -\frac{4\pi i}{h}\frac{e}{c}\sum_\alpha \varphi_\alpha \frac{\partial\psi}{\partial x_\alpha}
   -\frac{4\pi^2}{h^2}\left(\frac{e^2}{c^2}\sum_\alpha \varphi_\alpha^2 + m_0^2c^2\right)\psi
   =0
  \]
and further on the expression
  \[
    iS_\alpha=\psi\frac{\partial\psi^*}{\partial x_\alpha}-\psi^*\frac{\partial\psi}{\partial x_\alpha}
                       +\frac{4\pi i}{h}\frac{e}{c}\varphi_\alpha \psi\psi^*\ ,
  \]
in which $\psi$ is Schr\"{o}dinger's function, $\psi^*$ the complex 
conjugate value and $\varphi_\alpha$ the four-potential 
of the electromagnetic field. Given $S_\alpha$, one calculates the radiation from {\em classical}
electrodynamics.\label{Pauli-classical} If now in the wave equation one replaces $\varphi_\alpha$ by the potential of
an incident plane wave, the terms that are proportional to the amplitude of this wave can be
considered infinitely small in the first order, and one can apply the approximation methods
of perturbation theory. This now is a point where one needs to be especially careful. It is 
all-important to know what one will take as the unperturbed field $\psi$, which must correspond 
to a solution of the wave equation for the free particle (corresponding to $\varphi_\alpha=0$).\endnote{The printed text 
reads `$\varphi_\alpha\omega$'.} One finds that in order to agree with the observations, it is necessary to 
take {\em two} infinitely extended monochromatic wave trains as being already present in
the unperturbed solution, of which one corresponds to the initial state, the other to
the final state of the Compton process. In my opinion this assumption, on which the theories 
of the Compton effect by Schr\"{o}dinger, Gordon and Klein are based, is unsatisfactory and this 
defect is corrected only by Dirac's quantum electrodynamics.\footnote{Compare below Schr\"{o}dinger's
contribution and the ensuing discussion ({\em eds.}).} But if one makes this assumption, the  
current distribution of the unperturbed solution corresponds to that of an infinitely
extended diffraction grating [un r\'{e}seau infiniment \'{e}tendu] that moves with a constant speed, and the
action of the radiation on this grating leads to a sharp modified ray.

If one considers a bound electron in an atom,
one has to replace one
component of the solution $\psi$ in the unperturbed charge and current distribution by the eigenfunction  
of the atom in the stationary state considered, and
the other component by a solution corresponding to the final state of the
Compton process (belonging to the continuous spectrum of the atom), which
at great distance from the atom behaves more or less as a
plane wave. One thus has a moving grating that first of all depends only on the finite extension 
of the atom and in the second place has
components no longer moving with the same speed at all. This gives rise to a
lack of sharpness of the shifted ray
of the scattered radiation.

But one can show that, from the point of view of quantum mechanics, 
this explanation for the lack of sharpness of the shifted ray is just another
form of the explanation given in the first instance and which relies on the 
different directions of the initial velocities of the electrons in the atom. 
For according to quantum mechanics if
  \[
    \psi=f(x,y,z)e^{2\pi ivt}
  \]
is the eigenfunction corresponding to a
given stationary state of the atom, the function\endnote{Brackets in the exponent added.}
  \[
    \varphi (p_x,p_y,p_z)=\int\!\int\!\int f(x,y,z)
    e^{-\frac{2\pi i}{\lambda}(p_x x + p_y y + p_z z)}dxdydz\ ,
  \]
which one obtains by decomposing $f$ in plane
waves according to Fourier can be interpreted in the sense that  
$|\varphi(p)|^2 dp_xdp_ydp_z$ denotes the probability that in the given
stationary state the components of the momentum of the electron lie between $p_x,p_y,p_z$ and $p_x+dp_x$, etc. Now,
if through the resulting velocity distribution
of the electrons in the atom one calculates the broadening of the shifted line
according to the first point of view, for light of sufficiently short wavelength with respect 
to which the electron can be considered free in the atom (and it is only under these conditions that the 
procedure is legitimate), one finds exactly the same result as with the other
method described.\footnote{Pauli was possibly the first to introduce the probability interpretation
of the wave function in momentum space, in a letter to Heisenberg of 19 October 1926 (Pauli, 1979, pp.~347--8). 
Cf.\ the footnote on p.~\pageref{Paulifootnote} ({\em eds}.).}\\

{\sc Mr Compton.}~---~Jauncey has calculated the broadening of the
modified ray using essentially the method that Mr Pauli has just
described. Jauncey assumed, however, that the
velocities of the electron are the ones given by Bohr's theory of
orbital motions. The broadening thus obtained is larger than that 
found experimentally.\\

{\sc Mrs Curie.}~---~In his very interesting report, Professor Compton
has dwelt on emphasising the reasons that
lead one to adopt the theory of a collision between a quantum and a free
electron. Along the same line of thought, I think it is useful to point out the following two views:

First, the existence of collision electrons seems to play a fundamental role in the
biological effects produced on living tissues by very high-frequency 
radiation, such as the most penetrating $\gamma$-rays emitted by 
radioelements. If one assumes that the biological effect may be attributed
to the ionisation produced in the cells subjected to radiation, this
effect cannot depend directly on the $\gamma$-rays, but is due to the emission of secondary $\beta$-rays
that accompanies the passage of the $\gamma$-rays through matter. Before
the discovery of the collision electrons,
only a single mechanism was known for the production of these secondary
rays, that consisting in the total absorption of a quantum of radiation
by the atom, with the emission of a photoelectron. The absorption coefficient
$\tau$ relating to this process varies with the wavelength $\lambda$ of
the primary $\gamma$-radiation, as well as with the density [$\rho$] of the absorbing
matter and the atomic number $N$ of the atoms composing it, according 
to the well-known relation of Bragg and Peirce
$\frac{\tau}{\rho}=\Lambda N^3\lambda^3$, where $\Lambda$
is a coefficient that has a constant value for frequencies higher than that of the K discontinuity. 
If this relation valid in the domain of X-rays can be applied to high-frequency $\gamma$-rays, the resulting value
of $\frac{\tau}{\rho}$ for the light elements is so weak that the emission of photoelectrons 
appears unable to explain the biological effects of radiation on the living tissues traversed.\footnote{It 
is true that several authors have recently contested the legitimacy of extending the absorption law of Bragg and 
Peirce to X-rays.}

The issue appears altogether different if one takes into consideration the emission of collision electrons
in these tissues, following Compton's theory. For a collimated primary beam of $\gamma$-rays, the 
fraction of electromagnetic energy converted into kinetic energy of the electrons
per unit mass of the absorbing matter is given by the coefficient
  \[
    \frac{\sigma_a}{\rho}=\frac{\alpha}{(1-2\alpha)^2}\frac{\sigma_0}{\rho}\ ,
  \]
where $\frac{\sigma_0}{\rho}$ is
the scattering coefficient per unit mass
valid for medium frequency X-rays, according to the theory of J.~J.~Thomson, and is close to $0.2$, 
while $\alpha$ is Compton's parameter $\alpha=\frac{h\nu}{mc^2}$ ($h$ Planck's constant, $\nu$
primary frequency, $m$ rest mass of the electron, $c$ speed of light).
Taking $\alpha=1.2$, a value suitable for an
important group of $\gamma$-rays (equivalent potential 610 kilovolts), one
finds $\frac{\sigma_a}{\rho}=0.02$, that is, 2 per cent of the primary energy is converted to
energy of the electron per unit
mass of absorbing matter, whence a possibility of
interpreting the observed biological effects. To this direct production
of collision electrons along the trajectory of the primary beam is added,
in an extended medium, a supplementary production, from the fact that to each of these electrons corresponds a 
scattered quantum, with a smaller value than the primary quantum, and that this scattered
quantum can in turn be subject to the Compton effect in the medium through
which it propagates, with production of a new collision electron and of an
even smaller quantum. This process, indefinitely repeatable and called the `multiple Compton effect' seems in fact 
to have been observed by certain authors.\footnote{[B.] Rajewsky, {\em Fortschritte
auf dem Gebiet der Roentgenstrahlung}, {\bf 35} (1926), 262.} Not only is the number of collision 
electrons thereby multiplied, but, further, the primary quantum, reduced by successive
collisions takes on values for which the absorption with
emission of photoelectrons becomes more and more probable.

These facts have an important repercussion on the technique of X-ray therapy. Certain authors 
had, in fact, denied the usefulness of producing very high-voltage apparatus providing X-rays
of very high frequency and very high penetrating power, whose use is
otherwise convenient owing to
the uniformity of irradiation they allow one to attain. If these rays 
had been devoid of efficacy, one would have had to give up on their use. Such is 
not the case if one adopts the point of view of the
Compton effect, and it is then legitimate to direct the technique towards the use of high voltages.

Another interesting point of view to examine is that of the emission
of $\beta$-rays by radioactive bodies. Professor Compton has pointed
out that among the $\beta$-rays of secondary origin, some could be 
collision electrons produced by the scattering of the primary $\gamma$-rays 
on the electrons contained in the matter they traverse.

It is in an effect of this type that Thibaud thinks one may
find the explanation for the appearance of the magnetic
spectra of the secondary $\gamma$-rays. These spectra are composed of lines
that may be attributed to groups of photoelectrons of the same speed,
each of which is emitted by absorption in a thin metallic envelope
of a group of homogeneous $\gamma$-rays emitted by a radioelement contained
in this envelope. Each line of photoelectric origin is accompanied by a band
beginning at the line itself and extending towards the
region of low velocities. Thibaud thinks that this band could be due 
to photoelectrons expelled from the screen by those $\gamma$-rays that, in this same screen,
had suffered the Compton effect with reduction of frequency.
This interpretation appears plausible; however, in order to prove it, it would be necessary to study the structure 
of the band and find in the same spectrum the
band that may be attributed to the collision electrons corresponding to the scattered
$\gamma$-rays.

An analogous problem arises regarding the emission of $\beta$-rays by radioactive bodies with negligible thickness, 
so as to eliminate, as far as possible, the secondary effects due to the supports
and envelopes. One then observes a magnetic spectrum that may be 
attributed to the radioelement alone and consisting either of a continuous
band, or of the superposition of a continuous spectrum and a line spectrum.
The latter has received a satisfactory interpretation in some recent papers 
(L.~Meitner, Ellis, Thibaud, etc.).

A line is due to a group of photoelectrons with the same speed expelled
from the levels of the radioactive atoms by a group of homogeneous $\gamma$-rays produced in their
nuclei. This effect is called `internal conversion', since one assumes that
the quantum emitted by an atomic nucleus is reabsorbed in the electron
cloud [enveloppe \'{e}lectronique] of the same atom. The great majority of observed lines find their explanation in this hypothesis.

The interpretation of the continuous spectrum appears to present more
difficulties. Some authors attribute it only to
the primary $\beta$-rays, while others consider the possibility of a
secondary origin and invoke the Compton effect as a possible cause 
of its production (L.~Meitner). This would be 
an `internal' Compton effect, such that a $\gamma$-ray emitted from the nucleus 
of an atom would experience a collision with one of the weakly bound
electrons at the periphery of the same atom. If that were the case,
the velocity distribution of the emitted collision electrons would not be
arbitrary, but would have to conform to the predictions of Compton's theory.

I have closely examined this problem, which has a very complex
appearance.\footnote{[M.]~Curie, {\em 
Le Journal de Physique et le Radium}, {\bf 7} (1926), 97.} 
Each group of homogeneous $\gamma$-rays is 
accompanied by scattered $\gamma$-rays, so that in the diffraction 
spectrum of the $\gamma$-rays, each line should experience a broadening
of $0.0485$ \AA\ units. The experiments on the diffraction of
$\gamma$-rays are difficult and not very numerous; so
far the broadening effect has not been reported.

Each homogeneous group of $\gamma$-rays must correspond to a group
of collision electrons, whose velocity varies continuously from zero
to an upper limit derived from Compton's theory and which in the 
magnetic spectrum corresponds to a band bounded
sharply on the side of the large velocities. The same group of $\gamma$-rays may
correspond to further groups of photoelectrons expelled from the different levels
K, L, etc.\ of the atom through internal absorption of the scattered $\gamma$-rays. 
For each group of photoelectrons, the velocity of emission lies 
between two well-defined limits. The upper
limit corresponds to the surplus energy of the primary $\gamma$-rays with 
respect to the extraction work $W$ characteristic of the given level; the lower 
limit corresponds to the surplus energy, with respect to the same work, of the 
$\gamma$-rays scattered in the direction opposite to that of the primary rays, and having
experienced because of that the highest loss of
frequency. In the magnetic spectrum, each group of photoelectrons will
be represented by a band equally well bounded on the side of the large
and of the small velocities, with the same difference 
between the extreme energies for each band.

It is easy to see that in the same magnetic spectrum the different bands
corresponding  to the same group of $\gamma$-rays 
may partially overlap, making it difficult to analyse the spectrum 
comparing the distribution of $\beta$-rays with that predicted by theory.
For substances emitting several groups
of $\gamma$-rays, the difficulty must become considerable,
unless there are large differences in their relative effectiveness in producing
the desired effect. Let us also point out that the continuous spectrum due to the Compton
effect may be superposed with a continuous spectrum independent of this
effect (that may be attributed for instance to the primary $\beta$-rays).

Examination of the experimental data available so far does not yet allow one 
to draw conclusions convincingly. Most of the spectra are 
very complex, and their precise study with respect to the energy 
distributions of the $\beta$-rays will require very detailed work. 
In certain simple spectra such as that of the $\beta$-rays of RaD, 
one observes lines of photoelectric origin that may be attributed to
a single group of monochromatic $\gamma$-rays. These lines form the upper edge
of bands extending towards low velocities
and probably arising from photoelectrons produced by the scattered 
$\gamma$-rays. In certain magnetic spectra obtained from the $\beta$-rays
of mesothorium 2 in the region of low velocities, one notices in the
continuous spectrum a gap that might correspond, for the group of primary
$\gamma$-rays with 58 kilovolts, to the separation between the band due to the collision 
electrons and that due to the
photoelectrons of the scattered $\gamma$-rays.\footnote{D.~K.~Yovanovitch and A.~Prola, 
{\em Comptes Rendus}, {\bf 183} (1926), 878.} \\

{\sc Mr Schr\"{o}dinger},\label{Schr-Compton} at the invitation of Mr Ehrenfest, draws on
the blackboard in coloured chalk the system of four wave trains by which he has 
tried  to represent the Compton effect in an anschaulich way [d'une fa\c{c}on intuitive]\footnote{For
discussions of the notion of Anschaulichkeit, see sections~\ref{foundmat}, \ref{Schr-conflict} and \ref{visualisability} 
({\em eds}.).} 
({\em Ann.\ d. Phys.} 4th series, vol.~{\bf 82} (1927), 257).\footnote{Schr\"{o}dinger 
(1928, p.~x) later remarked on a mistake pointed out to him by Ehrenfest in the figure as published in the original
paper ({\em eds}.).}\\

{\sc Mr Bohr.}\label{Bohr-disc2}~---~The simultaneous consideration of two systems of waves
has not the aim of giving a causal theory in the classical sense, but one
can show that it leads to a symbolic analogy. This has been studied 
in particular by Klein. Furthermore, it has been
possible to treat the problem in more depth through the way Dirac has formulated Schr\"{o}dinger's
theory. We find here an even more advanced renunciation of Anschaulichkeit [intuitivit\'{e}],
a fact very characteristic of the symbolic methods
in quantum theory.\\

{\sc Mr Lorentz.}~---~Mr Schr\"{o}dinger has shown how one can explain
the Compton effect in wave mechanics. In this explanation one considers
the waves associated with the electron ($e$) and the photon ($ph$), before
(1) and after (2) the encounter. It is natural to think that, of these four 
systems of waves $e_1$, $ph_1$, $e_2$ and $ph_2$, the latter two are produced
by the encounter. But they are not {\em determined} by $e_1$ and $ph_1$,
because one can for example choose arbitrarily the direction of $e_2$.
Thus, for the problem to be well-defined, it is not sufficient to know
$e_1$ and $ph_1$; another piece of data is necessary, just as in the case of the collision of two elastic balls
one must know not only their initial velocities but also a parameter
that determines the greater or lesser eccentricity of the collision,
for instance the angle between the relative velocity and the common 
normal at the moment of the encounter. Perhaps one could
introduce into the explanation given by Mr Schr\"{o}dinger something
that would play the role of this accessory parameter.\\

{\sc Mr Born.}~---~I think it is easy to understand why three 
of the four waves have to be given in order for the process to be determined; it suffices
to consider analogous circumstances in the classical theory. If the motions
of the two particles approaching each other are given, the effect of the 
collision is not yet determined; it can be made determinate 
by giving the position of closest approach or an equivalent piece of data.
But in wave mechanics such microscopic data are not available. That is why it
is necessary to prescribe the motion of one of the particles after the collision,
if one wants the motion of the second particle after the
collision to be determined. But there is nothing surprising in this, everything being exactly
as in classical mechanics. The only difference is that in the old theory one
introduces microscopic quantities, such as the radii of the atoms that
collide, which are eliminated from subsequent
calculations, while in the new theory one avoids the introduction of these 
quantities.\\


\newpage

\renewcommand{\enoteheading}{\section*{Notes to the translation}}
\addcontentsline{toc}{section}{\it Notes to the translation}
\theendnotes

\setcounter{endnote}{0}
\setcounter{equation}{0}

\chapter*{The new dynamics of quanta$^{\scriptsize\hbox{a}}$}\markboth{{\it L.~de~Broglie}}{{\it The new dynamics of quanta}}
\addcontentsline{toc}{chapter}{The new dynamics of quanta ({\em L.~de~Broglie\/})}
\begin{center}{\sc By Mr Louis de BROGLIE}\footnotetext[1]{Our translation of the title (`La
nouvelle dynamique des quanta') reflects de Broglie's frequent use of the word
`quantum' to refer to a (pointlike) particle, an association that would be
lost if the title were translated as, for example, `The new quantum dynamics'
(\textit{eds}.).}
\end{center}

\

\begin{center}
\par
\Needspace{5\baselineskip}
{\sc I. --- Principal points of view}\addcontentsline{toc}{section}{I. --- 
Principal points of view}\footnote[2]{On beginning this
exposition, it seems right to underline that Mr Marcel Brillouin was the true
precursor of wave Mechanics, as one may realise by referring to the following
works: \textit{C. R.} \textbf{168} (1919), 1318; \textbf{169} (1919), 48;
\textbf{171} (1920), 1000. --- \textit{Journ. Physique} \textbf{3} (1922),
65.}
\end{center}
1. \textit{First works of Mr Louis de Broglie} [1]. --- In his first works on
the Dynamics of Quanta, the author of the present report started with the
following idea: taking the existence of elementary corpuscles of matter and
radiation as an experimental fact, these corpuscles are supposed to be endowed
with a periodicity. In this way of seeing things, one no longer conceives of
the `material point' as a static entity pertaining to only a tiny region of
space, but as the centre of a periodic phenomenon spread all around it.

Let us consider, then, a completely isolated material point and, in a system
of reference attached to this point, let us attribute to the postulated
periodic phenomenon the appearance of a stationary wave defined by the
function%
\[
u(x_{0},\ y_{0},\ z_{0},\ t_{0})=f(x_{0},\ y_{0},\ z_{0})\cos2\pi\nu_{0}%
t_{0}\ .
\]

In another Galilean system $x$, $y$, $z$, $t$, the material point will have a
rectilinear and uniform motion with velocity $v=\beta c$. Simple application
of the Lorentz transformation shows that, as far as the phase is concerned, in
the new system the periodic phenomenon has the appearance of a plane wave
propagating in the direction of motion whose frequency and phase velocity are%
\[
\nu=\frac{\nu_{0}}{\sqrt{1-\beta^{2}}}\ ,\ \ \ V=\frac{c^{2}}{v}=\frac
{c}{\beta}\ .
\]

The appearance of this phase propagation with a speed superior to $c$, as an
immediate consequence of the theory of Relativity, is quite striking.

There exists a noteworthy relation between $v$ and $V$. The formulas giving
$\nu$ and $V$ allow us in fact to define a refractive index of the vacuum, for
the waves of the material point of proper frequency $\nu_{0}$, by the
dispersion law%
\[
n=\frac{c}{V}=\sqrt{1-\frac{\nu_{0}^{2}}{\nu^{2}}}\ .
\]
One then easily shows that%
\[
\frac{1}{v}=\frac{1}{c}\frac{\partial(n\nu)}{\partial\nu}\ ,
\]
that is, that the velocity $v$ of the material point is equal to the group
velocity corresponding to the dispersion law.

With the free material point being thus defined by wave quantities, the
dynamical quantities must be related back to these. Now, since the frequency
$\nu$ transforms like an energy, the obvious thing to do is to assume the
quantum relation%
\[
W=h\nu\ ,
\]
a relation that is valid in all systems, and from which one derives the
undulatory definition of the proper mass $m_{0}$%
\[
m_{0}c^{2}=h\nu_{0}\ .
\]

Let us write the function representing the wave in the system $x,y,z,t$ in the form%
\[
u(x, y, z, t)=f(x, y, z, t)\cos\frac{2\pi}{h}\varphi(x, y, z, t)\ .
\]
Denoting by $W$ and $p$ the energy and momentum, one easily shows that one
has\footnote{These are the relativistic guidance equations of de Broglie's
early pilot-wave theory of 1923--24, for the special case of a free particle
(\textit{eds.}).}$^{,}$\footnote{The vector `$\overrightarrow{\mathrm{grad\ }%
\varphi}$' is the vector whose components are $\partial\varphi/\partial x$,
$\partial\varphi/\partial y$, $\partial\varphi/\partial z$.}%
\[
W=\frac{\partial\varphi}{\partial t}\ ,\ \ \ \overrightarrow{p}%
=-\overrightarrow{\mathrm{grad\ }\varphi}\ .
\]

The function $\varphi$ is then none other than the Jacobi
function.\footnote{Usually called the Hamilton-Jacobi function (\textit{eds.}%
).} One deduces from this that, in the case of uniform rectilinear motion, the
principles of least action and of Fermat are identical.\label{identical-principles}

To look for a generalisation of these results, let us now assume that the
material point moving in a field derived from a potential function
$F(x, y, z, t)$ is represented by the function%
\[
u(x, y, z, t)=f(x, y, z, t)\cos\frac{2\pi}{h}\varphi(x, y, z, t)\ ,
\]
where $\varphi$ is the Jacobi function of the old Dynamics. This assimilation
of the phase into the Jacobi function then leads us to assume the following
two relations, which establish a general link between mechanical quantities
and wave quantities:%
\[
W=h\nu=\frac{\partial\varphi}{\partial t}\ ,\ \ \ \overrightarrow{p}%
=\frac{h\nu}{V}=-\overrightarrow{\mathrm{grad\ }\varphi}\ .
\]

One then deduces that, for the waves of the new Mechanics, the space occupied
by the field has a refractive index\label{refractive-index}%
\[
n=\sqrt{\left(  1-\frac{F}{h\nu}\right)  ^{2}-\frac{\nu_{0}^{2}}{\nu^{2}}}\ .
\]
Hamilton's equations show in addition that, here again, the velocity of the
moving body is equal to the group velocity.\footnote{In the case of motion of
a point charge in a magnetic field, space behaves like an anisotropic medium
(see \textit{Thesis}, p.~39).}

These conceptions lead to an interpretation of the stability conditions
introduced by quantum theory. If, indeed, one considers a closed trajectory,
the phase must be a single-valued function along this curve, and as a result
one is led to write the Planck condition\endnote{The integral sign is printed
as `$\int_{0}$' in the original.}%
\[
\oint(p\cdot dl)=k\cdot h\ \ \ \ (k\ \mathrm{\operatorname{integer}})\ .
\]

The Sommerfeld conditions for quasi-periodic motions may also be derived. The
phenomena of quantum stability thus appear to be analogous to phenomena of
resonance, and the appearance of whole numbers here becomes as natural as in
the theory of vibrating strings or plates. Nevertheless, as we shall see, the
interpretation that has just been recalled still constitutes only a first approximation.

The application of the new conceptions to corpuscles of light leads to
difficulties if one considers their proper mass to be finite. One avoids these
difficulties by assuming that the properties of the corpuscles of light are
deduced from those of ordinary material points by letting the proper mass tend
to zero. The two speeds $v$ and $V$ then both tend to $c$, and in the limit
one obtains the two fundamental relations of the theory of light quanta%
\[
h\mathrm{\nu}=W\ ,\ \ \ \frac{h\mathrm{\nu}}{c}=p\ ,
\]
with the aid of which one can account for Doppler effects, radiation pressure,
the photoelectric effect and the Compton effect.

The new wave conception of Mechanics leads to a new statistical Mechanics,
which allows us to unify the kinetic theory of gases and the theory of
blackbody radiation into a single doctrine. This statistics coincides with
that proposed independently by Mr Bose [2]; Mr Einstein [3] has shown its
scope and clarified its significance. Since then, numerous papers [4] have
developed it in various directions.

Let us add a few remarks. First, the author of this report has always assumed
that the material point occupies a well-defined position in space.\label{always-trajectories} As a
result, the amplitude $f$ should contain a singularity or at the very least
have abnormally high values in a very small region. But, in fact, the form of
the amplitude plays no role in the results reviewed above. Only the phase
intervenes: hence the name \textit{phase waves} originally given to the waves
of the new Mechanics.

On the other hand, the author, after having reduced the old forms of Dynamics
to geometrical Optics, realised clearly that this was only a first stage. The
existence of diffraction phenomena appeared to him to require the construction
of a new Mechanics `which would be to the old Mechanics (including that of
Einstein) what wave Optics is to geometrical Optics'.\label{oldp86}\footnote{\textit{Revue
G\'{e}n\'{e}rale des Sciences}, 30 November 1924, p.~633.} It is Mr
Schr\"{o}dinger who has had the merit of definitively constructing the new doctrine.

\ 

\noindent 2. \textit{The work of Mr E. Schr\"{o}dinger} [5]. --- Mr Schr\"{o}dinger's
fundamental idea seems to have been the following: the new Mechanics must
begin from wave equations, these equations being constructed in such a way
that in each case the phase of their sinusoidal solutions should be a solution
of the Jacobi equation in the approximation of geometrical Optics.

Instead of considering waves whose amplitude contains a singularity, Mr
Schr\"{o}dinger systematically looks at waves of classical type, that is to
say, waves whose amplitude is a continuous function. For him, the waves of the
new Mechanics are therefore represented by functions $\Psi$ that one can
always write in the canonical form%
\[
\Psi=a\cos\frac{2\pi}{h}\varphi\ ,
\]
$a$ being a continuous function and $\varphi$ being \textit{in the first
approximation} a solution of the Jacobi equation. We may understand the words
`in the first approximation'\ in two different ways: first, if the conditions
that legitimate the use of geometrical Optics are realised, the phase
$\varphi$ will obey the equation called the equation of geometrical Optics,
and this equation will have to be identical to that of Jacobi; second, one
must equally recover the Jacobi equation if one makes Planck's constant tend
to zero, because we know in advance that the old Dynamics must then become valid.

Let us first consider the case of the motion of a single material point in a
static field derived from the potential function $F(x, y, z)$. In his first
Memoir Schr\"{o}dinger shows that the wave equation, at least in the
approximation of Newton's Mechanics, is in this case%
\[
\triangle\Psi+\frac{8\pi^{2}m_{0}}{h^{2}}(E-F)\Psi=0\ .
\]
It is also just this equation that one arrives at beginning from the
dispersion law noted in the first section.

Having obtained this equation, Mr Schr\"{o}dinger used it to study the
quantisation of motion at the atomic scale (hydrogen atom, Planck oscillator,
etc.). He made the following fundamental observation: in the problems
considered in micromechanics, the approximations of geometrical Optics are no
longer valid at all. As a result, the interpretation of the quantum conditions
proposed by L. de Broglie shows only that the Bohr-Sommerfeld formulas
correspond to the approximation of the old Dynamics. To resolve the problem of
quantisation rigorously, one must therefore consider the atom as the seat of
stationary waves satisfying certain conditions. Schr\"{o}dinger assumed, as is
very natural, that the wave functions must be finite, single-valued and
continuous over all space. These conditions define a set of fundamental
functions (Eigenfunktionen) for the amplitude, which represent the various
stable states of the atomic system being considered. The results obtained have
proven that this new quantisation method, to which Messrs L\'{e}on Brillouin,
G. Wentzel and Kramers [6] have made important contributions, is the correct one.

For Mr Schr\"{o}dinger, one must look at continuous waves, that is to say,
waves whose amplitude does not have any singularities. How can one then
represent the `material point'? Relying on the equality of the velocity of the
moving body and the group velocity, Schr\"{o}dinger sees the material point as
a group of waves (Wellenpaket\endnote{The French uses `Wellenpacket'
throughout.}) of closely neighbouring frequencies propagating in
directions contained within the interior of a very narrow cone. The material
point would then not be really pointlike; it would occupy a region of space
that would be at least of the order of magnitude of its wavelength. Since, in
intra-atomic phenomena, the domain where motion takes place has dimensions of
the order of the wavelengths, there the material point would no longer be
defined at all; for Mr Schr\"{o}dinger, the electron in the atom is in some
sense `smeared out' [`fondu'], and one can no longer speak of its position or
velocity. This manner of conceiving of material points seems to us to raise
many difficulties; if, for example, the quantum of ultraviolet light occupies
a volume whose dimensions are of the order of its wavelength, it is quite
difficult to conceive that this quantum could be absorbed by an atom of
dimensions a thousand times smaller.

Having established the wave equation for a material point in a static field,
Mr Schr\"{o}dinger then turned to the Dynamics of many-body systems [la
Dynamique des syst\`{e}mes]. Still limiting himself to the
Newtonian\footnote{That is, nonrelativistic (\textit{eds}.).} approximation,
and inspired by Hamilton's ideas, he arrived at the following statement: Given
an isolated system whose potential energy is $F(q_{1}, q_{2}, ..., q_{n})$,
the kinetic energy is a homogeneous quadratic form in the momenta $p_{k}$ and
one may write%
\[
2T=\sum_{kl}m^{kl}p_{k}p_{l}\ ,
\]
the $m^{kl}$ being functions of the $q$. If $m$ denotes the determinant
$\left\vert m^{kl}\right\vert $ and if $E$ is the constant of energy in the
classical sense, then according to Schr\"{o}dinger one must begin with the
wave equation%
\[
m^{+\frac{1}{2}}\sum_{kl}\frac{\partial}{\partial q_{k}}\left[  m^{-\frac
{1}{2}}m^{kl}\frac{\partial\Psi}{\partial q^{l}}\right]  +\frac{8\pi^{2}%
}{h^{2}}(E-F)\Psi=0\ ,
\]
which describes the propagation of a wave in the configuration space
constructed by means of the variables $q$. Setting%
\[
\Psi=a\cos\frac{2\pi}{h}\varphi\ ,
\]
and letting $h$ tend to zero, in the limit one indeed recovers the Jacobi
equation%
\[
\frac{1}{2}\sum_{kl}m^{kl}\frac{\partial\varphi}{\partial q^{k}}\frac
{\partial\varphi}{\partial q^{l}}+F=E\ .
\]
To quantise an atomic system, one will here again determine the fundamental
functions of the corresponding wave equation.

We cannot recall here the successes obtained by this method (papers by Messrs
Schr\"{o}dinger, Fues,\endnote{Mis-spelt as `Fuess'.} Manneback [7],
etc.), but we must insist on the difficulties of a conceptual type that it
raises. Indeed let us consider, for simplicity, a system of $N$ material
points each possessing three degrees of freedom. The configuration space is in
an essential way formed by means of the coordinates of the points and yet Mr
Schr\"{o}dinger assumes that in atomic systems material points no longer have
a clearly defined position. It seems a little paradoxical to construct a
configuration space with the coordinates of points that do not exist.\label{paradoxical}
Furthermore, if the propagation of a wave in space has a clear physical
meaning, it is not the same as the propagation of a wave in the abstract
configuration space, for which the number of dimensions is determined by the
number of degrees of freedom of the system. We shall therefore have to return
later to the exact meaning of the Schr\"{o}dinger equation for many-body systems.

By a transformation of admirable ingenuity, Mr Schr\"{o}dinger has shown that
the quantum Mechanics invented by Mr Heisenberg and developed by Messrs Born,
Jordan, Pauli, etc., can be translated into the language of wave Mechanics. By
comparison with Heisenberg's matrix elements, he was able to derive the
expression for the mean charge density of the atom from the functions $\Psi$,
an expression to which we shall return later.

The Schr\"{o}dinger equations are not relativistic. For the case of a single
material point, various authors [8] have given a more general wave equation
that is in accord with the principle of Relativity. Let $e$ be the electric
charge of the point, $\mathcal{V}$ and $\overrightarrow{A}$ the two
electromagnetic potentials. The equation that the wave $\Psi$, written in
complex form, must satisfy is\footnote{This is the complex, time-dependent
Klein-Gordon equation in an external electromagnetic field (\textit{eds}.).}%
\[
\square\Psi+\frac{4\pi i}{h}\frac{e}{c}\left[  \frac{\mathcal{V}}{c}%
\frac{\partial\Psi}{\partial t}+\sum_{xyz}A_{x}\frac{\partial\Psi}{\partial
x}\right]  -\frac{4\pi^{2}}{h^{2}}\left[  m_{0}^{2}c^{2}-\frac{e^{2}}{c^{2}%
}(\mathcal{V}^{2}-A^{2})\right]  \Psi=0\ .
\]
As Mr O. Klein [9] and then the author [10] have shown, the theory of the
Universe with five dimensions allows one to give the wave equation a more
elegant form in which the imaginary terms, whose presence is somewhat shocking
for the physicist, have disappeared.

We must also make a special mention of the beautiful Memoirs in which Mr De
Donder [11] has connected the formulas of wave Mechanics to his general theory
of Einsteinian Gravity.

\ 

\noindent 3. \textit{The ideas of Mr Born} [12]. --- Mr Born was struck by the fact that
the continuous wave functions $\Psi$ do not allow us to say \textit{where} the
particle whose motion one is studying is and, rejecting the concept of the
Wellenpaket, he considers the waves $\Psi$ as giving only a statistical
representation of the phenomena. Mr Born seems even to abandon the idea of the
determinism of individual physical phenomena: 
the Quantum Dynamics, he wrote in his letter to \textit{Nature}, 
`would then be a singular fusion of mechanics and statistics~....~. 
A knowledge of $\Psi$ enables us to follow the course of a physical 
process in so far as it is quantum mechanically determinate: not in 
a causal sense, but in a statistical one'.\endnote{We follow the original 
English, which is a translation by Oppenheimer, from Born (1927, 
p.~355). De Broglie translates 'mechanics' as `dynamique' and includes the 
words `La Dynamique des Quanta' in the quotation, where Born has `it' 
(referring to `quantum mechanics').}

These conceptions were developed in a mathematical form by their author, in
Memoirs of fundamental interest. Here, by way of example, is how he treats the
collision of an electron and an atom. He writes the Schr\"{o}dinger equation
for the electron-atom system, and he remarks that before the collision, the
wave $\Psi$ must be expressed by the product of the fundamental
function\footnote{That is, eigenfunction (\textit{eds.}).} representing the
initial state of the atom and the plane wave function corresponding to the
uniform rectilinear motion of the electron. During the collision, there is an
interaction between the electron and the atom, an interaction that appears in
the wave equation as the mutual potential energy term. Starting from the
initial form of $\Psi$, Mr Born derives by methods of successive approximation
its final form after the collision, in the case of an elastic collision, which
does not modify the internal state of the atom, as well as in the case of an
inelastic collision, where the atom passes from one stable state to another
taking energy from or yielding it to the electron. According to Mr Born, the
final form of $\Psi$ determines the probability that the collision may produce
this or that result.

The ideas of Mr Born seem to us to contain a great deal of truth, and the
considerations that shall now be developed show a great analogy with them.

\

\begin{center}
\par
\Needspace{5\baselineskip}
{\sc II. --- Probable meaning of the continuous waves $\Psi$ 
[13]}\addcontentsline{toc}{section}{II. --- Probable meaning of the continuous waves $\Psi$}
\end{center}
4. \textit{Case of a single material point in a static field}. --- The body of
experimental discoveries made over forty years seems to require the idea that
matter and radiation possess an atomic structure. Nevertheless, classical
optics has with immense success described the propagation of light by means of
the concept of continuous waves and, since the work of Mr Schr\"{o}dinger,
also in wave Mechanics one always considers continuous waves which, not
showing any singularities, do not allow us to define the material point. If
one does not wish to adopt the hypothesis of the `Wellenpaket', whose
development seems to raise difficulties, how can one reconcile the existence
of pointlike elements of energy with the success of theories that consider the
waves $\Psi$? What link must one establish between the corpuscles and the
waves? These are the chief questions that arise in the present state of wave Mechanics.

To try to answer this, let us begin by considering the case of a single
corpuscle carrying a charge $e$ and moving in an electromagnetic
field\footnote{Here we leave aside the case where there also exists a
gravitational field. Besides, the considerations that follow extend without
difficulty to that case.} defined by the potentials $\mathcal{V}$ and
$\overrightarrow{A}$. Let us suppose first that the motion is one for which
the old Mechanics (in relativistic form) is sufficient. If we write the wave
$\Psi$ in the canonical form%
\[
\Psi=a\cos\frac{2\pi}{h}\varphi\ ,
\]
the function $\varphi$ is then, as we have seen, the Jacobi function, and the
velocity of the corpuscle is defined by the formula of Einsteinian Dynamics%
\begin{equation}
\overrightarrow{v}=-c^{2}\frac{\overrightarrow{\mathrm{grad\ }\varphi}%
+\frac{e}{c}\overrightarrow{A}}{\frac{\partial\varphi}{\partial t}%
-e\mathcal{V}}\ . \tag{I}%
\end{equation}

We propose to assume by induction that this formula is still valid when the
old Mechanics is no longer sufficient, that is to say when $\varphi$ is no
longer a solution of the Jacobi equation.\footnote{Mr De Donder assumes
equation (I) as we do, but denoting by $\varphi$ not the phase of the wave,
but the classical Jacobi function. As a result his theory and ours diverge as
soon as one leaves the domain where the old relativistic Mechanics is
sufficient.} If one accepts this hypothesis, which appears justified by
its\endnote{The French reads `ces' [these] rather than `ses'
[its].} consequences, the formula (I) completely determines the
motion of the corpuscle \textit{as soon as one is given its position at an
initial instant}. In other words, the function $\varphi$, just like the Jacobi
function of which it is the generalisation,\label{oldp90} determines a whole class of
motions, and to know which of these motions is actually described it suffices
to know the initial position.

Let us now consider a whole cloud of corpuscles, identical and without
interaction, whose motions, determined by (I), correspond to the same function
$\varphi$ but differ in the initial positions. Simple reasoning shows that if
the density of the cloud at the initial moment is equal to%
\[
Ka^{2}\left(  \frac{\partial\varphi}{\partial t}-e\mathcal{V}\right)  \ ,
\]
where $K$ is a constant, it will subsequently remain constantly given by this
expression. We can state this result in another form. Let us suppose there be
only a single corpuscle whose initial position we ignore; from the preceding,
the \textit{probability for its presence} [sa \textit{probabilit\'{e} de
pr\'{e}sence}] at a given instant in a volume $d\tau$ of space will be%
\begin{equation}
\pi\,d\tau=Ka^{2}\left(  \frac{\partial\varphi}{\partial t}-e\mathcal{V}%
\right)  \,d\tau\ . \tag{II}%
\end{equation}

In brief, in our hypotheses, each wave $\Psi$ determines a `class of
motions',\ and each one of these motions is governed by equation (I) when one
knows the initial position of the corpuscle. If one ignores this initial
position, the formula (II) gives the probability for the presence of the
corpuscle in the element of volume $d\tau$ at the instant $t$. The wave $\Psi$
then appears as both a \textit{pilot wave} (F\"{u}hrungsfeld of Mr Born) and a
\textit{probability wave}.\label{oldp9192} Since the motion of the corpuscle seems to us to be
strictly determined by equation (I), it does not seem to us that there is any
reason to renounce believing in the determinism of individual physical
phenomena,\footnote{Here, that is, of the motion of individual corpuscles.}
and it is in this that our conceptions, which are very similar in other
respects to those of Mr Born, appear nevertheless to differ from them markedly.

Let us remark that, if one limits oneself to the Newtonian approximation, in
(I) and (II) one can replace: $\frac{\partial\varphi}{\partial t}%
-e\mathcal{V}$ by $m_{0}c^{2}$, and one obtains the simplified forms%
\begin{equation}
\overrightarrow{v}=-\frac{1}{m_{0}}\left(  \overrightarrow{\mathrm{grad\ }%
\varphi}+\frac{e}{c}\overrightarrow{A}\right)  \ , \tag{I$^\prime$}%
\end{equation}%
\begin{equation}
\pi=\mathrm{const}\cdot a^{2}\ . \tag{II$^\prime$}%
\end{equation}
There is one case where the application of the preceding ideas is done in a
remarkably clear form: when the initial motion of the corpuscles is uniform
and rectilinear in a region free of all fields. In this region, the cloud of
corpuscles we have just imagined may be represented by the homogeneous plane
wave\endnote{`$v$' is misprinted as `$\mathcal{V}$'.}%
\[
\Psi=a\cos\frac{2\pi}{h}W\left(  t-\frac{vx}{c^{2}}\right)  \ ;
\]
here $a$ is a constant, and this means that a corpuscle has the same
probability to be at any point of the cloud. The question of knowing how this
homogeneous plane wave will behave when penetrating a region where a field is
present is analogous to that of determining the form of an initially plane
light wave that penetrates a refracting medium. In his Memoir `Quantenmechanik
der Stossvorg\"{a}nge', Mr Born has given a general method of successive
approximation to solve this problem, and Mr Wentzel [14] has shown that one
can thus recover the Rutherford formula for the deflection of $\beta$-rays by
a charged centre.

We shall present yet another observation\label{anotherobservation} on the Dynamics of the material point
such as results from equation (I): for the material point one can always write
the equations of the Dynamics of Relativity even when the approximation of the
old mechanics is not valid, on condition that one attributes to the body a
variable proper mass $M_{0}$ given by the formula\label{fordiscussion3}%
\[
M_{0}=\sqrt{m_{0}^{2}-\frac{h^{2}}{4\pi^{2}c^{2}}\frac{\square a}{a}}\ .
\]

\ 

\noindent 5. \textit{The interpretation of interference}.\label{alsooldp9192} --- The new Dynamics allows us
to interpret the phenomena of wave Optics in exactly the way that was
foreseen, a long time ago now, by Mr Einstein.\footnote{Cf.\ chapter~\ref{guiding-fields-in-3-space} 
(\textit{eds.}).} In the case of light,
the wave $\Psi$ is indeed the light wave of the classical
theories.\footnote{We then consider $\Psi$ as the `light variable' without at
all specifying the physical meaning of this quantity.}$^{,}$\footnote{By
`classical theories' de Broglie seems to mean scalar wave optics. In the
general discussion (p.~\pageref{p192DRAFT1}), de Broglie states that the
physical nature of $\Psi$ for photons is unknown (\textit{eds.}).} If we
consider the propagation of light in a region strewn with fixed obstacles, the
propagation of the wave $\Psi$ will depend on the nature and arrangement of
these obstacles, but the frequency $\frac{1}{h}\frac{\partial\varphi}{\partial
t}$ will not vary (no Doppler effect). The formulas (I) and (II) will then
take the form\label{fordiscussion2}%
\[
\overrightarrow{v}=-\frac{c^{2}}{h\mathrm{\nu}}\overrightarrow{\mathrm{grad\ }%
\varphi}\ ;\ \ \ \ \pi=\mathrm{const\cdot}a^{2}\ .
\]
The second of these formulas shows immediately that the bright and dark
fringes predicted by the new theory will coincide with those predicted by the
old.\label{forMinEss2} To record the fringes, for example by photography, one can do an
experiment of short duration with intense irradiation, or an experiment of
long duration with feeble irradiation (Taylor's experiment); in the first case
one takes a mean in space, in the second case a mean in time, but if the light
quanta do not act on each other the statistical result must evidently be the same.\label{forMinEss2B}

Mr Bothe [15] believed he could deduce, from certain experiments on the
Compton effect in a field of interference, the inexactitude of the first
formula written above, the one giving the velocity of the quantum, but in our
opinion this conclusion can be contested.

\ 

\noindent 6. \textit{The energy-momentum tensor of the waves} $\Psi$. --- In one of his
Memoirs [16], Mr Schr\"{o}dinger gave the expression for the energy-momentum
tensor in the interior of a wave $\Psi$.\footnote{Cf. Schr\"{o}dinger's
report, section II (\textit{eds.}).} Following the ideas expounded here, the
wave $\Psi$ represents the motion of a cloud of corpuscles; examining the
expression given by Schr\"{o}dinger and taking into account the relations (I)
and (II), one then perceives that it decomposes into one part giving the
energy and momentum of the particles, and another that can be interpreted as
representing a state of stress existing in the wave around the particles.
These stresses are zero in the states of motion consistent with the old
Dynamics; they characterise the new states predicted by wave Mechanics, which
thus appear as `constrained states'\ of the material point and are intimately
related to the variability of the proper mass $M_{0}$. Mr De Donder has also
drawn attention to this fact, and he was led to denote the amplitude of the
waves that he considered by the name of `internal stress potential'.

The existence of these stresses allows one to explain how a mirror reflecting
a beam of light suffers a radiation pressure, even though according to
equation (I), because of interference, the corpuscles of light do not
`strike'\ its surface.\footnote{Cf. Brillouin's example in the discussion at
the end of de Broglie's lecture (\textit{eds.}).}

\ 

\noindent 7. \textit{The dynamics of many-body systems}.\label{deBsection7} --- We must now examine how
these conceptions may serve to interpret the wave equation proposed by
Schr\"{o}dinger for the Dynamics of many-body systems. We have pointed out
above the two difficulties that this equation raises. The first, relating to
the meaning of the variables that serve to construct the configuration space,
disappears if one assumes that the material points always have a quite
definite position. The second difficulty remains. It appears to us certain
that if one wants to \textit{physically} represent the evolution of a system
of $N$ corpuscles, one must consider the propagation of $N$ waves in space,
each of the $N$ propagations being determined by the action of the $N-1$
corpuscles connected to the other waves.\label{otherwaves}\footnote{Cf.\ section~\ref{deB-1927-Solvay-report} 
(\textit{eds.}).} Nevertheless, if one focusses one's
attention only on the corpuscles, one can represent their states by a point in
configuration space, and one can try to relate the motion of this
representative point to the propagation of a fictitious wave $\Psi$ in
configuration space.\label{andagain} It appears to us very probable that the wave\footnote{The
amplitude $a$ is time-independent because de Broglie is assuming the
time-independent Schr\"{o}dinger equation. Later in his report, de Broglie
applies his dynamics to a non-stationary wave function as well, for the case
of an atomic transition (\textit{eds.}).}%
\[
\Psi=a(q_{1},\ q_{2},\ ...,\ q_{n})\cos\frac{2\pi}{h}\varphi(t,\ q_{1}%
,\ ...,\ q_{n})\ ,
\]
a solution of the Schr\"{o}dinger equation, is only a fictitious wave which,
in the \textit{Newtonian approximation}, plays for the representative point of
the system in configuration space the same role of pilot wave and of
probability wave that the wave $\Psi$ plays in ordinary space in the case of a
single material point.\label{singlematpoint}

Let us suppose the system to be formed of $N$ points having for rectangular
coordinates%
\[
x_{1}^{1},\ x_{2}^{1},\ x_{3}^{1},\ ...,\ x_{1}^{N},\ x_{2}^{N},\ x_{3}%
^{N}\ .
\]
In the configuration space formed by means of these coordinates, the
representative point of the system has for [velocity] components along the
axis $x_{i}^{k}$%
\[
v_{x_{i}^{k}}=-\frac{1}{m_{k}}\frac{\partial\varphi}{\partial x_{i}^{k}}\ ,
\]
$m_{k}$ being the mass of the $k$th corpuscle. This is the relation that
replaces (I$^{\prime}$) for many-body systems. From this, one deduces that the
probability for the presence of the representative point in the element of
volume $d\tau$ of configuration space is%
\[
\pi\ d\tau=\mathrm{const\cdot}a^{2}\ d\tau\ .
\]

This new relation replaces relation (II$^{\prime}$) for many-body systems. It
fully accords,\label{fullyaccords} it seems to us, with the results obtained by Mr Born for the
collision of an electron and an atom, and by Mr Fermi [17] for the collision
of an electron and a rotator.\footnote{Cf. the remarks by Born and Brillouin
in the discussion at the end of de Broglie's lecture, and the de Broglie-Pauli
encounter in the general discussion at the end of the conference 
(pp.~\pageref{Pauli-deB-beginning}~ff.) (\textit{eds.}).}

Contrary to what happens for a single material point, it does not appear easy
to find a wave $\Psi$ that would define the motion of the system taking
Relativity into account.\label{relativityaccount}

\ 

\noindent 8. \textit{The waves }$\Psi$\textit{ in micromechanics}. --- Many authors
think it is illusory to wonder what the position or the velocity of an
electron in the atom is at a given instant. We are, on the contrary, inclined
to believe that it is possible to attribute to the corpuscles a position and a
velocity even in atomic systems, in a way that gives a precise meaning to the
variables of configuration space.

This leads to conclusions that deserve to be emphasised. Let us consider a
hydrogen atom in one of its stable states. According to Schr\"{o}dinger, in
spherical coordinates\footnote{$r$, radius vector; $\theta$, latitude;
$\alpha$, longitude.} the corresponding function $\Psi_{n}$ is of the form%
\[
\Psi_{n}=F(r,\theta)[A\cos m\alpha+B\sin m\alpha]%
\begin{array}
[c]{c}%
\sin\\
\cos
\end{array}
\frac{2\pi}{h}W_{n}t\ \ \ \ (m\ \mathrm{\operatorname{integer}})
\]
with%
\[
W_{n}=m_{0}c^{2}-\frac{2\pi^{2}m_{0}e^{4}}{n^{2}h^{2}}\ .
\]

If we then apply our formula (I$^{\prime}$), we conclude that the electron is
motionless in the atom, a conclusion which would evidently be inadmissible in
the old Mechanics. However, the examination of various questions and notably
of the Zeeman effect has led us to believe that, in its stable states, the H
atom must rather be represented by the function%
\[
\Psi_{n}=F(r,\theta)\cos\frac{2\pi}{h}\left(  W_{n}t-\frac{mh}{2\pi}%
\alpha\right)\ ,
\]
which, being a linear combination of expressions of the type written above, is
equally acceptable.\footnote{In his memoir, `Les moments de rotation et le
magn\'{e}tisme dans la m\'{e}canique ondulatoire' (\textit{Journal de Physique}
\textbf{8} (1927), 74), Mr L\'{e}on Brillouin has implicitly assumed the
hypothesis that we formulate in the text.} If this is true the electron will
have, from (I$^{\prime}$), a uniform circular motion of speed%
\[
v=\frac{1}{m_{0}r}\frac{mh}{2\pi}\ .
\]
It will then be motionless only in states where $m=0$.

Generally speaking, the states of the atom at a given instant can always be
represented by a function%
\[
\Psi=\sum_{n}c_{n}\Psi_{n}\ ,
\]
the $\Psi_{n}$ being Schr\"{o}dinger's Eigenfunktionen. In particular, the
state of transition $i\rightarrow j$ during which the atom emits the frequency
$\nu_{ij}$ would be given by (this appears to be in keeping with
Schr\"{o}dinger's ideas)%
\[
\Psi=c_{i}\Psi_{i}+c_{j}\Psi_{j}\ ,
\]
$c_{i}$ and $c_{j}$ being two functions of time that change very slowly
compared with the trigonometric factors of the $\Psi_{n}$, the first varying
from 1 to 0 and the second from 0 to 1 during the transition. Writing the
function $\Psi$ in the canonical form $a\cos\frac{2\pi}{h}\varphi$, which is
always possible, formula (I$^{\prime}$) will give the velocity of the electron
during the transition, if one assumes the initial position to be given. So it
does not seem to be impossible to arrive in this way at a visual
representation of the transition.\footnote{In this example, de Broglie is
applying his dynamics to a case where the wave function $\Psi$ has a
time-dependent amplitude $a$ (\textit{eds.}).}

Let us now consider an ensemble of hydrogen atoms that are all in the same
state represented by the same function%
\[
\Psi=\sum_{n}c_{n}\Psi_{n}\ .
\]
The position of the electron in each atom is unknown to us, but if, in our
imagination, we superpose all these atoms, we obtain a \textit{mean atom}
where the probability for the presence of one of the electrons in an element
of volume $d\tau$ will be given by the formula (II),\footnote{In this section
on atomic physics (`micromechanics') de Broglie considers the non-relativistic
approximation, using the limiting formula (I$^{\prime}$) --- except in this
paragraph where he reverts to the relativistic formulas (I) and (II), for the
purpose of comparison with the relativistic formulas for charge and current
density obtained by other authors (\textit{eds.}).} $K$ being determined by
the fact that the total probability for all the possible positions must be
equal to unity. The charge density $\rho$ and the current density
$\overrightarrow{J}=\rho\overrightarrow{v}$ in the mean atom are then, from
(I) and (II),\label{p91DRAFT1}%
\begin{align*}
\rho &  =Kea^{2}\left(  \frac{\partial\varphi}{\partial t}-e\mathcal{V}%
\right)  \ ,\\
\overrightarrow{J}  &  =-Kec^{2}a^{2}\left(  \overrightarrow{\mathrm{grad\ }%
\varphi}+\frac{e}{c}\overrightarrow{A}\right)
\end{align*}
and these formulas coincide, apart from notation, with those of Messrs Gordon,
Schr\"{o}dinger and O. Klein [18].\label{fordiscussion1}

Limiting ourselves to the Newtonian approximation, and for a moment denoting
by $\Psi$ the wave written in \textit{complex} form, and by $\bar{\Psi}$
the conjugate function, it follows that%
\[
\rho=\mathrm{const\cdot}a^{2}=\mathrm{const\cdot}\Psi\bar{\Psi}\ .
\]

This is the formula to which Mr Schr\"{o}dinger was led in reformulating the
matrix theory; it shows that the electric dipole moment of the mean atom
during the transition $i\rightarrow j$ contains a term of frequency $\nu_{ij}%
$, and thus allows us to interpret Bohr's frequency relation.

Today it appears certain that one can predict the mean energy radiated by an
atom by using the Maxwell-Lorentz equations, on condition that one introduces
in these equations the mean quantities $\rho$ and $\rho\overrightarrow{v}$
which have just been defined.\footnote{Cf. Schr\"{o}dinger's report, p.~\pageref{forpageDEB13}, 
and the ensuing discussion, and section~\ref{Schr-deB} (\textit{eds.}).} 
One can thus give the correspondence principle an entirely precise meaning, as Mr
Debye [19] has in fact shown in the particular case of motion with one degree
of freedom. It seems indeed that classical electromagnetism can from now on
retain only a statistical value; this is an important fact, whose meaning one
will have to try to explore more deeply.

To study the interaction of radiation with an ensemble of atoms, it is rather
natural to consider a `mean atom',\ immersed in a `mean light'\ which one
defines by a homogeneous plane wave of the vector potential. The density
$\rho$ of the mean atom is perturbed by the action of the light and one
deduces from this the scattered radiation. This method, which gives good mean
predictions, is related more or less directly to the theories of scattering by
Messrs Schr\"{o}dinger and Klein [20], to the theory of the Compton effect by
Messrs Gordon and Schr\"{o}dinger [21], and to the Memoirs of Mr Wentzel [22]
on the photoelectric effect and the Compton effect, etc. The scope of this
report does not permit us to dwell any further on this interesting work.

\ 

\noindent 9. \textit{Conclusions and remarks}. --- So far we have considered the
corpuscles as `exterior'\ to the wave $\Psi$, their motion being only
determined by the propagation of the wave. This is, no doubt, only a
provisional point of view: a true theory of the atomic structure of matter and
radiation should, it seems to us, \textit{incorporate} the corpuscles in the
wave phenomenon by considering singular solutions of the wave equations.\label{Draft1p.96} One
should then show that there exists a correspondence between the singular waves
and the waves $\Psi$, such that the motion of the singularities is connected
to the propagation of the waves $\Psi$ by the relation (I).\footnote{Cf.\ 
section~\ref{Structure} (\textit{eds.}).} In the case of no
[external] field, this correspondence is easily established, but it is not so
in the general case.

We have seen that the quantities $\rho$ and $\rho\overrightarrow{v}$ appearing
in the Maxwell-Lorentz equations must be calculated in terms of the functions
$\Psi$, but that does not suffice to establish a deep link between the
electromagnetic quantities and those of wave Mechanics. To establish this
link,\footnote{The few attempts made till now in this direction, notably by Mr
Bateman (\textit{Nature} \textbf{118} (1926), 839) and by the author
(\textit{Ondes et mouvements}, Chap.~VIII, and \textit{C.~R}.\ \textbf{184}
(1927), 81) can hardly be regarded as satisfactory.} one should probably begin
with singular waves, for Mr Schr\"{o}dinger has very rightly remarked that the
potentials appearing in the wave equations are those that result from the
discontinuous structure of electricity and not those that could be deduced
from the functions $\Psi$.

Finally, we point out that Messrs Uhlenbeck and Goudsmit's hypothesis of the
magnetic electron, so necessary to explain a great number of phenomena, has
not yet found its place in the scope of wave Mechanics.

\

\begin{center}
\par
\Needspace{5\baselineskip}
{\sc III. --- Experiments showing preliminary direct evidence for the new
Dynamics of the electron}\addcontentsline{toc}{section}{III. --- Experiments 
showing preliminary direct evidence for the new Dynamics of the electron}
\end{center}
10. \textit{Phenomena whose existence is suggested by the new conceptions}.~--- 
The ideas that have just been presented lead one to consider the motion of
an electron as guided by the propagation of a certain wave. In many usual
cases, the old Mechanics remains entirely adequate as a first approximation;
but our new point of view, as Elsasser\endnote{Consistently mis-spelt
throughout the text as `Elsaesser'.} [23] pointed out already in 1925,
necessarily raises the following question: `Could one not observe electron
motions that the old Mechanics would be incapable of predicting, and which
would therefore be characteristic of wave Mechanics? In other words, for
electrons, could one not find the analogue of the phenomena of diffraction and
interference?'\footnote{This is not a quotation: these words do not appear in
the cited 1925 paper by Elsasser. Further, it was de Broglie who first
suggested electron diffraction, in a paper of 1923 (see section~\ref{first-papers}) 
(\textit{eds}.).}

These new phenomena, if they exist, must depend on the wavelength of the wave
associated with the electron motion. For an electron of speed $v$, the
fundamental formula%
\[
p=\frac{h\nu}{V}%
\]
gives%
\[
\lambda=\frac{V}{\nu}=\frac{h}{p}=\frac{h\sqrt{1-\beta^{2}}}{m_{0}v}\;.
\]

If $\beta$ is not too close to 1, it suffices to write%
\[
\lambda=\frac{h}{m_{0}v}\;.
\]
Let $\mathcal{V}$ be the potential difference, expressed in volts, that is
capable of imparting the speed $v$ to the electron; numerically, for the
wavelength in centimetres, one will have\footnote{Here we have adopted the
following values:%
\begin{align*}
h &  =6.55\times10^{-27}\;\mbox{\footnotesize erg-seconds}\ ,\\
m_{0} &  =9\times10^{-28}\;\mathrm{gr\ ,}\\
e &  =4.77\times10^{-10}\;\mathrm{e.s.u.}\ .
\end{align*}
}%
\[
\lambda=\frac{7.28}{v}=\frac{12.25}{\sqrt{\mathcal{V}}}\times10^{-8}\;.
\]

To do precise experiments, it is necessary to use electrons of at least a few
volts: from which one has an upper limit for $\lambda$ of a few angstroms. One
then sees that, even for slow electrons, the phenomena being sought are
analogous to those shown by X-rays and not to those of ordinary light. As a
result, it will be difficult to observe the diffraction of a beam of electrons
by a small opening, and if one wishes to have some chance of obtaining
diffraction by a grating, one must either consider those natural
three-dimensional gratings, the crystals, or use ordinary gratings under a
very grazing incidence, as has been done recently for X-rays. On making slow
electrons pass through a crystalline powder or an amorphous substance, one
could also hope to notice the appearance of rings analogous to those that have
been obtained and interpreted in the X-ray domain by Messrs Hull, Debye and
Scherrer, Debierne, Keesom and De Smedt, etc.

The exact theoretical prediction of the phenomena to be observed along these
lines is still not very advanced. Let us consider the diffraction of a beam of
electrons with the same velocity by a crystal; the wave $\Psi$ will propagate\label{willpropagate}
following the general equation, in which one has to insert the potentials
created by the atoms of the crystal considered as centres of force. One does
not know the exact expression for these potentials but, because of the regular
distribution of atoms in the crystal, one easily realises that the scattered
amplitude will show maxima in the directions predicted by Mr von Laue's
theory. Because of the role of pilot wave played by the wave $\Psi$, one must
then observe a selective scattering of the electrons in these directions.\label{selective}

Using his methods, Mr Born has studied\label{hasstudied} another problem: that of the collision
of a narrow beam of electrons with an atom. According to him, the curve giving
the number of electrons that have suffered an inelastic collision as a
function of the scattering angle must show maxima and minima; in other words,
these electrons will display rings on a screen placed normally to the
continuation of the incident beam.

It would still be premature to speak of agreement between theory and
experiment; nevertheless, we shall present experiments that have revealed
phenomena showing at least broadly the predicted character.

\ 

\noindent 11. \textit{Experiments by Mr E.G. Dymond} [24]. --- Without feeling obliged
to follow the chronological order, we shall first present Mr Dymond's experiments:

A flask of purified helium contained an `electron gun', which consisted of a
brass tube containing an incandescent filament of tungsten and in whose end a
slit was cut. This gun discharged a well-collimated beam of electrons into the
gas, with a speed determined by the potential difference (50 to 400 volts)
established between the filament and the wall of the tube. The wall of the
flask had a slit through which the electrons could enter a chamber where the
pressure was kept low by pumping and where, by curving their trajectories, a
magnetic field brought them onto a Faraday cylinder.

Mr Dymond first kept the orientation of the gun fixed and measured the speed
of the electrons thus scattered by a given angle. He noticed that most of the
scattered electrons have the same energy as the primary electrons; they have
therefore suffered an elastic collision. Quite a large number of electrons
have a lower speed corresponding to an energy loss from about 20 to 55 volts:
this shows that they made the He atom pass from the normal state $1^{1}S$ to
the excited state $2^{1}S$. One also observes a lower proportion of other
values for the energy of the scattered electrons; we shall not discuss the
interpretation that Mr Dymond has given them, because what interests us most
here is the variation of the number of scattered particles with the scattering
angle $\theta$. To determine this number, Mr Dymond varied the orientation of
the gun inside the flask, and for different scattering angles collected the
electrons that suffered an energy loss equivalent to 20 to 55 volts; he
constructed a series of curves of the angular distribution of these electrons
for different values of the tension $\mathcal{V}$ applied to the electron gun.
The angular distribution curve shows a very pronounced maximum for a low value
of $\theta$, and this maximum appears to approach $\theta=0$ for increasing
values of $\mathcal{V}$.

Another, less important, maximum appears towards $\theta=50%
{{}^\circ}%
$ for a primary energy of about a hundred volts, and then moves for increasing
values of $\mathcal{V}$ towards increasing $\theta$. Finally, a very sharp
maximum appears for a primary energy of about 200 volts at $\theta=30%
{{}^\circ}%
$, and then seems independent of $\mathcal{V}$. These facts are summarised in
the following table given by Dymond:\endnote{For clarity, the presentation of
the table has been slightly altered.}

\begin{center}
\begin{tabular}
[c]{ccc}
$\mathcal{V}$ (volts) &   \  & Positions of the maxima ($%
{{}^\circ}%
$)\\
\begin{tabular}[t]{c}
48.9\\
72.3\\
97.5\\
195\\
294\\
400
\end{tabular}
&
\begin{tabular}[t]{c}
....\\
....\\
....\\
....\\
....\\
....
\end{tabular}
&
\begin{tabular}[t]{rcc}
24    \, \ &  --- &  ---\\
8     \, \ &  --- &  ---\\
5     \, \ &  --- &  50\\
$<2.5$     &  30  &  59\\
$<2.5$     &  30  &  69\\
$<2.5$     &  30  &  70
\end{tabular}
\end{tabular}
\end{center}

The above results\label{aboveresults} must very probably be interpreted with the aid of the new
Mechanics and are to be related to Mr Born's predictions. Nevertheless, as Mr
Dymond very rightly says, `the theoretical side of the problem is however not yet
sufficiently advanced to give detailed information on the phenomena to be expected, so that
the results above reported cannot be said to substantiate the wave mechanics except in the most 
general way'.\endnote{We have used the original English (Dymond 1927, p.~441).}

\ 

\noindent 12. \textit{Experiments by Messrs C. Davisson and L. H. Germer}. --- In 1923,
Messrs Davisson and Kunsman [25] published peculiar results on the scattering
of electrons at low speed. They directed a beam of electrons, accelerated by a
potential difference of less than 1000 volts, onto a block of platinum at an
incidence of 45$%
{{}^\circ}%
$ and determined the distribution of scattered electrons by collecting them in
a Faraday cylinder. For potentials above 200 volts, one observed a steady
decrease in scattering for increasing values of the deviation angle, but for
smaller voltages the curve of angular variation showed two maxima. By covering
the platinum with a deposit of magnesium, one obtained a single small maximum
for electrons of less than 150 volts. Messrs Davisson and Kunsman attributed
the observed phenomena to the action of various layers of intra-atomic
electrons on the incident electrons, but it seems rather, according to
Elsasser's opinion, that the interpretation of these phenomena is a matter for
the new Mechanics.

Resuming analogous experiments with Mr Germer [26], Mr Davisson obtained very
important results this year, which appear to confirm the general predictions
and even the formulas of Wave Mechanics.

The two American physicists sent homogeneous beams of electrons onto a crystal
of nickel, cut following one of the 111 faces of the regular octahedron
(nickel is a cubic crystal). The incidence being normal, the phenomenon
necessarily had to show the ternary symmetry around the direction of the
incident beam. In a cubic crystal cut in this manner, the face of entry is cut
obliquely by three series of 111 planes, three series of 100 planes, and six
series of 110 planes. If one takes as positive orientation of the normals to
these series of planes the one forming an acute angle with the face of entry,
then these normals, together with the direction of incidence, determine
distinguished azimuths, which Messrs Davisson and Germer call azimuths (111),
(100), (110), and for which they studied the scattering; because of the
ternary symmetry, it evidently suffices to explore a single azimuth of each type.

Let us place ourselves at one of the distinguished azimuths and let us
consider only the distribution of Ni atoms on the face of entry of the
crystal, which we assume to be perfect. These atoms form lines perpendicular
to the azimuth being considered and whose equidistance $d$ is known from
crystallographic data. The different directions of scattering being identified
in the azimuthal plane by the angle $\theta$ of co-latitude, the waves
scattered by the atoms in the face of entry must be in phase in directions
such that one has%
\[
\theta=\arcsin\left(  \frac{n\lambda}{d}\right)  =\arcsin\left(  \frac{n}%
{d}\frac{12.25}{\sqrt{\mathcal{V}}}\cdot 10^{-8}\right)
\;\;\;\;\;(n\;\operatorname{integer})\ .
\]
One must then expect to observe maxima in these directions, for the scattering
of the electrons by the crystal.

Now here is what Messrs Davisson and Germer observed. By gradually varying the
voltage $\mathcal{V}$ that accelerates the electrons one observes, in the
neighbourhood of \textit{certain} values of $\mathcal{V}$, very distinct
scattering maxima in directions whose co-latitude is accurately given by the
above formula (provided one sets in general $n=1$, and sometimes $n=2$). There
is direct numerical confirmation of the formulas of the new Dynamics;\label{confirmation} this is
evidently a result of the highest importance.

However, the explanation of the phenomenon is not complete: one must explain
why the scattering maxima are observed only in the neighbourhood of certain
particular values of $\mathcal{V}$, and not for all values of $\mathcal{V}$.
One interpretation naturally comes to mind: we assumed above that only the
face of entry of the crystal played a role, but one can assume that the
electron wave penetrates somewhat into the crystal and, further, in reality
the face of entry will never be perfect and will be formed by several parallel
111 planes forming steps. In these conditions, it is not sufficient to
consider the interference of the waves scattered by a single reticular plane
at the surface, one must take into account the interference of the waves
scattered by several parallel reticular planes. In order for there to be a
strong scattering in a direction $\theta$, $\theta$ and $\mathcal{V}$ must
then satisfy not only the relation written above, but also another relation
which is easy to find; the scattering must then be \textit{selective}, that is
to say, occur with [significant] intensity only for certain values of
$\mathcal{V}$, as experiment shows. Of course, the theory that has just been
outlined is a special case of Laue's general theory.

Unfortunately, as Messrs Davisson and Germer have themselves remarked, in
order to obtain an exact prediction of the facts in this way, it is necessary
to attribute to the separation of the 111 planes next to the face of entry a
smaller value (of about 30\%) than that provided by Crystallography and by
direct measurements by means of X-rays. It is moreover not unreasonable to
assume that the very superficial reticular planes have a spacing different
from those of the deeper planes, and one can even try to connect this idea to
our current conceptions concerning the equilibrium of crystalline gratings.

If one accepts the preceding hypothesis, the scattering must be produced by a
very small number of reticular planes in the entirely superficial layer of the
crystal; the concentration of electrons in preferred directions must then be
much less pronounced than in the case of scattering by a whole unlimited
spatial grating. Is it nevertheless sufficient in order to explain the `peaks'
observed by Davisson and Germer? To this question, Mr Patterson has recently
provided an affirmative answer, by showing that the involvement of just two
superficial reticular planes already suffices to predict exactly the
variations of the selective reflection of electrons observed in the
neighbourhood of%
\[
\theta=50%
{{}^\circ}%
\;,\;\;\;\;\mathcal{V}=54\;\mathrm{volts}\;.
\]

To conclude, we can do no better than quote the conclusion of Mr Patterson [27]: 
`The agreement of these results with calculation seems to 
indicate that the phenomenon can be explained as a diffraction
of waves in the \textit{outermost layers} of the crystal surface. 
It also appears [....] that a complete analysis of the results
of such experiments will give valuable information as to the
conditions prevailing in the actual surface, and that a new
method has been made available for the investigation of the 
structure of crystals in a region which has up to the present 
almost completely escaped observation'.\endnote{We use here 
the original English text (Patterson 1927, p.~47). De Broglie 
changes `of these results' to `des r\'{e}sultats exp\'{e}rimentaux', 
omits the italics and translates `valuable' by `exacts'.}

\ 

\noindent 13. \textit{Experiments by Messrs G.~P.~Thomson and A.~Reid} [28]. --- Very
recently, Messrs Thomson and Reid have made the following results known: if a
narrow pencil of homogeneous cathode rays passes normally through a celluloid
film, and is then received on a photographic plate placed parallel to the film
at $10$ cm behind it, one observes rings around the central spot. With rays of
13\thinspace000 volts, a photometric examination has revealed the existence of
three rings. By gradually increasing the energy of the electrons, one sees the
rings appear around 2500 volts, and they have been observed up to
16\thinspace500 volts. The radii of the rings decrease when the energy
increases and, it seems, approximately in inverse proportion to the speed,
that is, to our wavelength $\lambda$.

These observations are very interesting, and again confirm the new conceptions
in broad outline. Is it a question here of an atomic phenomenon analogous to
those observed by Dymond, or else of a phenomenon of mutual interference
falling into one of the categories studied by Debye and Scherrer, Hull,
Debierne, Keesom and De Smedt? We are unable to say, and we limit ourselves to
remarking that here the electrons used are relatively fast; this is
interesting from the experimental point of view, because it is much easier to
study electrons of a few thousand volts than electrons of about a hundred volts.


\newpage

\par
\Needspace{5\baselineskip}
\begin{center}

{\bf Bibliography}\addcontentsline{toc}{section}{Bibliography}\footnote{The style 
of the bibliography has been slightly modernised and uniformised with that used in the
other reports ({\em eds.}).}\\
\hfill\\ 
\end{center}
\noindent [1] Louis de Broglie, \textit{C.\ R}., \textbf{177} (1923), 507, 548 and 630;
\textbf{179} (1924), 39, 676 and 1039. Doctoral thesis, November 1924 (Masson,
publisher), \textit{Annales de Physique} (10), \textbf{III} [(1925)], 22.
\textit{J.\ de Phys.\ }(6), \textbf{VII} [(1926)],~1.

\noindent [2] S.~N.~Bose, \textit{Zts.\ f.\ Phys.}, \textbf{27} (1924), 384.

\noindent [3] A.~Einstein, \textit{Berl.\ Ber.\ }(1924), 261; (1925), [3].

\noindent [4] P.~Jordan, \textit{Zts.\ f.\ Phys.}, \textbf{33} (1925), 649.

\noindent E.~Schr\"{o}dinger, \textit{Physik.\ Zts.}, \textbf{27} (1926), 95.

\noindent P.~Dirac, \textit{Proc.\ Roy.\ Soc.\ A}, \textbf{112} (1926), 661.

\noindent E.~Fermi, \textit{Zts.\ f.\ Phys.}, \textbf{36} (1926), 902.

\noindent L.~S.~Ornstein and H.~A.~Kramers, \textit{Zts.\ f.\ Phys.}, \textbf{42} (1927), 481.

\noindent [5] E.~Schr\"{o}dinger, \textit{Ann.\ der Phys.}, \textbf{79} (1926), 361, 489
and 734; \textbf{80} (1926), 437; \textbf{81} (1926), 109.
\textit{Naturwissensch.}, 14th year [(1926)], 664. \textit{Phys. Rev.}, 
\textbf{28} (1926), 1051.

\noindent [6] L.~Brillouin, \textit{C.\ R.}, \textbf{183} (1926), 24 and 270. \textit{J.\ de
Phys.\ }[(6)], \textbf{VII} (1926), 353.

\noindent G.~Wentzel, \textit{Zts.\ f.\ Phys.}, \textbf{38} (1926), 518.

\noindent H.~A.~Kramers, \textit{Zts.\ f.\ Phys.}, \textbf{39} (1926), 828.

\noindent [7] E.~Fues, \textit{Ann.\ der Phys.}, \textbf{80} (1926), 367; \textbf{81}
(1926), 281.

\noindent C.~Manneback, \textit{Physik.\ Zts.}, \textbf{27} (1926), 563; \textbf{28}
(1927), 72.

\noindent [8] L.~de Broglie, \textit{C.\ R.}, \textbf{183} (1926), 272. \textit{J.\ de
Phys.\ }(6), \textbf{VII} (1926), 332.

\noindent O.~Klein, \textit{Zts.\ f.\ Phys.}, \textbf{37} (1926), 895.

\noindent [V.] Fock, \textit{Zts.\ f.\ Phys.}, \textbf{38} (1926), 242.

\noindent W.~Gordon, \textit{Zts.\ f.\ Phys.}, \textbf{40} (1926), 117.

\noindent L.~Flamm, \textit{Physik.\ Zts.}, \textbf{27} (1926), 600.

\noindent [9] O.~Klein, \textit{loc.~cit.\ }in [8].

\noindent [10] L.~de Broglie, \textit{J.\ de Phys.\ }(6), \textbf{VIII} (1927), 65.

\noindent (See also L.~Rosenfeld, \textit{Acad.\ Roy.\ Belg.\ }(5), \textbf{13}, n.~6.)

\noindent [11] Th.~De Donder, \textit{C.\ R.}, \textbf{182} (1926), 1512; \textbf{183}
(1926), 22 (with Mr Van~den Dungen); \textbf{183} (1926), 594; \textbf{184}
(1927), 439 and 810. \textit{Acad.\ Roy.\ Belg.}, sessions of 9 October 1926, of
8 January, 5 February, 5 March and 2 April 1927.

\noindent [12] M.~Born, \textit{Zts.\ f.\ Phys.}, \textbf{37} (1926), 863; \textbf{38}
(1926), 803; \textbf{40} (1926), 167. \textit{Nature}, \textbf{119} (1927), 354.

\noindent [13] L.~de Broglie, \textit{C.\ R.}, \textbf{183} (1926), 447, and \textbf{184}
(1927), 273. \textit{Nature}, \textbf{118} (1926), 441. \textit{J.\ de Phys.\ }(6), 
\textbf{VIII} (1927), 225. \textit{C.\ R.}, \textbf{185} (1927), 380.

\noindent (See also E.~Madelung, \textit{Zts.\ f.\ Phys.}, \textbf{40} (1926), 322.)

\noindent [14] G.~Wentzel, \textit{Zts.\ f.\ Phys.}, \textbf{40} (1926), 590.

\noindent [15] W.~Bothe, \textit{Zts.\ f.\ Phys.}, \textbf{41} (1927), 332.

\noindent [16] E.~Schr\"{o}dinger, \textit{Ann.\ der Phys.}, \textbf{82} (1927), 265.

\noindent [17] E.~Fermi, \textit{Zts.\ f.\ Phys.}, \textbf{40} (1926), 399.

\noindent [18] W.~Gordon and E.~Schr\"{o}dinger, \textit{loc.~cit.\ }in [8] and [16].

\noindent O.~Klein, \textit{Zts.\ f.\ Phys.}, \textbf{41} (1927), 407.

\noindent [19] P.~Debye, \textit{Physik.\ Zts.}, \textbf{27} (1927), 170.

\noindent [20] E.~Schr\"{o}dinger and O.~Klein, \textit{loc.~cit.\ }in [5] and [18].

\noindent [21] W.~Gordon, \textit{loc.~cit.\ }in [8].

\noindent E.~Schr\"{o}dinger, \textit{Ann.\ der Phys.}, \textbf{82} (1927), 257.

\noindent [22] G.~Wentzel, \textit{Zts.\ f.\ Phys.}, \textbf{43} (1927), 1; \textbf{41}
(1927), 828.

\noindent (See also G.~Beck, \textit{Zts.\ f.\ Phys.}, \textbf{41} (1927), 443; 
J.~R.~Oppenheimer, \textit{Zts.\ f.\ Phys.}, \textbf{41} (1927), 268.)

\noindent [23] [W.] Elsasser, \textit{Naturwissensch.}, \textbf{13} (1925), 711.

\noindent [24] E.~G.~Dymond, \textit{Nature}, \textbf{118} (1926), 336. \textit{Phys.\ 
Rev.}, \textbf{29} (1927), 433.

\noindent [25] C.~Davisson and C.~H.~Kunsman, \textit{Phys.\ Rev.}, \textbf{22} (1923), 242.

\noindent [26] C.~Davisson and L.~H.~Germer, \textit{Nature}, \textbf{119} (1927), 558.

\noindent [27] A.~L.~Patterson, \textit{Nature}, \textbf{120} (1927), 46.

\noindent [28] C.~P.~Thomson and A.~Reid, \textit{Nature}, \textbf{119} (1927), 890.


\newpage

\section*{Discussion of Mr de~Broglie's report}\markboth{{\it L.~de~Broglie}}{{\it Discussion}}
\addcontentsline{toc}{section}{Discussion of Mr de~Broglie's report}

{\sc Mr Lorentz.}~---~I should like to see clearly how, in the
first form of your theory, you recovered Sommerfeld's quantisation conditions.
You obtained a single condition, applicable only to the case where the orbit
is closed: the wave must, after travelling along the orbit, finish in phase
when it comes back to the initial point. But in most cases the trajectory is
not closed; this happens, for example, for the hydrogen atom when one takes
relativity into account; the trajectory is then a rosette, and never comes
back to its initial point.

How did you find the quantisation conditions applicable to these multiperiodic problems?\\

{\sc Mr de Broglie.}~---~The difficulty is resolved by
considering pseudo-periods, as I pointed out in my Thesis (chap.~III, p.~41).
When a system is multiperiodic, with partial periods $\tau_{1}$, $\tau_{2}$,
..., $\tau_{n}$, one can prove that one can find quasi-periods $\tau$ that are
nearly exactly whole multiples of the partial periods:%
\[
\tau=m_{1}\tau_{1}+\varepsilon_{1}=m_{2}\tau_{2}+\varepsilon_{2}...=m_{n}%
\tau_{n}+\varepsilon_{n}\;,
\]
the $m_{1}$, $m_{2}$, ..., $m_{n}$ being integers and the $\varepsilon_{1}$,
$\varepsilon_{2}$, ..., $\varepsilon_{n}$ as small as one likes. The
trajectory then never comes back to its initial point, but at the end of a
quasi-period $\tau$ it comes back as closely as one likes to the initial
position. One will then be led to write that, at the end of a quasi-period,
the wave finishes in phase; now, there is an infinite number of quasi-periods,
corresponding to all kinds of systems of values of the integers $m_{1}$,
$m_{2}$, ..., $m_{n}$. In order that the wave finishes in phase after any one
of these quasi-periods, it is necessary that one have\endnote{The last
equality is misprinted as `$\int_{\tau_{n}}p_{n}dq=n_{n}h_{n}$'.}%
\[
\int_{\tau_{1}}p_{1}dq_{1}=n_{1}h\;,\;\;\;\;\int_{\tau_{2}}p_{2}dq_{2}%
=n_{2}h\;,\;\;\;\;...\;,\;\;\;\;\int_{\tau_{n}}p_{n}dq_{n}=n_{n}h\;,
\]
which gives exactly Sommerfeld's conditions.\footnote{Darrigol (1993, pp.~342--3, 
364--5) shows that this derivation is faulty: the condition that the
wave should finish in phase after any quasi-period does \textit{not} imply the
$n$ separate conditions listed above.
}\\

{\sc Mr Born.}~---~The definition\label{Bornobjection} of the trajectory of a particle
that Mr de Broglie has given seems to me to present difficulties in the case
of a collision between an electron and an atom. In an elastic collision, the
speed of the particle must be the same after the collision as before. I should
like to ask Mr de Broglie if that follows from his formula.\\

{\sc Mr de Broglie.}~---~That follows from it, indeed.\\

{\sc Mr Brillouin.}~---~It seems to me that no serious objection
can be made to the point of view of L. de Broglie. Mr Born can doubt\label{candoubt} the real
existence of the trajectories calculated by L. de Broglie, and assert that one
will never be able to observe them, but he cannot prove to us that these
trajectories do not exist. There is no contradiction between the point of view
of L. de Broglie and that of other authors, since, in his report (\S 8, 
p.~\pageref{p91DRAFT1}\endnote{The original reads `p.~18', which is presumably
in reference to de Broglie's Gauthier-Villars `preprint', in all
likelihood circulated before the conference (preprints of other lectures
were circulated as mimeographs). Cf. chapter~\ref{HistEss}, p.~\pageref{forpageDEB22}}) 
L.~de Broglie shows us that his formulas\footnote{That is, de Broglie's equations
for the mean charge and current density to be used in semiclassical radiation
theory (\textit{eds.}).} are in exact agreement with those of Gordon, at
present accepted by all physicists.\\

{\sc Mr Pauli.}~---~I should like to make a small remark\label{smallremark} on what
seems to me to be the mathematical basis of Mr de Broglie's viewpoint
concerning particles in motion on definite trajectories. His conception is
based on the principle of conservation of charge:%
\begin{equation}
\frac{\partial\rho}{\partial t}+\frac{\partial s_{1}}{\partial x}%
+\frac{\partial s_{2}}{\partial y}+\frac{\partial s_{3}}{\partial
z}=0\;\;\;\;\mathrm{or}\;\;\;\;\sum_{k=1}^{4}\frac{\partial s_{k}}{\partial
x_{k}}=0\;, \tag{a}%
\end{equation}
which is a consequence of the wave equation, when one sets%
\[
is_{k}=\psi\frac{\partial\psi^{\ast}}{\partial x_{k}}-\psi^{\ast}%
\frac{\partial\psi}{\partial x_{k}}+\frac{4\pi i}{h}\frac{e}{c}\Phi_{k}%
\psi\psi^{\ast}\;.
\]
Mr de Broglie introduces, in place of the complex function $\psi$, the two
real functions $a$ and $\varphi$ defined by%
\[
\psi=ae^{\frac{2\pi i}{h}\varphi}\;,\;\;\;\;\psi^{\ast}=ae^{-\frac{2\pi i}%
{h}\varphi}\;.
\]

Substituting these expressions into the expression for $s_{k}$ yields:%
\[
s_{k}=\frac{4\pi}{h}a^{2}\left(  \frac{\partial\varphi}{\partial x_{k}}%
+\frac{e}{c}\Phi_{k}\right)  \;.
\]

From this follow the expressions given by Mr de Broglie for the velocity
vector, defined by%
\begin{equation}
v_{1}=\frac{s_{1}}{\rho}\;,\;\;\;\;v_{2}=\frac{s_{2}}{\rho}\;,\;\;\;\;v_{3}%
=\frac{s_{3}}{\rho}\;. \tag{b}%
\end{equation}

Now if in a field theory there exists a conservation principle of the form
(a), it is always formally possible to introduce a velocity vector (b),
depending on space and time, and to imagine furthermore corpuscles that move
following the current lines of this vector. Something similar\label{PaulionSlater} 
was already proposed in optics by Slater; according to him, light quanta should always
move following the lines of the Poynting vector. Mr de Broglie now introduces
an analogous representation for material particles.

In any case, I do not believe that this representation may be developed in a
satisfactory manner; I intend to return to this during the general
discussion.\footnote{See pp.~\pageref{Pauli-deB-beginning}~ff.\ (\textit{eds.}).}\\

{\sc Mr Schr\"{o}dinger.}~---~\label{SchrdeBdisc}If I have properly understood Mr de
Broglie, the \textit{velocity} of the particles must have its analogue in a
vector field composed of the three \textit{spatial} components of the
\textit{current in a four-dimensional space}, after division of these by the
component \textit{with respect to time} (that is, the charge density). I
should like simply to recall now that there exist still other vector
quantities of a field, which can be made to correspond with the velocity of
the particles, such as the components of the \textit{momentum density} (see
\textit{Ann.\ d.\ Phys.}\endnote{`\textit{Ann.\ de Phys.}' in the
original.} \textbf{82}, 265). Which of the two analogies is the more convincing?\\

{\sc Mr Kramers.}~---~The fact that with independent particles in
motion one cannot construct an energy-momentum tensor having the properties
required by Maxwell's theory constitutes nevertheless a difficulty.\\

{\sc Mr Pauli.}~---~The quotient of the momentum by the energy
density which Mr Schr\"{o}dinger considers would in fact lead in a
relativistic calculation to other particle trajectories than would the
quotient of the densities of current and of charge.\\

{\sc Mr Lorentz.}~---~In using his formulas for the velocity of
the electron, has Mr de Broglie not calculated this velocity in particular
cases, for example for the hydrogen atom?\\

{\sc Mr de Broglie.}~---~When one applies the formula for
the velocity to a wave function representing a stable state of the hydrogen
atom according to Mr Schr\"{o}dinger, one finds circular orbits. One does not
recover the elliptical orbits of the old theory (see my report, \S 8).\\

{\sc Mr Ehrenfest.}~---~Can\label{deBDisc-beginning} the speed of an electron in a
stationary orbit be zero?\\

{\sc Mr de Broglie.}~---~Yes, the speed of the electron
can be zero.\\

{\sc Mr Schr\"{o}dinger.}\label{forpage89}~---~Mr de Broglie says that in the case
of the hydrogen atom his hypothesis leads to \textit{circular} orbits. That is
true for the particular solutions of the wave equation that one obtains when
one separates the problem in polar coordinates in space; perhaps it is still
true for the solutions that one obtains by making use of parabolic or
elliptical coordinates. But in the case of a degeneracy (as he considers it
here) it is, in reality, not at all the particular solutions which have a
significance, but only an \textit{arbitrary} linear combination, with constant
coefficients, of all the particular solutions belonging to the same
eigenvalue, because there is no means of distinguishing between them, all
linear combinations being equally justified in principle. In these conditions,
much more complicated types of orbit will certainly appear. But I do not
believe that in the atomic domain one may still speak of `orbits'.\\

{\sc Mr Lorentz.}~---~Does one know of such more complicated orbits?\\

{\sc Mr Schr\"{o}dinger.}~---~No, one does not know of them; but I
simply wanted to say that if one finds circular orbits, that is due to a
fortuitous choice of particular solutions that one considers, and this choice
cannot be motivated in a way that has no arbitrariness.\\

{\sc Mr Brillouin.}~---~\label{Brill-beginning}Perhaps it is not superfluous to give some
examples that illustrate well the meaning of Mr L. de Broglie's formulas, and
that allow one to follow the motion of the particles guided by the phase wave.
If the wave is plane and propagates freely, the trajectories of the particles
are the rays normal to the wave surface. Let us suppose that the wave is
reflected by a plane mirror, and let $\theta$ be the angle of incidence; the
wave motion in%
  \begin{figure}
    \centering
      \resizebox{\textwidth}{!}{\includegraphics[0mm,0mm][220.12mm,100.00mm]{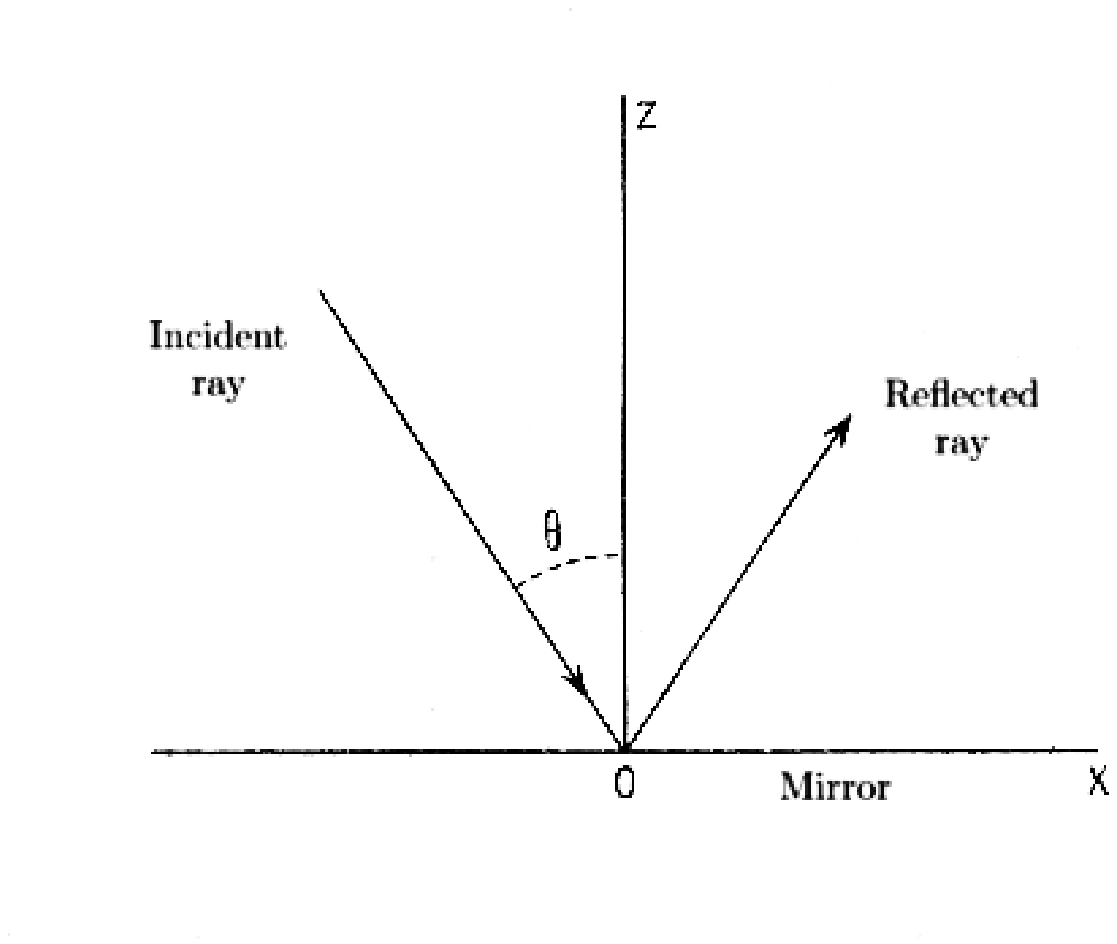}}
    \unnumberedcaption{Fig.~1.}
  \end{figure} 
front of the mirror is given by a superposition of the incident wave%
\[
\psi_{1}=a_{1}\cos2\pi\left(  \frac{t}{T}-\frac{x\sin\theta-z\cos\theta
}{\lambda}\right)
\]
and the reflected wave%
\[
\psi_{2}=a_{1}\cos2\pi\left(  \frac{t}{T}-\frac{x\sin\theta+z\cos\theta
}{\lambda}\right)  \;,
\]
which gives%
\[
\psi=2a_{1}\cos\frac{2\pi z\cos\theta}{\lambda}\cos2\pi\left(  \frac{t}%
{T}-\frac{x\sin\theta}{\lambda}\right)  \;.
\]

This wave is put in L.~de~Broglie's canonical form%
\[
\psi=a\cos\frac{2\pi}{h}\varphi
\]
with%
\[
a=2a_{1}\cos\frac{2\pi z\cos\theta}{\lambda}\;\;\;\;\mathrm{and}%
\;\;\;\;\varphi=h\left(  \frac{t}{T}-\frac{x\sin\theta}{\lambda}\right)  \;.
\]

Let us then apply L.~de~Broglie's formulas, in the simplified form given on
page~\pageref{fordiscussion2} (\S 5);\endnote{This is `p.~117' in the original.} and let
us suppose that it is a light wave guiding the photons; the velocity of these
is%
\[
\overrightarrow{v}=-\frac{c^{2}}{h\mathrm{\nu}}\overrightarrow{\mathrm{grad\,}%
\varphi}\;.
\]

We see that the projectiles move parallel to the mirror, with a speed
$v_{x}=c\sin\theta$, less than $c$. Their energy remains equal to
$h\mathrm{\nu}$, because their mass has undergone a variation, according to
the following formula (report by L. de Broglie, p.~~\pageref{fordiscussion3}):\endnote{Again, 
`p.~117' in the original.}%
\[
M_{0}=\sqrt{m_{0}^{2}-\frac{h^{2}}{4\pi^{2}c^{2}}\frac{\square a}{a}}=\frac
{h}{2\pi c}\sqrt{-\frac{\square a}{a}}=\frac{h\mathrm{\nu}}{c^{2}}\cos
\theta\;.
\]

The mass of the photons, which is zero in the case where the wave propagates
freely, is then assumed to take a non-zero value in the whole region where
there is interference or deviation of the wave.

Let us draw a diagram for the case of a limited beam of light falling on a%
  \begin{figure}
    \centering
      \resizebox{\textwidth}{!}{\includegraphics[0mm,0mm][220.18mm,100.53mm]{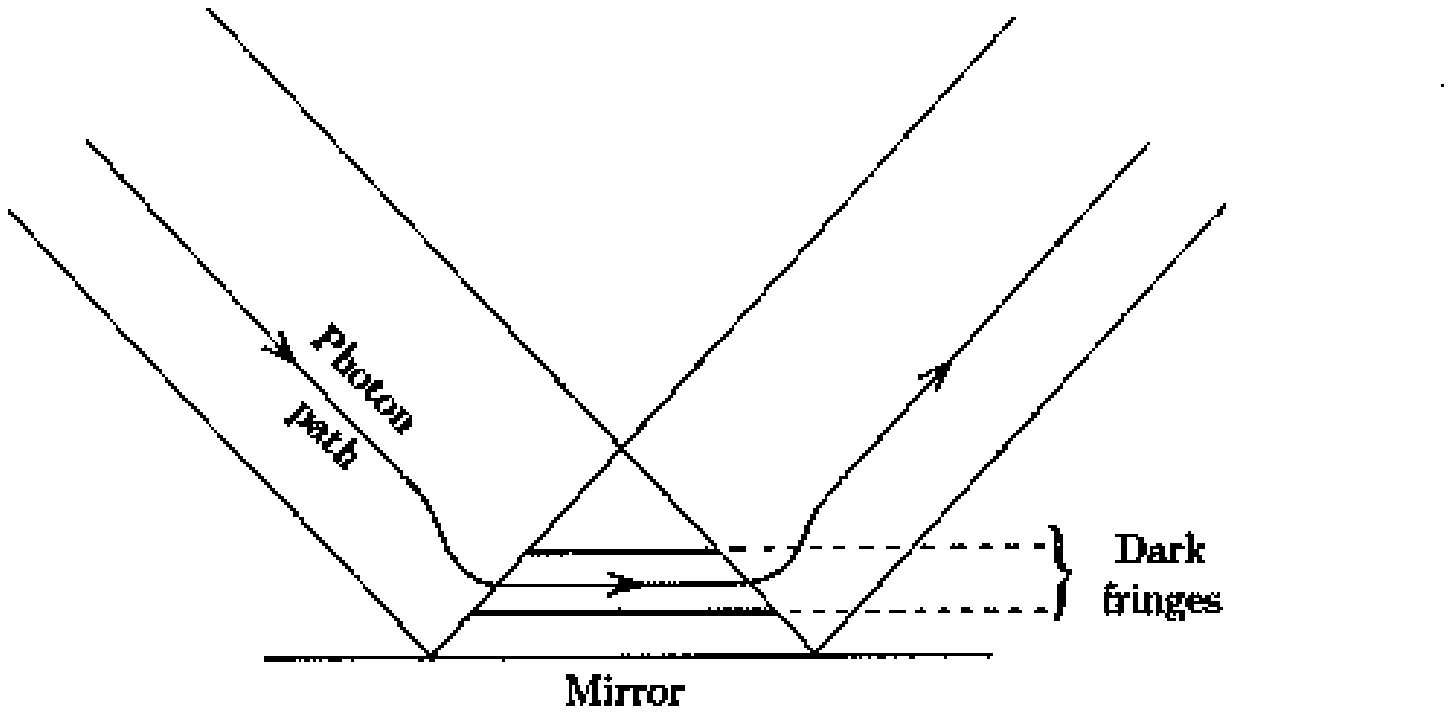}}
    \unnumberedcaption{Fig.~2.}\label{Brill-figure}
  \end{figure} 
plane mirror; the interference is produced in the region of overlap of two
beams. The trajectory of a photon will be as follows: at first a rectilinear
path in the incident beam, then a bending at the edge of the interference
zone, then a rectilinear path parallel to the mirror, with the photon
travelling in a bright fringe and avoiding the dark interference fringes;
then, when it comes out, the photon retreats following the direction of the
reflected light beam.

No photon actually strikes the mirror, nevertheless the mirror suffers
classical radiation pressure; it is in order to explain this fact that L. de
Broglie assumes the existence of special stresses in the interference zone;
these stresses, when added to the tensor of momentum flux transported by the
photons, reproduce the classical Maxwell tensor; there is then no difference
in the mechanical effects produced by the wave during its reflection by the mirror.

These remarks show how L.~de~Broglie's system of hypotheses preserves the
classical formulas, and avoids a certain number of awkward paradoxes. One thus
obtains, for example, the solution to a curious problem posed by G. N. Lewis\label{Lewisparadox}
(\textit{Proc.\ Nat.\ Acad.\ } \textbf{12} (1926), 22 and 439), which was the
subject of discussions between this author and R. C. Tolman and S. Smith
(\textit{Proc.\ Nat.\ Acad.\ } \textbf{12} (1926), 343 and 508).

Lewis assumed that the photons always follow the path of a light ray of
geometrical optics, but that they choose, among the different rays, only those
that lead from the luminous source to a bright fringe situated on an absorbing
body. He then considered a source S whose light is reflected by two mirrors AA%
\'{}
and BB%
\'{}%
; the light beams%
  \begin{figure}
    \centering
      \resizebox{\textwidth}{!}{\includegraphics[0mm,0mm][220.58mm,90.83mm]{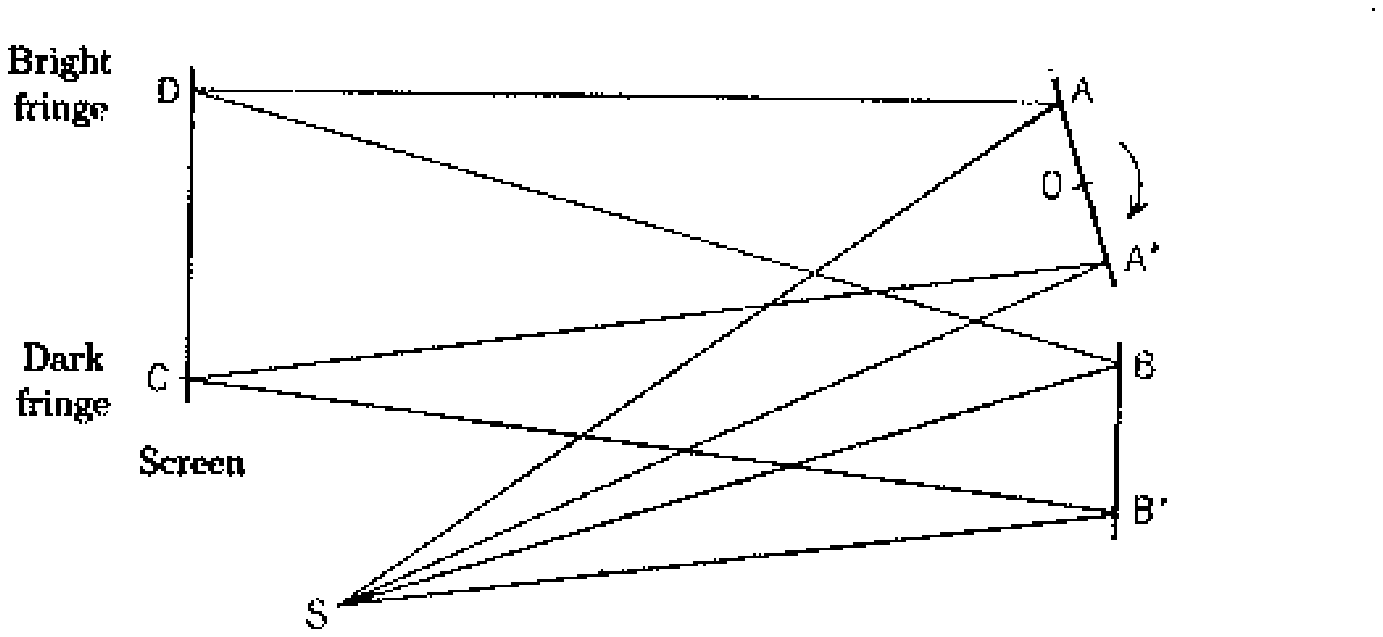}}
    \unnumberedcaption{Fig.~3.}
  \end{figure} 
overlap, producing interference [zones] in which one places a screen CD; the
dimensions are assumed to be such that there is a bright fringe on one of the
edges D of the screen and a dark fringe on the other edge C. Following the
hypothesis of Lewis, the photons would follow only the paths SBD and SAD,
which end at the bright fringe D; no photon will take the path SA%
\'{}%
C or SB%
\'{}%
C. All the photons come to strike the mirror AA%
\'{}
on the edge A, so one could predict that this mirror would suffer a torque; if
one made it movable around an axis O, it would tend to turn in the direction
of the arrow.

This paradoxical conclusion is entirely avoided by L. de Broglie, since his
system of hypotheses preserves the values of radiation pressure. This example
shows clearly that there is a contradiction between the hypothesis of
rectilinear paths for the photons (following the light rays) and the necessity
of finding photons only where a bright interference fringe is produced, no
photon going through the regions of dark fringes.\label{Brill-end}\\

{\sc Mr Lorentz} draws attention to a case where the classical
theory and the photon hypothesis lead to different results concerning the
ponderomotive forces produced by light. Let us consider reflection by the
hypotenuse face of a glass prism, the angle of incidence being larger than the
angle of total reflection. Let us place a\endnote{Misprinted as `au' instead
of `un'.} second prism behind the first, at a distance of the order of
magnitude of the wavelength, or only a fraction of this length. Then, the
reflection will no longer be total. The light waves that penetrate the layer
of air reach the second prism before their intensity is too much weakened, and
there give rise to a beam transmitted in the direction of the incident rays.

If, now, one calculates the Maxwell stresses on a plane situated in the layer
of air and parallel to its surfaces, one finds that, if the angle of incidence
exceeds a certain value (60$%
{{}^\circ}%
$ for example), there will be an \textit{attraction} between the two prisms.
Such an effect can never be produced by the motion of corpuscles, this motion
always giving rise to a [positive] pressure as in the kinetic theory of gases.

What is more, in the classical theory one easily sees the origin of the
`negative pressure'. One can distinguish two cases, that where the electric
oscillations are in the plane of incidence and that where this is so for the
magnetic oscillations. If the incidence is very oblique, the oscillations of
the incident beam that I have just mentioned are only slightly inclined with
respect to the normal to the hypotenuse face, and the same is true for the
corresponding oscillations in the layer of air.

One then has approximately, in the first case an electric field such as one
finds between the electrodes of a capacitor, and in the second case a magnetic
field such as exists between two opposite magnetic poles.

The effect would still remain if the second prism were replaced by a glass
plate, but it must be very difficult to demonstrate this experimentally.


\newpage

\renewcommand{\enoteheading}{\section*{Notes to the translation}}
\addcontentsline{toc}{section}{\it Notes to the translation}
\theendnotes

\setcounter{endnote}{0}
\setcounter{equation}{0}

\chapter*{Quantum mechanics$^{\scriptsize\hbox{a}}$}\markboth{{\it M.~Born and 
W.~Heisenberg}}{{\it Quantum mechanics}}
\addcontentsline{toc}{chapter}{Quantum mechanics ({\em M.~Born and 
W.~Heisenberg\/})}
\begin{center}{\sc by Messrs Max BORN and Werner HEISENBERG}\footnotetext[1]{Our translation
follows the German typescript in AHQP-RDN, document M-0309. Discrepancies 
between the typescript and the published version are reported in the endnotes.
The published version is reprinted in Heisenberg (1984, ser.~B, vol.~2, 
pp.~58--99) ({\em eds.}).}\end{center}

\

\begin{center}

\par
\Needspace{5\baselineskip}
{\sc Introduction}\addcontentsline{toc}{section}{Introduction}
\end{center}
Quantum mechanics is based on the intuition that the essential difference 
between atomic physics and classical physics is the occurrence of discontinuities (see in particular 
[1,4,58--63]).\footnote[2]{Numbers in square brackets 
refer to the bibliography at the end.} Quantum mechanics should thus 
be considered a direct continuation of the quantum theory founded 
by Planck, Einstein and Bohr. Bohr in particular stressed
repeatedly, already before the birth of quantum mechanics, that the 
discontinuities must lead to the introduction of new kinematical and 
mechanical concepts, so that indeed classical mechanics and its 
corresponding conceptual scheme should be abandoned [1,4]. 
Quantum mechanics tries to introduce the new concepts through a 
precise analysis of what is `observable in principle'. In fact, this does not mean 
setting up the principle that a sharp division between `observable' and 
`unobservable' quantities is possible and necessary. As soon as a conceptual 
scheme is given, one can infer from the observations to other facts that are actually not 
observable directly, and the boundary between `observable' and `unobservable'\endnote{Here 
and in a number of places in the following, the French edition omits quotation marks present 
in the German typescript. They are tacitly restored in this edition.} quantities becomes altogether 
indeterminate. But if the conceptual scheme itself is still unknown, it will be 
expedient to enquire only about the observations themselves, without drawing conclusions from them, 
because otherwise wrong concepts and prejudices taken over from before will block the way to 
recognising the physical relationships [Zusammenh\"{a}nge]. At the same time the new 
conceptual scheme provides the anschaulich content of the new theory.\footnote{For the notion 
of Anschaulichkeit, see the comments in sections~\ref{foundmat}, \ref{Schr-conflict} and 
\ref{visualisability} ({\em eds.}).} 
From a theory that is anschaulich in this sense, one can thus demand only 
that it is consistent in itself and that it allows one to predict unambiguously\label{unambiguously} the results 
for all experiments conceivable in its domain. Quantum mechanics is meant as
a theory that is in this sense anschaulich and complete for the micromechanical 
processes [46].\endnote{The French edition gives `[47]'.}

Two kinds of discontinuities are characteristic of atomic physics: the
existence of corpuscles (electrons, light quanta) on the one hand, and 
the occurrence of discrete stationary states (discrete\endnote{[diskrete] --- 
[d\'{e}termin\'{e}es]} energy values, momentum values etc.)\
on the other. Both kinds of discontinuities can be introduced in the 
classical theory only through artificial auxiliary assumptions. For quantum 
mechanics, the existence of discrete stationary 
states and energy values is just as natural as the existence of discrete 
eigenoscillations in a classical oscillation problem [4]. 
The existence of corpuscles will perhaps later turn out to be reducible 
just as easily to discrete stationary states of the wave processes
(quantisation of the electromagnetic waves on the one hand, and of the de Broglie waves on the other) [4], [54]. 

The discontinuities, as the notion of `transition probabilities' already shows, 
introduce a statistical element into atomic physics. This statistical element 
forms an {\em essential} part of the foundations of quantum mechanics 
(see in particular [4,30,38,39,46,60,61,62]);\endnote{The French edition omits `[60]'.} according to the latter, 
for instance, in many cases the course of an experiment is determinable   
from the initial conditions only statistically, at least if in fixing the initial conditions one
takes into account only the experiments conceivable in principle up to now. This 
consequence of quantum mechanics is empirically testable. 
Despite its statistical character, the theory nevertheless accounts for the 
apparently fully causal determination of macroscopic\endnote{[makroskopische] --- [microscopiques]} 
processes. In particular, the 
principles of conservation of energy and momentum hold exactly also in quantum mechanics.
There seems thus to be no empirical argument against accepting fundamental indeterminism
for the microcosm.

\par
\Needspace{5\baselineskip}
\begin{center} {\sc I. --- The mathematical methods of quantum 
mechanics}\footnote{Section \ref{beyond} contains additional material on 
the less familiar aspects of the formalisms presented here ({\em eds.}).}\addcontentsline{toc}{section}{I. 
--- The mathematical methods of quantum mechanics}\\
\end{center}
The phenomenon for whose study the mathematical formalism of quantum mechanics
was first developed is the spontaneous radiation of an excited atom. After
innumerable attempts to explain the structure of the line spectra with classical
mechanical models had proved inadequate, one returned to the direct description
of the phenomenon on the basis of its simplest empirical laws (Heisenberg [1]). 
First among these is Ritz's combination principle, according to
which the frequency of each spectral line of an atom
appears as the difference of two terms $\nu_{ik}=T_i-T_k$; thus the set of all 
lines of the atom will be best described by specifying a quadratic 
array [Schema], and since each line possesses besides its 
frequency also an intensity and a phase, one will write in each position of the array 
an elementary oscillation function with complex amplitude:
  \begin{equation}
    \left(\begin{array}{ccc}
            q_{11}e^{2\pi i\nu_{11}t}   &   q_{12}e^{2\pi i\nu_{12}t}  &  \ldots  \\
            q_{21}e^{2\pi i\nu_{21}t}   &   q_{22}e^{2\pi i\nu_{22}t}  &  \ldots  \\
            \ldots\ldots\ldots          &   \ldots\ldots\ldots         &  \ldots      
          \end{array}\right)\ .
    \label{eq1}
  \end{equation}
This array is understood as representing a coordinate $q$ 
as a function of time in a similar way as the totality of terms of the Fourier series 
  \[
    q(t)=\sum_n q_n e^{2\pi i\nu_n t},\quad\nu_n=n\nu_0
  \]
in the classical theory; except that now because of the two indices the sum no longer makes sense.
The question arises of which expressions correspond to functions of the 
classical coordinate, for instance to the square $q^2$. Now, such arrays 
ordered by two indices occur as {\em matrices} in mathematics in the theory of quadratic forms 
and of linear transformations; the composition of two linear transformations,
  \[
   x_k=\sum_l a_{kl}y_l\ , \qquad y_l=\sum_j b_{lj}z_j\ ,
  \]
to form a new one,
  \[
   x_k=\sum_j c_{kj}z_j\ ,
  \]
then corresponds to the composition or multiplication of the matrices
  \begin{equation}
    ab=c,\quad\mbox{that is,}\quad \sum_l a_{kl}b_{lj}=c_{kj}\ .
    \label{eq2}
  \end{equation}
This multiplication in general is {\em not} commutative. It is natural 
to apply this recipe to the array of the atomic oscillations (Born 
and Jordan [2], Dirac [3]); it is immediately evident that because of Ritz's
formula $\nu_{ik}=T_i-T_k$ no new frequencies appear, just as in the classical 
theory in the multiplication of two Fourier series, and herein 
lies the first justification for the procedure. By repeated application 
of additions and multiplications one can define arbitrary matrix functions. 

The analogy with the classical theory leads further to allowing as  
representatives of real quantities only those matrices that are `Hermitian', 
that is, whose elements go over to the complex conjugate numbers under 
permutation of the indices. The discontinuous nature of the atomic 
processes here is put into the theory from the start as empirically established.
However, this does not establish yet the connection with quantum theory and its 
characteristic constant $h$. This is also achieved, by carrying over the 
content\endnote{[durch sinngem\"{a}sse \"{U}bertragung] --- [par une extension logique]} 
of the Bohr-Sommerfeld quantum conditions in a form given by Kuhn and 
Thomas, in which they are written as relations between the Fourier coefficients 
of the coordinates $q$ and momenta $p$. In this way one obtains the matrix equation
  \begin{equation}
    pq-qp=\frac{h}{2\pi i}\cdot 1\ ,
    \label{eq3}
  \end{equation}
where $1$ means the unit matrix. The matrix $p$ thus does not `commute' with $q$. 
For several degrees of freedom the commutation relation (\ref{eq3}) holds
for every pair of conjugate quantities, while the $q_k$ commute with each other, 
the $p_k$ with each other, and also the $p_k$ with the non-corresponding $q_k$.

In order to construct the new mechanics (Born, Heisenberg and Jordan [4]), 
one carries over as far as possible the notions of the classical theory. 
It is possible to define the differentiation of a matrix with respect to 
time and that of a matrix function with respect to an argument matrix. 
One can thus carry over to the matrix theory the canonical equations
  \[
    \frac{dq}{dt}=\frac{\partial H}{\partial p}\ ,\qquad
    \frac{dp}{dt}=-\frac{\partial H}{\partial q}\ ,
  \]
where one should understand $H(p,q)$ 
as the same function of the matrices $p,q$ 
that occurs in the classical theory as a function of the numbers $p,q$. 
(To be sure,\endnote{In the French edition, the parenthetical remark
is given as a footnote.} ambiguity can occur because of the 
noncommutativity of the multiplication; for example, $p^2q$ is different 
from $pqp$.) This procedure was tested in simple examples (harmonic and 
anharmonic oscillator). Further, one can prove the theorem of conservation of 
energy, which for non-degenerate systems (all terms $T_k$ 
different from each other, or: all frequencies $\nu_{ik}$ different 
from zero) here takes the form: for the solutions $p,q$ of the 
canonical equations the Hamiltonian function $H(p,q)$ becomes a 
diagonal matrix $W$. It follows immediately that the elements of 
this diagonal matrix represent the terms $T_n$ of Ritz's formula 
multiplied by $h$ (Bohr's frequency condition). It is particularly 
important to realise that conversely the requirement
  \[
    H(p,q)=W\qquad\mbox{(diagonal matrix)}
  \]
is a complete substitute for the canonical equations of motion, and leads to unambiguously 
determined solutions even if one allows for degeneracies (equality of terms, 
vanishing frequencies).

By a matrix with elements that are harmonic functions of time, one can of course 
represent only quantities (coordinates) that correspond to time-periodic 
quantities of the classical theory. Therefore cyclic coordinates (angles), 
which increase proportionally to time, cannot be treated at present.\footnote{This 
point is taken up again shortly after eq.~(\ref{eq10}).  
({\em eds.}).} Nevertheless, one easily manages to subject rotating systems to the matrix method
by representing the Cartesian components of the angular 
momentum with matrices [4].\endnote{The typescript does not give the reference number, only the brackets.
The French edition omits the reference entirely. The mentioned results are to be found in section 4.1 
of Born, Heisenberg and Jordan (1926 [4]).} One obtains thereby expressions for the 
energy\footnote{Angular momentum is of course responsible for a 
characteristic splitting of the energy terms ({\em eds.}).} that 
differ characteristically from the
corresponding classical ones; for instance the modulus\endnote{Word omitted in the French edition.}
of the total angular momentum is not equal to $\frac{h}{2\pi}j$ $(j=0,1,2,\ldots)$, but to 
$\frac{h}{2\pi}\sqrt{j(j+1)}$, in accordance with empirical rules that Land\'{e} and others 
had derived from the term splitting in the Zeeman effect.\footnote{For Land\'{e}'s work on the
anomalous Zeeman effect, see Mehra and Rechenberg (1982a, sec.~IV.4, esp.\ \mbox{pp.~467--76} 
and \mbox{482--5}) ({\em eds.}).} 
Further, one obtains for the changes in the angular quantum numbers [Rotationsquantenzahlen]
the correct selection rules and intensity formulas, as had already  
been arrived at earlier by correspondence arguments and 
confirmed by the Utrecht\label{utrecht} observations.\footnote{For the `Utrecht observations' see 
Mehra and Rechenberg (1982a, sec.~VI.6, \mbox{pp.~647--8}) and Mehra and Rechenberg (1982b, sec.~III.4, 
esp.\ \mbox{pp.~154--61}) ({\em eds.}).}

Pauli [6], avoiding angular variables, even managed to work out 
the hydrogen atom with matrix mechanics, at least with regard to the 
energy values and some aspects of the intensities.

Asking for the most general coordinates for which the 
quantum mechanical laws are valid leads to the generalisation of the notions
of canonical variables and canonical transformations known from the classical 
theory. Dirac [3] has noted that the content of the expressions such as 
$\frac{2\pi i}{h}(p_kq_l-q_lp_k)-\delta_{kl}$, which appear in the commutation 
relations of the type (\ref{eq3}) corresponds\endnote{[sind sinngem\"{a}sse 
\"{U}bertragungen] --- [sont des extensions logiques]} to that  of the Poisson 
brackets, whose vanishing in classical mechanics characterises a system 
of variables as canonical. Therefore also in quantum mechanics one will 
denote as canonical every system of matrix pairs $p,q$ that satisfy the commutation 
relations, and as a canonical transformation every transformation that leaves these relations 
invariant. One can write these with the help of an arbitrary matrix 
$S$  in the form\footnote{Cf.\ p.~\pageref{unitarity} above ({\em eds.}).}  
  \begin{equation}
    P=S^{-1}pS,\quad Q=S^{-1}qS\ ,
    \label{eq4}
  \end{equation}
and in a certain sense this is the most general canonical transformation. 
Then for an arbitrary function one has
  \[
    f(P,Q)=S^{-1}f(p,q)S\ .
  \]
Now one can also carry over the main idea of the Hamilton-Jacobi 
theory [4]. Indeed, if the Hamiltonian function $H$ is given as a 
function of any known canonical matrices $p_0, q_0$, then the solution 
of the mechanical problem defined by $H$ reduces to finding a matrix $S$ that satisfies the equation
  \begin{equation}
    S^{-1}H(p_0,q_0)S=W\ .
    \label{eq5}
  \end{equation}
This is an analogue of the Hamilton-Jacobi differential equation of classical mechanics.

Exactly as there, also here perturbation theory can be treated most clearly 
with the help of equation (\ref{eq5}). If $H$ is given as a power 
series in some small parameter
  \[
    H=H_0+\lambda H_1+\lambda^2 H_2+\ldots
  \]
and the mechanical problem is solved for $\lambda=0$, that is, $H_0=W_0$ is 
known as a diagonal matrix, then the solution to (\ref{eq5}) can be obtained easily as a power series
  \[
    S=1+\lambda S_1+\lambda^2 S_2+\ldots
  \]
by successive approximations. Among the numerous applications of this procedure,
only the derivation of Kramers' dispersion formula shall be mentioned here, 
which results if one assumes that the light-emitting and the scattering systems are weakly coupled 
and if one calculates the perturbation on the latter ignoring the 
backreaction [4].\footnote{In other words, one considers just the scattering system under an external 
perturbation (Born, Heisenberg and Jotrdan [4], section 2.4, in particular eq.~(32)). See also Mehra 
and Rechenberg (1982c, ch.~III, esp.\ \mbox{pp.~93--4} and \mbox{103--9}) ({\em eds.}).}

The theory of the canonical transformations leads to a deeper conception, 
which later became essential in understanding the physical meaning of the formalism.

To each matrix $a=(a_{nm})$ one can associate a quadratic (more precisely: 
Hermitian) form\footnote{$\bar{\varphi}$ denotes the complex number conjugate to $\varphi$.}
  \[
    \sum_{nm}a_{nm}\varphi_n\bar{\varphi}_m
  \]
of a sequence of variables $\varphi_1,\varphi_2\ldots$, or also a linear transformation of 
the sequence of variables $\varphi_1,\varphi_2\ldots$, into another one 
$\psi_1,\psi_2\ldots$\endnote{Misprint in the French edition: 
summation index `$n$' in the equation.}
  \begin{equation}
    \psi_n=\sum_m a_{nm}\varphi_m\ ,
    \label{eq6}
  \end{equation}
where provisionally
the meaning of the variables $\varphi_n$ and $\psi_n$ 
shall be left unspecified; we shall return to this.

A transformation (\ref{eq6}) is called `orthogonal' if it maps the identity form into itself
  \begin{equation}
    \sum_n\varphi_n\bar{\varphi}_n=\sum_n\psi_n\bar{\psi}_n\ .
    \label{eq7}
  \end{equation}
Now these orthogonal transformations of the auxiliary variables $\varphi_n$ 
immediately turn out to be essentially identical to the canonical 
transformations of the $q$ and $p$ matrices; 
the Hermitian character and the commutation relations are preserved. Further, one can replace 
the matrix equation (\ref{eq5}) by the equivalent requirement [4]: the form
  \[
    \sum_{nm}H_{nm}(q_0,p_0)\varphi_n\bar{\varphi}_m
  \]
is to be transformed orthogonally into a sum of squares
  \begin{equation}
    \sum_n w_n\psi_n\bar{\psi}_n\ .
    \label{eq8}
  \end{equation}

The fundamental problem of mechanics is thus none other than the principal axes problem 
for surfaces of second order in infinite-dimensional space, occurring everywhere in 
pure and applied mathematics and variously studied. As is well known, this is equivalent 
to asking for the values of the parameter $W$ for which the linear equations
  \begin{equation}
    W\varphi_n=\sum_m H_{nm}\varphi_m
    \label{eq9}
  \end{equation}
have a non-identically vanishing solution. The values $W=W_1,W_2,\ldots$ 
are called eigenvalues of the form $H$; they are the energy values (terms) 
of the mechanical system. To each eigenvalue 
$W_n$ corresponds an eigensolution $\varphi_k=\varphi_{kn}$.\footnote{This 
is the notation used by Born and Heisenberg: the $n$th eigensolution is 
represented by an infinite vector with components labelled by $k$ ({\em eds.}).} 
The set of these eigensolutions evidently again forms a matrix and it is easy 
to see that this is identical with the transformation matrix $S$ appearing 
in (\ref{eq5}).\footnote{This point is made more explicit after eq.~(\ref{eq17}). 
See also the relevant contributions by Dirac and by Kramers in the general discussion, 
pp.~\pageref{Dirac-Schr} and \pageref{Kramers-Schr} ({\em eds.}).}

The eigenvalues, as is well known, are invariant under orthogonal transformations of the
$\varphi_k$,\endnote{[orthogonale Transformationen der $\varphi_k$] --- [transformations 
orthogonales $\varphi_k$]} and since these correspond to the canonical substitutions 
of the $p$ and $q$ matrices, one recognises immediately the canonical 
invariance of the energy values $W_n$.

While the quantum theoretical matrices do not belong to the class of matrices 
(finite and bounded infinite\endnote{[beschr\"{a}nkte unendliche] --- [partiellement 
infinies]}) investigated by the mathematicians 
(especially by Hilbert and his school), one can nevertheless carry over the main aspects of the 
known theory to the more general case. The precise formulation of these 
theorems\endnote{[S\"{a}tze] --- [principes]} has been recently given by J.~von Neumann [42] in a paper to 
which we shall have to return.\endnote{[noch zur\"{u}ckzukommen haben] --- [n'avons pas \`{a} revenir]}

The most important result that is achieved in this way is 
the theorem that a form cannot always be decomposed into a sum of squares (\ref{eq8}), but 
that there also occur invariant integral components
  \begin{equation}
    \int W\psi (W)\overline{\psi(W)}dW\ ,
    \label{eq10}
  \end{equation}
where the sequence of variables $\psi_1,\psi_2,\ldots$ has to be complemented 
by the continuous distribution $\psi(W)$.

In this way the continuous spectra appear in the theory in the 
most natural way. But this implies by no means that in this domain the classical theory comes again into its own.
Also here the characteristic discontinuities of quantum theory remain; 
also in the continuous spectrum a (spontaneous) state transition consists 
of a `jump' of the system from a point $W'$ to another one $W''$ with 
emission of a wave $q(W,W')e^{2\pi i\nu t}$ with the frequency 
$\nu=\frac{1}{h}(W'-W'')$.

The main defect of matrix mechanics consists in its clumsiness, even helplessness, in the 
treatment of non-periodic quantities, such as angular variables or coordinates that attain
infinitely large values (e.g.\ hyperbolic trajectories). To overcome this difficulty  
two essentially different routes have been taken, the operator calculus of Born and Wiener 
[21], and the so-called\endnote{Word omitted in the French edition.} q-number theory of Dirac [7].

The latter starts from the idea that a great part of the matrix relations 
can be obtained without an explicit representation of the matrices, simply on 
the basis of the rules for operating with the matrix symbols. These depart from the rules for numbers
only in that the multiplication is generally not commutative. Dirac therefore considers  
abstract quantities, which he calls q-numbers (as opposed to the ordinary c-numbers) 
and with which he operates according to the rules of the 
noncommutative algebra. It is therefore a kind of hypercomplex number 
system. 
The commutation relations are of course preserved. The theory acquires an extraordinary resemblance 
to the classical one; for instance, one can 
introduce angle and action variables 
$w,J$ and expand any q-number into a Fourier series with respect to the $w$; the 
coefficients are functions of the $J$ and turn out to be
identical to the matrix elements if one replaces the $J$ by integer multiples of $h$. By his method Dirac has 
achieved important results, for instance worked out the hydrogen atom independently of 
Pauli [7] and determined the intensity of radiation 
in the Compton effect [12].  A drawback of this formalism --- apart from the quite 
tiresome dealing with the noncommutative algebra --- is the 
necessity to replace at a certain point of the calculation certain q-numbers with ordinary 
numbers (e.g.\ $J=hn$), in order to obtain results comparable with experiment. 
Special `quantum conditions' which had disappeared from matrix mechanics are thus needed again.

The operator calculus differs from the q-number method in that it does not introduce 
abstract hypercomplex numbers, but concrete, constructible mathematical objects 
that obey the same laws, namely operators or functions in the space 
of infinitely many variables. The method is by Eckart [22] and was then developed 
further by many others following on from Schr\"{o}dinger's wave 
mechanics, especially by Dirac [38] and Jordan [39] and in an impeccable mathematical 
form by J.\ von Neumann [42]; it rests roughly on the following idea.

A sequence of variables $\varphi_1,\varphi_2,\ldots$ can be interpreted as 
a point in an infinite-dimensional space. If the 
sum of squares $\sum_n|\varphi_n|^2$ converges, then it represents a measure of distance, a 
Euclidean metric [Massbestimmung], in this space; this metric space of infinitely
many dimensions is called for short a Hilbert space. The canonical transformations of matrix 
mechanics correspond thus to the rotations of the Hilbert space. Now, however, one can also fix
a point in this space other than by the specification of discrete coordinates 
$\varphi_1,\varphi_2,\ldots\ $. Take for instance a complete, normalised orthogonal system of functions 
$f_1(q),f_2(q),\ldots\ $, that is one for which\endnote{The overbar
is missing in the original typescript (only here), but is included
in the French edition.} 
  \begin{equation}
    \int f_n(q)\overline{f_m(q)}dq=\delta_{nm}=\left\{\begin{array}{cl}
                        1   &  \mbox{for $n=m$}   \\
                        0   &  \mbox{for $n\neq m$\ ;}
                                      \end{array}\right.
    \label{eq11}
  \end{equation}
the variable $q$ can range here over an arbitrary, also multi-dimensional domain. 
If one then sets (Lanczos [23])\endnote{The typescript
reads: `Lanczos [\ ]', the reference number is added in the French edition.}
  \begin{equation}
    \left\{\begin{array}{lcl}
                \varphi(q)          & = &    \sum_n\varphi_nf_n(q)     \\[0.7ex]
                H(q',q'')    & = &    \sum_{nm}H_{nm}f_n(q')\overline{f_m(q'')}\ ,                      
              \end{array}\right.
    \label{eq12}
  \end{equation}
the linear equations (\ref{eq9})\endnote{The typescript consistently 
gives this reference as `(q)', the French edition as `(\ref{eq9})'.} turn into the integral 
equation\endnote{Equation number missing in the French edition.} 
  \begin{equation}
    W\varphi(q')=\int H(q',q'')\varphi(q'')dq''\ .
    \label{eq13}
  \end{equation}
This relation established through (\ref{eq12}) means thus nothing but a 
change of the coordinate system in the Hilbert space, given by the orthogonal 
transformation matrix $f_n(q)$ with one discrete and one continuous index.

One sees thus that the preference for `discrete' coordinate 
systems in the original version of the matrix theory is by no means something 
essential. One can just as well use `continuous matrices' such as 
$H(q',q'')$. Indeed, the specific representation of a point in
the Hilbert space by projection onto certain orthogonal coordinate 
axes does not matter at all; rather, one can summarise equations 
(\ref{eq9}) and (\ref{eq13}) in the more general equation
  \begin{equation}
     W\varphi=H\varphi\ ,
    \label{eq14}
  \end{equation}
where $H$ denotes a linear operator which transforms the point $\varphi$ of the Hilbert 
space into another. The equation requires to find those points $\varphi$ which under the 
operation $H$ only suffer a displacement along the line joining them to the 
origin.\endnote{[eine Verschiebung l\"{a}ngs ihrer Verbindungslinie mit dem Nullpunkt] --- 
[un d\'{e}placement de leur droite de jonction avec 
l'origine]} The points satisfying this condition determine an orthogonal  
system of axes, the principal axes frame of the operator $H$; 
the number of axes is finite or infinite, in the latter case distributed discretely or 
continuously, and the eigenvalues $W$ are the lengths of the principal axes. The linear 
operators in the Hilbert space are thus the general concept that can serve to 
represent a physical quantity mathematically. The calculus with operators proceeds obviously  
according to the same rules as the one with Dirac's q-numbers; they\endnote{[sie] --- [ces r\`{e}gles]} 
constitute a realisation of this abstract notion. So far we have analysed the 
situation with the example of the Hamiltonian function, but 
the same holds for any quantum mechanical quantity. 
Any coordinate $q$ can be written, instead of as a matrix with discrete indices $q_{nm}$, also
as a function of two continuous variables $q(q',q'')$ by projection onto an orthogonal system 
of functions, or, more generally, can also be considered 
as a linear operator in the Hilbert space; then it has eigenvalues that are invariant, 
and eigensolutions with respect to each orthogonal coordinate system. 
The same holds for a momentum $p$ and every function of $q$ and $p$,
indeed for every quantum mechanical `quantity'. While in the classical theory
physical quantities are represented by variables that can take numerical values from 
an arbitrary value range, a physical quantity in quantum theory is represented by a linear 
operator and the stock of values that it can take by the eigenvalues of the corresponding 
principal axes problem in the Hilbert space.

In this view, Schr\"{o}dinger's wave mechanics [24] appears formally as a special case. 
The simplest operator whose characteristic values are all the real numbers, is in fact the 
multiplication of a function $F(q)$ by the real number $q$; one writes it simply $q$.
Then, however, the eigenfunctions are `improper' 
functions; for according to (\ref{eq14}) they must have the property of being everywhere 
zero except if $W=q$. Dirac [38] has 
introduced for the representation of such improper functions the `unit function' $\delta(s)$, which should always be 
zero when $s\neq 0$, but for which nonetheless $\int_{-\infty}^{+\infty}\delta(s)ds=1$
should hold. Then one can write down the (normalised) eigenfunctions 
  \begin{equation}
    \varphi(q,W)=q\delta(W-q)
    \label{eq15}
  \end{equation}
belonging to the operator $q$.

The conjugate to the operator $q$ is the differential operator
  \begin{equation}
    p=\frac{h}{2\pi i}\frac{\partial}{\partial q}\ ;
    \label{eq16}
  \end{equation}
indeed, the commutation relation (\ref{eq3}) holds, which means just the trivial 
identity
  \[
    (pq-qp){\cal F}(q)=\frac{h}{2\pi i}\left\{\frac{d}{dq}(q{\cal F})
    -q\frac{d{\cal F}}{dq}\right\}=\frac{h}{2\pi i}{\cal F}(q)\ .
  \]

If one now constructs a Hamiltonian 
function out of $p,q$ (or out of several such conjugate pairs), 
then equation (\ref{eq14}) becomes a differential equation for the quantity 
$\varphi(q)$:
  \begin{equation}
    H(q,\frac{h}{2\pi i}\frac{\partial}{\partial q})\varphi(q)=
    W\varphi(q)\ .
    \label{eq17}
  \end{equation}
This is Schr\"{o}dinger's wave equation, which appears here as a special case of the 
operator theory. The most important point about this formulation of the quantum laws 
(apart from the great advantage of connecting to known mathematical methods) is the
replacement of all `quantum conditions', such as were still necessary in Dirac's theory of
q-numbers, by the simple requirement that the eigenfunction $\varphi(q)=\varphi(q,W)$
should be everywhere finite in the domain of definition of the variables $q$; 
from this, in the event, a discontinuous spectrum of eigenvalues $W_n$ (along with continuous ones) 
arises automatically.
But Schr\"{o}dinger's eigenfunction $\varphi(q,W)$ is actually nothing but 
the transformation matrix $S$ of equation (\ref{eq5}), which one can indeed also write 
in the form
  \[
    HS=SW\ ,
  \]
analogous to (\ref{eq17}).

Dirac [38] has made this state of affairs even clearer by 
writing the operators $q$ and $p$ and thereby also $H$ as integral operators, as in 
(\ref{eq13}); then one has to set
  \begin{equation}
    \begin{array}{rclcl}
             q{\cal F}(q')  & = &  {\displaystyle  \int q''\delta(q'-q''){\cal F}(q'')dq''  }
                            & = &  q'{\cal F}(q')\ ,                          \\
             p{\cal F}(q')  & = & {\displaystyle \int\frac{h}{2\pi i}\delta'(q'-q'')
                                   {\cal F}(q'')dq''    }
                            & = &  {\displaystyle  \frac{h}{2\pi i}\frac{d{\cal F}}{dq'}\ ,   }
           \end{array}
    \label{eq18}
  \end{equation}
where, however, the occurrence of the derivative of the singular 
function $\delta$ has to be taken into the bargain. Then Schr\"{o}dinger's 
equation (\ref{eq17}) takes the form (\ref{eq13}).

The direct passage to the matrix representation in the strict
sense takes place by inverting the formulas (\ref{eq12}), in which one identifies
the orthogonal system $f_n(q)$ with the eigenfunctions 
$\varphi(q,W_n)$ belonging to the discrete spectrum.
If $T$ is an arbitrary operator (constructed from $q$ and 
$p=\frac{h}{2\pi i}\frac{\partial}{\partial q}$), define the 
corresponding matrix $T_{nm}$ by the coefficients of the
expansion
  \begin{equation}
    T\varphi_n(q)=\sum_mT_{nm}\varphi_m(q)
    \label{eq19}
  \end{equation}
or \addtocounter{equation}{-1}
  \begin{subequations}
  \begin{align}
    T_{nm}=\int\overline{\varphi_m(q)}T\varphi_n(q)dq\ ;
    \label{eq19a}
    \end{align}
  \end{subequations}
then one easily sees that equation (\ref{eq17}) is equivalent 
to (\ref{eq9}).

The further development of the formal theory has taken place
in close connection with its physical interpretation, to which 
we therefore turn first.

\par
\Needspace{5\baselineskip}
\begin{center} {\sc II. --- Physical interpretation}\addcontentsline{toc}{section}{II. --- Physical interpretation}\\
\end{center}
The most noticeable\label{noticeable} defect of the original matrix mechanics consists in 
the fact that at first it appears to give information not about actual phenomena, 
but rather only about possible states and processes. It allows one to calculate
the possible stationary states of a system; further it makes a statement about the nature of the harmonic oscillation 
that can manifest itself as a light wave in a quantum jump.
But it says nothing\label{nothing} about when a given state is present, or when a change is to 
be expected. The reason for this is clear: matrix mechanics 
deals only with closed periodic systems, and in these there are indeed no changes. 
In order to have true processes,\endnote{[Vorg\"{a}nge] --- [ph\'{e}nom\`{e}nes]} as long 
as one remains in the domain of matrix mechanics,  one must direct one's attention
to a {\em part} of the system; this is no longer closed and enters into interaction
with the rest of the system. The question is what matrix
mechanics can tell us about this.

Imagine, for instance, two systems 1 and 2 weakly coupled to each other
(Heisenberg [35], Jordan [36]).\footnote{The form of the result as 
given here is similar to that in Heisenberg [35]. For further details,
see the discussion in section~\ref{resonance} ({\em eds.}).} For the total 
system conservation of energy then holds; that is, $H$ is a diagonal 
matrix. But for a subsystem, for instance 1, $H^{(1)}$ 
is not constant, the matrix has elements off the diagonal.\endnote{Both 
the manuscript and the French edition read `$H_1$' and `$H_2$' in this 
paragraph and two paragraphs later, and `$H^{(1)}$' in the intervening 
paragraph. We have uniformised the notation.} The energy 
exchange can now be interpreted in two ways: for one, the periodic
elements of the matrix of $H^{(1)}$ (or of $H^{(2)}$) represent
a slow beating, a continuous oscillation of the energy to and fro; 
but at the same time, one can also describe the process with the concepts of the 
discontinuum theory and say that system 1 performs
quantum jumps and carries over the energy that is thereby 
freed to system 2 as quanta, and vice versa. But one can now show 
that these two apparently very different views do not
contradict each other at all. This rests on a mathematical 
theorem that states the following:

Let $f(W_n^{(1)})$ be any function of the energy values $W_n^{(1)}$ of
the isolated subsystem 1; if one forms the same function of the matrix
$H^{(1)}$ that represents the energy of
system 1 in the presence of the coupling to system 2, then $f(H^{(1)})$
is a matrix that does not consist only of diagonal elements 
$f(H^{(1)})_{nn}$. But these represent the time-averaged value of the 
quantity $f(H^{(1)})$. The effect of the coupling is thus measured by 
the difference\endnote{The French edition consistently reads `$\overline{\delta f_n}$'.}
  \[
    \delta\overline{f_n}= f(H^{(1)})_{nn}- 
    f(W^{(1)}_n)\ .
  \]
The first part of the said theorem now states that
$\delta\overline{f_n}$ can be brought into the form\endnote{The right-hand side of this equation
reads `$\sum_m \{f(W_n)-f(W_m)\}\Phi_{nm}$' in both the typescript and the French edition, but it should be as shown 
(see above, p.~\pageref{Heisenberg-correction}).}
  \begin{equation}
    \delta\overline{f_n}=\sum_m \{f(W_m)-f(W_n)\}\Phi_{nm}\ .
    \label{eq20}
  \end{equation}
This can be interpreted thus: the time average of the change in
$f$ due to the coupling is the arithmetic mean, with certain 
weightings $\Phi_{nm}$, of all possible jumps
of $f$ for the isolated system.

These $\Phi_{nm}$ will have to be called `transition probabilities'. 
The second part of the theorem determines the $\Phi_{nm}$ through 
the features of the coupling. Namely, if $p^0_1,q^0_1,p^0_2,q^0_2$ are coordinates satisfying the evolution
equations of the uncoupled systems, for which therefore $H^{(1)}$ and $H^{(2)}$ on 
their own are diagonal matrices, one can then think of 
the energy, including the interaction, as expressed as a function of 
these quantities. Then the solution of the mechanical problem according 
to (\ref{eq5})\endnote{The French edition gives `(\ref{eq2})'.} reduces 
to constructing a matrix $S$ that satisfies the equation
  \[
    S^{-1}H( p^0_1,q^0_1,p^0_2,q^0_2)S=W\ .
  \]
Denoting the states of system 1 by $n_1$, those of system 2 by $n_2$,
a state of the total system is given by 
$n_1n_2$,\endnote{Both the typescript and the French edition read (only here) `$n_1,n_2$'.} 
and to each transition $n_1n_2\rightarrow m_1m_2$ corresponds an element of 
$S$, $S_{n_1n_2,m_1m_2}$. Then the result is:\endnote{Both the typescript and 
the French edition read `$\Phi_{nm}$'.}
  \begin{equation}
    \Phi_{n_1m_1}=\sum_{n_2m_2}|S_{n_1n_2,m_1m_2}|^2\ .
    \label{eq21}
  \end{equation}
The squares of the elements of the $S$-matrix thus determine the
transition probabilities. The individual sum term 
$|S_{n_1n_2,m_1m_2}|^2$ 
in (\ref{eq21}) obviously means
that component of the transition probability for the jump $n_1\rightarrow m_1$
of system 1 that is induced by the jump $n_2\rightarrow m_2$
of system 2. 

By means of these results the contradiction between the two views from which 
we started is removed. Indeed, for the mean values, which alone may be observed, the conception of 
continuous beating always leads to the same result as the conception of quantum jumps.

If one asks the question {\em when} a quantum jump occurs, the theory provides no answer.
At first it seemed as if there were a gap here which might 
be filled with further probing. But soon it became apparent 
that this is not so, rather, that it is a failure of principle, 
which is deeply anchored in the nature of the possibility of 
physical knowledge [physikalisches Erkenntnisverm\"{o}gen].

One sees that quantum mechanics yields mean values correctly, 
but cannot predict the occurrence of an individual event. Thus 
determinism, held so far to be the foundation of the exact 
natural sciences, appears here to go no longer unchallenged. 
Each further advance in the interpretation of the formulas has 
shown that the system of quantum mechanical formulas can be 
interpreted consistently only from the point of view of a 
fundamental indeterminism, but also, at 
the same time, that the totality of empirically 
ascertainable facts can be reproduced by the system of the theory.

In fact, almost all observations in the field of atomic physics 
have a statistical character; they are countings, for instance 
of atoms in a certain state. While the determinateness
of an individual process is assumed by classical physics, in fact it plays practically
no role, because the microcoordinates that exactly determine an atomic process can 
never all be given; therefore by averaging they are eliminated 
from the formulas, which thereby
become statistical statements. It has become apparent that quantum mechanics
represents a merging of mechanics and statistics, in which the unobservable microcoordinates are eliminated.

The clumsiness\label{clumsiness} of the matrix theory in the description of 
processes developing in time can be avoided by making use of the more general 
formalisms\endnote{Singular in the French edition.} we have described above. In the general equation
(\ref{eq14}) one can easily introduce time explicitly by invoking the theorem of 
classical mechanics that energy $W$ and time $t$ behave as canonically conjugate 
quantities; in quantum mechanics it corresponds to having 
a commutation relation
  \[
    Wt-tW=\frac{h}{2\pi i}\ .
  \]
Thus for $W$ one can posit the operator 
$\frac{h}{2\pi i}\frac{\partial}{\partial t}$.
Equation (\ref{eq14}) then reads
  \begin{equation}
    \frac{h}{2\pi i}\frac{\partial\varphi}{\partial t}=
   H\varphi\ ,
    \label{eq22}
  \end{equation}
and here one can consider $H$ as depending explicitly on time. A special case of this is the 
equation \addtocounter{equation}{-1}
  \begin{subequations}
    \begin{align}
    \left\{H(q,\frac{h}{2\pi i}\frac{\partial}{\partial q})
    -\frac{h}{2\pi i}\frac{\partial}{\partial t}\right\}\varphi(q)=0\ , 
    \label{eq22a}
    \end{align}
  \end{subequations}
given by Schr\"{o}dinger [24],\endnote{The typescript includes the square 
brackets but no reference number. The French edition omits the reference entirely.} 
which stands to (\ref{eq17}) in the same relation as (\ref{eq22}) to (\ref{eq14}), 
as well as the form: \addtocounter{equation}{-1}
  \begin{subequations}
     \begin{align}\addtocounter{equation}{1}
    \vspace{-2em}
\frac{h}{2\pi i}\frac{\partial\varphi(q')}{\partial t}=\int H(q',q'')\varphi(q'')dq''\ ,
    \label{eq22b}
    \end{align}
  \end{subequations}
much used by Dirac, which relates to the integral 
formula (\ref{eq13}). Essentially, the introduction of time as a numerical variable 
reduces to thinking of the system under consideration\label{consideration} as coupled to 
another one and neglecting the reaction on the latter. But this 
formalism is very convenient and leads to a further development of the statistical 
view,\footnote{See the discussion in section~\ref{stateprob} ({\em eds.}).} 
namely, if one considers the case where
an explicitly time-dependent 
perturbation $V(t)$ is added to a time-independent energy function $H^0$, so that one has the equation
  \begin{equation}
    \frac{h}{2\pi i}\frac{\partial\varphi}{\partial t}=\left\{ H^0+V(t)\right\}\varphi\ 
    \label{eq23}
  \end{equation}
(Dirac [37], Born [34]).\endnote{Only brackets in 
the typescript, references omitted in the French edition.} Now if $\varphi^0_n$ are the eigenfunctions of
the operator $H^0$, which for the sake of simplicity we assume
to be discrete, the desired quantity $\varphi$ can be expanded in terms 
of these:
  \begin{equation}
    \varphi(t)=\sum_n c_n(t)\varphi^0_n\ .
    \label{eq24}
  \end{equation}
The $c_n(t)$ are then the coordinates of $\varphi$ in the Hilbert space
with respect to the orthogonal system $\varphi^0_n$; they can be calculated from
the differential equation (\ref{eq23}), if their initial values $c_n(0)$ 
are given. The result can be expressed as:
  \begin{equation}
    c_n(t)=\sum_m S_{nm}(t)c_m(0)\ ,
    \label{eq25}
  \end{equation}
where $S_{nm}(t)$ is an orthogonal matrix 
depending\endnote{[abh\"{a}ngige] --- [ind\'{e}pendante]} 
on $t$ and determined by $V(t)$.

The temporal process is thus represented by a rotation of the Hilbert space
or by a canonical transformation (\ref{eq4}) with the time-dependent
matrix $S$.

Now how is one to interpret this?

From the point of view of Bohr's theory\label{BHviewpoint} a system can always be in only  
{\em one} quantum state. To each of these belongs an eigensolution $\varphi^0_n$ of the unperturbed system.
If now one wishes to calculate what happens to 
a system that is initially in a certain state, 
say the $k$th, one has to choose $\varphi=\varphi^0_k$ as the initial condition for equation (\ref{eq23}), i.e.\ 
$c_n(0)=0$ for $n\neq k$, and $c_k(0)=1$. 
But then, after the perturbation is over, $c_n(t)$ will have become equal to $S_{nk}(t)$, and 
the solution consists of a superposition of eigensolutions. According to Bohr's 
principles it makes no sense to say a system is simultaneously in several states. 
The only possible interpretation seems to be statistical: 
the superposition of several eigensolutions expresses that through the perturbation the initial
state can go over to any other quantum state, and it 
is clear that as measure for the transition probability one has to take
the quantity
  \[
    \Phi_{nk}=|S_{nk}(t)|^2\ ;
  \]
because then one obtains again equation
(\ref{eq20}) for the average change of any state function.

This interpretation is supported by the fact that one establishes the validity of Ehrenfest's 
adiabatic theorem (Born [34]); one can show that under an infinitely slow action, one has
  \[
    \Phi_{nn}\rightarrow 1,\qquad\Phi_{nk}\rightarrow 0\qquad  (n\neq k)\ ,
  \]
that is, the probability of a jump tends to zero.

But this assumption also leads immediately to an interpretation of the $c_n(t)$
themselves: the $|c_n(t)|^2$ must be the state probabilities\label{mustbe}  [Zustandswahrscheinlichkeiten].

Here, however,\label{BH-beginning} one runs into a difficulty of principle that is of great importance, 
as soon as one starts from an initial state for which not 
all the $c_n(0)$ except one vanish. Physically, this case occurs if a system is 
given for which one does not know exactly the quantum 
state in which it is, but knows only the probability $|c_n(0)|^2$ for each quantum 
state. As a matter of fact, the phases [Arcus] of the complex quantities $c_n(0)$ still 
remain indefinite; if one sets $c_n(0)=|c_n(0)|e^{i\gamma_n}$, then
the $\gamma_n$ denote some phases whose meaning needs to be established.
The probability distribution at the end of the perturbation  according 
to (\ref{eq25}) is then
  \begin{equation}
    |c_n(t)|^2=\Big| \sum_m S_{nm}(t)c_m(0)\Big|^2
    \label{eq26}
  \end{equation}
and not
  \begin{equation}
    \sum_m|S_{nm}(t)|^2|c_m(0)|^2\ ,
    \label{eq27}
  \end{equation}
as one might suppose from the usual probability calculus.

Formula (\ref{eq26}), following Pauli\label{Pauli}, can be 
called the theorem of the {\em interference of probabilities}; its deeper meaning has become clear only
through the wave mechanics of de Broglie and Schr\"{o}dinger, which we shall 
presently discuss. Before this, however, it should be noted 
that this `interference' does not represent a contradiction\label{Born-contradiction} with the rules 
of the probability calculus, that is, with the assumption that the $|S_{nk}|^2$ 
are quite usual probabilities.\footnote{The notation $|S_{nm}|^2$ would probably be clearer, at least 
according to the reading of this passage proposed in section~\ref{on-interference} ({\em eds.}).} In fact, 
the composition rule (\ref{eq27}) follows from the concept of probability for the problem 
treated here when and only when the relative number, that is, the probability 
$|c_{n}|^2$ of the atoms in the state $n$, has been {\em established} beforehand 
{\em experimentally}.\endnote{The French edition reads 
`$(c_{nk})^2$' instead of `$|c_{n}|^2$' and `$nk$' instead of 
`$n$'.} In this case the phases $\gamma_n$  
are unknown in principle,\endnote{The French edition reads `$p_{nk}$' instead of `$\gamma_n$'.} 
so that (\ref{eq26}) then naturally goes over to 
(\ref{eq27}) [46].

It should be noted further that the formula (\ref{eq26}) 
goes over to the expression (\ref{eq27}) if the perturbation function 
proceeds totally irregularly as a function of time. That is for instance 
the case when the perturbation is produced by `white light'.\footnote{Compare also Born's 
discussion in Born (1926c [34]) ({\em eds.}).} Then, on average, 
the surplus terms in (\ref{eq26}) drop out and one obtains (\ref{eq27}). In
this way it is easy to derive the Einstein coefficient\label{forpage315} $B_{nm}$ for the 
probability per unit radiation of the quantum jumps induced by light 
absorption (Dirac [37], Born [30]). But, in general, according to (\ref{eq26})
the knowledge of the probabilities $|c_n(0)|^2$ is by no means sufficient to 
calculate the course of the perturbation, rather one has to know 
also the phases $\gamma_n$.

This circumstance recalls vividly the behaviour of light in interference
phenomena. The intensity of illumination on a screen is by no means always 
equal to the sum of the light intensities of the individual
beams of rays that impinge on the screen, or, as
one can well say, it is by no means equal to the sum of 
the light quanta that move in the individual beams; instead it depends 
essentially on the phases of the waves. Thus at this point an analogy between the quantum
mechanics of corpuscles and the wave theory of light becomes apparent.\label{BH-end}

As a matter of fact this connection was found by de Broglie in
a quite different way. It is not our purpose to discuss this. It is enough to formulate 
the result of de Broglie's considerations, and their further development by 
Schr\"{o}dinger, and to put it in relation to quantum mechanics.

The dual nature of light --- waves, light quanta --- corresponds 
to the analogous dual nature of material particles; these also behave 
in a certain respect like waves. Schr\"{o}dinger has set up the laws of 
propagation of these waves [24] and has arrived at equation [(\ref{eq17})],\endnote{Both
the typescript and the French edition give `(\ref{eq11})', but this should evidently be
either `(\ref{eq17})' or `(\ref{eq22a})'.}
here derived in a different way. His view, however, that these waves exhaust the
essence of matter and that particles are nothing but wave packets,
not only stands in contradiction with the principles of Bohr's empirically 
very well-founded theory, but also leads to impossible conclusions; here therefore 
it shall be left to one side. Instead we attribute a dual 
nature to matter also: its description requires both corpuscles (discontinuities) 
and waves (continuous processes). From the viewpoint of the statistical
approach to quantum mechanics it is now clear why these can be reconciled: 
the waves are probability waves. Indeed, it is not the probabilities themselves, rather
certain `probability amplitudes' that propagate continuously  and obey differential or 
integral equations, as in classical continuum physics; but additionally there are
discontinuities, corpuscles whose frequency is governed by the square of these 
amplitudes.

The most definite support for this conception is given by collision phenomena for material
particles (Born [30]). Already Einstein [16], when he deduced from 
de Broglie's daring theory the possibility of 
`diffraction' of material particles,\label{for-first-papers}\footnote{Note that the first prediction of such
diffraction appears in fact to have been made by de Broglie in 1923; cf.\ section~\ref{first-papers} 
({\em eds}.).} tacitly assumed that it is the particle number that is determined by the intensity of
the waves. The same occurs in the interpretation given by
Elsasser [17] of the experiments by Davisson and 
Kunsman [18,19] on the reflection of electrons by crystals; also here
one assumes directly that the {\em number} of electrons is a maximum in the
diffraction maxima.
The same holds for Dymond's [20] experiments on the diffraction
of electrons by helium atoms.

The application of wave mechanics to the calculation of collision processes
takes a form quite analogous to the theory of diffraction of light by small 
particles. One has to find the solution to Schr\"{o}dinger's wave equation 
(\ref{eq17}) that goes over at infinity to a given incident 
plane wave; this solution behaves everywhere at infinity like an 
outgoing\endnote{The adjective is omitted in the French edition.} spherical wave.
The intensity of this spherical wave in any direction compared to the intensity of the 
incoming wave determines the relative number of particles
deflected in this direction from a parallel ray. As a measure of the intensity 
one has to take a `current vector'\endnote{[`Strahlvektor'] --- [`vecteur radiant']} which can be
constructed from the solution $\varphi(q,W)$, and
which is formed quite analogously to the Poynting vector of the electromagnetic theory 
of light, and which measures the number of particles crossing a unit surface in unit time.

In this way Wentzel [31] and Oppenheimer [32] have derived wave 
mechanically the famous Rutherford law for the scattering of $\alpha$-particles 
by heavy nuclei.\footnote{Cf. Born (1969, Appendix XX). The current 
vector, as defined there, is the usual 
${\bf j}=\frac{h}{2\pi i}\frac{1}{2m}(\psi^*{\bf \nabla}\psi-\psi{\bf\nabla}\psi)$ ({\em eds.}).}

If one wishes to calculate the probabilities of excitation and ionisation of atoms [30], then one 
must introduce the coordinates of the atomic electrons as variables on an equal footing with those 
of the colliding electron. The waves then propagate no longer in three-dimensional space but in multi-dimensional
configuration space. From this one sees that the quantum mechanical waves are indeed something
quite different from the light waves of the classical theory.

If one constructs the current vector just defined for a solution of the generalised Schr\"{o}dinger equation 
(\ref{eq22}), which describes time evolution, 
one sees that the time derivative of the integral
  \[
    \int |\varphi|^2dq'\ ,
  \]
ranging over an arbitrary domain of the independent numerical variables $q'$, 
can be transformed into the surface integral of the current vector over the boundary 
of that domain. From this it emerges that 
  \[
    |\varphi|^2
  \]
has to be interpreted as particle density or, better, as probability density. The solution 
$\varphi$ itself is called `probability amplitude'.

The amplitude $\varphi(q',W')$ belonging to a stationary state thus yields via 
$|\varphi(q',W')|^2$ the probability that for given energy $W'$ the coordinate $q'$ is\label{falls} 
in some given element $dq'$.\endnote{The absolute square is missing in the German typescript, 
but is added in the French edition.} But this can be generalised immediately. In fact,
$\varphi(q',W')$ is the projection of the principal
axis $W'$ of the operator $H$ onto the principal axis $q'$ of the 
operator $q$. One can therefore say in general (Jordan [39]): if two physical
quantities are given by the operators $q$ and $Q$ and if one knows the principal axes of the former, for 
instance, according to magnitude and 
direction,\endnote{[des einen, etwa, nach Gr\"{o}sse und Richtung] --- [de l'un, 
par exemple en grandeur et en direction]} then from the equation
  \[
    Q\varphi(q',Q')=Q'\varphi(q',Q')
  \]
one can determine the principal axes $Q'$ of $Q$\endnote{[von $Q$] --- [et $Q$]} 
and their projections $\varphi(q',Q')$ on the axes of $q$. Then $|\varphi(q',Q')|^2dq'$ 
is the probability that for given $Q'$ the value of $q'$ falls in a given interval $dq'$.

If conversely one imagines the principal axes of $Q$ as given, 
then those of $q$ are obtained\endnote{The French edition has a prime on `$q$'.} through 
the inverse rotation; from this one easily recognises that 
$\overline{\varphi(Q',q')}$ is the corresponding amplitude,\endnote{The overbar 
is missing in the German typescript, but is added in the French edition.} 
so that $|\varphi(Q',q')|^2dQ'$ means the probability, given $q'$, to find\label{find} 
the value of $Q'$ in $dQ'$. If for instance one takes for $Q$ the operator
$p=\frac{h}{2\pi i}\frac{\partial}{\partial q}$, then one has the equation
  \[
    \frac{h}{2\pi i}\frac{\partial\varphi}{\partial q'}=p'\varphi\ ,
  \]
thus
  \begin{equation}
    \varphi=Ce^{\frac{2\pi i}{h}q'p'}\ .
    \label{eq28}
  \end{equation}
This is therefore the probability amplitude for a pair of conjugate quantities. 
For the probability density one obtains $|\varphi|^2=C$, that is, for given $q'$ every value $p'$ is equally probable.

This is an important result, since it allows one to retain the concept of `conjugate quantity'
even in the case where the differential definition fails, namely when
the quantity $q$ has only a discrete spectrum or even when it is only capable of taking
finitely many values. The latter for instance is the case for angles with quantised 
direction [richtungsgequantelte Winkel],\footnote{This was a standard term referring to the 
fact that in the presence of an external magnetic field, the projection of the angular momentum 
in the direction of the field has to be quantised (quantum number $m$). Therefore, the direction 
of the angular momentum with respect to the magnetic field can be said to be quantised. Cf. Born
(1969, p.~121) ({\em eds.}).} say for the magnetic electron, or in the 
Stern-Gerlach experiments. One can then, as Jordan does, call by definition a quantity $p$  
conjugate to $q$, if the corresponding probability amplitude has the expression (\ref{eq28}).

As the amplitudes are the elements of the rotation matrix of one orthogonal system into
another, they are composed according to the matrix rule:
  \begin{equation}
    \varphi(q',Q')=\int\psi(q',\beta')\chi(\beta',Q')d\beta'\ ;
    \label{eq29}
  \end{equation}
in the case of discrete spectra, instead of the integral one has finite or infinite sums. 
This is the general formulation of the theorem of the interference of probabilities.
As an application, let us look again at formula (\ref{eq24}). Here $c_n(t)$ was
the amplitude for the probability that the system at time $t$ has energy $W_n$;
$\varphi_n^0(q')$ is the amplitude for the probability that for given energy
$W_n$ the coordinate $q'$ has a given value. Thus
  \[
    \varphi(q',t)=\sum_n c_n(t)\varphi_n^0(q')
  \]
expresses the amplitude for the probability that $q'$ at time $t$ has a given value.

Alongside the concept of the relative state probability $|\varphi(q',Q')|^2$, there also occurs
the concept of the transition probability,\endnote{Throughout this paragraph, the French edition translates 
`\"{U}bergang' as `transformation' instead of `transition'.} namely, every time
one considers a system as depending on an external 
parameter, be it time or any property of a weakly coupled external system.
Then the system of principal axes of any quantity becomes dependent on this parameter; it
experiences a rotation, represented by an orthogonal transformation
$S(q',q'')$, in which the parameter enters (as in formula (\ref{eq25})). The 
quantities $|S(q',q'')|^2$ are the `transition 
probabilities';\endnote{`$S$' missing in the French edition.} 
in general, however, they are not independent, instead the `transition 
amplitudes' are composed according to the interference rule.

\

\par
\Needspace{5\baselineskip}
\begin{center} {\sc III. --- Formulation of the principles and 
delimitation of their scope}\addcontentsline{toc}{section}{III. --- 
Formulation of the principles and delimitation of their scope}\\
\end{center}
After the general concepts of the
theory have been developed through analysis of empirical findings, the dual task arises,  first of giving
a system of principles as simple as possible and connected directly to the observations, from which the 
entire theory can be deduced as from a mathematical system of axioms, and second of
critically scrutinising experience to assure oneself that no 
observation conceivable by today's means stands in contradiction
to the principles.

Jordan [39] has formulated such a `system of axioms', which takes the following
statements as fundamental:\endnote{[das folgende S\"{a}tze zugrunde legt] --- [qui est 
\`{a} la base des th\'{e}or\`{e}mes suivants]. Note that `Satz' can indeed 
mean both `statement' and `theorem'.}

\

\noindent 1) One requires for each pair of quantum mechanical
quantities $q,Q$ the existence of a probability amplitude $\varphi(q',Q')$, 
such that $|\varphi|^2$ gives the probability\endnote{The French edition omits 
absolute bars.} that for given $Q'$ the value of 
$q'$ falls in a given infinitesimal interval.

\

\noindent 2) Upon permutation of $q$ and $Q$, the corresponding amplitude should be $\overline{\varphi(Q',q')}$.

\

\noindent 3) The theorem (\ref{eq29}) of the composition of probability amplitudes.

\

\noindent 4) To each quantity $q$ there should belong a 
canonically conjugate one $p$, defined by the amplitude (\ref{eq28}). This is 
the only place where the quantum constant $h$
appears.\endnote{The `$h$' is present in the French edition 
but not in the typescript.}

\

Finally one also takes as obvious that, if the quantities $q$ and $Q$ 
are identical, the amplitude $\varphi(q',q'')$ becomes equal to the 
`unit matrix' $\delta(q'-q'')$, that is, always to zero, except when
$q'=q''$. This assumption and the multiplication theorem 
3) together characterise the amplitudes thus defined as the coefficients of an 
orthogonal transformation; one obtains the orthogonality conditions simply 
by stating that the composition of the amplitude belonging to $q,Q$ with 
that belonging to $Q,q$ must yield the identity.

One can then reduce all given quantities, including the 
operators,  to amplitudes by writing them as integral operators as in formula 
(\ref{eq13}). The noncommutative operator multiplication is then a 
consequence of the axioms and loses all the strangeness attached
to it in the original matrix theory.

Dirac's method [38] is completely equivalent to Jordan's formulation,\label{completely-equivalent} except in
that he does not arrange the principles in axiomatic form.\footnote{There are nevertheless
some differences between the approaches of Dirac and Jordan. Cf.\ Darrigol (1992, pp.~343--4) ({\em eds}.).}

This theory now indeed summarises all of quantum mechanics in a system in 
which the simple concept of the calculable probability [berechenbare 
Wahrscheinlichkeit]\endnote{In the typescript, this is typed over an (illegible)
previous alternative. Jordan in his habilitation lecture (1927f [62]) uses the term `angebbare 
Wahrscheinlichkeit' (`assignable probability' in Oppenheimer's translation (Jordan 1927g)).} 
for a given event plays the main role.\endnote{[in dem der einfache Begriff der berechenbaren Wahrscheinlichkeit 
f\"{u}r ein bestimmtes Ereignis die Hauptrolle spielt] --- [dans lequel la simple notion de la 
probabilit\'{e} calculable joue le r\^{o}le principale pour un \'{e}v\'{e}nement d\'{e}termin\'{e}]}
It also has some shortcomings, however. One formal shortcoming is the occurrence of improper functions, like the 
Dirac $\delta$, which one needs for the representation of the unit matrix 
for continuous ranges of variables. More serious is the circumstance
that the amplitudes are not directly measurable quantities, rather, only the 
squares of their moduli; the phase factors are indeed 
essential for how different phenomena are connected [f\"{u}r den 
Zusammenhang der verschiedenen Erscheinungen wesentlich], but are only 
indirectly determinable, exactly as phases in optics 
are deduced indirectly by combining measurements of intensity.
It is, however, a tried and proven 
principle, particularly in quantum mechanics, that one should introduce as far as 
possible only directly observable quantities as fundamental concepts of a theory. 
This defect\endnote{[\"{U}berstand] --- [d\'{e}faut]. The word `\"{U}berstand' may be characterising 
the phases as some kind of surplus structure, but it is quite likely a mistyping of `\"{U}belstand', 
which can indeed be translated as `defect', as in the French version.} is related 
mathematically to the fact that 
the definition of probability in terms of the amplitudes does not express 
the invariance under orthogonal transformations of the Hilbert space 
(canonical transformations).

These gaps in the theory have been filled by von Neumann [41,42]. There
is\endnote{[Es gibt] --- [Cet auteur donne]} an invariant definition of the 
eigenvalue spectrum for arbitrary operators, and of the relative probabilities, 
without presupposing the existence of eigenfunctions or indeed using 
improper functions. Even though this theory has not yet been elaborated
in all directions, one can however say with certainty 
that a mathematically irreproachable grounding of quantum mechanics
is possible.

Now the second question has to be answered: is this theory in accord with 
the totality\label{totality} of our experience? In particular, given that the individual 
process is only statistically determined, how can the usual
deterministic order be preserved in the composite macroscopic 
phenomena?\endnote{[Wie kann insbesondere bei der nur statistischen 
Bestimmtheit des Einzelvorgangs in den zusammengesetzten makroskopischen 
Erscheinungen die gewohnte deterministische Ordnung aufrecht erhalten 
werden?] --- [En particulier comment, vu la d\'{e}termination uniquement 
statistique des processus individuels dans les ph\'{e}nom\`{e}nes macroscopiques 
compliqu\'{e}s, l'ordre d\'{e}terministe auquel nous sommes accoutum\'{e}s peut-il 
\^{e}tre conserv\'{e}?]}

The most important step in testing the new conceptual system in this 
direction consists in the determination of the boundaries within which the 
application of the old (classical) words and concepts is allowed, such as 
`position, velocity, momentum, energy of a particle (electron)' 
(Heisenberg [46]). It now turns out that all these quantities
can be {\em individually} exactly measured and defined, as in the classical 
theory, but that for simultaneous measurements of canonically conjugate 
quantities (more generally: quantities whose operators do not commute) 
one cannot get below a characteristic limit of indeterminacy [Unbestimmtheit].\footnote{Here
and in the following, the choice of translation reflects the characteristic terminology of the original. Born 
and Heisenberg use the terms `Unbestimmtheit' (indeterminacy) and `Ungenauigkeit' (imprecision), while 
the standard German terms today are `Unbestimmtheit' and `Unsch\"{a}rfe' (unsharpness) ({\em eds.}).}
To determine this, according to Bohr [47]\endnote{This reference is to a supposedly forthcoming 
`\"{U}ber den begrifflichen Aufbau der Quantentheorie'. Yet, no such published or unpublished work 
by Bohr is extant. Some pages titled `Zur Frage des begrifflichen Aufbaus der Quantentheorie' 
are contained in the folder `Como lecture II' in the Niels Bohr archive, microfilmed in AHQP-BMSS-11, 
section 4. See also Bohr (1985, p.~478). We wish to thank 
Felicity Pors, of the Niels Bohr archive, for correspondence on this point.} one can start quite generally
from the empirically given dualism between waves and corpuscles. One has
essentially the same phenomenon already in every diffraction of light 
by a slit. If a wave impinges perpendicularly on an (infinitely long) slit 
of width $q_1$, then the light distribution as a function of the deviation 
angle $\varphi$ is given according to Kirchhoff by the square of the 
modulus of the quantity
  \[
    a\int_{-\frac{q_1}{2}}^{+\frac{q_1}{2}}e^{\frac{2\pi i}{\lambda}\sin\varphi q}dq\,=\,
    2a\frac{\sin\left({\displaystyle\frac{\pi q_1}{\lambda}\sin\varphi}\right)}{{\displaystyle\frac{\pi q_1}{\lambda}\sin\varphi}}\ ,
  \]
and thus ranges over a domain whose order of magnitude is given by\endnote{The French edition incorrectly reads 
`$\sin\varphi_1=\frac{q_1}{\lambda}$'.} 
$\sin\varphi_1=\frac{\lambda}{q_1}$ and gets ever larger with 
decreasing slit width $q_1$. 
If one considers this process from the point of view of the corpuscular 
theory, and if the association given by de Broglie of frequency and
wavelength with energy and momentum of the light quantum is valid, 
  \[
    h\nu=W,\quad\frac{h}{\lambda}=P\ ,
  \]
then the momentum component perpendicular to the direction of the slit is
  \[
    p=P\sin\varphi=\frac{h}{\lambda}\sin\varphi\ .
  \]
One sees thus that after the passage through the slit the light quanta have 
a distribution whose amplitude is given by
  \[
    e^{\frac{2\pi i}{\lambda}\sin\varphi\cdot q}=
    e^{\frac{2\pi i}{h}p\cdot q}\ ,
  \]
precisely as quantum mechanics requires for two 
canonically conjugate variables; further, the width of the domain of the 
variable $p$ that contains the greatest number of light 
quanta is
  \[
    p_1=P\sin\varphi_1=\frac{P\lambda}{q_1}=\frac{h}{q_1}\ .
  \] 
By general considerations of this kind one arrives at the insight 
that the imprecisions (average errors) of two canonically
conjugate variables $p$ and $q$ always stand in the relation
  \begin{equation}
    p_1 q_1\geq h\ .
    \label{eq30}
  \end{equation}
The narrowing of the range of one variable, which forms the essence 
of a measurement, widens unfailingly the range of the other. The same 
follows immediately from the mathematical formalism of 
quantum mechanics on the basis of formula (\ref{eq28}). The actual meaning of Planck's 
constant $h$ is thus that it is the universal measure of the indeterminacy 
that enters the laws of nature through the dualism of waves and corpuscles.

That quantum mechanics is a mixture of strictly mechanical and 
statistical principles can be considered a consequence of this
indeterminacy. Indeed, in the classical theory one may fix the state of 
a mechanical system by, for instance, measuring the initial values 
of $p$ and $q$ at a certain instant. In quantum mechanics such a 
measurement of the initial state is possible only with the accuracy 
(\ref{eq30}). Thus the values of $p$ and $q$ are known also at later times only statistically.

The relation between the old and the new theory can therefore be 
described thus:

In classical mechanics one assumes the possibility of determining exactly the initial 
state; the further development is then determined by the
laws themselves.

In quantum mechanics, because of the imprecision relation, the result 
of each measurement can be expressed by the choice of appropriate initial 
values for probability functions;\label{BHprobfunc} the quantum mechanical laws\label{laws} determine the
change (wave-like propagation) of these probability functions. The
result of future experiments however remains in general indeterminate and 
only the expectation\endnote{[Erwartung] --- [attente]} of the result is 
statistically constrained. Each new experiment replaces the probability 
functions valid until now with new ones, which correspond to the result of
the observation; it separates  
the physical quantities into known and unknown (more precisely and less
precisely known) quantities in a way characteristic of the experiment.

That in this view certain laws, like the principles
of conservation of energy and momentum, are strictly valid, follows
from the fact that they are relations between commuting quantities 
(all quantities of the kind $q$ or all quantities of the kind 
$p$).\footnote{A similar but more explicit phrasing is used by Born 
(1926e, lecture 15): assuming that $H(p,q)=H'(p)+H''(q)$, the time derivatives 
not only of $H$ but also of all components of momentum and of angular momentum have 
the form $f(q)+g(p)$ with suitable functions $f$ and $g$. Born states that since all $q$ 
commute with one another and all $p$ commute with one another, 
the expressions $f(q)+g(p)$ will vanish under 
the same circumstances as in classical mechanics ({\em eds.}).}

The transition from micro- to macromechanics results naturally 
from the imprecision relation because of the smallness of Planck's constant $h$. 
The fact of the propagation,\endnote{[Ausbreitung] --- [extension]} the `melting away' 
of a `wave or probability packet' is crucial to this. For some simple mechanical 
systems (free electron in a magnetic or electric field  
(Kennard [50]), harmonic oscillator (Schr\"{o}dinger [25])),
the quantum mechanical propagation of the wave packet agrees with
the propagation of the system trajectories that would occur in the 
classical theory if the initial conditions were known only with the
precision restriction (\ref{eq30}). Here the purely classical treatment 
of $\alpha$ and $\beta$ particles, for instance
in the discussion of Wilson's photographs, 
immediately finds its justification. But in general the statistical laws 
of the propagation of a `packet' for the classical and the quantum theory 
are different; one has particularly extreme examples of this in the cases 
of `diffraction' or `interference' of material rays, as in
the already mentioned experiments of Davisson, Kunsman and Germer [18,19] 
on the reflection of electrons by metallic crystals.

That the totality of experience can be fitted into the system of this theory 
can of course be established only by calculation and discussion of all the 
experimentally accessible cases. Individual experimental setups,\endnote{[einzelne 
Versuchsanordnungen] --- [des essais isol\'{e}s]} in which the suspicion of a contradiction with 
the precision limit (\ref{eq30}) might arise, have been discussed [46,47]; 
every time the reason for the impossibility of fixing exactly all 
determining data could be exhibited
intuitively [anschaulich aufgewiesen].

There remains only to survey the most important consequences of the theory 
and their experimental verification.

\

\par
\Needspace{5\baselineskip}
\begin{center} {\sc IV. --- Applications of quantum mechanics}\addcontentsline{toc}{section}{IV. --- 
Applications of quantum mechanics}\\
\end{center}
In this section we shall briefly discuss those applications of quantum mechanics that stand in
close relation\label{closerelation} to questions of principle. Here the Uhlenbeck-Goudsmit 
theory of the magnetic electron shall be mentioned first. Its formulation
and the treatment of the anomalous Zeeman effects with the matrix calculus
raise no difficulties [11]; the treatment with the method of eigenoscillations
succeeds only with the help of the general Dirac-Jordan theory
(Pauli [45]).\label{help} Here, two three-dimensional 
wave functions are associated with each electron. It becomes natural 
thereby to look for an analogy between matter waves and polarised light waves,
which in fact can be carried through to a certain extent (Darwin [49], 
Jordan [53]). What is common to both phenomena is that the number of terms 
is finite, so the representative matrix is also finite (two arrangements [Einstellungen]
for the electron, two directions of polarisation for light). Here the 
definition of the conjugate quantity by means of differentiation thus fails; one must resort 
to Jordan's definition by means of the
probability amplitudes (formula (\ref{eq28})).

From among the other applications, the quantum mechanics of many-body 
problems shall be mentioned [28,29,40]. In a system that contains a 
number of similar particles [gleicher Partikel],\endnote{Again, the terminology has changed 
both in German and in English. The term `similar particles' for `identical particles' is used
for instance by Dirac (1927a [37]).} there occurs between them a kind of `resonance' and from that results a 
decomposition of the system of terms into subsystems that do not combine (Heisenberg, Dirac [28,37]). 
Wigner has systematically investigated this phenomenon 
by resorting to group theoretic methods, and has set up the totality of 
the non-combining systems of terms [40]; Hund has managed to derive the majority of these results by 
comparatively elementary means [48]. A special role is played by the `symmetric' and `antisymmetric'
subsystems of terms; in the former every eigenfunction remains unchanged under permutation of 
arbitrary similar particles, in the latter it changes sign under permutation of any two particles. 
In applying this theory to the spectra of atoms with several electrons it turns out that the Pauli 
equivalence rule\footnote{That is, the Pauli principle applied to electrons that are `equivalent' 
in the sense of having the same quantum numbers $n$ and $l$; cf.\ Born (1969, p.~178). ({\em eds.}).} 
allows only the antisymmetric 
subsystem.\endnote{[nur das antisymmetrische Teilsystem zul\"{a}sst] --- [ne permet pas 
le syst\`{e}me antisym\'{e}trique]} On the basis of this insight one can 
establish quantum mechanically the systematics 
of the line spectra and of the electron grouping throughout the whole periodic system of elements.

If one has a large number of similar particles, which are to be
given a statistical treatment (gas theory), one obtains different statistics depending on 
whether one chooses the corresponding wave function according to the one or the other subsystem. 
The symmetric system is characterised by the fact that no new
state arises under permutation of the 
particles from\endnote{[aus] --- [dans]} a state described by a symmetric eigenfunction; thus all permutations that belong
to the same set of quantum numbers (lie in the same `cell') together always have the 
weight 1. This corresponds to the Bose-Einstein statistics [56,16]. In the antisymmetric 
term system two quantum numbers may never become equal, because otherwise the
eigenfunction vanishes; a set of quantum numbers  corresponds therefore either 
to no proper function at all or at most to one, thus the weight of a state is 
0 or 1. This is the Fermi-Dirac statistics [57,37].

Bose-Einstein statistics holds for light quanta, as 
emerges from the validity of Planck's radiation formula. Fermi-Dirac statistics certainly holds 
for (negative) electrons, as emerges from the above-mentioned systematics 
of the spectra on the basis of Pauli's 
equivalence rule, and with great likelihood also for the 
positive elementary particles (protons); one can infer this from 
observations of band spectra [28,43] and in particular from the 
specific heat of hydrogen at low temperatures [55].\endnote{The French edition gives `[56]'.}
The assumption of Fermi-Dirac statistics for the positive and negative elementary particles 
of matter has the consequence that Bose-Einstein statistics holds for all neutral structures, 
e.g.\ molecules (symmetry of the eigenfunctions under permutation of an even number\endnote{Both 
the German version followed here and the French version (`a whole number of particles') seem rather infelicitous.} 
of particles of matter). Within quantum mechanics, in which a many-body problem is treated in 
configuration space, the new statistics of Bose-Einstein and Fermi-Dirac has a 
perfectly legitimate place, unlike in the classical theory, where an arbitrary modification of 
the usual statistics is impossible; nevertheless the restrictions made on the form of the  
eigenfunctions appear as an arbitrary additional assumption. In particular, the example of light quanta 
indicates that the new statistics is related in an essential way to the wave-like properties
of matter and light. If one decomposes the electromagnetic oscillations of a cavity 
into spatial harmonic components, each of these behaves like a harmonic oscillator as regards time 
evolution; it now turns out that under quantisation of this system of oscillators a solution 
results that behaves exactly like a system of light quanta obeying Bose-Einstein
statistics [4]. Dirac has used this fact for a consistent treatment of electrodynamical
problems [51,52], to which we shall return briefly.

The corpuscular structure of light thus appears here
as quantisation of light waves, such as vice versa
the wave nature of matter manifests itself in the 
`quantisation' of the corpuscular motion.
Jordan has shown [54] that one can
proceed analogously with electrons; one has then
to decompose the Schr\"{o}dinger function of a cavity 
into fundamental and harmonics and to quantise each of these 
as a harmonic oscillator, in such a way in fact
that Fermi-Dirac statistics is obtained.
The new quantum numbers,
which express the `weights' in the 
usual many-body theory, have thus only the values 0 and 1.
Therefore one has again here a case of finite matrices, which can be treated only with Jordan's general theory. 
The existence of electrons thus plays the same role in the formal elaboration of the theory as that of light 
quanta; both are discontinuities no different in kind from the stationary states of a quantised system. 
However, if the material particles stand in interaction with each other, the development of this idea might
run into difficulties of a deep nature.

The results of Dirac's investigations [51,52] of quantum electrodynamics 
consist above all in a rigorous derivation of Einstein's transition probabilities for spontaneous 
emission.\footnote{As opposed to the induced emission 
discussed on p.~\pageref{forpage315} ({\em eds.}).} Here the 
electromagnetic field (resolved into quantised harmonic oscillations) 
and the atom are considered as a coupled system and quantum mechanics 
is applied in the form of the integral equation (\ref{eq13}).
The interaction energy appearing therein is obtained by
carrying over classical 
formulas. In this connection, the nature of
absorption and scattering of light by atoms is clarified. Finally, Dirac [52] has
managed to derive a dispersion formula with damping term; this includes also the quantum mechanical 
interpretation of Wien's experiments on the decay\label{decay} in luminescence of canal 
rays.\footnote{See above, p.~\pageref{Wien} ({\em eds.}).} 
His method consists in considering the process
of the scattering of light by atoms as a collision of light quanta. However, since one can 
indeed attribute energy and momentum to the light quantum but not easily a spatial position, 
there is a failure of the wave mechanical collision theory 
(Born [30]), in which one presupposes knowledge of the interaction between the collision partners as
a function of the relative position. It is thus necessary to use the momenta as independent variables, 
and an operator equation of  matrix character instead of 
Schr\"{o}dinger's wave equation. Here one has a case where the use of
the general points of view which we have emphasised in this 
report cannot be avoided. At the same time, the theory 
of Dirac reveals anew the deep analogy between electrons and light quanta.

\

\par
\Needspace{5\baselineskip}
\begin{center} {\sc Conclusion}\addcontentsline{toc}{section}{Conclusion} \end{center}
By way of summary, we wish to emphasise that while we consider the last-mentioned enquiries, which 
relate to a quantum mechanical treatment of the electromagnetic field, as not yet completed [unabgeschlossen], 
we consider {\em quantum mechanics} to be a closed theory [geschlossene Theorie], whose fundamental 
physical and mathematical assumptions are no longer susceptible of any modification. Assumptions about the physical
meaning of quantum mechanical quantities that
contradict Jordan's or equivalent postulates will
in our opinion also contradict experience. (Such
contradictions\label{contradictions} can arise for example if the square of the modulus of the 
eigenfunction is interpreted as charge density.\footnote{See Schr\"{o}dinger's report, especially his section I, and 
section~\ref{Schr-radiation} above ({\em eds.}).}) On the
question of the `validity of the law of causality' we have this opinion: as long 
as one takes into account only experiments  that lie in the domain of our currently acquired physical 
and quantum mechanical experience, the assumption of indeterminism in principle, here taken 
as fundamental, agrees with experience. The further development of 
the theory of radiation will change nothing in this state of affairs, because the
dualism between corpuscles and waves, which in quantum mechanics appears as part of a contradiction-free, 
closed theory [abgeschlossene Theorie], holds in quite a similar way for radiation. The relation between 
light quanta and electromagnetic waves must be just as statistical as that between de Broglie waves and 
electrons. The difficulties still standing at present in the way of a complete theory of radiation thus 
do {\em not} lie in the dualism between light quanta and waves --- which is entirely intelligible 
--- instead they appear only when one attempts to arrive at a relativistically
invariant, closed formulation of the electromagnetic laws; all questions for which such a formulation 
is unnecessary can be treated by Dirac's method [51,52]. However, the first steps also towards overcoming 
these relativistic difficulties have already been made.


\newpage

\par
\Needspace{5\baselineskip}
\begin{center}

{\bf Bibliography}\addcontentsline{toc}{section}{Bibliography}\footnote{The style 
of the bibliography has been both modernised and uniformised. Amendments and fuller 
details (when missing) are given in square brackets, mostly without commentary. 
Amendments in the French edition of mistakes in the typescript (wrong initials, 
spelling of names etc.) are taken over also mostly without commentary. Mistakes 
occurring only in the French edition are endnoted ({\em eds.}).}\\
\hfill\\ 
\end{center}
\noindent [1] W.~Heisenberg, \"{U}ber quanten[theoretische]
Umdeutung kinematischer und mechanischer Beziehungen, 
{\em Z.\ f.\ Phys.}, {\bf 33} (1925), 879. 

\noindent [2] M.~Born and P.~Jordan, Zur 
Quantenmechanik, I, {\em Z.\ f.\ Phys.}, {\bf 34} (1925), 858.

\noindent [3] P.~Dirac, The fundamental equations of quantum 
mechanics, {\em Proc.\ Roy.\ Soc.\ A}, {\bf 109} (1925), 
642.

\noindent [4] M.~Born, W.~Heisenberg and P.~Jordan, 
Zur Quantenmechanik, II, {\em Z.\ f.\ Phys.}, {\bf 35} (1926),
557.

\noindent [5] N.~Bohr, Atomtheorie und Mechanik, 
{\em Naturwiss.}, {\bf 14} (1926), 1.

\noindent [6] W.~Pauli, \"{U}ber das Wasserstoffspektrum
vom Standpunkt der neuen Quantenmechanik, {\em Z.\ f.\ Phys.}, {\bf 36} (1926), 336.

\noindent [7] P.~Dirac, Quantum mechanics and a
preliminary investigation of the hydrogen atom, {\em Proc.\ Roy.\ Soc.\ A}, {\bf 110}  (1926), 
561.

\noindent [8]  G.~E.~Uhlenbeck and  S.~Goudsmit,
Ersetzung der Hypothese vom unmechanischen Zwang durch eine
Forderung bez\"{u}glich des inneren Verhaltens jedes einzelnen 
Elektrons, {\em Naturw.}, {\bf 13} (1925), 953.\endnote{Both 
the typescript and the French edition add `(Magnetelektron)'. The French 
edition reads `{\em Nature}'.}

\noindent [9] { L.~H.~Thomas}, The motion of the spinning 
electron, {\em Nature}, {\bf 117} (1926), 514.

\noindent [10] { J.~Frenkel}, Die Elektrodynamik des rotierenden
Elektrons, {\em Z.\ f.\ Phys.}, {\bf 37} (1926),  243.

\noindent [11] { W.~Heisenberg} and { P.~Jordan}, Anwendung
der Quantenmechanik auf das Problem der anomalen Zeemaneffekte, {\em Z.\ f.\ 
Phys.}, {\bf 37} (1926),  263.

\noindent [12] { P.~Dirac}, Relativity quantum mechanics with
an application to Compton scattering, {\em Proc.\ Roy.\ Soc.\ }[{\em A}], {\bf 111} (1926), 405.

\noindent [13] { P.~Dirac}, The elimination of the nodes in
quantum mechanics, {\em Proc.\ Roy.\ Soc.\ }[{\em A}], {\bf 111} (1926), 281.

\noindent [14] { P.~Dirac}, On quantum algebra, 
{\em Proc.\ Cambridge Phil.\ Soc.}, {\bf 23} (1926), 412.

\noindent [15] { P.~Dirac}, The Compton effect in
wave mechanics, {\em Proc.\ Cambridge Phil.\ Soc.}, {\bf 23} (1926),  500.

\noindent [16] { A.~Einstein}, Quantentheorie des einatomigen
idealen Gases, II, {\em Berl.\ Ber.\ }(1925), [3].\endnote{Both the typescript and the French 
edition read `p.~5'.}

\noindent [17] { W.~Elsasser}, Bemerkungen zur Quanten[mechanik]
freier Elektronen, {\em Naturw.}, {\bf 13} (1925),  711.

\noindent [18] { C.~Davisson} and { [C.] Kunsman},  
[The s]cattering of low speed electrons by [Platinum] and [Magnesium], 
{\em Phys.\ Rev.}, {\bf 22} (1923),  [242].\endnote{Typescript 
and published volume read `Pt' and `Mg', as well as `243'.}

\noindent [19] { C.~Davisson} and { L.~Germer}, The scattering
of electrons by a single crystal of Nickel, {\em Nature}, {\bf 119} (1927), 558.

\noindent [20] { E.~G.~Dymond}, On electron scattering in Helium,
{\em Phys.\ Rev.}, {\bf 29} (1927), 433.

\noindent [21] { M.~Born} and { N.~Wiener}, Eine neue 
Formulierung der Quantengesetze f\"{u}r periodische und [nichtperiodische]
Vorg\"{a}nge, {\em Z.\ f.\ Phys.}, {\bf 36} (1926),  174.

\noindent [22] { C.~Eckart}, Operator calculus and the solution
of the equation[s] of quantum dynamics, {\em Phys.\ Rev.}, {\bf 28} (1926),  711.

\noindent [23] { K.~Lanczos}, \"{U}ber eine feldm\"{a}ssige
Darstellung der neuen Quantenmechanik, {\em Z.\ f.\ Phys.}, {\bf 35} (1926), 812.

\noindent [24] { E.~Schr\"{o}dinger}, Quantisierung als 
Eigenwertproblem, I to IV, {\em Ann.\ d.\ Phys.}, {\bf 79} (1926), 
 361; {\em ibid.},  489; {\em ibid.}, {\bf 80} (1926),  437; 
{\em ibid.}, {\bf 81} (1926),  109.

\noindent [25] { E.~Schr\"{o}dinger}, Der stetige \"{U}bergang von 
der Mikro- zur Makromechanik, {\em Naturw.}, {\bf 14} (1926), 664.

\noindent [26] { E.~Schr\"{o}dinger}, \"{U}ber das Verh\"{a}ltnis der
Hei\-sen\-berg-Born-Jor\-dan\-schen Quantenmechanik zu der meinen, {\em Ann.\ d.\ Phys.},  
{\bf 79} (1926), 734.

\noindent [27] { P.~Jordan}, Bemerkung \"{u}ber einen Zusammenhang
zwischen Duanes Quantentheorie der Interferenz und den de Broglieschen Wellen,
{\em Z.\ f.\ Phys.}, {\bf 37} (1926), 376.

\noindent [28] { W.~Heisenberg}, Mehrk\"{o}rperproblem und 
Resonanz in der Quantenmechanik, I and II, {\em Z.\ f.\ Phys.}, {\bf 38} (1926), 
411; {\em Ibid.\/} {\bf 41} (1927), 239.

\noindent [29] { W.~Heisenberg}, \"{U}ber die Spektren von 
Atomsystemen mit zwei Elektronen, {\em Z.\ f.\ Phys.}, {\bf 39} (1926), 
499.\endnote{In the French edition: `409'.}

\noindent [30] { M.~Born}, Zur Quantenmechanik der 
Stossvorg\"{a}nge, {\em Z.\ f.\ Phys.}, {\bf 37} (1926),  863; 
[Quantenmechanik der Stossvorg\"{a}nge], {\em ibid.}, {\bf 38} (1926),  803.

\noindent [31] { G.~Wentzel}, Zwei Bemerkungen \"{u}ber die
Zerstreuung korpuskularer Strahlen als Beugungserscheinung, {\em Z.\ f.\ Phys.}, {\bf 40} (1926),  
590.

\noindent [32] { J.~R.~Oppenheimer}, Bemerkung zur Zerstreuung
der $\alpha$-Teilchen, {\em Z.\ f.\ Phys.}, {\bf 43} (1927),  413.

\noindent [33] { W.~Elsasser}, Diss.\ G\"{o}ttingen, 
[Zur Theorie der Stossprozesse bei Wasserstoff], {\em Z.\ f.\ Phys.},   
[{\bf 45} (1927), 522].\endnote{This is indeed the (abridged) published 
version of Elsasser's G\"{o}ttingen dissertation.}

\noindent [34] { M.~Born}, Das Adiabatenprinzip in der Quantenmechanik,
{\em Z.\ f.\ Phys.}, {\bf 40} ([1926]),  167.\endnote{In the French 
edition: `Das Adiabatenprinzip in den Quanten', as well as `1927' (the latter as in 
the typescript).}

\noindent [35] { W.~Heisenberg},  Schwankungserscheinungen und
Quantenmechanik, {\em Z.\ f.\ Phys.}, {\bf 40} (1926),  501.

\noindent [36] { P.~Jordan}, \"{U}ber quantenmechanische Darstellung
von Quantenspr\"{u}ngen, {\em Z.\ f.\ Phys.}, {\bf 40} ([1927]),  661.

\noindent [37] { P.~Dirac}, On the theory of quantum mechanics, 
{\em Proc.\ Roy.\ Soc.\ A}, {\bf 112} (1926),  661.

\noindent [38] { P.~Dirac}, The physical interpretation of
quantum dynamics, {\em Proc.\ Roy.\ Soc.\ A}, {\bf 113} ([1927]),  621.

\noindent [39] { P.~Jordan}, \"{U}ber eine neue Begr\"{u}ndung der
Quantenmechanik, {\em Z.\ f.\ Phys.}, {\bf 40} ([1927]), [809]; 
Second Part, {\em ibid.}, {\bf 44} (1927),  1.

\noindent [40] { E.~Wigner}, \"{U}ber nichtkombinierende Terme in
der neueren Quantentheorie, First Part, {\em Z.\ f.\ Phys.}, {\bf 40} (1926),  
492; Second Part, {\em ibid.}, {\bf 40} (1927),  883.

\noindent [41] { D.~Hilbert, [J.].~v.~Neumann} and 
{ L.~Nordheim}, [\"{U}ber die Grundlagen der Quantenmechanik,] {\em Mathem.\ Ann.}, 
{\bf 98} [1928], 1.

\noindent [42] { [J.]~v.~Neumann}, [Mathematische Begr\"{u}ndung 
der Quantenmechanik,] {\em G\"{o}tt.\ Nachr.}, 
20 May 1927, [1].

\noindent [43] { F.~Hund}, Zur Deutung der Molekelspektr[en],
{\em Z.\ f.\ Phys.}, {\bf 40} ([1927]),  742; {\em ibid.}, {\bf 42} (1927), 
93; {\em ibid.}, {\bf 43} (1927),  805.

\noindent [44] { W.~Pauli}, \"{U}ber Gasentartung und 
Paramagnetismus, {\em Z.\ f.\ Phys.}, {\bf 41} (1927),  81.

\noindent [45] { W.~Pauli}, Zur Quantenmechanik des magnetischen 
Elektrons, {\em Z.\ f.\ Phys.}, {\bf 43} (1927),  601.

\noindent [46] { W.~Heisenberg}, \"{U}ber den anschaulichen Inhalt 
der quantentheoretischen Kinematik und Mechanik, {\em Z.\ f.\ Phys.}, 
{\bf 43} (1927),  172.\endnote{In both the typescript and the  
French edition the title of the paper is given as `\"{U}ber den 
anschaulichen Inhalt der Quantenmechanik'.}

\noindent [47] { N.~Bohr}, \"{U}ber den begrifflichen Aufbau der
Quantentheorie, forthcoming [im Erscheinen].

\noindent [48] { F.~Hund}, Symmetriecharaktere von Termen bei
Systemen mit gleichen Partikeln in der Quantenmechanik, {\em Z.\ f.\ Phys.}, 
{\bf 43} (1927),  788.

\noindent [49] { [C.~G.] Darwin}, The electron as a vector wave, {\em Nature}, 
{\bf 119} (1927),  282.

\noindent [50] { E.~Kennard}, Zur Quantenmechanik einfacher
Bewegungstypen, {\em Z.\ f.\ Phys.}, {\bf 44} (1927),  326.

\noindent [51] { P.~Dirac}, The quantum theory of emission and
absorption of radiation, {\em Proc.\ Roy.\ Soc.\ A}, {\bf 114} (1927),  243.

\noindent [52] { P.~Dirac}, The quantum theory of dispersion, {\em Proc.\ Roy.\ Soc.\ A}, {\bf 114} (1927),  
710.

\noindent [53] { P.~Jordan}, \"{U}ber die Polarisation der
Lichtquanten, {\em Z.\ f.\ Phys.}, {\bf 44} (1927),  292.

\noindent [54] { P.~Jordan}, Zur Quantenmechanik der Gasentartung,
{\em Z.\ f.\ Phys.}, forthcoming [im Erscheinen] [{\bf 44} (1927), 473].

\noindent [55] { D.~Dennison}, A note on the specific heat of [the]
Hydrogen [molecule], {\em Proc.\ Roy.\ Soc.\ A}, [{\bf 115}] (1927),  483.

\

\par
\Needspace{5\baselineskip}
\noindent On statistics also:

\noindent [56] { N.~S.~Bose}, Plancks Gesetz und 
Lichtquantenhypothese, {\em Z.\ f.\ Phys.}, {\bf 26} (1924),  178.

\noindent [57] { E.~Fermi}, [Sulla quantizzazione del gas perfetto monatomico], 
{\em Lincei Rend.}, {\bf 3} (1926),  145.

\

\par
\Needspace{5\baselineskip}
\noindent General surveys:

\noindent [58] { M.~Born} ({\em Theorie des Atombaus\/}?)  
[{\em Probleme der Atomdynamik\/}], Lectures given at the Massachusetts 
Institute of Technology (Springer, [1926]).\endnote{Date given as `1927' 
in typescript and volume.}

\noindent [59] { W.~Heisenberg}, [\"{U}ber q]uantentheoretische Kinematik 
und Mechanik, {\em Mathem.\ Ann.}, {\bf 95} (1926),  683.

\noindent [60] { W.~Heisenberg}, Quantenmechanik,
{\em Naturw.}, {\bf 14} (1926),  [989].

\noindent [61] { M.~Born}, Quantenmechanik und Statistik,
{\em Naturw.}, {\bf 15} (1927),  238.\endnote{In the French edition: `288'.}

\noindent [62] { P.~Jordan}, Kausalit\"{a}t und Statistik
in der modernen Physik, {\em Naturw.}, {\bf 15} (1927), 105.

\noindent [63] { P.~Jordan}, Die Entwicklung der neuen
Quantenmechanik, {\em Naturw.}, {\bf 15} (1927),  614, 
636.\endnote{The French edition omits `614'.}


\newpage

\section*{Discussion of Messrs Born and Heisenberg's report\footnotemark}\markboth{{\it M.~Born and W.~Heisenberg}}{{\it Discussion}}
\addcontentsline{toc}{section}{Discussion of Messrs Born and Heisenberg's report}

{\sc Mr Dirac.}\label{forpage65}\footnotetext{The two discussion contributions by Dirac 
follow his manuscript in AHQP, microfilm 36, section 10. Deviations in the French edition (which may or may not be due to 
Dirac) are reported in endnote, as well as interesting variants or cancellations in the manuscript, and punctuation has 
been slightly altered ({\em eds.}).}~---~I should like 
to point out the exact nature of the correspondence between the matrix mechanics and 
the classical mechanics. In classical mechanics one can work out a problem by two methods: (1) by taking all the variables 
to be numbers and working out the motion, e.g.\ by Newton's laws, which means one is calculating the motion resulting 
from one particular set of numerical values for the initial coordinates and momenta, and (2) by considering the 
variables to be functions of the $J$'s (action variables)\endnote{The manuscript includes also `and $w$'s' and `and 
angle', both cancelled.} and using the general transformation theory of dynamics and thus determining 
simultaneously the motion resulting from all possible initial conditions.\endnote{The French edition breaks up and 
rearranges this sentence.} The matrix theory corresponds to this second classical method. It gives 
information about all the states of the system simultaneously. A difference between the matrix method and the second 
classical method arises since in the latter one requires to treat simultaneously only states having nearly the same 
$J$'s (one uses, for instance, the operators $\frac{\partial}{\partial J}$), while in the matrix theory one must treat 
simultaneously states whose $J$'s differ by finite amounts.

To get results comparable with experiment when one uses the second classical method,\endnote{The French edition omits 
the temporal clause.} one must substitute numerical values for the $J$'s in the functions of the $J$'s 
obtained from the general treatment.  One has to do the same in the matrix theory. This gives rise to a difficulty  
since the results of the general treatment are now matrix elements, each referring in general to two different sets of 
$J$'s. It is only the diagonal elements, for which these two sets of $J$'s coincide, that have a direct physical 
interpretation.\\

{\sc Mr Lorentz.}~---~I was very surprised to see that the matrices satisfy equations of motion. In theory that is
very beautiful, but to me it is a great mystery, which, I hope, will
be clarified. I am told that by all these considerations one has come to construct matrices that 
represent what one can observe in the atom, for instance the frequencies
of the emitted radiation. Nevertheless, the fact that the coordinates, the potential energy, and so on,
are now represented by matrices indicates that the quantities have lost their original meaning
and that one has made a huge step in the direction of abstraction.

Allow me to draw attention to another point that has struck me. Let us consider the elements of the matrices
representing the coordinates of a particle in an atom, a hydrogen atom for instance, and satisfying
the equations of motion. One can then change the phase of each element of the matrices without these ceasing 
to satisfy the equations of motion; one can, for instance, change the time. But one can go even further and 
change the phases, not arbitrarily, but by multiplying each element by a factor of the form $e^{i(\delta_m-\delta_n)}$,
and this is quite different from a change of time origin.\footnote{This corresponds of course to the choice of a 
phase factor $e^{i\delta_n}$ for each stationary state. This point (among others) had been raised in the correspondence
between Lorentz and Ehrenfest in the months preceding the conference. See Lorentz to Ehrenfest, 4 July 1927, AHQP-EHR-23
(in Dutch) ({\em eds.}). }

Now these matrix elements ought to represent emitted radiation. If the emitted radiation were what
is at the basis of all this, one could expect to be able to change all phases in an arbitrary way. The 
above-mentioned fact then leads us very naturally to the idea that it is not the radiation that is the
fundamental thing: it leads us to think that behind the emitted oscillations are hidden some true oscillations, of 
which the emitted oscillations are difference oscillations.

In this way then, in the end there would be oscillations of which the emitted oscillations are differences, 
as in Schr\"{o}dinger's theory,\footnote{See section \ref{Schr-radiation} ({\em eds.}).} and it seems to me that this is 
contained in the matrices. This circumstance indicates the existence of a simpler wave substrate.\\

{\sc Mr Born.}~---~Mr Lorentz is surprised that the matrices satisfy the equations of motion; with regard
to this I would like to note the analogy with complex numbers. Also here we have a case where in an
extension of the number system the formal laws are preserved almost completely. Matrices are some kind of
hypercomplex numbers, which are distinguished from the ordinary numbers by the fact that the law
of commutativity no longer holds.\\

{\sc Mr Dirac.}~---~The arbitrary phases occurring in the matrix method correspond exactly\endnote{The French edition 
reads `trouvent une analogie'.} to the arbitrary phases in the second classical method, where the variables 
are functions of the $J$'s and $w$'s (action and angle variables). There are arbitrary\endnote{In the manuscript this 
replaces the cancelled word `unknown'.} phases in the $w$'s, which may have different values for each 
different set of  values for the $J$'s. This is completely analogous\endnote{In the manuscript this replaces `corresponds 
exactly'.} to the matrix theory, in which each arbitrary phase is associated with a row and column, and 
therefore with a set of values for the $J$'s.\\

{\sc Mr Born.}~---~The phases $\alpha_n$ which Mr Lorentz has just mentioned are associated with the different 
energy levels, quite like in classical mechanics. I do not think there is anything mysterious hiding behind this.\\

{\sc Mr Bohr.}\label{Bohr-BH}~---~The issue of the meaning of the arbitrary phases, raised by Mr Lorentz, is of very great
importance, I think, in the discussion of the consistency of the methods of quantum theory. Although the
concept of phase is indispensable in the calculations, it hardly enters the interpretation of the observations.


\newpage

\renewcommand{\enoteheading}{\section*{Notes to the translation}}
\addcontentsline{toc}{section}{\it Notes to the translation}
\theendnotes

\setcounter{endnote}{0}
\setcounter{equation}{0}

\chapter*{Wave mechanics$^{\scriptsize\hbox{a}}$}\markboth{{\it E.~Schr\"{o}dinger}}{{\it Wave mechanics}}
\addcontentsline{toc}{chapter}{Wave mechanics ({\em E.~Schr\"{o}dinger\/})}
\begin{center}
{\sc By Mr E. SCHR\"{O}DINGER}\footnotetext[1]{Our translation follows Schr\"{o}dinger's German 
typescript in AHQP-RDN, document M-1354. 
Discrepancies between the typescript and the French edition are endnoted. Interspersed in the German 
text, Schr\"{o}dinger provided his own summary of the paper (in French). We translate this in the 
footnotes. The French version of this report is also reprinted in 
Schr\"{o}dinger (1984, vol.~3, pp.~302--23)  ({\em eds.}).}
\end{center}

\

\begin{center}
\par
\Needspace{5\baselineskip}
{\sc Introduction}\addcontentsline{toc}{section}{Introduction}
\end{center}
Under this name at present two theories are being carried on, which are indeed closely related 
but not identical. The first, which follows on directly from the famous doctoral thesis by L.\ de Broglie, 
concerns waves in three-dimensional space. Because of the strictly relativistic treatment 
that is adopted in this version from the outset, we shall refer to it as the {\em four-dimensional} wave 
mechanics. The other theory is more remote from Mr de Broglie's original ideas,
insofar as it is based on a wave-like process in the space of {\em position coordinates} ($q$-space) 
of an arbitrary mechanical system.\endnote{Here and in the following, the French edition omits some italics, 
which are quite characteristic of Schr\"{o}dinger's writing style and which we tacitly restore.} We shall 
therefore call it the {\em multi-dimensional} wave mechanics. Of course this use of the $q$-space is to be seen
only as a mathematical tool, as it is often applied
also in the old mechanics; ultimately, in this version also, the process to be described is one in space and time. 
In truth, however, a complete unification of the two conceptions has not yet been achieved.
Anything over and above the motion of a single electron could be treated so far only in the {\em multi}-dimensional 
version; also, this is the one that provides the mathematical solution to the problems posed
by the Heisenberg-Born matrix mechanics. For these reasons I shall place it first, hoping in this way also to 
illustrate better the characteristic difficulties of the as such more beautiful four-dimensional 
version.\footnote{Summary of the introduction: Currently there are in fact {\em two} [theories of] wave mechanics, 
very closely related to each other but not identical, that is, the relativistic or {\em four-dimensional} theory, which 
concerns waves in ordinary space, and the {\em multi-dimensional} theory, which originally concerns waves in the 
configuration space of an arbitrary system. The former, until now, is able to deal only with the case of a single 
electron, while the latter, which provides the solution to the matrix problems of Heisenberg-Born, comes up against 
the difficulty of being put in relativistic form. We start with the latter.}

\begin{center} 
\par
\Needspace{5\baselineskip}
{\sc I. --- Multi-dimensional theory}\addcontentsline{toc}{section}{I. --- Multi-dimensional theory}
\end{center}
Given a system whose configuration is described by the generalised position coordinates 
$q_1,q_2,\ldots,q_n$, classical mechanics considers its task as being that of determining the 
$q_k$ as functions of time, that is, of exhibiting {\em all} 
systems of functions $q_1(t),q_2(t),\ldots,q_n(t)$ that correspond to a dynamically
possible motion of the system. Instead, according to wave
mechanics the solution to the problem of motion is  not given by a system of $n$ functions of the single variable 
$t$, but by a {\em single} function $\psi$ of the $n$ variables $q_1,q_2,\ldots,q_n$ and perhaps of time (see below). 
This is determined by a {\em partial} differential equation with 
$q_1,q_2,\ldots,q_n$ (and perhaps $t$) as 
{\em independent} variables. This change of role of the $q_k$, which from dependent become independent variables,
appears to be the crucial point. More later on the meaning of the function $\psi$, which is still controversial. We 
first describe how it is determined, thus what corresponds to the equations of motion of the old mechanics.

First let the system be a {\em conservative} one. We start from its Hamiltonian function
  \[
    H=T+V\ ,
  \]
that is, from the total energy
expressed as a function of the $q_k$ and the canonically-conjugate momenta
$p_k$. We take $H$ to be a {\em homogeneous} quadratic function of the $q_k$ and of unity and replace in it each 
$p_k$ by 
$\frac{h}{2\pi}\frac{\partial\psi}{\partial q_k}$
and unity by $\psi$. We call the function 
of the $q_k$, $\frac{\partial\psi}{\partial q_k}$ and $\psi$ thus obtained $L$ (because in wave mechanics 
it plays the role of a Lagrange function). Thus
  \begin{equation}
    L=T\left(q_k, \frac{h}{2\pi}\frac{\partial\psi}{\partial q_k}\right)
      +V\psi^2\ .
    \label{Sch1}
  \end{equation}
Now we determine $\psi(q_1,q_2,\ldots,q_n)$ by the requirement that under variation of $\psi$,
  \begin{equation}
    \delta\int Ld\tau =0\quad\mbox{with}\quad
    \int\psi^2 d\tau =1\ .
    \label{Sch2}
  \end{equation}
The integration is to be performed over the whole of $q$-space  
(on whose perhaps infinitely distant boundary, $\partial\psi$ must 
disappear). However, $d\tau$ is not simply the product of 
the $dq_k$, rather the `rationally measured' volume element in 
$q$-space:
  \begin{equation}
    d\tau = dq_1dq_2\ldots dq_n
    \Big|\pm\frac{\partial^2T}{\partial p_1\ldots\partial p_k}\Big|^{-\frac{1}{2}}
    \label{Sch3}
  \end{equation}
(it is the volume element of a {\em Riemann}ian $q$-space, whose {\em metric},
as for instance also in Hertz's mechanics,\label{Hertz}\footnote{For Schr\"{o}dinger's interest in Hertz's work on 
mechanics, see Mehra and Rechenberg (1987, pp.~522--32) ({\em eds.}).} is determined by the kinetic en\-er\-gy).~--- 
Performing the variation, taking the normalisation constraint with the multiplier [Factor] $-E$, yields the Euler 
equation
  \begin{equation}
    \Delta\psi+\frac{8\pi^2}{h^2}(E-V)\psi=0
    \label{Sch4}
  \end{equation}
($\Delta$ stands for the analogue of the Laplace operator in the generalised
Riemannian sense). As is well known,
  \[
    \int Ld\tau =E
  \]
for a function that satisfies the Euler equation (\ref{Sch4}) and the constraint 
in (\ref{Sch2}).

Now, it turns out that equation (\ref{Sch4}) in general does not have, for every $E$-value, 
a solution $\psi$  that is single-valued and always 
finite and continuous together with its first and second derivatives; 
instead, in all special cases examined so far, this is the case precisely 
for the $E$-values that Bohr's theory would describe
as stationary energy levels of the system (in the case of discrepancies, 
the recalculated values explain the facts of experience {\em better} than the
old ones). The word `stationary' used by Bohr is thus
given a very pregnant meaning by the variation problem (\ref{Sch4}).

We shall refer to these values as eigenvalues, $E_k$, and to 
the corresponding solutions $\psi_k$ as eigenfunctions.\footnote{As a rule, 
in certain domains of the energy axis\endnotemark\,  
the eigenvalue spectrum is continuous, so that the index $k$ is replaced by
a continuous parameter. In the notation we shall generally not take this into 
account.}\endnotetext{[Energiegerade] --- [s\'{e}rie des \'{e}nergies]} We shall number 
the eigenvalues always in increasing order
and shall number repeatedly those with multiple eigensolutions. The $\psi_k$ form a 
normalised complete
orthogonal system in the $q$-space, with respect to which every well-behaved function of the $q_k$ can be 
expanded in a series. Of course this does not mean that every
well-behaved function solves the homogeneous equation (\ref{Sch4}) and thus the
variation problem, because (\ref{Sch4}) is indeed an equation {\em system}, 
each single eigensolution $\psi_k$ satisfying
a different element of the system, namely the one with
$E=E_k$.\footnote{Summary of the above: Wave mechanics demands that events in a mechanical system that 
is in motion be described not by giving $n$ generalised coordinates $q_1,q_2\ldots q_n$ as functions of the time 
$t$, but by giving a {\em single} function [$\psi$] of the $n$ variables $q_1,q_2\ldots q_n$ and maybe of the 
time $t$. The system of equations of motion of classical mechanics corresponds in wave mechanics to a {\em single} 
partial differential equation, eq.~(\ref{Sch4}), which can be obtained by a certain variational procedure. $E$ is 
a Lagrange multiplier, $V$ is the potential energy, a function of the coordinates; $h$ is Planck's constant, 
$\Delta$ denotes the Laplacian in $q$-space, generalised in the sense of Riemann. One finds in specific cases 
that finite and continuous solutions, `eigenfunctions' $\psi_k$ of eq.~(\ref{Sch4}), exist only for certain 
`eigenvalues' $E_k$ of $E$. The set of these functions forms a complete orthogonal system in the coordinate 
space. The eigenvalues are precisely the `stationary energy levels' of Bohr's theory.}

One can take the view\label{Schr-content} that one should 
be content in principle with what has been said so far and its very diverse special
applications. The single stationary states of Bohr's theory would then in a way 
be described by the eigenfunctions $\psi_k$, which {\em do not
contain time at all\/}.\footnote{Cf.\ section~\ref{Time-in-quantum-theory} ({\em eds}.).} One would find that one can derive much 
more from them that is worth knowing, in particular, one can 
form from them, by fixed general rules, quantities that can be aptly taken to be 
{\em jump probabilities} between the single stationary states. Indeed, 
it can be shown for instance that the integral 
  \begin{equation}
    \int q_i\psi_k\psi_{k'}d\tau\ ,
    \label{Sch5}
  \end{equation}
extended to the whole of $q$-space,
yields precisely the matrix element bearing the indices $k$ and $k'$ of the 
`matrix $q$' in the Heisenberg-Born theory; similarly, the elements of all 
matrices occurring there can be calculated from the wave mechanical eigenfunctions.

The theory as it stands, restricted to conservative systems, could treat 
already even the {\em interaction} between two or more systems, 
by considering these as one single system, with the addition of a suitable
term in the potential energy depending on the coordinates of all subsystems. 
Even the interaction of a material system with the radiation field is not out of
reach, if one imagines the system together with certain ether oscillators 
(eigenoscillations of a cavity) as a single conservative
system, positing suitable interaction terms.

On this view\label{Sch-Campbell}  the {\em time variable} would play absolutely no role
in an isolated system --- a possibility to which N. Campbell ({\em Phil.\
Mag.}, [{\bf 1}] (1926), [1106]) has recently pointed.  
Limiting our attention to an isolated system, we would not perceive the passage of time in it any more
than we can notice its possible progress in space, an assimilation of time to the spatial coordinates that 
is very much in the spirit of relativity. What we would notice would be
merely a sequence of discontinuous transitions, so to speak a cinematic image, 
but without the possibility of
comparing the time intervals between the transitions.
Only secondarily, and in fact with increasing precision 
the more extended the system, would a {\em statistical} definition of time result from 
{\em counting} the transitions taking place (Campbell's `radioactive clock').
Of course then one cannot understand the jump probability in the usual way 
as the probability of a transition calculated relative to unit time.
Rather, a {\em single} jump probability is then utterly meaningless; only
with {\em two} possibilities for jumps, the probability that the one 
may happen {\em before} the other is equal to {\em its} jump probability 
divided by the sum of the two.

I consider this view the only one that would make it possible
to hold on to `quantum jumps' in a coherent way.
Either all changes in nature are discontinuous or not a single one. The first view
may have many attractions; for the time being however, it still poses great difficulties. 
If one does not wish to be so radical and give up in principle the use of the
{\em time variable} also for the single atomistic system, then it is very natural 
to assume that it is contained hidden also 
in equation (\ref{Sch4}). One will conjecture that equation system 
(\ref{Sch4}) is the {\em amplitude} equation of an {\em oscillation} 
equation, from which time has been eliminated by setting\footnote{Schr\"{o}dinger introduces the time-dependent 
equation in his fourth paper on quatisation (1926g). There (p.~112), Schr\"{o}dinger 
leaves the sign of time undetermined, settling on the same convention as in (\ref{Sch6}) --- the opposite of today's
convention --- on pp.~114-15. As late as Schr\"{o}dinger (1926h, p.~1065), one reads that `the most general
solution of the wave-problem will be (the real part of) [eq.~(27) of that paper]'. Instead the wave function
is characterised as `essentially complex' in Schr\"{o}dinger (1927c, fn.~3 on p.~957) ({\em eds}.).}
  \begin{equation}
    \psi\sim e^{2\pi i\nu t}\ .
    \label{Sch6}
  \end{equation} 
  $E$ must then be proportional to a power of $\nu$, and it is natural 
to set $E=h\nu$. Then the following is the oscillation equation that 
leads to (\ref{Sch4}) with the ansatz
(\ref{Sch6}):\footnote{Recall that Schr\"{o}dinger does not in fact 
set $m=1$, but absorbs the mass in the definition of $\Delta$ ({\em eds.}).}
  \begin{equation}
    \Delta\psi -\frac{8\pi^2}{h^2}V\psi 
    -\frac{4\pi i}{h}\frac{\partial\psi}{\partial t}=0\ .
   \label{Sch7}
  \end{equation}    
Now {\em this} is satisfied not just by a single\endnote{Bracket added in the French edition.}
  \[
    \psi_k e^{2\pi i\nu_k t}\quad (\nu_k=\frac{E_k}{h})\ ,
  \]
but by an arbitrary linear combination
  \begin{equation}
    \psi=\sum_{k=1}^{\infty}c_k\psi_k e^{2\pi i\nu_k t}
    \label{Sch8}
  \end{equation}
with arbitrary (even complex) constants $c_k$ . If one considers {\em this}
$\psi$ as the description\endnote{[als Beschreibung] --- [comme la d\'{e}finition]} of a certain  
sequence of phenomena in the system, then this
is now given by a (complex) function of the $q_1,q_2,\ldots,q_n$ {\em and} 
of time, a function which can even be given arbitrarily at $t=0$ (because of
the completeness\endnote{[Vollst\"{a}ndigkeit] --- [perfection]} and orthogonality of the $\psi_k$); 
the oscillation equation  
(\ref{Sch7}), or its solution (\ref{Sch8}) with suitably chosen $c_k$, 
then governs the temporal development. Bohr's stationary 
states correspond to the eigenoscillations of the structure
({\em one} $c_k=1$, all others $=0$).

There now seems to be no obstacle to assuming that equation (\ref{Sch7}) 
is valid immediately also for non-conservative systems (that is, $V$
may contain time explicitly). Then, however,\endnote{[freilich] --- [\'{e}videmment]} 
the solution no longer has the simple form (\ref{Sch8}). A particularly interesting
application hereof is the perturbation of an atomic system by an electric alternating field.
This leads to a theory of {\em dispersion}, but we must forgo here a more detailed
description of the same. --- From (\ref{Sch7}) there {\em always} follows 
  \begin{equation}
    \frac{d}{dt}\int d\tau\psi\psi^*=0\ .
    \label{Sch9}
  \end{equation}
(An asterisk shall always denote the complex conjugate.\endnote{Printed as a footnote in 
the French edition.}) Instead of the earlier normalisation condition
(\ref{Sch2}), one can thus require
  \begin{equation}
    \int d\tau\psi\psi^*=1\ ,
    \label{Sch10}
  \end{equation}
which in the conservative case, equation (\ref{Sch8}), means  
  \begin{equation}
    \sum_{k=1}^\infty c_k c_k^*=1\ .
    \label{Sch11}
  \footnote{Summary of the above: Even limiting oneself 
to what has been said up to now, it would be possible to derive much of interest from these results, for instance 
the transition probabilities, formula (\ref{Sch5}) yielding precisely the matrix element $q_i(k,k')$ for the same 
mechanical problem formulated according to the Heisenberg-Born theory. Although we have restricted ourselves so 
far to conservative systems, it would be possible to treat in this way also the mutual action between several 
systems and even between a material system and the radiation field; one would only have to add all relevant systems
to the system under consideration. {\em Time} does not appear at all in our considerations and one could imagine 
that the only events that occur are sudden transitions from one quantum state of the total system to another 
quantum state, as Mr N.~Campbell has recently thought. Time would be defined only statistically by counting the 
quantum jumps (Mr Campbell's `radioactive clock'). --- Another, less radical, point of view is to assume that time 
is hidden already in the family of equations (\ref{Sch4}) parametrised by $E$, this family being the {\em amplitude} 
equation of an {\em oscillation} equation, from which time has been eliminated by the ansatz (\ref{Sch6}). 
Assuming $h\nu=E$ one arrives at eq.~(\ref{Sch7}), which, because it no longer contains the frequency $\nu$, is solved 
by the {\em series} (\ref{Sch8}), where the $c_k$ are arbitrary, generally complex, constants. Now $\psi$ is a function 
of the $q_1,q_2\ldots q_n$ {\em as well as} of time $t$ and, by a suitable choice of the $c_k$, it can be adjusted to an 
arbitrary initial state. Nothing prevents us now from making the time appear also in the function $V$ --- this is the 
theory of non-conservative systems, one of whose most important applications is the theory of dispersion. --- The 
important relation (\ref{Sch9}), which follows from eq.~(\ref{Sch7}), allows one in all cases to normalise $\psi$ 
according to eq.~(\ref{Sch10}).}
  \end{equation}  

What does the $\psi$-function mean now,\label{mean} that is, {\em how does the system described by it
really look like in three dimensions\/}? Many physicists today are of
the opinion that it does not describe\endnote{[sie nicht .... beschreibe] --- [qu'ils ne d\'{e}crivent pas]} the occurrences 
in an individual 
system,\endnote{[Einzelsystem] --- [syst\`{e}me d\'{e}termin\'{e}]} but only the processes in an ensemble of very many
like constituted systems that do not sensibly influence one another\endnote{This clause is omitted in the French 
edition.} and are all under the very same conditions. I shall skip
this point of view, since others are presenting it.\footnote{See 
the report by Messrs Born and  Heisenberg.\endnotemark}\endnotetext{[von anderer Seite vertreten] --- [d\'{e}fendue par 
d'autres]. Footnote only in the French edition.} I myself have so far found useful
the following perhaps somewhat naive but quite concrete idea [daf\"{u}r recht greifbare Vorstellung]. The classical 
system of material points does not really
exist, instead there exists something that continuously 
fills the entire space and of which one would obtain a `snapshot' if one dragged the classical system, with the camera 
shutter open,  through {\em all} its configurations, 
the representative point in $q$-space spending in each volume 
element $d\tau$ a time that is proportional to the {\em instantaneous} value of $\psi\psi^*$.
(The value of $\psi\psi^*$ for only {\em one} value of the
argument $t$ is thus in question.) Otherwise stated: the real system is a superposition
of the classical one in all its possible states, using $\psi\psi^*$ as
`weight function'.

The systems to which the theory is applied consist classically of several\endnote{[aus einer 
Anzahl] --- [d'un grand nombre]} charged point masses. In the interpretation
just discussed\endnote{[Durch die eben besprochene Deutung] --- [Ainsi que nous venons de le voir]} the charge of every 
single one of these is distributed continuously across space, the individual point mass with charge $e$ yielding
to the charge in the three-dimensional volume element $dx\,dy\,dz$ the
contribution\endnote{The equation number is missing in the French edition, and the following sentence is printed 
as a footnote.}
  \begin{equation}
    e\int'\psi\psi^*d\tau\ .
   \label{Sch12}
  \end{equation}
The prime on the integration sign means: one has to integrate
only over the part of the $q$-space corresponding to a 
position of the distinguished point mass within $dxdydz$. ---
Since $\psi\psi^*$ in general depends on time, these charges fluctuate; only
in the special case of a conservative system oscillating with a single
eigenoscillation are they distributed 
permanently, so to speak statically.

It must now be emphasised that by the claim that there {\em are}\endnote{[es {\em gebe}] --- [sont donn\'{e}es]} 
these charge densities (and the current densities arising from their fluctuation), we can mean at best 
{\em half} of what classical electrodynamics would mean by that. Classically, charge and current densities 
are (1) {\em application points}, (2) {\em source points} of the electromagnetic field.
As {\em application points} they are completely out of the question here; the assumption that these charges
and currents act, say, according to Coulomb's or Biot-Savart's law directly on one another, or are directly 
affected in such a way by external fields, this assumption is
either superfluous or wrong (N.B. {\em de facto} wrong), because the changes in the
$\psi$ function and thereby in the charges are indeed to be determined through the
oscillation equation (\ref{Sch7}) --- thus we must not think of them as determined also in 
another way, by forces acting on them. An external electric field 
is to be taken account of in (\ref{Sch7}) in the potential function $V$, an external 
magnetic field in a similar way to be discussed below, --- this is the way their 
application to the charge distribution is expressed in the present theory.

Instead, our spatially distributed charges prove themselves excellently 
as {\em source points} of the field, at least for the external action of the system, in particular with
respect to its {\em radiation}. Considered as source points in the sense of the usual electrodynamics, they yield 
largely\endnote{[weitgehend] --- [tout \`{a} fait]} correct information\label{forpageDEB13} about its frequency, intensity and 
polarisation.\footnote{See the discussion after the report, as well as section~\ref{Schr-radiation} ({\em eds}.).} 
In most cases, the charge is in practice confined to a region that is small compared to the wavelengths of the 
emitted light. The radiation is then determined by 
the resulting {\em dipole moment} [{\em elektrisches Moment\/}] of the charge distribution. According to the principles 
determined above, this is calculated from the classical dipole 
moment of an arbitrary configuration by performing an average using $\psi\psi^*$ 
  \begin{equation}
    M_{\mbox{\scriptsize qu}}=\int M_{\mbox{\scriptsize cl}}\psi\psi^*d\tau\ . 
    \label{Sch13}
  \end{equation}
A glance at (\ref{Sch8}) shows that in $M_{\mbox{\scriptsize qu}}$ the 
{\em differences}
of the $\nu_k$ will appear as emission frequencies; since the $\nu_k$ are the
spectroscopic term values, our picture provides an {\em understanding\endnote{[{\em Verst\"{a}ndnis] --- [interpr\'{e}tation]}} of
Bohr's frequency condition}. The integrals that appear as 
amplitudes of the different partial oscillations
of the dipole moment represent according to the remarks 
on (\ref{Sch5}) the elements of Born and Heisenberg's `dipole
moment matrix'. By evaluating these integrals one obtained the
correct polarisations and intensities of the emitted light in many special cases, 
in particular intensity zero in all cases where a line allowed by the frequency 
condition is missing according to experience 
({\em understanding\/\endnote{[{\em Verst\"{a}ndnis}] --- [signification]} of the selection principle}).  
--- Even though all these results, if one so wishes,
can be detached from the picture of the fluctuating charges
and be represented in a more abstract form, yet they put quite beyond doubt that the picture is tremendously 
useful for one who has the need for Anschaulichkeit!\endnote{No exclamation mark in the French 
edition.}$^{,}$\footnote{Summary of the above: The physical meaning of the function 
$\psi$ appears to be that the system of charged point particles imagined by classical mechanics does not in fact 
exist, but that there is a continuous distribution of electric charge, whose density can be calculated at each 
point of ordinary space using $\psi$ or rather $\psi\psi^*$, the square of the absolute value of $\psi$. According 
to this idea, the quantum (or: real) system is a superposition of all the possible configurations of the classical 
system, the real function $\psi\psi^*$ in $q$-space occurring as `weighting function'. Since $\psi\psi^*$ in 
general contains time, fluctuations of charge must occur. What we mean by the {\em existence} of these continuous 
and fluctuating charges is not at all that they should act on each other according to Coulomb's or Biot-Savart's 
law --- the {\em motion} of these charges is already completely governed by eq.~(\ref{Sch7}). But what we mean is 
that they are the {\em sources} of the electric fields and magnetic fields proceeding from the atom, above all the 
sources of the observed radiation. In many a case one has obtained wonderful agreement with experiment by calculating 
the radiation of these fluctuating charges using classical electrodynamics. In particular, they yield a complete and 
general explanation of Bohr's `frequency condition' and of the spectral `polarisation and selection rules'.} 

In no way should one claim that the provisional attempt of a classical-electrodynamic coupling of the field to the 
charges generating the field is already the last word
on this issue. There are  internal\endnote{[innere] --- [intimes]} reasons for doubting this.
First, there is a serious difficulty in the question of the
{\em reaction} of the emitted radiation on the emitting system, which is not yet
expressed by the wave equation (\ref{Sch7}), according to which 
also such wave forms of the system that continuously emit radiation 
could and would in fact always persist unabated. Further, one should consider the following. We always observe the radiation emitted by an atom only 
through its action on another atom or molecule. Now, from the wave mechanical standpoint we can consider two charged 
point masses that belong to the {\em same} atom, neither as acting directly on each other in their pointlike form
(standpoint of classical mechanics), nor are we allowed to think this of their `smeared out'
wave mechanical charge distributions (the wrong move taunted above). Rather, we have to take account of their 
classical potential energy, considered as a function in $q$-space, in the coefficient $V$ of the wave equation 
(\ref{Sch7}). But then, when we have two {\em different} atoms,  it will surely not be correct in principle to 
insert the fields
generated by the {\em spread-out} charges of the first at the position of the second in the wave equation for 
the latter. And yet we do this when we calculate the radiation of an atom in the way described above  and now treat
wave mechanically the behaviour of another atom in this radiation field. I say 
this way of calculating the interaction between the charges of different atoms can be at most approximate, but 
not correct in principle. For {\em within one} system it is certainly wrong. But if we bring the two atoms 
closer together, then the distinction between the charges of one and those of the other 
gradually disappears, it is actually never a distinction of principle.\endnote{[eine prinzipielle] --- [essentielle]} --- 
The coherent wave mechanical route would surely be to combine both 
the emitting and the receiving system into a single one and to describe them through a
{\em single} wave equation with appropriate coupling terms, however large the
distance between emitter and receiver may be. Then one could be completely silent about the processes 
in the radiation field. But what would be the correct coupling terms? 
Of course not the usual Coulomb potentials, as soon as
the distance is equal to several wavelengths!\endnote{No exclamation mark in the French edition.} (One realises 
from here that without important amendments the {\em entire} theory
in reality can only be applied to very
{\em small} systems.) Perhaps one should use the {\em retarded} potentials. But these are not
functions in the (common) $q$-space, instead they are something much more complicated. 
Evidently we encounter here the provisional limits of the theory and must be happy to possess in the 
procedure depicted above an approximate treatment that appears to be very useful.\footnote{Summary 
of the above: However, there are reasons to believe that our fluctuating and purely classically radiating 
charges do not provide the last word on this question. Since we observe the radiation of an atom only by 
its effect on another atom or molecule (which we shall thus also treat quite naturally by the methods of 
wave mechanics), our procedure reduces to substituting into the wave equation of {\em one} system the 
potentials that would be produced according to the classical laws by the extended charges of another system. 
This way of accounting for the mutual action of the charges belonging to two different systems cannot be 
absolutely correct, since for the charges belonging to the same system it is not. The correct method of 
calculating the influence of a radiating atom on another atom would be perhaps to treat them as {\em one} 
total system according to the methods of wave mechanics. But that does not seem at all possible, since the 
retarded potentials, which should no doubt occur, are not simply functions of the configuration of the systems, 
but something much more complicated. Evidently, at present these are the limits of the method!}

\

\begin{center}
\par
\Needspace{5\baselineskip}
{\sc II. --- Four-dimensional theory}\addcontentsline{toc}{section}{II. --- Four-dimensional theory}
\end{center}
If one applies the {\em multi}-dimensional version of  
wave mechanics to a single electron of mass $m$ and charge $e$ moving in a space with the electrostatic potential
$\varphi$ and to be described by the three rectangular
coordinates $x,y,z$, then the wave equation (\ref{Sch7}) becomes
  \begin{equation}
      \frac{1}{m}\left(\frac{\partial^2\psi}{\partial x^2}+
                               \frac{\partial^2\psi}{\partial y^2}+
                               \frac{\partial^2\psi}{\partial z^2}\right)-
     \frac{8\pi^2}{h^2}e\varphi\psi-
    \frac{4\pi i}{h}\frac{\partial\psi}{\partial t}=0\ .
    \label{Sch14}
  \end{equation}
(N.B.  The factor 
$\frac{1}{m}$ derives from the fact that, given the way of determining the metric of the $q$-space through the kinetic 
energy, $x\sqrt{m}$, $y\sqrt{m}$, $z\sqrt{m}$ should be used as coordinates rather than 
$x,y,z$.\endnote{Printed as a footnote in the French edition.}) It now turns out that the present equation is nothing 
else but the ordinary three-dimensional wave equation for de Broglie's `phase waves' of the electron, except that the 
equation in the above form is shortened or truncated
in a way that one can call `neglecting the influence of relativity'. 

In fact, in the electrostatic field de Broglie gives the following expression\footnote{Cf.\ the formula for the 
refractive index on p.~\pageref{refractive-index} of de~Broglie's report ({\em eds.}).} for the wave velocity $u$ of his phase 
waves, depending on the potential $\varphi$ (i.e.\ on position)
and on the frequency $\nu$:\endnote{In the French edition this equation number 
is given to the following equation (unnumbered in the typescript).}
  \begin{equation}
    u=c\frac{h\nu}{\sqrt{(h\nu-e\varphi)^2-h^2\nu_0^2}}
    \qquad
    \left(\nu_0=\frac{mc^2}{h}\right)\ .
   \label{Sch15}
  \end{equation}
If one inserts this into the ordinary three-dimensional wave equation
  \[
    \frac{\partial^2\psi}{\partial x^2}+\frac{\partial^2\psi}{\partial y^2}+
    \frac{\partial^2\psi}{\partial z^2}-\frac{1}{u^2}
    \frac{\partial^2\psi}{\partial t^2}=0\ ,
  \]
and uses (\ref{Sch6}) to eliminate the frequency $\nu$ from
the equation, one has\endnote{Bracket printed as a footnote in the French edition, 
with the addition: `$\Delta$ stands for the Laplacian'.} ($\Delta=\frac{\partial^2}{\partial x^2}
+\frac{\partial^2}{\partial y^2}
+\frac{\partial^2}{\partial z^2}$)
  \begin{equation}
    \Delta\psi-\frac{1}{c^2}\frac{\partial^2\psi}{\partial t^2}+
    \frac{4\pi ie\varphi}{hc^2}\frac{\partial\psi}{\partial t}+
    \frac{4\pi^2}{c^2}\left(\frac{e^2\varphi^2}{h^2}-\nu_0^2
    \right)\psi=0\ .    
    \label{Sch16}
  \end{equation}
Now if one considers that in the case of  `slow electron motion' (a) the occurring frequencies are always very nearly 
equal to the rest frequency $\nu_0$, so that in order of magnitude the derivative with respect to time in (\ref{Sch16}) 
is equal to a multiplication by $2\pi i\nu_0$, and that (b) $\frac{e\varphi}{h}$ 
in this case\endnote{[$\frac{e\varphi}{h}$] --- [$e\frac{e\varphi}{h}$]} is always small 
with respect to $\nu_0$; and if one then sets in equation (\ref{Sch16})
  \begin{equation}
    \psi=e^{2\pi i\nu_0 t}\tilde{\psi}\ ,
    \label{Sch17}
  \end{equation} 
and disregarding squares of small quantities, one obtains for $\tilde{\psi}$ exactly equation (\ref{Sch14}) derived from 
the multi-dimensional version of wave mechanics. As claimed, this is thus indeed the `classical 
approximation' of the wave equation holding 
for de Broglie's phase waves.\footnote{That is, the nonrelativistic approximation ({\em eds.}).} The transformation (\ref{Sch17})
here shows us that, considered from de Broglie's point of view, the multi-dimensional
theory is committed to a so to speak truncated view of the frequency, in that it subtracts 
once and for all from all frequencies the rest frequency $\nu_0$ (N.B.  In calculating the charge 
density from $\psi\psi^*$,\endnote{[die Ladungsdichte aus $\psi\psi^*$] --- [la densit\'{e} de charge 
$\psi\psi^*$]} the additional factor is of course irrelevant since it has modulus\endnote{[Betrag] --- [valeur]} 
1.\endnote{Bracket printed as a footnote in the French edition.})\footnote{Summary of the 
above: The three-dimensional wave 
equation, eq.~(\ref{Sch14}), obtained by applying the multi-dimensional theory to a single electron in an 
electrostatic potential field $\varphi$, is none other than the nonrelativistic approximation of the wave equation 
that results from Mr L.~[d]e~Broglie's ideas for his `phase waves'. The latter, eq.~(\ref{Sch16}), is obtained by 
substituting into the ordinary wave equation expression (\ref{Sch15}), which Mr [d]e~Broglie has derived for the 
phase velocity $u$ as a function of the frequency $\nu$ and of the potential $\varphi$ (that is, of the coordinates 
$x,y,z$, on which $\varphi$ will depend) and by eliminating from the resulting formula the frequency $\nu$ by means 
of (\ref{Sch6}).}

Let us now keep to the form (\ref{Sch16}) of the wave equation. It still requires an important generalisation. 
In order to be truly relativistic it must be invariant with respect to Lorentz transformations. But if we perform 
such a transformation on our electric field, hitherto assumed to be static, then
it loses this feature and a magnetic field appears by itself next to it. In this way one derives almost 
unavoidably the form of the wave equation in an 
{\em arbitrary} electromagnetic field. The result can be put in the following transparent form, which makes the 
complete equivalence [Gleichberechtigung] of time and the three spatial coordinates fully explicit :
  \begin{equation}
    \begin{array}{l}  {\displaystyle
    \left[\left(\frac{\partial}{\partial x}+
    \frac{2\pi ie}{hc}a_x\right)^2+
     \left(\frac{\partial}{\partial y}+
     \frac{2\pi ie}{hc}a_y\right)^2\right.    }   \\[0.5em]
         {\displaystyle
         \quad\left.+\left(\frac{\partial}{\partial z}+
         \frac{2\pi ie}{hc}a_z\right)^2+
          \left(\frac{1}{ic}\frac{\partial}{\partial t}+
                \frac{2\pi ie}{hc}i\varphi\right)^2-
          \frac{4\pi^2\nu_0^2}{c^2}\right]\psi=0\ .  }
    \end{array}
    \label{Sch18}
  \end{equation}
(N.B.\ $a$ is the vector potential.\endnote{The rest of the bracket is printed as a footnote in the French edition.} 
In evaluating the squares one has to take account of the {\em order} of the factors, since one is dealing with operators, 
and further of Maxwell's relation:
  \begin{equation}
    \frac{\partial a_x}{\partial x}+\frac{\partial a_y}{\partial y}+
    \frac{\partial a_z}{\partial z}+
    \frac{1}{ic}\frac{\partial (i\varphi)}{\partial t}=0\ .)
    \label{Sch19}
  \end{equation}
This wave equation is of very manifold interest. First, as shown by 
Gordon,\footnote{W.~Gordon, {\em Zeitschr.\  
f.\ Phys.}, {\bf 40} (1926), 117.} it can be derived in a way very similar to what we have seen above for the 
amplitude equation of conservative systems, from a variational principle, which now obtains in four dimensions, 
and where time plays a perfectly symmetrical role with respect to the three spatial
coordinates. Further: if one adds to the {\em Lagrange function} of Gordon's
variational principle the well-known Lagrange 
function of the Maxwell field {\em in vacuo}
(that is, the half-difference of the squares of
the magnetic and the electric field strenghts) and varies in the spacetime integral of the new Lagrange function thus 
obtained not only $\psi$, but also the potential components $\varphi,a_x,a_y,a_z$, one obtains as the {\em five} Euler 
equations along with the wave equation 
(\ref{Sch18}) also the four retarded potential equations for 
$\varphi,a_x,a_y,a_z$.\footnote{E.~Schr\"{o}dinger, {\em Ann.\ d.\ Phys.}, {\bf 82} (1927), [265].\endnotemark}\endnotetext{Both 
typescript and French edition give `365' as page number.} (One could also say: 
Maxwell's {\em second} quadruple of equations, while the first, as is well-known, holds identically in the 
potentials.\endnote{The French edition adds this to the footnote.}) It contains as {\em charge and current 
density} quadratic forms in $\psi$ and its first derivatives\endnote{[treten darin 
in $\psi$ und seinen ersten Ableitungen quadratische Formen auf] --- [y 
figurent dans $\psi$ et ses premi\`{e}res d\'{e}riv\'{e}es des formes quadratiques]} that agree completely with the 
rule which we had given in the {\em multi}-dimensional theory for calculating the true charge distribution from the 
$\psi$-function. Second, one can further define\footnote{E. Schr\"{o}dinger, 
{\em loc.\ cit.}\endnotemark}\endnotetext{Footnote only in the French edition.} a
{\em stress-energy-momentum tensor} of the {\em charges}, whose ten components are also
quadratic forms of $\psi$ and its first derivatives, and which together
with the well-known Maxwell tensor obeys the laws of conservation of energy
and of momentum (that is, the sum of the two tensors has a vanishing divergence).\footnote{Summary of the 
above: In order to generalise equation (\ref{Sch16}) so that it may apply to an arbitrary electromagnetic field, 
one subjects it to a Lorentz transformation, which automatically makes a magnetic field appear. One arrives at 
eq.~(\ref{Sch18}), in which time enters in a perfectly symmetrical way with the spatial coordinates. Gordon has 
shown that this equation derives from a four-dimensional variational principle. By adding to Gordon's Lagrangian 
the well-known Lagrangian of the free field and by varying along with $\psi$ also the four components of the 
potential, one derives from a single variational principle besides eq.~(\ref{Sch18}) also the laws of electromagnetism 
with certain homogeneous quadratic functions of $\psi$ and its first derivatives as charge and current densities. These 
agree well with what was said in the previous chapter regarding the calculation of the fluctuating charges using the 
$\psi$ function. --- One finds a definition of the stress-energy-momentum tensor, which, added to Maxwell's tensor, 
satisfies the conservation laws.}

But I shall not bother you here with the rather complex mathematical development of these issues, since the view still 
contains a serious inconsistency. Indeed, according
to it, it would be the {\em same} potential components $\varphi,a_x,a_y,a_z$ which {\em on the
one hand} act to modify the wave equation (\ref{Sch18}) (one could say: 
they act {\em on} the charges {\em as movers\/}\endnote{[{\em bewegend}] --- [{\em par le mouvement}]}) and which
{\em on the other hand} are determined in turn, via the retarded potential equations, {\em by} these same 
charges, which occur as {\em sources} in the latter equations. (That is: the wave equation
(\ref{Sch18}) determines the $\psi$ function, from 
the latter one derives the charge and current densities, which 
as sources determine the potential components.) --- In reality,
however, one operates {\em otherwise} in the application of
the wave equation (\ref{Sch18}) to the hydrogen electron, and one {\em must}
operate otherwise to obtain the correct result: one substitutes in the wave equation (\ref{Sch18}) 
the {\em already given} potentials of the nucleus and of possible external fields (Stark and Zeeman 
effect). From the solution for $\psi$ thus 
obtained one derives the fluctuating charge densities discussed above, which 
one in fact\endnote{[allerdings] --- [certainement]} has to use for the determination from sources 
of the {\em emitted radiation}; but one must {\em not} add {\em a posteriori} 
to the field of the nucleus and the possible external fields also the fields produced by these charges at the 
position of the atom itself in equation (\ref{Sch18})~---\endnote{[(\ref{Sch18})] --- [(\ref{Sch8})]} something 
totally wrong would result.

Clearly this is a painful lacuna. The pure field theory is not enough, it has to be supplemented by performing a 
kind of {\em individualisation}\label{indiv} of the charge densities coming from the single point charges of the
classical model, where however each single `individual' may
be spread over the whole of space, so that they overlap.
In the wave equation for the single
individual one would have to take into account only the fields produced by the other individuals but not its self-field. 
These remarks, however, are only meant to characterise the general nature of the required supplement, not to constitute a 
programme to be taken completely literally.\footnote{Summary of the above: However, these last developments 
run into a great difficulty. From their direct application would follow the logical necessity of taking into account 
in the wave equation, for instance in the case of the hydrogen atom, not only the potential arising from the nucleus, 
but also the potentials arising from the fluctuating charges; which, apart from the enormous mathematical complications 
that would arise, would give completely wrong results. The field theory (`Feldtheorie') appears thus inadequate; it should 
be supplemented by a kind of individualisation of the electrons, despite these being extended over the whole of space.}

We wish to present also the remarkable special
result yielded by the relativistic form (\ref{Sch18}) of the wave equation for the hydrogen atom. One would at first expect and hope to find the well-known Sommerfeld formula 
for the fine structure of terms. Indeed one {\em does} obtain a fine structure and one {\em does} obtain
Sommerfeld's formula, however the result contradicts experience, because
it is exactly what one would find in the Bohr-Sommerfeld theory, if
one were to posit the radial as well as the azimuthal
quantum number as {\em half-integers} [halbzahlig], that is, half of an odd integer. --- Today this result is not as 
disquieting as when it was first 
encountered.\footnote{E.~Schr\"{o}dinger, {\em Ann.\ d.\ Phys.}, {\bf 79} (1926), [361], p.~372.} In fact, it is 
well-known that the extension of Bohr's theory through the Uhlenbeck-Goudsmit electron spin [Elektronendrall], required 
by many other facts of experience, has to be supplemented in turn by the move to secondary quantum `half'-numbers 
[`halbe' Nebenquantenzahlen] in order to obtain good 
results. How the spin is represented in wave
mechanics is still uncertain.
Very promising suggestions\footnote{C.~G.~Darwin, {\em Nature}, {\bf 119} (1927), 282,  {\em Proc.\ Roy.\ Soc.\ A}, 
{\bf 116} (1927), 227.}
point in the direction that instead of the {\em scalar} $\psi$ a {\em vector} should be introduced. 
We cannot discuss here this latest turn in the theory.\footnote{Summary of the above: For the hydrogen 
atom the relativistic equation (\ref{Sch18}) yields a result that, although disagreeing with experience, is 
rather remarkable, that is: one obtains the same fine structure as the one that would result from the 
Bohr-Sommerfeld theory by assuming the radial and azimuthal quantum numbers to be `integral and a half', that is, 
half an odd integer. The theory has evidently to be completed by taking into account what in Bohr's theory is called 
the spin of the electron. In wave mechanics this is perhaps expressed (C.~G.~Darwin) by a polarisation of the $\psi$ 
waves, this quantity having to be modified from a scalar to a vector.}

\

\begin{center}
\par
\Needspace{5\baselineskip}
{\sc III. --- The many-electron problem}\addcontentsline{toc}{section}{III. --- The many-electron problem}
\end{center}
The attempts\footnote{See in particular A.~Unsold, {\em
Ann.\ d.\ Phys.}, {\bf 82} (1927), 355.} to derive numerical results by
means of approximation methods for the atom with {\em several} 
electrons, whose amplitude equation (\ref{Sch4}) or wave
equation (\ref{Sch7}) cannot be solved directly, have led to the remarkable
result that actually, despite the multi-dimensionality of
the original equation,  in this procedure one always needs to calculate only
with the well-known three-dimensional eigenfunctions of hydrogen; indeed 
one has to calculate certain {\em three}-dimensional {\em charge distributions} that
result from the hydrogen eigenfunctions according to the principles presented above, and one has 
to calculate according the principles of classical electrostatics the self-potentials and
interaction potentials of these charge distributions; these {\em
constants} then enter as coefficients in a system
of equations that in a simple way determines {\em in principle} the behaviour 
of the many-electron atom. Herein, I think, lies a hint that with 
the furthering of our understanding
`in the end everything will indeed become intelligible in three dimensions again'.\endnote{The
French edition omits the inverted commas.}
For this reason I want to elaborate a little on what has just been said.

Let
  \[
    \psi_k(x,y,z)\quad\mbox{and}\quad E_k\ ;\quad
    (k=1,2,3,\ldots)
  \]
be the normalised eigenfunctions (for simplicity assumed as {\em real}) and 
corresponding eigenvalues of the {\em one}-electron atom
with $Z$-fold positive nucleus, which for brevity we shall call the
hydrogen problem. They satisfy the three-dimensional
amplitude equation (compare equation (\ref{Sch4})):
   \begin{equation}
      \left\{\begin{array}{ccl}  {\displaystyle
              \frac{1}{m}\left(\frac{\partial^2\psi}{\partial x^2}+
                               \frac{\partial^2\psi}{\partial y^2}+
                               \frac{\partial^2\psi}{\partial z^2}\right)+
              \frac{8\pi^2}{h^2}\left( E+\frac{Ze^2}{r}\right)\psi } & = & 0    
                                                                 \\[2em]
              (r=\sqrt{x^2+y^2+z^2})\ . & &
            \end{array}\right.
    \label{Sch20}
  \end{equation}
If only one eigenoscillation is present, 
one has the {\em static} charge dis\-tri\-bu\-tion\endnote{The equation number is missing in the printed volume.}
  \begin{equation}
    \rho_{kk}=-e\psi_k^2\ .
    \label{Sch21}
  \end{equation}
If one imagines {\em two} being excited with maximal strength,
one adds to $\rho_{kk}+\rho_{ll}$ a charge distribution oscillating 
with frequency $|E_k-E_l|/h$, whose {\em amplitude} distribution is given by
  \begin{equation}
    2\rho_{lk}=-2e\psi_k\psi_l\ .
    \label{Sch22}
  \end{equation}
The spatial integral of $\rho_{kl}$ vanishes when $k\neq l$
(because of the orthogonality of the $\psi_k$) and it is $-e$ for $k=l$. The charge 
distribution resulting from the presence of two eigenoscillations together 
has thus at every instant the sum zero. --- One can now form the electrostatic potential energies
  \begin{equation}
    p_{k,l;k',l'} =   \int\!\ldots\!\int dx\,dy\,dz\,dx'\,dy'\,dz'\,  
                            \frac{\rho_{kl}(x,y,z)\rho_{k'l'}(x',y',z')}{r'}\ ,
    \label{Sch23}
  \end{equation}
where $r'=\sqrt{(x-x')^2+(y-y')^2+(z-z')^2}$ and
the indices $k,l,k',l'$ may exhibit arbitrary degeneracies (to be sure, in the case $k=k',l=l'$, $p$ is
{\em twice} the potential self-energy of the charge distribution $\rho_{kl}$; but that is of no importance). 
{\em It is the constants $p$ that control also the many-electron atom}.

Let us sketch this. Let the classical model now consist 
of $n$ electrons and a $Z$-fold positively charged nucleus at
the origin. We shall use the wave equation in the form (\ref{Sch7}). It becomes 
$3n$-dimensional,\endnote{[$3n$-dimensional] --- [tridimensionelle]}
say thus  
  \begin{equation} 
    \frac{1}{m}(\Delta_1+\Delta_2+\ldots+\Delta_n)\psi-
    \frac{8\pi^2}{h^2}(V_n+V_e)\psi-
    \frac{4\pi i}{h}\frac{\partial\psi}{\partial t}=0\ .
    \label{Sch24}
  \end{equation}
Here
  \begin{equation}
    \Delta_\sigma=\frac{\partial^2}{\partial x_\sigma^2}+
                           \frac{\partial^2}{\partial y_\sigma^2}
                          +\frac{\partial^2}{\partial z_\sigma^2}\ ;
                           \quad \sigma=1,2,3,\ldots,n\ .
    \label{Sch25}
  \end{equation}
We have considered the potential energy function
as decomposed in two parts, $V_n+V_e$; $V_n$ should 
correspond to the 
interaction of all $n$ electrons with the nucleus, $V_e$ to 
their interaction with one another, therefore\footnote{Analogously to eq.~(\ref{Sch12}),
the prime on the summation sign should be interpreted as meaning that the sum is to be taken over all 
pairs with $\sigma\neq\tau$ ({\em eds}.).}
  \begin{equation}
    V_n=-Ze^2\sum_{\sigma=1}^{n}\frac{1}{r_\sigma}\ ,
    \label{Sch26}
  \end{equation}
  \begin{equation}
    V_e=+e^2\sideset{}{'}\sum_{(\sigma,\tau)}\frac{1}{r_{\sigma\tau}}
    \label{Sch27}
  \end{equation}
  \[
    \left[ r_\sigma=\sqrt{x_\sigma^2+y_\sigma^2+z_\sigma^2},\;\; 
    r_{\sigma\tau}=\sqrt{(x_\sigma-x_\tau)^2+(y_\sigma-y_\tau)^2+(z_\sigma- 
    z_\tau)^2}\right].
  \]
As the starting point for an approximation procedure we choose now the eigensolutions 
of equation (\ref{Sch24}) with $V_e=0$, that is with the interaction between the electrons disregarded. 
The eigenfunctions are then {\em products} of hydrogen
eigenfunctions, and the eigenvalues are {\em sums} of the corresponding eigenvalues of hydrogen. As a matter of fact, 
one easily shows that\endnote{Misprint in the French edition: the $E_k$ are not in the exponent.}
  \begin{equation}   
      \psi_{k_1\ldots k_n}  = 
      \psi_{k_1}(x_1\,y_1\,z_1)\ldots
                                                 \psi_{k_n}(x_n\,y_n\,z_n)
      e^{\frac{2\pi it}{h}(E_{k_1}+\ldots+E_{k_n})}
    \label{Sch28}
  \end{equation} 
always satisfies equation (\ref{Sch24}) (with $V_e=0$).
And if one takes all possible sequences of numbers [Zahlenkombinationen] for the $k_1,k_2,\ldots,k_n$, then
these products of $\psi_k$ form a
complete orthogonal system in the $3n$-dimensional 
$q$-space --- one has thus integrated the approximate equation completely.

One now aims to solve the full\endnote{[komplet[t]e] --- [complexe]} equation (\ref{Sch24})
(with $V_e\neq 0$) by {\em expansion} with respect to this complete orthogonal system, that is one makes 
{\em this} ansatz:
  \begin{equation}
    \psi=\sum_{k_1=1}^\infty\ldots\sum_{k_n=1}^\infty
             a_{k_1\ldots k_n}\psi_{k_1\ldots k_n}\ .
    \label{Sch29}
  \end{equation}
But of course the coefficients $a$ cannot be {\em constants}, otherwise the above sum would again be only 
a solution of the truncated equation with $V_e=0$. It turns out,
however, that it is enough to consider the $a$ as functions of {\em time} alone (`method of the variation of 
constants').\footnote{P.~A.~M.~Dirac, {\em Proc.\ Roy.\ Soc.\ A},
{\bf 112} (1926), [661] p.~674.} Substituting (\ref{Sch29}) into 
(\ref{Sch24}) one finds that the following conditions on the time dependence of the $a$
must hold:\endnote{[dass f\"{u}r die Abh\"{a}ngigkeit der $a$ von der Zeit folgende Forderungen bestehen] --- [que
pour que $a$ d\'{e}pende du temps les conditions suivantes doivent \^{e}tre satisfaites]}$^,$\endnote{Two 
misprints in the French edition: the $E_k$ are not in the exponent, and the $k_i$ run to $n$.}
  \begin{multline}
    \frac{da_{k_1\ldots k_n}}{dt}  = 
                \frac{2\pi i}{h}\sum_{l_1=1}^\infty
                \ldots\sum_{l_n=1}^\infty
                v_{k_1\ldots k_n , l_1\ldots l_n}a_{l_1\ldots l_n}       
               e^{\frac{2\pi it}{h}
               (E_{l_1\ldots l_n}-E_{k_1\ldots k_n})}    \\                                                                            
    [k_1\ldots k_n=1,2,3\ldots\ ]. 
    \label{Sch30}
  \end{multline}
Here we have set for brevity
  \begin{equation}
    E_{k_1}+\ldots+E_{k_n}=E_{k_1\ldots k_n}\ .
    \label{Sch31}
  \end{equation}
The $v$ are {\em constants}, indeed they are {\em prima facie} $3n$-tuple integrals ranging
over the whole of $q$-space (Additional explanation:\endnote{Printed as a footnote in the French edition.} 
Where do these $3n$-tuple integrals come from? They derive from the fact that after substituting
(\ref{Sch29}) into (\ref{Sch24}) one replaces the latter equation by the mathematically equivalent condition
that its left-hand side shall be orthogonal to all functions of the complete orthogonal system in $R_{3n}$. 
The system  (\ref{Sch30}) expresses {\em this} condition.) Writing this out 
one has
  \begin{equation}
    \begin{split}
      v_{k_1\ldots k_n,l_1\ldots l_n} &=    \\[0.5em]
      & \!\!\!\!\!\!\!\!\!\!\!\!\!\int^{\mbox{\scriptsize $3n$-fold}}\!\!\!\!\!\!\!\!\!\ldots\int 
      dx_1\ldots dz_nV_e\psi_{k_1}(x_1,y_1,z_1)
              \ldots\psi_{k_n}(x_n,y_n,z_n)   \\
      & \qquad\psi_{l_1}(x_1,y_1,z_1)\ldots\psi_{l_n}(x_n,y_n,z_n)\ .
    \end{split}
    \label{Sch32}
  \end{equation}
If one now considers the simple  structure of $V_e$
given in (\ref{Sch27}), one recognises that the $v$ can be reduced to sextuple integrals, in
fact each of them is a finite sum of some of the Coulomb potential energies defined in (\ref{Sch23}). 
Indeed, if in the finite sum representing $V_e$, we focus on an individual term, for example
$e^2/r_{\sigma\tau}$,
this contains only the six variables $x_\sigma,\ldots,z_\tau$. One can thus immediately perform in (\ref{Sch32}) 
precisely $3n-6$ integrations on this term, yielding (because of the orthogonality
and normalisation of the $\psi_k$) the factor 1, if 
$k_\rho=l_\rho$ for all indices $\rho$ that coincide
neither with $\sigma$ nor with $\tau$, and yielding 
instead the factor 0 if even just for a single $\rho$ different from $\sigma$ and $\tau$ one has: $k_\rho\neq l_\rho$. 
(One sees thus that very many terms disappear.) For the 
non-vanishing terms, it is easy to see that they coincide
with one of the $p$ defined in (\ref{Sch23}). QED\footnote{Summary of the above: Calling $\psi_k$ and 
$E_k$ the eigenfunctions and eigenvalues of the problem for one electron, charge $-e$, in the field of a nucleus 
$+Ze$ (hydrogen problem), let us form the charge distributions (\ref{Sch21}) and (\ref{Sch22}), the former 
corresponding to the existence of a single normal mode, the latter to the cooperation of two of them. Taken as 
charge densities in ordinary electrostatics, each of these would have a certain potential energy and there would 
even be a certain mutual potential energy between two of them, assumed to coexist. These are the constants 
$p_{kl;k'l'}$ in (\ref{Sch23}). --- With these givens, let us attack the problem of the 
$n$-electron atom. Dividing the potential energy in the wave equation (\ref{Sch24}) for this problem into two 
terms and neglecting at first the term $V_e$, due to the mutual action between the electrons, the eigensolutions 
would be given by (\ref{Sch28}), that is, by the products of $n$ hydrogen functions. From these products, taken in 
all combinations, form the series (\ref{Sch29}), which will yield the {\em exact} solution of equation (\ref{Sch24}), 
provided that the coefficients $a_{k_1 k_2\ldots k_n}$ are functions of time satisfying the equations (\ref{Sch30}); 
(see the abbreviation (\ref{Sch31})). The coefficients $v$ in (\ref{Sch30}) are {\em constants}, defined originally 
by the $3n$-tuple integrals (\ref{Sch32}), which however, thanks to the simple form of $V_e$ (see (\ref{Sch27})), 
reduce to sextuple integrals, namely precisely to the constants $p_{k,l;k',l'}$ (see
(\ref{Sch23})).}

Let us now have a somewhat closer look at the equation system (\ref{Sch30}), whose coefficients,
as we have just seen, have such a relatively simple structure, and which determines the varying amplitudes 
of our ansatz\endnote{[unseres Ansatzes] --- [de notre expression fondamentale]} 
(\ref{Sch29}) as 
functions of time. We can allow ourselves to introduce a somewhat simpler symbolic notation, by letting 
the {\em string} of indices  
$k_1,k_2,\ldots,k_n$ be represented by the {\em single} index $k$, and similarly $l_1,l_2,\ldots,l_n$ by $l$. One then 
has
  \begin{equation}
    \frac{da_k}{dt}=\frac{2\pi i}{h}\sum_{l=1}^\infty v_{kl}a_le^{\frac{2\pi it}{h}(E_l-E_k)}\ .
    \label{Sch33}
  \end{equation}
(One must not confuse,
however,  $E_l,E_k$ with the {\em single}\endnote{[den {\em einzelnen}] --- [les {\em diverses}]} eigenvalues 
of the hydrogen problem, which were earlier denoted in the same way.\endnote{Bracket printed as footnote 
in the French edition.}) 
This is now a system of {\em infinitely many} differential equations, which we cannot solve
directly: so, practically nothing seems to have been gained.
In turn, however, we have as yet also {\em neglected
nothing}: with {\em exact} solutions $a_k$ of (\ref{Sch33}), 
(\ref{Sch29}) would be an {\em exact} solution of 
(\ref{Sch24}). This is precisely where I want to place the main emphasis, 
greater than on the practical implementation of the approximation procedure, 
which shall be sketched below only for the sake of completeness. {\em In principle}
the equations (\ref{Sch33}) determine the solution of the many-electron problem 
exactly;\endnote{[bestimmen die L\"{o}sung exact] --- [d\'{e}terminent la solution]} --- 
and they no longer contain anything multi-dimensional; their coefficients are simple Coulomb energies of
charge distributions that already
occur in the hydrogen problem. Further, the equations (\ref{Sch33}) determine the solutions of the many-electron problem 
according to (\ref{Sch29}) as a combination 
of products of the hydrogen eigenfunctions. While these products 
(denoted above by $\psi_{k_1k_2\ldots k_n}$)\endnote{Misprint in the 
French edition: `$\psi(k_1,k_2,\ldots,k_n)$'.} are still functions on the $3n$-dimensional $q$-space,
any two of them yield in the calculation of the three-dimensional charge distributions in the many-electron atom,
as is easily seen, a charge distribution which if it is not identically zero
corresponds again to a hydrogen distribution  
(denoted above by $\rho_{kk}$ or $\rho_{kl}$).

These considerations are the analogue of the construction 
of the higher atoms from hydrogen trajectories in Bohr's theory. They reinforce the
hope that by delving more deeply one will be able to interpret and {\em understand} 
the results of the multi-dimensional theory in three dimensions.\footnote{Summary of the above: Although 
the system of eqs.\ (\ref{Sch30}) (abbreviated to (\ref{Sch33})) does not admit a direct solution, the number of 
equations as well as the number of unknown functions being infinite, it seems to me very interesting that the solution 
to the multi-dimensional problem is provided in principle by a system of equations whose coefficients have such simple 
meanings in three dimensions. Further, one realises that the charge distribution that corresponds to the solution 
(\ref{Sch29}) of the $n$-electron problem turns out to be the superposition of the distributions $\rho_{kk}$ and 
$\rho_{kl}$ that occur already in the hydrogen problem. The hope of interpreting and of {\em understanding} the 
multi-dimensional theory in three dimensions is thus strengthened.} 

Now, as far as the approximation method is concerned, it consists in fact of considering the contribution 
$V_e$ made to the potential energy function $V$ by
the interaction of the electrons with one another, 
to be as far as possible {\em small} as compared to the action of the nucleus. The $v_{kl}$ 
are then considered small compared to the eigenvalue 
differences $E_l-E_k$, except if $E_l=E_k$. The $a_l$
will then
vary {\em slowly} by comparison to the powers of $e$ 
appearing on the
right-hand side of equation (\ref{Sch33}), {\em as long as the latter are not equal to} 1, and all those terms on the 
right-hand side for which this is not the case will
yield only small fluctuations of short period of the $a_k$ and can be neglected in the
approximation.\endnote{Both the typescript and the French edition read `$c_l$' and `$c_k$' instead of `$a_l$' and 
`$a_k$'.} Thereby, first, the sums on the right become {\em finite},
because in fact always only a finite number of eigenvalues coincide. Second, the infinitely many
equations separate into groups; each group contains only a finite number of $a_l$ and can be integrated very
easily.\endnote{Again, both the typescript and the French edition read `$c_k$' instead of `$a_k$'.} This is the first 
step of the approximation procedure, which in theory can be continued
indefinitely, but becomes more and more cumbersome. 
We shall not enter into details. 

One can also transform the untruncated system of differential equations (\ref{Sch33}) at a {\em single} stroke 
into a system of ordinary linear equations (with
infinitely many unknowns!) by setting
  \begin{equation}
    a_l=c_l e^{\frac{2\pi it}{h}(E-E_l)}\ ,
    \label{Sch34}
  \end{equation}
where the quantity $E$ and the quantities $c_l$ are unknown {\em constants}. Substituting
into (\ref{Sch33}) one finds
  \begin{equation}
    (E-E_k)c_k=\sum_{l=1}^\infty v_{kl}c_l\ ;\quad (k=1,2,3,\ldots )\ .
    \label{Sch35}
  \end{equation}
This equation system coincides with the Heisenberg-Born `principal axes problem'. 
If the $v_{kl}$ are very small quantities, then, if not {\em all} $c_l$ are to be
very small, $E$ must be close to {\em one} of the $E_l$, 
let us say to $E_k$. In the first approximation then only 
$c_k$, and all those $c_l$ for which $E_l=E_k$, are different 
from zero. The problem thus separates in the first approximation into 
a denumerable set of {\em finite} principal axes problems.\footnote{Summary of the above: One can embark 
on the solution of the system of equations (\ref{Sch33}) by an approximation method. Positing (\ref{Sch34}), the 
{\em constants} $E$ and $c_l$ have to satisfy the system (\ref{Sch35}) of ordinary linear homogeneous equations, 
whose number as well as that of the unknown constants, however, is infinite. It is only by assuming all 
coefficients $v_{kl}$ to be {\em small} that one can conclude that $E$ has to be very close to one of the values 
$E_l$, for instance $E_k$, and that [$c_l$] approximately vanishes, unless $E_l$ is 
equal to $E_k$. Since there is only a finite number of $E_l$ that coincide with $E_k$, the problem reduces in the 
first approximation to a problem of a finite number of `principal axes', or rather to an infinity of such finite 
problems. --- As a matter of fact, the equations (\ref{Sch35}) coincide with the problem of an infinite number of 
principal axes, which the Heisenberg-Born mechanics reduces to.}


\newpage

\section*{Discussion of Mr Schr\"{o}dinger's report}\markboth{{\it E.~Schr\"{o}dinger}}{{\it Discussion}}
\addcontentsline{toc}{section}{Discussion of Mr Schr\"{o}dinger's report}

{\sc Mr Schr\"{o}dinger.}~---~It would seem that my description in terms of a snapshot was not very
fortunate, since it has been misunderstood.\label{misunderstood} Perhaps the following explanation is clearer. 
The interpretation of Born is well-known, who takes $\psi\psi^*d\tau$ to be the probability for the system being in the volume 
element $d\tau$ of the configuration space. Distribute a very large number $N$ of systems in the
configuration space, taking the above probability as `frequency function'. Imagine these 
systems as superposed in real space, dividing however by $N$ the charge of 
each point mass in each system. In this way, in the limiting case where $N=\infty$ one obtains
the wave mechanical picture of the system.\\

{\sc Mr Bohr.}\label{Bohr-Schr}~---~You have said that from the charge distribution $\psi\psi^*d\tau$ and the classical
laws you obtain the frequency and intensity of light, but do the remarks about difficulties you made
later indicate that what you had obtained was not correct?\\

{\sc Mr Schr\"{o}dinger.}~---~The difficulty I mentioned is the following. If one expands the general
solution as a series with respect to the eigenfunctions
  \[
    \psi=\sum_kc_k\psi_k
  \]
and if one calculates the intensity of the radiation resulting from $\psi_k$
and $\psi_l$ together, one finds that it becomes proportional to $c_k^2c_l^2$. However, according to the
old theory, only the square of the amplitude corresponding to the `initial level' should appear here; that
of the `final level' should be replaced by 1.\\ 

{\sc Mr Bohr.}~---~Has Dirac not found the solution to the difficulty?\\ 

{\sc Mr Schr\"{o}dinger.}~---~Dirac's results are certainly very interesting and point the way toward a solution, 
if they do not contain it already. Only, we should first come to an understanding in
physical terms [nous
devrions d'abord nous entendre en langage physique]. I find it still impossible, for the time being, to
see an answer to a physical question in the assertion that certain quantities obey a noncommutative
algebra, especially when these quantities are meant to represent numbers of atoms. The relation
between the continuous spatial densities, described earlier, and the observed intensities and
polarisations of the spectral rays is [too natural]\footnote{The French here reads `trop peu naturelle', 
which has the exact opposite meaning. The context would seem, however, to justify the amendment ({\em eds}.).} 
for me to deny all meaning to these densities
only because some difficulties appear that are not yet resolved.\\

{\sc Mr Born.}~---~It seems to me that interpreting the quantity $\psi\psi^*$ as a charge 
density leads to difficulties in the case of quadrupole moments. The latter in fact 
need to be taken into account in order to obtain the radiation, not only for theoretical
reasons, but also for experimental reasons.

For brevity let us set 
  \[
    e^2\psi\psi^*=e^2|\psi|^2=\Psi
  \]
and let us consider, for example, the case of two particles; $\Psi$ becomes a function of $x_1$ and $x_2$,
where for brevity $x_1$ stands for all the coordinates of the first particle; $x_2$ has a similar meaning. The
electric density is then, according to Schr\"{o}dinger,
  \[
    \rho(x)=\int\Psi(x,x_2)dx_2+\int\Psi(x_1,x)dx_1\ .
  \]
In wave mechanics the quadrupole moment
  \[
    \int\!\int x_1x_2\Psi(x_1,x_2)dx_1dx_2
  \]
cannot, as far as I can tell, be expressed using the function $\rho(x)$. I would like to know how one 
can, in this case, reduce the radiation of the quadrupole to the motion of a charge distribution $\rho(x)$
 in the usual three-dimensional space.\\

{\sc Mr Schr\"{o}dinger.}~---~I can assure you that the calculation of the dipole moments
is perfecly correct and rigorous and that this objection by Mr Born is unfounded. Does the
agreement between wave mechanics and matrix mechanics extend to the possible
radiation of a quadrupole?\label{quadrupole} That is a question I have not examined. Besides, we do not
possess observations on this point that could allow us to use a possible disagreement
between the two approaches to decide between them.\\

{\sc Mr Fowler} asks for explanations regarding the method for solving
the equations in the case of the many-electron problem.\\  

{\sc Mr De Donder.}~---~Equation (\ref{Sch24}) of Mr Schr\"{o}dinger's report can be extended to the case
in which the $n$ charged particles are {\em different} and where the external actions as
well as the interactions can be described, in spacetime, by a gravitational field [champ 
gravifique].\footnote{Th.~De Donder, L'\'{e}quation 
fondamentale de la Chimie quantique, {\em Comptes Rendus Acad.\ Sci.\ Paris}, session of 10 October 1927,
pp.~698--700. See esp.\ eq.\ (10).} The quantum equation thus obtained is the sum of the quantum equations
for the $n$ particles taken separately, each of the equations being divided by the (rest) mass of the corresponding
particle. Thus, for instance, the quantum equation for the nucleus will not enter if one assumes, as a first approximation,
that the mass of the nucleus is infinitely large with respect to that of an electron.

When there is interaction, the problem is much more complex. One can, as Mr Schr\"{o}dinger indicates, consider
the action of the nucleus as an {\em external} action acting on the electrons of the cloud [couronne], and
the (electrostatic) actions {\em between} the electrons in this cloud as a {\em perturbation}; but
that is only a first approximation. In order to account for relativistic and electromagnetic effects
I have assumed that the molecular systems have an {\em additive} character.\footnote{For more 
details, one can consult our note: `L'\'{e}quation de quantification des mol\'{e}cules comprenant $n$ particules
\'{e}lectris\'{e}es', published after this meeting, in the {\em Bull.\ Ac.\ R.\ Belg., Cl.\ des Sciences}, session
of 5 November 1927.} One can thus recover, as a special case, the above-mentioned method
of quantisation by Schr\"{o}dinger.\\

{\sc Mr Born.}~---~In G\"{o}ttingen we have embarked on a systematic calculation of the matrix elements that appear
in perturbation theory, with the aim of collecting them in tables up to the principal quantum
number $10$. Part of these calculations, which are very extended, has already been
done. My coworker Mr Biem\"{u}ller has used them to calculate the lower terms 
of the helium atom according to the usual perturbation method up to perturbations of the second order. 
The agreement of the ground term with the empirical value, despite the defects 
of the procedure, is hardly worse than in the recently published paper by
Kellner [{\em Zeitschr.\ f.\ Phys.}, {\bf 44} (1927), 91], who has applied a more precise method (Ritz's procedure).\\

{\sc Mr Lorentz.}~---~Do you see the outcome of this long labour as satisfactory?\\

{\sc Mr Born.}~---~The calculation has not attained yet the precision of the measurements. The calculations
we have done applying the ordinary perturbation method [m\'{e}thode des perturbations ordinaires] consist of a series 
expansion with respect to the inverse of the nuclear charge $Z$, of the form
  \[
    E=Z\left(a+\frac{b}{Z}+\frac{c}{Z^2}+\ldots\right)\ .
  \]
The three terms shown have been calculated. Nevertheless, in the case of helium ($Z=2$) the precision is not
yet as good as in the calculations done by Kellner using Ritz's approximation method.\\

{\sc Mr Lorentz.}~---~But you hope however to improve your results.\\

{\sc Mr Born.}~---~Yes, only the convergence of the series is very slow.\\

{\sc Mr Heisenberg.}~---~On the subject of this approximation method, Mr Schr\"{o}dinger says at the end of his 
report that the discussion he has given
reinforces the hope that when our knowledge will be deeper it
will be possible to explain and to understand in three
dimensions the results provided by the multi-dimensional theory. I see nothing in Mr Schr\"{o}dinger's calculations
that would justify this hope. What Mr Schr\"{o}dinger does in his very beautiful 
approximation method,
is to replace the $n$-dimensional differential equations by an infinity of linear equations. That reduces the problem,
as Mr Schr\"{o}dinger himself states, to a problem with ordinary matrices, in which the coefficients can be interpreted
in three-dimensional space. The equations are thus `three-dimensional' exactly in the same sense as in the usual
matrix theory. It thus seems to me that, in the classical sense, we are just as far from understanding the theory in 
three dimensions as we are in the matrix theory.\\

{\sc Mr Schr\"{o}dinger.}~---~I would not know how to express more precisely my hope of a possible formulation in a
three-dimensional space. Besides, I do not believe that one would obtain simpler
calculational methods in this way, and it is probable that one will always do calculations using the multi-dimensional 
wave equation. But then one will be able to grasp its physical meaning better. I am not precisely searching for a 
three-dimensional partial differential equation. Such a simple formulation is surely impossible. If I am not
satisfied with the current state of the problem, it is because I do not understand yet the physical meaning of its
solution.

What Mr Heisenberg has said is mathematically unexceptionable, but the point in question is that of the physical 
interpretation. This is indispensable\label{indispensable} for the further development of the theory. Now, this
development is necessary. For one must agree that all current ways of formulating the results of the new quantum
mechanics only correspond to the classical mechanics of actions at a distance. As soon as light crossing times become 
relevant in the system, the new mechanics fails, because the classical potential energy function no longer exists.

Allow me, to show that my hope of achieving a three-di\-men\-sion\-al conception is not quite utopian, to recall 
what Mr Fowler has told us on the topic of Mr Hartree's approximation method.\footnote{See the discussion after 
Bragg's report, p.~\pageref{Hartree} ({\em eds}.).} It is true that this method abstracts
from what one calls the `exchange terms' (which correspond, for instance, to the distance 
between the ortho and para terms of neutral helium). But, abstracting from that, it already achieves the three-dimensional
aim I tend to. Should one declare {\em a priori} impossible that Hartree's method might be modified or developed
in such a way as to take into account the exchange terms while working with a satisfactory
three-dimensional model?\\

{\sc Mr Born.}~---~Regarding the question of knowing whether it is possible to describe a many-electron problem by a 
field equation in three dimensions, I would like to point out the following. The number of quantum 
numbers of an atom rises by three with each additional electron; it is thus equal to $3n$
for $n$ electrons. It seems doubtful that there should be an ordinary, three-dimensional eigenvalue problem, whose
eigenvalues have a range of size $\infty^{3n}$ [dont la valeur caract\'{e}ristique ait une
multitude de $\infty^{3n}$ dimensions].\footnote{The French text here appears to make little sense, but Born 
is possibly referring to the dimension of the space of solutions ({\em eds}.).} 
Instead, it follows from recent papers by Dirac
and Jordan\footnote{Cf.\ section IV of Born and Heisenberg's report ({\em eds}.).} that one can build on a 
three-dimensional oscillation equation if one considers the eigenfunction itself not as an ordinary number, but as one 
of Dirac's q-numbers, that is, if one quantises again its amplitude as a function of time. An $n$-quanta oscillation 
with this amplitude then yields together with the three spatial quantum numbers the necessary range [multitude] of 
quantum numbers. From this point of view the number of electrons in a system appears itself as a quantum number, that 
is, the electrons themselves appear as discontinuities of the same nature as the stationary states.\\

{\sc Mr Schr\"{o}dinger.}~---~Precisely the structure of the periodic system is already
contained 
in the physics [m\'{e}canique] of the three-dimensional hydrogen problem. The degrees of 
degeneracy 1, 4, 9, 16, etc., multiplied by 2, yield precisely the periodic numbers [nombres de p\'{e}riodes]. The factor 
2 that I have just mentioned derives from the spin [giration (spin)]. From the point of view of wave mechanics, the 
apparently mysterious `Pauli action' of the first two electrons on the third (which they 
prevent from also following an orbit with quantum number 1) means strictly speaking nothing other than the non-existence 
of a third eigenfunction with principal quantum number 1. This non-existence is precisely a property of the 
three-dimensional model, or of the three-dimensional equation. The multi-dimensional equation
has too many eigenfunctions; it is this [elle] that makes the `Pauli exclusion' (Pauliverbot) necessary to eliminate 
this defect.\footnote{The French text refers to the four-dimensional equation (`l'\'{e}quation 
\`{a} quatre dimensions') as having too many solutions. This reading could 
be correct, in the sense that the exclusion principle was first introduced
in the context of the relativistic (four-dimensional) Bohr-Sommerfeld theory
of the atom, but the above reading seems much more natural in context. Note that Schr\"{o}dinger 
throughout his report uses `vierdimensional' and `vieldimensional', which could be easily confused, 
for `four-dimensional' and `many-dimensional', respectively ({\em eds}.).}


\newpage

\renewcommand{\enoteheading}{\section*{Notes to the translation}}
\addcontentsline{toc}{section}{\it Notes to the translation}
\theendnotes

\setcounter{endnote}{0}
\setcounter{equation}{0}

\chapter*{General discussion of the new ideas presented$^{\scriptsize\hbox{a}}$}\markboth{{\it General 
discussion}}{{\it Causality, determinism, probability}}
\addcontentsline{toc}{chapter}{General discussion of the new ideas presented}
\begin{center}
\par
\Needspace{5\baselineskip}
\textsc{Causality, determinism, probability}
\footnotetext[1]{As mentioned in section~\ref{editing}, the Bohr archives contain a copy 
of the galley proofs of 
the general discussion, dated 1 June 1928.\endnotemark A few of the contributions in these proofs seem to have 
been still largely unedited: they contain some gaps and incomplete sentences, some more colloquial formulations, 
and in at least one case a sentence that was dropped from the published volume. We reproduce in endnotes the most 
substantial examples of these alternative versions. For most of the discussion contributions by Dirac, we have 
followed his manuscript version.\endnotemark For Bohr's discussion contributions, 
we have used material from Bohr (1985) and from notes taken by Richardson\endnotemark (also mentioned
in section~\ref{editing}). See our notes for further details ({\em eds.}).}
\addtocounter{endnote}{-2}\endnotetext{Microfilmed in AHQP-BMSS-11, section 5.}
\addtocounter{endnote}{1}\endnotetext{AHQP-36, section 10.} 
\addtocounter{endnote}{1}\endnotetext{These notes are to be found in the Richardson collection in Houston, 
included with the copy of Born and Heisenberg's report (microfilmed in AHQP-RDN, document M-0309).}
\addcontentsline{toc}{section}{Causality, determinism, probability}
\end{center}
\textsc{Mr Lorentz}.~---~I should like to draw attention to the difficulties
one encounters in the old theories.

We wish\label{Lorentzwish} to make a representation of the phenomena, to form an image of them in
our minds. Until now, we have always wanted to form these images by means of
the ordinary notions of time and space. These notions are perhaps innate; in
any case, they have developed from our personal experience, by our daily
observations. For me, these notions are clear and I confess that I should be
unable to imagine physics without these notions. The image that I wish to form
of phenomena must be absolutely sharp and definite, and it seems to me that we
can form such an image only in the framework of space and time.

For me, an electron is a corpuscle that, at a given instant, is present at a
definite point in space, and if I had the idea that at a following moment the
corpuscle is present somewhere else, I must think of its trajectory, which is
a line in space. And if the electron encounters an atom and penetrates it, and
after several incidents leaves the atom, I make up a theory in which the
electron preserves its individuality; that is to say, I imagine a line
following which the electron passes through the atom. Obviously, such a theory
may be very difficult to develop, but \textit{a priori} it does not seem to me impossible.

I imagine that, in the new theory, one still has electrons. It is of course
possible that in the new theory, once it is well-developed, one will have to
suppose that the electrons undergo transformations. I happily concede that the
electron may dissolve into a cloud. But then I would try to discover on which
occasion this transformation occurs. If one wished to forbid me such an
enquiry by invoking a principle, that would trouble me very much. It seems to
me that one may always hope one will do later that which we cannot yet do at
the moment. Even if one abandons the old ideas, one may always preserve the
old classifications [d\'{e}nominations]. I should like to preserve this ideal
of the past, to describe everything that happens in the world with distinct
images. I am ready to accept other theories, on condition that one is able to
re-express them in terms of clear and distinct images.

For my part, despite not having yet become familiar with the new ideas that I
now hear expressed,\footnote{In fact, Lorentz had followed the recent
developments rather closely. In particular, he had corresponded extensively  
with Ehrenfest and with Schr\"{o}dinger, and had even
delivered seminars and lectures on wave mechanics and on matrix mechanics at
Leiden, Cornell and Caltech. See section~\ref{scientific} (\textit{eds.}).} 
I could visualise these ideas thus. Let us take the case
of an electron that encounters an atom; let us suppose that the electron
leaves the atom and that at the same time there is emission of a light
quantum. One must consider, in the first place, the systems of waves that
correspond to the electron and to the atom before the collision. After the
collision, we will have new systems of waves. These systems of waves can be
described by a function $\psi$ defined in a space with a large number of
dimensions and satisfying a differential equation. The new wave mechanics will
work with this equation and will determine the function $\psi$ before and
after the collision.

Now, there are phenomena that teach us that there is something else in
addition to the waves, namely corpuscles; one can, for example, perform an
experiment with a Faraday cylinder; one must then take into account the
individuality of the electrons and also of the photons. I think I would find
that, to explain the phenomena, it suffices to assume that the expression
$\psi\psi^{\ast}$ gives the probability that the electrons and the photons
exist in a given volume; that would suffice to explain the experiments. But
the examples given by Mr Heisenberg teach me that I will have thus attained
everything that experiment allows me to attain. However, I think that this
notion of probability should be placed at the end\label{attheend}, and as a conclusion, of
theoretical considerations, and not as an \textit{a priori} axiom, though I
may well admit that this indeterminacy corresponds to experimental
possibilities. I would always be able to keep my deterministic faith for the
fundamental phenomena, of which I have not spoken. Could a deeper mind\label{deepermind} not be
aware of the motions of these electrons? Could one not keep determinism by
making it an object of belief? Must one necessarily elevate indeterminism to 
a principle?

\


\textsc{Mr Bohr} expounds his point of view with respect to the problems of
quantum theory.
  \begin{quote}
    The original published proceedings add `(see the preceding article)'. In the proceedings, the article 
    preceding the general discussion is a French translation of the German version of Bohr's Como lecture 
    (Bohr 1928) (published in {\em Naturwissenschaften}). As described in section~\ref{editing}, this article 
    was included at Bohr's request, to replace his remarks made at this point in the general discussion. 
    (In our translation of the proceedings, we have omitted this well-known article.)

    The extant notes relating to Bohr's remarks at this point are particularly fragmentary. Kalckar's introduction 
    to volume 6 of Bohr's {\em Collected Works} (Bohr 1985) describes the corresponding part of notes (taken by 
    Kramers and by Verschaffelt) in the Bohr archives as too incomplete to warrant reproduction in that volume, 
    but provides the following summary and comparison with the printed versions of the Como lecture: 
    `The notes cover the wave-corpuscle aspects of light 
    and matter (corresponding to the first sections of the printed lecture). The $\gamma$-ray microscope 
    is analysed, although the notes are somewhat incomplete here (as in many other places), and the 
    r\^{o}le of the finite wave trains is discussed in connection with the momentum measurement through 
    the Doppler effect (as in the printed versions). After some questions~.... Bohr continues by 
    discussing the significance of the phase and comments on the Stern-Gerlach experiment and the 
    inobservability of the phase in a stationary state~....~' (Bohr 1985, p.~37).   

    Further details of what Bohr said at this point may be obtained from notes on the general discussion 
    taken by Richardson.\endnote{Included in AHQP-RDN, document M-0309.}\label{werepr} Below, we reproduce the relevant 
    parts of these notes, and comment on their relation to Bohr's paper translated in the proceedings. 

    The first part of Richardson's notes relating to Bohr reads as follows:
  \end{quote}
  \begin{align*}
    E  & = h\nu \qquad\qquad e^{i2\pi(\tau_x x+\tau_y y+\tau_z z -\nu t)}  \\
    p  & = h\tau
  \end{align*}
Int[er]f[eren]ce. $\qquad\qquad$ ?$h\rightarrow\infty$~[?] 
  \[
    \left.
    \begin{array}{lcl}
      \Delta x \Delta\tau_x   & \sim & 1  \\
      \Delta t \Delta\nu      & \sim & 1
    \end{array}
    \right\}
    \qquad\qquad
    \begin{array}{lcl}
      \Delta x \Delta p_x   & \sim & h  \\
      \Delta t \Delta E     & \sim & h
    \end{array}
  \]

  \begin{quote}
    \noindent This corresponds to part of section~2 of Bohr's paper translated in the proceedings. There Bohr introduces the concepts 
    of energy and momentum for plane waves, and the idea that waves of limited extent in spacetime are obtained through the 
    `interference' (that is, superposition) of different plane waves, the resulting waves satisfying (at best) the given 
    relations. (As a consequence, a group of waves has no well-defined phase, a point Bohr takes up again below.) This is 
    used to justify Bohr's idea of complementarity between a causal picture (in the sense of energy-momentum conservation 
    for elementary processes) and a spacetime picture. 
      \begin{figure}
        \centering
         \resizebox{\textwidth}{!}{\includegraphics[0mm,0mm][220.30mm,160.98mm]{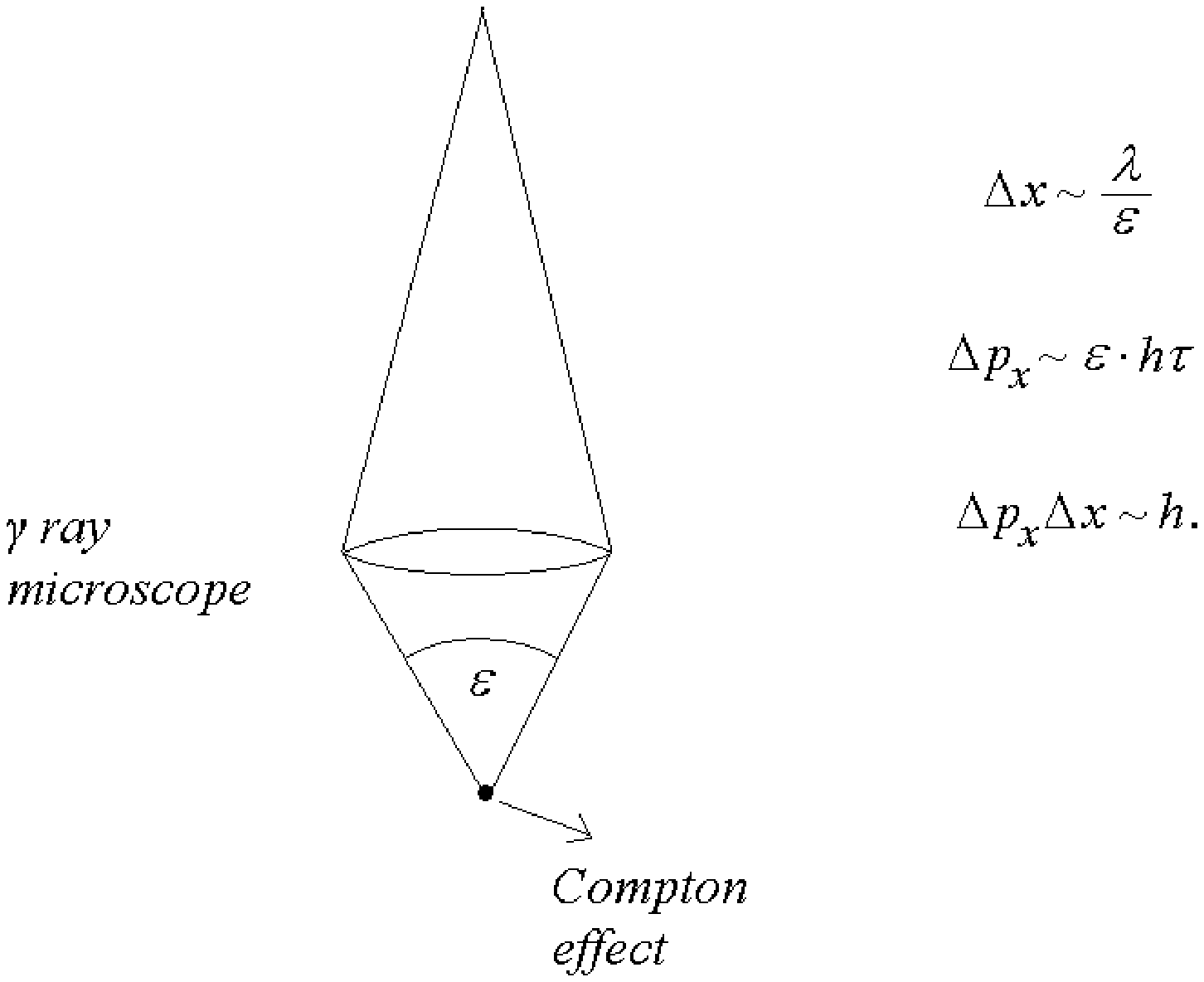}}
        \unnumberedcaption{Fig.~A.}
      \end{figure} 
    
    Richardson's notes then continue as shown in Fig.~A.
    The $\gamma$-ray microscope is discussed in section~3 of Bohr (1928) (the section on 
    measurement, which also discusses momentum measurements based on the Doppler effect). Bohr appears 
    to have inserted a discussion of these experiments as an illustration of the uncertainty-type relations above.

    The next part of Richardson's notes returns to section~2 of the paper, and is reproduced in Fig.~B.
      \begin{figure}
       \centering
         \resizebox{\textwidth}{!}{\includegraphics[-50mm,0mm][220.30mm,160.98mm]{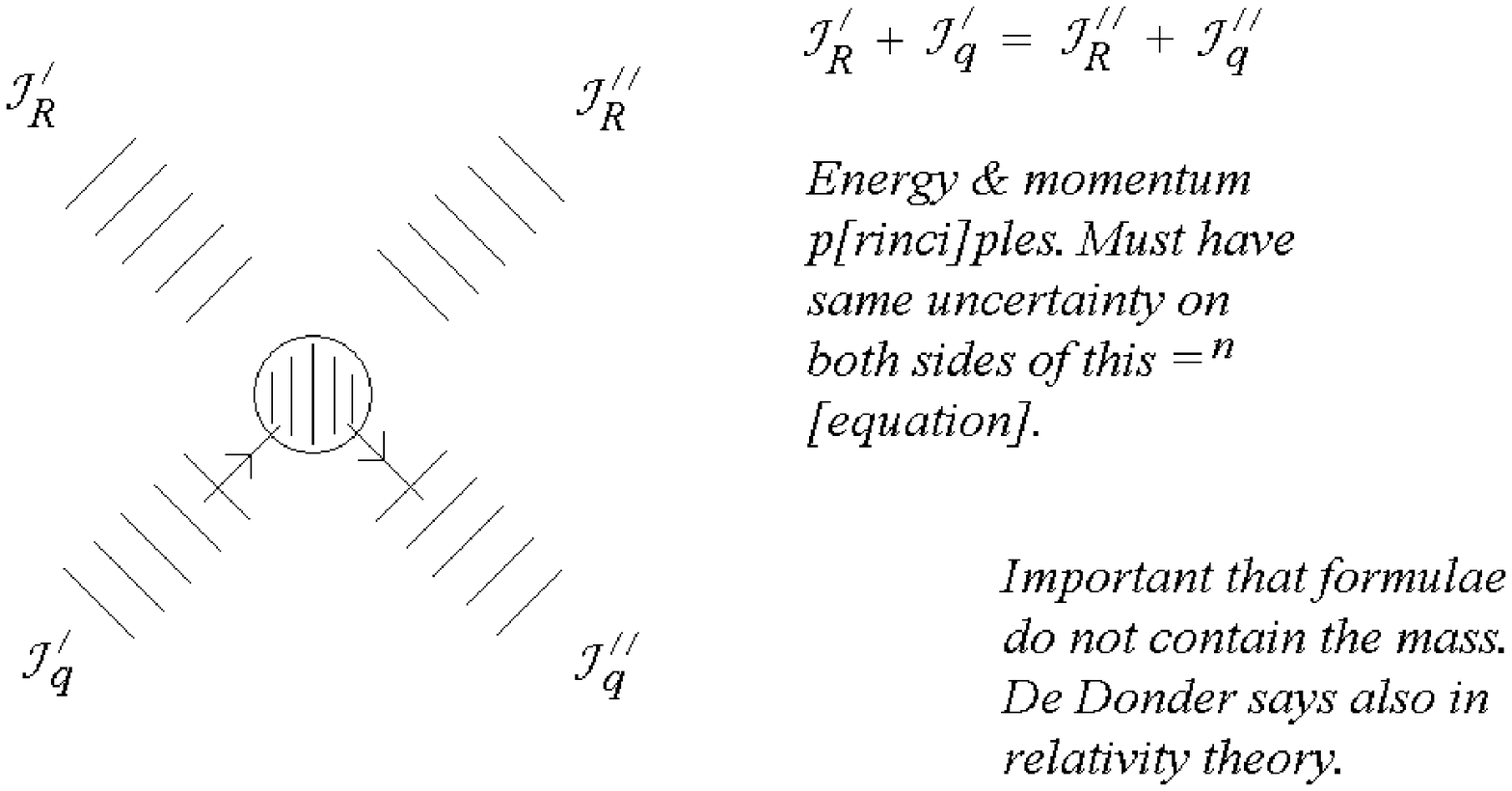}}
        \unnumberedcaption{Fig.~B.}
      \end{figure} 
    This corresponds in fact to the subsequent paragraphs of section~2, in which Bohr applies the notion of 
    complementarity to resolve the perceived paradoxes related to the scattering of radiation by free electrons 
    (note the extended --- as opposed to pointlike --- region of scattering in the diagram, and see Bohr's 
    contribution to the discussion of Compton's report, p.~\ref{forpage170}) as well as the perceived paradoxes 
    related to collisions (cf. sec\-tion~\ref{BornBohr}). 
    Possibly, ${\cal I}$ stands for `Impuls' (that is, momentum), $R$ for radiation, $q$ for charge.

    The next part of Richardson's notes, shown in Fig.~C,
      \begin{figure}
       \centering
         \resizebox{\textwidth}{!}{\includegraphics[0mm,-10mm][220.30mm,140.98mm]{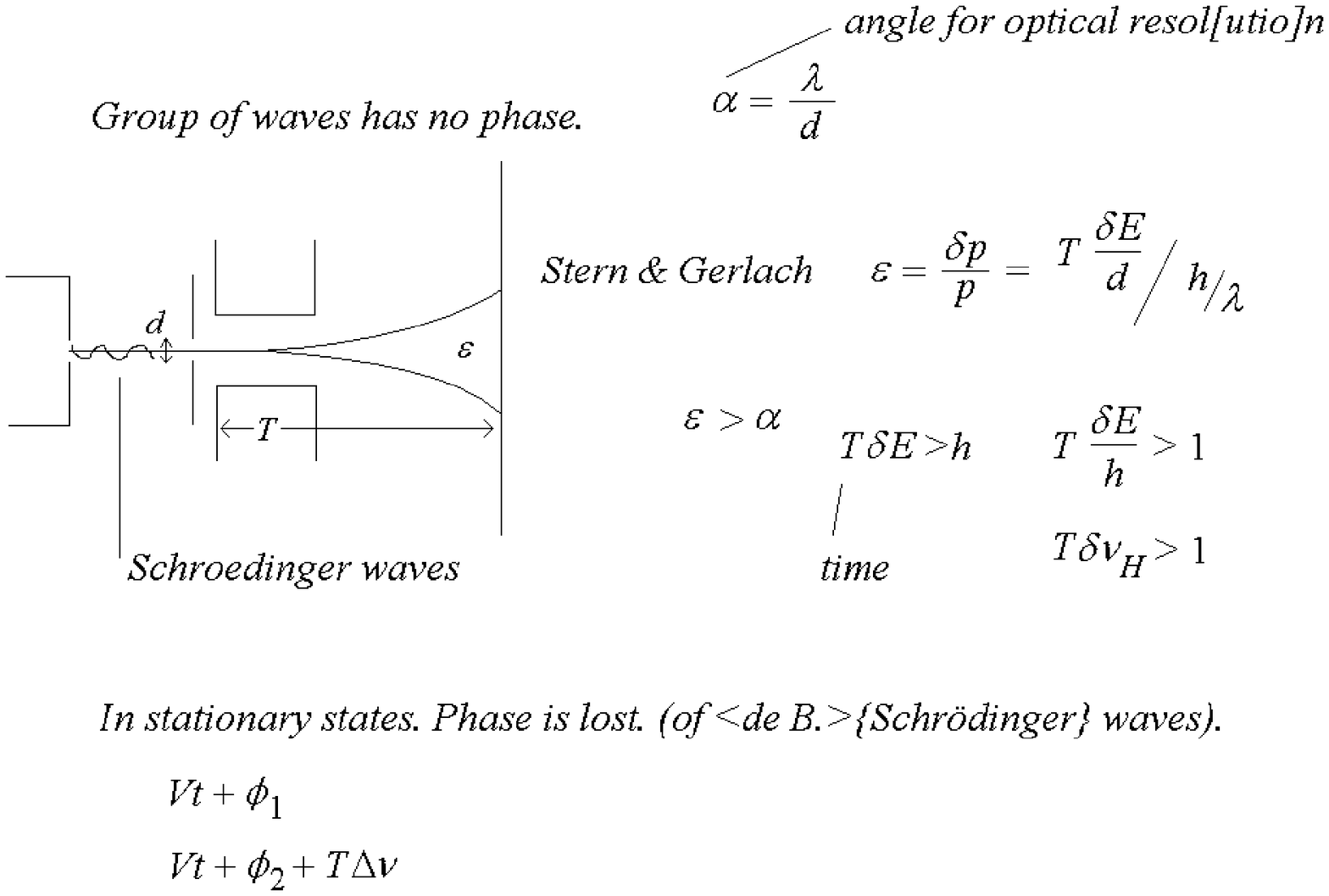}}
        \unnumberedcaption{Fig.~C.}
      \end{figure} 
    instead relates to part of section~6 of Bohr's paper (sections~4 and 5 of the paper are, respectively, a review 
    of the correspondence principle and of matrix mechanics, and a discussion and critique of wave mechanics). In 
    section~6 of the published paper, Bohr raises the following puzzle. According to Bohr, in any observation that 
    distinguishes between different stationary states one has to disregard the past history of the atom, but, 
    paradoxically, the theory assigns a phase to a stationary state. However, since the system will not be strictly 
    isolated, one will work with a group of waves, which (as mentioned in section~2) has no well-defined phase. Bohr 
    then illustrates this with the Stern-Gerlach experiment. The condition for distinguishability of the eigenstates 
    of the hydrogen atom is that the angular spreading of the beam should be greater than that given by diffraction 
    at the slit ($\varepsilon>\alpha$), which translates into the time-energy uncertainty relation. As Bohr mentions, 
    Heisenberg (1927) uses this as an illustration of the uncertainty relation, while Bohr uses it as an illustration 
    of how knowledge of the phase is lost. (This section also discusses the limit of high quantum numbers.)

    The final section~7 of the paper (`The problem of elementary particles') has no parallel in Richardson's notes. 
    The part of the notes relating to Bohr's remarks at this point concludes instead with the following (explicitly 
    labelled `Bohr'):
  \end{quote}

    \par
    \Needspace{5\baselineskip}
    \noindent 1.~[blank]
    
    \noindent 2. Stationary states, past lost $\because$ [because] phase indetermination --- Stern \&\ Gerlach's Exp[erimen]t.
    
    \noindent 3. Schroedinger's $\psi$, prob[abilit]y of electron at a given place at a given time [?], 
    uncertainty $\Delta\nu\Delta t\sim 1$
      \[
        \gamma\times\frac{v}{c}
      \]

\ 

\textsc{Mr Brillouin}.~---~Mr Bohr insists on the uncertainty of simultaneous
measurements of position and momentum; his point of view is closely connected
to the notion of \textit{cells in phase space} introduced by Planck a very
long time ago. Planck assumed that if the representative point of a system is
in a cell (of size $\Delta p\Delta q=h$) one cannot distinguish it from
another point in the same cell. The examples brought by Mr Bohr aptly make
precise the physical meaning of this quite abstract notion.

\ 

\textsc{Mr De Donder}.~---~The considerations that Mr Bohr has just developed
are, I think, in close relation with the following fact: in the Einsteinian
Gravitation\footnote{Th.~De Donder, {\em Th\'{e}orie des
champs gravifiques} (\textit{M\'{e}morial des sciences math\'{e}matiques}, part
14, Paris, 1926). See esp. equations (184), (184%
\'{}%
) and (188), (188%
\'{}%
). One can also consult our lectures: \textit{The Mathematical Theory of
Relativity} (Massachusetts Institute of Technology), Cambridge, Mass., 1927.
See esp. equations (23), (24) and (28), (29).} of a continuous system or of a
pointlike system, there appear not the masses and charges of the particles,
but entities $\tau^{(m)}$ and $\tau^{(e)}$ in \textit{four} dimensions; note
that these \textit{generalised} masses and charges, localised in spacetime,
\textit{are conserved} along their worldlines.

\ 

\textsc{Mr Born}.\label{Bornbeginning}~---~Mr Einstein has considered the following problem: A
radioactive sample emits $\alpha$-particles in all directions; these are made
visible by the method of the Wilson cloud [chamber]. Now, if one associates a
spherical wave with each emission process, how can one understand that the
track of each $\alpha$ particle appears as a (very nearly) straight line? In
other words: how can the corpuscular character of the phenomenon be reconciled
here with the representation by waves?

To do this, one must appeal to the notion of `reduction of the probability
packet'\label{page172} developed by Heisenberg.\footnote{Born is referring here in particular 
to Heisenberg's uncertainty paper (Heisenberg 1927) ({\em eds}.).} The description\label{thedescription} of the emission 
by a spherical wave is valid only for as long as one does not observe ionisation;
as soon as such ionisation is shown by the appearance of cloud droplets, in
order to describe what happens afterwards one must `reduce' the wave packet in
the immediate vicinity of the drops. One thus obtains a wave packet in the
form of a ray, which corresponds to the corpuscular character of the phenomenon.

Mr Pauli\label{MrPauli}\footnote{Cf.\ Pauli's letter to Bohr, 17 October 1927, discussed in
section~\ref{QM-without-collapse} (\textit{eds.}).} has asked me
if it is not possible to describe the process without the reduction of wave
packets, by resorting to a multi-dimensional space, whose number of dimensions
is three times the number of all the particles present ($\alpha$-particles and
atoms hit by the radiation).

This is in fact possible and can even be represented in a very anschaulich manner 
[d'une mani\`{e}re fort intuitive] by means of an appropriate simplification, but this does
not lead us further as regards the fundamental questions. Nevertheless, I
should like to present this case here as an example of the multi-dimensional
treatment of such problems.
  \begin{figure}
    \centering
      \resizebox{\textwidth}{!}{\includegraphics[0mm,0mm][200.00mm,150.45mm]{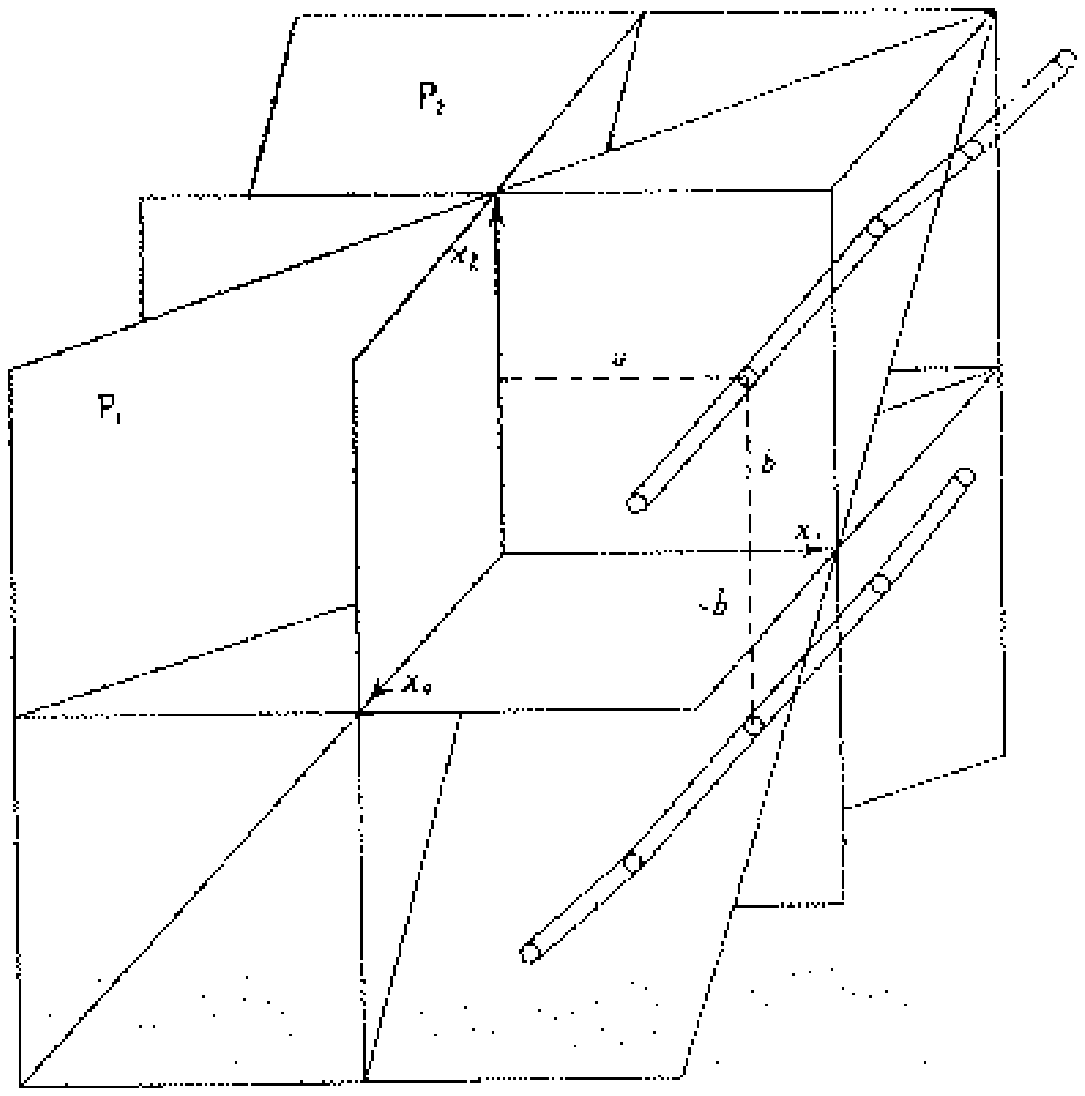}}
    \unnumberedcaption{Fig.~1.}
    \label{Bornfigure}
  \end{figure} 
I assume, for simplicity, that there are only two atoms that may be hit. One
then has to distinguish two cases: either the two atoms 1 and 2 lie on the
same ray starting from the origin (the place where the preparation is), or
they do not lie on the same ray. If we represent by $\varepsilon$ the
probability that an atom will be hit, we have the following probability
diagram:\footnote{In the following tables, the probability for the number of
particles hit to equal 1 should be read as the probability for \textit{each
case} in which the number of particles hit equals 1 (\textit{eds.}).}

\

\par
\Needspace{8\baselineskip}
\noindent I. The points 1 and 2 are located on the same ray starting from the origin.

\begin{center}%
\begin{tabular}
[c]{cc}%
{\small Number of particles hit} & {\small Probability}\\
${\small 0}$ & ${\small 1-\varepsilon}$\\
${\small 1}$ & ${\small 0}$\\
${\small 2}$ & ${\small \varepsilon}$%
\end{tabular}

\end{center}

\par
\Needspace{8\baselineskip}
\noindent II. The points 1 and 2 are not on the same ray.

\begin{center}%
\begin{tabular}
[c]{cc}%
{\small Number of particles hit} & {\small Probability}\\
${\small 0}$ & ${\small 1-2\varepsilon}$\\
${\small 1}$ & ${\small \varepsilon}$\\
${\small 2}$ & ${\small 0}$%
\end{tabular}

\end{center}

This is how one should express the probability of events in the case of
rectilinear propagation.

To make possible a graphical representation of the phenomenon, we will
simplify it further by assuming that all the motions take place following only
a single straight line, the axis $x$. We must then distinguish the two cases
where the atoms lie on the same side and on either side of the origin. The
corresponding probabilities are the following:

\

\par
\Needspace{8\baselineskip}
\noindent I. The points 1 and 2 are located on the same side.

\begin{center}%
\begin{tabular}
[c]{cc}%
{\small Number of particles hit} & {\small Probability}\\
${\small 0}$ & $\frac{1}{2}$\\
${\small 1}$ & ${\small 0}$\\
${\small 2}$ & $\frac{1}{2}$%
\end{tabular}

\end{center}

\

\par
\Needspace{8\baselineskip}
\noindent II. The points 1 and 2 are located on different sides.

\begin{center}%
\begin{tabular}
[c]{cc}%
{\small Number of particles hit} & {\small Probability}\\
${\small 0}$ & ${\small 0}$\\
${\small 1}$ & $\frac{1}{2}$\\
${\small 2}$ & ${\small 0}$%
\end{tabular}

\end{center}

Now, these relations can be represented by the motion of a wave packet in a
space with three dimensions $x_{0}$, $x_{1}$, $x_{2}$. To the initial state
there corresponds:%
\[
\begin{array}{llll}
\mbox{In case I, the point}  & x_0=0, & x_1=a & x_2=b \\
\mbox{In case II, the point} & x_0=0, & x_1=a & x_2=-b 
\end{array}
\]
where $a$ and $b$ are positive numbers. The wave packet at first fills the
space surrounding these points and subsequently moves parallel to the axis
$x_{0}$, dividing itself into two packets of the same size going in opposite
directions. Collisions are produced when $x_{0}=x_{1}$ or $x_{0}=x_{2}$, that
is to say, on two planes of which one, P$_{1}$, is parallel to the axis
$x_{2}$ and cuts the plane $x_{0}x_{1}$ following the bisector of the positive
quadrant, while the second, P$_{2}$, is parallel to the axis $x_{1}$ and cuts
the plane $x_{0}x_{2}$ following the bisector of the positive quadrant. As
soon as the wave packet strikes the plane P$_{1}$, its trajectory receives a
small kink in the direction $x_{1}$; as soon as it strikes P$_{2}$ the
trajectory receives a kink in the direction $x_{2}$ (Fig.~1).

Now, one immediately sees in the figure that the upper part of the wave
packet, which corresponds to case I, strikes the planes P$_{1}$, P$_{2}$ on
the same side of the plane $x_{1}x_{2}$, while the lower part strikes
them\endnote{French edition: `les' is misprinted as `le'.} on different sides. The
figure then gives an anschaulich representation of the cases indicated in the
above diagram. It allows us to recognise immediately whether, for a given size
of wave packet, a given state, that is to say a given point $x_{0},x_{1},x_{2}$, 
can be hit or not.

To the `reduction'\label{tothereduction} of the wave packet corresponds the choice of one of the two
directions of propagation $+x_{0}$, $-x_{0}$, which one must take as soon as
it is established that one of the two points 1 and 2 is hit, that is to say,
that the trajectory of the packet has received a kink.

This example serves only to make clear that a complete description of the
processes taking place in a system composed of several molecules is possible
only in a space of several dimensions.

\ 

\textsc{Mr Einstein}.\label{Einstein-disc}\footnote{The extant manuscript in the Einstein 
archives\endnotemark consists of the first four paragraphs only, which we have
translated here (footnoting significant differences from the published
French) ({\em eds}.).}\endnotetext{AEA 16-617.00 (in German, with transcription and archival comments).}~---~Despite 
being conscious of the fact that I have not entered
deeply enough into the essence of quantum mechanics, nevertheless I want to
present here some general remarks.\footnote{The published French has: `I must
apologise for not having gone deeply into quantum mechanics. I should
nevertheless want to make some general remarks' ({\em eds}.).}

One can take two positions towards the theory with respect to its postulated
domain of validity, which I wish to characterise with the aid of a simple example.

Let S be a screen provided with a small opening O (Fig.~2), and P a
hemispherical photographic film
  \begin{figure}
    \centering
      \resizebox{\textwidth}{!}{\includegraphics[0mm,0mm][220.58mm,110.88mm]{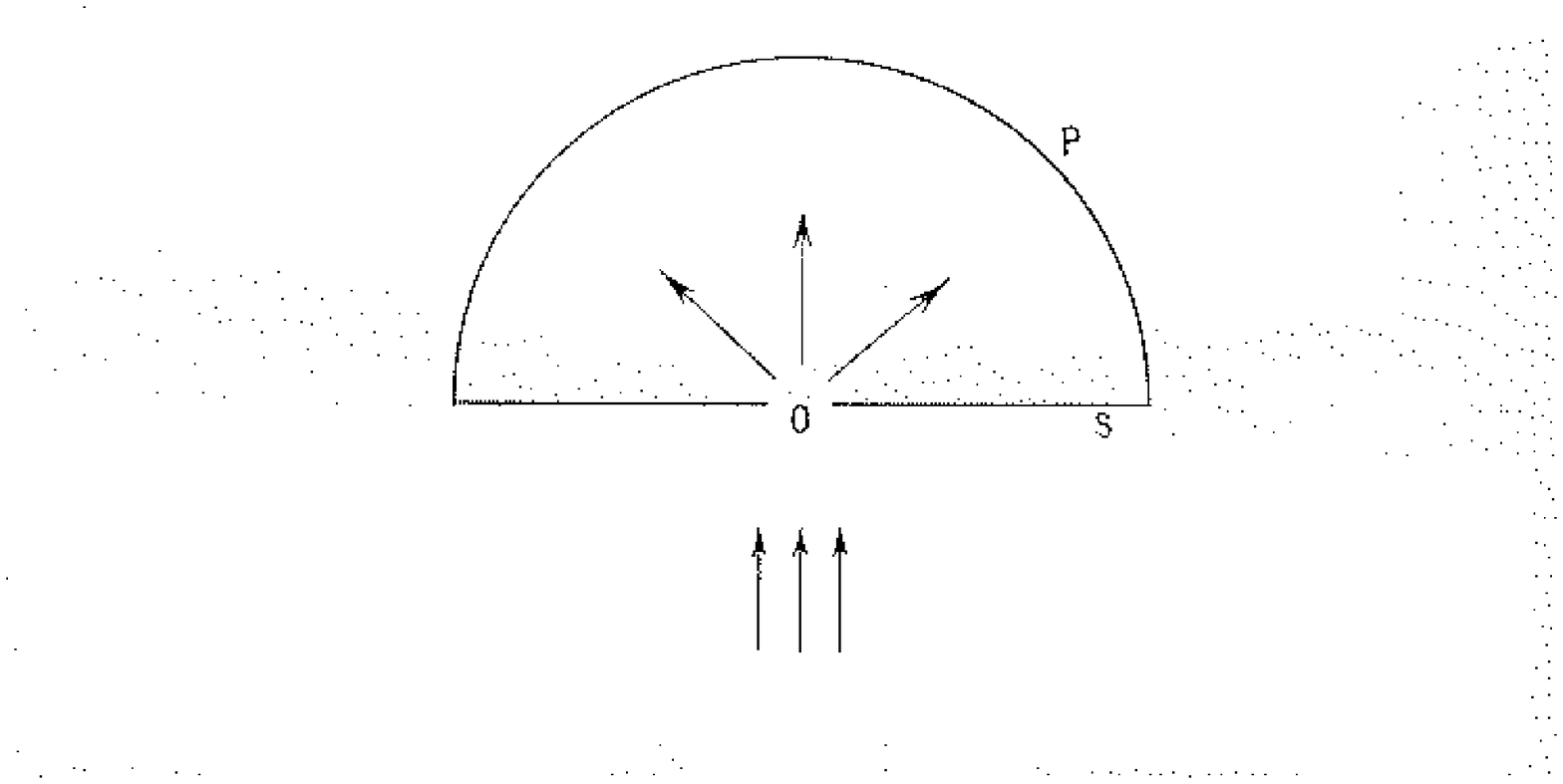}}
    \unnumberedcaption{Fig.~2.}
    \label{Einsteinfigure}
  \end{figure} 
of large radius. Electrons impinge on S in the direction of the arrows. Some
of these go through O, and because of the smallness of O and the speed of the
particles, are dispersed uniformly over the directions of the hemisphere, and
act on the film.

Both ways of conceiving the theory now have the following in common. There are
de Broglie waves, which impinge approximately normally on S and are diffracted
at O. Behind S there are spherical waves, which reach the screen P and whose
intensity at P is responsible [massgebend] for what happens at P.\footnote{In the 
published French, the German expression `ist massgebend' is misrendered as `donne 
la mesure' [gives the measure] instead of as `is responsible'. This is of some 
significance for the interpretation of Einstein's remarks as a form of the later 
EPR argument; see section~\ref{1927-incompleteness} (\textit{eds.}).}

We can now characterise the two points of view as follows.

\ 

\noindent 1. \textit{Conception I}. --- The de Broglie-Schr\"{o}dinger waves do not
correspond to a single electron, but to a cloud of electrons extended in
space. The theory gives no information about individual processes, but only
about the ensemble of an infinity of elementary processes.

\ 

\noindent 2. \textit{Conception II}. --- The theory claims to be a complete theory of
individual processes. Each particle directed towards the screen, as far as can
be determined by its position and speed, is described by a packet of de
Broglie-Schr\"{o}dinger waves of short wavelength and small angular width.
This wave packet is diffracted and, after diffraction, partly reaches the film
P in a state of resolution [un \'{e}tat de r\'{e}solution].

\

According to the first, purely statistical, point of view $\left\vert
\psi\right\vert ^{2}$ expresses the probability that there exists at the point
considered \textit{a particular} particle of the cloud, for example at a given
point on the screen.

According to the second, $\left\vert \psi\right\vert ^{2}$ expresses the
probability that at a given instant \textit{the same} particle is present at a
given point (for example on the screen). Here, the theory refers to an
individual process and claims to describe everything that is governed by laws.

The second conception\label{secondconception} goes further than the first, in the sense that all the
information resulting from I results also from the theory by virtue of II, but
the converse is not true.\footnote{The French has `I' and `II' exchanged in
this sentence, which is illogical (\textit{eds.}).} It is only by virtue of II that the
theory contains the consequence that the conservation laws are valid for the
elementary process; it is only from II that the theory can derive the result
of the experiment of Geiger and Bothe, and can explain the fact that in the
Wilson [cloud] chamber the droplets stemming from an $\alpha$-particle are
situated very nearly on continuous lines.

But on the other hand, I have objections to make to conception II.\label{Einstein-loc} The
scattered wave directed towards P does not show any privileged direction. If
$\left\vert \psi\right\vert ^{2}$ were simply regarded as the probability that
at a certain point a given particle is found at a given time, it could happen
that \textit{the same} elementary process produces an action \textit{in two or
several} places on the screen. But the interpretation, according to which
$\left\vert \psi\right\vert ^{2}$ expresses the probability that \textit{this}
particle is found at a given point, assumes an entirely peculiar mechanism of
action at a distance, which prevents the wave continuously distributed in
space from producing an action in \textit{two} places on the screen.

In my opinion, one can remove this objection only in the following way, that
one does not describe the process solely by the Schr\"{o}dinger wave, but that
at the same time one localises the particle during the propagation.\label{locprop} I think
that Mr de Broglie is right to search in this direction.\label{searchdirection} If one works solely
with the Schr\"{o}dinger waves, interpretation II of $\left\vert
\psi\right\vert ^{2}$ implies to my mind a contradiction with the postulate of relativity.

I should also like to point out briefly two arguments which seem to me to
speak against the point of view II.\label{Einstein-config} This [view] is essentially tied\label{essentiallytied} to a
multi-dimensional representation (configuration space), since only this mode
of representation makes possible the interpretation of $\left\vert
\psi\right\vert ^{2}$ peculiar to conception II. Now, it seems to me that
objections of principle are opposed to this multi-dimensional representation.
In this representation, indeed, two configurations of a system that are
distinguished only by the permutation of two particles of the same species are
represented by two different points (in configuration space), which is not in
accord with the new results in statistics. Furthermore, the feature of forces\label{featureofforces}
of acting only at small \textit{spatial} distances finds a less natural
expression in configuration space than in the space of three or four dimensions.

\ 

\textsc{Mr Bohr}.\footnote{These remarks by Bohr do not appear in the
published French. We have reproduced them from Bohr's {\em Collected Works}, vol.~6 
(Bohr 1985, p.~103), which contains a reconstruction of Bohr's remarks from notes by
Verschaffelt (held in the Bohr archive). The tentative interpolations in square brackets 
are by the editor of Bohr (1985), J.~Kalckar (\textit{eds.}).}~---~I feel myself in a very difficult position
because I don't understand what precisely is the point which Einstein wants to
[make]. No doubt it is my fault.

....

As regards general problem I feel its difficulties. I would put problem in
other way. I do not know what quantum mechanics is. I think we are dealing
with some mathematical methods which are adequate for description of our
experiments. Using a rigorous wave theory we are claiming something which the
theory cannot possibly give. [We must realise] that we are away from that
state where we could hope of describing things on classical theories.
Understand same view is held by Born and Heisenberg. I think that we actually
just try to meet, as in all other theories, some requirements of nature, but
difficulty is that we must use words which remind of older theories. The whole
foundation\label{Bohrfoundation} for causal spacetime description is taken away by quantum theory,
for it is based on assumption of observations without interference. ....
excluding interference means exclusion of experiment and the whole meaning of
space and time observation .... because we [have] interaction [between object
and measuring instrument] and thereby we put us on a quite different
standpoint than we thought we could take in classical theories. If we speak of
observations we play with a statistical problem. There are certain features
complementary to the wave pictures (existence of individuals). ....

....

The saying that spacetime is an abstraction might seem a philosophical
triviality but nature reminds us that we are dealing with something of
practical interest. Depends on how I consider theory. I may not have
understood, but I think the whole thing lies [therein that the] theory is
nothing else [but] a tool for meeting our requirements and I think it does.

\ 

\textsc{Mr Lorentz}.~---~To represent the motion of a system of $n$ material
points, one can of course make use of a space of 3 dimensions with $n$ points
or of a space of $3n$ dimensions where the systems will be represented by a
single point. This must amount to exactly the same thing; there can be no
fundamental difference. It is merely a question of knowing which of the two
representations is the most suitable, which is the most convenient.

But I understand that there are cases where the matter is difficult. If one
has a representation in a space of $3n$ dimensions, one will be able to return
to a space of $3$ dimensions only if one can reasonably separate the $3n$
coordinates into $n$ groups of $3$, each corresponding to a point, and I could
imagine that there may be cases where that is neither natural nor simple. But,
after all, it certainly seems to me that all this concerns the form rather
than the substance of the theory.

\ 

\textsc{Mr Pauli}.~---~I am wholly\label{Pauliwholly} of the same opinion as Mr Bohr, when he
says that the introduction of a space with several dimensions is only a
technical means of formulating mathematically the laws of mutual action
between several particles, actions which certainly do not allow themselves to
be described simply, in the ordinary way, in space and time. It may perfectly
well be that this technical means may one day be replaced by another, in the
following fashion. By Dirac's method one can, for example, quantise the
characteristic vibrations of a cavity filled with blackbody radiation, and
introduce a function $\psi$ depending on the amplitudes of these
characteristic vibrations of unlimited number. One can similarly use, as do
Jordan and Klein, the amplitudes of ordinary four-dimensional material waves
as arguments of a multi-dimensional function $\phi$. This gives, in the
language of the corpuscular picture, the probability that at a given instant
the numbers of particles of each species present, which have certain
kinematical properties (given position or momentum), take certain values. This
procedure also has the advantage that the defect of the ordinary
multi-dimensional method, of which Mr Einstein has spoken and which appears
when one permutes two particles of the same species, no longer exists. As
Jordan and Klein have shown, making suitable assumptions concerning the
equations that this function $\phi$ of the amplitudes of material waves in
ordinary space must satisfy,\endnote{The French text has `$\psi$' instead of `$\phi$', 
and `doit satisfaire dans l'espace ordinaire' instead of the other way round. 
Note that $\phi$ is a functional of `material' waves which themselves
propagate in ordinary space.} one arrives exactly at the same results as
by basing oneself on Schr\"{o}dinger's multi-dimensional theory.

To sum up, I wish then to say that Bohr's point of view, according to which
the properties of physical objects of being defined and of being describable
in space and time are complementary, seems to be more general than a special
technical means. But, independently of such a means, one can, according to
this idea, declare in any case that the mutual actions of several particles
certainly cannot be described in the ordinary manner in space and time.\label{ordinary}

To make clear the state of things of which I have just spoken, allow me to
give a special example. Imagine two hydrogen atoms in their ground state at a
great distance from each other, and suppose one asks for their energy of
mutual action. Each of the two atoms has a perfectly isotropic distribution of
charge, is neutral as a whole, and does not yet emit radiation. According to
the ordinary description of the mutual action of the atoms in space and time,
one should then expect that such a mutual action does not exist when the
distance between the two neutral spheres is so great that no notable
interpenetration takes place between their charge clouds. But when\label{whenonetreats} one treats
the same question by the multi-dimensional method, the result is quite
different, and in accordance with experiment.

The classical analogy to this last result would be the following: Imagine
inside each atom a classical oscillator whose moment $p$ varies periodically.
This moment produces a field at the location of the other atom whose
periodically variable intensity is of order $\mathcal{E}\sim\frac{p}{r^{3}}$,
where $r$ is the distance between the two atoms. When two of these oscillators
act on each other, a polarisation occurs with the following potential energy,
corresponding to an attractive force between the atoms,%
\[
-\frac{1}{2}\alpha\mathcal{E}^{2}\sim\frac{1}{2}\alpha p^{2}\frac{1}{r^{6}%
}\;,
\]
where $\alpha$ represents the polarisability of the atom.

In speaking of these oscillators, I only wanted to point out a classical
analogy with the effect that one obtains as a result of multi-dimensional wave
mechanics. I had found this result by means of matrices, but Wang has derived
it directly from the wave equation in several dimensions. In a paper by
Heitler and London, which is likewise concerned with this problem, the authors
have lost sight of the fact that, precisely for a large distance between the
atoms, the contribution of polarisation effects to the energy of mutual
action, a contribution which they have neglected, outweighs in order of
magnitude the effects they have calculated.

\ 

\textsc{Mr Dirac}.\label{Dirac-contribution}\footnote{Here we mostly follow the English 
version from Dirac's manuscript.\endnotemark (The French translation may have been done 
from a typescript or fairer copy.) We generally follow the French paragraphing, and we
uniformise Dirac's notation. Interesting variants, cancellations and additions will be noted, as 
will significant deviations from the published French (\textit{eds.}).}\endnotetext{AHQP-36, 
section 10.}~---~I should like to express my ideas on a few questions.

The first is the one that has just been discussed and I have not much to add
to this discussion. I shall just mention the explanation that the quantum
theory would give of Bothe's experiment.\endnote{The French adds:
`d\'{e}crite par M. Compton'.} The difficulty arises from\endnote{The
French reads: `provient uniquement de'.} the inadequacy of the
3-dimensional wave picture. This picture cannot distinguish between the case
when there is a probability $p$ of a light-quant being in a certain small
volume, and the case when there is a probability $\frac{1}{2}p$ of two
light-quanta being in the volume, and no probability for only one. But the
wave function in many-dimensional space does distinguish between these cases.
The theory of Bothe's experiment in many-dimensional space would show that,
while there is a certain probability for a light-quantum appearing in one or
the other of the counting chambers, there is no probability of two appearing simultaneously.

At present the general theory of the wave function in many-dimensional space
necessarily involves the abandonment of relativity.\label{abandonment}\endnote{Dirac's
manuscript omits `At present'.} One might, perhaps, be able to bring
relativity into the general quantum theory in the way Pauli has mentioned 
of quantising 3-dimensional waves, but this would not lead to greater
Anschaulichkeit\endnote{The French reads `intuitivit\'{e}'.} in the
explanation of results such as Bothe's.

I shall now show how Schr\"{o}dinger's expression for the electric density
appears naturally in the matrix theory.\label{Dirac-Schr} This will show the exact signification
of this density and the limitations which must be imposed on its use. Consider
an electron moving in an arbitrary field, such as that of an H atom. Its
coordinates $x,y,z$ will be matrices. Divide the space up into a number of
cells, and form that function of $x,y,z$ that is equal to 1 when the electron
is in a given cell and 0 otherwise. This function of the matrices $x,y,z$ will
also be a matrix.\footnote{The published version has: `Divide the space up
into a large number of small cells, and consider the function of three
variables $\xi$, $\eta$, $\zeta$ that is equal to $1$ when the point $\xi$,
$\eta$, $\zeta$ is in a given cell and equal to $0$ when the point is
elsewhere. This function, applied to the matrices $x$, $y$, $z$, gives another
matrix' (\textit{eds.}).} There is one such matrix for each cell whose matrix
elements will be functions of the coordinates $a,b,c$ of the cell, so that it
can be written $A(a,b,c)$.

Each of these matrices represents a quantity that if measured experimentally
must have either the value 0 or 1. Hence each of these matrices has the
characteristic values 0 and 1 and no others. If one takes the two matrices
$A(a,b,c)$ and $A(a^{\prime},b^{\prime},c^{\prime})$, one sees that they must
commute,\endnote{Instead of `commute' the French has `permuter leurs valeurs'.} 
since one can give a numerical value to both simultaneously; for
example, if the electron is known to be in the cell $a,b,c$, it will
certainly not be in the cell $a^{\prime},b^{\prime},c^{\prime}$, so that if
one gives the numerical value 1 to $A(a,b,c)$, one must at the same time give
the numerical value 0 to $A(a^{\prime},b^{\prime},c^{\prime})$.

We can transform each of the matrices $A$ into a diagonal matrix $A^{\ast}$ by
a transformation\endnote{The French reads: `transformation canonique'.}
of the type%
\[
A^{\ast}=BAB^{-1}.
\]

Since all the matrices $A(a,b,c)$ commute,\endnote{Instead of `commute' the
French has `changent de valeur'.} they can be transformed simultaneously
into diagonal matrices by a transformation of this type. The diagonal elements
of each matrix $A^{\ast}(a,b,c)$ are its characteristic values, which are the
same as the characteristic values of $A(a,b,c)$, that is, 0 and 1.

Further, no two $A^{\ast}$ matrices, such as $A^{\ast}(a,b,c)$ [and] $A^{\ast
}(a^{\prime},b^{\prime},c^{\prime})$, can both have 1 for the same diagonal
element, as a simple argument shows that $A^{\ast}(a,b,c)+A^{\ast}(a^{\prime
},b^{\prime},c^{\prime})$ must also have only the characteristic values 0 and
1. We can without loss of generality assume that each $A^{\ast}$ has just one
diagonal element equal to 1 and all the others zero. By transforming back, by
means of the formula
\[
A(a,b,c)=B^{-1}A^{\ast}(a,b,c)B\ ,
\]
we now find that the matrix elements of $A(a,b,c)$ are of the form%
\[
A(a,b,c)_{mn}=B_{m}^{-1}B_{n}\ ,
\]
i.e.\ a function of the row multiplied by a function of the column.

It should be observed that the proof of this result is quite independent of
equations of motion and quantum conditions. If we take these into account, we
find that $B_{m}^{-1}$ and $B_{n}$ are apart from constants just
Schr\"{o}dinger's eigenfunctions $\bar{\psi}_{m}$ and $\psi_{n}$ at the
point $a,b,c$.

Thus Schr\"{o}dinger's density function $\bar{\psi}_{m}(x,y,z)\psi
_{m}(x,y,z)$ is a\endnote{Dirac's manuscript reads `the'.} diagonal
element of the matrix $A$ referring to a cell about the point $x,y,z$. The
true quantum expression for the density is the whole matrix. Its diagonal
elements give only the average density, and must not be used when the density
is to be multiplied by a dynamical variable represented by a matrix.

I should now like to express my views on determinism\label{Dirac-determinism} and the nature of the
numbers appearing in the calculations of the quantum theory, as they appear to
me after thinking over Mr Bohr's remarks of yesterday.\endnote{In Dirac's
manuscript, the words `determinism and' are cancelled and possibly reinstated.
They appear in the French, which also omits `of yesterday'.} In the
classical theory one starts from certain numbers describing completely the
initial state of the system, and deduces other numbers that describe
completely the final state. This deterministic theory applies only to an
isolated system.

But, as Professor Bohr has pointed out, an isolated system is by definition
unobservable. One can observe the system only by disturbing it and observing
its reaction to the disturbance. Now since physics is concerned only with
observable quantities the deterministic classical theory is
untenable.\endnote{%
$<$%
therefore unsatisfactory%
$>$
$<$%
untenable%
$>$%
, the latter seems reinstated. The French has `ind\'{e}fendable'.}

In the quantum theory\label{inthequantumtheory} one also begins with certain numbers and deduces others
from them. Let us inquire into the distinguishing
cha\-rac\-te\-ris\-tics\endnote{Instead of `the distinguishing characteristics' the
French has `l'essence physique'.} of these two sets of numbers. The
disturbances that an experimenter applies to a system to observe it are
directly under his control, and are acts of freewill by him.\label{Dirac-freewill} \emph{It is only
the numbers that describe these acts of freewill that can be taken as initial
numbers for a calculation in the quantum theory}. Other numbers describing the
initial state of the system are inherently unobservable, and do\endnote{%
$<$%
do%
$>$%
, \{would\} appears above the line, \{can\} below. The French reads `ne
figurent pas'.} not appear in the quantum theoretical treatment.

Let us now consider the final numbers obtained as the result of an experiment.
It is essential that the result of an experiment shall be a permanent record.
The numbers that describe such a result must help to not only describe the
state of the world at the instant the experiment is ended, but also help to
describe the state of the world at any subsequent time. These numbers describe
what is common to all the events in a certain chain of causally connected
events, extending indefinitely into the future.

Take as an example a Wilson cloud expansion experiment. The causal chain here
consists of the formation of drops of water round ions, the scattering of
light by these drops of water, and the action of this light on a photographic
plate, where it leaves a permanent record. The numbers that form the result of
the experiment describe all of the events in this chain equally well and help
to describe the state of the world at any time after the chain began.

One could perhaps extend the chain further into the past.\endnote{This
sentence does not appear in the French.} In the example one could,
perhaps, ascribe the formation of the ions to a $\beta$-particle, so that the
result of the experiment would be numbers describing the track of a $\beta
$-particle. In general one tries with the help of theoretical considerations
to extend the chain as far back into the past as possible, in order that the
numbers obtained as the result of the experiment may apply as directly as
possible to the process under investigation.\endnote{In the French, this
sentence appears at the beginning of the paragraph.}

This view of the nature of the results of experiments fits in very well with
the new quantum mechanics. According to quantum mechanics the state of the
world at any time is describable by a wave function $\psi$, which normally
varies according to a causal law, so that its initial value determines its
value at any later time.\label{Dirac-state} It may however happen that at a certain time $t_{1}$,
$\psi$ can be expanded in the form%
\[
\psi=\sum_{n}c_{n}\psi_{n}\ ,
\]
where the $\psi_{n}$'s are wave functions of such a nature that they cannot
interfere with one another at any time subsequent to $t_{1}$. If such is the
case, then the world at times later than $t_{1}$ will be described not by
$\psi$ but by one of the $\psi_{n}$'s. The particular $\psi_{n}$ that it shall
be must be regarded as chosen by nature.\endnote{This sentence does not
appear in the French.} One may say that nature chooses which $\psi_{n}$
it is to be, as the only information given by the theory is that the
probability of any $\psi_{n}$ being chosen is $\left\vert c_{n}\right\vert
^{2}$.\endnote{Dirac's manuscript has `$c_{n}^{2}$'.} The value of the
suffix $n$ that labels the particular $\psi_{n}$ chosen may be the result of
an experiment, and the result of an experiment must always be such a number.
It is a number describing an irrevocable choice of nature,\label{Dirac-choice} which must affect
the whole of the future course of events.\footnote{The last two sentences
appear differently in the published version: `The choice, once made, is
irrevocable and will affect the whole future state of the world. The value of
$n$ chosen by nature can be determined by experiment and \textit{the results
of all experiments} are numbers describing such choices of nature'.
\par
Dirac's notes contain a similar variant written in the margin: `The value of
$n$ chosen by nature may be determined by experiment. The result of every
experiment consists of numbers determining one of these choices of nature, and
is permanent since such a choice is irrevocable and affects the whole future
state of the world' (\textit{eds.}).}

As an example take the case of a simple collision problem. The wave packet
representing the incident electron gets scattered in all directions. One must
take for the wave function after the process not the whole scattered wave, but
once again a wave packet moving in a definite direction. From the results of
an experiment,\label{fromresults} by tracing back a chain of causally connected events one could
determine in which direction the electron was scattered and one would thus
infer that nature had chosen this direction. If, now,\label{ifnow} one arranged a mirror to
reflect the electron wave scattered in one direction $d_{1}$ so as to make it
interfere with the electron wave scattered in another direction $d_{2}$, one
would not be able to distinguish between the case when the electron is
scattered in the direction $d_{2}$ and when it is scattered in the direction
$d_{1}$ and reflected back into $d_{2}$. One would then not be able to trace
back the chain of causal events so far, and one would not be able to say that
nature had chosen a direction as soon as the collision occurred, but only
[that] at a later time nature chose where the electron should appear.
The\endnote{%
$<$%
Thus
$<$%
a possibility%
$>$
\{the existence\} of%
$>$
.} interference between the $\psi_{n}$'s compels nature to postpone her choice.\label{postpone}

\ 

\textsc{Mr Bohr}.\footnote{Again, these remarks do not appear in the published
French and we have reproduced them from Bohr's {\em Collected Works} (Bohr 1985, 
p.~105) (\textit{eds.}).}~---~Quite see that one must go into details of
pictures, if one wants to control or illustrate general statements. I think
still that you may simpler put it in my way. Just this distinction between
observation and definition allows to let the quantum mechanics appear as
generalisation. What does mean: get records which do not allow to work
backwards. Even if we took all molecules in photographic plate one would have
closed system. If we tell of a record we give up definition of plate. Whole
point lies in that by observation we introduce something which does not allow
to go on.

.... 

\ 

\textsc{Mr Born}.~---~I should like to point out, with regard to the
considerations of Mr Dirac, that they seem closely related to the ideas
expressed in a paper by my collaborator J.\endnote{The French has `F.'.}
von Neumann, which will appear shortly. The author of this paper shows that
quantum mechanics can be built up using the ordinary probability calculus,
starting from a small number of formal hypotheses; the probability amplitudes
and the law of their composition do not really play a role there.

\ 

\textsc{Mr Kramers}.\label{Kramers-Schr}~---~I think the most elegant way to arrive at the results
of Mr Dirac's considerations is given to us by the methods he presented in his
memoir in the \textit{Proc.\ Roy.\ Soc.}, ser.\ {\em A}, vol.~\textbf{113}, p.~621. Let
us consider a function of the coordinates $q_{1},q_{2},q_{3}$ of an
electron, that is equal to $1$ when the point considered is situated in the
interior of a certain volume $V$ of space and equal to zero for every exterior
point, and let us represent by $\psi(q,\alpha)$ and $\overline{\psi}%
(\alpha,q)$ the transformation functions that allow us to transform a physical
quantity $F$, whose form is known as a matrix $(q^{\prime},q^{\prime\prime})$,
into a matrix $(\alpha^{\prime},\alpha^{\prime\prime})$, $\alpha_{1},\alpha_{2},\alpha_{3}$ 
being the first integrals of the equation of
motion. The function $f$, written as a matrix $(q^{\prime},q^{\prime\prime})$,
will then take the form $f(q^{\prime})\delta(q^{\prime}-q^{\prime\prime}\,)$,
where $\delta(q^{\prime}-q^{\prime\prime}\,)$ represents Dirac's unit matrix.
As a matrix $(\alpha^{\prime},\alpha^{\prime\prime})$, $f$ will then take the
form%
\begin{align*}
f(\alpha^{\prime},\alpha^{\prime\prime})  &  =\int\bar{\psi}%
(\alpha^{\prime},q^{\prime})dq^{\prime}f(q^{\prime})\delta(q^{\prime
}-q^{\prime\prime})dq^{\prime\prime}\psi(q^{\prime\prime},\alpha^{\prime
\prime})\\
&  =\int_{V}\bar{\psi}(\alpha^{\prime},q^{\prime})dq^{\prime}%
\psi(q^{\prime},\alpha^{\prime\prime})\;,
\end{align*}
the integral having to be extended over the whole of the considered volume.
The diagonal terms of $f(\alpha^{\prime},\alpha^{\prime\prime})$, which may be
written in the form%
\[
f(\alpha)=\int\psi\bar{\psi}dq\;,
\]
will directly represent, in accordance with Dirac's interpretation of the
matrices, the probability that, for a state of the system characterised by
given values of $\alpha$, the coordinates of the electron are those of a point
situated in the interior of $V$. As $\psi$ is nothing other than the solution
of Schr\"{o}dinger's wave equation, we arrive at once at the interpretation of
the expression $\psi\bar{\psi}$ under discussion.

\ 

\textsc{Mr Heisenberg}.\label{Heisenberg-determinism}~---~I do not agree with Mr Dirac when he says that, in
the described experiment, nature makes a choice. Even if you place yourself
very far away from your scattering material, and if you measure after a very
long time, you are able to obtain interference by taking two mirrors. If
nature had made a choice, it would be difficult to imagine how the
interference is produced. Evidently, we say that this choice of nature can
never be known before the decisive experiment has been done; for this reason,
we can make no real objection to this choice, because the expression `nature
makes a choice' then implies no physical observation. I should rather say,\label{Heisenberg-determinism2} as
I did in my last paper, that the
\textit{observer himself} makes the choice,\footnote{From Heisenberg's publication record, it is
clear that he is here referring to his uncertainty paper, which had appeared
in May 1927. There we find the statement that `all perceiving is a
choice from a plenitude of possibilities' (Heisenberg 1927, p.~197). When 
Heisenberg says, in his above comment on Dirac, that
the observer `makes' the choice, he seems to mean this in the sense of the
observer \textit{bringing about} the choice (\textit{eds.}).} because it is only at the moment
where the observation is made that the `choice' has become a physical reality
and that the phase relationship in the waves, the power of interference, is destroyed.

\ 

\textsc{Mr Lorentz}.~---~There is then, it seems to me, a fundamental
difference of opinion on the subject of the meaning of these choices made by nature.

To admit the possibility that nature makes a choice means, I think, that it is
impossible for us to know in advance how phenomena will take place in the
future. It is then indeterminism that you wish to erect as a principle.
According to you there are events that we cannot predict, whereas until now we
have always assumed the possibility of these predictions.

\

\begin{center}
\par
\Needspace{5\baselineskip}
\textsc{Photons}\markboth{{\it General discussion}}{{\it Photons}}
\addcontentsline{toc}{section}{Photons}
\end{center}
\textsc{Mr Kramers}.~---~During the discussion of Mr de Broglie's report, Mr
Brillouin explained to us how radiation pressure is exerted in the case of
interference and that one must assume an auxiliary stress. But how is
radiation pressure\label{Kramersquestion} exerted in the case where it is so weak that there is only
one photon in the interference zone? And how does one obtain the auxiliary
tensor in this case?

\ 

\textsc{Mr de Broglie}.~---~The proof of the existence of these stresses can
be made only if one considers a cloud of photons.

\ 

\textsc{Mr Kramers}.~---~And if there is only one photon, how can one account
for the sudden change of momentum suffered by the reflecting object?

\ 

\textsc{Mr Brillouin}.~---~No theory\label{notheory} currently gives the answer to Mr Kramers' question.

\ 

\textsc{Mr Kramers}.~---~No doubt one would have to imagine a complicated
mechanism, that cannot be derived from the electromagnetic theory of waves?

\ 

\textsc{Mr de Broglie}.~---~The dualist representation by corpuscles and
associated waves does not constitute a definitive picture of the phenomena. It
does not allow one to predict the pressures exerted on the different points of
a mirror during the reflection of a single photon. It gives only the mean
value of the pressure during the reflection of a cloud of photons.

\ 

\textsc{Mr Kramers}.~---~What advantage\label{advantage} do you see in giving a precise value
to the velocity $v$ of the photons?

\ 

\textsc{Mr de Broglie}.~---~This allows one to imagine the trajectory followed
by the photons and to specify the meaning of these entities; one can thus
consider the photon as a material point having a position and a velocity.

\ 

\textsc{Mr Kramers}.~---~I do not very well see, for my part, the advantage
that there is, for the description of experiments, in making a picture where
the photons travel along well-defined trajectories.

\ 

\textsc{Mr Einstein}.~---~During reflection\label{duringreflection} on a mirror, Mr L.~de~Broglie
assumes that the photons move parallel to the mirror with a speed $c\sin
\theta$; but what happens if the incidence is normal? Do the photons then have
zero speed, as required by the formula ($\theta=0$)?

\ 

\textsc{Mr Piccard}.~---~Yes. In the case of reflection, one must assume that
the component of the velocity of the photons parallel to the mirror is
constant. In the interference zone, the component normal to the mirror
disappears. The more the incidence increases, the more the photons are slowed
down. One thus indeed arrives at stationary photons in the limiting case of
normal incidence.\footnote{Note that here the wave train is tacitly assumed to
be limited longitudinally. Cf. our discussion of the de Broglie-Pauli
encounter, section~\ref{elastic-inelastic} (\textit{eds.}).}

\ 

\textsc{Mr Langevin}.~---~In this way then, in the interference zone, the
photons no longer have the speed of light; they do not then always have the
speed $c$?

\ 

\textsc{Mr de Broglie}.~---~No, in my theory the speed of photons is equal to
$c$ only outside any interference zone, when the radiation propagates freely
in the vacuum. As soon as there are interference phenomena, the speed of the
photons becomes smaller than $c$.

\ 

\textsc{Mr De Donder}.~---~I should like to show how the research of Mr L. de
Broglie is related to mine on some points.

By identifying the ten equations of the gravitational field and the four
equations of the electromagnetic field with the fourteen equations of the wave
mechanics of L.~Rosenfeld, I have obtained\footnote{\textit{Bull.\ Ac.\ Roy.\ de
Belgique}, \textit{Cl.\ des Sc.\ } (5) \textbf{XIII}, ns. 8--9, session of 2
August 1927, 504--9. See esp. equations (5) and (8).} a \textit{principle of
correspondence} that clarifies and generalises that of O.~Klein.\footnote{\textit{Zeitschr.\ 
f.\ Phys.\ } \textbf{41}, n. 617 (1927). See esp. equations (18), p.~414.}

In my principle of correspondence, there appear \textit{the quantum current}
and \textit{the quantum tensor}. I will give the formulas for them later on;
let it suffice to remark now that the example of correspondence that Mr de
Broglie has expounded is in harmony with my principle.

Mr L.~Rosenfeld\footnote{L.~Rosenfeld, `L'univers \`{a} cinq dimensions et la
m\'{e}canique ondulatoire (quatri\`{e}me communication)', \textit{Bull.\ Ac.\ Roy.\ Belg.}, 
\textit{Cl.\ des Sc.}, October 1927. See esp. paragraphs 4 and 5.}
has given another example. Here, the mass is \textit{conserved} and, moreover,
one resorts to the quantum current. We add that this model of quantisation is
also included, as a particular case, in our principle of correspondence.

Mr Lorentz has remarked, with some surprise, that the continuity equation for
charge is preserved in Mr de Broglie's example. Thanks to our principle of
correspondence, and to Rosenfeld's compatibility\endnote{Misprinted as
`comptabilit\'{e}', despite having been corrected in the galley proofs.}
theorem, one can show that it will always be so for the total current
(including the quantum current) and for the theorem of energy and momentum.
The four equations that express this last theorem are satisfied by virtue of
the two generalised quantum equations of de Broglie-Schr\"{o}dinger.

One further small remark, to end with. Mr de Broglie said that
\textit{relativistic} systems do not exist yet. I have given the theory of
\textit{continuous} or \textit{holonomic} systems.\footnote{\textit{C.\ R.\ 
Acad.\ Sc.\ Paris}, 21 February 1927, and \textit{Bull.\ Ac.\ Roy.\ Belgique},
\textit{Cl.\ des Sc.}, 7 March 1927.} But Mr de Broglie gives another meaning
to the word \textit{system}; he has in mind \textit{interacting} systems, such
as the Bohr atom, the system of three bodies, etc. I have remarked
recently\footnote{\textit{Bull.\ Ac.\ Roy.\ Belgique}, \textit{Cl.\ des Sc.}, 2
August 1927. See esp. form.~(22).} that the quantisation of these systems
should be done by means of a $(ds)^{2}$ taken in a \textit{configuration}
space with $4n$ dimensions, $n$ denoting the number of particles. In a paper
not yet published, I have studied particular systems called \textit{additive}.

\ 

\textsc{Mr Lorentz}.~---~The stresses of which you speak and which you call
quantum, are they those of Maxwell?

\ 

\textsc{Mr De Donder}.~---~Our quantum stresses must contain the Maxwell
stresses as a particular case; this results from the fact that our principle
of correspondence is derived (in part, at least) from Maxwell's equations, and
from the fact that these quantum stresses here formally play the same role as
the stresses of electrostriction\footnote{For more details, see our Note:
`L'\'{e}lectrostriction d\'{e}duite de la gravifique einsteinienne',
\textit{Bull.\ Ac.\ Roy.\ Belgique}, \textit{Cl.\ des Sc.}, session of 9 October
1926, 673--8.} in Einsteinian Gravity. Let us recall, on this subject, that
our principle of correspondence is also derived from the fundamental equations
of Einsteinian Gravity. Mr de Broglie has, by means of his calculations, thus
recovered the stresses of radiation.

\

\begin{center}
\par
\Needspace{5\baselineskip}
\textsc{Photons and electrons}\markboth{{\it General discussion}}{{\it Photons and electrons}}
\addcontentsline{toc}{section}{Photons and electrons}
\end{center}
\textsc{Mr Langevin} makes a comparison between the old and modern statistics.

Formerly, one decomposed the phase space into cells, and one evaluated the
number of representative points attributing an individuality to each
constituent of the system.

It seems today that one must modify this method by suppressing the
individuality of the constituents of the system, and substituting instead the
individuality of the states of motion. By assuming that any number of
constituents of the system can have the same state of motion, one obtains the
statistics of Bose-Einstein.

One obtains a third statistics, that of Pauli-Fermi-Dirac,\endnote{In the
printed text, the word `Dirac' is misplaced to later in the paragraph.}
by assuming that there can be only a single representative point in each cell
of phase space.

The new type of representation seems more appropriate to the conception of
photons and particles: since one attributes a complete identity of nature to
them, it appears appropriate to not insist on their individuality, but to
attribute an individuality to the states of motion.

In the report of Messrs Born and Heisenberg, I see that it results from
quantum mechanics that the statistics of Bose-Einstein is suitable for
molecules, that of Pauli-Dirac for electrons and protons. This means that for
photons\endnote{Misprinted as `protons'.} and molecules there is
superposition, while for protons and electrons there is impenetrability.
Material particles are then distinguished from photons\endnote{Again misprinted as
`protons'.} by their impenetrability.\endnote{The version 
of this contribution in the galley proofs reads as follows:
  
    \ 

    \noindent \textsc{Mr Langevin} makes a comparison between the old and modern statistics.
    
    \

    \noindent Formerly, one decomposed the phase space, into cells and one evaluated the
    representative points.
    
    \

    \noindent It seems that one must modify this method by suppressing the
    individuality of the representative points and~[blank]

    \

    \noindent Third method: that of Pauli.

    \

    \noindent     This type of representation seems more appropriate to the conception of
    photons and particles~[blank] attribute identity of
    nature, attribute at the same time individuality representing a state.

    \

    \noindent     In the report of Messrs Born and Heisenberg, I see that it results from
    quantum mechanics that the statistics of Bose-Einstein is suitable for
    molecules, that of Pauli-Dirac, instead, is suitable for electrons. 
    This means that for~[blank] there is superposition, while for photons 
    and electrons there is impenetrability.

    \
  }

\ 

\textsc{Mr Heisenberg}.~---~There is no reason, in quantum mechanics, to
prefer one statistics to another. One may always use different statistics,
which can be considered as complete solutions of the problem of quantum
mechanics. In the current state of the theory, the question of interaction has
nothing to do with the question of statistics.

We feel nevertheless that Einstein-Bose statistics could be the more suitable
for light quanta, Fermi-Dirac statistics for positive and negative
electrons.\footnote{That is, for protons and electrons (\textit{eds.}).} The
statistics could be connected with the difference between radiation and
matter, as Mr Bohr has pointed out. But it is difficult to establish a link
between this question and the problem of interaction. I shall simply mention
the difficulty created by electron spin.

\ 

\textsc{Mr Kramers} reminds us of Dirac's research on statistics, which has
shown that Bose-Einstein statistics can be expressed in an entirely different
manner. The statistics of photons, for example, is obtained by considering a
cavity filled with blackbody radiation as a system having an infinity of
degrees of freedom. If one quantises this system according to the rules of
quantum mechanics and applies Boltzmann statistics, one arrives at Planck's
formula, which is equivalent to Bose-Einstein statistics applied to photons.

Jordan has shown that a formal modification of Dirac's method allows one to
arrive equally at a statistical distribution that is equivalent to Fermi
statistics. This method is suggested by Pauli's exclusion principle.

\ 

\textsc{Mr Dirac}\label{page187}\footnote{On this criticism 
by Dirac, cf.\ Kragh (1990, pp.~128--30) ({\em eds.}).} points out 
that this modification, considered 
from a general point of view, is quite artificial. Fermi statistics is not
established on exactly the same basis as Einstein-Bose statistics, since the
natural method of quantisation for waves leads precisely to the latter
statistics for the particles associated with the waves. To obtain Fermi
statistics, Jordan had to use an unusual method of quantisation for waves,
chosen specially so as to give the desired result. There are mathematical
errors in the work of Jordan that have not yet been redressed.

\ 

\textsc{Mr Kramers}.~---~I willingly grant that Jordan's treatment does not
seem as natural as the manner by which Mr Dirac quantises the solution of the
Schr\"{o}dinger equation. However, we do not yet understand why nature
requires this quantisation, and we can hope that one day we will find the
deeper reason for why it is necessary to quantise in one way in one case and
in another way in the other.

\ 

\textsc{Mr Born}.~---~An essential difference between Debye's old theory, in
which the characteristic vibrations of the blackbody cavity are treated like
Planck oscillators, and the new theory is this, that both yield quite exactly
Planck's radiation formula (for the mean density of radiation), but that the
old theory leads to inexact values for the local fluctuations of radiation,
while the new theory gives these values exactly.

\ 

\textsc{Mr Heisenberg}.~---~According to the experiments, protons and
electrons both have an angular momentum and obey the laws of the statistics of
Fermi-Dirac; these two points seem to be related. If one takes two particles
together, if one asks, for example, which statistics one must apply to a gas
made up of atoms of hydrogen, one finds that the statistics of Bose-Einstein
is the right one, because by permuting two H atoms, we permute one positive
electron and one negative electron,\footnote{That is, we permute the two
protons, and also the two electrons (\textit{eds.}).} so that we change the
sign of the Schr\"{o}dinger function \textit{twice}. In other words,
Bose-Einstein statistics is valid for all gases made up of neutral molecules,
or more generally, composed of systems whose charge is an even multiple of
$e$. If the charge of the system is an odd multiple of $e$, the statistics of
Fermi-Dirac applies to a collection of these systems.

The He nucleus does not rotate and a collection of He nuclei obeys the laws of
Bose-Einstein statistics.

\ 

\textsc{Mr Fowler} asks if the fine details of the structure of the bands of
helium agree better with the idea that we have only symmetric states of
rotation of the nuclei of helium than with the idea that we have only
antisymmetric states.

\ 

\textsc{Mr Heisenberg}.~---~In the bands of helium, the fact that each second
line disappears teaches us that the He nucleus is not endowed with a spinning
motion. But it is not yet possible to decide experimentally, on the basis of
these bands, if the statistics of Bose-Einstein or that of Fermi-Dirac must be
applied to the nucleus of He.

\ 

\textsc{Mr Schr\"{o}dinger}.~---~You have spoken of experimental evidence in
favour of the hypothesis that the proton is endowed with a spinning motion
just like the electron, and that protons obey the statistical law of
Fermi-Dirac. What evidence are you alluding to?

\ 

\textsc{Mr Heisenberg}.~---~The experimental evidence is provided by the work
of Dennison\footnote{\textit{Proc.\ Roy.\ Soc.\ }\textit{A\/} \textbf{114} (1927),
483.} on the specific heat of the hydrogen molecule, work which is based on
Hund's research concerning the band spectra of hydrogen.

Hund found good agreement between his theoretical scheme and the experimental
work of Dilke, Hopfield and Richardson, by means of the hypotheses mentioned
by Mr Schr\"{o}dinger. But for the specific heat, he found a curve very
different from the experimental curve. The experimental curve of the specific
heat seemed rather to speak in favour of Bose-Einstein statistics. But the
difficulty was elucidated in the paper by Dennison, who showed that the
systems of `symmetric' and `antisymmetric' terms (with regard to protons) do
not combine in the time necessary to carry out the experiment. At low
temperature, a transition takes place about every three months. The ratio of
statistical weights of the systems of symmetric and antisymmetric terms is
$1:3$, as in the helium atom. But at low temperatures the specific heat must
be calculated as if one had a mixture of two gases, an `ortho' gas and a
`para' gas. If one wished to perform experiments on the specific heat with a
gas of hydrogen, kept at low temperature for several months, the result would
be totally different from the ordinary result.

\ 

\textsc{Mr Ehrenfest} wishes to formulate a question that has some relation to
the recent experiments by Mr Langmuir on the disordered motion of electrons in
the flow of electricity through a gas.

In the well-known Pauli exclusion (Pauliverbot), one introduces (at least in
the language of the old quantum theory) a particular incompatibility relation
between the quantum motions of the different particles of a single system,
without speaking explicitly of the role possibly played by the forces acting
between these particles. Now, suppose that through a small opening one allows
particles that, so to speak, do not exert forces on each other, to pass from a
large space into a small box bounded by quite rigid walls with a complicated
shape, so that the particles encounter the opening and leave the box only at
the end of a sufficiently long time. Before entering the box, if the particles
have almost no motion relative to one another, the Pauli exclusion intervenes.
After their exit, will they have very different energies, independently of the
weakness of the mutual action between the particles? Or else what role do
these forces play in the production of Pauli's incompatibility (choice of
antisymmetric solutions of the wave equation)?

\ 

\textsc{Mr Heisenberg}.~---~The difficulty with Mr Ehrenfest's experiment is
the following: the two electrons must have different energies. If the energy
of interaction of the two electrons is very small, the time $\tau_{1}$
required for the electrons to exchange an appreciable amount of energy is very
long. But to find experimentally which state, symmetric or antisymmetric, the
system of the two electrons in the box is in, we need a certain time $\tau
_{2}$ which is at least $\sim1/\nu$, if $h\nu$ is the [energy] difference
between the symmetric and antisymmetric states. Consequently, $\tau_{1}%
\sim\tau_{2}$ and the difficulty disappears.

\ 

\textsc{Mr Richardson}.~---~The evidence for a nuclear spin is much more
complete than Mr Heisenberg has just said. I have recently had occasion to
classify a large number of lines in the visible bands of the spectrum of the
$\mathrm{H}_{2}$ molecule. One of the characteristic features of this spectrum
is a rather pronounced alternation in the intensity of the successive lines.
The intensities of the lines of this spectrum were recently measured by
MacLennan, Grayson-Smith and Collins. Unfortunately, a large number of these
lines overlap with each other, so that the intensity measurements must be
accepted only with reservations.

But nevertheless, I think one can say, without fear of being mistaken, that
all the bands that are sufficiently well-formed and sufficiently free of
influences of the lines on each other (so that one can have confidence in the
intensity measurements) have lines, generally numbered 1, 3, 5,~..., that are
intrinsically three times more intense than the intermediate lines, generally
numbered 2, 4, 6, ...\,. By intrinsic intensity, I mean that which one obtains
after having taken into account the effects on the intensity of temperature
and quantum number (and also, of course, the effects of overlap with other
lines, where it is possible to take this into account). In other words, I wish
to say that the constant $c$ of the intensity formula%
\[
\mathcal{J}=c\left(  m+\frac{1}{2}\right)  e^{\frac{-\left(  m+\frac{1}%
{2}\right)  ^{2}h^{2}}{8\pi KkT}}\;,
\]
where $m$ is the number of the line and $K$ the moment of inertia of the
molecule, is three times bigger for the odd-numbered lines than for the
even-numbered ones. This means that the ratio $3:1$ applies, with an accuracy
of about $5\%$, for at least five different vibration states of a
three-electron state of excitation. It also applies to another state, which is
probably $3^{1}P$ if the others are $3^{3}P$. It is also shown, but in a less
precise way, that it applies to two different vibration states of a state of
excitation with four electrons.

At present, then, there is a great deal of experimental evidence that this
nuclear spin persists through the different states of excitation of the
hydrogen molecule.

\ 

\textsc{Mr Langmuir}.~---~The question has often been raised of a similarity
in the relation between light waves and photons on the one hand, and de
Broglie waves and electrons on the other. How far can this analogy be
developed? There are many remarkable parallels, but also I should like to see
examined if there are no fundamental differences between these relations.
Thus, for example, an electron is characterised by a constant charge. Is there
a constant property of the photon that may be compared with the charge of the
electron? The speed of the electron is variable; is that of the photon also?
The electromagnetic theory of light has suggested a multitude of experiments,
which have added considerably to our knowledge. The wave theory of the
electron explains the beautiful results of Davisson and Germer. Can one hope
that this theory will be as fertile in experimental suggestions as the wave
theory of light has been?\endnote{The galley proofs contain the following version of this
contribution:

\

\noindent    \textsc{Mr Langmuir} would like to see established clearly a parallel between
    electrons and photons. What characterises an electron? A well-defined charge.
    What characterises the photon? Its velocity, perhaps? What is the analogy, what  
    are the differences? Electron: de Broglie waves; photon: electromagnetic waves.
    For certain respects, this parallelism is clear, but perhaps it can be pursued
    to the end? What are the suggestions in the way of experiments?

\
  }

\ 

\textsc{Mr Ehrenfest}.~---~When one examines a system of plane waves of
elliptically polarised light, placing oneself in differently moving coordinate
systems, these waves show the same degree of ellipticity whatever system one
places oneself in. Passing from the language of waves to that of photons, I
should like to ask if one must attribute an elliptical polarisation (linear or
circular in the limiting cases) to each photon? If the reply is affirmative,
in view of the invariance of the degree of ellipticity in relativity, one must
distinguish as many species of photons as there are degrees of ellipticity.
That would yield, it seems to me, a new difference between the photon and the
spinning electron. If, on the other hand, one wishes above all to retain the
analogy with the electron, as far as I can see one comes up against two difficulties:

\

\noindent 1. How then must one describe linearly polarised light in the language of
photons? (It is instructive, in this respect, to consider the way in which the
two linearly polarised components, emitted perpendicularly to the magnetic
field by a flame showing the Zeeman effect, are absorbed by a second flame
placed in a magnetic field with antiparallel orientation.)

Mr Zeeman, to whom I posed the question, was kind enough to perform the
experiment about a year ago, and he was able to notice that the absorption is
the same in parallel and antiparallel fields, as one could have predicted, in
fact, by considerations of continuity.

\

\noindent 2. For electrons, which move always with a speed less than that of light, the
universality of the spin may be expressed as follows, that one transforms the
corresponding antisymmetric tensor into a system of coordinates carried with
the electron in its translational motion (`at rest'). But photons always move
with the speed of light!

\ 

\textsc{Mr Compton}.~---~Can light be elliptically polarised when the photon
has an angular momentum?

\ 

\textsc{Mr Ehrenfest}.~---~Because the photons move with the speed of light, I
do not really understand what it means when one says that each photon has a
universal angular momentum just like an electron.

Allow me to remind you of yet another property of photons. When two photons
move in directions that are not exactly the same, one can say quite
arbitrarily that one of the photons is a radio-photon and the other a $\gamma
$-ray photon, or inversely. That depends quite simply on the moving system of
coordinates to which one refers the pair of photons.

\ 

\textsc{Mr Lorentz}.~---~Can you make them identical by such a transformation?

\ 

\textsc{Mr Ehrenfest}.~---~Perfectly. If they move in different directions.
One can then give them the same colour by adopting a suitable frame of
reference. It is only in the case where their worldlines are exactly parallel
that the ratio of their frequencies remains invariant.

\ 

\textsc{Mr Pauli}.~---~The fact that the spinning electron can take two
orientations in the field allowed by the quanta seems to invite us at first to
compare it to the fact that there are, for a given direction of propagation of
the light quanta, two characteristic vibrations of blackbody radiation,
distinguished by their polarisation. Nevertheless there remain essential
differences between the two cases. While in relativity one describes waves by
a (real) sextuple vector $F_{ix}=-F_{xi}$, for the spinning electron one has
proposed the following two modes of description for the associated de Broglie
waves: 1. One describes these waves by two complex functions $\psi_{\alpha}$,
$\psi_{\beta}$ (and so by four real functions); but these functions transform
in a way that is hardly intuitive during the change from one system of
coordinates to another. That is the route I followed myself. Or else: 2.
Following the example of Darwin, one introduces a \textit{quadruple} vector
with generally complex components (and so eight real functions in total). But
this procedure has the inconvenience that the vector involves a redundancy
[ind\'{e}termination], because all the verifiable results depend on only
\textit{two} complex functions.

These two modes of description are mathematically equivalent, but
independently of whether one decides in favour of one or the other, it seems
to me that one cannot speak of a \textit{simple} analogy between the
polarisation of light waves and the polarisation of de Broglie waves
associated with the spinning electron.

Another essential difference between electrons and light quanta is this, that
between light quanta there does not exist direct (immediate) mutual action,
whereas electrons, as a result of their carrying an electric charge, exert
direct mutual actions on each other.

\ 

\textsc{Mr Dirac}.\footnote{Again, here we follow Dirac's original English
(\textit{eds.}).}~---~I should like to point out an important failure in the
analogy between the spin of electrons and the polarisation of photons. In the
present theory of the spinning electron one assumes that one can specify the
direction of the spin axis of an electron at the same time as its position, or
at the same time as its momentum. Thus the spin variable of an electron
commutes\endnote{The French renders `commute' throughout with
`changer'.} with its coordinate and with its momentum variables. The case
is different for photons. One can specify a direction of polarisation for
plane monochromatic light waves, representing photons of given momentum, so
that the polarisation variable commutes with the momentum variables. On the
other hand, if the position of a photon is specified, it means one has an
electromagnetic disturbance confined to a very small volume,\endnote{The
French adds: `\`{a} un instant donn\'{e}'.} and one cannot give a
definite polarisation, i.e. a definite direction for the electric vector, to
this disturbance. Thus the polarisation variable of a photon does not commute
with its coordinates.

\ 

\textsc{Mr Lorentz}.~---~In these different theories, one deals with the
probability $\psi\psi^{\ast}$. I should like to see quite clearly how this
probability can exist when particles move in a well-defined manner following
certain laws. In the case of electrons, this leads to the question of motions
in the field $\psi$ (de Broglie). But the same question arises for light
quanta. Do photons allow us to recover all the classical properties of waves?
Can one represent the energy, momentum and Poynting vector by photons? One
sees immediately that, when one has an energy density and energy flow, if one
wishes to explain this by photons then the number of photons per unit volume
gives the density, and the number of photons per second that move across a unit
surface gives the Poynting vector.

The photons will then have to move with a speed different from that of light.\label{Lordiff}
If one wished to assign always the same speed $c$ to the photons, in some
cases one would have to assume a superposition of several photon currents. Or
else one would have to assume that the photons cannot be used to represent all
the components of the energy-momentum tensor. Some of the terms must be
continuous in the field. Or else the photons are smeared out [fondus].

A related question is to know whether the photons can have a speed different
from that of light and whether they can even be at rest. That would altogether
displease me. Could we speak of these photons and of their motion in a field
of radiation?

\ 

\textsc{Mr de Broglie}.~---~When I tried to relate the motion of the photons
to the propagation of the waves $\psi$ of the new mechanics, I did not worry
about putting this point of view in accord with the electromagnetic conception
of light waves, and I considered only waves $\psi$ of scalar character, which
one has normally used until now.

\ 

\textsc{Mr Lorentz}.~---~One will need these waves for photons also. Are they
of a different nature than light waves? It would please me less to have to
introduce two types of waves.

\ 

\textsc{Mr de Broglie}.\label{p192DRAFT1}~---~At present one does not know at all the physical
nature of the $\psi$-wave of the photons. Can one try to identify it with the
electromagnetic wave? That is a question that remains open. In any case, one
can provisionally try to develop a theory of photons by associating them with
waves $\psi$.

\ 

\textsc{Mr Lorentz}.~---~Is the speed of the wave equal to that of light?

\ 

\textsc{Mr de Broglie}.~---~In my theory, the speed of photons is equal to
$c$, except in interfering fields. In general, I find that one must assign to
a moving corpuscle a proper mass $M_{0}$ given by the formula%
\[
M_{0}=\sqrt{m_{0}^{2}-\frac{h^{2}}{4\pi^{2}c^{2}}\frac{\square a}{a}}\;,
\]
the function $\frac{\square a}{a}$ being calculated at the point where the
moving body is located at the given moment ($a$ is the amplitude of the wave
$\psi$). For photons, one has%
\[
m_{0}=0\;.
\]
Thus, when a photon moves freely, that is to say, is associated with an
ordinary plane wave, $M_{0}$ is zero and, to have a finite energy, the photon
must have speed $c$. But, when there is interference, $\frac{\square a}{a}$
becomes different from zero, $M_{0}$ is no longer zero and the photon, to
maintain the same energy, must have a speed less than $c$, a speed that can
even be zero.

\ 

\textsc{Mr Lorentz}.~---~The term $\frac{\square a}{a}$ must be negative,
otherwise the mass would become imaginary.

\ 

\textsc{Mr de Broglie}.~---~In the corpuscular conception of light, the
existence of diffraction phenomena occuring at the edge of a screen requires
us to assume that, in this case, the trajectory of the photons is curved. The
supporters of the emission theory\label{oldp193} said that the edge of the screen exerts a
force on the corpuscle. Now, if in the new mechanics as I develop it, one
writes the Lagrange equations for the photon, one sees appear on the
right-hand side of these equations a term proportional to the gradient of
$M_{0}$.

This term represents a sort of force of a new kind, which exists only when the
proper mass varies, that is to say, where there is interference. It is this
force that will curve the trajectory of the photon when its wave $\psi$ is
diffracted by the edge of a screen.

Furthermore, for a cloud of photons the same Lagrange equations lead one to
recover the internal stresses pointed out by Messrs Schr\"{o}dinger and De
Donder.\footnote{Cf.\ Schr\"{o}dinger (1927b) and De Donder's comments
above (\textit{eds.}).} One finds, indeed, the relations%
\[
\frac{\partial}{\partial x^{k}}\left[  T^{ik}+\Pi^{ik}\right]  =0\;,
\]
where the tensor $T^{ik}$ is the energy-momentum tensor of the corpuscles%
\[
T^{ik}=\rho_{0}u^{i}u^{k}\;.
\]

The tensor $\Pi^{ik}$, which depends on derivatives of the amplitude of the
wave $\psi$ and is zero when this amplitude is constant, represents stresses
existing in the cloud of corpuscles, and these stresses allow us to recover
the value of the radiation pressure in the case of reflection of light by a mirror.

The tensor $T^{ik}+\Pi^{ik}$ is certainly related to the Maxwell tensor but,
to see clearly how, one would have to be able to clarify the relationship
existing between the wave $\psi$ of the photons and the electromagnetic light wave.

\ 

\textsc{Mr Pauli}.\label{Pauli-deB-beginning}\footnote{Cf.\ section~\ref{elastic-inelastic}
(\textit{eds.}).}~---~It seems to me that, concerning the statistical results
of scattering experiments, the conception of Mr de Broglie is in full
agreement with Born's theory in the case of elastic collisions, but that it is
no longer so when one also considers inelastic collisions. I should like to
illustrate this by the example of the rotator, which was already mentioned by
Mr de Broglie himself. As Fermi\footnote{\textit{Zeitschr.\ f.\ Phys.\ }\textbf{40} 
(1926), 399.} has shown, the treatment by wave mechanics of the
problem of the collision of a particle that moves in the $(x,y)$ plane and of
a rotator situated in the same plane, may be made clear in the following
manner.\footnote{See section~\ref{elastic-inelastic} for a discussion of Fermi's
argument (\textit{eds.}).} One introduces a configuration space of three
dimensions, of which two coordinates correspond to the $x$ and $y$ of the
colliding particle, while as third coordinate one chooses the angle $\varphi$
of the rotator. In the case where there is no mutual action between the
rotator and the particle, the function $\psi$ of the total system is given
by\endnote{`$h$' misprinted as `$\lambda$'.}%
\[
\psi(x,y,\varphi)=Ae^{2\pi i\left[  \frac{1}{h}(p_{x}x+p_{y}y+p_{\varphi
}\varphi)-\nu t\right]  }\;,
\]
where one has put%
\[
p_{\varphi}=m\frac{h}{2\pi}\;\;\;\;\;(m=0,1,2,...)\;.
\]

In particular, the sinusoidal oscillation of the coordinate $\varphi$
corresponds to a stationary state of the rotator. According to Born, the
superposition of several partial waves of this type, corresponding to
different values of $m$ and by consequence of $p_{\varphi}$,\endnote{`$p_{\varphi}$' 
misprinted as `$\varphi$'.} means that there
is a probability different from zero for several stationary states of the
rotator, while according to the point of view of Mr de Broglie, in this case
the rotator no longer has a constant angular velocity and can also execute
oscillations in certain circumstances.

Now, in the case of a finite energy of interaction between the colliding
particle and the rotator, if we study the phenomenon of the collision by means
of the wave equation in the space $(x,y,\varphi)$, according to Fermi the
result can be interpreted very simply. Indeed, since the energy of interaction
depends on the angle $\varphi$ in a periodic manner and vanishes at large
distances from the rotator, that is to say from the axis $\varphi$, in the
space $(x,y,\varphi)$ we are dealing simply with a wave that falls on a
grating and, in particular, on a grating that is unlimited in the direction of
the axis $\varphi$. At large distances from the grating, waves come out only
in fixed directions in configuration space, characterised by integral values
of the difference $m^{\prime}-m^{\prime\prime}$. Fermi has shown that the
different spectral orders correspond simply to the different possible ways of
transferring the energy of the colliding particle to the rotator, or
conversely. Thus to each spectral order of the grating corresponds a given
stationary state of the rotator after the collision.

It is, however, an essential point\label{essentialpoint} that, in the case where the rotator is in a
stationary state before the collision, the incident wave is unlimited in the
direction of the axis. For this reason, the different spectral orders of the
grating will always be superposed at each point of configuration space. If we
then calculate, according to the precepts of Mr de Broglie, the angular
velocity of the rotator after the collision, we must find that this velocity
is not constant. If one had assumed that the incident wave is
limited\footnote{The French reads `illimit\'{e}e' [unlimited], which we
interpret as a misprint. Pauli seems to be saying that if, on the other hand,
the incident wave had been taken as limited, then \textit{before} the
collision the rotator could not have been in a stationary state and its
angular velocity could not have been constant (\textit{eds.}).} in the
direction of the axis $\varphi$, it would have been the same before the
collision. Mr de Broglie's point of view does not then seem to me compatible
with the requirement of the postulate of the quantum theory, that the rotator
is in a stationary state both before and after the collision.

To me this difficulty\label{Paulidifficulty} does not appear at all fortuitous or inherent in the
particular example of the rotator; in my opinion, it is due directly to the
condition assumed by Mr de Broglie, that in the individual collision process
the behaviour of the particles should be completely determined and may at the
same time be described completely by ordinary kinematics in spacetime. In
Born's theory, agreement with the quantum postulate is realised thus, that the
different partial waves in configuration space, of which the general solution
of the wave equation after the collision is composed, are applicable
[indiqu\'{e}es] \textit{separately} in a statistical way. But this is no
longer possible in a theory that, in principle, considers it possible to avoid
the application of notions of probability to \textit{individual} collision 
processes.

\ 

\textsc{Mr de Broglie}.~---~Fermi's problem is not of the same type as that
which I treated earlier; indeed, he makes configuration space play a part, and
not ordinary space.

The difficulty\label{pointedout} pointed out by Mr Pauli has an analogue in classical optics.
One can speak of the beam diffracted by a grating in a given direction only if
the grating and the incident wave are laterally limited, because otherwise all
the diffracted beams will overlap and be bathed in the incident wave. In
Fermi's problem, one must also assume the wave $\psi$ to be limited laterally
in configuration space.

\ 

\textsc{Mr Lorentz}.~---~The question is to know what a particle should do
when it is immersed in two waves at the same time.

\ 

\textsc{Mr de Broglie}.~---~The whole question is to know if one has the right
to assume the wave $\psi$ to be limited laterally in configuration space. If
one has this right, the velocity of the representative point of the system
will have a constant value, and will correspond to a stationary state of the
rotator, as soon as the waves diffracted by the $\varphi$-axis will have
separated from the incident beam.

One can say that it is not possible to assume the incident beam to be limited
laterally, because Fermi's configuration space is formed by the superposition
of \textit{identical} layers of height $2\pi$ in the direction of the
$\varphi$-axis; in other words, two points of configuration space lying on the
same parallel to the $\varphi$-axis and separated by a whole multiple of
$2\pi$ represent the \textit{same} state of the system. In my opinion, this
proves above all the artificial character of configuration spaces, and in
particular of that which one obtains here by rolling out along a line the
cyclic variable $\varphi$.\label{Pauli-deB-end}

\ 

\textsc{Mr De Donder}.~---~In the course of the discussion of Mr L.~de
Broglie's report, we explained how we obtained our Principle of
Correspondence; thanks to this principle, one will have\footnote{I adopt here
L.~Rosenfeld's notation, so as to facilitate the comparison with his formulas,
given later.}%
\begin{align*}
\rho_{(e)}u^{a}+\Lambda^{a}  &  =\sqrt{-g}K^{2}\frac{c}{e}\sum_{n}\frac
{-h}{2i\pi}g^{an}\left(  \psi\bar{\psi}_{.n}-\bar{\psi}\psi
_{.n}\right)  -2\frac{e}{c}\Phi^{a}\psi\bar{\psi}\;,\\
\rho_{(m)}u^{a}u^{b}+\Pi^{ab}  &  =\sqrt{-g}\sum_{\alpha}\sum_{\beta}%
\gamma^{a\alpha}\gamma^{b\beta}\left(  \psi_{.\alpha}\bar{\psi}_{.\beta}
+\bar{\psi}_{.\alpha}\psi_{.\beta}\right)  -\gamma^{ab}L\\
(a,\;b,\;n  &  =1,...,4;\;\alpha,\;\beta=0,1,...,4)\ .
\end{align*}

The first relation represents \textit{the total current} ($\equiv$ electronic
current $+$ quantum current) as a function of $\psi$ and of the potentials
$g^{an}$, $\Phi^{a}$. Recall that one has set%
\begin{align*}
L  &  \equiv\sum_{\alpha}\sum_{\beta}\gamma^{\alpha\beta}\psi_{.\alpha
}\overline{\psi_{.\beta}}+k^{2}\left(  \mu^{2}-\frac{1}{2\chi}\right)
\psi\overline{\psi}\;,\\
\gamma^{ab}  &  \equiv g^{ab}\;,\;\;\;\gamma^{0a}\equiv-\alpha\Phi
^{a}\;,\;\;\;\gamma^{00}\equiv\alpha^{2}\Phi^{a}\Phi_{a}-\frac{1}{\xi}\;,\\
\xi\alpha^{2}  &  \equiv2\chi\;,\;\;\;\chi\equiv\frac{8\pi G}{c^{2}%
}\;,\;\;\;G\equiv6.7\times10^{-8}\ \mathrm{c.g.s.}%
\end{align*}

We have already mentioned the examples (or models) of correspondence found
respectively by L.~de Broglie and L.~Rosenfeld. To be able to show clearly a
new solution to the problem relating to \textit{photons} that Mr L.~de Broglie
has just posed, I am going to display the formulas concerning the two
above-mentioned models.\footnote{L. Rosenfeld, `L'univers \`{a} cinq dimensions
et la m\'{e}canique ondulatoire', \textit{Bull.\ Ac.\ Roy.\ Belgique}, \textit{Cl.\
des Sc.}, October 1927. See respectively the formulas (*38%
\'{}%
), (*31), (*27), (21), (1), (8), (35), (28), (29), (*35).}

\

\vspace*{-0.2em}
\hspace*{-0.5em}\begin{minipage}[t]{14.5em}
\begin{center}
{\em Model of L.~de Broglie.}
\end{center}
Quantum current $\Lambda_{a}\equiv0$.

\vspace{27.3em}

Charge density $\rho_{(e)}=2K^{2}A^{\prime2}\mu^{\prime}$, \\
where we have put 
  \[
    \mu^{\prime2}=\mu^{2}+\frac{\square A^{\prime}}{K^{2}A^{\prime}}\ ,
  \]
which, retaining the charge $e$, reduces to substituting
for the mass $m_{0}$ \textit{the modified mass of L.~de Broglie}:
  \[
    M_{0}\equiv\sqrt{m_{0}^{2}+\frac{h^{2}}{4\pi^{2}c^{2}}\frac{\square A^{\prime}}{A^{\prime}}}\ . 
  \]
\end{minipage}\hspace{0.5em}
\begin{minipage}[t]{15.5em}
\begin{center}
{\em Model of L.~Rosenfeld.}
\end{center}
Quantum current $\Lambda_{a}=2K^{2}A^{\prime2}C_{.a}$,\\
where $A^{\prime}$ is the modulus of $\psi$ and where
the potential $C\equiv S^{\prime}-S$. The function
$S$ satisfies the \textit{classical} Jacobi
equation; the function $S^{\prime}$ satisfies
the \textit{modified} Jacobi equation;
one then has
    \begin{align*}
      \gamma^{\alpha\beta}S_{.\alpha}S_{.\beta}                       & = \mu^{2}-\frac{1}{2\chi}\ ,\\
      \gamma^{\alpha\beta}S_{.\alpha}^{\prime}S_{.\beta}^{\prime}     & = 
      \mu^{2}-\frac{1}{2\chi}+\frac{\square A^{\prime}}{K^{2}A^{\prime}}\ .
    \end{align*}
The quantum potential $C$ produces \textit{the difference} between \textit{physical} quantisation
and \textit{geometrical} quantisation.

Recall that $\mu\equiv\frac{m_{0}c^{2}}{e}$, where $m_{0}$ and $e$
are respectively the mass (at rest) and charge of the particle under consideration. We have also put
  \[
    k\equiv iK\equiv i\frac{2\pi}{h}\frac{e}{c}\ .
  \]
Charge density $\rho_{(e)}=2K^{2}A^{\prime2}\mu$.\\
Here then \textit{one retains}, at the same time, the mass $m_0$ and the charge $e$.
\end{minipage}


\

Let us respectively apply these formulas to the problem of the photon pointed
out by Mr L. de Broglie. The \textit{proper} mass $m_{0}$ of the photon is
\textit{zero}; in the model of Mr L. de Broglie, this mass must be replaced by
the \textit{modified} mass $M_{0}$; on the contrary, in the model of Mr L.~Rosenfeld, 
one uses only the \textit{proper} mass $m_{0}\equiv0$. In the two
models, the charge density $\rho_{(e)}$ is zero. Finally, in the first model,
the speed of the photon \textit{must} vary; in contrast, in the second model,
one can assume that this speed is always that of light. These conclusions
obviously speak in favour of the model of L.~Rosenfeld, and, in consequence,
also in favour of the \textit{physical} existence of our quantum current
$\Lambda^{a}$ ($a=1,2,3,4$). This current will probably play a dominant
role in still unexplained optical phenomena.\footnote{On this subject, Mr L.~Brillouin 
has kindly drawn my attention to the experiments by Mr~F.~Wolfers: `Sur un nouveau 
ph\'{e}nom\`{e}ne en optique: interf\'{e}rences par diffusion' 
(\textit{Le Journal de Physique et le Radium\/} (VI) \textbf{6}, n.~11, November
1925, 354--68).}

\ 

\textsc{Mr Lorentz}.~---~Let us take an atom of hydrogen and let us form the
Schr\"{o}dinger function $\psi$.\endnote{`$\psi$' missing in the original,
with a space instead.} We consider $\psi\psi^{\ast}$ as the probability for the
presence of the electron in a volume element. Mr Born has mentioned all the
trajectories in the classical theory: let us take them with all possible
phases,\footnote{The `phases' of classical trajectories seems to be meant in
the sense of action-angle variables (\textit{eds.}).} but let us now take the
$\psi$ corresponding to a single value $W_{n}$ of energy and then let us form
$\psi\psi^{\ast}$. Can one say that this product $\psi_{n}\psi_{n}^{\ast}$
represents the probability that the electrons move with the given energy
$W_{n}$? We think that the electron cannot escape from a certain sphere. The
atom is limited, whereas $\psi$ extends to infinity. That is disagreeable.\endnote{Here 
the galley proofs include an additional sentence: 

\

\noindent If one took the integral 
extended over the whole of this space, the exterior part would be comparable.

\
}

\ 

\textsc{Mr Born}.~---~The idea that $\psi\psi^{\ast}$ represents a probability
density has great importance in applications. If, for example, in the
classical theory an electron had two equilibrium positions separated by a
considerable potential energy, then classically, for a sufficiently weak total
energy only one oscillation could ever take place, around one of the two
equilibrium positions. But according to quantum mechanics, each eigenfunction
extends from one domain into the other; for this reason there always exists a
probability that a particle, which at first vibrates in the neighbourhood of
one of the equilibrium positions, jumps to the other. Hund has made important
applications of this to molecular structure. This phenomenon probably also
plays a role in the explanation of metallic conduction.

\ 

\textsc{Mr de Broglie}.~---~In the old theory of the motion of an electron in
the hydrogen atom, an electron of total energy%
\[
W=\frac{m_{0}c^{2}}{\sqrt{1-\beta^{2}}}-\frac{e^{2}}{r}%
\]
cannot escape from a sphere of radius%
\[
R=-\frac{e^{2}}{W-m_{0}c^{2}}%
\]
because the value of the term $\frac{m_{0}c^{2}}{\sqrt{1-\beta^{2}}}$ has
$m_{0}c^{2}$ as a lower limit.

In my conception one must take%
\[
W=\frac{M_{0}c^{2}}{\sqrt{1-\beta^{2}}}-\frac{e^{2}}{r}\ ,
\]
as the expression for the energy, where $M_{0}$ is the variable proper mass
which I have already defined. Calculation shows that the proper mass $M_{0}$
diminishes when $r$ increases, in such a way that an electron of energy $W$ is
no longer at all constrained to be in the interior of a sphere of radius $R$.

\ 

\textsc{Mr Born}.~---~Contrary to Mr Schr\"{o}dinger's opinion, that it is
nonsense to speak of the location and motion of an electron in the atom, Mr
Bohr and I are of the opinion that this manner of speaking always
has a meaning when one can specify an experiment allowing us to measure the
coordinates and the velocities with a certain approximation.

\ 

\begin{quote}
Again in Richardson's notes on the general discussion (cf.\ p.~\pageref{werepr}), 
the following text together with Fig.~D (both labelled `Bohr'), and a similar 
figure with the shaded region labelled `B', appear immediately after notes on 
De Donder's lengthy exposition just above, and clearly refer to remarks Bohr made 
on the topic being addressed here: 
\end{quote}

  \begin{figure}
    \centering
     \resizebox{\textwidth}{!}{\includegraphics[-40mm,0mm][220.30mm,130.98mm]{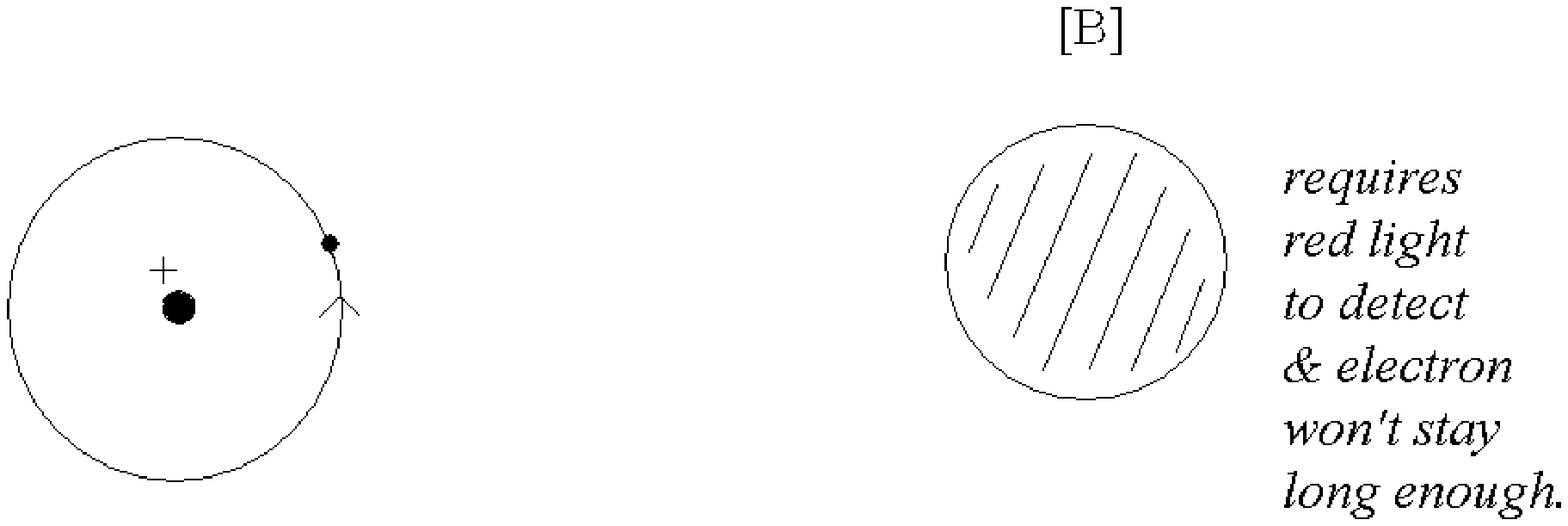}}
    \unnumberedcaption{Fig.~D.}
  \end{figure} 

B[ohr] says it has no point to worry about the paradox that the electron in the
atom is in a fixed path (ellipse or circle) and the probability that it should
be found in a given place is given by the product $\psi\bar{\psi}$ which is a
continuous function of space extending from zero to $\infty$. He says if we
take a region such as B a long way from the atom in
order to find if the electron is there we must illuminate it with long light
waves and the frequency of these is so low that the electron is out of the
region by reason of its motion in the stationary state before it has been
illuminated long enough for the photoelectric act to occur. I am really not
sure if this is right. But, anyway, it is no objection to pulling it out with
an intense \textit{static} electric field \& this appears to be what is
happening in the W experiments.

\

\textsc{Mr Pauli}.~---~One can indeed determine the location of the electron
outside the sphere, but without modifying its energy to the point where an
ionisation of the atom occurs.

\ 

\textsc{Mr Lorentz}.\label{forjustabove}~---~I should 
like to make a remark on the subject of wave packets.\footnote{Cf.\ also the 
discussion of the Lorentz-Schr\"{o}dinger correspondence in 
section~\ref{Schr-packets} (\textit{eds.}).}

When Mr Schr\"{o}dinger drew attention to the analogy between mechanics and
optics, he suggested the idea of passing from corpuscular mechanics to wave
mechanics by making a modification analogous to that which is made in the
passage from geometrical optics to wave optics.\endnote{The original
mistakenly reads `geometrical mechanics' and `corpuscular optics'.} The
wave packet gave a quite striking picture of the electron, but in the atom the
electron had to be completely smeared out [fondu], the packet having the
dimensions of the atom. When the dimensions of the wave packet become
comparable to those of the trajectories of the classical theory, the material
point would start to spread; having passed this stage, the electron will be
completely smeared out.

The mathematical difficulty of constructing wave packets in the atom is due to
the fact that we do not have at our disposal wavelengths sufficiently small or
sufficiently close together. The frequencies of stable waves in the atom
(eigenvalues) are more or less separated from each other; one cannot have
frequencies very close together corresponding to states differing by very
little, because the conditions at infinity would not be satisfied. To
construct a packet, one must superpose waves of slightly different
wavelengths; now, one can use only eigenfunctions $\psi_{n}$, which are
sharply different from each other. In atoms, then, one cannot have wave
packets. But there is a difficulty also for free electrons, because in reality
a wave packet does not, in general, retain its shape in a lasting manner.
Localised [limit\'{e}s] wave packets do not seem able to maintain themselves;
spreading takes place. The picture of the electron given by a wave packet is
therefore not satisfying, except perhaps during a short enough time.

What Mr Bohr does is this: after an observation he again localises [limite]
the wave packet so as to make it represent what this observation has told us
about the position and motion of the electron; a new period then starts during
which the packet spreads again, until the moment when a new observation allows
us to carry out the reduction again. But I should like to have a picture of
all that during an unlimited time.\endnote{The version in the galley proofs reads as follows.
(Note that in the case of this and the preceding contribution by Lorentz in the galley 
proofs, the published version was clearly not edited by him, since he had died
at the beginning of February.)
 
    \

    \noindent \textsc{Mr Lorentz}.~---~I should like to make a remark on the subject of wave packets.

    \

    \noindent When Mr Schr\"{o}dinger drew attention to the analogy between mechanics and
    optics, he suggested the idea of passing from geometrical mechanics to wave
    mechanics by making a modification analogous to that which is made in the
    passage from corpuscular optics to wave optics. The wave packet was a quite
    striking picture, but in the atom the electron is completely smeared out, 
    the packet being of the dimensions of the atom~[blank], material point that 
    would start to spread~[blank], passed, these electrons are completely smeared out.

    \

    \noindent Mathematical difficulty, wave packets in the atom, more or less distinguished frequencies
    (eigenvalues), but you could not have frequencies very close together by states differing 
    by much or little [par des \'{e}tats tant soit peu diff\'{e}rants],
    because one would not have the conditions at infinity. To
    construct a packet, one must superpose waves of slightly different
    wavelengths; now, one can use only eigenfunctions $\psi_{n}$, which are
    sharply different from each other. Thus one does not have the waves with which one 
    could build a packet. In atoms, then, one cannot have the wave packets; 
    it is the same for free electrons. All these wave packets will end up dissolving.

    \

    \noindent In reality a wave packet does not last; wave packets that would remain localised
    [limit\'{e}s] do not seem to maintain themselves; spreading
    takes place; the picture is therefore not satisfying~[blank], short enough time 
    perhaps [blank].

    \
 
    \noindent What Mr Bohr does is this~[small blank] after an observation we have again localised 
    [limit\'{e}]~[blank]; a new period starts~[blank]. But I should like to have 
    a picture of all that during an indefinite time.

    \
  }

\ 

\textsc{Mr Schr\"{o}dinger}.~---~I see no difficulty at all in the fact that
on orbits of small quantum number one certainly cannot construct wave packets
that move in the manner of the point electrons of the old mechanics.

The fact that this is impossible is precisely the salient point of the wave
mechanical view, the basis of the absolute powerlessness of the old mechanics
in the domain of atomic dimensions. The original picture\label{Schr-picture} was this, that what
moves is in reality not a point but a domain of excitation of finite
dimensions, in particular at least of the order of magnitude of a few
wavelengths. When such a domain of excitation propagates along a trajectory
whose dimensions and radii of curvature are large compared with the dimensions
of the domain itself, one can abstract away the details of its structure and
consider only its progress along the trajectory. This progress takes place
following exactly the laws of the old mechanics. But if the trajectory shrinks
until it becomes of the order of magnitude of a few wavelengths, as is the
case for orbits of small quantum number, all its points will be continually
inside the domain of excitation and one can no longer reasonably speak of the
propagation of an excitation along a trajectory, which implies that the old
mechanics loses all meaning.

That is the original idea. One has since found that the naive identification
of an electron, moving on a macroscopic orbit, with a wave packet encounters
difficulties and so cannot be accepted to the letter. The main difficulty is
this, that with certainty the wave packet spreads in all directions when it
strikes an obstacle, an atom for example. We know today, from the interference
experiments with cathode rays by Davisson and Germer, that this is part of the
truth, while on the other hand the Wilson cloud chamber experiments have shown
that there must be something that continues to describe a well-defined
trajectory after the collision with the obstacle. I regard the compromise
proposed from different sides, which consists of assuming a combination of
waves and point electrons, as simply a provisional manner of resolving the difficulty.

\ 

\textsc{Mr Born}.~---~Also in the classical theory, the precision\label{Bornprecision} with which
the future location of a particle can be predicted depends on the accuracy of
the measurement of the initial location. It is then not in this that the
manner of description of quantum mechanics, by wave packets, is different from
classical mechanics. It is different because the laws of propagation of
packets are slightly different in the two cases.

\newpage

\renewcommand{\enoteheading}{\section*{Notes to the translation}}
\addcontentsline{toc}{section}{\it Notes to the translation}
\theendnotes

\backmatter

\end{document}